\pdfoutput=1
\documentclass[rmp,aps,nofootinbib,longbibliography,twocolumn,10pt]{revtex4-1}
\usepackage[T1]{fontenc}
\usepackage[colorlinks=true,linkcolor=blue,citecolor=blue,urlcolor=blue]{hyperref}
\usepackage{amsmath}
\usepackage{mathtools}

\usepackage{graphics}
\usepackage{epsfig}
\usepackage{amssymb}
\usepackage{graphicx}
\usepackage{graphics}
\usepackage[english]{babel}
\usepackage{amsmath}        %% To split one equation
\usepackage{amsfonts}       %% To type simbol complex space
\usepackage{amsbsy,amssymb}     %% For bold mathematical symbols
\usepackage{xcolor}
\definecolor{red}{rgb}{0.7,0,0}%darkred MIT
\definecolor{green}{rgb}{0.,0.35,0.}%darkgreen
\definecolor{blue}{rgb}{0.2,0.2,0.7}
\usepackage{times}
\usepackage{latexsym}
\usepackage{fancyhdr}
\usepackage{verbatim}
\usepackage{epsf}
\usepackage{subfigure}
\usepackage{bm}
\usepackage{epsfig}
\usepackage{psfrag}
\usepackage{dcolumn}
\usepackage{makeidx}
\usepackage[T1]{fontenc}
\def\inbar{\,\vrule height1.5ex width.4pt depth0pt}
\def\IR{\relax{\rm I\kern-.18em R}}
\def\IC{\relax\hbox{$\inbar\kern-.3em{\rm C}$}}

\definecolor{red}{rgb}{0.7,0,0}%darkred MIT
\definecolor{green}{rgb}{0.,0.35,0.}%darkgreen
\definecolor{blue}{rgb}{0.2,0.2,0.7}

\newcommand{\beq}{\begin{equation}}
\newcommand{\eeq}{\end{equation}}

\newcommand{\be}{\begin{equation}}
\newcommand{\ee}{\end{equation}}
\newcommand{\bea}{\begin{eqnarray}}
\newcommand{\eea}{\end{eqnarray}}

\newcommand{\gdD}[1]{g}
\newcommand{\UdD}[1]{U}

\newcommand{\ba}{{\bf a}}

\def\ee{\mathord{\rm e}}

\def\be{\begin{equation}}
\def\ba{\begin{align}}
\def\enda{\end{align}}
\def\bi{\begin{itemize}}
\def\ei{\end{itemize}}

\def\Rb87{^{87}\rm{Rb}}                 % Rb 87
\def\Li6{^{6}\rm{Li}}                   % Li 6
 \usepackage{color}
\usepackage[normalem]{ulem} %usage \sout{...}

\definecolor{color_dl}{rgb}{1 0 0} % Dirk's color
\begin{document}

\title{Attosecond physics at the nanoscale}
\author{M. F. Ciappina$^{1,2}$, J. A. P\'erez-Hern\'andez$^{3}$, A. S. Landsman$^{4}$, W. Okell$^{1}$, S. Zherebtsov$^{1}$, B. F\"org$^{1}$, J. Sch\"otz$^{1}$, J. L. Seiffert$^{5}$, T. Fennel$^{5}$, T. Shaaran$^{6}$, T. Zimmermann$^{7,8}$, A. Chac\'on$^{9}$, R. Guichard$^{10}$, A. Za\"ir$^{11}$, J. W. G. Tisch$^{11}$, J. P. Marangos$^{11}$, T. Witting$^{11}$, A. Braun$^{11}$, S. A. Maier$^{11}$,  L. Roso$^{3}$, M. Kr\"uger$^{1,12,13}$, P. Hommelhoff$^{1,12,14}$, M. F. Kling$^{1,15}$, F. Krausz$^{1,15}$ and M. Lewenstein$^{9,16}$}
\affiliation{
\mbox{$^1$ Max-Planck-Institut f\"ur Quantenoptik, Hans-Kopfermann-Str. 1, D-85748 Garching, Germany}
\mbox{$^2$ Institute of Physics of the ASCR, ELI-Beamlines project, Na Slovance 2, 18221
Prague, Czech Republic}
\mbox{$^3$ Centro de L\'aseres Pulsados (CLPU), Parque Cient\'{\i}fico, E-37008 Villamayor, Salamanca, Spain}
\mbox{$^4$ Max-Planck-Institut f\"ur Physik komplexer Systeme, N\"othnitzer Str. 38, D-01187 Dresden, Germany}
\mbox{$^5$ Physics Department, University of Rostock, D-18051 Rostock, Germany}
\mbox{$^6$ Max-Planck-Institut f\"ur Kernphysik, Saupfercheckweg 1, D-69117 Heidelberg, Germany}
\mbox{$^7$Seminar for Applied Mathematics, ETH Zurich, CH-8093 Zurich, Switzerland}
\mbox{$^{8}$Max Planck-POSTECH Center for Attosecond Science, Pohang, Gyeongbuk 
37637, Korea}
\mbox{$^{9}$ ICFO - Institut de Ci\`encies Fot\`oniques, The Barcelona Institute of Science and Technology, 08860 Castelldefels (Barcelona), Spain}
\mbox{$^{10}$ Department of Physics and Astronomy, University College London, Gower Street, London WC1E 6BT, United Kingdom}
\mbox{$^{11}$ Blackett Laboratory, Imperial College, London SW7 2AZ, United Kingdom}
\mbox{$^{12}$ Department f\"ur Physik, Friedrich-Alexander-Universit\"at Erlangen-N\"urnberg, Staudtstrasse 1, D-91058 Erlangen, Germany}
\mbox{$^{13}$ Department of Physics of Complex Systems, Weizmann Institute of Science, 76100 Rehovot, Israel}
\mbox{$^{14}$ Max-Planck-Insitut f\"ur die Physik des Lichts, G{\"u}nther-Scharowsky-Str. 1 Blg. 24, D-91058 Erlangen, Germany}
\mbox{$^{15}$ Fakult\"at f\"ur Physik, Ludwig-Maximilians-Universit\"at M\"unchen, Am Coulombwall 1, D-85748 Garching, Germany}
\mbox{$^{16}$ ICREA - Instituci\'o Catalana de Recerca i Estudis Avan\c{c}ats, Lluis Companys 23, 08010 Barcelona, Spain}}

\date{\today}

\begin{abstract}

Recently two emerging areas of research, attosecond and nanoscale physics, have started to come together. Attosecond physics deals with phenomena occurring when ultrashort laser pulses,
with duration on the femto- and sub-femtosecond time scales, interact with atoms, molecules
or solids. The laser-induced electron dynamics occurs natively on a timescale down to a few hundred or even tens of attoseconds (1 attosecond=1 as=10$^{-18}$ s), which is comparable with the optical field. For comparison, the revolution of an electron on a 1s orbital of a hydrogen atom is $\sim152$ as. On the other hand, the second branch involves the manipulation and engineering of mesoscopic systems, such as solids, metals and dielectrics, with nanometric precision. Although nano-engineering is a vast and well-established research field on its own, the merger with intense laser physics is relatively recent.

In this report on progress we present a comprehensive experimental and theoretical
overview of physics that takes place when short and intense laser pulses
interact with nanosystems, such as metallic and dielectric nanostructures. In particular we elucidate how the spatially inhomogeneous laser induced fields at a nanometer
scale modify the laser-driven electron dynamics. Consequently, this has important impact on pivotal processes such as above-threshold ionization and high-order harmonic generation. The
deep understanding of the coupled dynamics between these spatially inhomogeneous fields and matter configures a promising way to new avenues of research and applications. Thanks to the maturity that attosecond physics has reached, together with the tremendous advance in material engineering and manipulation techniques, the age of atto-nano physics has begun, but it is in the initial stage. We present thus some of the open questions, challenges and prospects for experimental confirmation of theoretical predictions, as well as experiments aimed at characterizing the induced fields and the unique electron dynamics initiated by them with high temporal and spatial resolution.

\end{abstract}
\maketitle
\tableofcontents

%% SECTION I: Introduction %%%%%%%%%%%%%%%%%%%%%%%%%%%%%%%%%

\section{Introduction}

This report on progress presents a new emerging field of atomic, molecular, and optical physics: atto-nanophysics. It is an area that combines the traditional and already very mature attosecond physics with the equally well developed nanophysics. In the introduction we give just general motivations and description of this new area, restricting ourselves to vary basic (mostly review style) references. Extensive set of references concerning the new area is included in the bulk of the report.  

Attosecond physics has traditionally focused on atomic and small molecular targets~\cite{Scrinzi06,Krausz09}.  For such targets the electron excursion amplitude induced by the ultrafast laser pulse is small compared to the wavelength of the driving laser. Hence,  the spatial dependence of the laser field can be safely neglected.  In the presence of such spatially homogeneous laser fields the time-dependent processes occurring on the attosecond time scale have been extensively investigated~\cite{Krausz01,baltuska_attosecond_2003}.  This subject has now reached maturity based upon well-established theoretical developments and the understanding of various nonlinear phenomena (cf.~\cite{Salieres-adv,Joachain2, MaciejChapter}), as well as the formidable advances in experimental laser techniques.  Nowadays, measurements with attosecond precision are routinely performed in several facilities around the world. 

At the same time, bulk matter samples have been scaled in size to nanometer dimensions, paving the way to study light-matter interaction in a completely new regime. When a strong laser interacts, for instance, with a metallic structure, it can couple with the plasmon modes inducing the ones corresponding to  collective oscillations of free charges. These free charges, driven by the field, generate spots of few nanometers size of highly enhanced near-fields, which exhibit  unique temporal and spatial characteristics. The near-fields in turn  induce appreciable changes in the local field strength at a scale of the order of tenths of nanometers, and in this way modify the field-induced electron dynamics. In other words, in this regime, the spatial scale on which  the electron dynamics takes place is of the same order as the field variations. Moreover,  the near-fields change on a sub-cycle timescale as the free charges respond almost instantaneously to the driving laser. As a consequence, we face an unprecedented scenario: the possibility to study and manipulate strong-field induced phenomena by rapidly changing fields, which are not spatially homogeneous.

This report on progress is devoted and focused on the experimental and theoretical consequences of spatially inhomogeneous laser driven strong fields in atoms, molecules and nano-structures. We begin with a brief subsection about attosecond physics. The purpose here is not to describe the subject in detail (for recent review articles on this topic we refer the reader to, e.g.~\cite{Scrinzi06,Krausz09}), but rather to give a general overview of the strong field processes driven by intense ultrashort laser pulses in optical to mid-IR frequencies.   Such pulses are instrumental to all phenomena described here, including high-order harmonic generation (HHG), above-threshold ionization (ATI) and non-sequential double ionization (NSDI). 

The following subsection indicates how our understanding of these strong field processes, relatively well known and studied for atomic gas targets, is affected in the presence of nanoscale condensed matter targets.  The emergent field of attosecond physics at the nanoscale marries very fast attosecond processes (1 as = $10^{-18}$ s), with very short nanometric spatial scales (1 nm = $10^{-9}$ m), bringing a unique and sometimes unexpected perspective on important underlying strong field phenomena.

Section II is quite extended and includes a short description of various experimental techniques and methods used in atto-nanophysics,  from generation of nano-plasmonic fields, design of nano-structures, to more general techniques of super intense laser physics: generation of few-cycle phase stabilized laser pulses, generation of attosecond pulses via HHG, and combining both on attosecond streaking. In this Section we present as an excellent example a case study of the Imperial College attosecond beamline and its applications of attostreaking at surfaces. 

Section III is devoted to the discussion of electron emission imaging from isolated nanoparticles, a subject which has grown in importance and maturity in the last 5 years or so. Similarly hot subject: attosecond control of electrons at nanoscale needle tips is the subject of Section IV.  Section V describes specific aspect of attosecond streaking in nanolocalized plasmonic fields, originating both from isolated nanospheres as well as from nanoantennas.  In Section VI we turn to the discussion of experiments on extreme XUV generation by atoms in plasmonic nanofields. 

Theoretical approaches are summarized in Secton VII, while selected theoretical predictions concerning HHG are presented in Section VIII.  Section IX is exclusively devoted to theoretical predictions concerning ATI driven by spatially inhomogeneous fields, while in Section X we briefly mention other processes of interest, such as multielectron effects and multielectron ionization. 

We conclude in a short Section XI, stressing the explosive character of the recent development of the atto-nanophysics, and quoting examples of very recent breakthrough papers. 

\subsection{Strong field phenomena driven by spatially homogeneous fields}

A common way of initiating electronic dynamics in atoms or molecules is to expose these systems to an intense and
coherent electromagnetic radiation.  This interplay results in a variety of widely studied and important phenomena, which we simply list and shortly describe in this subsection. To put the relevant laser parameters into context, it is useful to compare them with an atomic reference. In the present context, laser fields are considered intense when their strength is not much smaller or even comparable to the Coulomb field experienced by an atomic electron. The Coulomb field in an hydrogen atom is approximately $5\times10^9$ V cm$^{-1}$ ($\approx 514$ V nm$^{-1}$), corresponding to an equivalent intensity of $3.51\times10^{16}$ W cm$^{-2}$ -- this last value actually defines the atomic unit of intensity. With regard to time scales, we note that in the Bohr model of hydrogen atom, the electron takes about 150 as to orbit around the proton, defining the characteristic time for electron dynamics inside atoms and molecules~\cite{Krausz07}.  Finally, the relevant laser sources are typically in the near-IR regime, and hence laser frequencies are much below the ionisation potential.  In particular, an 800 nm source corresponds to a photon energy of $0.057$ au (1.55 eV), which is much below the ionisation potential of hydrogen, given by $1/2$ au (13.6 eV).  At the same time, laser intensities are in the $10^{13} - 10^{15}$ W cm$^{-2}$ range: high enough to ionize some fraction of the sample, but low enough to avoid space charge effects.

While the physics of interactions of atoms and molecules with intense laser pulses is quite complex, much can be understood using theoretical tools developed over the past decades, starting with the seminal work by Keldysh in the 1960's~\cite{Keldysh,PPT1966,Reiss,ADK1986,FaisalBook}.  According to the Keldysh theory, an electron can be freed from an atomic or molecular core either via tunnel or multiphoton ionization. These two regimes are characterized by the Keldysh parameter:

\begin{equation}
\gamma=\omega_0\frac{\sqrt{2I_p}}{E_0}=\sqrt{\frac{I_p}{2U_p}},
\end{equation}
where $I_p$ is the ionization potential, $U_p$ is the ponderomotive energy, defined as $U_p=E_0^2/ 4 \omega_0^2$ where $E_0$ is the peak laser electric field and $\omega_0$ the laser carrier frequency. The adiabatic tunnelling regime is then characterized by $\gamma\ll 1$, whereas the multiphoton ionization regime by $\gamma \gg 1$. In the multiphoton regime ionisation rates scale as laser intensity $I^N$, where $N$ is the order of the process, i.e.~the number of photon necessary to overpass the ionization potential. 

Many experiments take place in an intermediate or {\it cross-over} region, defined by $\gamma\sim 1$ ~\cite{AttoTunnelTime}.  Another way to interpret $\gamma$ is to note that $\gamma=\tau_T/\tau_L$, where $\tau_T$ is the Keldysh time (defined as $\tau_T = \frac{\sqrt{2I_p}}{E_0}$) and $\tau_L$ is the laser period.  Hence  $\gamma$ serves as a measure of non-adiabaticity by comparing the response time of the electron wavefunction to the period of the laser field.  

When laser intensities approach $10^{13}\sim10^{14}$ W cm$^{-2}$, the usual perturbative scaling  observed in the multiphoton regime ($\gamma\gg 1$) does not hold, and the emission process becomes dominated by tunnelling ($\gamma<1$). In this regime a strong laser field bends the binding potential of the atom creating a penetrable potential barrier. The ionization process is governed thus by electrons tunnelling through this potential barrier, and subsequently interacting "classically" with the strong laser field far from the parent ion~\citep{corkum93, schafer93,Lewenstein94}.  

This concept of tunnel ionization underpins many important theoretical advances, which have received spectacular experimental confirmation with the development of intense ultra-short lasers and attosecond sources over the past two decades.  On a fundamental level, theoretical and experimental progress opened the door to the study of basic atomic and molecular processes on the attosecond time scale.  On a practical level, this led to the development of attosecond high frequency extreme ultraviolet and X-ray sources, which promise many important applications, such  fine control of atomic and molecular reactions among others. The very fact that we deal here with sources that produce pulses of attosecond duration is remarkable. Attosecond XUV pulses allow in principle  to capture all processes underlying structural dynamics and chemical reactions, including electronic motion coupled to nuclear dynamics. They allow also to address basic unresolved and controversial questions in quantum mechanics, such as for instance  the duration of the strong field ionization process or the tunnelling time~\cite{AttoTunnelTime, pazourek15}.

As was already mentioned, among the variety of phenomena which take place when atomic systems are driven by coherent and intense electromagnetic radiation, the most notable examples are HHG, ATI and NDSI.  All these processes present similarities and differences, which we describe briefly below~\cite{Joachain1, Joachain2, MaciejChapter}
 
HHG takes place whenever an atom or molecule interacts with an intense laser field of frequency $\omega_0$, producing radiation of higher multiples of the fundamental frequency $K\omega_0$, where in the simplest case of rotationally symmetric target $K$ is an odd integer. HHG spectra present very distinct characteristics: there is a sharp decline in conversion efficiency followed by a plateau in which the harmonic intensity hardly varies with the harmonic order $K$, and eventually an abrupt cutoff.  For an inversion symmetric medium (such as all atoms and some molecules), only odd harmonics of the driving field have been observed because of dipole selection rules and the central symmetric character of the potential formed by the laser pulse and the atomic field. The discovery of this plateau region in HHG has made it possible to generate coherent XUV radiation using table-top lasers. The above mentioned features characterize a highly nonlinear process~\cite{AnneHHG}. Furthermore, HHG spectroscopy (i.e.~the measurement and interpretation of the HHG emission from a sample) has been widely applied to studying the ultrafast dynamics of molecules interacting with strong laser fields (see, e.g.~\cite{marangos2016Review}).

Conceptually, HHG is easily understood using the three-step model~\citep{kulander,corkum93,Lewenstein94,muller,kuchiev1987}: (i) tunnel ionization due to the intense and low frequency laser field; (ii) acceleration of the free electron by the laser electric field, and (iii) re-collision with the parent ion. The kinetic energy gained by the electron in its journey, under the presence of the laser oscillatory electric field, is converted into a high energy photon and can be easily calculated starting from semiclassical assumptions. 

HHG has received special attention because it underpins the creation of attosecond pulses and, simultaneously, it exemplifies a special challenge from a theoretical point of view due to the complex intertwining between the Coulomb and external laser fields. Additionally, HHG is a promising way to provide a coherent table-top sized short wavelength light sources in the extreme-ultraviolet (XUV) and soft x-ray regions of the spectrum. Nonlinear atom-electron dynamics triggered by focusing intense laser pulses onto noble gases generates broadband high photons whose energy reaches the soft X-ray region. This nonlinear phenomenon requires laser intensities in the range of $10^{14}$ W cm$^{-2}$, routinely available from Ti:sapphire femtosecond laser amplifiers~\cite{Krausz00}.

Another widely studied phenomenon is the above-threshold ionization (ATI).  In fact, and from an historical viewpoint, it was the first one to be considered as a strong nonperturbative laser-matter interaction process~\cite{VanderWiel,AgostiniATI}.  Conceptually, ATI is similar to HHG, except the electron does not recombine with the parent atom in the step (iii), but rather is accelerated away by the laser field, eventually registering at the detector.   Hence, ATI is a much more likely process than HHG, although the latter has opened a venue for a larger set of applications and technological developments. Nevertheless, ATI is an essential tool for laser pulse characterization, in particular in a few-cycle pulses regime.  Unlike in HHG, where macroscopic effects, such as phase matching, often have to be incorporated to reliably reproduce the experiment, single atom simulations are generally enough for ATI modeling. 

In an ordinary ATI experiment, the energy and/or angular distribution of photoelectrons is measured. The ATI spectrum in energy presents a series of peaks given by the formula $E_p=(m+s)\omega_0-I_p$, where $m$ is the minimum number of laser photons needed to exceed the atomic binding energy $I_p$ and $s$ is commonly called the number of `above-threshold' photons carried by the electron. This picture changes dramatically when few-cycle pulses are used to drive the media and the ATI energy spectra becomes much richer structurally speaking~\citep{Milosevic06}.

In this case, we can clearly distinguish two different regions, corresponding to direct and rescattered electrons.  The low energy region, given by $E_k\lesssim2 U_p$, corresponds to direct electrons or electrons which never come back to the vicinity of the parent atom.

On the other hand, the high energy part of the ATI spectrum $2 U_p\lesssim E_k\lesssim 10 U_p$ is dominated by the rescattered electrons, i.e.~the electrons that reach the detector after being rescattered by the remaining ion-core~\cite{Paulusplateau}. The latter are strongly influenced by the absolute phase of a few-cycle pulse and as a consequence they are used routinely for laser pulse characterization~\cite{paulus_measurement_2003}. These two energy limits for both the direct and rescattered electrons, i.e.~$2U_p$ and $10 U_p$ can be easily obtained invoking purely classical arguments~\citep{Becker02Chapter,Milosevic06,salieres2001}.

Most of the ATI and HHG experiments use as an interacting media multielectronic atoms and molecules, and recently condensed and bulk matter.  Nevertheless, one often assumes that only one valence electron is active and hence determines all the significant features of the strong field laser-matter interaction.  The first observations of two-electron effects in ionization by strong laser pulses go back to the famous Anne L'Huillier's `knee'~\cite{Anne-knee}. This paper and later the influential Paul Corkum's work~\cite{corkum93} stimulated the discussion about sequential versus non-sequential ionization, and about a specific mechanism of the latter (shake-off, rescattering, etc.). In the last twenty years, and more recently as well, there has been a growing interest in electron correlations, both in single- and multi-electron ionization regimes, corresponding to lower and higher intensities, respectively (cf. ~\cite{Walker94,olgaHHG,Corkumcollective}).

One notable example where electron correlation plays an instrumental role is the so-called non-sequential double ionization (NSDI)~\cite{Walker94}. It stands in contrast to sequential double (or multiple) ionization, i.e.~when the process undergoes a sequence of single ionization events, with no correlation between them. NSDI has attracted considerable interest, since it gives direct experimental access to electron-electron correlation -- something that is famously difficult to analyse both analytically and numerically (for recent review see, e.g.~\cite{bergues15}).  

\subsection{Introduction to atto-nano physics}

%The interaction of ultra-short laser pulses with nanosystems (which includes clusters of atoms and molecules, as well as metal and dielectric nanoparticles and nanotips) presents a unique opportunity to study strong-field-induced electron dynamics in both their natural time and spatial scales. 

%Thanks to recent advances in experimental and engineering techniques (both in laser technology and in fabrication of plasmonic nanostructures), there has been considerable activity in this area (for a recent compilation see e.g.~\cite{Hommelhoff15}).  

The interaction of ultra-short strong laser pulses with extended systems has recently received much attention and led to an advance in our understanding of the attosecond to few-femtosecond electronic and nuclear dynamics. For instance, the interaction of clusters with strong ultrafast laser fields has long been known  to lead to the formation of nanoplasmas in which there is a high degree of charge localisation  and ultrafast dynamics, with the emission of energetic (multiple keV) electrons  and highly charged -up to Xe$^{40+}$- ions with high energy (MeV scale)~\cite{Shao1996,DitmireNature,Ditmire1997,Smith1998,Tisch1997}. Most recently use of short pulses ($\sim10$ fs)  has succeeded in isolating the electron dynamics from the longer timescale  ion dynamics (which are essentially frozen) revealing a higher degree of fragmentation  anisotropy in both electrons and ions compared to the isotropic distributions found from longer pulses ($\sim100$ fs)~\cite{Skopalova2010}.

Likewise, interactions of intense lasers with nano-particles, such as micron scale liquid droplets,  leads to hot plasma formation. An important role is found for enhanced local fields on the surface of  these droplets driving this interaction via ``field hot-spots''~\cite{Symes2004,Gumbrell2001,Donnelly2001,Mountford1998,Sumeruk2007Plasmas,Sumeruk2007}.

Furthermore, studies of driving bound and free charges in larger molecules, e.g.~collective electron dynamics in fullerenes~\cite{li_etal2015}, and in graphene-like structures~\cite{baum2015}, proton migration in hydrocarbon molecules~\cite{kuebel2015}, and charge migration in proteins~\cite{calegari2012,calegari2014} could be included in this category. In turn, laser-driven broad-band electron wavepackets have been used for static and dynamic diffraction imaging of molecules~\cite{blaga2012,xu2014,mick2015}, obtaining structural information with sub-nanometer resolution.

Tailored ultra-short and intense fields have also been used to drive electron dynamics and electron or photon emission from (nanostructured) solids (for a recent compilation see e.g.~\cite{Hommelhoff15}). The progress seen in recent years has been largely driven by advances in experimental and engineering techniques (both in laser technology and in nanofabrication). Among the remarkable achievements in just the latest years are the demonstration of driving electron currents and switching the conductivity of dielectrics with ultrashort pulses~\cite{schiffrin2013, schultze2013controlling}, controlling the light-induced electron emission from nanoparticles~\cite{Suessmann15,Zherebtsov11} and nanotips~\cite{Kruger11,Herink12,Piglosiewicz2014}, and the sub-cycle driven photon emission from solids~\cite{Ghimire2011,Trang2015,Huber2014,Vampa2015}. Furthermore, the intrinsic electron propagation and photoemission processes have been investigated on their natural, attosecond timescales~\cite{schultze2010delay,neppl2012attosecond,cavalieri2007attosecond,locher2015,okell_temporal_2015}.

\begin{figure}
	\centering	\includegraphics[width=1\columnwidth]{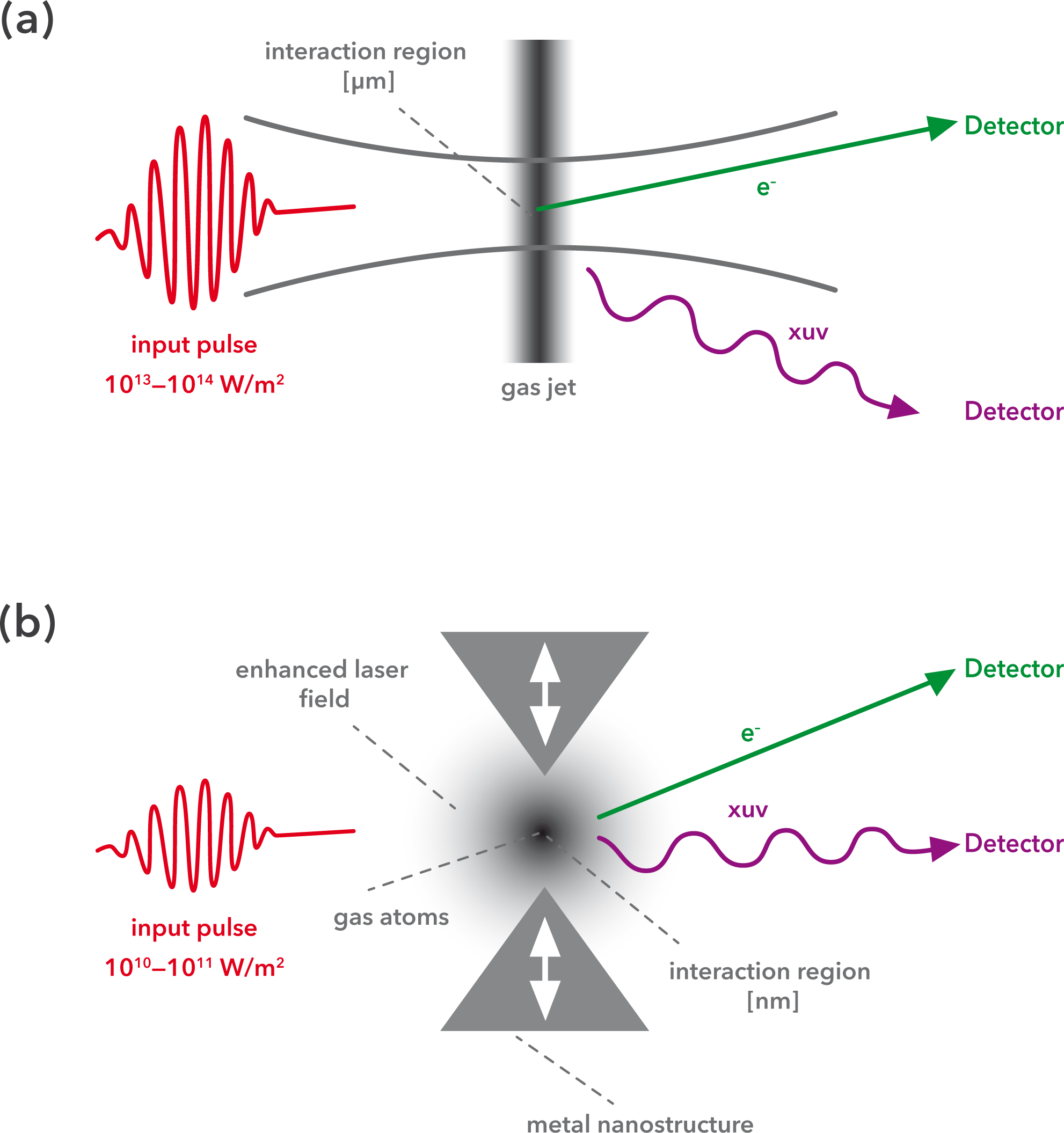}
	\caption{Sketch of conventional (a) and plasmonic-enhanced (b) strong field processes.}
	\label{Intro-fig:sketch}
\end{figure}

A key feature of light-nanostructure interaction is the enhancement of the electric near-field by several orders of magnitude, and its local confinement on a sub-wavelength scale~\cite{stockmanreview}.  From a theoretical viewpoint, this field localisation presents a unique challenge:  we have at our disposal strong fields that  change on a comparable spatial scale of the oscillatory electron dynamics that are initiated by those same fields.  As will be shown throughout this contribution, this singular property entails profound consequences in the underlying physics of the conventional strong field phenomena.  In particular, it violates one of the main assumptions that modelling of strong-field interactions is based upon: the spatial homogeneity of laser fields in the volume of the electronic dynamics under scrutiny. 

Interestingly, an exponential growing attraction in strong field phenomena induced
by plasmonic-enhanced fields was triggered by the controversial work of Kim et al.~\cite{Kim08}.
These authors claimed to observe efficient HHG from bow-tie metallic nanostructures.
Although the interpretation of the outcomes was incorrect, this paper definitively 
stimulated a constant interest in the plasmonic-enhanced HHG and ATI~\cite{Sivis13,Park11,Kovacev13NJP,Sivis12A,Kim12Reply,Park13}.

 Within the convetional assumption, both the laser electric field, $E(\mathbf{r},t)$, and the corresponding vector potential, $A(\mathbf{r},t)$, are spatially homogeneous in the region where the electron moves and only their time dependence is considered, i.e.~$E(\mathbf{r},t)=E(t)$ and $A(\mathbf{r},t)=A(t)$. This is a valid assumption considering the usual electron excursion (estimated classically using $\alpha=E_0/\omega_0^2$) is bounded roughly by a few nanometers in the near-IR, for typical laser intensities, and several tens of nanometers for mid-IR sources (note that $\alpha\propto \lambda_0^2$, where $\lambda_0$ is the wavelength of the driving laser and $E_0=\sqrt{I}$, where $I$ is the laser intensity)~\cite{Krausz00}. Hence, electron excursion is very small relative to the spatial variation of the field in the absence of local (or nanoplasmonic) field enhancement (see Fig.~\ref{Intro-fig:sketch}(a)). On the contrary, the fields generated using surface plasmons are spatially dependent on a nanometric region (cf. Fig.~\ref{Intro-fig:sketch}(b)).  As a consequence, all the standard theoretical tools in the strong field ionization toolbox (ranging from purely classical to frequently used semiclassical and complete quantum mechanical descriptions) have to be re-examined.  In this review, we will therefore  focus on how the most important and basic processes in strong field physics, such as HHG and ATI, are modified in a new setting of strong field ultrafast phenomena on a nano-scale.  Note that the strong field phenomena driven by plasmonic fields could be treated theoretically within a particular flavour of a non-dipole approximation, but neglecting completely magnetic effects. We will give more details about this particular point throughout this report.

%Marcelo:  I would avoid putting something like that especially in the Intro... you may be suggesting that all of the theory that we are presenting was inspired by a faulty experiment.  Also, talking about ISI Web of Science and citations does not seem the right place for a review paper...

%The theoretical work was mainly sparked by the seminal experiment of Kim et al.~\cite{Kim08}, which, even when their outcomes where put under controversy later on~\cite{Sivis12A,Kim12Reply,Sivis13}, has initiated an intense and frenetic activity, including experimental developments as well. Since such a milestone paper, that has more than 600 citations according to ISI Web of Science$^{\textregistered}$, around a hundred theoretical papers has been published, most of them in well renowned journals.

%We begin with a summary of experimental work, including some of the seminal contributions, which sparked significant experimental and theoretical advances. Next we give an overview of the theoretical tools, with specific emphasis on the approaches developed by the authors, with reference to important contributions of others.  The concluding section will offer an outlook, perspectives and future directions on this novel subject, with both theoretical and experimental aims.  

% Imperial Section

\section{Experimental tools and techniques}

\subsection{Near-fields and nanoplasmonics}

In this section we will give a brief introduction to nanoplasmonics. Since nanoplasmonics constitutes a vast field of research, we limit our discussion to aspects that are relevant for the attosecond physics discussed in this review. 

The interaction of light with matter is naturally confined by the length scales involved -- the wavelength of the light and the length associated with the spatial structure of the matter. If the length scale of the structure is much smaller than the wavelength the confinement of the interaction reaches the nano-scale. Electromagnetic near-fields are excited that enable optics below Abbe's diffraction limit. This mechanism opened up the field of nano-optics, also called near-field optics (see, e.g.~\cite{maier_plasmonics:_2007,Sarid2010,Novotny2012} for exhaustive literature on the topic). Nano-optics has found a wide range of applications in microscopy and spectroscopy, among them scanning near-field microscopy (SNOM)~\cite{Wessel1985,Inouye1994,Hartschuh2008} and tip-enhanced Raman scattering (TERS)~\cite{Wessel1985,Stockle2000}.

\begin{figure}
\centering
\includegraphics[width=\columnwidth]{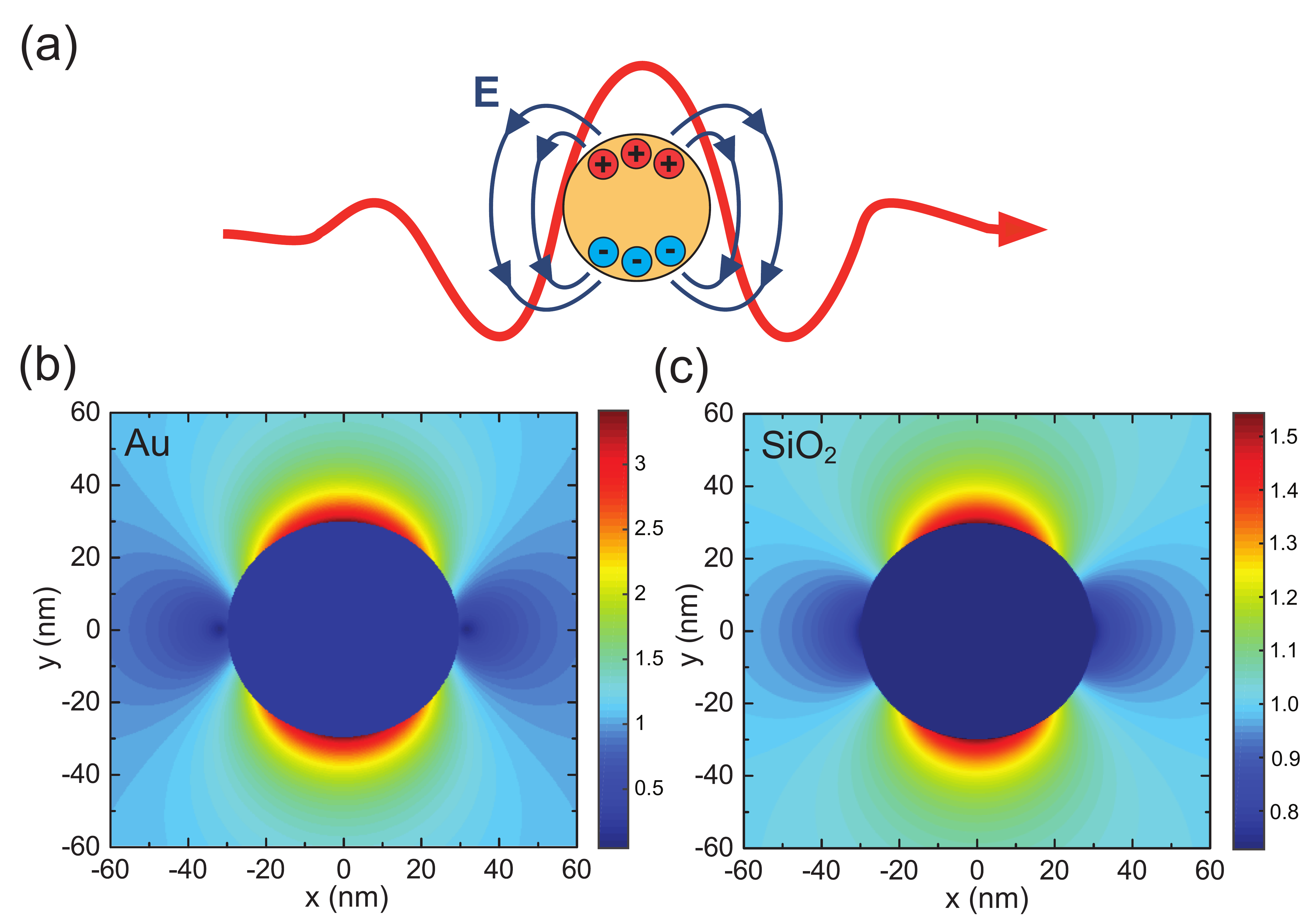}
\caption{
(a) Illustration of an optical near-field (blue) at a nanosphere excited by an external light pulse (red). Adapted from~\cite{Sussmann2014}. (b) Normalized local electric field strength $|{\bf E}({\bf r})| / |{\bf E}_0|$ at a gold nanosphere with a radius of 30\,nm at an excitation wavelength of 720\,nm ($\epsilon = -16.41 + 1.38 i$). (c) The same for a $\mathrm{SiO}_2$ nanosphere ($\epsilon = 2.12$).
\label{Figure1MK}}
\end{figure}

A prominent and illustrative example of a nano-scale structure used in nano-optics is a nanosphere. In the following we consider the interaction of light of wavelength $\lambda$ with such a sphere made from a linear, local, isotropic material, situated in vacuum. The essential assumption is that the sphere's radius $R$ is much smaller than $\lambda$. In linear optics, described by classical electrodynamics, the properties of the material of the sphere are entirely given by the complex dielectric constant $\epsilon(\lambda) = \epsilon_\mathrm{r}(\lambda) + i \epsilon_\mathrm{i}(\lambda)$, which is the square of the material's complex refractive index $n$. The dielectric constant describes the electronic response of the system to external electromagnetic fields. If we now expose the sphere to a homogeneous static electric field, we induce a collective displacement of electric charge along the field direction with respect to the ionic background. When moving from the static field to a linearly polarized light field, this displacement becomes oscillatory, leading to a time-dependent polarization (see Fig.~\ref{Figure1MK}(a)). The sphere now acts as a strongly confined source of light, in other words, as an optical nano-emitter.

For nanospheres with radii $R$ much smaller than the incident wavelength, the quasi-static approximation provides a simple approach to estimate the resulting local electric field. Omitting its time dependence in the quasi-static approximation and neglecting weak magnetic effects, the local field is given by~\cite{Jackson1999,maier_plasmonics:_2007}
\begin{equation}
{\bf E}({\bf r}) = 
  \begin{dcases}
         \frac{3}{\epsilon + 2} {\bf E}_0 &  , \ \ |{\bf r}| < R, \\
         {\bf E}_0 + \frac{3 {\bf \hat{r}} ({\bf \hat{r}} \cdot {\bf p}) - {\bf p}}{4 \pi \epsilon_0 |{\bf r}|^3} & , \ \ |{\bf r}| > R, \\
  \end{dcases}
	\label{eq:spherefield}
\end{equation}
where ${\bf r}$ is the spatial vector pointing from the centre of the sphere to the point of interest, ${\bf \hat{r}}$ is its unity vector, ${\bf E}_0$ is the spatially homogeneous incident field and $\epsilon_0$ is the vacuum permittivity. The (complex) dipole moment is given by
\begin{equation}
{\bf p} = 4 \pi \epsilon_0 R^3 \frac{\epsilon - 1}{\epsilon + 2} {\bf E}_0.
\label{eq:spheredipole}
\end{equation}
Figures~\ref{Figure1MK}(b) and~\ref{Figure1MK}(c) show the normalized local field strength $|{\bf E}({\bf r})| / |{\bf E}_0|$ for a gold nanosphere and a $\mathrm{SiO}_2$ nanosphere, respectively ($R = 30$\,nm, $\lambda = 720$\,nm). It is evident the optical near-field is spatially inhomogeneous. At the poles the field is strongly enhanced and rapidly decays with increasing distance from the surface, with a $1/e$ decay constant on the order of the radius $R$ of the sphere. The maximum local field enhancement $\xi = \mathrm{max}\,|{\bf E}({\bf r})| / |{\bf E}_0|$, here found at the sphere's poles, is independent of the sphere's radius and is given by
\begin{equation}
\xi = \left| 1 + 2\frac{\epsilon - 1}{\epsilon + 2}  \right|.
\label{eq:sphereenh}
\end{equation}
The enhancement factor for the gold nanosphere at its optical ``hotspots'' is $\xi = 3.41$. Inside the sphere, the field is uniformly screened and amounts only to a fraction of the strength of the incident field. The example of the nanosphere demonstrates the main characteristics of nano-optics, namely localization, enhancement and screening of electric fields at the nano-scale. In the context of this review, the induced spatial inhomogeneity and the strong enhancement attained at nanostructures is attractive in particular for driving and spatially confining nonlinear processes like low-order harmonic generation~\cite{Bouhelier2003,Neacsu2005,Wolf2016} or strong-field photoemission~\cite{Bormann2010,Schenk10}.

In general, the properties of the excited near-fields critically depend on the polarization of the incident light, the geometry of the nanostructure and on the (wavelength-dependent) dielectric constant of the material $\epsilon(\lambda)$. Depending on these factors, three effects can be distinguished that contribute to near-field excitation and field enhancement~\cite{Hartschuh2008,Martin1997,Martin2001}. The first effect is geometric in nature and benefits from sharp edges and protrusions of the nanostructure. Under light irradiation, surface charge is accumulated due to the discontinuity of the dielectric constant at the metal-vacuum boundary. This charge in conjunction with the sharp features of the nanostructure leads to strong local electric fields, similar to the electrostatic lightning rod effect. This effect mostly depends on geometry and can be observed for a wide range of materials and wavelength regimes. Prominent examples for nanostructures relying on the geometric effect are nanotips, nanotapers and nanorods. The second effect is observed at nanostructures that are odd multiples of half the wavelength in size. Antenna resonances in the optical domain are excited, leading to strong increase of field enhancement. The third effect is strongly coupled both to material properties and to geometry. Returning to the example of the nanosphere, Eq.~(\ref{eq:sphereenh}) predicts a resonance at $\epsilon_\mathrm{r} = -2$. This resonance condition, named Fr\"ohlich condition~\cite{Frohlich1955}, can be fulfilled satisfactorily by plasmonic metals in the visible domain ($\epsilon_\mathrm{r} < 0$ and $0 < \epsilon_\mathrm{i} \ll |\epsilon_\mathrm{r}|$), such as gold and silver. Such resonances are called localized surface plasmon resonances (LSPRs). Localized surface plasmons are excited that can lead to long-lived charge oscillations and also higher field enhancement than for other materials. Furthermore, propagating plasmon waves can be excited and observed, e.g.~at the shank of a nanotip~\cite{Berweger2012}. This shows that plasmonics can be confined to the nano-scale, enabling nanoplasmonics (see, e.g.~\cite{Sonnefraud2012,stockmanreview,Kauranen2012} for review articles on various aspects of nanoplasmonics).

Time-domain effects become very important if the incident light field is pulsed and broadband. Depending on the excitation spectrum in amplitude and phase, on the morphology of the nanostructure and on the wavelength-scaling of the dielectric constant, the induced near-field can be shaped in amplitude and phase and in its spatial behaviour. In particular, plasmonic materials typically feature long-lived plasmon oscillations that persist after the excitation pulse has ended, with lifetimes in the femtosecond domain (see, e.g.~\cite{Sonnichsen2002}).

Near-field optics is fully described by linear classical electrodynamics as defined by Maxwell's equations. Analytical modelling of near-fields with the quasistatic approximation or with Mie theory~\cite{Bohren1998}, however, is only possible for a few special cases such as nanospheres and nanoellipsoids~\cite{maier_plasmonics:_2007}, both assumed to be much smaller than the driving wavelength. For larger spheres and ellipsoids, higher-order modes start to contribute to the near-field and an analytical treatment becomes elusive. Numerical methods need to be applied in order to solve Maxwell's equations within the system's defining boundary conditions. Among those methods are the finite elements method (FEM), the finite-difference time-domain (FDTD) approach and the boundary element method (BEM)~\cite{Taflove2005}. In general, spectroscopic investigations of the optical response of nanostructures agree well with numerical simulations. Nanophotonic devices can be engineered using numerics and tailored to specific needs (see next subsection). In order to increase field enhancement, oftentimes nanostructure dimers are placed very close to each, for example in a bow-tie configuration~\cite{Sivis13}. Due to coupling of the modes of the dimers, the near-field in the gap between the dimers is strongly enhanced compared to that of a single nanostructure. At very small gap sizes, below 1\,nm, classical electrodynamics breaks down and quantum effects like electron tunnelling and nonlocal screening set in~\cite{Savage2012,Scholl2013}. In such a case, self-consistent theory approaches have to be applied, such as time-dependent density functional theory~\cite{Zuloaga2009,Marinica2015}, or quantum corrections to classical electrodynamics have to be introduced (see, e.g.~\cite{Esteban2012}). This limits the achievable field enhancement factor to lower values than the classical prediction, not only in dimer gaps and in other multiparticle systems~\cite{Ciraci2012}, but also very close to the surface of a single nanostructure~\cite{Zuloaga2010}.

Near-fields are accessible experimentally by various techniques. Nonlinear processes like second-harmonic generation~\cite{Neacsu2005} or nonlinear photoemission~\cite{Ropers07PRL,Thomas2013,Kruger14} or also attosecond streaking~\cite{SuessmannPRB11} can be employed as means to probe the magnitude of the local field enhancement. The spatial profile of near-fields can be resolved with the help of inelastic electron scattering processes where a tightly focused, high-energy electron beam is passing close to the nanostructure. Possible experimental observables are cathodoluminescence~\cite{Vesseur2007,Chaturvedi2009}, incoherent electron energy loss~\cite{Nelayah2007,Huth2013,Schroder2015a} or coherent electron energy gain~\cite{Barwick2009,Feist2015}. All these methods are particularly useful to probe plasmonic resonances and spatial structures like standing waves and optical hotspots.

\subsection{Design and manufacture of nano-scale targets}

\subsubsection{Design rules for plasmonic nanostructures}

Like many other resonating systems, plasmonic structures are characterized by a strong dispersive response to the probing field, with strong frequency dependence of the scattering and absorption cross-sections and the field enhancement at the vicinity of the nanoantennas~\cite{maier_plasmonics:_2007}. The peak of the extinction spectrum is known as the localized surface plasmon resonance (LSPR), and its resonance frequency and shape, so as the near-field enhancement and its time-dependent spectral properties are determined by the materials and the geometry (shape and size) of the nanostructures and their surrounding medium~\cite{fernandez-garcia_design_2014,choipra2016,anneoptexp2015}. The LSPR may be affected also by dipolar coupling between adjunct nanostructures and standing waves in periodically assembled structures~\cite{jain_plasmon_2006}.  
The plasma frequency--a material constant depending on its properties (namely: free electron density, electrons effective mass, and the effective electron dumping rate)--sets the upper limit for the frequency at which an LSPR is achievable for a specific material. For applications in the visible and near-infra red (NIR), a range of suitable plasmonic materials exists, (e.g.~aluminium, silver, gold, metal-nitrides, semiconductors and transparent conductive oxides (TCO)~\cite{naik_alternative_2013}), where the choice of the optimized material usually depends on its materials-compatibility with the overall fabrication process, and on its chemical and physical stability of the materials under operating conditions, e.g.~Ag and Al tend to oxidise in free atmosphere; TiN is stable at high temperatures and to the optical range of interest -Al can support LSPR at ultra-violet (UV) frequencies; Ag shows stronger resonance than Au in the visible due to electronic inter-band transitions in the latter~\cite{maier_plasmonics:_2007,fernandez-garcia_design_2014}; TCOs have negative real permittivity only for wavelength longer than 1.3 to 1.5\,$\mu$m~\cite{naik_alternative_2013}. The materials surrounding the nanostructures have a strong effect on the frequency of the LSPR which is red shifted with increase of the refractive index ($n$). Devices that are designed to operate at the visible optical range are usually fabricated on-top of glass-like substrate with $n$ ranging from 1.3 to 1.5, compare to $n=3.5$ to 4 for typical semiconductors, and use self-assembly monolayer as adhesion promoters rather than chromium of titanium layer~\cite{habteyes_metallic_2012}. Finally, after the selection of materials, the exact shape and dimensions of the nano-structures are designed to tune the plasmonic resonance frequency to the desired spectral range and functionality. The typical dimensions of gold structures designed to operate in the visible to NIR range are at the order of tens to few hundred of nanometers. The last designing step is usually accomplished by means of finite-elements or wave analysis numerical simulations~\cite{veronis_overview_2007}, covering a wide range of design parameters.

\subsubsection{Fabrication methods of arrays of plasmonic nanostructures}

\begin{figure}[htbp]
	\includegraphics[width=\columnwidth]{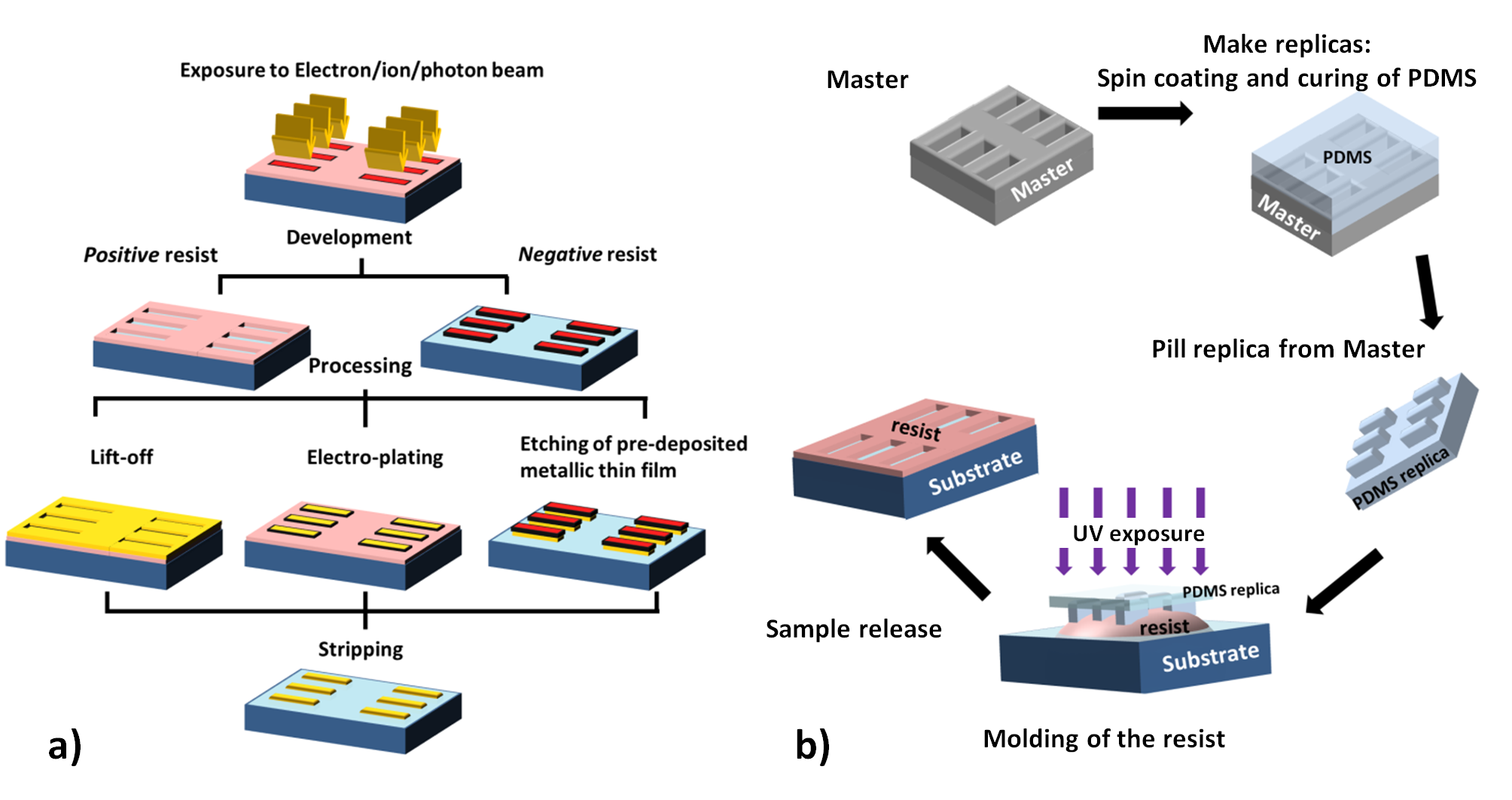}
	\caption{Schematic illustration of the nanofabrication process of plasmonic structures. (a) Complete  nanofabrication process: from direct lithography using a beam of photons or charged particles, through chemical development of the resist, post development processing, and stripping of the resists. (b) The moulding lithography process: The hard master is covered with PDMS which is then exposed to UV light, harden, and released from the master. The PDMS replica is pressed into a liquid resist, and the resist is cured by UV light shone through the transparent replica. After stripping, the sample is etched to remove any excess resist. The final outcome is a sample covered with a patterned resist that can be processed is a similar fashion to panel (a).}
	\label{fig:nanofab_fig1}
\end{figure}

Fabrication methods of metallic nanostructures are differed one from another by the resolution and critical dimension of the written features, accuracy of placement of structures, and speed and costs of the fabrication process. Here we restrict the discussion to methods that allow a precise placement of nanostructures in the substrate. In these techniques (as oppose to colloidal deposition, for example) direct deposition of nanostructured metals on the substrate is hardly feasible, and the patterning is usually done on electron- or photo-resists masks and then transferred into metallic nanostructures, by means of lift-off (where the resists is used a sacrifice layer), or by selective etching of the pre-deposited metallic layer under the patterned resist (which perform as a masking layer). Alternatively, electro-chemical deposition can be used to grow metal on the exposed section on the resist. The lithography process is illustrated in Fig.~\ref{fig:nanofab_fig1}(a).

Below we briefly describe four different methods for nanofabrication of plasmonic nanostructures:  Electron Beam Lithography (EBL), focused ion-beam (FIB), direct laser writing (DLW) and soft lithography.  

%\subsubsection{Electron Beam Lithography}

In {\bf Electron Beam Lithography (EBL)}, an electron beam emitted from thermionic or field-emission sources, with typical acceleration voltage of $10-100$ kV, is focused using electromagnetic and electrostatic lenses onto a thin layer of electron-sensitive resist. The primary electrons strike the resist and generate a cascade of secondary electrons with lower energies, which alter the chemical structure of the exposed area of the resist, changing the solubility. The desired pattering is achieved by a selective exposure of the resist using electron deflectors which direct the focused beam to the desired position on the sample. 
%Since the deflectors usually have a limited writing field at the order of few hundreds of micrometers, in order to write on larger areas, the sample is usually mounted on piezo-, or mechanical moving-stage. After exposure, the resist is developed in chemical bath, which, depending on the nature of the resist, removes only the exposed (or unexposed) resist for 'positive' (or 'negative') resists. 
The attainable resolution of EBL patterning is limited to $\sim 10$ nm due to scattering of electrons in the resists leading to unintended exposure (known as proximity effect) of the resist~\cite{del_campo_fabrication_2008}, and not by the wavelength of the electrons, which is at the order of 1 \AA\,for $10$\, keV electrons. 
%A common 'positive' resist with resolution below 50 nm is polymethylmetacrylate (PMMA); higher resolution of less than 10 nm is achievable using hybrid organic-inorganic~\cite{grenci_high_2015} or inorganic 'negative' resists such as hydrogen-silsesquioxane (HSQ)~\cite{manfrinato_resolution_2013}, though stripping of the exposed resists may require strong etchers. 
High resolution EBL resists usually requires relatively high electron exposure dose, which results in a slow patterning time. Nevertheless, due to its high resolution and its practically ultimate flexibility over the design and placement of structures, EBL is currently the workhorse of nanofabrication of plasmonic nanostructures.

\begin{figure}[htbp]
	\includegraphics[width=\columnwidth]{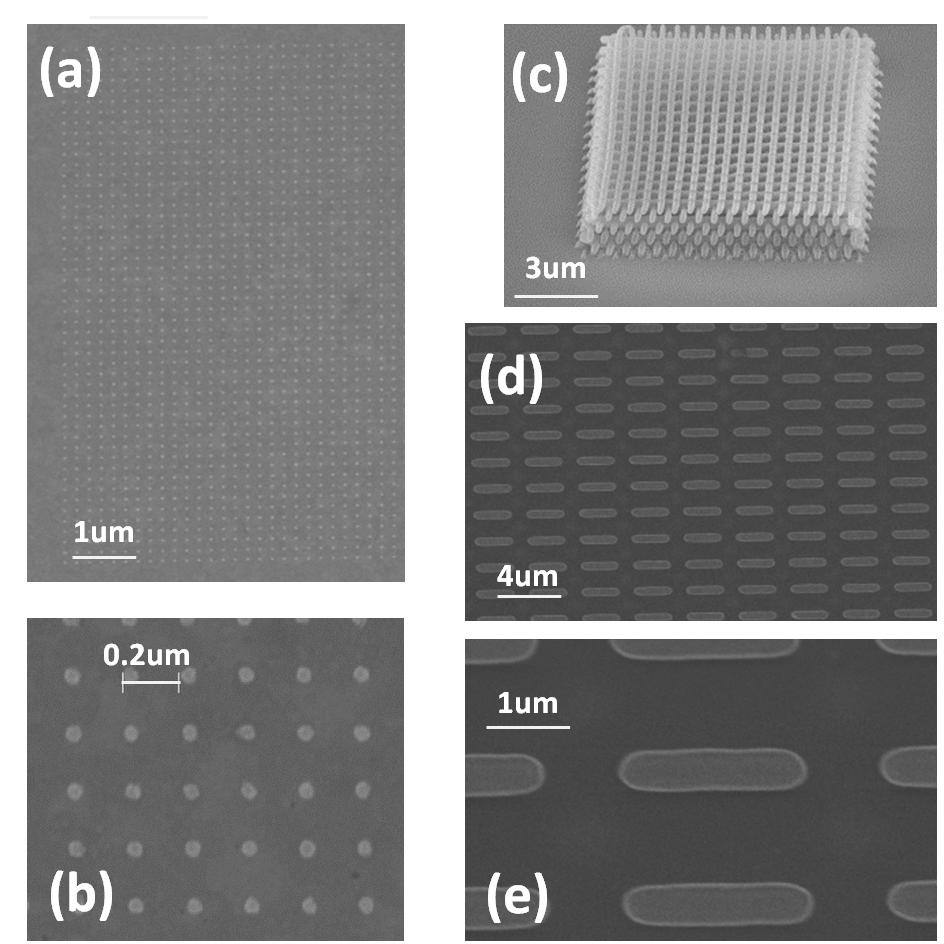}
	\caption{Scanning Electron Microscopy images of nanostructures. (a)-(b) Positive EBL writing: 50 nm thick Au discs with a diameter of 55 nm made by writing in positive resist (PMMA), followed by thermal evaporation and lift off process. (b): FIB, and (c)-(d): DLW.  (c): 3D woodpile structure made of polymeric photoresist coated with 25 nm of gold. (d)-(e): Au antennas with LSPR at the Near- and Mid-IR regime, fabricated by DLW of positive photo-resist (AZ Mir 701), followed by gold sputtering and lift off process.}
	\label{fig:nanofab_fig2}
\end{figure}

%\subsubsection{Focused ion-beam}

{\bf Focused ion-beam (FIB)} is similar to EBL, but with more capabilities: here, gallium ions are emitted from an ion source, and focused and controllably deflected into the sample. 
%The gallium ions can pattern the photo resists by the emission of secondary electrons, creating lift-off, or etching masks, just like EBL; alternatively, the heavy ions can directly knock-out atoms by physical sputtering of the pre-deposited metallic layer. 
Compared to EBL, FIB demonstrates a faster patterning of electron-resists (up to 100 times faster) due to the large number of secondary electrons that are emitted from each collision of the gallium ions in the resists~\cite{del_campo_fabrication_2008,cui_focused_2011}. With direct subtractive patterning (milling), re-deposition of the milled material is avoided by reaction with reactive gas, and the typically 5\,nm diameter beam can realise structures with feature sizes of $20-30$\,nm. 

%\subsubsection{Direct laser writing}

{\bf Direct laser writing (DLW)}~\cite{deubel_direct_2004}, is a mask-less photolithography method that allows a fast writing of flat and 3D structures at a resolution lower than that of electron/ion based techniques. A femtosecond  laser is coupled to a microscope and focused via objective with high numerical aperture onto a substrate covered with photosensitive material. High writing speeds of $10-50$ mm/sec~\cite{bagheri_fabrication_2015} can be achieved by combining sensitive photo-resists with fast deflection of the light-beam using galvanic mirrors. With these writing speeds, a cm$^2$ array of nanoantennae can be written by a single source in only few hours. Like many optical lithography systems, the resolution of DLW is diffraction-limited ($\sim 250$ nm for a UV source of $\lambda=405$ nm). However, a higher resolution of 150 nm can be achieved when an IR femtosecond laser (usually at 780 nm) is used to probe resists which support a non-linear absorption (namely, two-photon absorption (TPA)). 

%In TPA direct laser writing, the absorption happens only when a large number of photons are present simultaneously in the area in the resist, leading to a smaller area in which photons are absorbed and chemical modification of the resists occurs. The small absorption volume, enables DLW in 3D at a sub-micron resolution (Fig.~\ref{fig:nanofab_fig2}(c)). Recent developments in TPA DLW in the form of adding stimulated emission depletion (STED) to DLW, demonstrated structures with lateral sizes of 55 nm, and resolution of 120 nm~\cite{wollhofen_120_2013} --similar to the resolution of EBL with positive resists. 

%\subsubsection{Soft lithography: moulding lithography}

In contrast to the conventional fabrication methods portrayed above, which are based on exposure of the sample to a beam of photons or charged particles, {\bf soft-lithography} methods are based on physical contact of the stamp with the substrate. Moulding (embossing) lithography is a form of soft lithography that can be used to print large arrays of plasmonic nanostructures~\cite{qin_soft_2010}. 
Nanostructures with feature sizes of 20 nm and below can be reproducibly fabricated by this technique.

%The process starts with the fabrication of a 'master' --a hard material patterned by conventional methods (usually by EBL patterning followed by reactive etching of Si substrates). Then, the master is pressed into a soft material, followed by hardening of the material and release of the master form the solidified sample. In this way the pattern of a single master can be transferred to multiple arrays --allowing a rapid and cost effective fabrication of large arrays of nano-structures with high resolution that is not subject to scattering of diffraction, but limited by the graininess of the matter and the hardness of the master/stamp. 

%Nanostructures with feature sizes of 20 nm and below can be reproducibly fabricated by this technique. In moulding fabrication of plasmonic nano-structures, usually the nano-pattern is first replicated from the master to a flexible transparent stamp, usually poly(dimethylsiloxane) (PDMS) The replicated stamp is pressed into a UV curable liquid photoresist, exposed to UV light which solidifies the soft resist, and then released from the sample. After the release, excess polymer is etched down and the sample is ready for the next fabrication step of metal deposition and lift off, or chemical etching. The moulding nano-lithography process is illustrated in Fig.~\ref{fig:nanofab_fig1}(a).

\subsection{Few-cycle carrier envelope phase stabilised lasers}

The main enabling technology for strong field and attosecond physics is the ability to generate intense few-cycle laser pulses with stabilized waveforms \cite{Krausz00}.
%ii.	oscillators (as used in EUV nano-antenna expts) and amplifiers; explain different pulse energies and repetition rates [the conclusion in recent literature and in this paper is that oscillator based HHG in nanostructures does not work.]
% typical laser system for attoscience
%iii.	typical few-cycle system (Imperial) using hollow fibre pulse compression; briefly mention pulse characterisation techniques (minimal details)
A typical laser system for attosecond science consists of a Ti:sapphire chirped pulse amplification (OPA) system providing $<30$\,fs pulses at pulse energies around 1\,mJ or above with repetition rates of 1 to 5\,kHz. To achieve the  few-cycle pulse durations required for attosecond science these pulses are sent through a hollow core capillary (typically 250--400 $\mu$m inner diameter) for spectral broadening. Mostly an argon or neon gas fill is used. The hollow capillary pulse compression system can be used in static fill mode or in differential pumping/gradient pressure mode, where the gas is supplied at the exit side, whilst the entrance side is kept at vacuum. This improves coupling efficiency and beam quality \cite{robinson_generation_2006}. The broadened laser pulses exit the gas filled capillary with a positive chirp. Pulse compression to few-cycle duration is achieved with broadband chirped mirrors combined with thin glass wedges for fine-tuning. The current state of the art of these pulse compression systems are pulses below 4\,fs with 0.5 to 1\,mJ level pulse energies. Pulses with 3.8, 3.5, and 4\,fs durations and 0.4, 0.5, and 1\,mJ pulse energies at 1 and 4\,kHz repetition rate have been produced \cite{cavalieri_intense_2007,witting_characterization_2011,okell_carrier_envelope_2013,schweinberger_waveform_controlled_2012}. 

\begin{figure}[htbp]
	\includegraphics[width=\columnwidth]{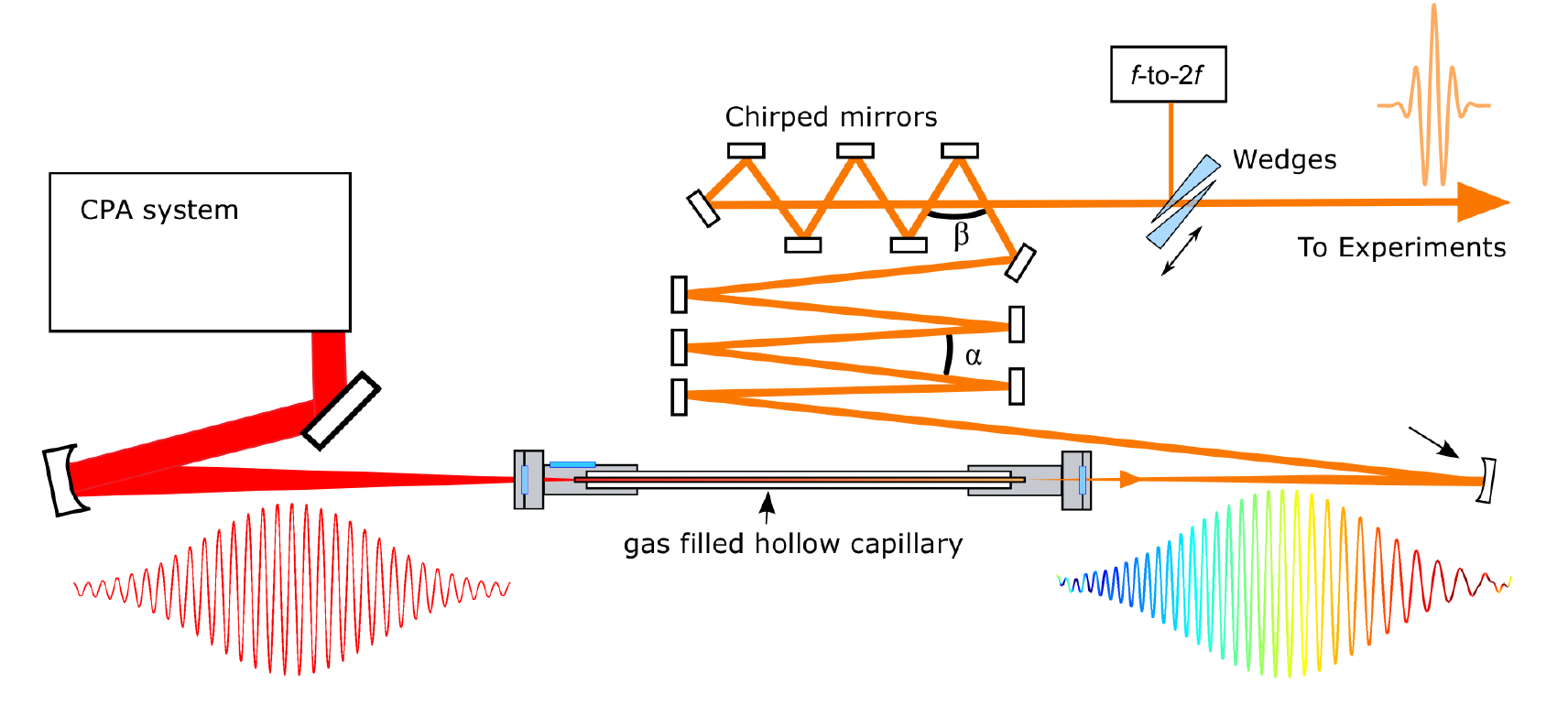}
	\caption{Typical few-cycle laser system with CPA, gas filled hollow capillary for spectral broadening, and compression with chirped mirrors and glass wedges.}
	\label{fig:cpahcfpc}
\end{figure}

Tailored electric field waveforms have been produced by a combination of discrete spectral bands derived from a hollow capillary waveguide \cite{wirth_synthesized_2011}. Recently the combination and careful compression led to optical waveforms with attosecond pulse durations \cite{hassan_optical_2016}.

Alternative approaches allowing the use of non-CEP stabilised laser systems~\cite{schmidt_cep_2011}.
Targeting higher repetition rates is a current area of research. Fibre lasers with postcompression similar to the scheme described above promise to deliver few-cycle laser pulses with repetition rates in the MHz range \cite{limpert_ultrafast_2011}. A viable alternative to achieve amplification of large bandwidths at high repetition rates is the optical parametric chirped pulse amplification (OPCPA) technology. Recently 6\,fs pulses at 300\,kHz repetition rates have been demonstrated~\cite{prinz_cep_stable_2015}.

At pulse durations of only a few or even a single optical cycle the phase between the carrier wave and the evelope, the carrier envelope phase (CEP) becomes an important parameter. The electric field of a laser field can be written as
\begin{equation}
	E(t) = E_0(t) \cos{[\omega t + \phi_{\text{CEP}}]}.
	\label{eq:CEPphase}
\end{equation}
The CEP $\phi_{\text{CEP}}$ becomes an important parameter in strong field driven interactions and in attosecond pulse generation~\cite{xu_route_1996,baltuska_phase_controlled_2003,paulus_measurement_2003,apolonski_controlling_2000,dietrich_determining_2000,luecking_long_term_2012}. 

\begin{figure}[htbp]
	\includegraphics[width=\columnwidth]{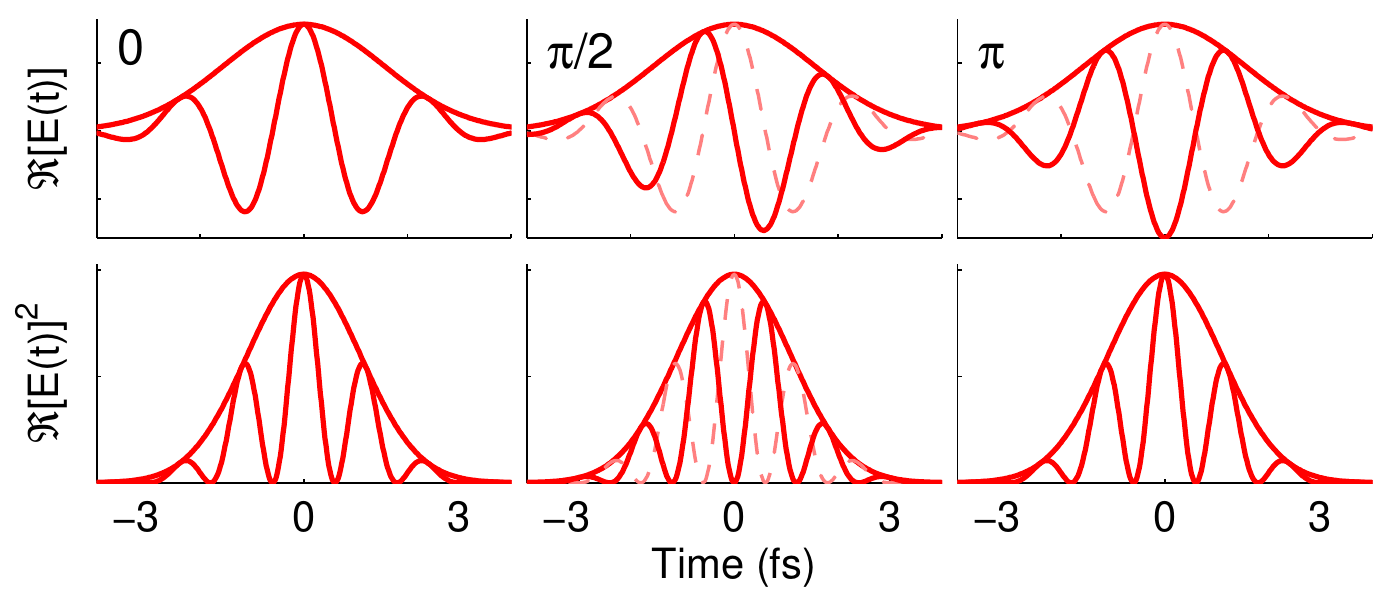}
	\caption{Illustration of the CEP for a few-cycle pulse. Top row: Electric field and envelope of three pulses with CEP of 0, $\pi/2$, and $\pi$. Bottom row: Field intensity and envelope for the same three pulses. The 'cos' waveforms for CEP $0$ and $\pi$ have a single strongest half-cycle, whilst the 'sin' waveform with CEP $\pi/2$ is characterized by two equally strong neighbouring half-cycles.}
	\label{fig:cpahcfpc1}
\end{figure}

%CEP control and measurement \cite{xu_route_1996}.
%phase controlled amplification \cite{baltuska_phase_controlled_2003}
%stereo ATI Paulus \cite{paulus2003}.
%feedback \cite{apolonski_controlling_2000} \cite{dietrich_determining_2000}

%feedforward methods \cite{luecking_long_term_2012}

An important aspect of few-cycle laser systems for attosecond science is the characterization of the generated few-cycle laser pulses. For a long time autocorrelation and especially the inteferometric autocorrelation has been popular. However, despite giving an estimate about the pulse duration, autocorrelation cannot recover the full temporal pulse shape \cite{chung_ambiguity_2001}. In the last decade a number of advanced metrology methods, able to recover the full complex electric field of ultrashort laser pulses, have been developed. Amongst them are frequency resolved optical gating (FROG) \cite{kane_characterization_1993}. FROG can be understood as a frequency resolved autocorrelation. The electric field of the unknown laser pulse is recovered with an iterative optimization algorithm. FROG has been employed to measure pulses down to near single cycle pulse durations \cite{akturk_measuring_2008}. A more recent development is the dispersion scan technique (d-scan). In d-scan an iterative algorithm recovers the electric field of ultrafast laser pulses from a series of second harmonic spectra for a varying amount of dispersion introduced into the unknown pulse \cite{miranda_simultaneous_2012}. Another attractive pulse characterization method is spectral phase interferometry for direct electric field reconstruction (SPIDER) \cite{iaconis_spectral_1998}. SPIDER relies on self referencing spectral shearing interferometry to characterize the electric field of ultrashort laser pulses. The spectral phase of an unknown laser pulse is recovered from the measurement interferogram with a direct algebraic reconstruction algorithm. As a one-dimensional data trace describes a one-dimensional laser field $E(x_0,y_0,t)$ SPIDER can be extended to multiple spatial dimensions to deliver spatio-temporal information. A variant of SPIDER, that uses spatial encoding of the phase and direct spectral filtering (SEA-F-SPIDER), has been employed to spatio-temporally characterize near-single cycle pulses \cite{witting_characterization_2011,balciunas_strong_field_2015}. Recent developments are direct field sampling techniques that employ a strong and short pulse and high harmonic generation to sample arbitrary electric field waveforms \cite{kim_petahertz_2013,wyatt_aries_2016}. An excellent review of ultrafast metrology can be found here \cite{walmsley_characterization_2009}.

\begin{figure}[htbp]
	\includegraphics[width=\columnwidth]{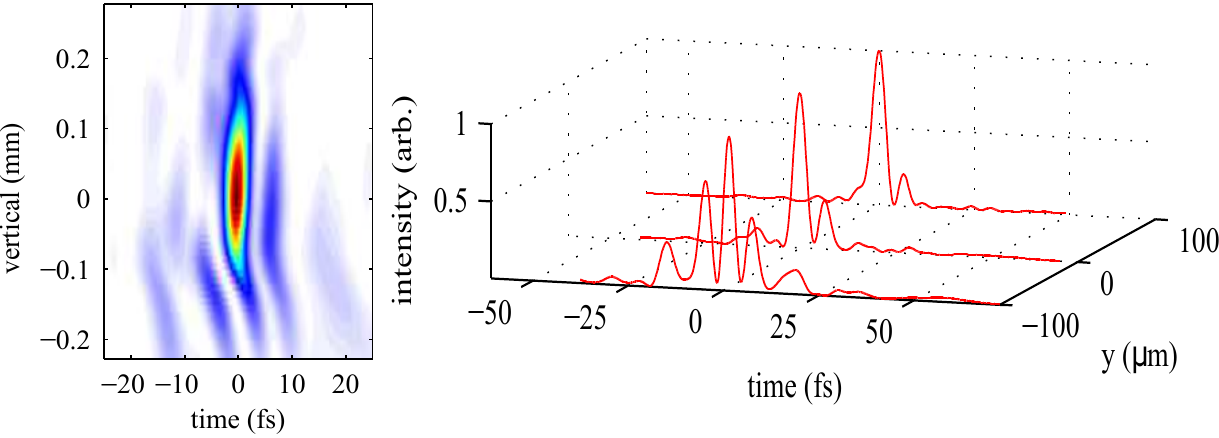}
	\caption{Spatio-temporal characterization of a 1.5-cycle laser pulse.}
	\label{fig:shortpulse2D}
\end{figure}

\subsection{Attosecond pulse generation from HHG}

The HHG process, in which high-order harmonics of a strong laser field are generated in its interaction with a gas-phase medium, was introduced in Section I. The short wavelength emission (VUV-XUV) from HHG can be of attosecond duration and hence HHG lies at the heart of attosecond science. The attosecond pulses can be emitted in pulse trains at repetition rates typically in the petahertz range \cite{antoine1996attosecond,mairesse2003attosecond} or as isolated pulses, i.e.~one attosecond pulse generated per laser pulse. A key feature of the attosecond pulse emission is its automatic synchronisation with the driving laser pulse. This permits pump-probe experiments using the attosecond pulse and the laser pulse.

Here we concentrate on the generation of isolated attosecond pulses that can be obtained by driving the HHG process with few-cycle CEP stabilised laser pulses, as described in the Section IIC. This method is known as amplitude gating \cite{Krausz01,goulielmakis2008single,kienberger2004atomic,frank2012invited}, because the highest energy emission from the HHG process is confined to the single, highest amplitude half-cycle of the drive laser pulse. This gating is only possible for a few-cycle pulse for which the field amplitude of neighbouring half-cycles is significantly lower. CEP stabilisation is crucial to ensure the peak of the carrier field coincides with the peak of the envelope \cite{baltuska_attosecond_2003,jones2000carrier}. The temporal confinement of the highest energy HHG emission to a single half cycle of the laser field leads to the formation of a spectral continuum at the short wavelength limit (the cutoff) of the HHG spectrum. This continuum region can be spectrally bandpass filtered using sub-micron thickness foil filters and multi-layer XUV mirrors to produce an isolated attosecond pulse.

Since the first measurement of an isolated sub-femtoscond pulse (650\,as at a photon energy of around 90\,eV) produced from HHG in Ne driven by a 7 fs near infra-red (NIR) pulse in 2001 \cite{Krausz01}, there has been considerable progress in attosecond pulse generation. Using shorter NIR drive pulses, attosecond pulses as short as 80 as have been generated \cite{goulielmakis2008single}. A range of techniques have also been developed to allow the generation of isolated attosecond pulses from multi-cycle rather than few-cycle pulses \cite{altucci2011single}. Longer drive laser pulses are technically less demanding to produce and can have higher energy than few-cycle pulses, with the potential to generate more intense attosecond pulses. Foremost amongst these techniques is polarisation gating  \cite{corkum1994subfemtosecond,sola2006controlling,sansone2006isolated}. Polarisation gating uses a drive pulse with a time varying ellipticity to confine the HHG emission to a short interval during which the pulse is approximately linearly polarised. Extensions of this technique, known as double optical gating (DOG) \cite{mashiko2008double} and generalised DOG (GDOG) \cite{feng2009generation}, where polarisation gating is combined with two-colour gating \cite{mashiko2008double}, have proven particularly effective for isolated attosecond pulse generation using multi-cycle drive laser pulses. In fact, the current record for the shortest attosecond pulse (67\,as) was obtained using DOG \cite{zhao2012tailoring}. Another method, known as ionisation gating \cite{ferrari2010high,JoseNat}, uses rapid field-ionisation of the generating medium on the rising edge of the laser pulse to confine the HHG emission to a single half cycle.

Isolated attosecond pulses can currently be produced over the spectral range 20-145\,eV (the extremes of this range are obtained in \cite{guggenmos2015chromium,huppert2015attosecond}) with durations well below 100\,as. Attosecond pulses are emitted with low divergence, spatially-coherent beams that by virtue of their short wavelength and excellent beam quality can be focused to relatively small spots ($<1\,\mu{}$m). For a linearly polarised drive laser pulse, the attosecond pulse is linearly polarised, but elliptical polarised attosecond pulses have also been predicted using polarisation gating \cite{henkel2013prediction}. Resonant HHG driven by elliptically polarised laser pulses has been shown to deliver quasi-circularly polarised ultrashort pulses in the extreme ultraviolet \cite{ferre2015table}, but this has not yet be extended to the attosecond domain. 

Attosecond pulse energies are typically in the picojoule-nanojoule range and this relatively low photon flux is frequently a limitation in gas-phase experiments, where target densities are typically low and interaction cross sections are often small. For condensed phase targets, including those at the nanoscale, one usually encounters the opposite problem. Due to the high target density, space-charge effects can distort the spectrum and spatial distribution of photoemitted electrons. This typically necessitates a reduction in the ionising photon flux to levels where space-charge effects are negligible. This often leads to low signal count rates that are not so dissimilar to those obtained in gas phase targets. In such circumstances, the use of high repetition rate lasers (hundreds of kHz to MHz) \cite{limpert_ultrafast_2011} is particularly advantageous. With nanoscale targets in particular, care must be taken not to damage the targets through excessive laser fluence. For example, nanoantennae (see Section VI) can easily be damaged by melting~\cite{Kovacev13NJP}.

We conclude this section on attosecond sources by noting that free electron lasers (FELs) are soon likely to provide another source of attosecond pulses with a brightness far exceeding current HHG-based system (see, for example, \cite{marangos2011introduction} and references therein). The unprecedented brightness of femtosecond x-ray pulses from FELS is already being used in coherent diffractive imaging experiments to image the dynamics of individual nanostructures, for example the transient melting of a single gold nanocrystal \cite{clark2015imaging} and the 3D imaging of lattice dynamics in individual gold nanocrystals \cite{scholz2013crystal}.

\subsection{Attosecond streaking}
Attosecond streaking \cite{goulielmakis2008single,kienberger2004atomic,baltuska_attosecond_2003,mashiko2008double,itatani2002attosecond,mairesse2005frequency,witting2012sub} is one of the most important techniques to measure electron dynamics with attosecond resolution. It features prominently in the nanoscale experiments described in this review. It is important to note that attosecond temporal resolution can also be obtained using HHG spectroscopy, which exploits the sub-cycle dynamics of the HHG process itself to interrogate the generating molecular system (for a comprehensive recent review see~\cite{marangos2016Review}). The intriguing possibility of applying HHG spectroscopy to molecules chemisorbed at surfaces is being considered by a number of groups as a way of understanding the ultrafast exchange of charge between the molecule and the surface. Such efforts would greatly benefit from localised nanoplasmonic field enhancement.

In an attosecond streaking experiment, an attosecond pulse (in the VUV-XUV range) generated by HHG and a synchronised laser pulse (typically a few-cycle pulse in the NIR range; usually the drive laser pulse for the HHG) propagating collinearly are focused on a target with a controllable delay between them. The photoelectron wavepacket produced by the attosecond pulse is accelerated in the laser field, which is known as the streaking field. This imprints sub-cycle timing information on the photoelectron spectrum. A streaking trace $S(E,\tau)$ is built up over multiple laser shots (usually $>10^5$) by recording the photoelectron energy spectrum for a range of delays, $\tau$, between the attosecond pulse and the streaking pulse. 

As will be described in Section V, different streaking regimes can be demarcated in terms of the time taken for the photoelectron to escape the streaking near field. For gas phase targets the streaking is in the so-called ponderomotive limit - the electron does not experience spatial variations in the streaking near field. In this regime, energy peaks in the photoelectron spectrum corresponding to different photoemission channels display modulations that follow the vector potential of the streaking field at the time of photoemission. In fact, the streaking trace can be treated as a frequency resolved optical gating (FROG) trace in which the streaking field acts as a pure phase gate function on the photoelectron wavepacket \cite{itatani2002attosecond}. FROG is a widely-used technique for characterising femtosecond laser pulses \cite{trebino1997measuring}. The extension of FROG to invert attosecond streaking traces is known as FROG for complete reconstruction of attosecond bursts (FROG-CRAB) \cite{mairesse2005frequency}. It can be performed using iterative algorithms \cite{kane1997simultaneous,kane1999recent} initially developed for the inversion of standard laser FROG traces to yield the full electric fields (phase and amplitude) of both the attosecond field and the streaking field. Attosecond streaking was initially used for the temporal characterisation of attosecond pulses. Its ability to fully retrieve the streaking field is particular attractive for the study of nanolocalised plasmonic fields (e.g. surrounding nanoantennas, nanotips and nanospheres) that are excited by ultrafast laser pulses (see Sections III-V). In such studies, it is the plasmonic field that acts as the streaking field. In principle, the time-dependent plasmonic field can be directly compared to the field of the excitation laser pulse by recording a streaking traces in a reference gas-phase atomic target.

\subsection{Case study: Imperial College Attosecond Beamline}
We now describe the attosecond beamline at Imperial College London~\cite{frank2012invited} which serves to illustrate the practical implementation of the concepts outlined above and give the reader an idea of the scientific ``tools'' required for attosecond physics at the nanoscale. Descriptions of other attosecond beamlines can be found in~\cite{fiess2010versatile,frassetto2014high,locher2014versatile,weber2015flexible,huppert2015attosecond}. The Imperial College beamline employs amplitude gating for the generation of isolated attosecond pulses using the sub-4 fs CEP-stabilised laser system described in Section IIC. It is capable of producing two synchronised attosecond pulses per laser drive pulse, one in the VUV spectral range ($\approx$20\,eV) and the other in the XUV range ($\approx$90\,eV) \cite{fabris2015synchronized}. This capability is targeted at pump-probe studies where both pulses are of attosecond duration, though such schemes are at the limits of current capability due to the limited pulse energy available for the pump step. VUV pulses are advantageous because of the high photo-ionisation cross section of many molecules in this spectral region \cite{kameta2002photoabsorption} and also because of the higher HHG photon flux possible in this energy range \cite{l1991theoretical}. Attosecond pulses in the $20-40$ eV range have previously been generated using polarisation gating \cite{mashiko2010tunable,feng2009generation}, though not synchronously with another attosecond pulse. In general, the ability to generate attosecond pulses from the VUV to XUV enables the study of electron dynamics over a wide energy range. This is likely to prove advantageous for the study of nanoscale systems, for example, in photoemission experiments in condensed phase system where electron mean free path is known to be strongly energy-dependent.

\begin{figure}[htbp]
	\includegraphics[width=\columnwidth]{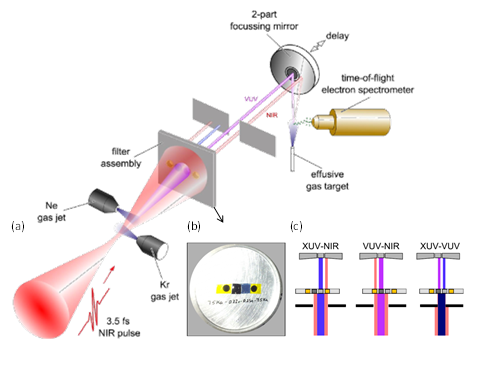}
	\caption{(a) The experimental setup for the generation of synchronised VUV and XUV attosecond pulses by high harmonic generation. NIR laser pulses at 1 kHz repetition-rate are focused into two in-line gas targets of Kr and Ne which are independently optimised for efficient VUV and XUV attosecond pulse generation, respectively. The NIR beam and generated radiation travels collinearly to a filter assembly comprising free-standing thin foil filters that allows different combinations of photon energies to be selected. For attosecond streaking measurements, the VUV and NIR beams, or XUV and NIR beams are selected with the appropriate filters and focused into an effusive gas target by a two-part mirror that allows a controllable time-delay to be introduced between the pulses. The photoelectron energy spectrum is measured as a function of delay using a time-of-flight electron spectrometer. (b) A photograph of the filter assembly. From left to right the filters are Kapton (NIR bandpass), indium (XUV bandpass), tin (VUV bandpass) and Kapton. (c) Moving the filter assembly across the beam allows different combinations of the beams to be selected.}
	\label{fig:imperial_beamline_1}
\end{figure}

Returning to our discussion of the beamline, the 3.5\,fs, 0.4\,mJ CEP stabilised laser pulses centred at a wavelength of 760\,nm and at 1 kHz repetition rate are introduced into the vacuum beamline through a thin optical window. As shown in Fig.~\ref{fig:imperial_beamline_1}, they are focused by a concave mirror into two closely-spaced in-line gas jets in which HHG occurs. This common path geometry minimises timing jitter between the pulses \cite{bothschafter2010collinear,brizuela2013efficient}. The VUV radiation is generated by HHG in a krypton gas target, the XUV radiation in a neon gas target. Kr proved to be the most efficient rare-gas for VUV harmonic generation. Meanwhile, the higher ionisation potential of Ne is better suited to XUV generation. 

The two pulses propagate collinearly with the NIR laser pulse to a filter assembly comprising different thin foil filters that provide spectral bandpass for the NIR (7.5\,$\mu$m Kapton foil), VUV (200\,nm Sn foil) and XUV (200\,nm Zr foil) pulses. By translating the filter assembly across the beam, different combinations of the pulses can be selected. Delay between the collinearly propagating beams is introduced using a two-part MoSi multilayer mirror assembly comprising a piezo-actuated central mirror inside an annular outer mirror \cite{drescher2001x}. One beam is reflected by the inner part of the two-part mirror, the other beam by the outer part. To characterise the XUV pulse by attosecond streaking, the XUV and NIR pulses were selected with the appropriate filters and focused by the two-part MoSi mirror into an effusive Ne target. Similarly, for streaking the VUV pulse, the VUV and NIR pulses were selected and focused into an effusive Ar target. Photoelectron energies were measured with a time-of-flight (TOF) electron spectrometer \cite{hemmers1998high} with a 0.02\,sr collection solid angle and an energy resolution of $\Delta{}E/E \approx 0.5\%$. At each delay value, the photoelectron spectrum was integrated for 3 minutes ($1.8\times10^5$\,shots at 1\,kHz pulse repetition rate).

\begin{figure}[htbp]
	\includegraphics[width=\columnwidth]{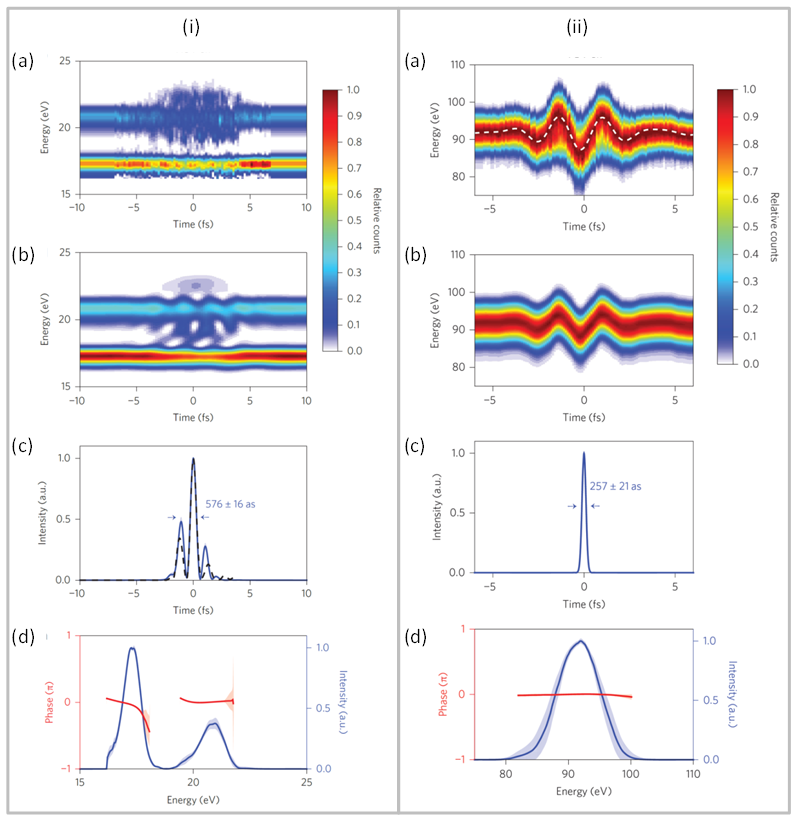}
	\caption{Attosecond streaking measurements of (i) VUV, and (ii) XUV, pulses that were generated synchronously by high harmonic generation in Kr and Ne gas targets, respectively. For both streaking measurements, the other gas target was present. (a) Measured streak trace. (b) Retrieved trace using the FROG-CRAB method. (c) Temporal intensity profile. The VUV intensity profile, (i), exhibits pre and post pulses, as expected due to the transmission of two neighbouring harmonic orders through the Sn spectral filter. The full-width-at-half-maximum VUV and XUV pulse duration were determined to be 576$\pm$16\,as and 257$\pm$21\,as, respectively. (d) Amplitude and phase of the retrieved spectrum. The shaded area represents one standard deviation about the mean.}
	\label{fig:imperial_beamline_2}
\end{figure}

The streak traces and FROG-CRAB retrievals for the VUV and XUV pulses are shown in Fig.~\ref{fig:imperial_beamline_2} (i) and (ii), respectively. The measured pulse durations were $576\pm16$\,as for the VUV pulse centred at 20\,eV and $257\pm21$as for the XUV pulse centred at 90\,eV. In separate measurements, the VUV pulse energy was determined to be $\approx0.5$\,nJ. This should be scalable to higher values by increasing the drive laser pulse energy above the 0.4\,mJ used in this experiment.

\subsection{Attosecond streaking at surfaces -- a stepping stone to nanoscale systems}
Attosecond streaking has enabled the complete characterisation of short-wavelength attosecond pulses and has permitted electron dynamics in matter to be resolved with attosecond precision \cite{drescher2002time,uphues2008ion,uiberacker2007attosecond,schultze2010delay}. Progress is being made to widen the scope of attosecond science. A natural extension is the study of condensed phase matter. The response of solids to electromagnetic fields is important in many areas of science and technology. For example, the study of the time evolution of electron-hole pair formation, charge density distributions, and electron propagation
in wide-bandgap semiconductors interacting with ultrafast laser fields is of relevance to the development of petahertz signal sampling and processing technologies \cite{schultze2013controlling,krausz2014attosecond}. 

In this section we review studies of electron dynamics at solid surfaces using attosecond streaking. This work addresses fundamental questions, such as how is the photoexcited electron affected by the periodic potential as it travels in the solid, and how do other electrons respond in these strongly correlated systems? Laser-assisted photoemission from a surface was first observed in~\cite{miaja2006laser}, where the cross-correlation between a 42\,eV XUV pulse and a NIR pulse was measured on a Pt surface. Subsequent condensed-phase attosecond streaking measurements were conducted on single crystal samples of tungsten \cite{cavalieri2007attosecond} and magnesium \cite{neppl2012attosecond}. These works provided the first experimental data on the time delay of photoemission from surfaces more than 100 years after Einstein's paper on the photoelectric effect~\cite{einstein1905erzeugung}.

In~\cite{cavalieri2007attosecond}, attosecond streaking on a W(110) surface was performed using 300 as XUV pulses centred at 91\,eV, with a 5\,fs streaking field (central wavelength of 750\,nm). To minimise the effect of surface contamination, the measurement chamber was maintained under ultrahigh vacuum (UHV) conditions and the tungsten crystal was cleaned before measurements were made by retracting it to a separate vacuum chamber where it underwent a number of heating cycles, some of which were in an oxygen environment. The attosecond streaking trace showed two pronounced peaks at $\approx$\,83\,eV and $\approx$\,56\,eV corresponding, respectively, to the 4f-state and valence-band photoemission. By comparing the relative phases of the characteristic streaking oscillations for these two peaks, a delay of $110\pm70$\,as was found between the emission of photoelectron originating from the localised 4f core states and those liberated from delocalised conduction band states.  The relatively large measurement error was reduced in subsequent experiments on W(110) \cite{neppl_attosecond_2012} and a smaller delay of $28\pm14$\,as was found. The larger delay in the initial measurement was attributed to surface impurities, despite the precautions taken.

The origin of this photoemission delay provoked considerable theoretical attraction~\cite{kazansky2009one,zhang2009attosecond,zhang2009four,krasovskii2011attosecond,zhang_streaking_2011,zhang_effect_2011}. Interestingly, no significant delay was found between photoemission from 2p-states and conduction band electrons in attosecond streaking measurements on Mg(0001) surfaces \cite{miaja2006laser}. These measurements used a 435\,as pulse at a higher photon energy of 118\,eV compared to the tungsten measurements. The quasi-synchronous release of the photoelectrons for Mg(0001) was explained in~\cite{miaja2006laser} in terms of a simple heuristic model in which the photoemission delay is $\tau_p = \lambda_{\text{mfp}}/v_i$, where $\lambda_{\text{mfp}}$ is the inelastic electron mean free path in the solid which provides a measure of the average travel distance to the surface, and $v_i = \sqrt{2E_i/m_e}$ is the initial electron velocity, where $E_i$ is the initial electron energy and $m_e$ the free electron mass. The electron mean free path was estimated to be 5.9 \AA \,for the 115\,eV valence band electrons and 4.8 \AA \, for the 68\,eV 2p electrons. By coincidence, the delay times are thus almost identical for the valence and 2p electrons at $\approx$92\,as.

Further theoretical work \cite{borisov_resonant_2013,liao_attosecond_2014} has examined the photoemission delays in tungsten and magnesium using quantum-mechanical models. In~\cite{zhang_streaking_2011}, the role of resonant and nonresonant processes in the origin of the delays is considered. Calculations indicate the valence band electrons can be either retarded or advanced with respect to the localised state electrons, depending on the interplay between the surface and resonant valence band emission. Modelling the W(001) experiment, their calculations revealed a strong surface state contribution ($\tau=0$) to the valence-band photoemission. Hence, the valence-band photoelectrons appear before the 4f electrons, with a difference in delay time $\Delta{}\tau = \tau_{4f}$. However, for Mg(0001), resonant processes were calculated to dominate the valence band emission, i.e.~bulk-type photoemission, so both 2f and valence bands are predicted to be photoemitted with delays. This leads to a smaller delay difference -- their calculations suggest $\approx$10--20\,as -- but still not in agreement with the quasi-synchronous photoemission observed experimentally. In~\cite{zhang_effect_2011} the Mg(0001) experiment was modelled. The relative photoemission delay between the valence band and 2p photoelectrons was found to be sensitive to the electron mean free path and screening of the streaking laser field inside the solid. The quasi-synchronous photoemission was reproduced in these calculations.

In addition to photoemission delays, the temporal structure of the photoemitted electron wavepacket can also be extracted from experimental attosecond streaking traces \cite{itatani2002attosecond}. For gas phase atoms, the photoelectron wavepacket can be taken as a ``perfect'' replica of the incident XUV pulse \cite{goulielmakis2008single,kienberger2004atomic}. In a solid, be it bulk or nanoscale, extra information is encoded in the temporal properties of the photoelectron wavepacket, for example connected with electron transport, and dispersion. In~\cite{okell_temporal_2015} attosecond streaking measurements were carried out on thin films of polycrystalline Au, a material used widely in plasmonics, and amorphous WO$_3$, a wide-bandgap (3.41eV) semiconductor \cite{nakamura_fundamental_1981}. These were the first streaking measurements made on solid samples that were not single crystals, and thus they represent a stepping stone towards attosecond streaking of nanoplasmonic fields, as described theoretically in Section V.

\begin{figure}[htbp]
	\includegraphics[width=\columnwidth]{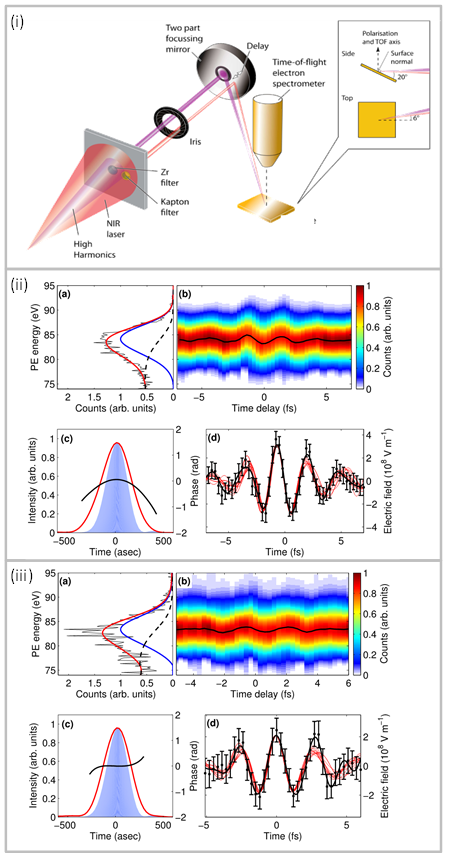}
	\caption{Attosecond streaking at surfaces provides a stepping stone to the nanoscale. (i) Experimental setup for attosecond streaking at surfaces. Few-cycle NIR and attosecond-duration XUV pulses are selected with thin foil filters and focused onto the sample (Au or WO$_3$ in this work) using a two-part mirror that provide a variable time delay. An iris allows the NIR intensity to be reduced to a level where the sample if not damaged but there is still sufficient streaking amplitude. Photoemitted electrons are detected with a time-of-flight electron spectrometer. The geometry of the interaction is shown in the inset. The pulses are focused onto the sample with an incidence angle of $20^{\circ}$. The laser polarization lies along the TOF axis. The incident beam is rotated in a horizontal plane by $6^{\circ}$. (ii) Streaking results from an amorphous, WO$_3$ surface. (iii) Streaking results from a polycrystalline Au surface. (a) Valence band photoelectron spectrum with no streaking field: raw data (solid black), Fourier filtered spectrum (red), secondary electron background (dashed black), and background subtracted and filtered spectrum (blue). (b) Streaking trace after Fourier filtering and background subtraction. (c) Photoelectron wavepacket intensity (red) and phase (black) retrieved from streaking trace using FROG-CRAB method. (d) Retrieved electric field at surface after bandpass filtering (black curve) and unfiltered data points. For comparison, the retrieved field from eight separate gas phase streaking measurements are shown (red curves). The peak field from each gas phase streak has been scaled to the peak field from the solid sample to aid comparison.}
	\label{fig:imperial_surfacestreaking}
\end{figure}

The experiments were conducted in a UHV ($<3\times10^{-9}$\,mbar) surface-science chamber on the Imperial College attosecond beamline that was described in Section IIF-IIG. This chamber contains a similar attosecond streaking set-up to that outlined in Section IIG, comprising a MoSi two-part mirror assembly and the same type of electron TOF spectrometer.  Streaking was conducted with a $248\pm15$\,as XUV pulse centred at 93\,eV (pulse duration determined from attosecond streaking in a Ne gas target) and a 3.5\,fs NIR streaking pulse which were focused onto the sample as shown in Fig.~\ref{fig:imperial_surfacestreaking}(i). The NIR intensity on the samples was $10^{10}$\,W/cm$^2$, well below the damage threshold of the samples material. At this intensity, above threshold photoemission from the NIR in the valence band region was negligible relative to photoemission from the XUV.

Figure~\ref{fig:imperial_surfacestreaking}(ii) shows the streaking results from a 20 nm tungsten sample that had been stored at ambient conditions and was not cleaned or prepared in any way prior to the streaking measurements. Separate analysis (XPS, XRD) revealed that the top 9 nm layer of this sample was amorphous WO$_3$. This is much greater than the electron mean free path in WO$_3$ which was estimated to be 0.5 nm based on \cite{tanuma_experimental_2005,liao_attosecond_2014,tanuma_calculations_2011}. Hence the measured photoelectron spectra are almost exclusively from the photoemission of WO$_3$. Separate streaking results for a 52 nm gold film are shown in Fig.~\ref{fig:imperial_surfacestreaking}(iii). Again, the sample was stored at ambient conditions and no sample cleaning or preparation was carried out.  XRD analysis revealed a polycrystalline surface.

The photoelectron wavepackets were retrieved using FROG-CRAB (see Fig.\ref{fig:imperial_surfacestreaking}(ii)-c, Fig.~\ref{fig:imperial_surfacestreaking}(iii)-c). For the WO3 and Au samples the wavepacket durations were measured to be $359^{+42}_{-25}$\,as and $319^{+43}_{-37}$\,as, respectively. The temporal broadening of the photoelectron wavepackets compared to the XUV pulse duration are $111^{+57}_{-42}$\,as and $71^{+58}_{-54}$\,as for the WO$_3$ and Au samples, respectively (details on the error analysis can be found in the supplement of~\cite{okell_temporal_2015}). Since the XUV pulse duration was measured independently by attosecond streaking in Ne atoms, these broadening measurements provide the first direct comparison of the electron wavepacket broadening inherent to photoemission at surfaces versus atomic ionization.

The broadening figures are consistent with a spread in escape times of free-electrons from within a mean free path of the surface (assuming perfect screening of the NIR field at the sample surface), in the spirit of the heuristic model of \cite{miaja2006laser}. The accuracy of this simple picture of free electron transport in the solid is likely to be a consequence of the XUV photon energy being much larger than the work function of the sample, leading to wavepackets with a free-electron-like character. At lower photon energies, the effective electron mass $m^*$ must be considered and the dispersion relation can depart significantly from that of a free electron \cite{lemell_simulation_2009}. It can exhibit rapid variations in group velocity with energy, which would increase the dispersion broadening of the electron wavepacket as it propagates through the solid (the free-electron group velocity dispersion at $\approx$90\,eV accounts for $<1$\,as of wavepacket broadening for the Au and WO$_3$ streaking measurements). Repeating these measurements with lower energy attosecond pulses is therefore an extremely interesting topic for future investigation.

From the streaking traces it was also possible to fully characterize the streaking near-field at the surface of each sample and compare it to the streaking fields retrieved from gas phase streaking measurements in Ne (see Fig.\ref{fig:imperial_surfacestreaking}(ii)-d, Fig.\ref{fig:imperial_surfacestreaking}(iii)-d). Though Au has a tendency to form rough surfaces which can enhance the excitation of local surface plasmons (LSPs) and surface plasmon polaritons (SPPs), AFM measurement of the gold sample used in this experiment revealed it was mostly plane with a 0.7 nm rms roughness. Therefore, no substantial plasmonic effects were expected, and indeed, as can be seen in Fig.~\ref{fig:imperial_surfacestreaking}(iii)-d, the retrieved near-field is in close agreement with the field recorded from the gas phase streaking measurements. However, by showing that attosecond streaking is possible on unprepared gold films, these experiments clearly demonstrate that streaking measurements in Au nanostructures (such as nanoantennas) should, in principle, be able to retrieve plasmonic fields with attosecond precision (see Section VB). Though such a measurement are likely to pose a significant signal-to-noise challenge, since the nanoplasmonic regions typically make up only a very small fraction of the sample area ionized by the XUV radiation.  This may necessitate the use of spatially-resolved electron detection, as discussed in Section V.B. In any case, it may be beneficial to use XUV high photon energies in order to minimise the wavepacket broadening and thus provide the highest temporal resolution possible.

%% MPQ Section 

\section{Waveform-controlled imaging of electron photoemission from isolated nanoparticles.}

\subsection{Introduction}

Application of the ultra-short waveform-controlled laser fields to nanostructured materials enables generation of localized near-fields with well-defined field evolution. The optical fields that can be tailored on sub-wavelength spatial and attosecond temporal scales have a high potential for control of ultrafast nonlinear processes at the nanoscale, with important implication for laser driven electron acceleration, XUV generation, and nanoscale electronics operating at optical frequencies. Recently, waveform-controlled enhanced electron acceleration in near-fields was observed in isolated nanoparticles, nanotips, and surface based nanostructures. Here we focus on studies of strong field induced waveform-controlled electron emission from isolated nanoparticles.

\subsection{Imaging of laser-induced electron emission from nanoparticles}

\begin{figure}
\centering
\includegraphics[width=\columnwidth]{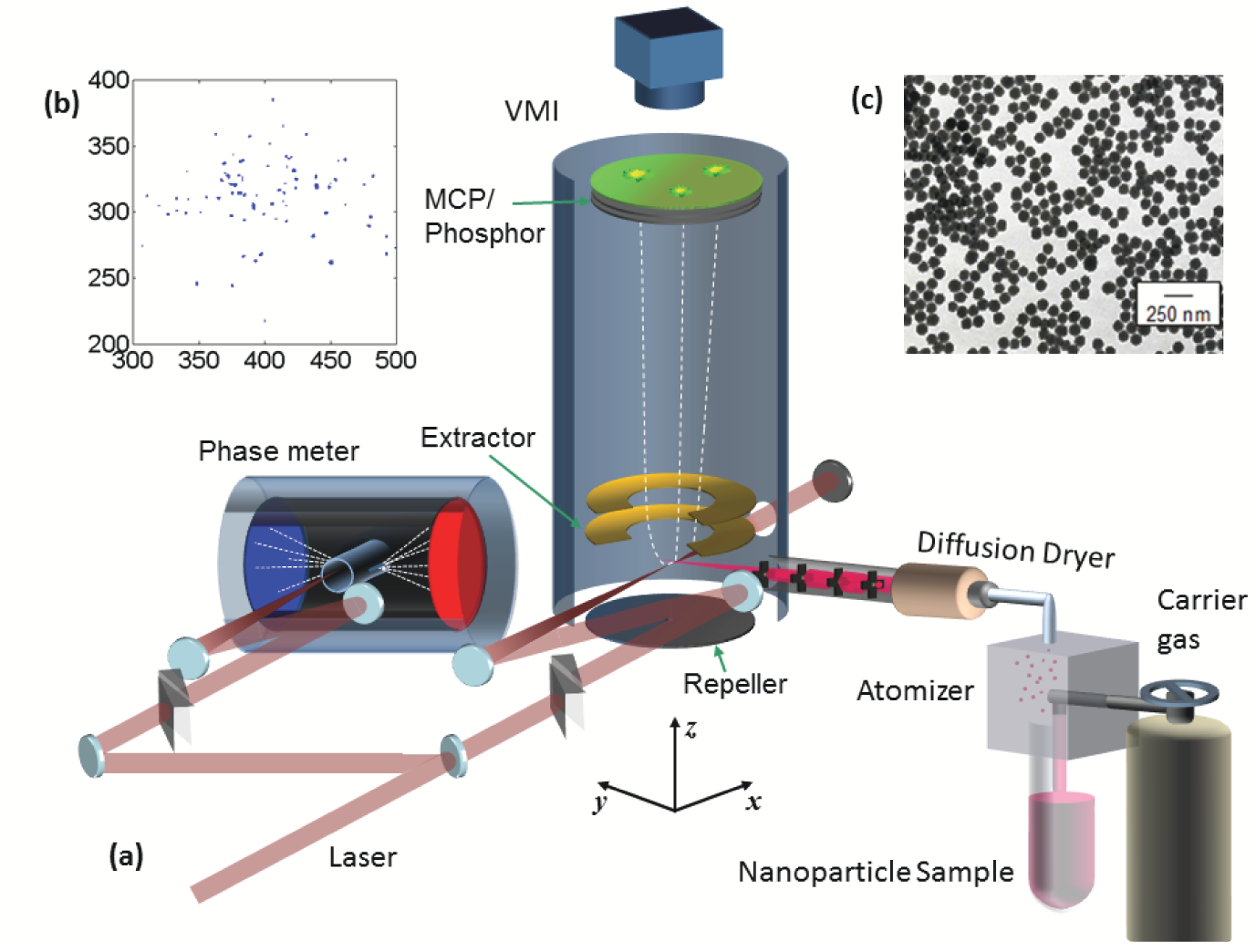}
\caption{
(a) Schematic of the VMI setup with an aerodynamic nanoparticle source and single-shot phase meter. The polarization of the laser was in the plane of the detector. (b) Single shot image recorded by the CMOS camera. (c) TEM image of 95 nm diameter SiO$_2$ nanoparticles. 
\label{Figure1MPQ}}
\end{figure}

In the experiments exploring the electron emission from isolated nanoparticles, reported in~\cite{Zherebtsov11,Zherebtsov12}, few-cycle laser fields (4 fs at 750 nm) have been employed. Such short fields have the advantage that the nanoparticles do not significantly expand during the interaction with the laser pulses, and purely electronic dynamics can be investigated. The pulses were generated from the output of an amplified laser system (25 fs pulse duration, 790 nm central wavelength)~\cite{Ahmad09} that was spectrally broadened in a capillary filled with 2.8 bar Ne gas and compressed by a chirped mirror compressor. The carrier-envelope phase (CEP) of the pulses was measured with a single-shot stereo-ATI phase meter~\cite{Wittmann09,Rathje12} using a small fraction of the laser beam ($\sim 15$ \%), see Fig.~\ref{Figure1MPQ}. The main part of the beam was focused into the center of the electrostatic optics of a velocity-map imaging (VMI) setup where it intersected with a nanoparticle beam. The electron emission distribution was projected onto a microchannel plate (MCP)/phosphor screen assembly and light flashes on the phosphor screen were recorded by a high-speed CMOS camera at the full repetition rate of the laser (1 kHz)~\cite{Suessmann11}. In order to enable storage of single-shot images at these high rates only pixels with brightness above threshold level were stored on the computer. The single shot detection significantly improved the experimental signal-to-noise ratio as it allows suppressing/identifying background contributions by selecting only the frames that contain nanoparticle signal. 

The SiO$_2$ nanoparticles were prepared by the groups of C.~Graf and E.~R{\"u}hl at Freie Universit{\"a}t (FU) Berlin using wet chemistry methods based on the St\"ober procedure~\cite{Stoeber68} and subsequent seeded growth process. This technique allowed producing particles with diameters in the range 50-550 nm with a polydispersity of less than 8\%~\cite{Zherebtsov11,SuessmannPhD}. After the synthesis the particles were purified by centrifugation/redispersion in ethanol. For size and shape characterization of the samples transmission electron microscopy (TEM) images were taken. Figure~\ref{Figure1MPQ}(c) shows a typical TEM image of the $95\pm6$ nm particles. The isolated nanoparticles were delivered into the interaction region by injection of the nanoparticle suspension into a carrier gas and focusing of the nanoparticle stream with an aerodynamic lens.

\subsection{Waveform controlled electron acceleration in near-field of a nanosphere}

\begin{figure}
\centering
\includegraphics[width=\columnwidth]{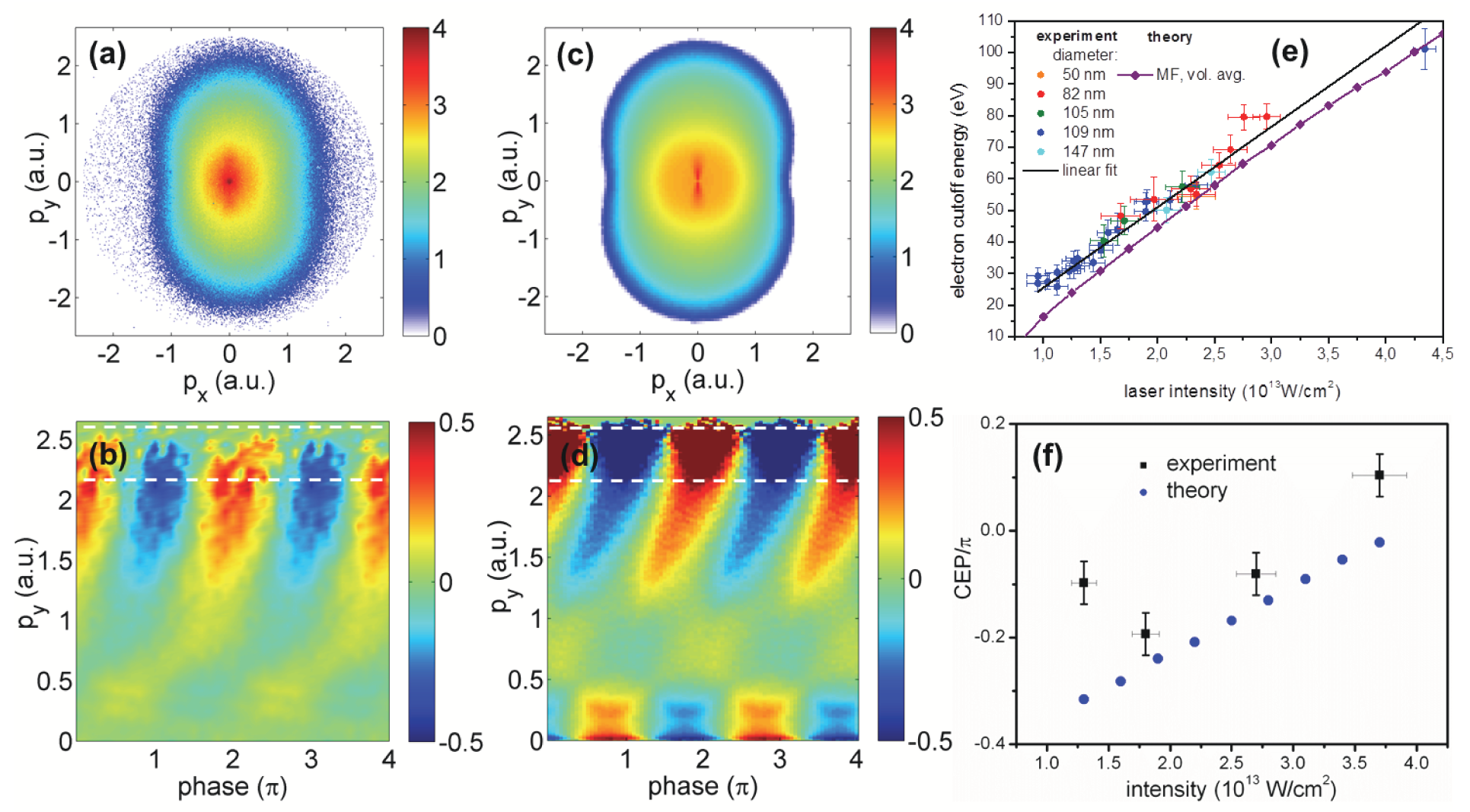}
\caption{(a) Photoelectron momentum map (projected along $p_z$) averaged over the CEP (log color scale) and (b) asymmetry of the electron emission as a function of the electron momentum and the CEP measured for 95 nm SiO$_2$ nanoparticles at $3.7\times10^{13}$ W cm${^{-2}}$.  (c)-(d) Photoelectron momentum and asymmetry map calculated for the same parameters as in (a)-(b). (e) Intensity dependence of the cutoff in the electron emission from SiO$_2$ nanoparticles with indicated diameters. (f) Dependence of the CEPs at the maximum asymmetry $\phi_{\mathrm{max}}$ of the electron emission from SiO$_2$ nanoparticles of 95 nm diameter on the laser intensity measured (black boxes) and calculated (blue filled circles). To obtain $\phi_{\mathrm{max}}$ the asymmetry maps were integrated over $p_y$ in the cutoff region (indicated by white dashed lines) and fitted with a function $f(\phi_{\mathrm{max}})=A \cos(\phi_{\mathrm{CEP}}-\phi_{\mathrm{max}})$.
\label{Figure2MPQ}}
\end{figure}

Figures~\ref{Figure2MPQ}(a)-~\ref{Figure2MPQ}(d) show typical results from the laser-induced electron emission from 95 nm diameter SiO$_2$ nanoparticles. The laser polarization is along the $p_y$ axis. The electron momentum distribution has an elliptical shape and is elongated along the polarization direction. Few-cycle laser pulses illuminating the nanoparticles offer a possibility to explore the CEP-dependence of the electron emission. The CEP dependence of the directional emission can be quantified with an asymmetry parameter 
\begin{equation}
A(p_y,\phi)=\frac{P_{\mathrm{up}}(p_y,\phi)-P_{\mathrm{down}}(p_y,\phi)}{P_{\mathrm{up}}(p_y,\phi)+P_{\mathrm{down}}(p_y,\phi)},
\end{equation}
where $P_{\mathrm{up}}(p_y,\phi)$ and $P_{\mathrm{down}}(p_y,\phi)$ are the angle integrated electron yields (within $[-25^{\circ},+25^{\circ}]$ angular range) in the up (positive $p_y$ momentum) and down (negative $p_y$ momentum) directions and $\phi$ is the CEP. The asymmetry parameter exhibits a pronounced CEP dependence with the largest amplitude near the highest recorded electron momentum. The cutoff of the CEP dependent electron emission is in agreement with the cutoff of the momentum map in Fig.~\ref{Figure2MPQ}(a) and is at about $50 U_p$, where $U_p$ is the ponderomotive potential of an electron in the driving laser field. The intensity dependence of the electron emission is illustrated in Figs.~\ref{Figure2MPQ}(e) and~\ref{Figure2MPQ}(f). For the studied intensity range $(1-4.5)\times10^{13}$ W cm$^{-2}$ the measurements show a nearly linear intensity dependence of the cutoff energy with an average scaled cutoff about $53.0 U_p$. The obtained cutoff is much higher than the modified classical atomic cutoff of $\sim 24 U_p$ that is expected for the dielectrically enhanced field near a nanosphere. The maximum asymmetry phase $\phi_{\mathrm{max}}$ increases with the laser intensity (except at the lowest intensity point). 

The mechanism of the enhanced electron acceleration was analyzed with quasi-classical trajectory-based simulations using the Mean-field-Mie-Monte-Carlo (M$^3$C) model~\cite{Zherebtsov12}. Results of these calculations performed for the same parameters as in the experiment are presented in Figs.~\ref{Figure2MPQ}(c)-~\ref{Figure2MPQ}(f). The simulations reproduce the main features of the experiment such as the overall shape of the momentum and asymmetry maps as well as the cutoff value. The simulation shows a similar increase of $\phi_{\mathrm{max}}$ with laser intensity as the experiment except for the lowest intensity point (Fig.~\ref{Figure2MPQ}(f)). The discrepancy at the lowest intensity may be ascribed to a deviation of the initial ionization mechanism from the pure tunneling regime assumed in the model.

\subsection{Effect of near-field deformation on electron photoemission from a nanosphere}

\begin{figure}
\centering
\includegraphics[width=\columnwidth]{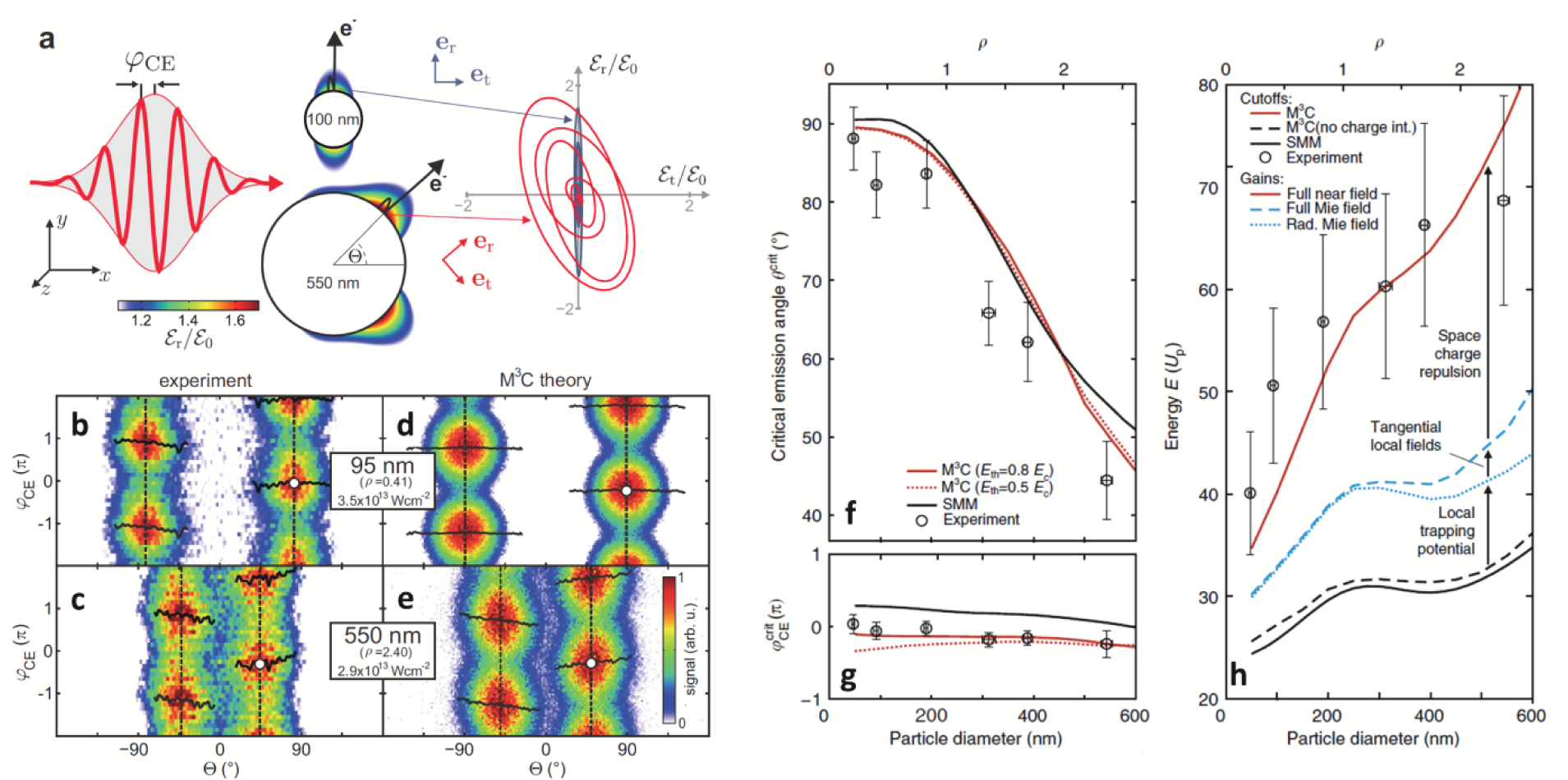}
\caption{(a) Peak radial field enhancement in the $x-y$ plane at $z=0$ obtained by Mie solution for SiO$_2$ spheres illuminated with a 4 fs linear polarized laser pulse centered at 720 nm and a CEP of $\phi_{\mathrm{CE}}=0$ (left). Field evolution in the local reference frame in the points of maximum field enhancement (right). (b)-(e) Measured (b)-(c) and simulated (d)-(e) angle and CEP-resolved electron yields of energetic electrons near-cutoff. The white dots indicate CEP values $\phi_{\mathrm{CE}}^{\mathrm{crit}}$ and emission angles $\theta^{\mathrm{crit}}$ of maximum upward emission. (f)-(g)-(h) Particle size dependence of the critical emission angle (f), critical phase (g), and cutoff energies (h). The symbols and lines indicate measured and calculated parameters. The simple man's model (SMM) simulations are described in detail in~\cite{Suessmann15,Seiffert15}.\label{Figure3MPQ}}
\end{figure}

The angle resolution provided by VMI detection offers a possibility for a more detailed visualization of the CEP-dependent photoemission. It was demonstrated recently that phase controlled electron photoemission provides a sensitive probe for localized fields~\cite{Suessmann15,Seiffert15}. In this work isolated nanospheres served as a test system for the generation of near-fields with adjustable polarization and spatial characteristics. Figure~\ref{Figure3MPQ}(a) shows the enhancement field distribution of the radial electric field as predicted by Mie theory for 100 and 550 nm diameter SiO$_2$ particles. For the particle much smaller than the wavelength of the incident field the near-field exhibits dipolar character and peaks along the laser polarization axis (see Section IIA). As the particle size becomes comparable to the laser wavelength the effect of field propagation becomes noticeable resulting in a shift of the region of maximal enhancement in propagation direction towards the rear side of the sphere.

The effect of field propagation on the phase controlled electron photoemission is illustrated in Figs.~\ref{Figure3MPQ}(b)-~\ref{Figure3MPQ}(e). For small nanospheres the electron yield peaks at the poles of the particle with the maximum signal at a critical CEP, $\phi_{\mathrm{CE}}^{\mathrm{crit}}$. For the large particles the electron yield shows similar phase dynamics and a significant shift of the critical emission angle to almost $45^{\circ}$. The size dependence of the main emission parameters is illustrated in Figs.~\ref{Figure3MPQ}(f)-~\ref{Figure3MPQ}(h). The experiment shows good agreement with the M$^3$C simulations, supporting proper description of a tunable directionality and attosecond control of electron dynamics in strongly deformed near-fields. Quantitative analysis of different many-particle contributions to the acceleration process (Fig.~\ref{Figure3MPQ}(h)) shows that the local trapping potential is only weakly size-dependent. That can be explained by the local character of this potential, being the latter determined mainly by the local electron density. On the other hand, contribution from the space-charge repulsion increases strongly with the particle size that indicates its sensitivity to the full electron distribution.

\begin{figure}
\centering
\includegraphics[width=\columnwidth]{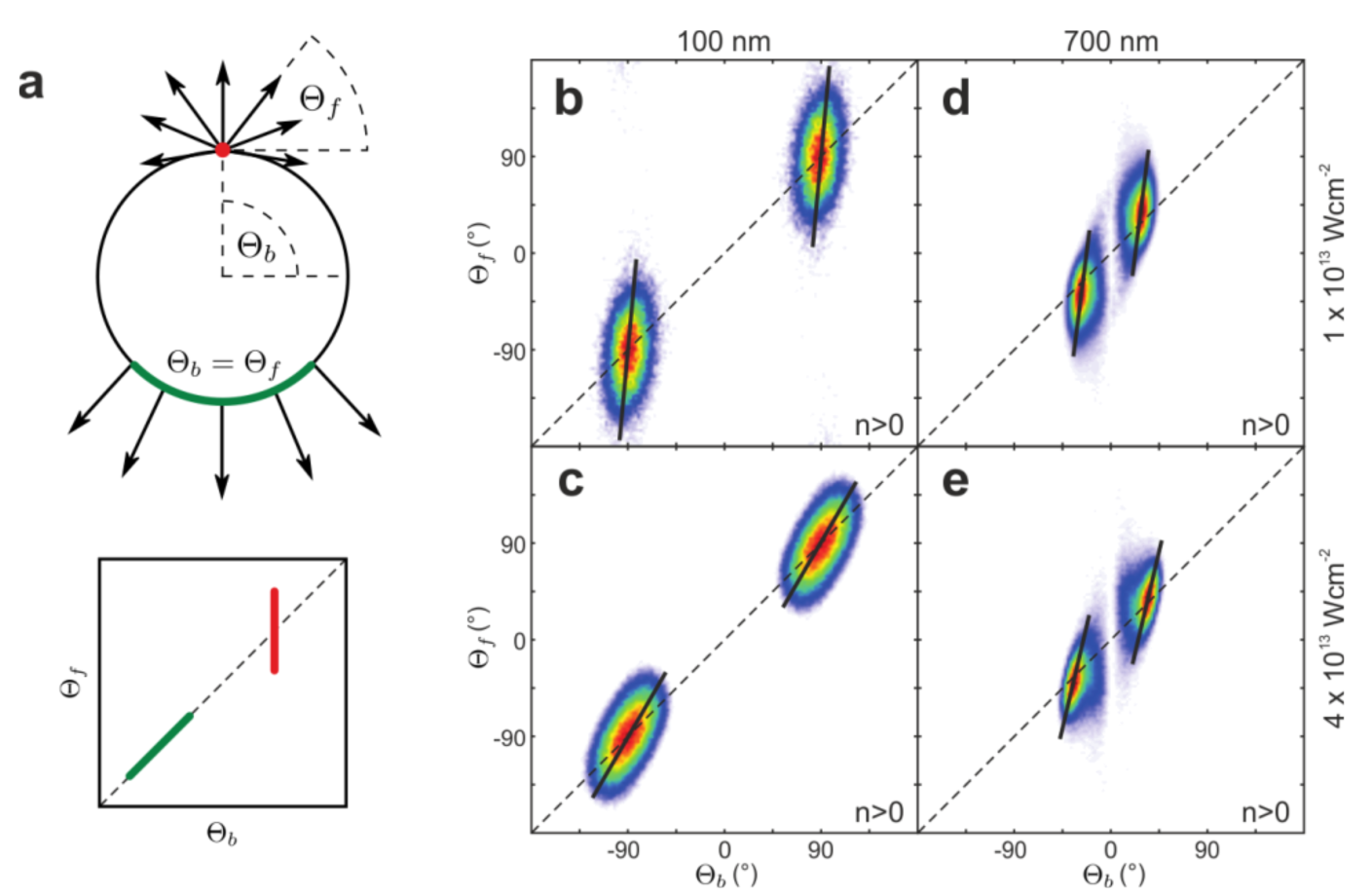}
\caption{(a) Schematic representation of correlation characteristics between birth angle $\Theta_b$ and final angle $\Theta_f$ of the unidirectional (red) and radial (green) emission. The birth and final angle are defined as projections of birth position and final momentum vector on the $x-y$ and $p_x-p_y$ planes respectively. (b)-(e) Correlation plots for energetic electrons ($E>E_c/2$) emitted from large and small nanospheres at two different intensities. Here $E_c$ denotes the cutoff energy of the electron emission. Only trajectories with electron collisions ($n>0$) were selected for the analysis. The dashed black line represents the case of radial emission.
\label{Figure4MPQ}}
\end{figure}

The trajectory based model allows correlation analysis between the electron emission position and its final momentum direction. Figure~\ref{Figure4MPQ}(a) illustrates two limiting cases of the emission from a sphere. The radial emission allows correlation of the final momentum direction to the initial birth angle. The analysis of energetic trajectories from small nanospheres, where the tangential field component is negligible, shows transition from unidirectional to radial emission with increase of the laser intensity (Figs.~\ref{Figure4MPQ}(b)-~\ref{Figure4MPQ}(c)). This intensity dependence can be attributed to the effect of the trapping potential that favors radial emission. For large nanoparticles the tangential and normal components of the driving field at the surface of the particle become comparable (Fig.~\ref{Figure3MPQ}(a)). The non-diagonal shape of the correlation plots reflects increased importance of the tangential field component for the acceleration process. 

% Peter and Michael contribution

\section{Attosecond control of electrons at nanoscale needle tips}

As discussed in Section I, hallmarks of attosecond physics include electric-field driven control of electron motion and the re-scattering plateau. First observed and understood in the context of atomic physics in the gas phase in the 1980s and 1990s (see e.g.~\cite{Scrinzi06,Krausz09,Milosevic06} and references therein), at solids and in particular at nanostructures they have been first observed and theoretically understood about two decades later~\cite{Kruger11, Wachter12, Kruger12N, Herink12, Piglosiewicz2014}. Attosecond physics phenomena at single nanostructures have been discussed in several original papers and review articles (see, e.g.~\cite{Hommelhoff15}), which is why here we only give a comprehensive overview of the field and refer the reader interested in the technical details to the more extensive review articles and original papers.

%\TV{About 50 years ago in his seminal work Keldysh has treated both atoms as well as solids~\cite{Keldysh1965}. While for atoms he has discussed ionization driven by oscillating fields, for solids he has treated oscillating field-driven population transfer from valence to conduction band, so processes taking place inside of a solid. This last point is of great current interest but outside of the scope of this review, which is why we do not discuss it here.
%\TV{The Keldysh theory, formulated about 50 years ago, beautifully applies to both atomic systems and solids~\cite{Keldysh1965}.} But there is one more aspect, namely that the treatment of {\it solid surfaces} bears great similarity to the {\it atomic} treatment. So t...}
%About 50 years ago in his seminal work Keldysh has treated both strong-field ionization of atoms as well as interband population transfer inside solids~\cite{Keldysh1965}. 
About 50 years ago in his seminal work Keldysh has come up with a theory that insightfully connects atomic tunnelling ionization in a strong laser field with ionization in a static electric field~\cite{Keldysh} (see Section IA). The same relation holds for {\it solid surfaces}, not discussed in Keldysh's pioneering work: DC field emission from solids (see~\cite{Fursey2005} and references therein) and optical tunnelling photoemission are closely linked, which is why the latter is consequently called optical field emission~\cite{Bunkin1965}. In addition, multiphoton emission, another limiting case of the Keldysh theory, may also arise at solids. DC field emission routinely requires sharp nanoscale needle tips in order to reach field strengths on the order of 1\,V\,nm$^{-1}$ on the tip surface by the virtue of the lightning rod effect. The optical counterpart of this DC field enhancement effect at nanotips, optical near-field enhancement, pushes laser-tip interactions into regimes of high intensity of up $10^{14}\,\mathrm{W\,cm}^{-2}$, corresponding to peak electric field strengths of $2.7$\,V\,\AA$^{-1}$. This is of great practical relevance: it enables strong-field physics experiments without the use of amplified laser systems. Initial studies focusing on the very nature of femtosecond laser-induced electron emission pointed to tunnelling photoemission~\cite{PeterH06, Hommelhoff2006a} or to multiphoton photoemission~\cite{Ropers07,Ropers07PRL,Barwick2007,Hilbert2009}, the two limiting regimes for oscillating fields of the Keldysh theory. Spectrally resolved measurement demonstrated above-threshold photoemission, the analogue of gas-phase above-threshold ionization, and a clear ponderomotive shift of above-threshold peaks -- hallmarks of the onset of a strong-field photoemission regime~\cite{Schenk10}.

\begin{figure}
	\centering	\includegraphics[width=0.72\columnwidth]{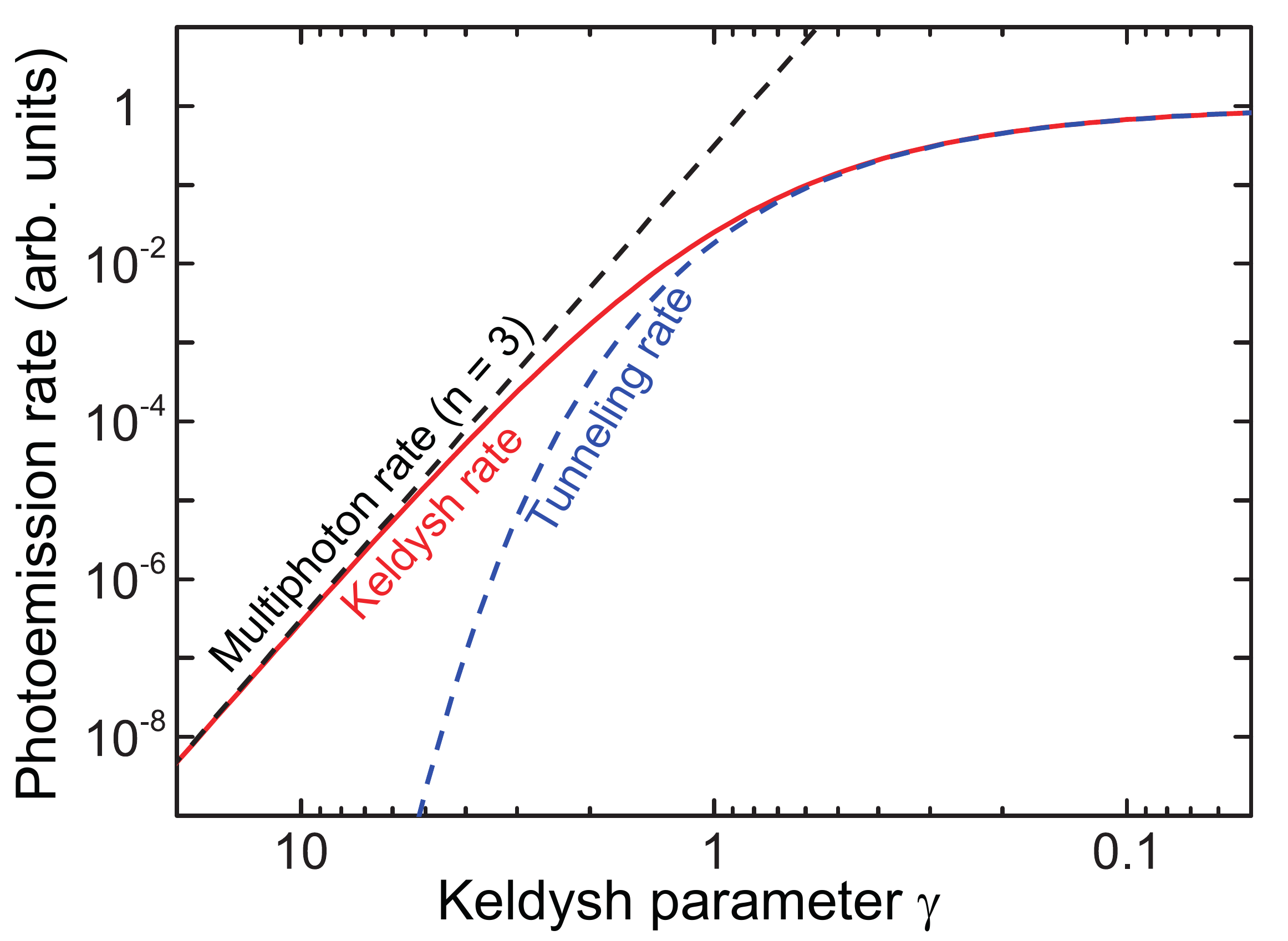}
	\caption{Illustration of the transition from multiphoton to tunnelling regime. Photoemission rate as a function of Keldysh parameter $\gamma$~\cite{Keldysh, Toth1991} (red full curve). The soft kink indicates the transition from multiphoton limit (dashed black line) to the tunnelling limit (dashed blue curve). In this calculation, the workfunction is 4.5\,eV and the photon energy 1.5\,eV. The theory curves are calculated with exponential accuracy.}
	\label{PH-fig:GeneralizedKeldyshRate}
\end{figure}

The transition from the multiphoton regime (Keldysh parameter $\gamma \gg 1$) to the tunnelling regime ($\gamma \ll 1$) was first reported in photoemission from a gold nanotip in a near-infrared laser field~\cite{Bormann2010}. Similar to initial work performed at a planar solid surface~\cite{Toth1991}, the authors observed a soft kink in the scaling of photocurrent with intensity. At low intensity, the multiphoton scaling dominates, with the current $j$ scaling as $j \propto I^p$, with $I$ the laser intensity and $p$ the minimum required number of photons for photoemission. Around an intensity corresponding to the intermediate regime of Keldysh parameter $\gamma \sim 1$, the scaling changes into a field-dependent tunnelling behavior ($j \propto \exp{-C / \sqrt{I}}$), featuring a much less steep slope (see Fig.~\ref{PH-fig:GeneralizedKeldyshRate} for an illustration). This transition has been observed in many more experiments for different wavelengths and materials, including plasmonic nanostructures (see, e.g.,~\cite{Dombi10,Keathley2013,Piglosiewicz2014}, and can be well modeled by strong-field theory~\cite{Yalunin2011}. The change of slope as a function of intensity appears to be more rapid than expected from this theory, which has been explained with an additional photocurrent contribution from the laser field penetrating into the metal surface~\cite{Bormann2010}. Also strong saturation of the photoemission yield at intensities slightly higher than the kink has been reported~\cite{Piglosiewicz2014}.

Tunnelling photoemission is prompt by definition and features sub-optical-cycle-resolved bursts of electron wavepackets~\cite{Hommelhoff2006a,Yalunin2011,Kruger12B}, even in the intermediate, non-adiabatic tunnelling regime around $\gamma$$\sim$$1$~\cite{Yudin2001}. Strongly delayed photoemission, on the other hand, is a sign of the formation of an excited non-equilibrium electron distribution inside the solid. A prominent example is thermally enhanced field emission where the laser pulses heat the electron gas~\cite{Kealhofer2012}. Pronounced electron-electron and electron-phonon scattering can also result in a photoemission delay, as encountered in a comparatively long laser pulse~\cite{Yanagisawa2011a} or, as it has been argued, when electrons return to the surface and undergo backscattering inside the metal~\cite{Yanagisawa2014}. While the first experiments resorted to verify prompt photoemission by measuring current additivity in an autocorrelation experiment~\cite{Hommelhoff2006a,Ropers07,Hilbert2009}, a recent study indicates that photoemission can be prompt up to a Keldysh parameter of $\sim$13~\cite{Juffmann2015}. This study employed a microwave cavity to streak photoelectrons from a nanotip, measuring their emission time with an accuracy of about 2\,fs. If identified, delays smaller than 2\,fs already are on a sub-optical-cycle level and might be interpreted with tunnelling time delays, a research subject of high current interest.

\begin{figure}
	\centering	\includegraphics[width=\columnwidth]{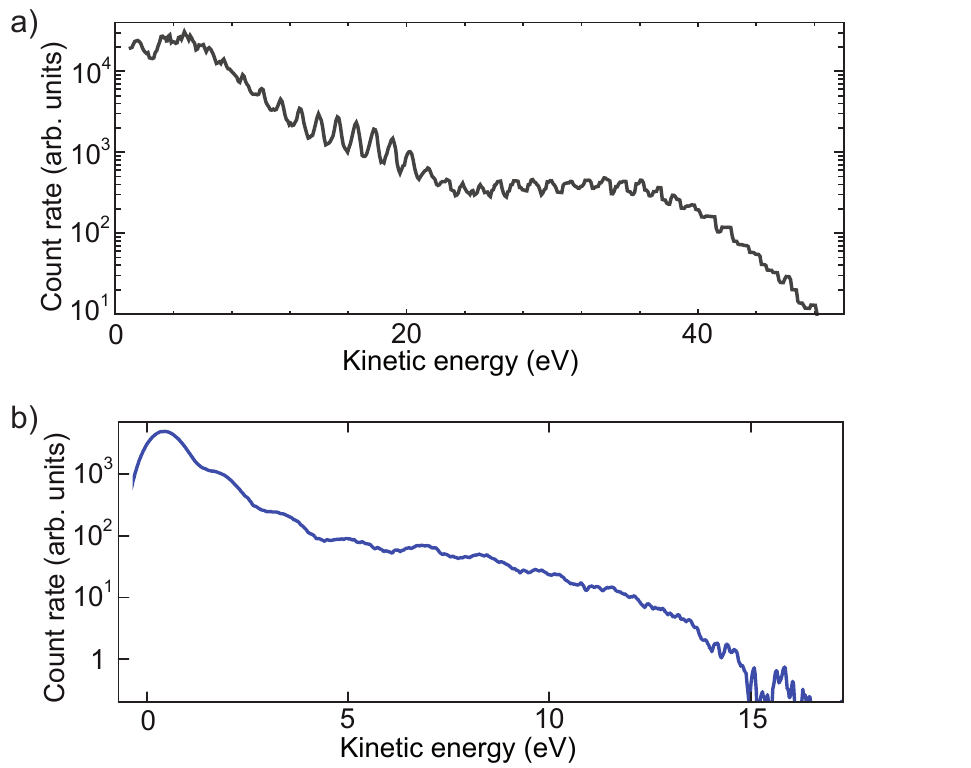}
	\caption{For comparison: photon-order resolved strong field spectra from xenon atoms (a) and from a tungsten needle tip (b). In both cases, the plateau and the cutoff are clearly discernible. Figure in (a) taken from \cite{Paulus2004} and modified.}
	\label{PH-fig:PlateauAtomsTip}
\end{figure}

As outlined above, the regimes of atomic gas-phase ionization can readily be transferred to photoemission from solid surfaces. This holds especially true for the electron dynamics following photoemission, as revealed by spectral features. The re-scattering plateau, its cutoff and a comparison to early work in atoms is shown in Fig.~\ref{PH-fig:PlateauAtomsTip}~\cite{Lindner2005,Kruger11}. Clearly, the overall shape of the spectra is very similar, with the direct part (exponential decrease of the count rate at small energies), the plateau part, which is terminated by the cutoff and a subsequent steep decrease in count rate. Both spectra exhibit peaks spaced by the photon energy, which is a clear sign of above-threshold ionization and photoemission. Because of the almost identical driver wavelengths of around 800\,nm in the two experiments, the photon energies are about equal, indicating that although the shape of the spectra are very similar, the energy scales differ. This is owed to the fact that the intensity driving electron re-scattering is very different (gas: $7 \times 10^{13}\,\mathrm{W\,cm}^{-2}$, tip: $4 \times 10^{11}\,\mathrm{W\,cm}^{-2}$ in the bare focus). This large discrepancy is partially lifted by optical near-field enhancement at the nanoscale needle tip, leading to an effective intensity of $\sim$$1.5 \times 10^{13}\,\mathrm{W\,cm}^{-2}$ at the tip's apex. In addition, the ionization potential (or workfunction in metals) is different: in xenon, the ionization potential is 12.1\,eV, while in tungsten it is 4.5\,eV.

\begin{figure}
  \centering
  \includegraphics[width=\columnwidth]{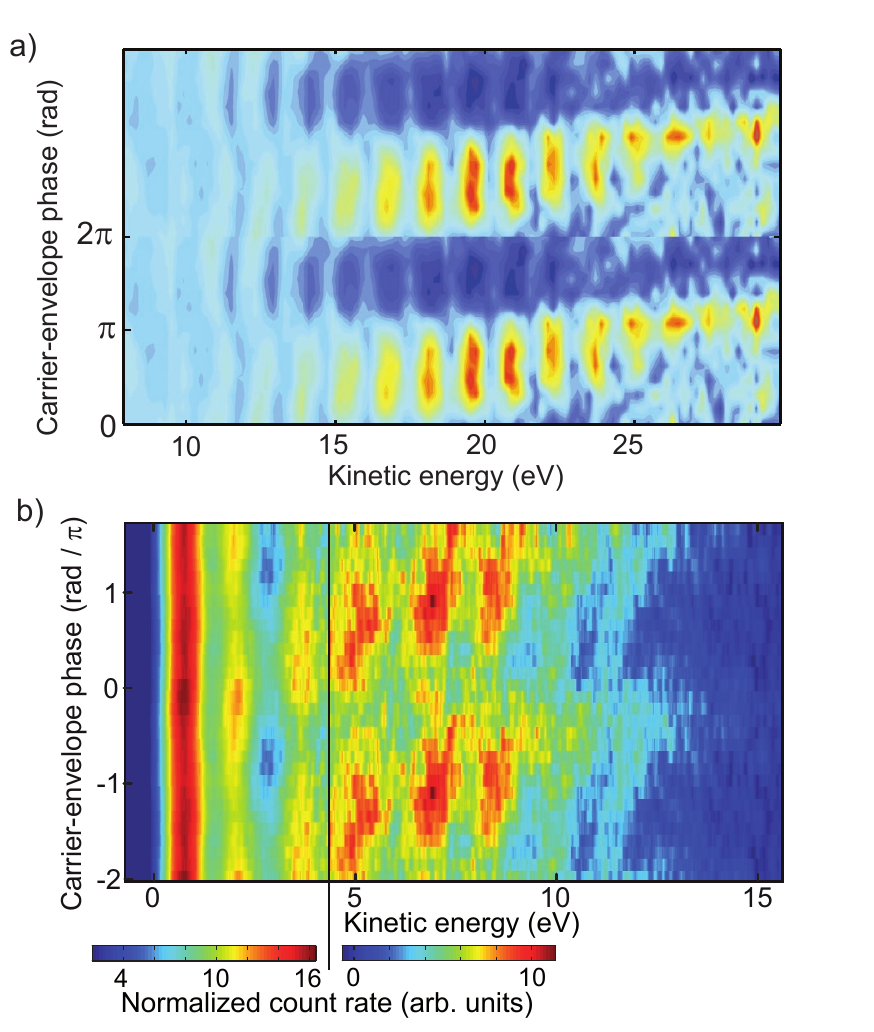}
  \caption{CEP-resolved spectra recorded in atomic argon gas (a) and at a tungsten tip (b). Data in (a) are taken from~\cite{Lindner2005}, data in (b) from~\cite{Kruger11}. In both cases only the range ${0 \dots 2 \pi}$ was measured.  The same data is shown twice on top of each other for clarity. }\label{PH-fig:CEP-AtomsTip}
\end{figure}

Accordingly, CEP resolved spectra of argon gas and tip look very similar, see Fig.~\ref{PH-fig:CEP-AtomsTip}. The similarity of these data is based in the understanding that both the electron emission mechanism as well as the external dynamics of the electron in the laser field are essentially identical. The Keldysh parameter in both cases leans towards optical-cycle resolved electron emission in the non-adiabatic tunnelling regime~\cite{Yudin2001}, after which the dynamics of the electron in the field of the laser, including the re-collision process of the electron with the parent matter, seems identical. 

For the general understanding of the process, the atomic physics picture view holds: Corkum's seminal three-step model~\cite{corkum93} fully applies and can explain the position of the cutoff, while the matter wave interference picture explains the photon orders, and their nonappearance for certain CEP values, provided that few-cycle laser pulses are used: Photon orders may not show up if electrons can only be accelerated to high enough energies during a single laser cycle. In that case, no other time-window exists from which electrons with sufficient energies can be released in order to interfere with those ionized by the first optical cycle~\cite{Kruger11, Kruger12B}. Intriguingly, a very simple model based on the propagation of Gaussian electron pulses suffices to explain the spectra, notably over the full dynamic range of several orders of magnitude in count rate~\cite{Kruger12N}. Because the solid surface breaks the symmetry, the number of electron trajectories that contribute are only half as many as in the case of atoms in the gas phase, making the system even simpler to describe. Hence it may be called a model system for strong-field processes at surfaces. We conclude that it is predominantly the dynamics of the single free active electron driven in the laser field that determines the shape of the spectra, in particular the plateau and cutoff regimes. The direct part and its behavior can be well modeled with extant theory such as PPT and ADK~\cite{PPT1966, ADK1986, Yalunin13, Bionta2014}. When, however, many photoelectrons are emitted per laser pulse and influence each other by Coulomb repulsion, the simple picture of direct photoemission and re-scattering might not be sufficient anymore, as a recent study suggests~\cite{Yanagisawa2014}. Here, the Coulomb repulsion itself leads to the formation of a plateau in electron spectra, whereas the low-energy part is formed by electrons that have been slowed down by scattering effects.

With this understanding, we can turn the perspective around and utilize the single active electron as a probe to measure the strong field that is driving it, namely the optical near-field at the nanostructure. Optical near-fields decay over a characteristic length not given by the driving wavelength, but given by the typical dimension of the structure provided the latter is much smaller than the former~\cite{Novotny2012, Novotny2011}. Typically, the sharpest nanostructures have kinks and edges with radii of curvature larger than 3\,nm, hence the decay length of the near-field is usually larger than $\sim$2\,nm~\cite{Thomas2013}. With electrons driven by 800\,nm laser light and effective field strengths in the range of $1\,\mathrm{V}$\,\AA$^{-1}$, the classical excursion length according to the three-step model equals 0.3\,nm. Therefore, in the re-collision scenario the electron only samples a field region extending the excursion length away from the nanostructure surface. For such small excursion lengths, the near-field can be considered constant for all but the sharpest nanostructures. Based on this idea, the field enhancement factor can be accurately measured, as has been done in~\cite{Thomas2013, Kruger14}. Similarly, electron acceleration at arrays of gold nanostructures~\cite{Dombi2013} and nanopillars~\cite{Nagel2013} was used to deduce the magnitude of the field enhancement, demonstrating how their shape affects the behaviour of optical near-fields through plasmon resonances. Also for non-plasmonic materials a strong shape dependence is expected that can lead to a dramatic increase of field enhancement~\cite{Thomas2015}.

Recent progress in needle tip-based optical cycle resolved physics experiments encompasses work at longer wavelengths than the typical 800\,nm of the initial work, extending up to 8\,$\mu$m ~\cite{Herink12, Park2012, Park13, Yalunin13, Piglosiewicz2014}. Longer wavelengths are of interest for several reasons. First, the typical time scales are prolonged, meaning that the tunnel barrier responsible for electron emission is established for longer times. Thereby, it is easy to reach deep into the tunnel regime (note the definition of the Keldysh parameter in terms of tunnel duration, see Section IA). Second, by a similar token, the electron, spending more time within a single laser cycle, is accelerated to larger kinetic energies, which could be of interest for source applications and the like. However, these are limited by the third point, namely that the increased classical excursion length of the classical electron trajectory can now easily overcome the optical field's decay length, which, as pointed out above, is given by the dimensions of the nanostructure and not the driving wavelength. The last point is closely related to the discussion of attosecond physics in inhomogeneous fields, treated in Section VIIIE and IXD. The observations reported in the aforementioned experimental works confirmed two outcomes of these effects, namely the suppression of re-scattering and electron emission followed by instantaneous acceleration within less than an optical cycle. Like in the near-infrared case~\cite{Kruger11}, the electron motion can be controlled with the CEP~\cite{Piglosiewicz2014} and the near-field can be investigated based on electron kinematics~\cite{Park13}. Wavelength-scaling studies recently included also the terahertz regime, which was explored as a means to streak photoemission~\cite{Wimmer2014} or drive field emission~\cite{Herink2014}.

Until recently, standard materials such as tungsten and gold have been used in the study of femtosecond laser driven electron emission from needle tips. It has been pointed out that this was not a bad choice, as the comparably large heat conductivity of both materials seems central to observing prompt electron emission mechanisms, such as multiphoton and tunneling processes~\cite{Kealhofer2012}. Nanotips made from highly doped silicon were also proven to support these processes~\cite{Swanwick2014}. In contrast, hafnium carbide, a material with an extremely high melting point of $\sim$4200\,K, which thus may also seem well suited as a prompt femtosecond electron emitter, displays a large thermal and thus non-instantaneous electron emission current contribution, due mainly to its poor heat conductivity~\cite{Kealhofer2012}. It will be interesting to see if strong-field effects can be observed at extremely well controlled and rugged modern materials, such as carbon nanotube electron emitters~\cite{Bionta2015}.

Based on the fundamental understanding gained in the last decade, the research field has enabled a range of applications of laser-driven nanotip photoemission, in particular as a source of ultrashort electron pulses. Crucial is the development and characterization of various electron source designs, a very active research area of recent years~\cite{Paarmann2012,Luneburg2013,Hoffrogge2014,Ehberger2015,Bormann2015,Vogelsang15,Schroder2015,Muller2016}. The spectrum of applications ranges from demonstrations of quantum optical phenomena with free electrons~\cite{Caprez2007,Feist2015} via the generation of X-ray pulses~\cite{Foreman2013} to ultrafast microscopy and low-energy electron diffraction experiments for fundamental material science~\cite{Quinonez2013,Gulde2014,Muller2014}. Notable is also the source development focussing on emitter arrays for, e.g., injecting high-brightness electron beams into accelerators~\cite{Ganter2008,Tsujino2008,Tsujino2009,Mustonen2011,Mustonen2012,Keathley2013,Swanwick2014,Hobbs2014}.

However, none of these applications so far goes beyond the femtosecond regime and makes direct use of the sub-optical-cycle nature of strong-field photoemission or the electric field control of electron motion. Due to the matter-wave dispersion of free electrons, attosecond dynamics are essentially limited to the vicinity of the nanostructure. Apart from probing the near-field and its structure as described above, attosecond control capabilities have been explored for detecting the CEP of few-cycle laser pulses~\cite{Hommelhoff2006a,Hommelhoff2006b,Kruger11,Schenk2011,Piglosiewicz2014}. Experimental efforts are currently underway towards attosecond electronics where the electric field switches and controls current between nanoelectrodes. Here, in an initial experiment the operation of a nanoscale vacuum tube diode in the femtosecond regime was reported, with the prospect of extending this time scale to the sub-optical-cycle regime~\cite{Higuchi2015}. Also it will be interesting to see if high-harmonic generation -- the recollision mechanism of attosecond science par excellence -- can be observed, without the use of an additional gas, and controlled by the CEP~\cite{Marcelo14}. Last, we note that the physics discussed here is also of current interest in the context of nanoplasmonics~\cite{Schertz2012, Dombi2013, Kusa2015}, optical control of photoemission sites on a nanostructure~\cite{Yanagisawa2009,Yanagisawa2010} and VMI~\cite{Bainbridge2014}.

\section{Attosecond streaking in nanolocalized plasmonic fields}

While the waveform controlled electron emission contains some spatial information about the near field distribution~\cite{Kruger11}, reconstruction of the time evolution relies to a large extent on model calculations. A pump-probe approach provides more direct access to the time-resolved near field dynamics. So far, ultrafast plasmonic near fields surrounding nanowires, nanoantennas, and nanotips have been fully characterized using femtosecond pulse characterization techniques~\cite{Dombi10,Vogelsang15,Hanke,Anderson10,Rewitz12}. Attosecond streaking measurements are expected to yield an even deeper understanding of the collective electron dynamics governing plasmon formation and decay, where transport and interaction effects on sub-cycle timescales are expected to be important.

\begin{figure}
\centering
\includegraphics[width=\columnwidth]{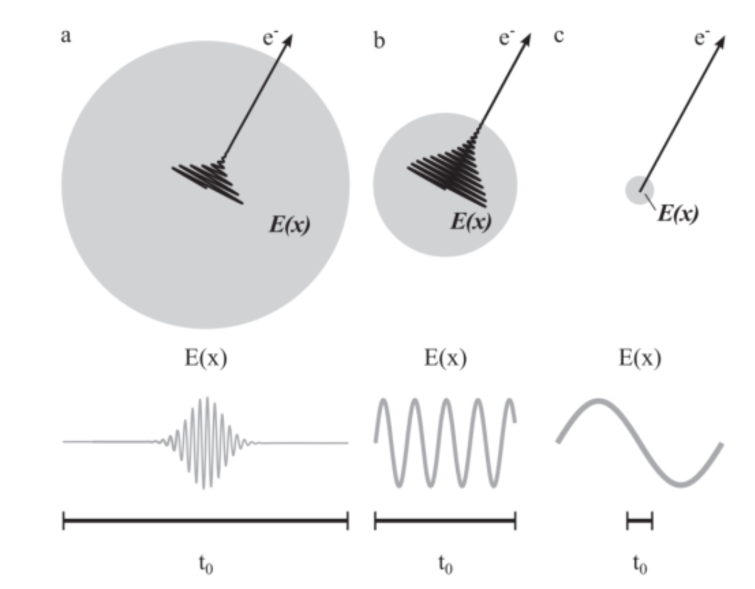}
\caption{Schematic illustration of the three attosecond streaking regimes (see text for details).
\label{Figure5MPQ}}
\end{figure}

As mentioned previously in Section IIE, attosecond streaking measurements can be used to trace attosecond electron dynamics in gas-phase samples and plain solid surfaces, and to fully characterize both the near-infrared (NIR) laser pulses (the streaking field) and the extreme ultraviolet (XUV) attosecond pulses. In attosecond streaking measurements on plasmonic nanostructures, the streaking field is replaced by a plasmonic field excited by (and typically enhanced with respect to) the incident NIR laser field, while the XUV acts as a probe by photoemitting an electron wavepacket that subsequently gets accelerated (streaked) in the plasmonic field~\cite{Stockman07}. In principle, similar information to standard streaking measurements can be obtained: the temporal structures of the streaking field and the XUV pulse, and information about attosecond electron dynamics taking place in the system. For plasmonic nanostructures, however, the situation is much more complex than in standard streaking measurements because the nanolocalized fields are spatially inhomogeneous~\cite{Stockman07}. The shift of the XUV photoemission is determined by the external field
\begin{equation}
\label{vf}
v_f(t_e)=v_0-\int_{t_e}^{\infty}\frac{e E(\mathbf{r},t)}{m} dt,
\end{equation}
where the field $E(\mathbf{r},t)$ has spatial and temporal dependence. Figure~\ref{Figure5MPQ} illustrates three different regimes of streaking in inhomogeneous fields as reported in~\cite{Stockman07,Skopalova11,Kelkensberg12}. In the ponderomotive limit the streaking field pulse duration $t_p$ is much shorter than the time it takes the electron to leave the near-field $t_0$ ($t_p\ll t_0$) and the electron does not experience spatial variation of the near-field (Fig.~\ref{Figure5MPQ}(a)). This corresponds to the case of conventional streaking in gas targets. Figure~\ref{Figure5MPQ}(c) illustrates the other, instantaneous limit when the electron leaves the localized field within a fraction of the optical cycle $T$ ($t_0\ll T$). This corresponds to quasi-electrostatic acceleration and the streaking field can be described by an electrostatic scalar potential. In contrast to conventional, ponderomotive streaking, in the instantaneous regime the electron streaking curve follows the electric field evolution. Finally, in the intermediate regime the electron transverses the field within several optical oscillations $t_0\approx T$ (Fig.~\ref{Figure5MPQ}(b)) and the streaking trace shows a phase-shift, which lies in-between the other two limits. Since the retrieval of the near-field in this case requires extensive modeling, the other two regimes are most desirable. 

\subsection{Attosecond streaking from an isolated nanosphere}

\begin{figure}
\centering
\includegraphics[width=\columnwidth]{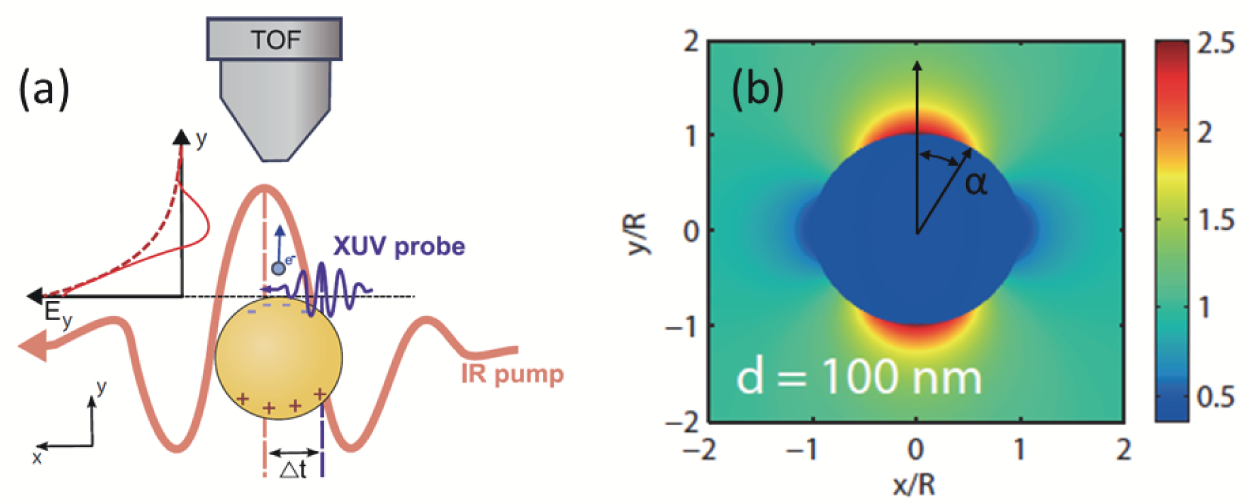}
\caption{Schematic of attosecond streaking on an isolated nanosphere (a). Amplitude of the field ($E_y$) distribution at an Au sphere of 100 nm diameter illuminated at 720 nm (b). The field is normalized to the incident field.
\label{Figure6MPQ}}
\end{figure}

Due to their simple shape and the possibility of an analytical description of the near-field of isolated nanospheres, they can be used as a reference system for tracing plasmonic excitations (see Section IIA). Isolated nanoparticles of well-defined size and shape can be produced by wet chemistry methods~\cite{Stoeber68,Sau04} and introduced into the interaction region by employing aerodynamic lenses~\cite{Zherebtsov11,Zherebtsov12} or optical trapping~\cite{Hansen05}. Figure~\ref{Figure6MPQ}(a) shows a schematic of a streaking experiment with isolated nanospheres~\cite{SuessmannPRB11}. The plasmonic oscillations are excited with a few-cycle NIR laser pulse and probed with photoemission induced by an attosecond XUV pulse. The electron emission is detected along the polarization direction with a time-of-flight (TOF) spectrometer. In the simulation an Au sphere of 100 nm diameter excited with a laser pulse of 5 fs duration (FWHM of electric field) centered at 720 nm with peak intensity of $1\times10^{13}$ W cm$^{-2}$ was considered~\cite{SuessmannPRB11}. 

The local field was calculated by finding the Mie solution at the central laser wavelength. The simulated near field in Fig.~\ref{Figure6MPQ}(b) exhibits symmetry relative to the polarization vector of the incident field, with the maximum field enhancement at the poles along the polarization vector. The non-resonant excitation leads to a maximum field enhancement factor of 2.5 on the sphere surface. The electric field quickly decays from the surface with a typical length scale of tenth of nanometers. 
For the photoemission step, the XUV pulse duration and bandwidth were taken as 250 as, and 7 eV, respectively. The central photon energy was 105 eV, giving a 100 eV initial electron energy. Electrons photoemitted by the XUV pulse were assumed to have initial velocity vectors parallel to the $y$-axis (along the TOF axis). The electron initial position is represented by the angle $\alpha$ in Fig.~\ref{Figure6MPQ}(b). The XUV pulse penetrates sufficiently deep into the nanoparticle for photoemission from the whole surface facing the TOF to be important. The relative emission from the back of the sphere was modeled using tabulated material data to calculate the XUV transmission through the sphere. To achieve good statistics approximately $1.5\times 10^5$ trajectories were initialized from the surface at each delay step.

\begin{figure}[h]
\centering
\includegraphics[width=\columnwidth]{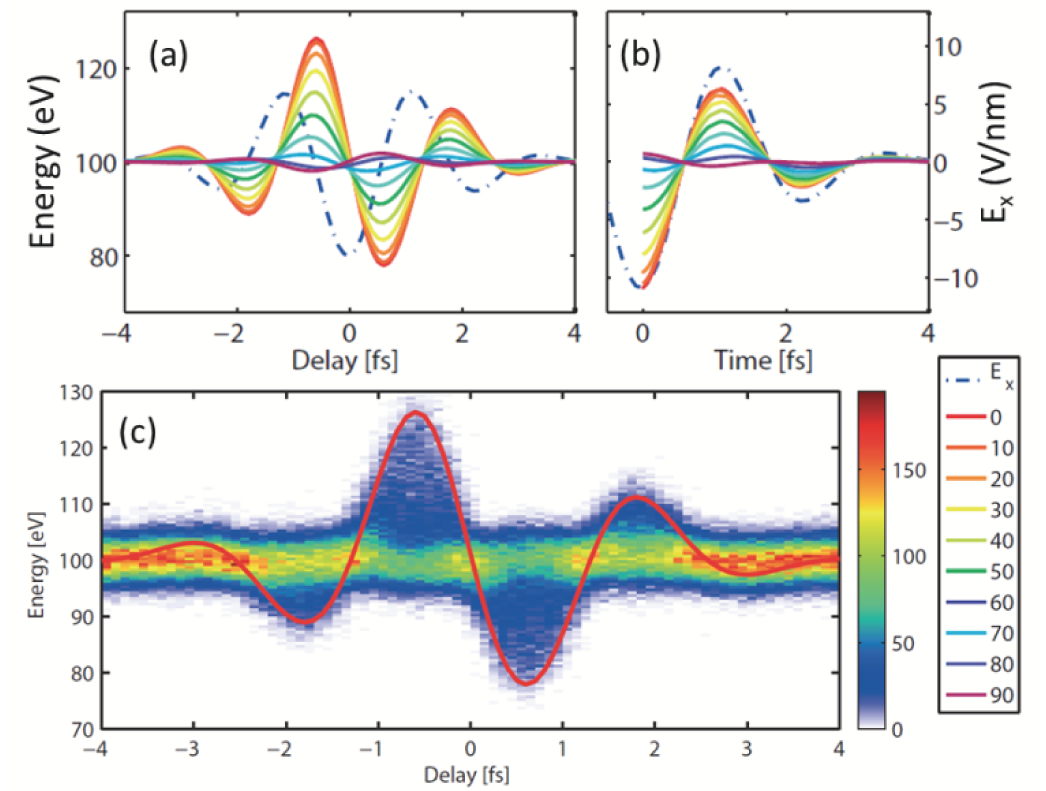}
\caption{(a) Simulated streaking waveform for electrons emitted at different positions on a sphere of 100 nm diameter. (b) Effective field for electrons emitted at time $t_e=0$ at the same positions as in (a). (c) Simulated streaking spectrogram. The red line indicates the streaking curve for electrons emitted at the particle pole. 
\label{Figure7MPQ}}
\end{figure}

Figure~\ref{Figure7MPQ}(a) shows streaking curves simulated for different electron initial positions. The plasmonic streaking field $E_x$ acting on the electrons emitted at $t_e=0$ is shown in Fig.~\ref{Figure7MPQ}(b), and the incident laser field is depicted by the blue dashed-dotted line. The electrons emitted at the poles show a streaking curve shifted in phase by $\sim \pi/2$ rad relative to the incident laser field, which is consistent with the ponderomotive picture of streaking. At larger values of $\alpha$, the streaking amplitude becomes smaller, but the phase of the streaking curve does not change significantly. For very large angles the phase shift relative to the laser field abruptly changes to approximately $\pi$ rad. This emission position dependence of the streaking traces and the fields accelerating the photoemitted electrons results from the dipolar character of the near field. 

The resulting streaking trace is shown in Fig.~\ref{Figure7MPQ}(c). Contributions from trajectories originating from different parts of the surface result in a blurred spectrogram in comparison to typical streaking measurements in an atomic gas. Trajectory analysis shows that the electrons emitted from the poles contribute to the largest energy shifts of the photoelectron spectra (red line in Fig.~\ref{Figure7MPQ}(c)). The case considered here is in the ponderomotive streaking regime, resulting in a simple phase shift of the plasmonic field with respect to the streaking trace (see Fig.~\ref{Figure5MPQ}(a)). Once this phase shift has been determined from theory, full characterization of the plasmonic field can be performed experimentally. An analytical solution of Eq.~(\ref{vf}) is generally not available, and in the intermediate streaking regime more complex streaking traces will be observed~\cite{Kelkensberg12,SuessmannPRB11}. Here, numerical simulations combined with appropriate feedback may be employed for the retrieval of the spatiotemporal evolution of the near fields.  

\subsection{Attosecond streaking from nanoantennas}

\begin{figure}[h]
\centering
\includegraphics[width=\columnwidth]{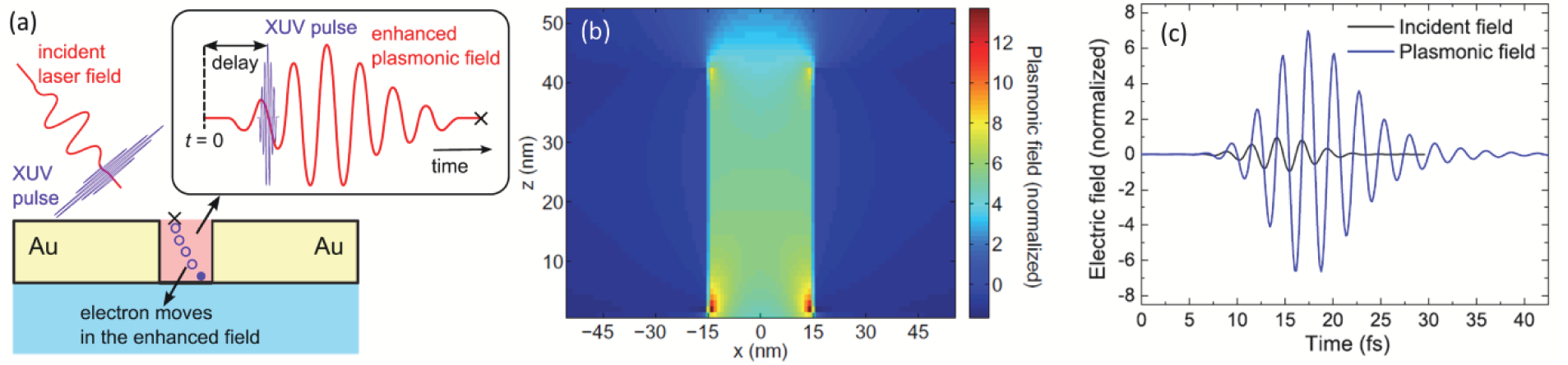}
\caption{(a) Schematic representation of the experiment. (b) Calculated plasmonic field $E_x (x,z)$ in the gap between the antennas at the time of the maximum of the plasmonic field. (c) Time evolution of the incident laser field (black line) and the plasmonic field response $E_x(t)$ at the point $x=10$ nm, $z=0$ nm. 
\label{Figure8MPQ}}
\end{figure}

We now turn our attention to streaking measurements on plasmonic nanostructures with more complex geometries. Plasmonic properties of surface based nanostructures have recently attracted attention due to their importance in applications ranging from chemical sensing~\cite{Liu11,Anker08} to the generation of XUV light~\cite{Kim08,Sivis13}. The possibility of tracing plasmonic fields of an array of Au nanoantennas with attosecond streaking has been studied numerically~\cite{Skopalova11}. Again, a few-cycle laser pulse excites the plasmonic field and a delayed attosecond pulse ionizes electrons that are then streaked in the plasmonic field (Fig.~\ref{Figure8MPQ}(a)). To calculate the time-dependent near-fields of the nanoantenna array three-dimensional finite-difference-time-domain (FDTD) simulations were performed for coupled antennas illuminated with a laser pulse of 5 fs duration polarized in $x$-direction~\cite{Skopalova11}. The dimensions and arrangement of the antenna elements were chosen such that the plasmon resonance of the nanostructure was centered at the carrier frequency of the incident laser pulse pulse (800 nm). The same time dependence for all points in space was assumed and the plasmonic field is presented as a decomposition of its spatial and temporal components $E(x,z,t)=E(x,z)E(t)$. This assumption was supported by the FDTD simulations and is needed for reconstruction of the electric field from the streaking process. The spatial distribution of the plasmonic field $E_x(x,z)$ exhibits maxima of the field enhancement near the corners of the gap (Fig.~\ref{Figure8MPQ}(b)). The time dependent evolution of the plasmonic field shows a resonance response with field oscillations lasting for more than 10 fs after the excitation pulse (Fig.~\ref{Figure8MPQ}(c)).

\begin{figure}[h]
\centering
\includegraphics[width=\columnwidth]{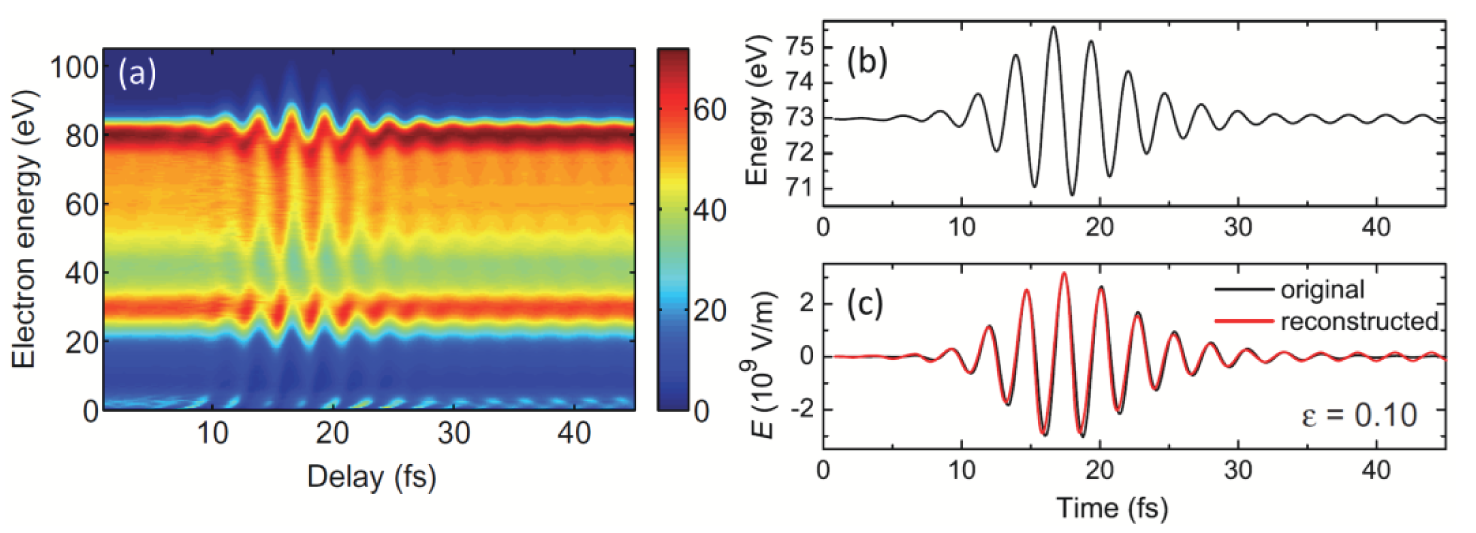}
\caption{(a) Simulated streaking spectrogram of Au nanoantenas. (b) Center of mass of the final electron energy as a function of the time delay. The spectra were integrated over the energy range 60-110 eV and the initial electron position was averaged over the $z$-direction. (c) Original field (black) and reconstructed field (red). The original field was normalized to have the same maximum as the reconstructed field.  
\label{Figure9MPQ}}
\end{figure}

The initial energy distribution of photoelectrons was simulated by convolution of 580 as Gaussian pulse at 90 eV photon energy with a spectrum obtained from narrow-line x-ray photoelectron spectroscopy measurements in Au and corrected for the energy-dependence of the ionization cross-section. Equation (\ref{vf}) was numerically solved for electrons emitted at different positions along the $x=15$ nm, $z=0-40$ nm surface, and the streaking spectrogram in Fig.~\ref{Figure9MPQ}(a) was obtained by averaging over these different initial positions. 

The spectrogram displays streaking of photoelectrons emitted from both the valence and 5p bands of Au, and resembles conventional streaking in gas indicating that the majority of electrons are streaked in the ponderomotive regime. To retrieve the plasmonic field evolution, the center-of-mass of the valence band was found as a function of time delay. The energy shift of the valence band (Fig.~\ref{Figure9MPQ}(b)) approximately follows the vector potential of the plasmonic field, allowing the electric field to be obtained by differentiating the center-of-mass curve as a function of the time delay. The reconstructed field is in close agreement with the original plasmonic field (Fig.~\ref{Figure9MPQ}(c)). It should be noted that the amplitude of the reconstructed waveform can be underestimated because of the finite XUV pulse duration. Similarly, for isolated rectangular nanoparticles it was found from simulations in~\cite{Borisov12} that the oscillations in the streaking spectrogram closely followed the plasmonic field.

Disentangling electrons from different emission positions indicates that not all of the electrons are streaked in the ponderomotive regime~\cite{Skopalova11}. In particular, high energy electrons emitted far from the substrate (i.e.~high initial $z$ position) and at short time-delays can escape fast enough to enter the intermediate streaking regime. This distorts the delay-energy relationship for these initial positions. To further understand the complex spatio-temporal structure of plasmonic near-fields, photoelectron emission microscopy (PEEM) setups aiming to combine nm spatial resolution with attosecond time resolution are currently being developed~\cite{Stockman07,Chew12,Mikkelsen09,Gong15}.

\subsection{Attosecond streaking at nanotapers}

While the numerous theoretical studies described above indicated that nano-localized fields can be characterized with attosecond precision using streaking, experimental implementation proved challenging. The linear XUV-induced photoemission process typically probes a much larger area than the nanoscale region of interest, and the streaking trace can be distorted because electrons emitted from different regions are streaked by different local fields. The absolute number of electrons emitted is also very low due to small sample sizes. However, a recent advance has been made in this area with the first streaking measurements performed on a nanostructure~\cite{Foerg2016}. By combining the measurements with a thorough analysis of the near-field spatial distribution and photoelectron trajectories, the authors were able to characterise the near-fields surrounding a gold nanotaper with attosecond precision. 

\begin{figure}[h]
\centering
\includegraphics[width=\columnwidth]{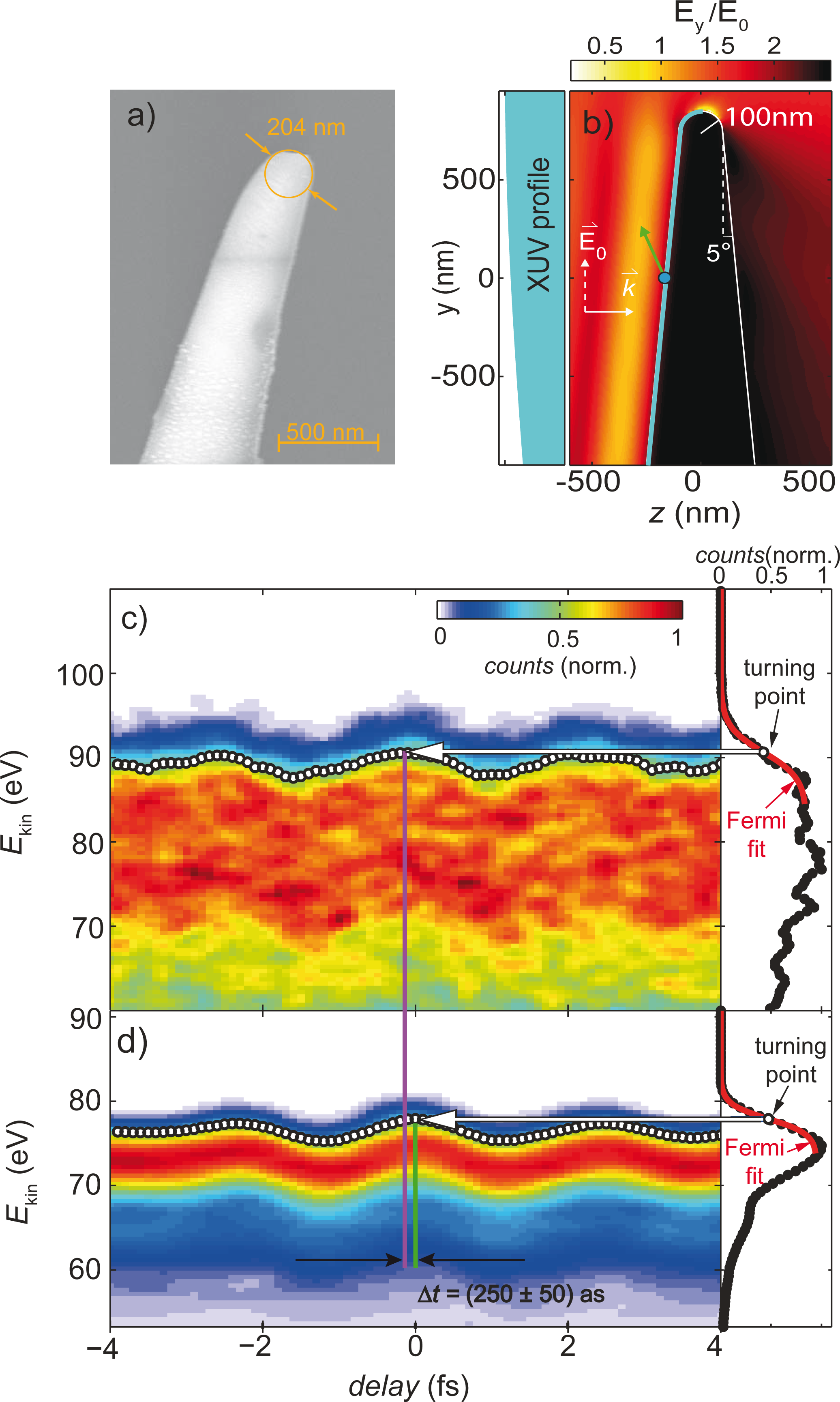}
\caption{(a) SEM image of nanotaper sample. (b) Normalized field strength of the field component parallel to the nanotaper axis, calculated using an FDTD method. The blue line shows the region of the sample illuminated by the XUV, and the spatial profile of the XUV focus is shown on the left. (c) Streaking measurement from the nanotaper sample. The energy shift of the streaking trace versus the time delay between the XUV and NIR pulses, shown by the white data points, was extracted by fitting a Fermi function (red) to the cut-off. (d) Reference streaking measurement in neon gas. The neon streaking trace is shifted in time by $\Delta t=(250\pm50)$ as relative to the nanotaper trace.   
\label{Figure1NanoTapers}}
\end{figure}

In the experiments, co-propagating 4.5 fs NIR laser pulses at 720 nm central wavelength, and isolated 220 as XUV pulses at 95 eV central energy, were generated and used to perform streaking measurements on a gold nanotaper. A scanning electron microscopy (SEM) image of the sample is shown in Fig.~\ref{Figure1NanoTapers}(b). The XUV focal spot (5 $\mu$m diameter) was centred on the tip apex (100 nm radius of curvature), although no significant XUV induced photoelectron signal was detected from the apex itself due to its small surface area. The XUV effectively probed the near fields surrounding the nanowire taper within a distance of 2.5 $\mu$m from the apex  and with a diameter tapering from 200 nm to 640 nm. The NIR polarization was aligned with the nanotaper axis.

Theoretical considerations indicated that the photoelectrons were streaked in the ponderomotive regime. The near-fields, calculated using an FDTD method, are shown in Fig. Fig.~\ref{Figure1NanoTapers}(b). The near-fields in the probed region have a high degree of spatial homogeneity in amplitude and phase, and are shifted in phase by 0.8 rad (corresponding to a temporal shift of 300 as) with respect to the incident NIR pulse. 

\begin{figure}[h]
\centering
\includegraphics[width=\columnwidth]{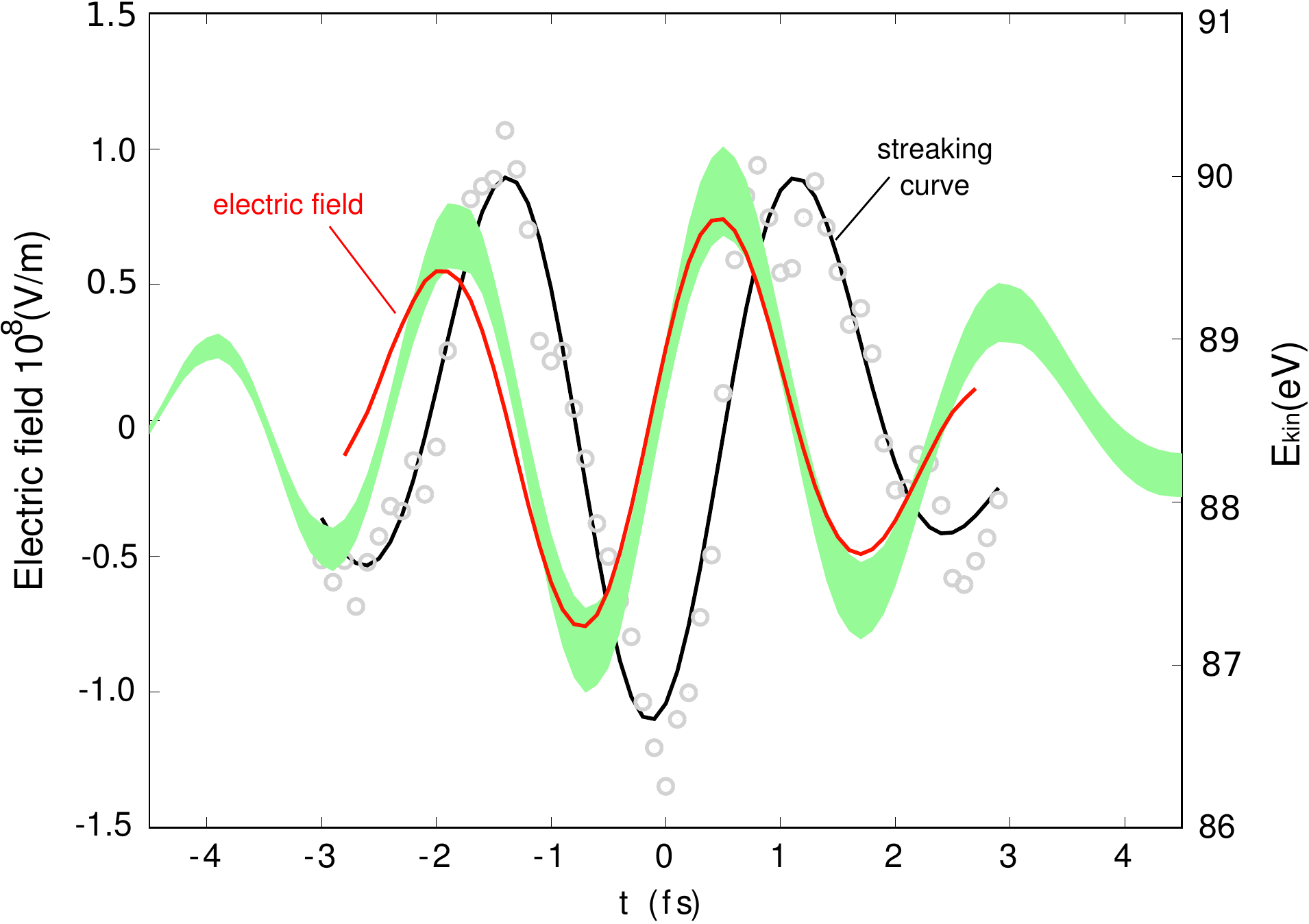}
\caption{Electric near-field (red) around the nanotaper, extracted from streaking measurements. The energy shift of the streaking measurement (data points), the Fourier filtered shift (streaking curve, black), and the calculated near-field (green shaded area) are also shown. 
\label{Figure2NanoTapers}}
\end{figure}

The experimental streaking trace from the nanotaper is shown in Fig.~\ref{Figure1NanoTapers}(c). The gas-phase streaking measurement in Fig.~\ref{Figure1NanoTapers}(d) gives the phase of the incident NIR pulse as a reference. The nanotaper streaking trace is shifted with respect to the gas phase streaking measurement by $\Delta t=(250\pm50)$ as. The measured shift was free from any significant contribution from photoemission time delays because the NIR field polarisation was parallel to the sample surface, resulting in a continuous electric field across the surface. The measured shift is in agreement with the theoretical value from Monte-Carlo simulations of photoelectron trajectories, confirming that the measurements successfully probed the near-fields around the nanotaper. The electric near-field retrieved from streaking measurements is shown in Fig.~\ref{Figure2NanoTapers}, and is in close agreement with the field expected from calculations (also shown in the same figure).

The experiments open the door to using the same approach to measure local near fields and attosecond plasmon dynamics in more complex nanostructures, such as ultrafast optoelectronic components. Future characterisation measurements of the electric field around the tip apex should furthermore yield a richer understanding of the physics discussed in Section IV, involving electron acceleration at nanoscale needle tips.

\section{Extreme-ultraviolet light generation in plasmonic nanofields}

\begin{figure}[h]
\centering
\includegraphics[width=\columnwidth]{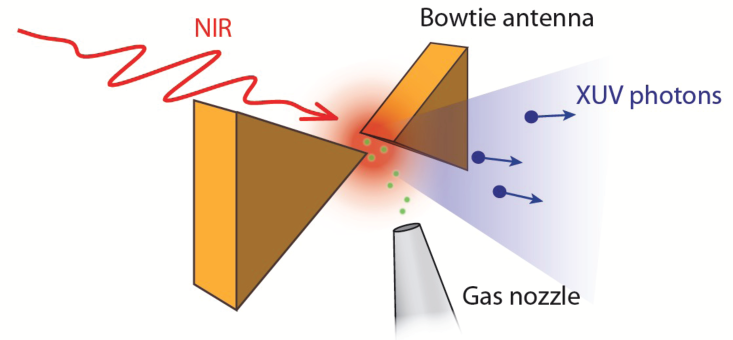}
\caption{Schematic representation of a bow-tie nanostructure for XUV generation. The structure is illuminated with a few cycle NIR laser field and gas is injected into the antenna gap.   
\label{Figure10MPQ}}
\end{figure}

The extreme local field enhancements that can be achieved by concentrating light into nanoscale volumes using plasmonic nanostructures have attracted significant interest from the ultrafast physics community. One of the applications that has generated the most excitement is the possibility to generate XUV light at high (MHz) repetition rates without need for an enhancement cavity. This work was initiated by~\cite{Kim08}, where pulses from a femtosecond oscillator (75 MHz repetition rate) were focused onto an array of bow-tie Au nanoantennas on a sapphire substrate and surrounded by argon gas. The bow-tie structures acted as resonant antennas concentrating the optical energy in the gaps between adjacent elements (see Fig.~\ref{Figure10MPQ}). The estimated intensity enhancement of more than 20 dB was sufficient enough to produce XUV radiation in the argon with wavelengths down to 47 nm.

The work of~\cite{Kim08} triggered a number of further experimental~\cite{Sivis13,Park11,Kovacev13NJP,Sivis12A,Kim12Reply,Park13} and theoretical~\cite{Marcelo12OE, Marcelo12A, Marcelo12AA, Marcelo12JMO, Jose13, Marcelo13AR, Marcelo13AP, Marcelo14,Husakou11A,Yavuz12,Stebbings11} efforts in a similar direction. Despite initial success in the observation of XUV light from bow-tie nanostructures the origin of the observed radiation remained debated. The study of~\cite{Sivis13} significantly deepened the understanding of the mechanisms responsible for the observed plasmon enhanced light emission. The authors compared the emission characteristics using nanostructures illuminated with low energy laser pulses from an oscillator to those obtained in conventional gas target with amplified high energy pulses. 

\begin{figure}[h]
\centering
\includegraphics[width=\columnwidth]{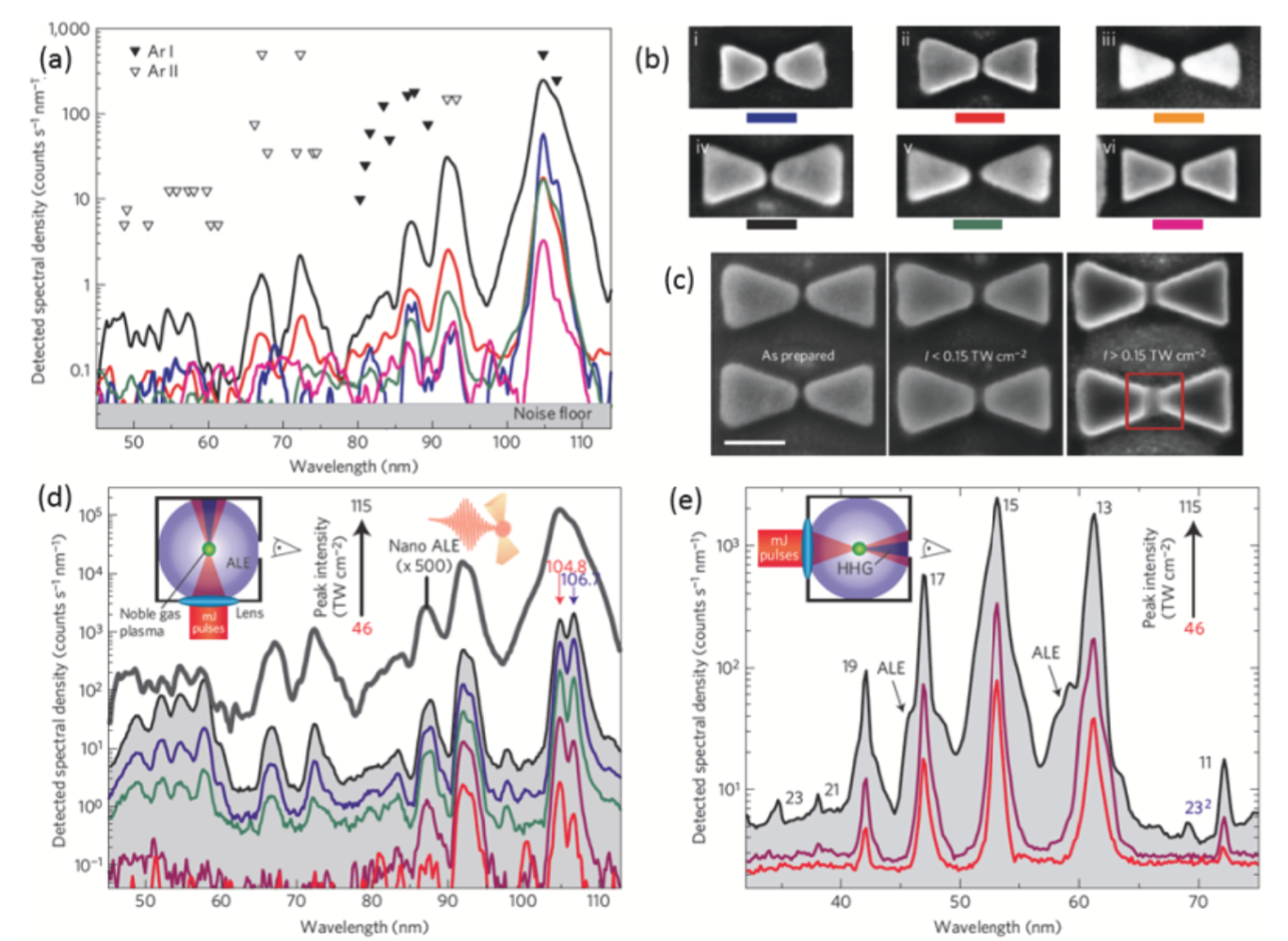}
\caption{(a) XUV spectra measured in bow-tie nanostructures exposed to Ar gas. Triangles indicate expected ALE transitions for neutral (filled) and singly ionized (open) Ar atoms. (b) SEM images of the bow-tie antennas used in the measurements presented in (a). (c) SEM images of nanoantennas {\it iv} after the preparation, after exposure for several hours to laser intensity up to 0.15 TW cm$^{-2}$, and after exposure for a few minutes to laser intensity up to 0.3 TW cm$^{-2}$ (from left to right respectively). (d) Intensity dependent spectra measured in Ar gas in the direction perpendicular to the laser beam propagation. For comparison a spectrum measured in bow-tie nanostructures is presented as a thick grey line. (e) Intensity dependent spectra measured along the laser beam propagation direction.   
\label{Figure11MPQ}}
\end{figure}

Spectra measured from different bow-tie nanoantenna samples are presented in Fig.~\ref{Figure11MPQ}(a). The authors identify the most pronounced features as atomic line emission (ALE) from neutral and singly ionized Ar atoms. The incoherent nature of these spectra is confirmed by the close agreement with ALE spectra measured in Ar gas illuminated with amplified pulses (Fig.~\ref{Figure11MPQ}(d)), detected in the direction perpendicular to the laser propagation. In contrast, the emission from the gas target in the laser propagation direction clearly shows high-order harmonic radiation (Fig.~\ref{Figure11MPQ}(e)).

Intensity dependent measurements on the nanostructures indicate local field intensities up to and beyond the damage threshold of the material. These intensities would in principle be sufficient for coherent high-harmonic generation (HHG) in Ar gas (Fig.~\ref{Figure11MPQ}(c)). The lack of high-harmonic emission in the measured spectra thus indicates that although the local intensities are clearly above the threshold for HHG, the small nanostructure generation volumes are insufficient for the coherent build-up of any noticeable HHG signal. A rough estimate using the actual experimental conditions indicates that the expected HHG signal from the nanostructure target is about $6\times10^{-3}$ smaller than the ALE.

Later work employed three dimensional tapered waveguides for XUV generation by adiabatically nanofocused SPPs~\cite{Park11}, where NIR pulses from a femtosecond oscillator were focused on the inlet of the waveguide with an intensity of $\sim10^{11}$ W cm$^{-2}$. These pulses excite an SPP wave that propagates inside the waveguide towards the exit. The parameters of the waveguide were optimized using FDTD simulations and a peak intensity enhancement factor of more than 20 dB relative to the incident field was obtained in a near cylindrical volume of diameter 240 nm and length 450 nm near the exit aperture. This volume is about three orders of magnitude larger than the generation volume (the volume containing an intensity enhancement of $>20$ dB) of a single bow-tie element used in previous work of~\cite{Kim08}. 

By back-filling the waveguide with Xe gas, XUV generation up to 70 eV photon energy was achieved~\cite{Park11} Compared to previous studies using bow-tie nanoantennas the three dimensional waveguide displays more than an order of magnitude higher XUV generation efficiency. In addition the waveguide fabricated on a cantilever microstructure is much less susceptible to thermal and optical damage. The origin of the observed radiation is, however, again disputed. Experimental investigations in~\cite{Sivis13PRL} indicate that while a sufficient intensity for HHG is achieved at the focus of the waveguide, the length of the guide is insufficient for a significant buildup of the signal. High-harmonic generation at MHz repetition rates in enhanced plasmonic fields remains an attractive prospect. However, efficient generation will require substantially higher gas pressures and larger interaction volumes. Meanwhile, the incoherent enhanced ALE that has been successfully generated in plasmonic near-fields could nevertheless find applications in areas such as near field imaging.

% Theory Section

\section{Theoretical approaches}

In the next subsections we describe the theoretical approaches we have developed to tackle strong field processes driven by spatially inhomogeneous laser fields. We put particular emphasis on the HHG and ATI, but we include at the end an incipient attempt to treat multielectronic phenomena. 

\subsection{HHG driven by spatially inhomogeneous fields}
Field-enhanced high-order-harmonic generation (HHG) using plasmonics fields, generated starting from engineered nanostructures or nanoparticles, requires no extra amplification stages due to the fact that, by exploiting surface plasmon resonances, the input driving electric field can be enhanced by more than 20 dB (corresponding to an increase in the intensity of several orders of magnitude). As a consequence of this enhancement, the threshold laser intensity for HHG generation in noble gases is largely exceeded and the pulse repetition rate remains unaltered. In addition, the high-harmonics radiation generated from each nanosystem acts as a pointlike source, enabling a high collimation or focusing of this coherent radiation by means of (constructive) interference. This fact opens a wide range of possibilities to spatially arrange nanostructures to enhance or shape the spectral and spatial properties of the harmonic radiation in numerous ways~\cite{Kim08,Park11,Kovacev13NJP}.

Due to the nanometric size of the so-called plasmonic 'hot spots', i.e.~the spatial region where the electric field reaches its highest intensity, one of the main theoretical assumptions, namely the spatial homogeneity of the driven electric field, should be removed (see Section IB). Consequently, both the analytical and numerical approaches to study laser-matter processes in atoms and molecules, in particular HHG, need to be modified to treat adequately this different scenario and allow now for a spatial dependence in the laser electric field. Several authors have addressed this problem recently~\cite{AlexisVG,Yavuz15,Husakou11A,Husakou11OE,Marcelo12OE,Marcelo12A,Marcelo12AA,Marcelo12AAA,Marcelo12JMO,Yavuz12,Jose13, Yavuz13,Marcelo13A,Marcelo13AR,Marcelo13AP,Marcelo13AA,Marcelo13LPL,Marcelo14,Marcelo14EPJD, Marcelo15CPC,Marcelo15,Husakou14A,Ebadi14A,Fetic12,Luo13,Feng13,Wang13,Lu13A,He13,Zhang13,Luo13JOSAB, Cao14,Cao14a,Feng15,Yu15}. As we will show below, this new characteristic affects considerably the electron dynamics and this is reflected on the observables, in the case of this subsection the HHG spectra. 

\subsection{Quantum approaches}

The dynamics of a single active atomic electron in a strong laser field takes place along the polarization direction of the field, when linearly polarized laser pulses are employed. It is then justifiable to model the HHG in a 1D spatial dimension by solving the time dependent Schr\"odinger equation (1D-TDSE)~\cite{Marcelo12A}:
\begin{eqnarray}
\label{tdse1d}
\mathrm{i} \frac{\partial \Psi(x,t)}{\partial t}&=&\mathcal{H}(t)\Psi(x,t) \\
&=&\left[-\frac{1}{2}\frac{\partial^{2}}{\partial x^{2}}+V_{\rm{a}}(x)+V_{\rm{l}}(x,t)\right]\Psi(x,t), \nonumber
\end{eqnarray}
where in order to model an atom in 1D, it is common to use soft core potentials, which are of the form:
\begin{equation}
V_{\rm{a}}(x)=-\frac{1}{\sqrt{x^2+b^2}},
\end{equation} 
where the parameter $b$ allows us to modify the ionization potential $I_p$ of the ground state, fixing it as close as possible to the value of the atom under consideration.
We consider the field to be linearly polarized along the $x$-axis and modify the interaction term $V_{\rm{l}}(x,t)$ in order to treat spatially
nonhomogeneous fields, while maintaining the dipole character.
Consequently we write
\begin{eqnarray}  
\label{vlaser1d}
V_{\rm{l}}(x,t)&=&-E(x,t)\,x
\end{eqnarray}
where $E(x,t)$ is the laser electric field defined as
\begin{equation}  
\label{electric}
E(x,t)=E_0\,f(t)\, (1+\varepsilon h(x))\,\sin(\omega t+\phi).
\end{equation}
In Eq.~(\ref{electric}), $E_0$, $\omega$ and $\phi$ are the peak
amplitude, the frequency of the laser pulse and the CEP, respectively. We refer to sin(cos)-like laser pulses where $\phi=0$ ($\phi=\pi/2$). The pulse envelope is given by $f(t)$ and $\varepsilon$ is a small parameter that characterizes the inhomogeneity strength. The function $h(x)$
represents the functional form of the spatial nonhomogeneous field and, in
principle, could take any form and be supported by the numerical algorithm (for details see e.g.~\cite{Marcelo12A,Marcelo12OE}). Most of the approaches use the simplest form for $h(x)$, i.e.~the linear term: $h(x)=x$. This choice is motivated by previous investigations~\cite{Husakou11A,Marcelo12A,Yavuz12}, but nothing prevents to use more general functional forms for $h(x)$.\footnote{The actual spatial dependence of the enhanced near-field in the surrounding of a metal nanostructure can be obtained by solving the Maxwell equations incorporating both the geometry and material properties of the nanosystem under study and the input laser pulse characteristics (see e.g.~\cite{Marcelo12OE}). The electric field retrieved numerically is then approximated using a power series  $h(x) =\sum_{i=1}^{N}b_{i}x^{i}$, where the coefficients $b_i$ are obtained by fitting the real electric field that results from a finite element simulation. Furthermore, in the region relevant for the strong field physics and electron dynamics and in the range of the parameters we are considering, the electric field can be indeed approximated by its linear dependence.}

The 1D-TDSE can be solved numerically by using the Crank-Nicolson scheme in order to obtain the time propagated electronic wavefunction $\Psi(x,t)$. Once $\Psi(x,t)$ is found, we can compute the harmonic spectrum by Fourier transforming the dipole acceleration of the active electron. One of the main advantages of the 1D-TDSE is that we are able to include any functional form for the spatial variation of the plasmonic field. For instance, we have implemented linear~\cite{Marcelo12A} and real (parabolic) plasmonic fields~\cite{Marcelo12OE}, as well as near-fields with exponential decay (evanescent fields)~\cite{Marcelo13AR}. 

An extension of the above described approach is to solve the TDSE in its full dimensionality and to include in the laser-electron potential the spatial variation of the laser electric field. For only one active electron we need to deal with 3 spatial dimensions and, due to the cylindrical symmetry of the problem, we are able to separate the electronic wavefunction in spherical harmonics, $Y_l^m$ and consider only terms with $m=0$ (see below). 

The 3D-TDSE in the length gauge can be written:  
\begin{eqnarray}
\label{tdse3d}
\nonumber
i\frac{\partial \Psi({\bf{r}},t)}{\partial t}&=&H\Psi({\bf{r}},t)\\
&=&\left [-\frac{\nabla^{2}}{2}+V_{SAE}({\bf{r}})+V_l({{\bf{r}},t})\right ]\Psi({\bf{r}},t),
\end{eqnarray}
where $V_{SAE}({\bf{r}})$ is the atomic potential in the single active electron (SAE) approximation and $V_l({{\bf{r}},t})$ the laser-electron coupling (see below).
The time-dependent electronic wave function $\Psi({\bf{r}},t)$, can be expanded in terms of spherical harmonics:
\begin{eqnarray}
\label{spherical}
\nonumber
\Psi({\bf{r}},t)&=&\Psi({r,\theta, \phi},t)\\
&\approx&\sum_{l=0}^{L-1}\sum_{m=-l}^{l}\frac{\Phi_{lm}(r,t)}{r}Y_{l}^{m}(\theta,\phi)
\end{eqnarray}
where the number of partial waves depends on each specific case. Here, in order to assure the numerical convergence, we have used up to $L\approx250$ in the most extreme case ($I\sim 5\times10^{14}$ W/cm$^{2}$). In addition, due to the fact that the plasmonic field is linearly polarized, the magnetic quantum number is conserved and consequently in the following we can consider only $m=0$ in Eq.~(\ref{spherical}). This property considerably reduces the complexity of the problem. In here, we consider $z$ as a polarization axis and we take into account that the spatial variation of the electric field is linear with respect to the position. As a result, the coupling
$V_{l}(\mathbf{r},t)$ between the atomic electron and the
electromagnetic radiation reads
\begin{equation}
\label{vlaserati} V_{l}(\mathbf{r},t)=\int ^\mathbf{r}
d\mathbf{r'}\cdot\mathbf{E}(\mathbf{r'},t)=E_0z(1+\varepsilon
z)f(t)\sin(\omega t+\phi)
\end{equation}
where $E_0$,  $\omega$ and $\phi$ are the laser electric field
amplitude, the central frequency and the CEP, respectively. As in previous investigations, the parameter $\varepsilon$ defines
the `strength'  of the inhomogeneity and has units of inverse
length (see also~\cite{Husakou11A,Yavuz12,Marcelo12A}). For modeling
short laser pulses in Eq.~(\ref{vlaserati}), we use a sin-squared
envelope $f(t)$ of the form $
f(t)=\sin^{2}\left(\frac{\omega t}{2 n_p}\right)$,
where $n_p$ is the total number of optical cycles. As a result,
the total duration of the laser pulse will be $T_p=n_p \tau_L$ where
$\tau_L=2\pi/\omega$ is the laser period. We focus our analysis on a hydrogen atom, i.e.~$V_{SAE}({\bf{r}})=-1/r$ in Eq.~(\ref{tdse3d}), and we also assume that before
switch on the laser ($t=-\infty$) the target atom is
in its ground state ($1s$), whose analytic form can be found in a
standard textbook. Within the SAE
approximation, however, our numerical scheme is tunable to treat
any complex atom by choosing the adequate effective (Hartree-Fock)
potential $V_{SAE}({\bf{r}})$, and finding the ground state by the means of numerical
diagonalization.

Next, we will show how the inhomogeneity modifies the equations which model 
the laser-electron coupling. Inserting Eq.~(\ref{spherical}) into Eq.~(\ref{tdse3d}) and considering that,
\begin{equation}
\cos \theta Y_{l}^{0}=c_{l-1}Y_{l-1}^{0}+c_{l}Y_{l+1}^{0}
\end{equation}
and
\begin{equation}
\label{square}
\cos^{2} \theta Y_{l}^{0}=c_{l-2}c_{l-1}Y_{l-1}^{0}+(c_{l-1}^{2}+c_{l}^{2})Y_{l}^{0}+c_{l}c_{l+1}Y_{l+2}^{0},
\end{equation}
where
\begin{equation}
c_{l}=\sqrt{\frac{(l+1)^2}{(2l+1)(2l+3)}},
\end{equation}
we obtain a set of coupled differential equations for each of the radial functions $\Phi_{l}(r,t)$:
\begin{eqnarray}
\label{diag}
i\frac{\partial\Phi_{l}}{\partial{t}}=\left [-\frac{1}{2}\frac{\partial^{2}}{\partial r^{2}}+\frac{l(l+1)}{2r^2}-\frac{1}{2} \right ]\Phi_{l}\nonumber\\
+\varepsilon r^{2}E(t)\left(c_{l}^{2}+c_{l-1}^{2}\right)\Phi_{l}\nonumber\\
+r E(t)\left(c_{l-1}\Phi_{l-1}+c_{l}\Phi_{l+1}\right)\nonumber\\
+\varepsilon r^{2}E(t)\left(c_{l-2}c_{l-1}\Phi_{l-2}+c_{l}c_{l+1}\Phi_{l+2}\right).
\end{eqnarray}
Equation (\ref{diag}) is solved using the Crank-Nicolson algorithm considering the additional term, i.e.~Eq.~(\ref{square}) due to the spatial inhomogeneity. As can be observed, the degree of complexity will increase substantially when a more complex functional form for the spatial inhomogeneous laser electric field is used. For instance, the incorporation of only a linear term couples the angular momenta $l,l\pm1,l\pm2$, instead of $l,l\pm1$, as in the case of conventional (spatial homogeneous) laser fields.

%Typically several hundreds of angular momenta $l$ should to be considered and we could recognize the time evolution of each of them as a 1D problem. We use a Crank-Nicolson method implemented on a splitting of the time-evolution operator that preserves the norm of the wave function for the time propagation, similar to the 1D-TDSE case. 

We have also made studies on helium because a majority of experiments in HHG are carried out in noble gases. Nonetheless, other atoms could be easily implemented by choosing the appropriate atomic model potential $V_{SAE}({\bf{r}})$. After time propagation of the electronic wavefunction, the HHG spectra can be computed in an analogous way as in the case of the 1D-TDSE. Due to the complexity of the problem, only simulations with nonhomogeneous fields with linear spatial variations along the laser polarization in the 3D-TDSE have been studied. This, however, is enough to confirm that even a small spatial inhomogeneity significantly modifies the HHG spectra (for details see~\cite{Jose13}). 

\subsection{Semiclassical approach}

An independent approach to compute high-harmonic spectra for atoms in intense laser pulses is the Strong Field Approximation (SFA) or Lewenstein model~\cite{Lewenstein94}.
The main ingredient of this approach is the evaluation of the time-dependent dipole moment $\mathbf{d}(t)$. Within the single active electron (SAE) approximation, it can be calculated starting from the ionization and recombination transition matrices combined with the classical action of the laser-ionized electron moving in the laser field. The SFA approximation has a direct interpretation in terms of the so-called three-step or simple man's model~\cite{Lewenstein94,corkum93} (see Section IA). 

Implicitly the Lewenstein model deals with spatially homogeneous electric and vector potential fields, i.e.~fields that do not experience variations in the region where the electron dynamics takes place. In order to consider spatial nonhomogeneous fields, the SFA approach needs to be modified accordingly, i.e.~the ionization and recombination transition matrices, joint with the classical action, now should take into account this new feature of the laser electric and vector potential fields (for details see~\cite{Marcelo12A,Marcelo13AA}).

\subsection{Classical framework}

Important information such as the HHG cutoff and the properties of the electron trajectories moving in the oscillatory laser electric field, can be obtained solving the classical Newton-Lorentz equation for an electron moving in a linearly polarized electric field. Specifically, we find the numerical solution of 
\begin{equation}
\label{newton}
\ddot{x}(t)=-\nabla_x V_{\rm{l}}(x,t),
\end{equation}
where $V_{\rm{l}}(x,t)$ is defined in Eq.~(\ref{vlaser1d}) with the laser electric field linearly polarized in the $x$ axis. For fixed values of ionization times $t_i$, it is possible to obtain the classical trajectories and to numerically calculate the times $t_r$ for which the electron recollides with the parent ion. In addition, once the ionization time $t_i$ is fixed, the full electron trajectory is completely determined (for more details about the classical model see~\cite{Marcelo15CPC}).

The following conditions are commonly set (the resulting model is also known as the simple man's model): i) the electron starts with zero velocity at the origin at time $t=t_i$, i.e., $x(t_i)=0$ and $\dot{x}(t_i)=0$; (ii) when the laser electric field reverses its direction, the electron returns to its initial position, i.e., recombines with the parent ion, at a later time, $t=t_r$, i.e. $x(t_r)=0$. $t_i$ and $t_r$ are known as ionization and recombination times, respectively. The electron kinetic energy at $t_r$ can be obtained from the usual formula $E_k(t_r)=\dot{x}(t_r)^2/2$, and, finding the value of $t_r$ (as a function of $t_i$ ) that maximizes this energy, we find that the HHG cutoff is given by $n_c\omega_0=3.17 U_p+I_p$, where $n_c$ is the harmonic order at the cutoff, $\omega_0$ is the laser frequency, $U_p$ is the ponderomotive energy and $I_p$ is the ionization potential of the atom or molecule under consideration. It is worth mentioning that the HHG cutoff will be extended when spatially inhomogeneous fields are employed. 

%We have solved numerically the Newton-Lorentz equation for an electron moving in a linearly polarized (in the $x$ axis) electric field with the same parameters used in the quantum models in order to compare the outcomes of both. Specifically, we find the numerical solution of 
%\begin{equation}
%\label{newton}
%\ddot{x}(t)=-\nabla_x V_{\rm{l}}(x,t),
%\end{equation}
%where $V_{\rm{l}}(x,t)$ is defined in Eq.~(\ref{vlaser1d}). For fixed values of ionization times $t_i$, it is possible to obtain the classical trajectories and to numerically calculate the times $t_r$ for which the electron recollides with the parent ion. In addition, once the ionization time $t_i$ is fixed, the full electron trajectory is completely determined (for more details about the classical model see~\cite{Marcelo15CPC}).

%%% Tomas Section

\subsection{Classical trajectory Monte Carlo (CTMC)}

In order to achieve quantitative accuracy for realistic systems, the 
classical framework may be coupled to accurate near-field, ionization, 
and scattering models into the classical trajectory Monte-Carlo (CTMC) 
scheme. Such scheme has several advantages in comparison with quantum 
simulations. First of all, even though the solution of the 3-D 
Schrödinger equation is possible in the SAE approximation under 
simplifying assumptions (see Sec.~VIIB above), a detailed description of 
complex geometry coupled with a realistic near-field is still out of 
reach for purely quantum methods. In addition, the (bulk) scattering and 
multi-electron effects complicate the quantum treatment to the extent 
that they are neglected in virtually every quantum calculation. 

%As detailed above, TDSE simulations have clarified a number of important
%phenomena in strong field ionization of nano-structures. Nevertheless,
%simulations of photoionization of nano-devices often rely on the classical
%trajectory Monte Carlo (CTMC) simulations for several reasons. First
%of all, even though the solution of the 3-D Schr\"odinger equation
%(corresponding to the single active electron approximation) is possible
%under simplifying assumptions (see Secs.~VIIB and IXB), a detailed
%description coupled with the realistic near-field is still out of
%reach for purely quantum methods. In addition, the scattering and
%multi-electron effects are often important which further complicate
%the quantum treatment. 

Similarly to the simple man's model of HHG, a CTMC simulation starts
with the ionization of electrons which is typically described stochastically
using methods based on a simplified quantum treatment, most often
the Fowler-Nordheim theory~\cite{Fowler1928} which is closely related
to the ADK theory of ionization of atoms~\cite{ADK1986}. Multiphoton
effects may be accounted for within the framework of Fowler-Norheim
theory by considering the response of the electron distribution function
of the nano-device to the laser field \cite{Yanagisawa2009,Yanagisawa2011,Yanagisawa2014}
or with more refined theoretical frameworks~\cite{Yalunin2011}.

After the ionization, the electrons are propagated using classical
equations of motion in the near field. For complex geometries, the
near field may be obtained in the time domain, e.g., with FDTD methods
or semi-analytically in the frequency domain, e.g., with the multiple
multipole programs (MMP)~\cite{Hafner1999}. For nanospheres, simpler
analytic Mie theory may be employed. For details see Sec. IIA. 

For low intensities or when only qualitative results are sought, the
electron-electron repulsion (space charge) may be neglected~\cite{Kruger11,Wachter12,Herink12,Dombi2013}.
For higher intensities and accuracy, the space charge effects may
be treated explicitly~\cite{Yanagisawa2014,Piglosiewicz2014} or
using the mean-field approximation~\cite{Suessmann15}. While the
mean-field approximation scales linearly with the number of electrons
ionized $N_{\text{el}}$, the explicit treatment leads to $N_{\text{el}}^{2}$
dependence. This unfavorable scaling may be alleviated without significantly
compromising the accuracy by using methods like fast multipole method
(FMM) or Barnes-Hut tree based methods with $N\log N$ scaling \cite{Winkel2012,Arnold2013,Bolten}.
In addition, the response of the nano-device on the ionized electron
cloud should be taken into account. Simple analytic formulas may be
used for planar and spherical geometries~\cite{Yanagisawa2014}.
For complex geometries, the electrostatic problem may be solved 
numerically~\cite{Zherebtsov11}. 

The description of the re-collision of an electron with a nano-device
ranges from a simple surface reflection (using several approximations)
\cite{Kruger11,Wachter12,Park13,Dombi2013} through approaches
relying on several empirical and fitted parameters~\cite{Yanagisawa2014}
to a propagation of electrons inside the nano-device using the Langevin
dynamics with stochastic events representing elastic and inelastic
scattering~\cite{Suessmann15,lemell_simulation_2009}. The probability $P$
of a particular scattering event in the time interval $(t,t+dt)$
is given by $P=v\cdot dt/\lambda_{m}$, where $v$ is the electron
velocity and $\lambda_{m}$ is the mean-free path. For elastic scattering,
mean free paths and scattering differential cross sections (DCSs)
may be obtained from quantum or DFT calculations for an electron interacting
with an isolated atom and combined with, e.g.~muffin-tin approximation
to model a solid state~\cite{Salvat2005}. The inelastic scattering
can be described as an interaction of the electron with a dielectric
medium defined by the complex\textendash wave vector $q$ and frequency
$\omega$-dependent\textendash dielectric function $\epsilon\left(q,\omega\right)$
which may be computed using the electron gas model~\cite{Lindhard1954,Mermin1970,Iafrate1980},
extended from experimentally measured optical energy loss function
via Drude models~\cite{Toekesi2001,Solleder2007,Da2014} or calculated
ab initio (typically employing TD-DFT). Alternatively, empirical formulas
may be used to describe some aspects of inelastic scattering~\cite{Lotz1967,Fernandez-Varea1993}.

\section{Selected results}
In the following sub-sections we present a brief summary of the results reported in several recent published works. In these articles, different noble gases (He, Ar and Xe) are used as atomic targets located in the vicinity of metal nanotips and nanoparticles and the HHG generated by them were studied and characterized. In addition, we include here predictions for the generation of coherent harmonic radiation directly from the metal surface of a nanotip.

\subsection{Spatially (linear) nonhomogeneous fields and electron confinement}
\begin{figure} [h]
\centering
\includegraphics[width=\columnwidth]{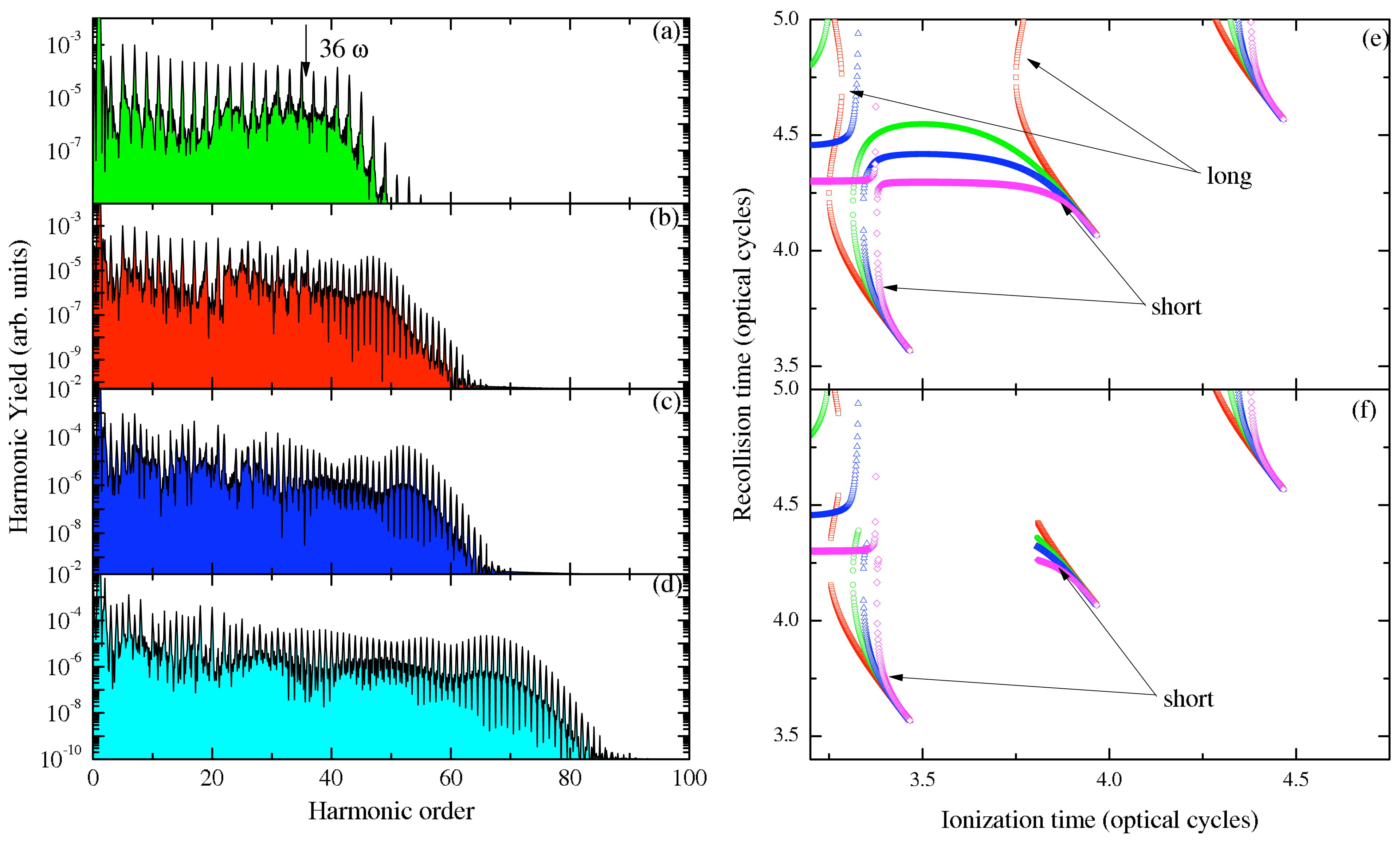}
\caption{HHG spectra for a model atom with a ground-state energy, $I_p=-0.67$ a.u. obtained using the 1D-TDSE approach. The laser parameters are $I=2\times10^{14}$ W$\cdot$cm$^{-2}$ and $\lambda=800$ nm. We have used a trapezoidal shaped pulse with two optical cycles turn on and turn off, and a plateau with six optical cycles, 10 optical cycles in total, i.e.~approximately 27 fs. The arrow indicates the cutoff predicted by the semiclassical model~\cite{Lewenstein94}. Panel (a): homogeneous case, (b): $\varepsilon=0.01$ (100 a.u), (c): $\varepsilon=0.02$ (50 a.u) and (d): $\varepsilon=0.05$ (20 a.u). The numbers in brackets indicate an estimate of the inhomogeneity region (for more details see e.g~\cite{Husakou11A,Marcelo12A}) . In panels (e) and (f) is shown the dependence of the semiclassical trajectories on the ionization and recollision times for different values of $\varepsilon$ and for the non confined case, panel (e) and the confined case, panel (f), respectively. Red squares: homogeneous case $\varepsilon=0$; green circles: $\varepsilon=0.01$; blue triangles: $\varepsilon=0.02$ and blue triangles: $\varepsilon=0.05$.}
\label{Figure1HHG}
\end{figure}

\begin{figure} [h]
\centering
\includegraphics[width=\columnwidth]{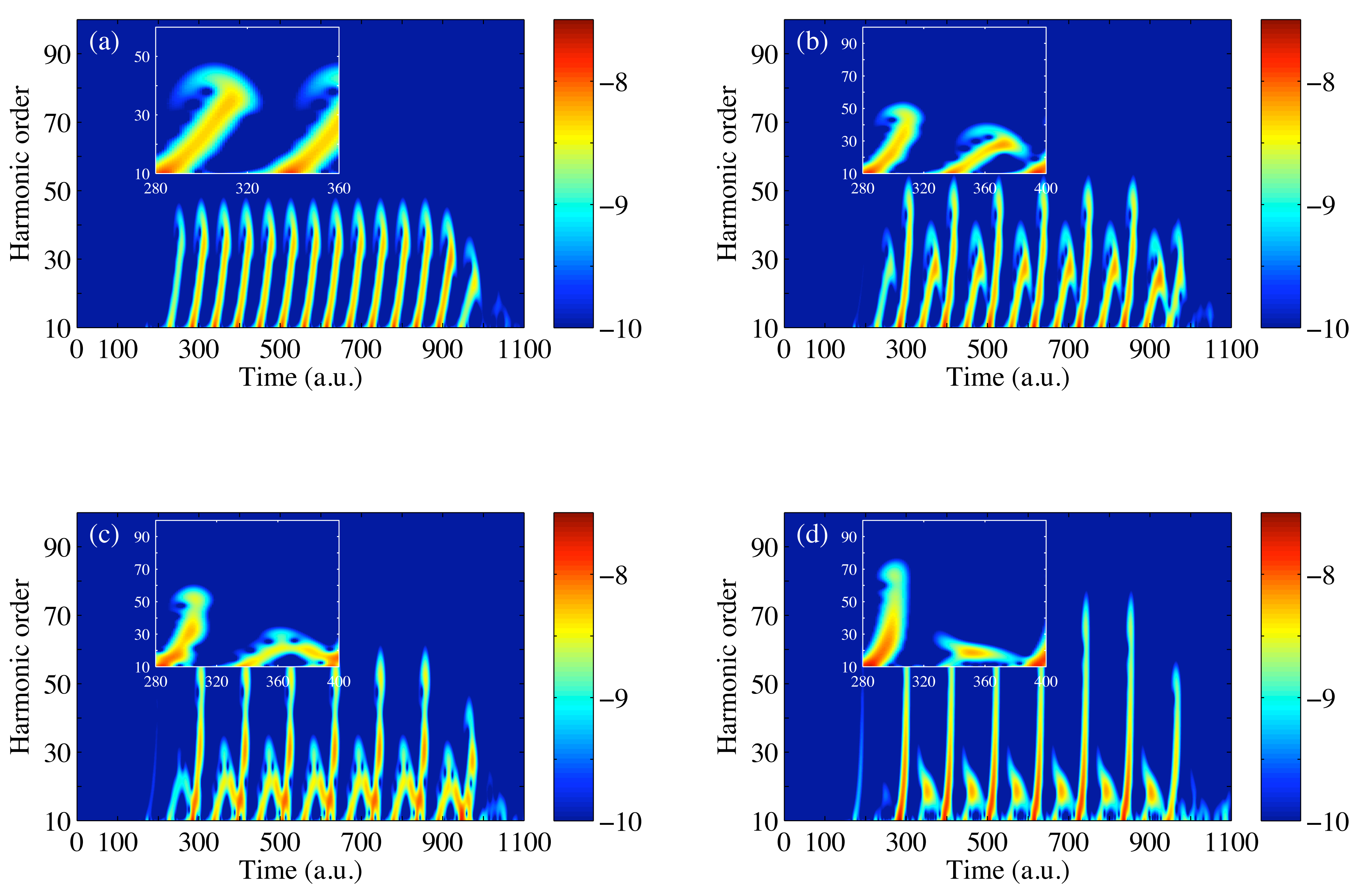}
\caption{Panels (a)-(d): Gabor analysis for the corresponding HHG spectra of panels (a)-(d) of Fig.~\ref{Figure1HHG}. The zoomed regions in all panels show a time interval during the laser pulse for which the complete electron trajectory, from birth time to recollision time, falls within the pulse plateau. In panels (a)-(d) the color scale is logarithmic.}
\label{Figure2HHG}
\end{figure}

In this sub-section we summarize the study carried out in~\cite{Marcelo12A} where it is shown that both the inhomogeneity of the local fields and the constraints in the electron movement, play an important role in the HHG process and lead to the generation of even harmonics and a significant increase in the HHG cutoff, more pronounced for longer wavelengths. In order to understand and characterize these new HHG features we employ two of the different approaches mentioned above: the numerical solution of the 1D-TDSE (see panels (a)-(d) in Fig.~\ref{Figure1HHG}) and the semiclassical approach known as Strong Field Approximation (SFA). Both approaches predict comparable results and describe satisfactorily the new features, but by employing the semiclassical arguments (see panels (e), (f) in Fig.~\ref{Figure1HHG}) behind the SFA and time-frequency analysis tools (Fig.~\ref{Figure2HHG}), we are able to fully explain the reasons of the cutoff extension. 

\newpage

\subsection{Spatially (linear) nonhomogeneous fields: the SFA approach}

In this subsection we summarize the work done in \cite{Marcelo12AA}. In this contribution, we perform a detailed analysis of high-order harmonic generation (HHG) in atoms within the strong field approximation (SFA) by considering spatially (linear) inhomogeneous monochromatic laser fields. We investigate how the
individual pairs of quantum orbits contribute to the harmonic spectra. To this end we have modified both the classical action and the saddle points equations by including explicitly the spatial dependence of the laser field. We show that in the case of a linear inhomogeneous field the electron tunnels with two different canonical momenta. One of these momenta leads to a higher cutoff and the other one develops a lower cutoff. Furthermore, we demonstrate that the quantum orbits have
a very different behavior in comparison to the conventional homogeneous field. A recent study supports our initial findings~\cite{carla2016}.

We also conclude that in the case of the inhomogeneous fields both odd and even harmonics are present in the HHG spectra. Within our extended SFA model, we show
that the HHG cutoff extends far beyond the standard semiclassical cutoff in spatially homogeneous fields. Our findings are in good agreement both with quantum-mechanical and classical models. Furthermore, our approach confirms the versatility of the SFA approach to tackle now the HHG driven by spatially (linear) inhomogeneous fields. 

\subsection{Real nonhomogeneous fields}

\begin{figure}[h]
\centering
\includegraphics[width=\columnwidth]{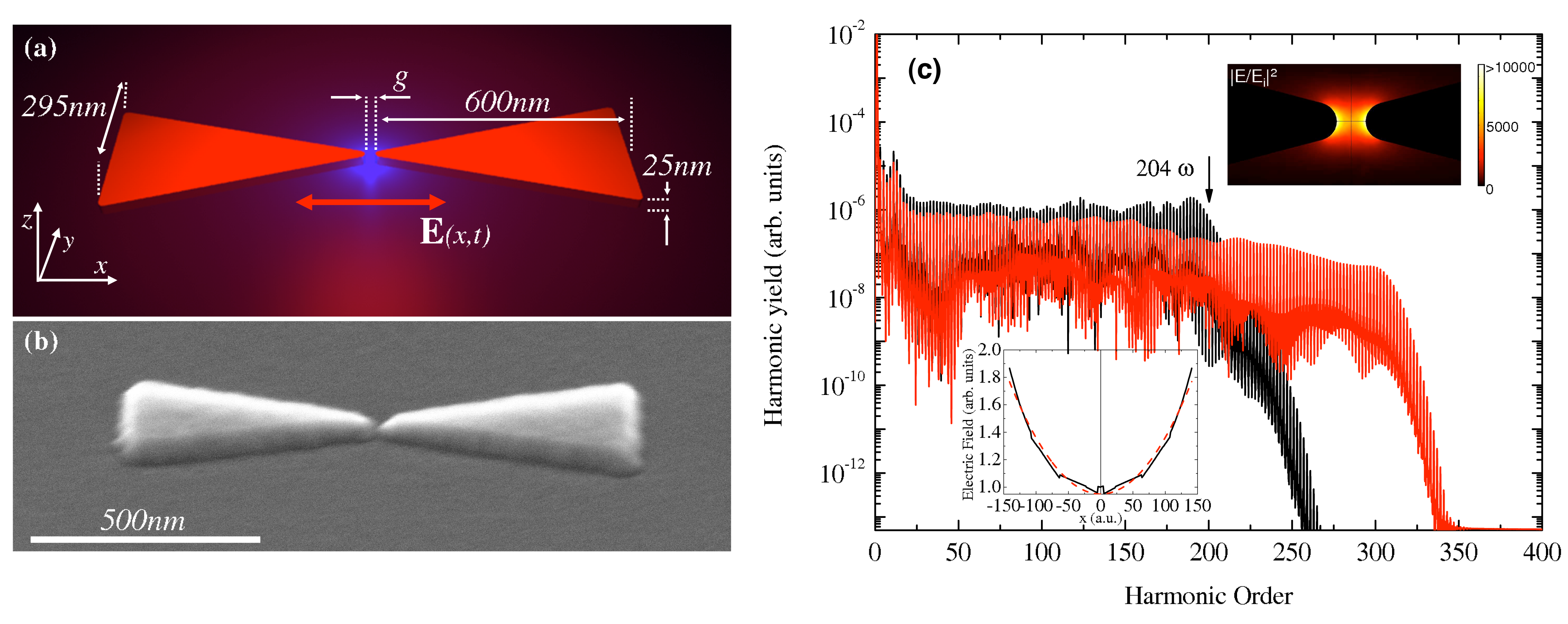}
\caption{(a) Schematic representation of the geometry of the
considered nanostructure. A gold bow-tie antenna resides on glass substrate
(refractive index $n = 1.52$) with superstate medium of air ($n = 1$). The
characteristic dimensions of the system and the coordinate system used in
the 1D-TDSE simulations are shown. (b) SEM image of a real gold bow-tie antenna. (c) High-order harmonic generation (HHG) spectra for a model of argon atoms ($I_p=-0.58$ a.u.), driven by a laser pulse with wavelength $\protect\lambda=1800$ nm and intensity $I=1.25\times10^{14}$ W$\cdot$cm$^{-2}$ at the center of the gap $x=0$. We have used a trapezoidal shaped pulse with three optical cycles turn on and turn off, and a plateau with four optical cycles (about 60 fs). The gold bow-tie nanostructure has a gap $g=15$ nm (283 a.u.). The black line indicates the homogeneous case while the red line indicates the nonhomogeneous case. The arrow indicates the cutoff predicted by the semiclassical model
for the homogeneous case~\cite{Lewenstein94}. The top left inset shows the
functional form of the electric field $E(x,t)$, where the solid lines are the
raw data obtained from the finite element simulations and the dashed line is a
nonlinear fitting. The top right inset shows the intensity enhancement in
the gap region of the gold bow-tie nanostructure. }
\label{Figure3HHG}
\end{figure}

In this sub-section we present numerical simulations of HHG in an argon model atom produced by the fields generated when a gold bow-tie nanostructure is illuminated by a short laser pulse of long wavelength $\lambda=1800$ nm (see~\cite{Marcelo12OE} for more details). The functional form of these fields is extracted from finite element simulations using both the complete geometry of the metal nanostructure and laser pulse characteristics (see Fig.~\ref{Figure3HHG}(a)).  We use the numerical solution of the TDSE in reduced dimensions to predict the HHG spectra. A clear extension in the harmonic cutoff position is observed. This characteristic could lead to the production of XUV coherent laser sources and open the avenue to the generation of shorter attosecond pulses. It is shown in Fig.~\ref{Figure3HHG}(c) that this new feature is a consequence of the combination of a spatial nonhomogeneous electric field, which modifies substantially the electron trajectories, and the confinement of the electron dynamics. Furthermore, our numerical results are supported by time-analysis
and classical simulations. A more pronounced increase in the harmonic cutoff, in addition to an appreciable growth in conversion efficiency,
could be attained by optimizing the nanostructure geometry and materials. These degrees of freedom could pave the way to tailor the
harmonic spectra according to specific requirements.

\subsection{Temporal and spatial synthesized fields}
In this sub-section we present a brief summary of the results published in~\cite{Jose13}. In short, numerical simulations of HHG in He atoms using a temporal and spatial synthesized laser field are considered using the full 3D-TDSE. This particular field provides a new route for the generation of photons at energies beyond the carbon K-edge using laser pulses at 800 nm, which can be obtained from conventional Ti:Sapphire laser sources.  The temporal synthesis is performed using two few-cycle laser pulses delayed in time~\cite{Jose09A}. On the other hand, the spatial synthesis is obtained by using a spatial nonhomogeneous laser field~\cite{Husakou11A,Yavuz12, Marcelo12A} produced when a laser beam is focused in the vicinity of a metal nanostructure or nanoparticle.

Focusing on the spatial synthesis, the nonhomogeneous spatial distribution of the laser electric field can be obtained experimentally by using the resulting field as produced after the interaction of the laser pulse with nanoplasmonic antennas~\cite{Husakou11A, Yavuz12, Marcelo12A,Kim08}, metallic nanowaveguides~\cite{Park11}, metal~\cite{Zherebtsov11,SuessmannSPIE11} and dielectric nanoparticles~\cite{SuessmannPRB11} or metal nanotips~\cite{PeterH06,Schenk10,Kruger11,Kruger12N,Kruger12B,Herink12}. 
%The main feature of these plasmonic fields is that they are no longer homogeneous in the region where the electron dynamics take place. 

The coupling between the atom and the laser pulse, linearly polarized along the $z$ axis, is modified in order to treat the spatially nonhomogeneous fields and can be written it as: $V_{\rm{l}}(z,t,\tau)=\tilde{E}(z,t,\tau)\,z$ with $\tilde{E}(z,t,\tau)=E(t,\tau)(1+\varepsilon z)$ and $E(t,\tau)=E_1(t)+E_2(t,\tau)$ the temporal synthesized laser field with $\tau$ the time delay between the two pulses (see e.g.~\cite{Jose09A} for more details). As in the 1D case the parameter $\varepsilon$ defines the strength of the nonhomogeneity and the dipole approximation is preserved because $\varepsilon\ll1$.

\begin{figure}[h]
\centering
\includegraphics[width=\columnwidth]{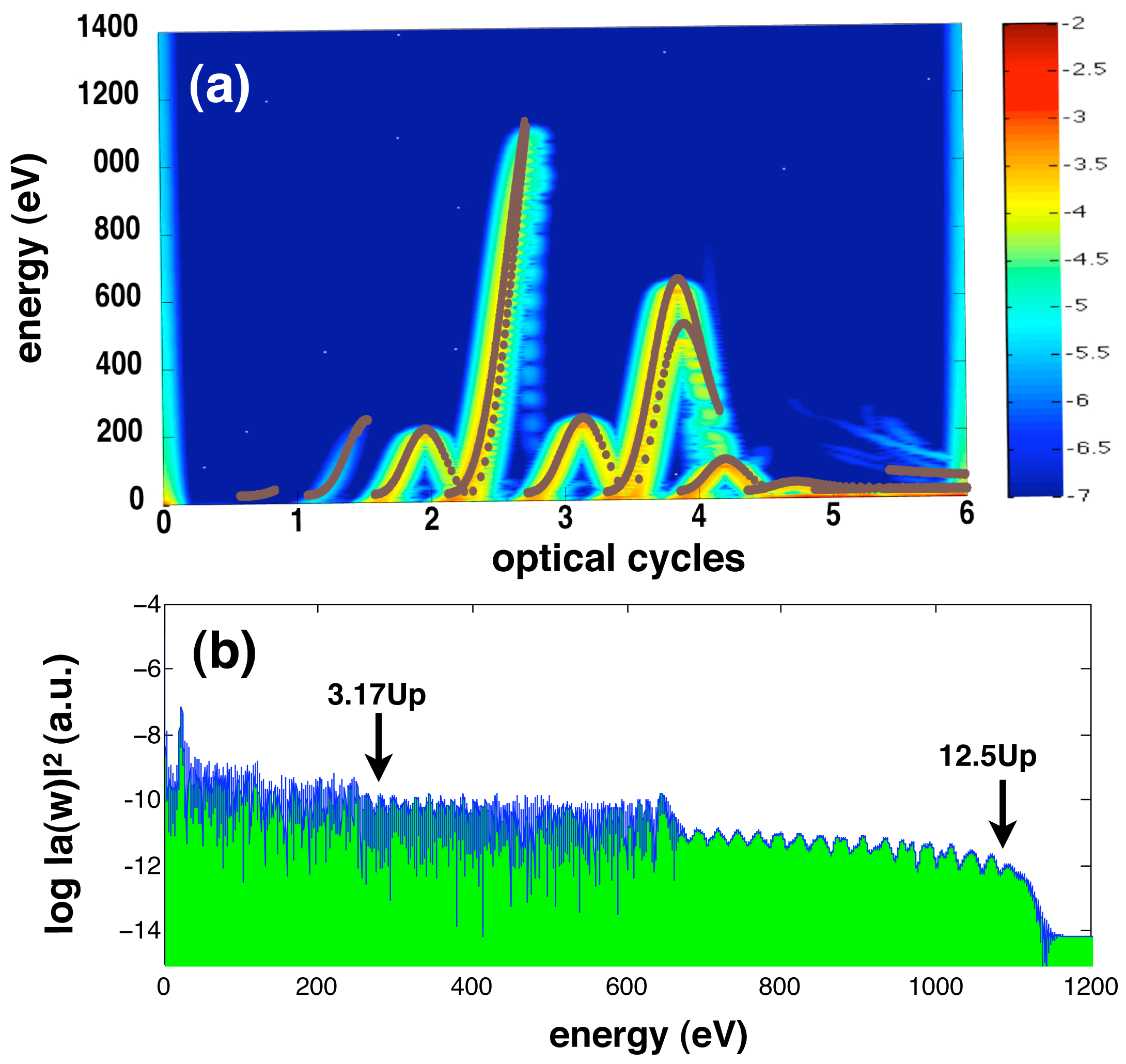}
\caption{(a) Time-frequency analysis obtained from the 3D-TDSE harmonic spectrum for a He atom driven by the spatially and temporally synthesized pulse described in the text with $\varepsilon=0.002$. The plasmonic enhanced intensity $I=1.4\times 10^{15}$ W cm$^{-2}$. Superimposed (in brown) are the classical rescattering energies; (b) 3D-TDSE harmonic spectrum for the same parameters used in (a).}.
\label{Figure4HHG}
\end{figure}

The linear functional form for the spatial non-homogeneity described above could be obtained engineering adequately the geometry of plasmonic nanostructures and by adjusting the laser parameters in such a way that the laser-ionized electron feels only a linear spatial variation of the laser electric field when in the continuum  (see e.g.~\cite{Marcelo12OE} and references therein). The harmonic spectrum then obtained in He for $\varepsilon=0.002$ is presented in Fig.~\ref{Figure4HHG}(b).  We can observe a considerable cut-off extension up to $12.5 U_p$ which is much larger when compared with the double pulse configuration employed alone (it leads only to a maximum of $4.5 U_p$~\cite{Jose09A}). This large extension of the cutoff is therefore a signature of the combined effect of the double pulse and the spatial nonhomogeneous character of the laser electric field. For this particular value of the laser peak intensity ($1.4\times 10^{15}$ W cm$^{-2}$) the highest photon energy is greater than 1 keV. Note that the quoted intensity is actually the plasmonic enhanced intensity, not the input laser intensity. The latter could be several orders of magnitude smaller, according to the plasmonic enhancement factor (see e.g.~\cite{Kim08,Park11}) and will allow the nanoplasmonic target to survive to the interaction.  In order to confirm the underlying physics highlighted by the classical trajectories analysis, we have retrieved the time-frequency distribution of the calculated dipole (from the 3D-TDSE) corresponding to the case of the spectra presented in Fig.~\ref{Figure4HHG}(b) using a wavelet analysis. The result is presented in Fig.~\ref{Figure4HHG}(a) where we have superimposed the calculated classical recombination energies (in brown) to show the excellent agreement between the two theoretical approaches. The consistency of the classical calculations with the full quantum approach is clear and confirms the mechanism of the generation of this $12.5 U_p$ cut-off extension. In addition, the HHG spectra exhibit a clean continuum as a result of the trajectory selection on the recombination time, which itself is a consequence of employing a combination of temporally and spatially synthesized laser field.

\subsection{Plasmonic near-fields}

This sub-section includes an overview of the results reported in~\cite{Marcelo13AR}. In this contribution it is shown how the HHG spectra from model Xe atoms are modified by using a plasmonic near enhanced field generated when a metal nanoparticle is illuminated by a short laser pulse. A setup combining a noble gas as a driven media and metal nanoparticles was also proposed recently in~\cite{Husakou14A,Husakou15}. 

\begin{figure}[h]
\centering
\includegraphics[width=\columnwidth]{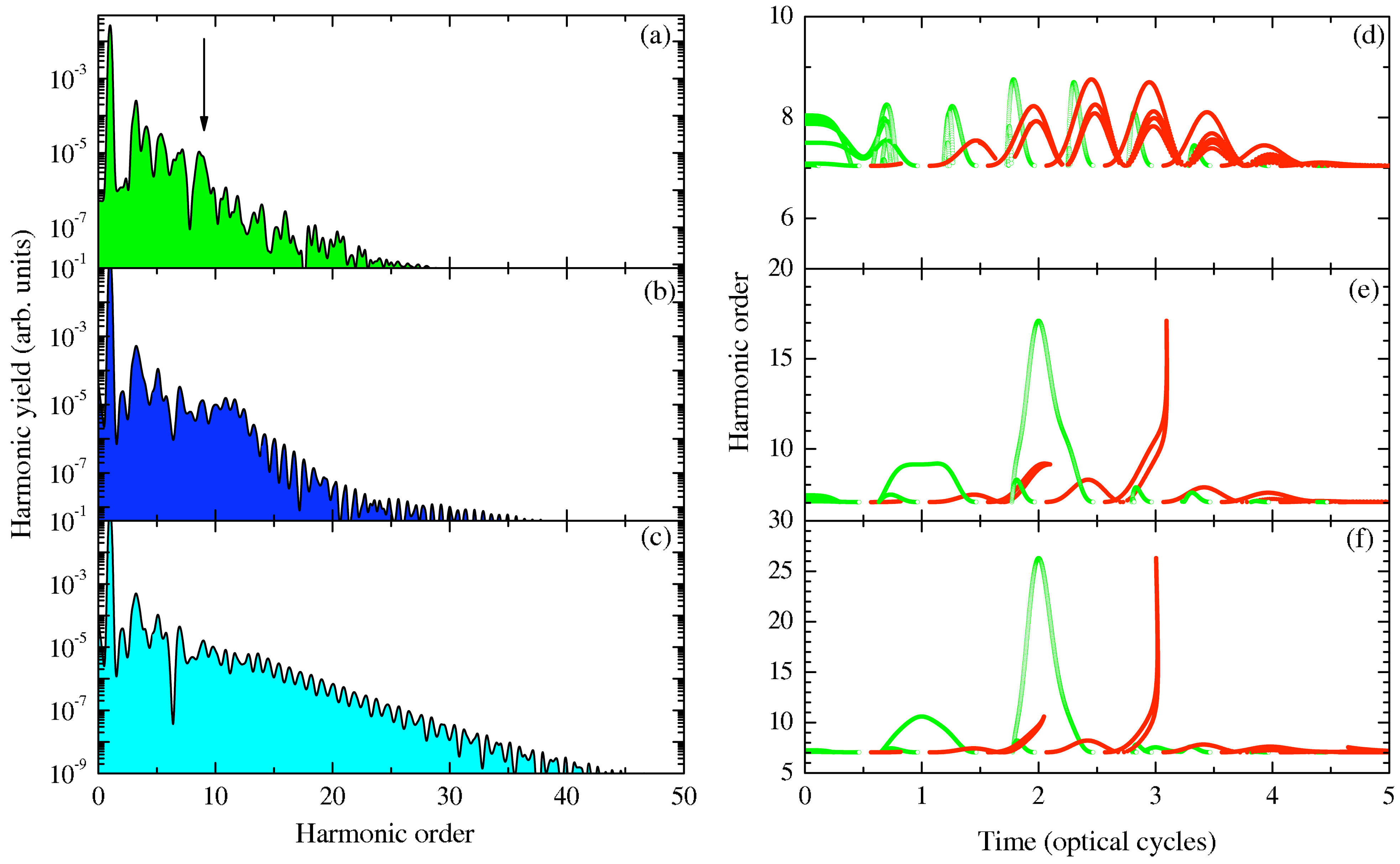}
\caption{HHG spectra for model Xe atoms, laser wavelength $\protect\lambda=720$ nm
and intensity $I=2\times10^{13}$ W$\cdot$cm$^{-2}$. We use a sin$^{2}$ pulse envelope with $n=5$. Panel (a) represents the homogeneous case, panel (b) $\protect\chi=50$ and panel (c) $\protect\chi=40$. The arrow in panel (a) indicates the cutoff predicted by the semiclassical approach~\cite{Lewenstein94}. Panels (d), (e), (f) show the corresponding total energy of the electron (expressed in harmonic order) driven by the laser field calculated from Newton-Lorentz equation and plotted as a function of the $t_i$ (green (light gray) circles) or the $t_r$ (red (dark gray) circles).}
\label{Figure5HHG}
\end{figure}

For our near-field we use the function given by~\cite{SuessmannSPIE11} to define the spatial nonhomogeneous laser electric field $E(x,t)$, i.e.
\begin{equation}
E(x,t)=E_{0}\,f(t)\,\exp (-x/\chi )\sin (\omega_0 t+\phi),
\end{equation}
where $E_{0}$, $\omega_0$, $f(t)$ and $\phi $ are the peak amplitude, the laser field frequency, the field envelope and the CEP, respectively. The functional form of the resulting laser electric field is extracted from attosecond streaking experiments and incorporated both in our quantum and classical approaches. In this specific case the spatial dependence of the plasmonic near-field is given by $\exp(-x/\chi )$ and it is a function of both the size and the material of the spherical nanoparticle. $E(x,t)$ is valid for $x$ outside of the metal nanoparticle, i.e.~$x\ge R_0$, where $R_0$ is its radius.  It is important to note that the electron motion takes place in the region $x\ge R_0$ with $(x+R_0)\gg0$. We consider the laser field having a sin$^{2}$ envelope: $f(t)=\sin ^{2}\left( \frac{\omega_0 t}{2n_{p}}\right)$, where $n_{p}$ is the total number of optical cycles, i.e.~the total pulse duration is $\tau_L =2\pi n_{p}/\omega_0$. The harmonic yield of the atom is obtained by Fourier transforming the acceleration $a(t)$ of the electronic wavepacket (see Section VIIB).

Figure~\ref{Figure5HHG},  panels (a), (b) and (c) show the harmonic spectra for model xenon atoms generated by a laser pulse with $I=2\times 10^{13}$ W cm$^{-2}$, $\lambda =720$ nm and a $\tau
_L=13$ fs, i.e.~$n_{p}=5$ (which corresponds to an intensity envelope of $\approx4.7$ fs FWHM)~\cite{SuessmannSPIE11}. 
In the case of a spatial homogeneous field, no harmonics beyond the $9^{th}$ order are observed. The spatial decay parameter $\chi$ accounts for the spatial nonhomogeneity induced by the nanoparticle and it varies together with its size and the kind of metal employed. Varying  the value of $\chi$ is therefore equivalent to choosing the type of nanoparticle used, which allows to overcome the semiclassically predicted cutoff limit and reach higher harmonic orders. For example, with $\chi =40$ and $\chi=50$ harmonics in the mid 20s (panel c) and well above the $9^{th}$ (a clear cutoff at $n_c\approx 15$ is achieved) (panel b), respectively, are obtained. A modification in the harmonic periodicity, related to the breaking of symmetry imposed by the induced nonhomogeneity, is also clearly noticeable.

Now, by the semiclassical simple man's (SM) model~\cite{corkum93,Lewenstein94} we will study the harmonic cut-off extension. This new effect may be caused by a combination of several factors (for details see~\cite{Marcelo12A,Marcelo12OE}). As is well known, the cutoff law is $n_{c}=(3.17U_{p}+I_{p})/\omega_0$, where $n_{c}$ is the harmonic order at the cutoff and $U_{p}$ the ponderomotive energy. We solve numerically Eq.~(\ref{newton}) for an electron moving in an electric field with the same parameters used in the TDSE-1D calculations, i.e.
\begin{equation}
\ddot{x}(t) =-\nabla _{x}V_{l}(x,t) = -E(x,t)(1-\frac{x(t)}{\chi }),
\end{equation}
and consider the SM model initial conditions: the electron starts at position zero at $t=t_{i}$ (the ionization time) with zero velocity, i.e. $x(t_{i})=0$ and $\dot{x}(t_{i})=0$. When the electric field reverses, the electron returns to its initial position (i.e.~the electron \textit{recollides} or recombines with the parent ion) at a later time $t=t_{r}$ (the recombination time), i.e.~$x(t_{r})=0$. The electron kinetic energy at the $t_{r}$ is calculated as usual from: $E_{k}(t_{r})=\frac{\dot{x}(t_{r})^{2}}{2}$ and finding the $t_r$ (as a function of $t_{i}$) that maximizes $E_k$, $n_c$ is also maximized.

Panels (d), (e) and (f) of Fig.~\ref{Figure5HHG} represent the behaviour of the harmonic order upon the $t_{i}$ and $t_{r}$, calculated from $n=(E_{k}(t_{i,r})+I_{p})/ \omega $ as for the cases (a), (b) and (c) of Fig.~\ref{Figure5HHG}, respectively. Panels (e) and (f) show how the nonhomogeneous character of the laser field strongly modifies the electron trajectories towards an extension of the $n_c$. This is clearly present at $n_{c}\sim 18\omega$ (28 eV) and $n_{c}\sim 27\omega$ (42 eV) for $\chi =50$ and $\chi =40$, respectively. These last two cutoff extensions are consistent with the quantum predictions presented in panels (b) and (c) of Fig.~\ref{Figure5HHG}.

Classical and quantum approaches predict cutoff extensions that could lead to the production of XUV coherent laser sources and open a direct route to the generation of attosecond pulses.  This effect is caused by the induced laser field spatial nonhomogeneity, which modifies substantially the electron trajectories. A more pronounced increment in the harmonic cutoff, in addition to an appreciable growth in the conversion efficiency, could be reached by varying both the radius and the metal material of the spherical nanoparticles. These new degrees of freedom could pave the way to extend the harmonic plateau reaching the XUV regime with modest input laser intensities.

\subsection{Metal nanotip photoemission}

In all the preceding subsections we use plasmonic enhanced fields as sources and atoms as active media.  On the contrary, in this subsection we predict that it is entirely possible to generate high-order harmonic radiation directly from metal nanotips. By employing available laser source parameters and treating the metal tip with a fully quantum mechanical model within the single-active electron (SAE) approximation, we are able to model the HHG process using a metal as active medium. As in previous cases we do not take into account any collective effect, such as propagation and phase matching.  Arguably, such collective effects could play a minor role in the generation of coherent radiation using nanosources due to the fact that radiation emission occurs at a sub-wavelength scale (see, e.g.~\cite{Kim08}). As was already discussed, the main physical mechanism behind the generation of high-order harmonics is the electron recollision step and consequently any reliable model should include it. It was already shown that the recollision mechanism is also needed to describe above-threshold photoemission (ATP) measurements and, considering these two laser-matter phenomena, i.e.~the photoemitted electrons and the high-frequency radiation, are physically linked, we could conclude that metal nanotips can be used as sources of coherent XUV radiation as well.  The theoretical model we use in this case has already been described in previous sections and employed for the calculation of electron photoemission from metal nanotips~\cite{Kruger12B,PeterH06}. As a consequence we do not repeat it here (for details we refer the reader to~\cite{Marcelo14}) and we only show and discuss briefly a couple of typical results. 

\begin{figure}[h]
\centering
\includegraphics[width=0.62\columnwidth]{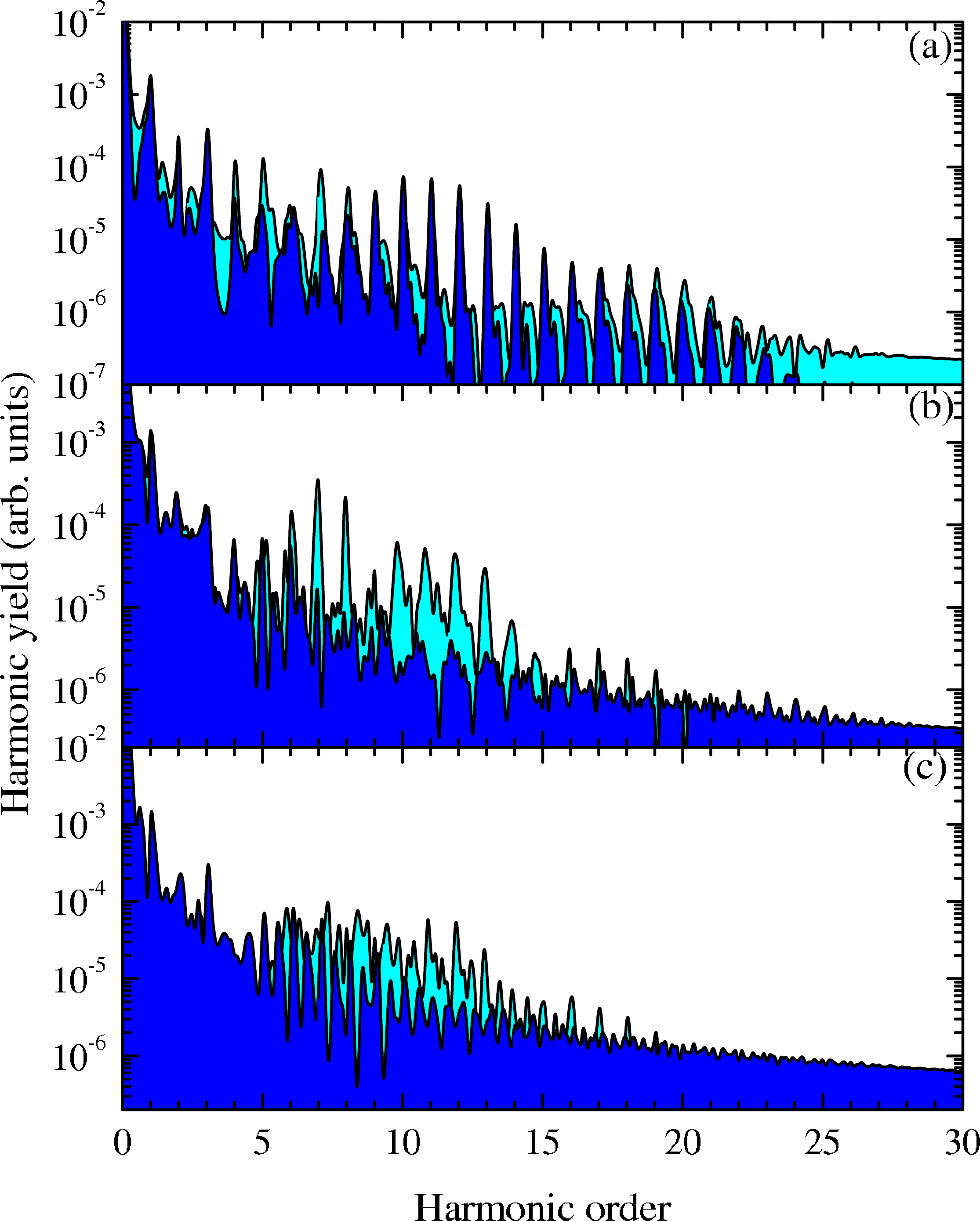}
\caption{(color online) Plots of the HHG spectra as a function of
harmonic order for a metal (Au) nanotip using a trapezoidal shaped
laser pulse with ten cycles of total time, $\lambda=685$ nm, and (a) $E_0 =
10$ GV m$^{-1}$, (b) $E_0 = 15$ GV m$^{-1}$, and (c) $E_0 = 20$ GV m$^{-1}$. In
all panels blue denotes $E_{dc} = -0.4$ GV m$^{-1}$ and cyan $E_{dc} =
2$ GV m$^{-1}$. Note that the harmonic yield scale ($y$ axis) is different in each panel.}
\label{Figure6HHG}
\end{figure}

In Fig.~\ref{Figure6HHG}, we show HHG spectra by using a long (ten-cycle) trapezoidal (two cycles of turn on and off and six cycles of constant amplitude, 23 fs of total time) laser pulse of $\lambda=685$ nm
(the corresponding photon energy is 1.81 eV). The different panels correspond to a set of values of the peak laser electric field $E_0$, namely, 10, 15, and 20 GV m$^{-1}$ for Figs.~\ref{Figure6HHG}(a), \ref{Figure6HHG}(b) and \ref{Figure6HHG}(c), respectively. For the three cases we have employed two values for the DC field, $E_{dc}$:  $-0.4$ GV m$^{-1}$ (blue -dark gray) and $2$ GV m$^{-1}$ (cyan -light gray).  Two main features can be observed:
(i) an increase of the relative yield in the plateau region for positive values of the $E_{dc}$ field.  This gain in conversion efficiency is important for ease
of experimental radiation detection; (ii) the occurrence of odd and even harmonics [see, e.g.~Fig.~\ref{Figure6HHG}(a)] is due to the broken symmetry at
the metal surface of the nanotip, in contrast to an atomic gas, which represents a centrosymmetric nonlinear medium. 

The main result of this subsection is that we show it is possible to generate
high-order harmonics directly from metal nanotips. Our predictions are based on 
a quantum mechanical approach, already successfully
applied to model the photoelectron spectra under similar
experimental conditions. As a consequence it appears perfectly feasible to obtain coherent harmonic radiation directly from these metal nanosources.  

\section{ATI driven by spatially inhomogeneous fields}

As was mentioned at the outset, ATI represents another key strong field phenomena. As a consequence, in the next subsections we summarize the theoretical work we have done in order to tackle the ATI driven by spatially inhomogeneous fields. As in the case of HHG, we include here results obtained using quantum, semiclassical and classical formalisms. 

\subsection{1D case}

Investigations carried out on ATI, generated by few-cycle driving laser pulses, have attracted much interest due to the sensitivity of the energy and angle-resolved photoelectron spectra to the absolute value of the CEP~\cite{Milosevic06,paulus2011}. This feature makes the ATI phenomenon a potential tool for laser pulse characterization. In order to characterize the CEP of a few-cycle laser pulse, the so-called backward-forward asymmetry of the ATI spectrum is measured and from the information collected the absolute CEP can be obtained~\cite{paulus2001,paulus2011}. Furthermore, nothing but the high energy region of the photoelectron spectrum appears to be strongly sensitive to the absolute CEP and consequently electrons with high kinetic energy are needed in order to characterize it~\cite{Milosevic06,paulus2001,paulus_measurement_2003}.

Nowadays, experiments have demonstrated that the electron spectra of ATI could be extended further by using plasmon field enhancement~\cite{Kim08,Zherebtsov11}. The strong confinement of the plasmonics spots and the distortion of the electric field by the surface plasmons induces a spatial inhomogeneity in the driving laser field, just before the interaction with the corresponding target gas. A related process employing solid state targets instead of atoms and molecules in gas phase is the so called above-threshold photoemission (ATP). This laser driven phenomenon has received special attention recently due to its novelty and the new physics involved. In ATP electrons are emitted directly from metallic surfaces or metal nanotips and they present distinct characteristics, namely higher energies, far beyond the usual cutoff for noble gases and consequently the possibility to reach similar electron energies with smaller laser intensities (see e.g.~\cite{PeterH06,Schenk10,Kruger11,Kruger14,Herink12}). Furthermore, the photoelectrons emitted from these nanosources are sensitive to the CEP and consequently it plays an important role in the angle and energy resolved photoelectron spectra~\cite{apolonski,Zherebtsov11,Kruger11}. 

Despite new developments, all numerical and semiclassical approaches to model the ATI phenomenon are based on the assumption that the external field is spatially homogeneous in the region where the electron dynamics take place. For a spatially inhomogeneous field, however, important modifications will occur to the strong field phenomena, as was already shown for the case of HHG.  These modifications occur because the laser-driven electric field, and consequently the force applied on the electron, will also depend on its position. 

From a theoretical viewpoint, the ATI process can be tackled using different approaches (for a summary see e.g.~\cite{Milosevic06} and references therein). In this subsection, we concentrate on extending one of the most and widely used approaches: the numerical solution of time-dependent Schr\"odinger Equation (TDSE) in reduced dimensions. 

In order to calculate the energy resolved photoelectron spectra, we use the same one-dimensional time-dependent Schr\"odinger
equation (1D-TDSE) employed for the computation of HHG (see Section VIIB). For calculating the energy-resolved photoelectron spectra $P(E)$ we use the
window function technique developed by Schafer~\cite{schaferwop1,schaferwop}. 
This tool has been widely used, both to calculate angle-resolved and
energy-resolved photoelectron spectra~\cite{schaferwop2} and it represents a
step forward with respect to the usual projection methods.

\begin{figure}[htb]
\centering
\includegraphics[width=0.45\textwidth]{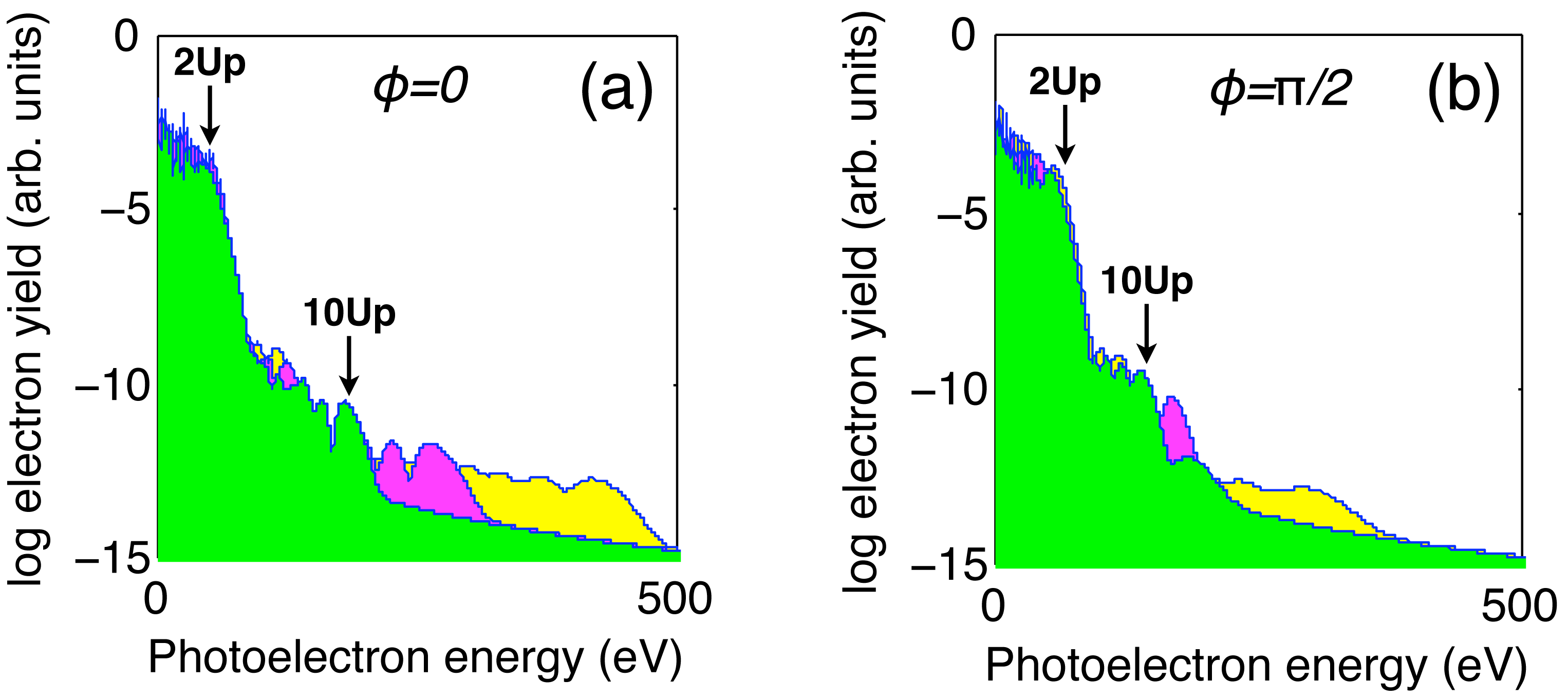}
\caption{1D-TDSE energy-resolved photoelectron spectra for a model atom with $I_{p}=-0.5$ a.u. and for the laser parameters, $I=3\times10^{14}$ W cm$^{-2}$, $\lambda=800$ nm and a sin-squared shaped pulse with a total duration of 4 cycles (10 fs ). In green for $\varepsilon=0$ (homogeneous case), in magenta for $\varepsilon=0.003$ and in yellow for $\varepsilon=0.005$. Panel (a) represent the case for $\phi=0$ (sin-like pulse) and panel (b) represents the case for $\phi=\pi/2$ (cos-like pulse). The arrows indicate the $2 U_p$ and $10 U_p$ cutoffs predicted by the classical model~\cite{Milosevic06}}
\label{Figure1ATI}
\end{figure}

In our simulations we employ as a driving field a four-cycle (total duration 10 fs) sin-squared laser pulse with an intensity  $I=3\times10^{14}$ W cm$^{-2}$ and wavelength $\lambda=800$ nm. We chose a linear inhomogeneous field and three different values for the parameter that characterizes the
inhomogeneity strength, namely $\varepsilon =0$ (homogeneous case), $\varepsilon=0.003$ and $\varepsilon=0.005$.  Figure \ref{Figure1ATI}(a)
shows the cases with $\phi =0$ (a sin-like laser pulse) meanwhile in Fig.~\ref{Figure1ATI}(b) $\phi =\pi/2$
(a cos-like laser pulse), respectively. In both panels
green represents the homogeneous case, i.e.~$\varepsilon =0$,
magenta is for $\varepsilon =0.003$ and yellow is for $\varepsilon =0.005$,
respectively.
For the homogeneous case, the spectra exhibits the usual distinct behavior,
namely the $2U_{p}$ cutoff ($\approx 36$ eV for our case) and the $10U_{p}$
cutoff ($\approx 180$ eV), where $U_{p}=E_{0}^{2}/4\omega^{2}$ is the
ponderomotive potential.  The former cutoff corresponds to those electrons
that, once ionized,  never return to the atomic core, while the latter one
corresponds to the electrons that, once ionized, return to the core and
elastically rescatter. It is well established using classical arguments that the maximum kinetic energies of the \textit{direct} and the \textit{rescattered} electrons are $E_{max}^{d}=2U_{p}$ and  $E_{max}^{r}=10U_{p}$,
respectively. In a quantum mechanical approach, however, it is possible to find electrons with energies beyond the 10$U_p$, although their yield drops several orders of magnitude~\cite{Milosevic06}. 
%Experimentally, both mechanisms contribute to the
%energy-resolved photoelectron spectra and consequently theoretical descriptions should also include both. 
The TDSE, which can be considered as an exact approach to the problem, is able to predict the $P(E)$ for the whole range of electron energies. In addition, the most energetic electrons, i.e.~those with $E_{k}\gg 2U_{p}$, are used to characterize the CEP of few-cycle pulses. As a result, a correct description
of the rescattering mechanism is needed.

For the spatial inhomogeneous case, the cutoff positions of both the \textit{direct} and the \textit{rescattered} electrons are extended towards larger energies.
For the \textit{rescattered} electrons, this extension  is very prominent. In
fact, for $\varepsilon =0.003$ and  $\varepsilon =0.005,$ it reaches  $\approx 260$ eV and  $\approx 420$ eV, respectively (see Fig.~\ref{Figure1ATI}(a)). 
Furthermore, it appears that the high energy region of $P(E)$, for instance,  the region between $200-400$ eV for $\varepsilon =0.005$
(Fig.~\ref{Figure1ATI} in yellow), is strongly sensitive to the CEP. This feature
indicates that the high energy region of the photoelectron spectra could resemble a new and better CEP characterization tool. It should be, however,
complemented by other well known and established CEP characterization tools, as, for instance, the forward-backward asymmetry (see~\cite{Milosevic06}). 
Furthermore, the utilization of nonhomogeneous fields would open the avenue for the
production of high energy electrons, reaching the keV regime, if a reliable control of the spatial and temporal shape of the laser electric field is attained. 

We now concentrate our efforts on explaining the extension of the
energy-resolved photoelectron spectra using classical arguments. From the
simple-man's model~\citep{corkum93,Lewenstein94} we can describe the physical origin of the
ATI process as follows: an atomic electron at a position $x=0$, is released
or \textit{born} at a given time, that we call \textit{ionization} time $t_{i}$, 
with zero velocity, i.e.~$\dot{x}(t_{i})=0$. This electron now moves
only under the influence of the oscillating laser electric field (the
residual Coulomb interaction is neglected in this model) and will reach
the detector either directly or through a rescattering process.  By using the
classical equation of motion, it is possible to calculate the maximum
energy of the electron for both direct and rescattered processes. The Newton
equation of motion for the electron in the laser field can be written as (see Eq.~(\ref{newton})):
\begin{eqnarray}
\ddot{x}(t) &=&-\nabla _{x}V_{\rm{l}}(x,t)  \notag  \label{newtonati} \\
&=&E(x,t)+\left[ \nabla _{x}E(x,t)\right] x  \notag \\
&=&E(t)(1+2\varepsilon x(t)),
\end{eqnarray}
where we have collected the time dependent part of the electric field in $E(t)$, i.e.~$E(t)=E_{0}f(t)\sin (\omega t+\phi )$ and 
specialized to the case $h(x)=x$. In the limit where $\varepsilon =0$ in Eq.~(\ref{newtonati}), we recover the spatial homogeneous case. For the
direct ionization, the kinetic energy of an electron released or born at
time $t_{i}$ is
\begin{equation}
\label{direct}
E_{d}=\frac{\left[ \dot{x}(t_{i})-\dot{x}(t_{f})\right] ^{2}}{2},
\end{equation}
where $t_{f}$ is the end time of the laser pulse. For the rescattering process, in which the electron returns to the core at a time $t_{r}$ and
reverses its direction, the kinetic energy  of the electron yields
\begin{equation}
\label{rescattered}
E_{r}=\frac{\left[ \dot{x}(t_{i})+\dot{x}(t_{f})-2\dot{x}(t_{r})\right] ^{2}}{2}.
\end{equation}

For homogeneous fields, Eqs.~(\ref{direct}) and (\ref{rescattered}) become $E_{d}=\frac{\left[ A(t_{i})-A(t_{f})\right] ^{2}}{2}$ and 
$E_{r}=\frac{\left[ A(t_{i})+A(t_{f})-2A(t_{r})\right] ^{2}}{2}$, with $A(t)$ being the laser vector potential $A(t)=-\int^{t} E(t')dt'$. For the case with $\varepsilon=0$, it
can be shown that the maximum value for $E_{d}$ is $2U_{p}$ while for $E_{r}$ it is $10U_{p}$~\cite{Milosevic06}. These two values appear as cutoffs in the energy resolved photoelectron spectrum as can be
observed in panels (a) and (b), in green, in Fig.~\ref{Figure1ATI} (see the respective arrows).

\begin{figure}[htb]
\centering
\includegraphics[width=0.47\textwidth]{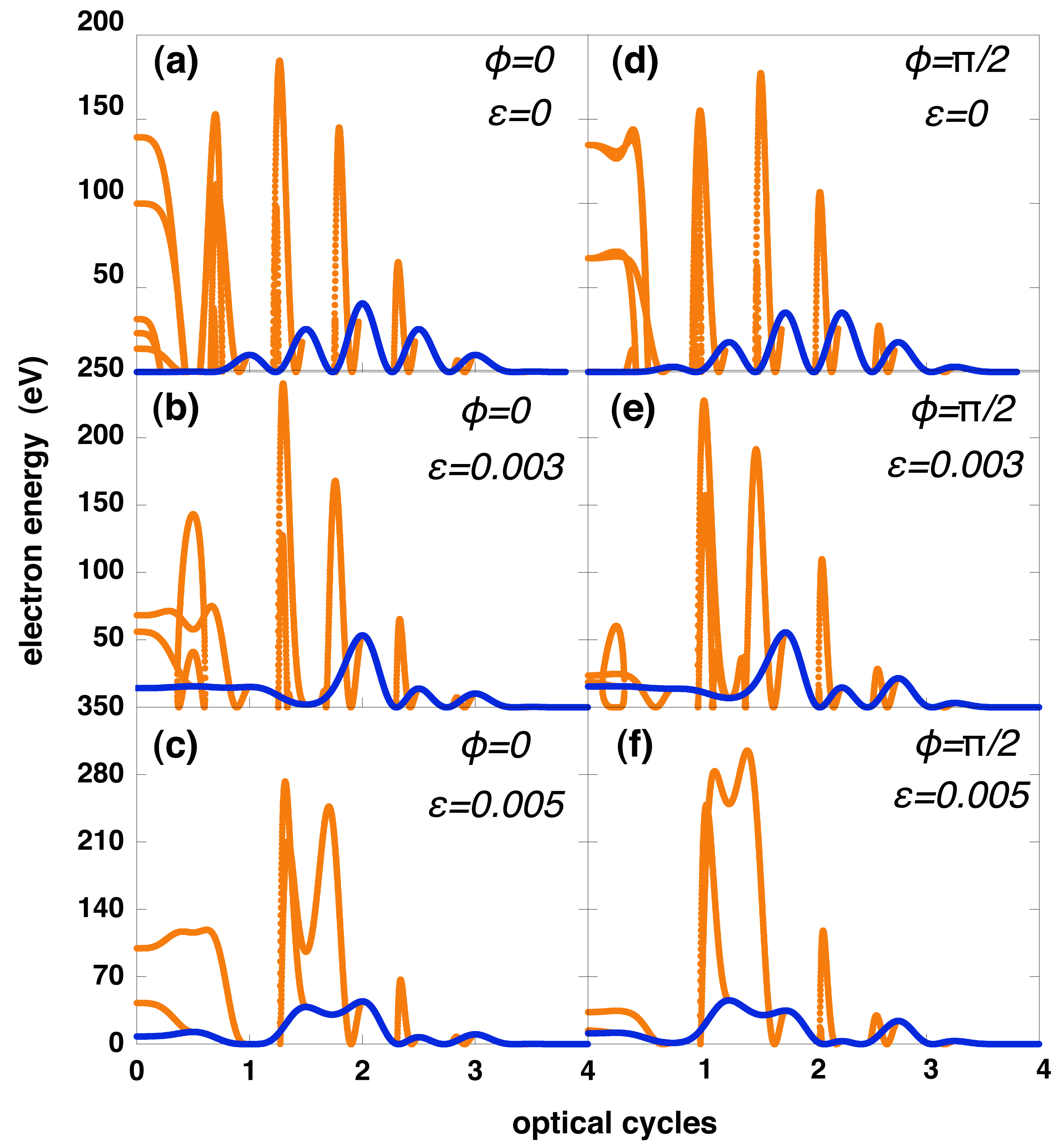}
\caption{Numerical solutions of Eq.~(\ref{newtonati}) plotted in terms of the direct (blue) and rescattered (orange) electron kinetic energy. The laser parameters are the same as in Fig.~\ref{Figure1ATI}.  Panels (a), (b) and (c) correspond to the case of sin-like pulses ($\phi=0$) and for $\varepsilon=0$ (homogeneous case), $\varepsilon=0.003$ and $\varepsilon=0.005$, respectively. Panels (d), (e) and (f) correspond to the case of cos-like pulses ($\phi=\pi/2$) and for $\varepsilon=0$ (homogeneous case), $\varepsilon=0.003$ and $\varepsilon=0.005$, respectively.}
\label{Figure2ATI}
\end{figure}

In Fig.~\ref{Figure2ATI}, we present the numerical solutions of Eq.~(\ref{newtonati}), which is plotted in terms of the kinetic energy of the direct and rescattered electrons. We employ the same laser parameters as in Fig.~\ref{Figure1ATI}. Panels (a), (b) and (c) correspond to the case of $\phi=0$ (sin-like pulses) and
for $\varepsilon=0$ (homogeneous case), $\varepsilon=0.003$ and $\varepsilon=0.005$, 
respectively. Meanwhile, panels (d), (e) and (f)
correspond to the case of $\phi=\pi/2$ (cos-like pulses) and for $\varepsilon=0$ (homogeneous case), 
$\varepsilon=0.003$ and $\varepsilon=0.005$, respectively.
From the panels (b), (c), (e) and (f) we can observe the strong
modifications that the nonhomogeneous character of the laser electric field
produces in the electron kinetic energy. These are related to the changes
in the electron trajectories (for details see e.g.~\cite{Yavuz12,Marcelo12A,Marcelo12OE}). In short, the electron trajectories are modified in such a way that now the electron ionizes at an earlier time and
recombines later, and in this way it spends more time in the continuum
acquiring energy from the laser electric field. Consequently, higher values
of the kinetic energy are attained. A similar behavior with the photoelectrons was observed recently in ATP using metal nanotips. According to the model presented in~\cite{Herink12} the localized fields modify the electron motion in such a way to allow sub-cycle dynamics. In our studies, however, we consider both direct and rescattered electrons (in~\cite{Herink12} only direct electrons are modeled) and the characterization of the dynamics of the photoelectrons is more complex. Nevertheless, the higher kinetic energy of the rescattered electrons is a clear consequence of the strong modifications of the laser electric field in the region where the electron dynamics takes place, as in the above mentioned case of ATP.

\subsection{3D case}

The logical extension to the numerical approach presented in the previous subsection is to use the three-dimensional time-dependent Schr\"odinger equation (3D-TDSE) to calculate angular electron momentum distributions and photoelectron spectra of atoms driven by spatially inhomogeneous fields. As in the 1D case the nonhomogeneous character of the laser electric field plays an important role on the ATI phenomenon. In addition, our 3D approach allows us to model in a reliable way the ATI process both in the tunneling and multiphoton regimes. We show that for the former, the spatial nonhomogeneous field causes significant modifications on the electron momentum distributions and photoelectron spectra, while its effects in the later appear to be negligible. Indeed, through the tunneling ATI process, one can obtain higher energy electrons as well as a high degree of
asymmetry in the momentum space map. In our study we consider near infrared laser fields with intensities in the mid- $10^{14}$ W cm$^{-2}$ range. We use a linear approximation for the plasmonic field, considered valid when the electron excursion is small compared with the inhomogeneity region. Indeed, our 3D simulations confirm that plasmonic fields could drive electrons with energies in the near-keV regime (see e.g.~\cite{Marcelo13A}). 

In order to obtain a more complete description of the ATI phenomenon driven by spatially
nonhomogeneous fields, we solve the 3D-TDSE) in the length gauge (see Section VIIB).  We then investigate the electron momentum distribution and energy-resolved photoelectron spectra $P(E)$, including the dynamics of both, direct and rescattered electrons.

\begin{figure}[htb]
\centering
\includegraphics[width=0.45\textwidth]{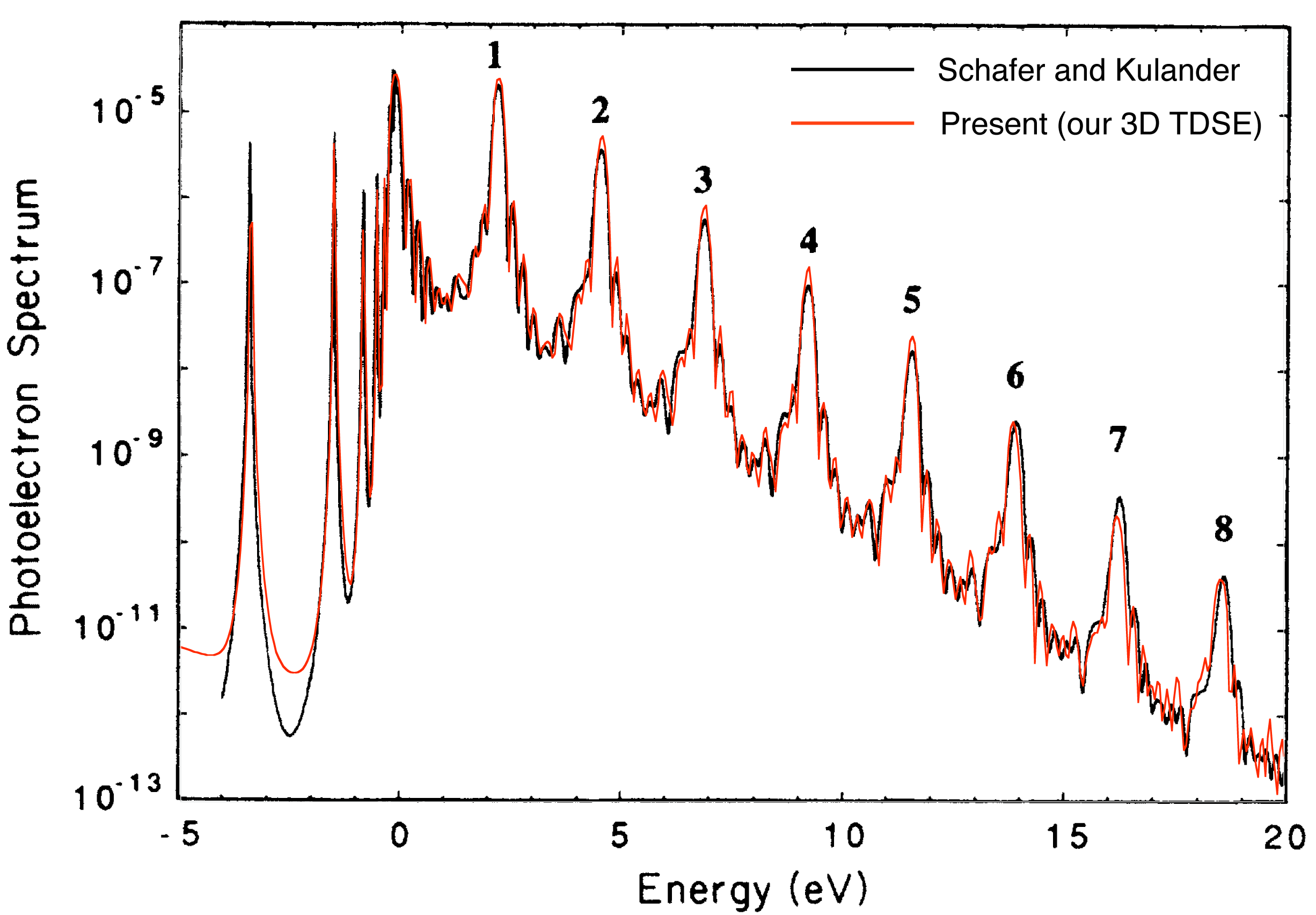}
\caption{(color online) Photoelectron spectrum resulting from our 3D TDSE simulations (in red) and superimposed (in black) with the ATI results calculated by Schafer and Kulander in Ref.~\cite{schaferwop1}. The laser wavelength is $\lambda=532$ nm and the intensity is $I=2\times10^{13}$ W cm$^{-2}$ (see Fig.~1 in~\cite{schaferwop1} for more details. The superimposed plot has been extracted from Fig.~1 of this last cited reference.}
\label{Figure3ATI}
\end{figure}

As in the 1D case, the ATI spectrum is calculated starting from the time propagated electron wave function, once the laser pulse has ceased. As a preliminary test and in order to assure the consistency of our numerical simulations, we have compared our calculations with the results previously obtained in~\cite{schaferwop1}. The comparison confirms the high degree of accuracy of our implementation as shown in Fig.~\ref{Figure3ATI}.
For computing the energy-resolved photoelectron spectra $P(E)$ and two-dimensional electron distributions $\mathcal{H}(k_z,k_r)$, where $k_z$ ($k_r$) is the electron momentum component parallel (perpendicular) to the polarization direction, we use the window function approach developed in~\cite{schaferwop1,schaferwop}.
%Note:  you already have the exact same sentence earlier in the text...
%This tool has been widely used, both to calculate angle-resolved and
%energy-resolved photoelectron spectra~\cite{schaferwop2}, and it represents a
%step forward with respect to the usual projection methods.

Experimentally speaking, both the direct and rescattered electrons contribute to the energy-resolved photoelectron spectra. It means that for tackling this problem both physical mechanisms should to be included in any theoretical model. In that sense, the 3D-TDSE, which can be considered as an exact approach to the ATI problem for atoms and molecules in the single active electron approximation (SAE), appears to be the most adequate tool to predict the $P(E)$ in the whole range of electron energies.

\begin{figure}[htb]
\centering
\includegraphics[width=0.47\textwidth]{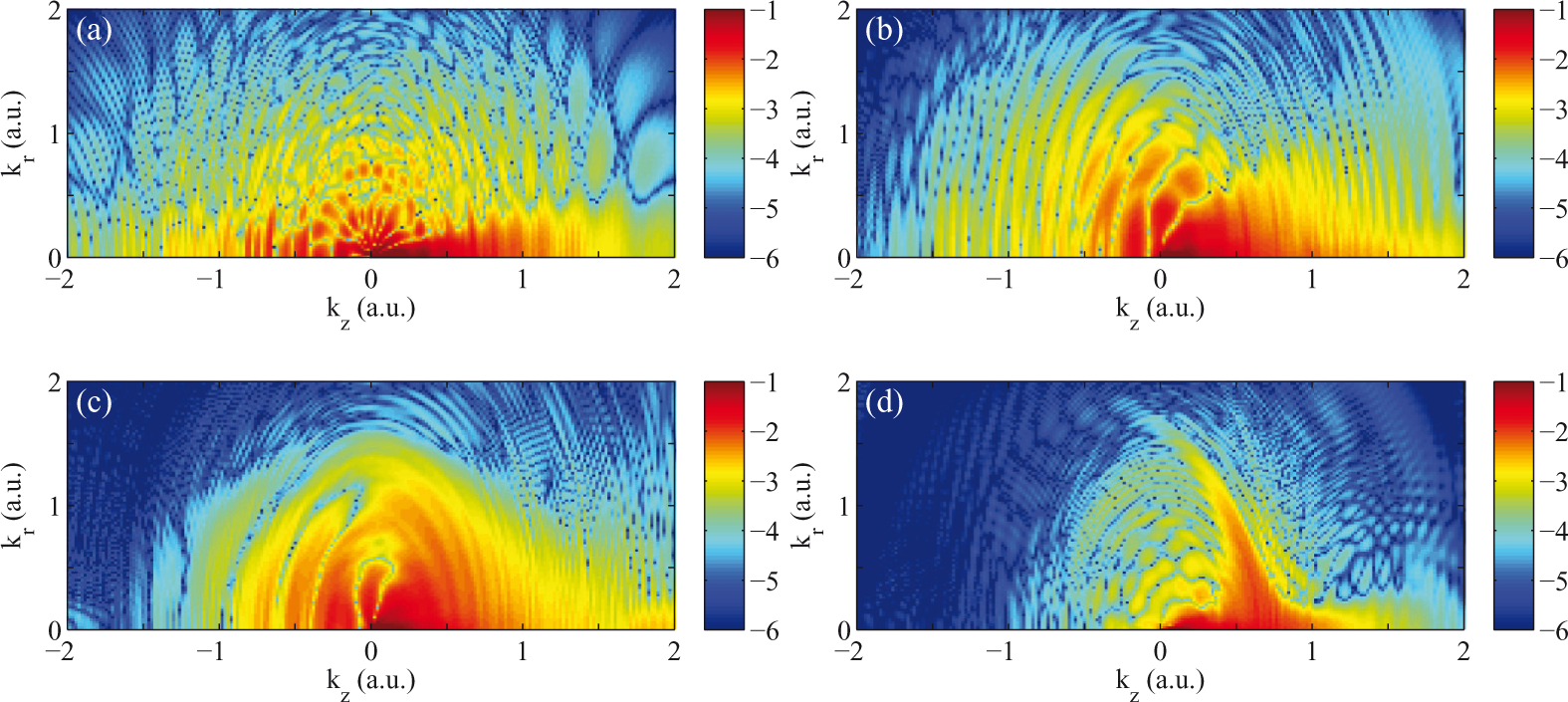}
\caption{(Color online) Two-dimensional electron momentum
distributions (logarithmic scale) in cylindrical coordinates ($k_z,k_r$)
using the exact 3D-TDSE calculation for an hydrogen atom.
The laser parameters are $I = 5.0544 \times 10^{14}$ W cm$^{-2}$ ($E_0=0.12$ a.u.) and $\lambda = 800$ nm. We have used a sin-squared shaped pulse with a total duration of four optical cycles
(10 fs) with $\phi=\pi/2$. (a) $\varepsilon = 0$
(homogeneous case), (b) $\varepsilon = 0.002$, (c) $\varepsilon = 0.003$ and (d) $\varepsilon = 0.005$.}
\label{Figure4ATI}
\end{figure}

In the following, we calculate two-dimensional electron momentum distributions for a
laser field intensity of $I=5.0544\times10^{14}$ W cm$^{-2}$
($E_0=0.12$ a.u). The results are depicted in Fig.~\ref{Figure4ATI} for
$\phi=\pi/2$. Here, panels (a), (b), (c) and (d) represent the
cases with $\varepsilon=0$ (homogeneous case), $\varepsilon=0.002$,
$\varepsilon=0.003$ and $\varepsilon=0.005$, respectively. 
By a simple inspection of Fig.~\ref{Figure4ATI} strong modifications produced by the spatial inhomogeneities in both the angular and low-energy structures can be appreciated (see~\cite{Marcelo13A} for more details).
\begin{figure}[htb]
\centering
\includegraphics[width=0.45\textwidth]{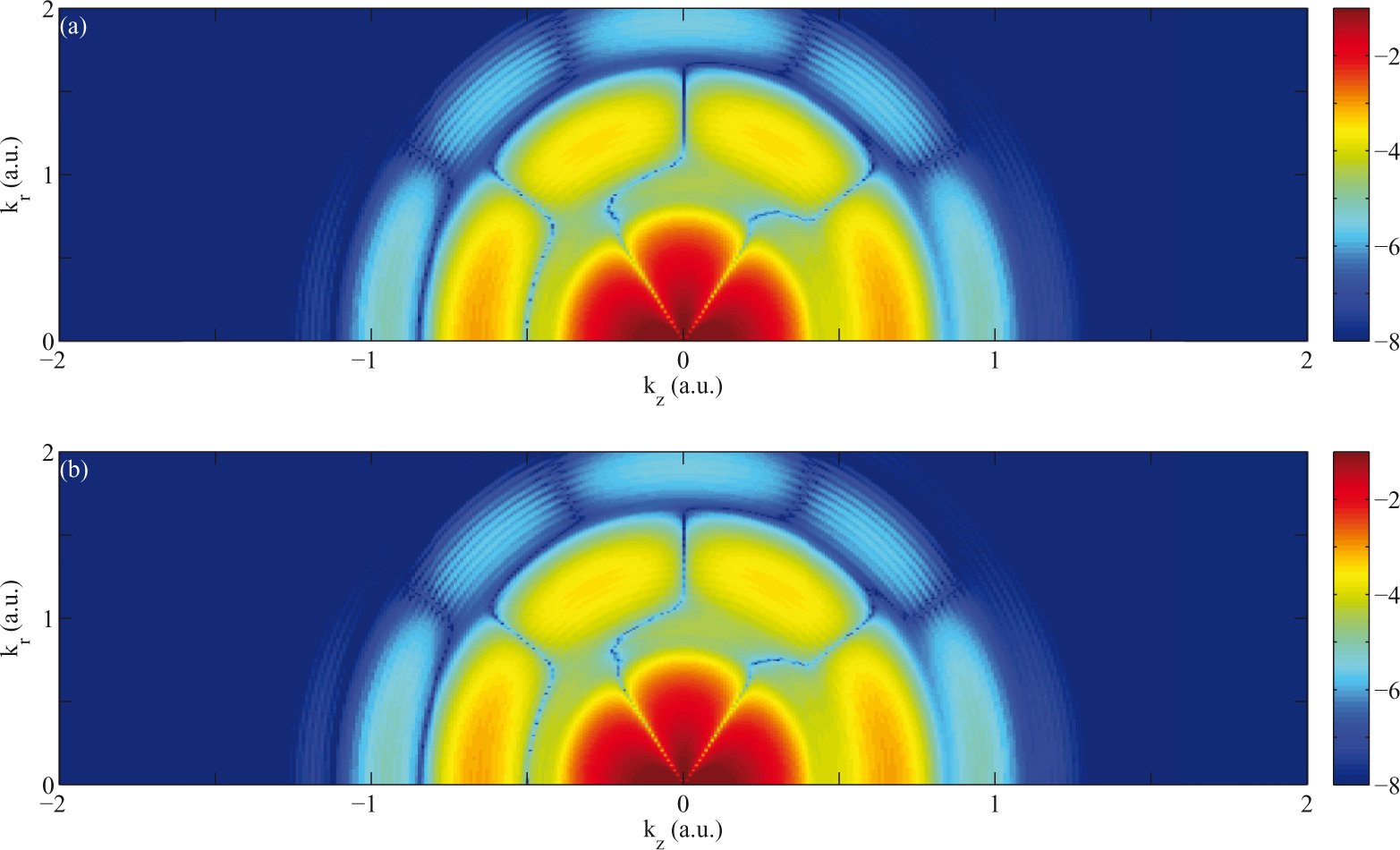}
\caption{(Color online) Two-dimensional electron momentum
distributions (logarithmic scale) in cylindrical coordinates ($k_z,k_r$)
using the exact 3D-TDSE calculation for an hydrogen atom.
The laser parameters are $E_0=0.05$ a.u. ($I=8.775\times10^{13}$ W cm$^{-2}$), $\omega=0.25$ a.u. ($\lambda=182.5$ nm) and $\phi=\pi/2$. We employ a laser pulse with 6 total cycles. Panel (a) corresponds to the homogeneous case ($\varepsilon=0$) and panel (b) is for $\varepsilon=0.005$.}
\label{Figure5ATI}
\end{figure}

However in the case of low intensity regime (i.e.~multiphoton regime, $\gamma\gg1$) the scenario changes radically. In order to study this regime we use a laser electric field with $E_0=0.05$ a.u. of peak amplitude ($I=8.775\times10^{13}$ W cm$^{-2}$), $\omega=0.25$ a.u. ($\lambda=182.5$ nm) and 6 complete optical cycles. The resulting Keldysh parameter $\gamma=5$ indicates the predominance of a multiphoton process~\cite{arbo1}. In  Fig.~\ref{Figure5ATI} we show the two-dimensional electron distributions for the two cases discussed above. For the homogeneous case our calculation is identical to the one presented in~\cite{arbo1}. We also notice the two panels present indistinguishable shape and magnitude.  Hence the differences introduced by the spatial inhomogeneity are practically imperceptible in the multiphoton ionization regime.

\subsection{SFA and quantum orbits}

As for the case of HHG driven by spatially inhomogeneous fields, ATI can also be modeled by using the SFA. In order to do so, it is necessary to modify the SFA ingredients, namely the classical action and the saddle point equations. The latter are more complex, but appear to be solvable for the case of spatially linear inhomogeneous fields (for details see~\cite{Marcelo13AA}). Within SFA it is possible to investigate how the individual pairs of quantum orbits contribute to the photoelectron spectra and the two-dimensional electron momentum distributions. We demonstrate that the quantum orbits have a very different behavior in the spatially inhomogeneous field when compared to the homogeneous field. In the case of inhomogeneous fields, the ionization and rescattering times differ between neighboring cycles, despite the field being nearly monochromatic. Indeed, the contributions from one cycle may lead to a lower cutoff, while another may develop a higher cutoff. As was shown both by our quantum mechanical and classical models, our SFA model confirms that the ATI cutoff extends far beyond the semiclassical cutoff, as a function of inhomogeneity strength. In addition, the angular momentum distributions have very different features compared to the homogeneous case. For the neighboring cycles, the electron momentum distributions do not share the same absolute momentum, and as a consequence they do not have the same yield. 

\subsection{Near-fields}

In this section we put forward the plausibility to perform ATI experiments by combining plasmonic enhanced near-fields and noble gases. The proposed experiment would take advantage of the plasmonic enhanced near-fields (also known as evanescent fields), which present a strong spatial nonhomogeneous character and the flexibility to use any atom or molecule in gas phase. A similar scheme was presented in Section VIIE, but now we are interested in generating highly energetic electrons, instead of coherent electromagnetic radiation. We employ 1D-TDSE (see Section VIIB) by including the actual functional form of metal nanoparticles plasmonic near-fields obtained from attosecond streaking measurements. We have chosen this particular nanostructure since its actual enhanced-field is known experimentally, while for the other nanostructures, like bow-ties~\cite{Kim08}, the actual plasmonic field is unknown. For most of the plasmonic nanostructures the enhanced field is theoretically calculated using the finite element simulation, which is based on an ideal system that may deviate significantly from actual experimental conditions. For instance,~\cite{Kim08} states an intensity enhancement of 4 orders of magnitude (calculated theoretically) but the maximum harmonic measured was the 17$^{\rm{th}}$, which corresponds to an intensity enhancement of only 2 orders of magnitude (for more details see~\cite{Marcelo12OE,Marcelo12JMO}). On the other hand, our numerical tools allow a treatment of a very general set of spatial nonhomogeneous fields such as those present in the vicinity of metal nanostructures~\cite{Kim08}, dielectric nanoparticles~\cite{Zherebtsov11}, or metal nanotips~\cite{Herink12}. The kinetic energy for the electrons both direct and rescattered can be classically calculated and compared to quantum mechanical predictions (for more details see e.g~\cite{Marcelo13LPL}).

\begin{figure}[htb]
\centering
\includegraphics[width=0.49\textwidth]{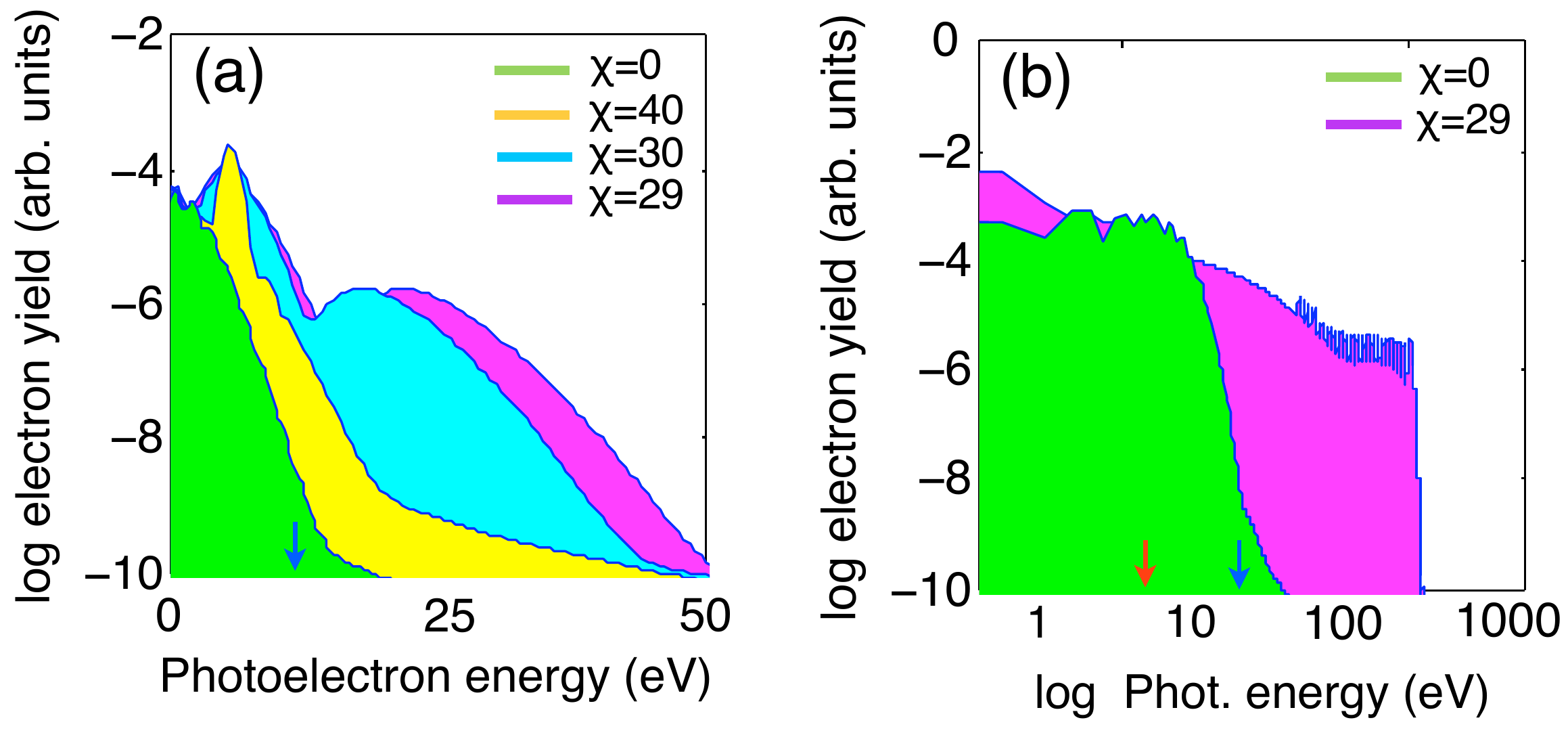}
\caption{Energy resolved photoelectron spectra for Xe atoms driven by an electric enhanced near-field. In panel (a) the laser intensity after interacting with the metal
nanoparticles is $I=2\times10^{13}$ W cm$^{-2}$. We employ $\phi=\pi/2$ (cos-like pulses) and the laser wavelength and number of cycles remain unchanged with respect to the input pulse, i.e. $\lambda=720$ nm and $n_p=5$ (13 fs in total). Panel (b) shows the output laser intensity of $I=5\times10^{13}$ W cm$^{-2}$ (everything else is the same as in panel (a)). The arrow indicates two conventional classical limits: $2 U_p$ (in red) at 5.24 eV and $10U_p$ (in blue) at 26.2 eV, respectively.}  
\label{Figure6ATI}
\end{figure}

We have employed the same parameters as the ones used in Section IXA, but now our aim is to compute the energy resolved photoelectron spectra. In Fig.~\ref{Figure6ATI} we present the photoelectron spectra calculated using 1D-TDSE for Xe atoms and for two different laser intensities, namely $I=2\times10^{13}$ W cm$^{-2}$ (Fig.~\ref{Figure6ATI}(a)) and $I=5\times10^{13}$ W cm$^{-2}$ (Fig.~\ref{Figure6ATI}(b)). In Fig.~\ref{Figure6ATI}(a) each curve presents different values of $\chi$: homogeneous case ($\chi\rightarrow\infty$), $\chi=40$, $\chi=35$ and $\chi=29$. For the homogeneous case there is a visible cutoff at $\approx 10.5$ eV confirming the well known ATI cutoff at $10U_{p}$, which corresponds to those electrons that once ionized return to the core and elastically rescatter. Here, $U_{p}$ is the ponderomotive potential given by $U_{p}=E_{p}^{2}/4\omega^{2}$. On the other hand, for this particular intensity, the cutoff at $2U_{p}$ ($\approx 2.1$ eV) developed by the direct ionized electrons is not visible in the spectrum. 

For the spatial nonhomogeneous cases the cutoff of the rescattered electron is far beyond the classical limit $10U_{p}$, depending on the $\chi$ parameter chosen. As it is depicted in Fig.~\ref{Figure6ATI}(a) the cutoff is extended as we decrease the value of $\chi$. For $\chi= 40$ the cutoff is at around $14$ eV, while for $\chi= 29$ it is around $30$ eV. The low energy region of the photoelectron spectra is sensitive to the atomic potential of the target and one needs to calculate TDSE in full dimensionality in order to model this region adequately. In this paper we are interested in the high energy region of the photoelectron spectra, which is very convenient because it is not greatly affected by the considered atom. Thus by employing 1D-TDSE the conclusions that can be taken from these highly energetic electrons are very reliable.

Figure~\ref{Figure6ATI}(b) shows the photoelectron spectra for the homogeneous case and for $\chi=29$ using a larger laser field intensity of $I=5\times10^{13}$ W cm$^{-2}$, while keeping all other laser parameters fixed. From this plot we observe that the nonhomogeneous character of the laser enhanced electric field introduces a highly nonlinear behavior. For this intensity with $\chi=29$ it is possible to obtain very energetic electrons reaching values of several hundreds of eV. This is a good indication that the nonlinear behavior of the combined system of the metallic nanoparticles and noble gas atoms could pave the way to generate keV electrons with tabletop laser sources. All the above quantum mechanical predictions can be directly confirmed by using classical simulations in the same way as for the case HHG (see Section VIIIE).

%All the above quantum mechanical predictions can be directly confirmed by using classical simulations in the same way as for the case HHG (see Section VIIIE). The only difference is the spatial functional form of the laser electric field. More specifically the Newton equation of motion now reads $\ddot{x}(t) =-E(x,t)(1-x(t)/\chi)$. Note that in the limit $\chi\rightarrow\infty$, we recover the conventional homogeneous case $\ddot{x}(t) =-E(t)$. 

Here we propose generation of high energy photoelectrons using near-enhanced fields by combining metallic nanoparticles and noble gas atoms. Near-enhanced fields present a strong spatial dependence at a nanometer scale and this behavior introduces substantial changes in the laser-matter processes. We have modified the 1D-TDSE to model the ATI phenomenon in noble gases driven by the enhanced near-fields of such nanostructure. We predict a substantial extension in the cutoff position of the energy-resolved photoelectron spectra, far beyond the conventional $10U_p$ classical limit. These new features are well reproduced by classical simulations. Our predictions would pave the way to the production of high energy photoelectrons reaching the keV regime by using a combination of metal nanoparticles and noble gases. In this kind of system each metal nanoparticle configures a laser nanosource with particular characteristics that allow not only the amplification of the input laser field, but also the modification of the laser-matter phenomena due to the strong spatial dependence of the generated coherent electromagnetic radiation.

\section{Other processes}

Most of the approaches applied to theoretically model HHG and ATI, both driven by spatial homogeneous and nonhomogeneous fields, 
are based on the hypothesis that the single active electron approximation (SAE) is good enough to describe both the harmonic emission and the laser ionized electron spectrum. For multielectronic processes, such as NSDI, clearly the SAE is insufficient. Studies of HHG considering two- and multi-electron effects 
have been performed by several authors~\cite{ASanpera1,Grobe1993HHG,Koval2007,Bandrauk2005,ShiLin2013,Santra2006}. 
From these contributions, we conclude that depending on the atomic target properties and the laser frequency-intensity regime, multielectronic effects could play an important role in HHG~\cite{Grobe1993HHG,ASanpera1,Koval2007}. We should mention, however, that all these theoretical approaches have been developed for spatially homogeneous fields and that,  to the best of our knowledge, studies of strong field phenomena driven by spatially inhomogeneous fields beyond the SAE have not been reported yet. 

We have started to investigate how plasmonic fields modify the electron dynamics in a prototypical two-electron systems, namely the He atom and the negative 
hydrogen ion (H$^-$). To this end we employ the numerical solutions of the reduced 1D$\times$1D-TDSE for both the two-active electron (TAE)
and the single active electron (SAE) models. From these models, we plan to trace out the analogies and differences in the HHG process from these two atomic systems, a priori very similar in their intrinsic structure~\cite{AlexisPRL}.

%%%%%%%%%%%%%%%%%%%%%%%%%%%%%%%%%%%%%%%%%%

%\input{Outlook}

%\section{Conclusions}
%%%  Conclusions

\section{Conclusions, Outlook and perspectives}

In this report on progress we have extensively reviewed, from both an experimental and theoretical viewpoints, the recent developments of the atto-nano physics. 

Nowadays, for the first time in the history of AMO physics we have at our disposal laser sources,  which, combined with nanostructures,  generate fields that exhibit spatial variation at a nanometric scale. This is the natural scale of the electron dynamics in atoms, molecules and bulk matter. Consequently, noticeable and profound changes occur in systems interacting with such spatially inhomogeneous fields. Using well known numerical techniques, based on solutions of Maxwell equations, one is able to model both the time and the spatial properties of these aser induced  plasmonic fields. This in the first  important step for the subsequent theoretical modelling of the strong-field physical processes driven by them. 

From a theoretical perspective, in the recent years there has been a profound and continuous activity in atto-nanophysics. Indeed,   all of the theoretical tools developed to tackle strong field processes driven by spatial homogeneous fields have beed generalized and  adapted to this new stage. Several open problems, however, still remain. For instance, the behaviour of complex systems, e.g.~multielectronic atoms and molecules, under the influence of spatial inhomogeneous fields is an unexplored area -- only few attempts to tackle this problem has been recently reported~\cite{Yavuz15,ilhan1,AlexisPRL}. In addition, and just to name another example, it was recently demonstrated that Rydberg atoms could be a plausible alternative as a driven media~\cite{rydberg}.

Several paths could be explored in the future. The manipulation and control of the plasmonic-enhanced fields appears as one them. From an experimental perspective this presents a tremendous challenge, considering the nanometric dimensions of the systems, although several experiments are planned in this direction, for instance combining metal nanotips and molecules in a gas phase. The possibility to tailor the electron trajectories at their natural scale is another avenue to be considered. By employing quantum control tools it would be possible, in principle theoretically, to drive the electron following a certain desired 'target',e.g.~a one which results with the largest possible velocity, now with a time and spatial dependent driving field. The spatial shape of this field could be, subsequently, obtained by engineering a nanostructure.

The quest for HHG from plasmonic nano-structures, joint with an explosive amount of theoretical work, started with the controversial report of a Korean group on HHG from bow-tie metal nano structures~\cite{Kim08}. Let us mention at the end of this report a very recent results of the same group, which clearly seems to be well justified and, as such, opens new perspectives and ways toward efficient HHG in nano-structures. In this recent preprint the authors demonstrate plasmonic HHG experimentally by devising a metal-sapphire nanostructure that provides a solid tip as the HHG emitter instead of gaseous atoms. The fabricated solid tips are made of monocrystalline sapphire surrounded by a gold thin-film layer, and intended to produce coherent extreme ultraviolet (EUV) harmonics by the inter- and intra-band oscillations of electrons driven by the incident laser. The metal-sapphire nanostructure enhances the incident laser field by means of surface plasmon polaritons (SPPs), triggering HHG directly from moderate femtosecond pulses of~0.1 TW cm$^{-2}$ intensities. Measured EUV spectra show odd-order harmonics up to~60 nm wavelengths without the plasma atomic lines typically seen when using gaseous atoms as the HHG emitter. This experimental outcome confirms that the plasmonic HHG approach is a promising way to realize coherent EUV sources for nano-scale near-field applications in spectroscopy, microscopy, lithography, and attosecond physics~\cite{Han2016}.  A new era of ultrafast physics is beginning!

%\cite{Han2016}
%
%
%
%
%
%Reference (preprint – but check maybe is on the arxives? Or is published? Ask them maybe?)
%
%
%
%
%
%High harmonic generation by strongly enhanced femtosecond pulses in
%
%2 metal-sapphire nanostructure
%
%3
%
%4 Seunghwoi Han1,5, Hyunwoong Kim1,5, Yong Woo Kim1, Young-Jin Kim2, Seungchul Kim3, In-Yong Park4
%
%5 and Seung-Woo Kim1,*
%
%6
%
%7 1Department of Mechanical Engineering, Korea Advanced Institute of Science and Technology (KAIST), 291
%
%8 Daehak-ro, Yuseong-gu, Daejeon 305-701, South Korea.
%
%9 2School of Mechanical and Aerospace Engineering, Nanyang Technological University (NTU), 50 Nanyang
%
%10 Avenue, Singapore 639798, Singapore.
%
%11 3Max Planck Center for Attosecond Science, Max Planck POSTECH/KOREA Res. Initiative, Pohang,
%
%12 Gyeongbuk 376-73, South Korea.
%
%13 4Division of Industrial Metrology, Korea Research Institute of Standards and Science (KRISS), Daejeon
%
%14 305-340, South Korea.
%
%15 5These authors equally contributed to this work.
%
%16 *Corresponding author (swk@kaist.ac.kr)

\section{Acknowledgements}
%%%  Acknowledgements

This work was supported by the project ELI--Extreme Light Infrastructure--phase 2 (CZ.02.1.01/0.0/0.0/15\_008/0000162 ) from European Regional Development Fund and in part by Max Planck POSTECH/KOREA Research Initiative Program [Grant No.~2011-0031558] through the National Research Foundation of Korea (NRF) funded by Ministry of Science, ICT \& Future Planning.  A.~S.~L.~acknowledges Max Planck Center for Attosecond Science. M.~K.~acknowledges financial support by the Minerva
Foundation and the Koshland Foundation. J.~A.~P.-H. and L.~R.~acknowledge to the Spanish Ministerio de Econom\'{\i}a y Competitividad (FURIAM Project No.~FIS2013-47741-R) and LaserLab IV Grant Agreement No.~654148. A.~B. is funded by the Imperial College London Junior Research Fellowship. A.~Z.~acknowledges support from UK-EPSRC project EP/J002348/1 'CADAM'. M.L. and A. C. acknowledge the support of Spanish  MINECO (Nacional Plan GRANT FOQUS FIS2013-46768-P, SEVERO OCHOA GRANT/SEV-2015-0522),  Fundaci\'o Cellex, Catalan AGAUR (SGR 2014 874) and 
ERC AdG OSYRIS.  P. H. acknowledges funding via the ERC Consolidator 
Grant NearFieldAtto, DFG-SPP 1840, DFG-SFB 953 and the Gordon and Betty 
Moore Foundation. M.F.C. acknowledges Dennis Luck for helping with the artwork. 

%\bibliography{biblio_marcelo}

\begin{thebibliography}{373}%
\makeatletter
\providecommand \@ifxundefined [1]{%
 \@ifx{#1\undefined}
}%
\providecommand \@ifnum [1]{%
 \ifnum #1\expandafter \@firstoftwo
 \else \expandafter \@secondoftwo
 \fi
}%
\providecommand \@ifx [1]{%
 \ifx #1\expandafter \@firstoftwo
 \else \expandafter \@secondoftwo
 \fi
}%
\providecommand \natexlab [1]{#1}%
\providecommand \enquote  [1]{``#1''}%
\providecommand \bibnamefont  [1]{#1}%
\providecommand \bibfnamefont [1]{#1}%
\providecommand \citenamefont [1]{#1}%
\providecommand \href@noop [0]{\@secondoftwo}%
\providecommand \href [0]{\begingroup \@sanitize@url \@href}%
\providecommand \@href[1]{\@@startlink{#1}\@@href}%
\providecommand \@@href[1]{\endgroup#1\@@endlink}%
\providecommand \@sanitize@url [0]{\catcode `\\12\catcode `\$12\catcode
  `\&12\catcode `\#12\catcode `\^12\catcode `\_12\catcode `\%12\relax}%
\providecommand \@@startlink[1]{}%
\providecommand \@@endlink[0]{}%
\providecommand \url  [0]{\begingroup\@sanitize@url \@url }%
\providecommand \@url [1]{\endgroup\@href {#1}{\urlprefix }}%
\providecommand \urlprefix  [0]{URL }%
\providecommand \Eprint [0]{\href }%
\providecommand \doibase [0]{http://dx.doi.org/}%
\providecommand \selectlanguage [0]{\@gobble}%
\providecommand \bibinfo  [0]{\@secondoftwo}%
\providecommand \bibfield  [0]{\@secondoftwo}%
\providecommand \translation [1]{[#1]}%
\providecommand \BibitemOpen [0]{}%
\providecommand \bibitemStop [0]{}%
\providecommand \bibitemNoStop [0]{.\EOS\space}%
\providecommand \EOS [0]{\spacefactor3000\relax}%
\providecommand \BibitemShut  [1]{\csname bibitem#1\endcsname}%
\let\auto@bib@innerbib\@empty
%</preamble>
\bibitem [{\citenamefont {Agostini}\ \emph {et~al.}(1979)\citenamefont
  {Agostini}, \citenamefont {Fabre}, \citenamefont {Mainfray}, \citenamefont
  {Petite},\ and\ \citenamefont {Rahman}}]{AgostiniATI}%
  \BibitemOpen
  \bibfield  {author} {\bibinfo {author} {\bibnamefont {Agostini},
  \bibfnamefont {P}}, \bibinfo {author} {\bibfnamefont {F.}~\bibnamefont
  {Fabre}}, \bibinfo {author} {\bibfnamefont {G.}~\bibnamefont {Mainfray}},
  \bibinfo {author} {\bibfnamefont {G.}~\bibnamefont {Petite}}, \ and\ \bibinfo
  {author} {\bibfnamefont {N.~K.}\ \bibnamefont {Rahman}}} (\bibinfo {year}
  {1979}),\ \bibfield  {title} {\enquote {\bibinfo {title} {Free-free
  transitions following six-photon ionization of xenon atoms},}\ }\href@noop {}
  {\bibfield  {journal} {\bibinfo  {journal} {Phys. Rev. Lett.}\ }\textbf
  {\bibinfo {volume} {42}},\ \bibinfo {pages} {1127}}\BibitemShut {NoStop}%
\bibitem [{\citenamefont {Ahmad}\ \emph {et~al.}(2009)\citenamefont {Ahmad},
  \citenamefont {Trushin}, \citenamefont {Major}, \citenamefont {Wandt},
  \citenamefont {Klingebiel}, \citenamefont {Wang}, \citenamefont {Pervak},
  \citenamefont {Popp}, \citenamefont {Siebold}, \citenamefont {Krausz},\ and\
  \citenamefont {Karsch}}]{Ahmad09}%
  \BibitemOpen
  \bibfield  {author} {\bibinfo {author} {\bibnamefont {Ahmad}, \bibfnamefont
  {I}}, \bibinfo {author} {\bibfnamefont {S.~A.}\ \bibnamefont {Trushin}},
  \bibinfo {author} {\bibfnamefont {Z.}~\bibnamefont {Major}}, \bibinfo
  {author} {\bibfnamefont {C.}~\bibnamefont {Wandt}}, \bibinfo {author}
  {\bibfnamefont {S.}~\bibnamefont {Klingebiel}}, \bibinfo {author}
  {\bibfnamefont {T.-J.}\ \bibnamefont {Wang}}, \bibinfo {author}
  {\bibfnamefont {V.}~\bibnamefont {Pervak}}, \bibinfo {author} {\bibfnamefont
  {A.}~\bibnamefont {Popp}}, \bibinfo {author} {\bibfnamefont {M.}~\bibnamefont
  {Siebold}}, \bibinfo {author} {\bibfnamefont {F.}~\bibnamefont {Krausz}}, \
  and\ \bibinfo {author} {\bibfnamefont {S.}~\bibnamefont {Karsch}}} (\bibinfo
  {year} {2009}),\ \bibfield  {title} {\enquote {\bibinfo {title} {Frontend
  light source for short-pulse pumped opcpa system},}\ }\href@noop {}
  {\bibfield  {journal} {\bibinfo  {journal} {Appl. Phys. B}\ }\textbf
  {\bibinfo {volume} {97}},\ \bibinfo {pages} {529--536}}\BibitemShut {NoStop}%
\bibitem [{\citenamefont {Akturk}\ \emph {et~al.}(2008)\citenamefont {Akturk},
  \citenamefont {D'Amico},\ and\ \citenamefont
  {Mysyrowicz}}]{akturk_measuring_2008}%
  \BibitemOpen
  \bibfield  {author} {\bibinfo {author} {\bibnamefont {Akturk}, \bibfnamefont
  {S}}, \bibinfo {author} {\bibfnamefont {C.}~\bibnamefont {D'Amico}}, \ and\
  \bibinfo {author} {\bibfnamefont {A.}~\bibnamefont {Mysyrowicz}}} (\bibinfo
  {year} {2008}),\ \bibfield  {title} {\enquote {\bibinfo {title} {Measuring
  ultrashort pulses in the single-cycle regime using frequency-resolved optical
  gating},}\ }\href@noop {} {\bibfield  {journal} {\bibinfo  {journal} {J. Opt.
  Soc. Am. B}\ }\textbf {\bibinfo {volume} {25}},\ \bibinfo {pages}
  {A63--A69}}\BibitemShut {NoStop}%
\bibitem [{\citenamefont {Altucci}\ \emph {et~al.}(2011)\citenamefont
  {Altucci}, \citenamefont {Tisch},\ and\ \citenamefont
  {Velotta}}]{altucci2011single}%
  \BibitemOpen
  \bibfield  {author} {\bibinfo {author} {\bibnamefont {Altucci}, \bibfnamefont
  {C}}, \bibinfo {author} {\bibfnamefont {J.~W.~G.}\ \bibnamefont {Tisch}}, \
  and\ \bibinfo {author} {\bibfnamefont {R.}~\bibnamefont {Velotta}}} (\bibinfo
  {year} {2011}),\ \bibfield  {title} {\enquote {\bibinfo {title} {Single
  attosecond light pulses from multi-cycle laser sources},}\ }\href@noop {}
  {\bibfield  {journal} {\bibinfo  {journal} {J. of Mod. Opt.}\ }\textbf
  {\bibinfo {volume} {58}},\ \bibinfo {pages} {1585--1610}}\BibitemShut
  {NoStop}%
\bibitem [{\citenamefont {Ammosov}\ \emph {et~al.}(1986)\citenamefont
  {Ammosov}, \citenamefont {Delone},\ and\ \citenamefont
  {Kra\u{\i}nov}}]{ADK1986}%
  \BibitemOpen
  \bibfield  {author} {\bibinfo {author} {\bibnamefont {Ammosov}, \bibfnamefont
  {M~V}}, \bibinfo {author} {\bibfnamefont {N.~B.}\ \bibnamefont {Delone}}, \
  and\ \bibinfo {author} {\bibfnamefont {V.~P.}\ \bibnamefont {Kra\u{\i}nov}}}
  (\bibinfo {year} {1986}),\ \bibfield  {title} {\enquote {\bibinfo {title}
  {Tunnel ionization of complex atoms and of atomic ions in an alternating
  electromagnetic field},}\ }\href@noop {} {\bibfield  {journal} {\bibinfo
  {journal} {Sov. Phys. JETP}\ }\textbf {\bibinfo {volume} {64}},\ \bibinfo
  {pages} {1191}}\BibitemShut {NoStop}%
\bibitem [{\citenamefont {Anderson}\ \emph {et~al.}(2010)\citenamefont
  {Anderson}, \citenamefont {Deryckx}, \citenamefont {Xu}, \citenamefont
  {Steinmeyer},\ and\ \citenamefont {Raschke}}]{Anderson10}%
  \BibitemOpen
  \bibfield  {author} {\bibinfo {author} {\bibnamefont {Anderson},
  \bibfnamefont {A}}, \bibinfo {author} {\bibfnamefont {K.~S.}\ \bibnamefont
  {Deryckx}}, \bibinfo {author} {\bibfnamefont {X.~G.}\ \bibnamefont {Xu}},
  \bibinfo {author} {\bibfnamefont {G.}~\bibnamefont {Steinmeyer}}, \ and\
  \bibinfo {author} {\bibfnamefont {M.~B.}\ \bibnamefont {Raschke}}} (\bibinfo
  {year} {2010}),\ \bibfield  {title} {\enquote {\bibinfo {title}
  {Few-femtosecond plasmon dephasing of a single metallic nanostructure from
  optical response function reconstruction by interferometric frequency
  resolved optical gating},}\ }\href@noop {} {\bibfield  {journal} {\bibinfo
  {journal} {Nano Lett.}\ }\textbf {\bibinfo {volume} {10}},\ \bibinfo {pages}
  {2519--2524}}\BibitemShut {NoStop}%
\bibitem [{\citenamefont {Anker}\ \emph {et~al.}(2008)\citenamefont {Anker},
  \citenamefont {Hall}, \citenamefont {Lyandres}, \citenamefont {Shah},
  \citenamefont {Zhao},\ and\ \citenamefont {Van~Duyne}}]{Anker08}%
  \BibitemOpen
  \bibfield  {author} {\bibinfo {author} {\bibnamefont {Anker}, \bibfnamefont
  {J~N}}, \bibinfo {author} {\bibfnamefont {W.~P.}\ \bibnamefont {Hall}},
  \bibinfo {author} {\bibfnamefont {O.}~\bibnamefont {Lyandres}}, \bibinfo
  {author} {\bibfnamefont {N.~C.}\ \bibnamefont {Shah}}, \bibinfo {author}
  {\bibfnamefont {J.}~\bibnamefont {Zhao}}, \ and\ \bibinfo {author}
  {\bibfnamefont {R.~P.}\ \bibnamefont {Van~Duyne}}} (\bibinfo {year} {2008}),\
  \bibfield  {title} {\enquote {\bibinfo {title} {Biosensing with plasmonic
  nanosensors},}\ }\href@noop {} {\bibfield  {journal} {\bibinfo  {journal}
  {Nat. Mater.}\ }\textbf {\bibinfo {volume} {7}},\ \bibinfo {pages}
  {442--453}}\BibitemShut {NoStop}%
\bibitem [{\citenamefont {Antoine}\ \emph {et~al.}(1996)\citenamefont
  {Antoine}, \citenamefont {L'Huillier},\ and\ \citenamefont
  {Lewenstein}}]{antoine1996attosecond}%
  \BibitemOpen
  \bibfield  {author} {\bibinfo {author} {\bibnamefont {Antoine}, \bibfnamefont
  {P}}, \bibinfo {author} {\bibfnamefont {A.}~\bibnamefont {L'Huillier}}, \
  and\ \bibinfo {author} {\bibfnamefont {M.}~\bibnamefont {Lewenstein}}}
  (\bibinfo {year} {1996}),\ \bibfield  {title} {\enquote {\bibinfo {title}
  {Attosecond pulse trains using high--order harmonics},}\ }\href@noop {}
  {\bibfield  {journal} {\bibinfo  {journal} {Phys. Rev. Lett.}\ }\textbf
  {\bibinfo {volume} {77}},\ \bibinfo {pages} {1234}}\BibitemShut {NoStop}%
\bibitem [{\citenamefont {Apolonski}\ \emph {et~al.}(2004)\citenamefont
  {Apolonski}, \citenamefont {Dombi}, \citenamefont {Paulus}, \citenamefont
  {Kakehata}, \citenamefont {Holzwarth}, \citenamefont {Udem}, \citenamefont
  {Lemell}, \citenamefont {Torizuka}, \citenamefont {Burgd\"orfer},
  \citenamefont {H\"ansch},\ and\ \citenamefont {Krausz}}]{apolonski}%
  \BibitemOpen
  \bibfield  {author} {\bibinfo {author} {\bibnamefont {Apolonski},
  \bibfnamefont {A}}, \bibinfo {author} {\bibfnamefont {P.}~\bibnamefont
  {Dombi}}, \bibinfo {author} {\bibfnamefont {G.~G.}\ \bibnamefont {Paulus}},
  \bibinfo {author} {\bibfnamefont {M.}~\bibnamefont {Kakehata}}, \bibinfo
  {author} {\bibfnamefont {R.}~\bibnamefont {Holzwarth}}, \bibinfo {author}
  {\bibfnamefont {Th.}\ \bibnamefont {Udem}}, \bibinfo {author} {\bibfnamefont
  {Ch.}\ \bibnamefont {Lemell}}, \bibinfo {author} {\bibfnamefont
  {K.}~\bibnamefont {Torizuka}}, \bibinfo {author} {\bibfnamefont
  {J.}~\bibnamefont {Burgd\"orfer}}, \bibinfo {author} {\bibfnamefont {T.~W.}\
  \bibnamefont {H\"ansch}}, \ and\ \bibinfo {author} {\bibfnamefont
  {F.}~\bibnamefont {Krausz}}} (\bibinfo {year} {2004}),\ \bibfield  {title}
  {\enquote {\bibinfo {title} {Observation of light-phase-sensitive
  photoemission from a metal},}\ }\href@noop {} {\bibfield  {journal} {\bibinfo
   {journal} {Phys. Rev. Lett.}\ }\textbf {\bibinfo {volume} {92}},\ \bibinfo
  {pages} {073902}}\BibitemShut {NoStop}%
\bibitem [{\citenamefont {Apolonski}\ \emph {et~al.}(2000)\citenamefont
  {Apolonski}, \citenamefont {Poppe}, \citenamefont {Tempea}, \citenamefont
  {Spielmann}, \citenamefont {Udem}, \citenamefont {Holzwarth}, \citenamefont
  {H\"ansch},\ and\ \citenamefont {Krausz}}]{apolonski_controlling_2000}%
  \BibitemOpen
  \bibfield  {author} {\bibinfo {author} {\bibnamefont {Apolonski},
  \bibfnamefont {A}}, \bibinfo {author} {\bibfnamefont {A.}~\bibnamefont
  {Poppe}}, \bibinfo {author} {\bibfnamefont {G.}~\bibnamefont {Tempea}},
  \bibinfo {author} {\bibfnamefont {C.}~\bibnamefont {Spielmann}}, \bibinfo
  {author} {\bibfnamefont {T.}~\bibnamefont {Udem}}, \bibinfo {author}
  {\bibfnamefont {R.}~\bibnamefont {Holzwarth}}, \bibinfo {author}
  {\bibfnamefont {T.~W.}\ \bibnamefont {H\"ansch}}, \ and\ \bibinfo {author}
  {\bibfnamefont {F.}~\bibnamefont {Krausz}}} (\bibinfo {year} {2000}),\
  \bibfield  {title} {\enquote {\bibinfo {title} {Controlling the phase
  evolution of few-cycle light pulses},}\ }\href@noop {} {\bibfield  {journal}
  {\bibinfo  {journal} {Phys. Rev. Lett.}\ }\textbf {\bibinfo {volume} {85}},\
  \bibinfo {pages} {740--743}}\BibitemShut {NoStop}%
\bibitem [{\citenamefont {Arb\'o}\ \emph {et~al.}(2008)\citenamefont {Arb\'o},
  \citenamefont {Miraglia}, \citenamefont {Gravielle}, \citenamefont
  {Schiessl}, \citenamefont {Persson},\ and\ \citenamefont
  {Burgd\"orfer}}]{arbo1}%
  \BibitemOpen
  \bibfield  {author} {\bibinfo {author} {\bibnamefont {Arb\'o}, \bibfnamefont
  {D~G}}, \bibinfo {author} {\bibfnamefont {J.~E.}\ \bibnamefont {Miraglia}},
  \bibinfo {author} {\bibfnamefont {M.~S.}\ \bibnamefont {Gravielle}}, \bibinfo
  {author} {\bibfnamefont {K.}~\bibnamefont {Schiessl}}, \bibinfo {author}
  {\bibfnamefont {E.}~\bibnamefont {Persson}}, \ and\ \bibinfo {author}
  {\bibfnamefont {J.}~\bibnamefont {Burgd\"orfer}}} (\bibinfo {year} {2008}),\
  \bibfield  {title} {\enquote {\bibinfo {title} {Coulomb-volkov approximation
  for near-threshold ionization by short laser pulses},}\ }\href@noop {}
  {\bibfield  {journal} {\bibinfo  {journal} {Phys. Rev. A}\ }\textbf {\bibinfo
  {volume} {77}},\ \bibinfo {pages} {013401}}\BibitemShut {NoStop}%
\bibitem [{\citenamefont {Arnold}\ \emph {et~al.}(2013)\citenamefont {Arnold},
  \citenamefont {Fahrenberger}, \citenamefont {Holm}, \citenamefont {Lenz},
  \citenamefont {Bolten}, \citenamefont {Dachsel}, \citenamefont {Halver},
  \citenamefont {Kabadshow}, \citenamefont {G\"ahler}, \citenamefont {Heber},
  \citenamefont {Iseringhausen}, \citenamefont {Hofmann}, \citenamefont
  {Pippig}, \citenamefont {Potts},\ and\ \citenamefont {Sutmann}}]{Arnold2013}%
  \BibitemOpen
  \bibfield  {author} {\bibinfo {author} {\bibnamefont {Arnold}, \bibfnamefont
  {A}}, \bibinfo {author} {\bibfnamefont {F.}~\bibnamefont {Fahrenberger}},
  \bibinfo {author} {\bibfnamefont {C.}~\bibnamefont {Holm}}, \bibinfo {author}
  {\bibfnamefont {O.}~\bibnamefont {Lenz}}, \bibinfo {author} {\bibfnamefont
  {M.}~\bibnamefont {Bolten}}, \bibinfo {author} {\bibfnamefont
  {H.}~\bibnamefont {Dachsel}}, \bibinfo {author} {\bibfnamefont
  {R.}~\bibnamefont {Halver}}, \bibinfo {author} {\bibfnamefont
  {I.}~\bibnamefont {Kabadshow}}, \bibinfo {author} {\bibfnamefont
  {F.}~\bibnamefont {G\"ahler}}, \bibinfo {author} {\bibfnamefont
  {F.}~\bibnamefont {Heber}}, \bibinfo {author} {\bibfnamefont
  {J.}~\bibnamefont {Iseringhausen}}, \bibinfo {author} {\bibfnamefont
  {M.}~\bibnamefont {Hofmann}}, \bibinfo {author} {\bibfnamefont
  {M.}~\bibnamefont {Pippig}}, \bibinfo {author} {\bibfnamefont
  {D.}~\bibnamefont {Potts}}, \ and\ \bibinfo {author} {\bibfnamefont
  {G.}~\bibnamefont {Sutmann}}} (\bibinfo {year} {2013}),\ \bibfield  {title}
  {\enquote {\bibinfo {title} {Comparison of scalable fast methods for
  long-range interactions},}\ }\href@noop {} {\bibfield  {journal} {\bibinfo
  {journal} {Phys. Rev. E}\ }\textbf {\bibinfo {volume} {88}},\ \bibinfo
  {pages} {063308}}\BibitemShut {NoStop}%
\bibitem [{\citenamefont {Bagheri}\ \emph {et~al.}(2015)\citenamefont
  {Bagheri}, \citenamefont {Weber}, \citenamefont {Gissibl}, \citenamefont
  {Weiss}, \citenamefont {Neubrech},\ and\ \citenamefont
  {Giessen}}]{bagheri_fabrication_2015}%
  \BibitemOpen
  \bibfield  {author} {\bibinfo {author} {\bibnamefont {Bagheri}, \bibfnamefont
  {S}}, \bibinfo {author} {\bibfnamefont {K.}~\bibnamefont {Weber}}, \bibinfo
  {author} {\bibfnamefont {T.}~\bibnamefont {Gissibl}}, \bibinfo {author}
  {\bibfnamefont {T.}~\bibnamefont {Weiss}}, \bibinfo {author} {\bibfnamefont
  {F.}~\bibnamefont {Neubrech}}, \ and\ \bibinfo {author} {\bibfnamefont
  {H.}~\bibnamefont {Giessen}}} (\bibinfo {year} {2015}),\ \bibfield  {title}
  {\enquote {\bibinfo {title} {Fabrication of {Square}-{Centimeter} {Plasmonic}
  {Nanoantenna} {Arrays} by {Femtosecond} {Direct} {Laser} {Writing}
  {Lithography}: {Effects} of {Collective} {Excitations} on {SEIRA}
  {Enhancement}},}\ }\href@noop {} {\bibfield  {journal} {\bibinfo  {journal}
  {ACS Photonics}\ }\textbf {\bibinfo {volume} {2}},\ \bibinfo {pages}
  {779--786}}\BibitemShut {NoStop}%
\bibitem [{\citenamefont {Bainbridge}\ and\ \citenamefont
  {Bryan}(2014)}]{Bainbridge2014}%
  \BibitemOpen
  \bibfield  {author} {\bibinfo {author} {\bibnamefont {Bainbridge},
  \bibfnamefont {A~R}}, \ and\ \bibinfo {author} {\bibfnamefont {W.~A.}\
  \bibnamefont {Bryan}}} (\bibinfo {year} {2014}),\ \bibfield  {title}
  {\enquote {\bibinfo {title} {Velocity map imaging of femtosecond laser
  induced photoelectron emission from metal nanotips},}\ }\href@noop {}
  {\bibfield  {journal} {\bibinfo  {journal} {New~J.~Phys.}\ }\textbf {\bibinfo
  {volume} {16}},\ \bibinfo {pages} {103031}}\BibitemShut {NoStop}%
\bibitem [{\citenamefont {Balciunas}\ \emph {et~al.}(2015)\citenamefont
  {Balciunas}, \citenamefont {Fourcade-Dutin}, \citenamefont {Fan},
  \citenamefont {Witting}, \citenamefont {Voronin}, \citenamefont {Zheltikov},
  \citenamefont {Gerome}, \citenamefont {Paulus}, \citenamefont {Baltuska},\
  and\ \citenamefont {Benabid}}]{balciunas_strong_field_2015}%
  \BibitemOpen
  \bibfield  {author} {\bibinfo {author} {\bibnamefont {Balciunas},
  \bibfnamefont {T}}, \bibinfo {author} {\bibfnamefont {C.}~\bibnamefont
  {Fourcade-Dutin}}, \bibinfo {author} {\bibfnamefont {G.}~\bibnamefont {Fan}},
  \bibinfo {author} {\bibfnamefont {T.}~\bibnamefont {Witting}}, \bibinfo
  {author} {\bibfnamefont {A.~A.}\ \bibnamefont {Voronin}}, \bibinfo {author}
  {\bibfnamefont {A.~M.}\ \bibnamefont {Zheltikov}}, \bibinfo {author}
  {\bibfnamefont {F.}~\bibnamefont {Gerome}}, \bibinfo {author} {\bibfnamefont
  {G.~G.}\ \bibnamefont {Paulus}}, \bibinfo {author} {\bibfnamefont
  {A.}~\bibnamefont {Baltuska}}, \ and\ \bibinfo {author} {\bibfnamefont
  {F.}~\bibnamefont {Benabid}}} (\bibinfo {year} {2015}),\ \bibfield  {title}
  {\enquote {\bibinfo {title} {A strong-field driver in the single-cycle regime
  based on self-compression in a kagome fibre},}\ }\href@noop {} {\bibfield
  {journal} {\bibinfo  {journal} {Nat. Comm.}\ }\textbf {\bibinfo {volume}
  {6}},\ \bibinfo {pages} {6117}}\BibitemShut {NoStop}%
\bibitem [{\citenamefont {Baltuska}\ \emph
  {et~al.}(2003{\natexlab{a}})\citenamefont {Baltuska}, \citenamefont {Udem},
  \citenamefont {Uiberacker}, \citenamefont {Hentschel}, \citenamefont
  {Goulielmakis}, \citenamefont {Gohle}, \citenamefont {Holzwarth},
  \citenamefont {Yakovlev}, \citenamefont {Scrinzi}, \citenamefont {Hansch},\
  and\ \citenamefont {Krausz}}]{baltuska_attosecond_2003}%
  \BibitemOpen
  \bibfield  {author} {\bibinfo {author} {\bibnamefont {Baltuska},
  \bibfnamefont {A}}, \bibinfo {author} {\bibfnamefont {Th.}\ \bibnamefont
  {Udem}}, \bibinfo {author} {\bibfnamefont {M.}~\bibnamefont {Uiberacker}},
  \bibinfo {author} {\bibfnamefont {M.}~\bibnamefont {Hentschel}}, \bibinfo
  {author} {\bibfnamefont {E.}~\bibnamefont {Goulielmakis}}, \bibinfo {author}
  {\bibfnamefont {Ch.}\ \bibnamefont {Gohle}}, \bibinfo {author} {\bibfnamefont
  {R.}~\bibnamefont {Holzwarth}}, \bibinfo {author} {\bibfnamefont {V.~S.}\
  \bibnamefont {Yakovlev}}, \bibinfo {author} {\bibfnamefont {A.}~\bibnamefont
  {Scrinzi}}, \bibinfo {author} {\bibfnamefont {T.~W.}\ \bibnamefont {Hansch}},
  \ and\ \bibinfo {author} {\bibfnamefont {F.}~\bibnamefont {Krausz}}}
  (\bibinfo {year} {2003}{\natexlab{a}}),\ \bibfield  {title} {\enquote
  {\bibinfo {title} {Attosecond control of electronic processes by intense
  light fields},}\ }\href@noop {} {\bibfield  {journal} {\bibinfo  {journal}
  {Nature}\ }\textbf {\bibinfo {volume} {421}},\ \bibinfo {pages}
  {611--615}}\BibitemShut {NoStop}%
\bibitem [{\citenamefont {Baltuska}\ \emph
  {et~al.}(2003{\natexlab{b}})\citenamefont {Baltuska}, \citenamefont
  {Uiberacker}, \citenamefont {Goulielmakis}, \citenamefont {Kienberger},
  \citenamefont {Yakovlev}, \citenamefont {Udem}, \citenamefont {Hansch},\ and\
  \citenamefont {Krausz}}]{baltuska_phase_controlled_2003}%
  \BibitemOpen
  \bibfield  {author} {\bibinfo {author} {\bibnamefont {Baltuska},
  \bibfnamefont {A}}, \bibinfo {author} {\bibfnamefont {M.}~\bibnamefont
  {Uiberacker}}, \bibinfo {author} {\bibfnamefont {E.}~\bibnamefont
  {Goulielmakis}}, \bibinfo {author} {\bibfnamefont {R.}~\bibnamefont
  {Kienberger}}, \bibinfo {author} {\bibfnamefont {V.~S.}\ \bibnamefont
  {Yakovlev}}, \bibinfo {author} {\bibfnamefont {T.}~\bibnamefont {Udem}},
  \bibinfo {author} {\bibfnamefont {T.~W.}\ \bibnamefont {Hansch}}, \ and\
  \bibinfo {author} {\bibfnamefont {F.}~\bibnamefont {Krausz}}} (\bibinfo
  {year} {2003}{\natexlab{b}}),\ \bibfield  {title} {\enquote {\bibinfo {title}
  {Phase-controlled amplification of few-cycle laser pulses},}\ }\href@noop {}
  {\bibfield  {journal} {\bibinfo  {journal} {IEEE Journal of Selected Topics
  in Quantum Electronics}\ }\textbf {\bibinfo {volume} {9}},\ \bibinfo {pages}
  {972--989}}\BibitemShut {NoStop}%
\bibitem [{\citenamefont {Bandrauk}\ and\ \citenamefont
  {Lu}(2005)}]{Bandrauk2005}%
  \BibitemOpen
  \bibfield  {author} {\bibinfo {author} {\bibnamefont {Bandrauk},
  \bibfnamefont {A~D}}, \ and\ \bibinfo {author} {\bibfnamefont {H.-Z.}\
  \bibnamefont {Lu}}} (\bibinfo {year} {2005}),\ \bibfield  {title} {\enquote
  {\bibinfo {title} {Harmonic generation in a 1d model of {H}$_2$ with single
  and double ionization},}\ }\href@noop {} {\bibfield  {journal} {\bibinfo
  {journal} {J. Phys. B}\ }\textbf {\bibinfo {volume} {38}},\ \bibinfo {pages}
  {2529}}\BibitemShut {NoStop}%
\bibitem [{\citenamefont {Barwick}\ \emph {et~al.}(2007)\citenamefont
  {Barwick}, \citenamefont {Corder}, \citenamefont {Strohaber}, \citenamefont
  {Chandler-Smith}, \citenamefont {Uiterwaal},\ and\ \citenamefont
  {Batelaan}}]{Barwick2007}%
  \BibitemOpen
  \bibfield  {author} {\bibinfo {author} {\bibnamefont {Barwick}, \bibfnamefont
  {B}}, \bibinfo {author} {\bibfnamefont {C.}~\bibnamefont {Corder}}, \bibinfo
  {author} {\bibfnamefont {J.}~\bibnamefont {Strohaber}}, \bibinfo {author}
  {\bibfnamefont {N.}~\bibnamefont {Chandler-Smith}}, \bibinfo {author}
  {\bibfnamefont {C.}~\bibnamefont {Uiterwaal}}, \ and\ \bibinfo {author}
  {\bibfnamefont {H.}~\bibnamefont {Batelaan}}} (\bibinfo {year} {2007}),\
  \bibfield  {title} {\enquote {\bibinfo {title} {Laser-induced ultrafast
  electron emission from a field emission tip},}\ }\href@noop {} {\bibfield
  {journal} {\bibinfo  {journal} {New~J.~Phys.}\ }\textbf {\bibinfo {volume}
  {9}},\ \bibinfo {pages} {142}}\BibitemShut {NoStop}%
\bibitem [{\citenamefont {Barwick}\ \emph {et~al.}(2009)\citenamefont
  {Barwick}, \citenamefont {Flannigan},\ and\ \citenamefont
  {Zewail}}]{Barwick2009}%
  \BibitemOpen
  \bibfield  {author} {\bibinfo {author} {\bibnamefont {Barwick}, \bibfnamefont
  {B}}, \bibinfo {author} {\bibfnamefont {D.~J.}\ \bibnamefont {Flannigan}}, \
  and\ \bibinfo {author} {\bibfnamefont {A.~H.}\ \bibnamefont {Zewail}}}
  (\bibinfo {year} {2009}),\ \bibfield  {title} {\enquote {\bibinfo {title}
  {Photon-induced near-field electron microscopy},}\ }\href@noop {} {\bibfield
  {journal} {\bibinfo  {journal} {Nature}\ }\textbf {\bibinfo {volume}
  {462}}~(\bibinfo {number} {7275}),\ \bibinfo {pages} {902--906}}\BibitemShut
  {NoStop}%
\bibitem [{\citenamefont {Batani}\ \emph {et~al.}(2001)\citenamefont {Batani},
  \citenamefont {Joachain}, \citenamefont {Martellucci},\ and\ \citenamefont
  {Chester}}]{Joachain2}%
  \BibitemOpen
  \bibfield  {author} {\bibinfo {author} {\bibnamefont {Batani}, \bibfnamefont
  {D}}, \bibinfo {author} {\bibfnamefont {C.~J.}\ \bibnamefont {Joachain}},
  \bibinfo {author} {\bibfnamefont {S.}~\bibnamefont {Martellucci}}, \ and\
  \bibinfo {author} {\bibfnamefont {A.~N.}\ \bibnamefont {Chester}}} (\bibinfo
  {year} {2001}),\ \href@noop {} {\emph {\bibinfo {title} {Atoms, Solids, and
  Plasmas in Super-Intense Laser Fields}}}\ (\bibinfo  {publisher} {Kluwer
  Academic/Plenum},\ \bibinfo {address} {New York})\BibitemShut {NoStop}%
\bibitem [{\citenamefont {Becker}\ \emph {et~al.}(2002)\citenamefont {Becker},
  \citenamefont {Grasbon}, \citenamefont {Kopold}, \citenamefont
  {Milo\u{s}evi\'c}, \citenamefont {Paulus},\ and\ \citenamefont
  {Walther}}]{Becker02Chapter}%
  \BibitemOpen
  \bibfield  {author} {\bibinfo {author} {\bibnamefont {Becker}, \bibfnamefont
  {W}}, \bibinfo {author} {\bibfnamefont {F.}~\bibnamefont {Grasbon}}, \bibinfo
  {author} {\bibfnamefont {R.}~\bibnamefont {Kopold}}, \bibinfo {author}
  {\bibfnamefont {D.~B.}\ \bibnamefont {Milo\u{s}evi\'c}}, \bibinfo {author}
  {\bibfnamefont {G.~G.}\ \bibnamefont {Paulus}}, \ and\ \bibinfo {author}
  {\bibfnamefont {H.}~\bibnamefont {Walther}}} (\bibinfo {year} {2002}),\
  \bibfield  {title} {\enquote {\bibinfo {title} {Above-threshold ionization:
  From classical features to quantum effects},}\ }in\ \href@noop {} {\emph
  {\bibinfo {booktitle} {Advances In Atomic, Molecular, and Optical
  Physics}}},\ \bibinfo {editor} {edited by\ \bibinfo {editor} {\bibfnamefont
  {B.}~\bibnamefont {Bederson}}\ and\ \bibinfo {editor} {\bibfnamefont
  {H.}~\bibnamefont {Walter}}}\ (\bibinfo  {publisher} {Academic Press},\
  \bibinfo {address} {San Diego})\ p.~\bibinfo {pages} {35}\BibitemShut
  {NoStop}%
\bibitem [{\citenamefont {Belshaw}\ \emph {et~al.}(2012)\citenamefont
  {Belshaw}, \citenamefont {Calegari}, \citenamefont {Duffy}, \citenamefont
  {Trabattoni}, \citenamefont {Poletto}, \citenamefont {Nisoli},\ and\
  \citenamefont {Greenwood}}]{calegari2012}%
  \BibitemOpen
  \bibfield  {author} {\bibinfo {author} {\bibnamefont {Belshaw}, \bibfnamefont
  {L}}, \bibinfo {author} {\bibfnamefont {F.}~\bibnamefont {Calegari}},
  \bibinfo {author} {\bibfnamefont {M.J.}\ \bibnamefont {Duffy}}, \bibinfo
  {author} {\bibfnamefont {A.}~\bibnamefont {Trabattoni}}, \bibinfo {author}
  {\bibfnamefont {L.}~\bibnamefont {Poletto}}, \bibinfo {author} {\bibfnamefont
  {M.}~\bibnamefont {Nisoli}}, \ and\ \bibinfo {author} {\bibfnamefont {J.B.}\
  \bibnamefont {Greenwood}}} (\bibinfo {year} {2012}),\ \bibfield  {title}
  {\enquote {\bibinfo {title} {Observation of ultrafast charge migration in an
  amino acid},}\ }\href@noop {} {\bibfield  {journal} {\bibinfo  {journal} {J.
  Phys. Chem. Lett.}\ }\textbf {\bibinfo {volume} {3}},\ \bibinfo {pages}
  {3751--3754}}\BibitemShut {NoStop}%
\bibitem [{\citenamefont {Bergues}\ \emph {et~al.}(2015)\citenamefont
  {Bergues}, \citenamefont {K\"ubel}, \citenamefont {Kling}, \citenamefont
  {Burger},\ and\ \citenamefont {Kling}}]{bergues15}%
  \BibitemOpen
  \bibfield  {author} {\bibinfo {author} {\bibnamefont {Bergues}, \bibfnamefont
  {B}}, \bibinfo {author} {\bibfnamefont {M.}~\bibnamefont {K\"ubel}}, \bibinfo
  {author} {\bibfnamefont {N.~G.}\ \bibnamefont {Kling}}, \bibinfo {author}
  {\bibfnamefont {C.}~\bibnamefont {Burger}}, \ and\ \bibinfo {author}
  {\bibfnamefont {M.~F.}\ \bibnamefont {Kling}}} (\bibinfo {year} {2015}),\
  \bibfield  {title} {\enquote {\bibinfo {title} {Single-cycle non-sequential
  double ionization},}\ }\href@noop {} {\bibfield  {journal} {\bibinfo
  {journal} {IEEE J. of Sel. Top. in Quant. Elect.}\ }\textbf {\bibinfo
  {volume} {21}},\ \bibinfo {pages} {8701009}}\BibitemShut {NoStop}%
\bibitem [{\citenamefont {Berweger}\ \emph {et~al.}(2012)\citenamefont
  {Berweger}, \citenamefont {Atkin}, \citenamefont {Olmon},\ and\ \citenamefont
  {Raschke}}]{Berweger2012}%
  \BibitemOpen
  \bibfield  {author} {\bibinfo {author} {\bibnamefont {Berweger},
  \bibfnamefont {S}}, \bibinfo {author} {\bibfnamefont {J.~M.}\ \bibnamefont
  {Atkin}}, \bibinfo {author} {\bibfnamefont {R.~L.}\ \bibnamefont {Olmon}}, \
  and\ \bibinfo {author} {\bibfnamefont {M.~B.}\ \bibnamefont {Raschke}}}
  (\bibinfo {year} {2012}),\ \bibfield  {title} {\enquote {\bibinfo {title}
  {Light on the tip of a needle: Plasmonic nanofocusing for spectroscopy on the
  nanoscale},}\ }\href@noop {} {\bibfield  {journal} {\bibinfo  {journal} {J.
  Phys. Chem. Lett.}\ }\textbf {\bibinfo {volume} {3}},\ \bibinfo {pages}
  {945--952}}\BibitemShut {NoStop}%
\bibitem [{\citenamefont {Bionta}\ \emph {et~al.}(2014)\citenamefont {Bionta},
  \citenamefont {Chalopin}, \citenamefont {Champeaux}, \citenamefont {Faure},
  \citenamefont {Masseboeuf}, \citenamefont {Moretto-Capelle},\ and\
  \citenamefont {Chatel}}]{Bionta2014}%
  \BibitemOpen
  \bibfield  {author} {\bibinfo {author} {\bibnamefont {Bionta}, \bibfnamefont
  {M~R}}, \bibinfo {author} {\bibfnamefont {B.}~\bibnamefont {Chalopin}},
  \bibinfo {author} {\bibfnamefont {J.P.}\ \bibnamefont {Champeaux}}, \bibinfo
  {author} {\bibfnamefont {S.}~\bibnamefont {Faure}}, \bibinfo {author}
  {\bibfnamefont {A.}~\bibnamefont {Masseboeuf}}, \bibinfo {author}
  {\bibfnamefont {P.}~\bibnamefont {Moretto-Capelle}}, \ and\ \bibinfo {author}
  {\bibfnamefont {B.}~\bibnamefont {Chatel}}} (\bibinfo {year} {2014}),\
  \bibfield  {title} {\enquote {\bibinfo {title} {Laser-induced electron
  emission from a tungsten nanotip: identifying above threshold photoemission
  using energy-resolved laser power dependencies},}\ }\href@noop {} {\bibfield
  {journal} {\bibinfo  {journal} {J. Mod. Opt.}\ }\textbf {\bibinfo {volume}
  {61}},\ \bibinfo {pages} {833--838}}\BibitemShut {NoStop}%
\bibitem [{\citenamefont {Bionta}\ \emph {et~al.}(2015)\citenamefont {Bionta},
  \citenamefont {Chalopin}, \citenamefont {Masseboeuf},\ and\ \citenamefont
  {Chatel}}]{Bionta2015}%
  \BibitemOpen
  \bibfield  {author} {\bibinfo {author} {\bibnamefont {Bionta}, \bibfnamefont
  {M~R}}, \bibinfo {author} {\bibfnamefont {B.}~\bibnamefont {Chalopin}},
  \bibinfo {author} {\bibfnamefont {A.}~\bibnamefont {Masseboeuf}}, \ and\
  \bibinfo {author} {\bibfnamefont {B.}~\bibnamefont {Chatel}}} (\bibinfo
  {year} {2015}),\ \bibfield  {title} {\enquote {\bibinfo {title} {First
  results on laser-induced field emission from a cnt-based nanotip},}\ }\href
  {\doibase 10.1016/j.ultramic.2014.11.027} {\bibfield  {journal} {\bibinfo
  {journal} {Ultramicroscopy}\ }\textbf {\bibinfo {volume} {159, Part 2}},\
  \bibinfo {pages} {152--155}},\ \bibinfo {note} {1st International Conference
  on Atom Probe Tomography and Microscopy}\BibitemShut {NoStop}%
\bibitem [{\citenamefont {Blaga}\ \emph {et~al.}(2012)\citenamefont {Blaga},
  \citenamefont {Xu}, \citenamefont {DiChiara}, \citenamefont {Sistrunk},
  \citenamefont {Zhang}, \citenamefont {Agostini}, \citenamefont {Miller},
  \citenamefont {DiMauro},\ and\ \citenamefont {Lin}}]{blaga2012}%
  \BibitemOpen
  \bibfield  {author} {\bibinfo {author} {\bibnamefont {Blaga}, \bibfnamefont
  {C~I}}, \bibinfo {author} {\bibfnamefont {J.}~\bibnamefont {Xu}}, \bibinfo
  {author} {\bibfnamefont {A.~D.}\ \bibnamefont {DiChiara}}, \bibinfo {author}
  {\bibfnamefont {E.}~\bibnamefont {Sistrunk}}, \bibinfo {author}
  {\bibfnamefont {K.}~\bibnamefont {Zhang}}, \bibinfo {author} {\bibfnamefont
  {P.}~\bibnamefont {Agostini}}, \bibinfo {author} {\bibfnamefont {T.~A.}\
  \bibnamefont {Miller}}, \bibinfo {author} {\bibfnamefont {L.~F.}\
  \bibnamefont {DiMauro}}, \ and\ \bibinfo {author} {\bibfnamefont {C.~D.}\
  \bibnamefont {Lin}}} (\bibinfo {year} {2012}),\ \bibfield  {title} {\enquote
  {\bibinfo {title} {Imaging ultrafast molecular dynamics with laser-induced
  electron diffraction},}\ }\href@noop {} {\bibfield  {journal} {\bibinfo
  {journal} {Nature}\ }\textbf {\bibinfo {volume} {483}},\ \bibinfo {pages}
  {194--197}}\BibitemShut {NoStop}%
\bibitem [{\citenamefont {Bohren}\ and\ \citenamefont
  {Huffman}(1998)}]{Bohren1998}%
  \BibitemOpen
  \bibfield  {author} {\bibinfo {author} {\bibnamefont {Bohren}, \bibfnamefont
  {C~F}}, \ and\ \bibinfo {author} {\bibfnamefont {D.~R.}\ \bibnamefont
  {Huffman}}} (\bibinfo {year} {1998}),\ \href@noop {} {\emph {\bibinfo {title}
  {Absorption and Scattering of Light by Small Particles}}}\ (\bibinfo
  {publisher} {Wiley-VCH})\BibitemShut {NoStop}%
\bibitem [{\citenamefont {Bolten}\ \emph {et~al.}(2016)\citenamefont {Bolten},
  \citenamefont {Fahrenberger}, \citenamefont {Halver}, \citenamefont {Heber},
  \citenamefont {Hofmann}, \citenamefont {Kabadshow}, \citenamefont {Lenz},
  \citenamefont {Pippig},\ and\ \citenamefont {Sutmann}}]{Bolten}%
  \BibitemOpen
  \bibfield  {author} {\bibinfo {author} {\bibnamefont {Bolten}, \bibfnamefont
  {M}}, \bibinfo {author} {\bibfnamefont {F.}~\bibnamefont {Fahrenberger}},
  \bibinfo {author} {\bibfnamefont {R.}~\bibnamefont {Halver}}, \bibinfo
  {author} {\bibfnamefont {F.}~\bibnamefont {Heber}}, \bibinfo {author}
  {\bibfnamefont {M.}~\bibnamefont {Hofmann}}, \bibinfo {author} {\bibfnamefont
  {I.}~\bibnamefont {Kabadshow}}, \bibinfo {author} {\bibfnamefont
  {O.}~\bibnamefont {Lenz}}, \bibinfo {author} {\bibfnamefont {M.}~\bibnamefont
  {Pippig}}, \ and\ \bibinfo {author} {\bibfnamefont {G.}~\bibnamefont
  {Sutmann}}} (\bibinfo {year} {2016}),\ \href {http://scafacos.github.com}
  {\enquote {\bibinfo {title} {{ScaFaCoS, C subroutine library}},}\
  }\BibitemShut {NoStop}%
\bibitem [{\citenamefont {Borisov}\ \emph {et~al.}(2012)\citenamefont
  {Borisov}, \citenamefont {Echenique},\ and\ \citenamefont
  {Kazansky}}]{Borisov12}%
  \BibitemOpen
  \bibfield  {author} {\bibinfo {author} {\bibnamefont {Borisov}, \bibfnamefont
  {A~G}}, \bibinfo {author} {\bibfnamefont {P.~M.}\ \bibnamefont {Echenique}},
  \ and\ \bibinfo {author} {\bibfnamefont {A.~K.}\ \bibnamefont {Kazansky}}}
  (\bibinfo {year} {2012}),\ \bibfield  {title} {\enquote {\bibinfo {title}
  {Attostreaking with metallic nano-objects},}\ }\href@noop {} {\bibfield
  {journal} {\bibinfo  {journal} {New. J. Phys.}\ }\textbf {\bibinfo {volume}
  {14}},\ \bibinfo {pages} {023036}}\BibitemShut {NoStop}%
\bibitem [{\citenamefont {Borisov}\ \emph {et~al.}(2013)\citenamefont
  {Borisov}, \citenamefont {S\'anchez-Portal}, \citenamefont {Kazansky},\ and\
  \citenamefont {Echenique}}]{borisov_resonant_2013}%
  \BibitemOpen
  \bibfield  {author} {\bibinfo {author} {\bibnamefont {Borisov}, \bibfnamefont
  {A~G}}, \bibinfo {author} {\bibfnamefont {D.}~\bibnamefont
  {S\'anchez-Portal}}, \bibinfo {author} {\bibfnamefont {A.~K.}\ \bibnamefont
  {Kazansky}}, \ and\ \bibinfo {author} {\bibfnamefont {P.~M.}\ \bibnamefont
  {Echenique}}} (\bibinfo {year} {2013}),\ \bibfield  {title} {\enquote
  {\bibinfo {title} {Resonant and nonresonant processes in attosecond streaking
  from metals},}\ }\href@noop {} {\bibfield  {journal} {\bibinfo  {journal}
  {Phys. Rev. B}\ }\textbf {\bibinfo {volume} {87}},\ \bibinfo {pages}
  {121110}}\BibitemShut {NoStop}%
\bibitem [{\citenamefont {Bormann}\ \emph {et~al.}(2010)\citenamefont
  {Bormann}, \citenamefont {Gulde}, \citenamefont {Weismann}, \citenamefont
  {Yalunin},\ and\ \citenamefont {Ropers}}]{Bormann2010}%
  \BibitemOpen
  \bibfield  {author} {\bibinfo {author} {\bibnamefont {Bormann}, \bibfnamefont
  {R}}, \bibinfo {author} {\bibfnamefont {M.}~\bibnamefont {Gulde}}, \bibinfo
  {author} {\bibfnamefont {A.}~\bibnamefont {Weismann}}, \bibinfo {author}
  {\bibfnamefont {S.~V.}\ \bibnamefont {Yalunin}}, \ and\ \bibinfo {author}
  {\bibfnamefont {C.}~\bibnamefont {Ropers}}} (\bibinfo {year} {2010}),\
  \bibfield  {title} {\enquote {\bibinfo {title} {Tip-enhanced strong-field
  photoemission},}\ }\href@noop {} {\bibfield  {journal} {\bibinfo  {journal}
  {Phys. Rev. Lett.}\ }\textbf {\bibinfo {volume} {105}},\ \bibinfo {pages}
  {147601}}\BibitemShut {NoStop}%
\bibitem [{\citenamefont {Bormann}\ \emph {et~al.}(2015)\citenamefont
  {Bormann}, \citenamefont {Strauch}, \citenamefont {Sch\"afer},\ and\
  \citenamefont {Ropers}}]{Bormann2015}%
  \BibitemOpen
  \bibfield  {author} {\bibinfo {author} {\bibnamefont {Bormann}, \bibfnamefont
  {R}}, \bibinfo {author} {\bibfnamefont {S.}~\bibnamefont {Strauch}}, \bibinfo
  {author} {\bibfnamefont {S.}~\bibnamefont {Sch\"afer}}, \ and\ \bibinfo
  {author} {\bibfnamefont {C.}~\bibnamefont {Ropers}}} (\bibinfo {year}
  {2015}),\ \bibfield  {title} {\enquote {\bibinfo {title} {An ultrafast
  electron microscope gun driven by two-photon photoemission from a nanotip
  cathode},}\ }\href@noop {} {\bibfield  {journal} {\bibinfo  {journal} {J.
  Appl. Phys.}\ }\textbf {\bibinfo {volume} {118}},\ \bibinfo {pages}
  {173105}}\BibitemShut {NoStop}%
\bibitem [{\citenamefont {Bothschafter}\ \emph {et~al.}(2010)\citenamefont
  {Bothschafter}, \citenamefont {Schiffrin}, \citenamefont {Yakovlev},
  \citenamefont {Azzeer}, \citenamefont {Krausz}, \citenamefont {Ernstorfer},\
  and\ \citenamefont {Kienberger}}]{bothschafter2010collinear}%
  \BibitemOpen
  \bibfield  {author} {\bibinfo {author} {\bibnamefont {Bothschafter},
  \bibfnamefont {E~M}}, \bibinfo {author} {\bibfnamefont {A.}~\bibnamefont
  {Schiffrin}}, \bibinfo {author} {\bibfnamefont {V.~S.}\ \bibnamefont
  {Yakovlev}}, \bibinfo {author} {\bibfnamefont {A.~M.}\ \bibnamefont
  {Azzeer}}, \bibinfo {author} {\bibfnamefont {F.}~\bibnamefont {Krausz}},
  \bibinfo {author} {\bibfnamefont {R.}~\bibnamefont {Ernstorfer}}, \ and\
  \bibinfo {author} {\bibfnamefont {R.}~\bibnamefont {Kienberger}}} (\bibinfo
  {year} {2010}),\ \bibfield  {title} {\enquote {\bibinfo {title} {Collinear
  generation of ultrashort uv and xuv pulses},}\ }\href@noop {} {\bibfield
  {journal} {\bibinfo  {journal} {Opt. Exp.}\ }\textbf {\bibinfo {volume}
  {18}},\ \bibinfo {pages} {9173--9180}}\BibitemShut {NoStop}%
\bibitem [{\citenamefont {Bouhelier}\ \emph {et~al.}(2003)\citenamefont
  {Bouhelier}, \citenamefont {Beversluis}, \citenamefont {Hartschuh},\ and\
  \citenamefont {Novotny}}]{Bouhelier2003}%
  \BibitemOpen
  \bibfield  {author} {\bibinfo {author} {\bibnamefont {Bouhelier},
  \bibfnamefont {A}}, \bibinfo {author} {\bibfnamefont {M.}~\bibnamefont
  {Beversluis}}, \bibinfo {author} {\bibfnamefont {A.}~\bibnamefont
  {Hartschuh}}, \ and\ \bibinfo {author} {\bibfnamefont {L.}~\bibnamefont
  {Novotny}}} (\bibinfo {year} {2003}),\ \bibfield  {title} {\enquote {\bibinfo
  {title} {Near-field second-harmonic generation induced by local field
  enhancement},}\ }\href@noop {} {\bibfield  {journal} {\bibinfo  {journal}
  {Phys. Rev. Lett.}\ }\textbf {\bibinfo {volume} {90}},\ \bibinfo {pages}
  {013903}}\BibitemShut {NoStop}%
\bibitem [{\citenamefont {Brabec}\ and\ \citenamefont
  {Krausz}(2000)}]{Krausz00}%
  \BibitemOpen
  \bibfield  {author} {\bibinfo {author} {\bibnamefont {Brabec}, \bibfnamefont
  {T}}, \ and\ \bibinfo {author} {\bibfnamefont {F.}~\bibnamefont {Krausz}}}
  (\bibinfo {year} {2000}),\ \bibfield  {title} {\enquote {\bibinfo {title}
  {Intense few-cycle laser fields: Frontiers of nonlinear optics},}\
  }\href@noop {} {\bibfield  {journal} {\bibinfo  {journal} {Rev. Mod. Phys.}\
  }\textbf {\bibinfo {volume} {72}},\ \bibinfo {pages} {545--591}}\BibitemShut
  {NoStop}%
\bibitem [{\citenamefont {Brizuela}\ \emph {et~al.}(2013)\citenamefont
  {Brizuela}, \citenamefont {Heyl}, \citenamefont {Rudawski}, \citenamefont
  {Kroon}, \citenamefont {Rading}, \citenamefont {Dahlstr{\"o}m}, \citenamefont
  {Mauritsson}, \citenamefont {Johnsson}, \citenamefont {Arnold},\ and\
  \citenamefont {L'Huillier}}]{brizuela2013efficient}%
  \BibitemOpen
  \bibfield  {author} {\bibinfo {author} {\bibnamefont {Brizuela},
  \bibfnamefont {F}}, \bibinfo {author} {\bibfnamefont {C.~M.}\ \bibnamefont
  {Heyl}}, \bibinfo {author} {\bibfnamefont {P.}~\bibnamefont {Rudawski}},
  \bibinfo {author} {\bibfnamefont {D.}~\bibnamefont {Kroon}}, \bibinfo
  {author} {\bibfnamefont {L.}~\bibnamefont {Rading}}, \bibinfo {author}
  {\bibfnamefont {J.~M.}\ \bibnamefont {Dahlstr{\"o}m}}, \bibinfo {author}
  {\bibfnamefont {J.}~\bibnamefont {Mauritsson}}, \bibinfo {author}
  {\bibfnamefont {P.}~\bibnamefont {Johnsson}}, \bibinfo {author}
  {\bibfnamefont {C.~L.}\ \bibnamefont {Arnold}}, \ and\ \bibinfo {author}
  {\bibfnamefont {A.}~\bibnamefont {L'Huillier}}} (\bibinfo {year} {2013}),\
  \bibfield  {title} {\enquote {\bibinfo {title} {Efficient high-order harmonic
  generation boosted by below-threshold harmonics},}\ }\href@noop {} {\bibfield
   {journal} {\bibinfo  {journal} {Sci. Rep.}\ }\textbf {\bibinfo {volume}
  {3}},\ \bibinfo {pages} {1410}}\BibitemShut {NoStop}%
\bibitem [{\citenamefont {Bunkin}\ and\ \citenamefont
  {Fedorov}(1965)}]{Bunkin1965}%
  \BibitemOpen
  \bibfield  {author} {\bibinfo {author} {\bibnamefont {Bunkin}, \bibfnamefont
  {F~V}}, \ and\ \bibinfo {author} {\bibfnamefont {M.~V.}\ \bibnamefont
  {Fedorov}}} (\bibinfo {year} {1965}),\ \bibfield  {title} {\enquote {\bibinfo
  {title} {Cold emission of electrons from the surface of a metal in a strong
  radiation field},}\ }\href@noop {} {\bibfield  {journal} {\bibinfo  {journal}
  {Sov.~Phys.~JETP}\ }\textbf {\bibinfo {volume} {21}},\ \bibinfo {pages}
  {896--899}}\BibitemShut {NoStop}%
\bibitem [{\citenamefont {Calegari}\ \emph {et~al.}(2014)\citenamefont
  {Calegari}, \citenamefont {Ayuso}, \citenamefont {Trabattoni}, \citenamefont
  {Belshaw}, \citenamefont {Camillis}, \citenamefont {Anumula}, \citenamefont
  {Frassetto}, \citenamefont {Poletto}, \citenamefont {Palacios}, \citenamefont
  {Decleva}, \citenamefont {Greenwood}, \citenamefont {Mart\'{\i}n},\ and\
  \citenamefont {Nisoli}}]{calegari2014}%
  \BibitemOpen
  \bibfield  {author} {\bibinfo {author} {\bibnamefont {Calegari},
  \bibfnamefont {F}}, \bibinfo {author} {\bibfnamefont {D.}~\bibnamefont
  {Ayuso}}, \bibinfo {author} {\bibfnamefont {A.}~\bibnamefont {Trabattoni}},
  \bibinfo {author} {\bibfnamefont {L.}~\bibnamefont {Belshaw}}, \bibinfo
  {author} {\bibfnamefont {S.~De}\ \bibnamefont {Camillis}}, \bibinfo {author}
  {\bibfnamefont {S.}~\bibnamefont {Anumula}}, \bibinfo {author} {\bibfnamefont
  {F.}~\bibnamefont {Frassetto}}, \bibinfo {author} {\bibfnamefont
  {L.}~\bibnamefont {Poletto}}, \bibinfo {author} {\bibfnamefont
  {A.}~\bibnamefont {Palacios}}, \bibinfo {author} {\bibfnamefont
  {P.}~\bibnamefont {Decleva}}, \bibinfo {author} {\bibfnamefont {J.~B.}\
  \bibnamefont {Greenwood}}, \bibinfo {author} {\bibfnamefont {F.}~\bibnamefont
  {Mart\'{\i}n}}, \ and\ \bibinfo {author} {\bibfnamefont {M.}~\bibnamefont
  {Nisoli}}} (\bibinfo {year} {2014}),\ \bibfield  {title} {\enquote {\bibinfo
  {title} {Ultrafast electron dynamics in phenylalanine initiated by attosecond
  pulses},}\ }\href@noop {} {\bibfield  {journal} {\bibinfo  {journal}
  {Science}\ }\textbf {\bibinfo {volume} {336}},\ \bibinfo {pages}
  {346}}\BibitemShut {NoStop}%
\bibitem [{\citenamefont {del Campo}\ and\ \citenamefont
  {Arzt}(2008)}]{del_campo_fabrication_2008}%
  \BibitemOpen
  \bibfield  {author} {\bibinfo {author} {\bibnamefont {del Campo},
  \bibfnamefont {A}}, \ and\ \bibinfo {author} {\bibfnamefont {E.}~\bibnamefont
  {Arzt}}} (\bibinfo {year} {2008}),\ \bibfield  {title} {\enquote {\bibinfo
  {title} {Fabrication approaches for generating complex micro- and
  nanopatterns on polymeric surfaces},}\ }\href@noop {} {\bibfield  {journal}
  {\bibinfo  {journal} {Chem. Rev.}\ }\textbf {\bibinfo {volume} {108}},\
  \bibinfo {pages} {911--945}}\BibitemShut {NoStop}%
\bibitem [{\citenamefont {Cao}\ \emph {et~al.}(2014)\citenamefont {Cao},
  \citenamefont {Jiang}, \citenamefont {Yu}, \citenamefont {Wang},
  \citenamefont {Bai},\ and\ \citenamefont {Lu}}]{Cao14}%
  \BibitemOpen
  \bibfield  {author} {\bibinfo {author} {\bibnamefont {Cao}, \bibfnamefont
  {X}}, \bibinfo {author} {\bibfnamefont {S.}~\bibnamefont {Jiang}}, \bibinfo
  {author} {\bibfnamefont {C.}~\bibnamefont {Yu}}, \bibinfo {author}
  {\bibfnamefont {Y.}~\bibnamefont {Wang}}, \bibinfo {author} {\bibfnamefont
  {L.}~\bibnamefont {Bai}}, \ and\ \bibinfo {author} {\bibfnamefont
  {R.}~\bibnamefont {Lu}}} (\bibinfo {year} {2014}),\ \bibfield  {title}
  {\enquote {\bibinfo {title} {Generation of isolated sub-10-attosecond pulses
  in spatially inhomogenous two-color fields},}\ }\href@noop {} {\bibfield
  {journal} {\bibinfo  {journal} {Opt. Exp.}\ }\textbf {\bibinfo {volume}
  {22}},\ \bibinfo {pages} {26153--26161}}\BibitemShut {NoStop}%
\bibitem [{\citenamefont {Caprez}\ \emph {et~al.}(2007)\citenamefont {Caprez},
  \citenamefont {Barwick},\ and\ \citenamefont {Batelaan}}]{Caprez2007}%
  \BibitemOpen
  \bibfield  {author} {\bibinfo {author} {\bibnamefont {Caprez}, \bibfnamefont
  {A}}, \bibinfo {author} {\bibfnamefont {B.}~\bibnamefont {Barwick}}, \ and\
  \bibinfo {author} {\bibfnamefont {H.}~\bibnamefont {Batelaan}}} (\bibinfo
  {year} {2007}),\ \bibfield  {title} {\enquote {\bibinfo {title} {Macroscopic
  test of the {A}haronov-{B}ohm effect},}\ }\href@noop {} {\bibfield  {journal}
  {\bibinfo  {journal} {Phys. Rev. Lett.}\ }\textbf {\bibinfo {volume} {99}},\
  \bibinfo {pages} {210401}}\BibitemShut {NoStop}%
\bibitem [{\citenamefont {Cavalieri}\ \emph
  {et~al.}(2007{\natexlab{a}})\citenamefont {Cavalieri}, \citenamefont
  {Goulielmakis}, \citenamefont {Horvath}, \citenamefont {Helml}, \citenamefont
  {Schultze}, \citenamefont {Fie{\ss}}, \citenamefont {Pervak}, \citenamefont
  {Veisz}, \citenamefont {Yakovlev}, \citenamefont {Uiberacker}, \citenamefont
  {Apolonski}, \citenamefont {Krausz},\ and\ \citenamefont
  {Kienberger}}]{cavalieri_intense_2007}%
  \BibitemOpen
  \bibfield  {author} {\bibinfo {author} {\bibnamefont {Cavalieri},
  \bibfnamefont {A~L}}, \bibinfo {author} {\bibfnamefont {E.}~\bibnamefont
  {Goulielmakis}}, \bibinfo {author} {\bibfnamefont {B.}~\bibnamefont
  {Horvath}}, \bibinfo {author} {\bibfnamefont {W.}~\bibnamefont {Helml}},
  \bibinfo {author} {\bibfnamefont {M.}~\bibnamefont {Schultze}}, \bibinfo
  {author} {\bibfnamefont {M.}~\bibnamefont {Fie{\ss}}}, \bibinfo {author}
  {\bibfnamefont {V.}~\bibnamefont {Pervak}}, \bibinfo {author} {\bibfnamefont
  {L.}~\bibnamefont {Veisz}}, \bibinfo {author} {\bibfnamefont {V.~S.}\
  \bibnamefont {Yakovlev}}, \bibinfo {author} {\bibfnamefont {M.}~\bibnamefont
  {Uiberacker}}, \bibinfo {author} {\bibfnamefont {A.}~\bibnamefont
  {Apolonski}}, \bibinfo {author} {\bibfnamefont {F.}~\bibnamefont {Krausz}}, \
  and\ \bibinfo {author} {\bibfnamefont {R.}~\bibnamefont {Kienberger}}}
  (\bibinfo {year} {2007}{\natexlab{a}}),\ \bibfield  {title} {\enquote
  {\bibinfo {title} {Intense 1.5-cycle near infrared laser waveforms and their
  use for the generation of ultra-broadband soft-x-ray harmonic continua},}\
  }\href@noop {} {\bibfield  {journal} {\bibinfo  {journal} {New J. of Phys.}\
  }\textbf {\bibinfo {volume} {9}},\ \bibinfo {pages} {242}}\BibitemShut
  {NoStop}%
\bibitem [{\citenamefont {Cavalieri}\ \emph
  {et~al.}(2007{\natexlab{b}})\citenamefont {Cavalieri}, \citenamefont
  {M{\"u}ller}, \citenamefont {Uphues}, \citenamefont {Yakovlev}, \citenamefont
  {Baltu{\v{s}}ka}, \citenamefont {Horvath}, \citenamefont {Schmidt},
  \citenamefont {Bl{\"u}mel}, \citenamefont {Holzwarth}, \citenamefont
  {Hendel}, \citenamefont {Drescher}, \citenamefont {Kleineberg}, \citenamefont
  {Echenique}, \citenamefont {Kienberger}, \citenamefont {Krausz},\ and\
  \citenamefont {Heinzmann}}]{cavalieri2007attosecond}%
  \BibitemOpen
  \bibfield  {author} {\bibinfo {author} {\bibnamefont {Cavalieri},
  \bibfnamefont {A~L}}, \bibinfo {author} {\bibfnamefont {N.}~\bibnamefont
  {M{\"u}ller}}, \bibinfo {author} {\bibfnamefont {Th.}\ \bibnamefont
  {Uphues}}, \bibinfo {author} {\bibfnamefont {V.~S.}\ \bibnamefont
  {Yakovlev}}, \bibinfo {author} {\bibfnamefont {A.}~\bibnamefont
  {Baltu{\v{s}}ka}}, \bibinfo {author} {\bibfnamefont {B.}~\bibnamefont
  {Horvath}}, \bibinfo {author} {\bibfnamefont {B.}~\bibnamefont {Schmidt}},
  \bibinfo {author} {\bibfnamefont {L.}~\bibnamefont {Bl{\"u}mel}}, \bibinfo
  {author} {\bibfnamefont {R.}~\bibnamefont {Holzwarth}}, \bibinfo {author}
  {\bibfnamefont {S.}~\bibnamefont {Hendel}}, \bibinfo {author} {\bibfnamefont
  {M.}~\bibnamefont {Drescher}}, \bibinfo {author} {\bibfnamefont
  {U.}~\bibnamefont {Kleineberg}}, \bibinfo {author} {\bibfnamefont {P.~M.}\
  \bibnamefont {Echenique}}, \bibinfo {author} {\bibfnamefont {R.}~\bibnamefont
  {Kienberger}}, \bibinfo {author} {\bibfnamefont {F.}~\bibnamefont {Krausz}},
  \ and\ \bibinfo {author} {\bibfnamefont {U}~\bibnamefont {Heinzmann}}}
  (\bibinfo {year} {2007}{\natexlab{b}}),\ \bibfield  {title} {\enquote
  {\bibinfo {title} {Attosecond spectroscopy in condensed matter},}\
  }\href@noop {} {\bibfield  {journal} {\bibinfo  {journal} {Nature}\ }\textbf
  {\bibinfo {volume} {449}},\ \bibinfo {pages} {1029--1032}}\BibitemShut
  {NoStop}%
\bibitem [{\citenamefont {Chac\'on}\ \emph
  {et~al.}(2015{\natexlab{a}})\citenamefont {Chac\'on}, \citenamefont
  {Ciappina},\ and\ \citenamefont {Lewenstein}}]{AlexisVG}%
  \BibitemOpen
  \bibfield  {author} {\bibinfo {author} {\bibnamefont {Chac\'on},
  \bibfnamefont {A}}, \bibinfo {author} {\bibfnamefont {M.~F.}\ \bibnamefont
  {Ciappina}}, \ and\ \bibinfo {author} {\bibfnamefont {M.}~\bibnamefont
  {Lewenstein}}} (\bibinfo {year} {2015}{\natexlab{a}}),\ \bibfield  {title}
  {\enquote {\bibinfo {title} {Numerical studies of light-matter interaction
  driven by plasmonic fields: The velocity gauge},}\ }\href@noop {} {\bibfield
  {journal} {\bibinfo  {journal} {Phys. Rev. A}\ }\textbf {\bibinfo {volume}
  {92}},\ \bibinfo {pages} {063834}}\BibitemShut {NoStop}%
\bibitem [{\citenamefont {Chac\'on}\ \emph
  {et~al.}(2015{\natexlab{b}})\citenamefont {Chac\'on}, \citenamefont
  {Ciappina},\ and\ \citenamefont {Lewenstein}}]{AlexisPRL}%
  \BibitemOpen
  \bibfield  {author} {\bibinfo {author} {\bibnamefont {Chac\'on},
  \bibfnamefont {A}}, \bibinfo {author} {\bibfnamefont {M.~F.}\ \bibnamefont
  {Ciappina}}, \ and\ \bibinfo {author} {\bibfnamefont {M.}~\bibnamefont
  {Lewenstein}}} (\bibinfo {year} {2015}{\natexlab{b}}),\ \href@noop {}
  {\enquote {\bibinfo {title} {Signatures of double-electron recombination in
  high-order harmonic generation driven by spatial inhomogeneous fields},}\
  }\bibinfo {note} {Submitted}\BibitemShut {NoStop}%
\bibitem [{\citenamefont {Chaturvedi}\ \emph {et~al.}(2009)\citenamefont
  {Chaturvedi}, \citenamefont {Hsu}, \citenamefont {Kumar}, \citenamefont
  {Fung}, \citenamefont {Mabon},\ and\ \citenamefont {Fang}}]{Chaturvedi2009}%
  \BibitemOpen
  \bibfield  {author} {\bibinfo {author} {\bibnamefont {Chaturvedi},
  \bibfnamefont {P}}, \bibinfo {author} {\bibfnamefont {K.~H.}\ \bibnamefont
  {Hsu}}, \bibinfo {author} {\bibfnamefont {A.}~\bibnamefont {Kumar}}, \bibinfo
  {author} {\bibfnamefont {K.~H.}\ \bibnamefont {Fung}}, \bibinfo {author}
  {\bibfnamefont {J.~C.}\ \bibnamefont {Mabon}}, \ and\ \bibinfo {author}
  {\bibfnamefont {N.~X.}\ \bibnamefont {Fang}}} (\bibinfo {year} {2009}),\
  \bibfield  {title} {\enquote {\bibinfo {title} {Imaging of plasmonic modes of
  silver nanoparticles using high-resolution cathodoluminescence
  spectroscopy},}\ }\href@noop {} {\bibfield  {journal} {\bibinfo  {journal}
  {ACS Nano}\ }\textbf {\bibinfo {volume} {3}},\ \bibinfo {pages}
  {2965--2974}}\BibitemShut {NoStop}%
\bibitem [{\citenamefont {Chew}\ \emph {et~al.}(2012)\citenamefont {Chew},
  \citenamefont {S{\"u}{\ss}mann}, \citenamefont {Sp{\"a}th}, \citenamefont
  {Wirth}, \citenamefont {Schmidt}, \citenamefont {Zherebtsov}, \citenamefont
  {Guggenmos}, \citenamefont {Oelsner}, \citenamefont {Weber}, \citenamefont
  {Kapaldo}, \citenamefont {Gliserin}, \citenamefont {Stockman}, \citenamefont
  {Kling},\ and\ \citenamefont {Kleineberg}}]{Chew12}%
  \BibitemOpen
  \bibfield  {author} {\bibinfo {author} {\bibnamefont {Chew}, \bibfnamefont
  {S~H}}, \bibinfo {author} {\bibfnamefont {F.}~\bibnamefont
  {S{\"u}{\ss}mann}}, \bibinfo {author} {\bibfnamefont {C.}~\bibnamefont
  {Sp{\"a}th}}, \bibinfo {author} {\bibfnamefont {A.}~\bibnamefont {Wirth}},
  \bibinfo {author} {\bibfnamefont {J.}~\bibnamefont {Schmidt}}, \bibinfo
  {author} {\bibfnamefont {S.}~\bibnamefont {Zherebtsov}}, \bibinfo {author}
  {\bibfnamefont {A.}~\bibnamefont {Guggenmos}}, \bibinfo {author}
  {\bibfnamefont {A.}~\bibnamefont {Oelsner}}, \bibinfo {author} {\bibfnamefont
  {N.}~\bibnamefont {Weber}}, \bibinfo {author} {\bibfnamefont
  {J.}~\bibnamefont {Kapaldo}}, \bibinfo {author} {\bibfnamefont
  {A.}~\bibnamefont {Gliserin}}, \bibinfo {author} {\bibfnamefont {M.~I.}\
  \bibnamefont {Stockman}}, \bibinfo {author} {\bibfnamefont {M.~F.}\
  \bibnamefont {Kling}}, \ and\ \bibinfo {author} {\bibfnamefont
  {U.}~\bibnamefont {Kleineberg}}} (\bibinfo {year} {2012}),\ \bibfield
  {title} {\enquote {\bibinfo {title} {Time-of-flight-photoelectron emission
  microscopy on plasmonic structures using attosecond extreme ultraviolet
  pulses},}\ }\href@noop {} {\bibfield  {journal} {\bibinfo  {journal} {Appl.
  Phys. Lett.}\ }\textbf {\bibinfo {volume} {100}},\ \bibinfo {pages}
  {051904}}\BibitemShut {NoStop}%
\bibitem [{\citenamefont {Choi}\ \emph {et~al.}(2016)\citenamefont {Choi},
  \citenamefont {Ciappina}, \citenamefont {P{\'e}rez-Hern{\'a}ndez},
  \citenamefont {Landsman}, \citenamefont {Kim}, \citenamefont {Kim},\ and\
  \citenamefont {Kim}}]{choipra2016}%
  \BibitemOpen
  \bibfield  {author} {\bibinfo {author} {\bibnamefont {Choi}, \bibfnamefont
  {S}}, \bibinfo {author} {\bibfnamefont {M.~F.}\ \bibnamefont {Ciappina}},
  \bibinfo {author} {\bibnamefont {P{\'e}rez-Hern{\'a}ndez}}, \bibinfo {author}
  {\bibfnamefont {A.~S.}\ \bibnamefont {Landsman}}, \bibinfo {author}
  {\bibfnamefont {Y.-J.}\ \bibnamefont {Kim}}, \bibinfo {author} {\bibfnamefont
  {S.~C.}\ \bibnamefont {Kim}}, \ and\ \bibinfo {author} {\bibfnamefont
  {D.~E.}\ \bibnamefont {Kim}}} (\bibinfo {year} {2016}),\ \bibfield  {title}
  {\enquote {\bibinfo {title} {Active tailoring of nano-antenna plasmonic field
  using few-cycle laser pulses},}\ }\href@noop {} {\bibfield  {journal}
  {\bibinfo  {journal} {Phys. Rev. A}\ }\textbf {\bibinfo {volume} {93}},\
  \bibinfo {pages} {021405(R)}}\BibitemShut {NoStop}%
\bibitem [{\citenamefont {Chung}\ and\ \citenamefont
  {Weiner}(2001)}]{chung_ambiguity_2001}%
  \BibitemOpen
  \bibfield  {author} {\bibinfo {author} {\bibnamefont {Chung}, \bibfnamefont
  {J-H}}, \ and\ \bibinfo {author} {\bibfnamefont {A.~M.}\ \bibnamefont
  {Weiner}}} (\bibinfo {year} {2001}),\ \bibfield  {title} {\enquote {\bibinfo
  {title} {Ambiguity of ultrashort pulse shapes retrieved from the intensity
  autocorrelation and the power spectrum},}\ }\href@noop {} {\bibfield
  {journal} {\bibinfo  {journal} {IEEE J. Select. Topics Quantum Electron.}\
  }\textbf {\bibinfo {volume} {7}},\ \bibinfo {pages} {656--666}}\BibitemShut
  {NoStop}%
\bibitem [{\citenamefont {Ciappina}\ \emph
  {et~al.}(2012{\natexlab{a}})\citenamefont {Ciappina}, \citenamefont
  {A{\'c}imovi{\'c}}, \citenamefont {Shaaran}, \citenamefont {Biegert},
  \citenamefont {Quidant},\ and\ \citenamefont {Lewenstein}}]{Marcelo12OE}%
  \BibitemOpen
  \bibfield  {author} {\bibinfo {author} {\bibnamefont {Ciappina},
  \bibfnamefont {M~F}}, \bibinfo {author} {\bibfnamefont {S.~S.}\ \bibnamefont
  {A{\'c}imovi{\'c}}}, \bibinfo {author} {\bibfnamefont {T.}~\bibnamefont
  {Shaaran}}, \bibinfo {author} {\bibfnamefont {J.}~\bibnamefont {Biegert}},
  \bibinfo {author} {\bibfnamefont {R.}~\bibnamefont {Quidant}}, \ and\
  \bibinfo {author} {\bibfnamefont {M.}~\bibnamefont {Lewenstein}}} (\bibinfo
  {year} {2012}{\natexlab{a}}),\ \bibfield  {title} {\enquote {\bibinfo {title}
  {Enhancement of high harmonic generation by confining electron motion in
  plasmonic nanostrutures},}\ }\href@noop {} {\bibfield  {journal} {\bibinfo
  {journal} {Opt. Exp.}\ }\textbf {\bibinfo {volume} {20}},\ \bibinfo {pages}
  {26261--26274}}\BibitemShut {NoStop}%
\bibitem [{\citenamefont {Ciappina}\ \emph
  {et~al.}(2012{\natexlab{b}})\citenamefont {Ciappina}, \citenamefont
  {Biegert}, \citenamefont {Quidant},\ and\ \citenamefont
  {Lewenstein}}]{Marcelo12A}%
  \BibitemOpen
  \bibfield  {author} {\bibinfo {author} {\bibnamefont {Ciappina},
  \bibfnamefont {M~F}}, \bibinfo {author} {\bibfnamefont {J.}~\bibnamefont
  {Biegert}}, \bibinfo {author} {\bibfnamefont {R.}~\bibnamefont {Quidant}}, \
  and\ \bibinfo {author} {\bibfnamefont {M.}~\bibnamefont {Lewenstein}}}
  (\bibinfo {year} {2012}{\natexlab{b}}),\ \bibfield  {title} {\enquote
  {\bibinfo {title} {High-order-harmonic generation from inhomogeneous
  fields},}\ }\href@noop {} {\bibfield  {journal} {\bibinfo  {journal} {Phys.
  Rev. A}\ }\textbf {\bibinfo {volume} {85}},\ \bibinfo {pages}
  {033828}}\BibitemShut {NoStop}%
\bibitem [{\citenamefont {Ciappina}\ \emph
  {et~al.}(2014{\natexlab{a}})\citenamefont {Ciappina}, \citenamefont
  {P{\'e}rez-Hern{\'a}ndez},\ and\ \citenamefont {Lewenstein}}]{Marcelo15CPC}%
  \BibitemOpen
  \bibfield  {author} {\bibinfo {author} {\bibnamefont {Ciappina},
  \bibfnamefont {M~F}}, \bibinfo {author} {\bibfnamefont {J.~A.}\ \bibnamefont
  {P{\'e}rez-Hern{\'a}ndez}}, \ and\ \bibinfo {author} {\bibfnamefont
  {M.}~\bibnamefont {Lewenstein}}} (\bibinfo {year} {2014}{\natexlab{a}}),\
  \bibfield  {title} {\enquote {\bibinfo {title} {Classstrong: Classical
  simulations of strong field processes},}\ }\href@noop {} {\bibfield
  {journal} {\bibinfo  {journal} {Comp. Phys. Comm.}\ }\textbf {\bibinfo
  {volume} {185}},\ \bibinfo {pages} {398--406}}\BibitemShut {NoStop}%
\bibitem [{\citenamefont {Ciappina}\ \emph {et~al.}(2015)\citenamefont
  {Ciappina}, \citenamefont {P{\'e}rez-Hern{\'a}ndez}, \citenamefont {Roso},
  \citenamefont {Za{\"i}r}, ,\ and\ \citenamefont {Lewenstein}}]{Marcelo15}%
  \BibitemOpen
  \bibfield  {author} {\bibinfo {author} {\bibnamefont {Ciappina},
  \bibfnamefont {M~F}}, \bibinfo {author} {\bibfnamefont {J.~A.}\ \bibnamefont
  {P{\'e}rez-Hern{\'a}ndez}}, \bibinfo {author} {\bibfnamefont
  {L.}~\bibnamefont {Roso}}, \bibinfo {author} {\bibfnamefont {A.}~\bibnamefont
  {Za{\"i}r}}, , \ and\ \bibinfo {author} {\bibfnamefont {M.}~\bibnamefont
  {Lewenstein}}} (\bibinfo {year} {2015}),\ \bibfield  {title} {\enquote
  {\bibinfo {title} {High-order harmonic generation driven by plasmonic fields:
  a new route towards the generation of uv and xuv photons?}}\ }\href@noop {}
  {\bibfield  {journal} {\bibinfo  {journal} {J. Phys.: Conf. Ser.}\ }\textbf
  {\bibinfo {volume} {601}},\ \bibinfo {pages} {012001}}\BibitemShut {NoStop}%
\bibitem [{\citenamefont {Ciappina}\ \emph
  {et~al.}(2012{\natexlab{c}})\citenamefont {Ciappina}, \citenamefont
  {P{\'e}rez-Hern{\'a}ndez}, \citenamefont {Shaaran}, \citenamefont {Biegert},
  \citenamefont {Quidant},\ and\ \citenamefont {Lewenstein}}]{Marcelo12AAA}%
  \BibitemOpen
  \bibfield  {author} {\bibinfo {author} {\bibnamefont {Ciappina},
  \bibfnamefont {M~F}}, \bibinfo {author} {\bibfnamefont {J.~A.}\ \bibnamefont
  {P{\'e}rez-Hern{\'a}ndez}}, \bibinfo {author} {\bibfnamefont
  {T.}~\bibnamefont {Shaaran}}, \bibinfo {author} {\bibfnamefont
  {J.}~\bibnamefont {Biegert}}, \bibinfo {author} {\bibfnamefont
  {R.}~\bibnamefont {Quidant}}, \ and\ \bibinfo {author} {\bibfnamefont
  {M.}~\bibnamefont {Lewenstein}}} (\bibinfo {year} {2012}{\natexlab{c}}),\
  \bibfield  {title} {\enquote {\bibinfo {title} {Above-threshold ionization by
  few-cycle spatially inhomogeneous fields},}\ }\href@noop {} {\bibfield
  {journal} {\bibinfo  {journal} {Phys. Rev. A}\ }\textbf {\bibinfo {volume}
  {86}},\ \bibinfo {pages} {023413}}\BibitemShut {NoStop}%
\bibitem [{\citenamefont {Ciappina}\ \emph
  {et~al.}(2014{\natexlab{b}})\citenamefont {Ciappina}, \citenamefont
  {P{\'e}rez-Hern{\'a}ndez}, \citenamefont {Shaaran},\ and\ \citenamefont
  {Lewenstein}}]{Marcelo14EPJD}%
  \BibitemOpen
  \bibfield  {author} {\bibinfo {author} {\bibnamefont {Ciappina},
  \bibfnamefont {M~F}}, \bibinfo {author} {\bibfnamefont {J.~A.}\ \bibnamefont
  {P{\'e}rez-Hern{\'a}ndez}}, \bibinfo {author} {\bibfnamefont
  {T.}~\bibnamefont {Shaaran}}, \ and\ \bibinfo {author} {\bibfnamefont
  {M.}~\bibnamefont {Lewenstein}}} (\bibinfo {year} {2014}{\natexlab{b}}),\
  \bibfield  {title} {\enquote {\bibinfo {title} {Coherent xuv generation
  driven by sharp metal tips photoemission},}\ }\href@noop {} {\bibfield
  {journal} {\bibinfo  {journal} {Eur. Phys. J. D}\ }\textbf {\bibinfo {volume}
  {68}},\ \bibinfo {pages} {172}}\BibitemShut {NoStop}%
\bibitem [{\citenamefont {Ciappina}\ \emph
  {et~al.}(2014{\natexlab{c}})\citenamefont {Ciappina}, \citenamefont
  {P{\'e}rez-Hern{\'a}ndez}, \citenamefont {Shaaran}, \citenamefont
  {Lewenstein}, \citenamefont {Kr{\"u}ger},\ and\ \citenamefont
  {Hommelhoff}}]{Marcelo14}%
  \BibitemOpen
  \bibfield  {author} {\bibinfo {author} {\bibnamefont {Ciappina},
  \bibfnamefont {M~F}}, \bibinfo {author} {\bibfnamefont {J.~A.}\ \bibnamefont
  {P{\'e}rez-Hern{\'a}ndez}}, \bibinfo {author} {\bibfnamefont
  {T.}~\bibnamefont {Shaaran}}, \bibinfo {author} {\bibfnamefont
  {M.}~\bibnamefont {Lewenstein}}, \bibinfo {author} {\bibfnamefont
  {M.}~\bibnamefont {Kr{\"u}ger}}, \ and\ \bibinfo {author} {\bibfnamefont
  {P.}~\bibnamefont {Hommelhoff}}} (\bibinfo {year} {2014}{\natexlab{c}}),\
  \bibfield  {title} {\enquote {\bibinfo {title} {High-order harmonic
  generation driven by metal nanotip photoemission: theory and simulations},}\
  }\href@noop {} {\bibfield  {journal} {\bibinfo  {journal} {Phys. Rev. A}\
  }\textbf {\bibinfo {volume} {89}},\ \bibinfo {pages} {013409}}\BibitemShut
  {NoStop}%
\bibitem [{\citenamefont {Ciappina}\ \emph
  {et~al.}(2013{\natexlab{a}})\citenamefont {Ciappina}, \citenamefont
  {P{\'e}rez-Hern{\'a}ndez}, \citenamefont {Shaaran}, \citenamefont {Roso},\
  and\ \citenamefont {Lewenstein}}]{Marcelo13A}%
  \BibitemOpen
  \bibfield  {author} {\bibinfo {author} {\bibnamefont {Ciappina},
  \bibfnamefont {M~F}}, \bibinfo {author} {\bibfnamefont {J.~A.}\ \bibnamefont
  {P{\'e}rez-Hern{\'a}ndez}}, \bibinfo {author} {\bibfnamefont
  {T.}~\bibnamefont {Shaaran}}, \bibinfo {author} {\bibfnamefont
  {L.}~\bibnamefont {Roso}}, \ and\ \bibinfo {author} {\bibfnamefont
  {M.}~\bibnamefont {Lewenstein}}} (\bibinfo {year} {2013}{\natexlab{a}}),\
  \bibfield  {title} {\enquote {\bibinfo {title} {Electron-momentum
  distributions and photoelectron spectra of atoms driven by an intense
  spatially inhomogeneous field},}\ }\href@noop {} {\bibfield  {journal}
  {\bibinfo  {journal} {Phys. Rev. A}\ }\textbf {\bibinfo {volume} {87}},\
  \bibinfo {pages} {063833}}\BibitemShut {NoStop}%
\bibitem [{\citenamefont {Ciappina}\ \emph
  {et~al.}(2013{\natexlab{b}})\citenamefont {Ciappina}, \citenamefont
  {Shaaran}, \citenamefont {Guichard}, \citenamefont {P{\'e}rez-Hern{\'a}ndez},
  \citenamefont {Roso}, \citenamefont {Arnold}, \citenamefont {Siegel},
  \citenamefont {Za{\"i}r},\ and\ \citenamefont {Lewenstein}}]{Marcelo13LPL}%
  \BibitemOpen
  \bibfield  {author} {\bibinfo {author} {\bibnamefont {Ciappina},
  \bibfnamefont {M~F}}, \bibinfo {author} {\bibfnamefont {T.}~\bibnamefont
  {Shaaran}}, \bibinfo {author} {\bibfnamefont {R.}~\bibnamefont {Guichard}},
  \bibinfo {author} {\bibfnamefont {J.~A.}\ \bibnamefont
  {P{\'e}rez-Hern{\'a}ndez}}, \bibinfo {author} {\bibfnamefont
  {L.}~\bibnamefont {Roso}}, \bibinfo {author} {\bibfnamefont {M.}~\bibnamefont
  {Arnold}}, \bibinfo {author} {\bibfnamefont {T.}~\bibnamefont {Siegel}},
  \bibinfo {author} {\bibfnamefont {A.}~\bibnamefont {Za{\"i}r}}, \ and\
  \bibinfo {author} {\bibfnamefont {M.}~\bibnamefont {Lewenstein}}} (\bibinfo
  {year} {2013}{\natexlab{b}}),\ \bibfield  {title} {\enquote {\bibinfo {title}
  {High energy photoelectron emission from gases using plasmonic enhanced
  near-fields},}\ }\href@noop {} {\bibfield  {journal} {\bibinfo  {journal}
  {Las. Phys. Lett.}\ }\textbf {\bibinfo {volume} {10}},\ \bibinfo {pages}
  {105302}}\BibitemShut {NoStop}%
\bibitem [{\citenamefont {Ciappina}\ \emph
  {et~al.}(2013{\natexlab{c}})\citenamefont {Ciappina}, \citenamefont
  {Shaaran},\ and\ \citenamefont {Lewenstein}}]{Marcelo13AP}%
  \BibitemOpen
  \bibfield  {author} {\bibinfo {author} {\bibnamefont {Ciappina},
  \bibfnamefont {M~F}}, \bibinfo {author} {\bibfnamefont {T.}~\bibnamefont
  {Shaaran}}, \ and\ \bibinfo {author} {\bibfnamefont {M.}~\bibnamefont
  {Lewenstein}}} (\bibinfo {year} {2013}{\natexlab{c}}),\ \bibfield  {title}
  {\enquote {\bibinfo {title} {High order harmonic generation in noble gases
  using plasmonic field enhancement},}\ }\href@noop {} {\bibfield  {journal}
  {\bibinfo  {journal} {Ann. Phys. (Berlin)}\ }\textbf {\bibinfo {volume}
  {525}},\ \bibinfo {pages} {97--106}}\BibitemShut {NoStop}%
\bibitem [{\citenamefont {Cirac{\`\i}}\ \emph {et~al.}(2012)\citenamefont
  {Cirac{\`\i}}, \citenamefont {Hill}, \citenamefont {Mock}, \citenamefont
  {Urzhumov}, \citenamefont {Fern{\'a}ndez-Dom{\'\i}nguez}, \citenamefont
  {Maier}, \citenamefont {Pendry}, \citenamefont {Chilkoti},\ and\
  \citenamefont {Smith}}]{Ciraci2012}%
  \BibitemOpen
  \bibfield  {author} {\bibinfo {author} {\bibnamefont {Cirac{\`\i}},
  \bibfnamefont {C}}, \bibinfo {author} {\bibfnamefont {R.~T.}\ \bibnamefont
  {Hill}}, \bibinfo {author} {\bibfnamefont {J.~J.}\ \bibnamefont {Mock}},
  \bibinfo {author} {\bibfnamefont {Y.}~\bibnamefont {Urzhumov}}, \bibinfo
  {author} {\bibfnamefont {A.~I.}\ \bibnamefont
  {Fern{\'a}ndez-Dom{\'\i}nguez}}, \bibinfo {author} {\bibfnamefont {S.~A.}\
  \bibnamefont {Maier}}, \bibinfo {author} {\bibfnamefont {J.~B.}\ \bibnamefont
  {Pendry}}, \bibinfo {author} {\bibfnamefont {A.}~\bibnamefont {Chilkoti}}, \
  and\ \bibinfo {author} {\bibfnamefont {D.~R.}\ \bibnamefont {Smith}}}
  (\bibinfo {year} {2012}),\ \bibfield  {title} {\enquote {\bibinfo {title}
  {Probing the ultimate limits of plasmonic enhancement},}\ }\href@noop {}
  {\bibfield  {journal} {\bibinfo  {journal} {Science}\ }\textbf {\bibinfo
  {volume} {337}}~(\bibinfo {number} {6098}),\ \bibinfo {pages}
  {1072--1074}}\BibitemShut {NoStop}%
\bibitem [{\citenamefont {Clark}\ \emph {et~al.}(2015)\citenamefont {Clark},
  \citenamefont {Beitra}, \citenamefont {Xiong}, \citenamefont {Fritz},
  \citenamefont {Lemke}, \citenamefont {Zhu}, \citenamefont {Chollet},
  \citenamefont {Williams}, \citenamefont {Messerschmidt}, \citenamefont
  {Abbey}, \citenamefont {Harder}, \citenamefont {Korsunsky}, \citenamefont
  {Wark}, \citenamefont {Reis},\ and\ \citenamefont
  {Robinson}}]{clark2015imaging}%
  \BibitemOpen
  \bibfield  {author} {\bibinfo {author} {\bibnamefont {Clark}, \bibfnamefont
  {J~N}}, \bibinfo {author} {\bibfnamefont {L.}~\bibnamefont {Beitra}},
  \bibinfo {author} {\bibfnamefont {G.}~\bibnamefont {Xiong}}, \bibinfo
  {author} {\bibfnamefont {D.~M.}\ \bibnamefont {Fritz}}, \bibinfo {author}
  {\bibfnamefont {H.~T.}\ \bibnamefont {Lemke}}, \bibinfo {author}
  {\bibfnamefont {D.}~\bibnamefont {Zhu}}, \bibinfo {author} {\bibfnamefont
  {M.}~\bibnamefont {Chollet}}, \bibinfo {author} {\bibfnamefont {G.~J.}\
  \bibnamefont {Williams}}, \bibinfo {author} {\bibfnamefont {M.~M.}\
  \bibnamefont {Messerschmidt}}, \bibinfo {author} {\bibfnamefont
  {B.}~\bibnamefont {Abbey}}, \bibinfo {author} {\bibfnamefont {R.~J.}\
  \bibnamefont {Harder}}, \bibinfo {author} {\bibfnamefont {A.~M.}\
  \bibnamefont {Korsunsky}}, \bibinfo {author} {\bibfnamefont {J.~S.}\
  \bibnamefont {Wark}}, \bibinfo {author} {\bibfnamefont {D.~A.}\ \bibnamefont
  {Reis}}, \ and\ \bibinfo {author} {\bibfnamefont {I.~K.}\ \bibnamefont
  {Robinson}}} (\bibinfo {year} {2015}),\ \bibfield  {title} {\enquote
  {\bibinfo {title} {Imaging transient melting of a nanocrystal using an x-ray
  laser},}\ }\href@noop {} {\bibfield  {journal} {\bibinfo  {journal} {Proc.
  Natl. Acad. Sci. USA}\ }\textbf {\bibinfo {volume} {112}},\ \bibinfo {pages}
  {7444--7448}}\BibitemShut {NoStop}%
\bibitem [{\citenamefont {Corkum}(1993)}]{corkum93}%
  \BibitemOpen
  \bibfield  {author} {\bibinfo {author} {\bibnamefont {Corkum}, \bibfnamefont
  {P~B}}} (\bibinfo {year} {1993}),\ \bibfield  {title} {\enquote {\bibinfo
  {title} {Plasma perspective on strong field multiphoton ionization},}\
  }\href@noop {} {\bibfield  {journal} {\bibinfo  {journal} {Phys. Rev. Lett.}\
  }\textbf {\bibinfo {volume} {71}},\ \bibinfo {pages} {1994}}\BibitemShut
  {NoStop}%
\bibitem [{\citenamefont {Corkum}\ \emph {et~al.}(1994)\citenamefont {Corkum},
  \citenamefont {Burnett},\ and\ \citenamefont
  {Ivanov}}]{corkum1994subfemtosecond}%
  \BibitemOpen
  \bibfield  {author} {\bibinfo {author} {\bibnamefont {Corkum}, \bibfnamefont
  {P~B}}, \bibinfo {author} {\bibfnamefont {N.~H.}\ \bibnamefont {Burnett}}, \
  and\ \bibinfo {author} {\bibfnamefont {M.~Yu.}\ \bibnamefont {Ivanov}}}
  (\bibinfo {year} {1994}),\ \bibfield  {title} {\enquote {\bibinfo {title}
  {Subfemtosecond pulses},}\ }\href@noop {} {\bibfield  {journal} {\bibinfo
  {journal} {Opt. Lett.}\ }\textbf {\bibinfo {volume} {19}},\ \bibinfo {pages}
  {1870--1872}}\BibitemShut {NoStop}%
\bibitem [{\citenamefont {Corkum}\ and\ \citenamefont
  {Krausz}(2007)}]{Krausz07}%
  \BibitemOpen
  \bibfield  {author} {\bibinfo {author} {\bibnamefont {Corkum}, \bibfnamefont
  {P~B}}, \ and\ \bibinfo {author} {\bibfnamefont {F.}~\bibnamefont {Krausz}}}
  (\bibinfo {year} {2007}),\ \bibfield  {title} {\enquote {\bibinfo {title}
  {Attosecond science},}\ }\href@noop {} {\bibfield  {journal} {\bibinfo
  {journal} {Nat. Phys.}\ }\textbf {\bibinfo {volume} {3}},\ \bibinfo {pages}
  {381--387}}\BibitemShut {NoStop}%
\bibitem [{\citenamefont {Da}\ \emph {et~al.}(2014)\citenamefont {Da},
  \citenamefont {Shinotsuka}, \citenamefont {Yoshikawa}, \citenamefont {Ding},\
  and\ \citenamefont {Tanuma}}]{Da2014}%
  \BibitemOpen
  \bibfield  {author} {\bibinfo {author} {\bibnamefont {Da}, \bibfnamefont
  {B}}, \bibinfo {author} {\bibfnamefont {H.}~\bibnamefont {Shinotsuka}},
  \bibinfo {author} {\bibfnamefont {H.}~\bibnamefont {Yoshikawa}}, \bibinfo
  {author} {\bibfnamefont {Z.~J.}\ \bibnamefont {Ding}}, \ and\ \bibinfo
  {author} {\bibfnamefont {S.}~\bibnamefont {Tanuma}}} (\bibinfo {year}
  {2014}),\ \bibfield  {title} {\enquote {\bibinfo {title} {Extended mermin
  method for calculating the electron inelastic mean free path},}\ }\href@noop
  {} {\bibfield  {journal} {\bibinfo  {journal} {Phys. Rev. Lett.}\ }\textbf
  {\bibinfo {volume} {113}},\ \bibinfo {pages} {063201}}\BibitemShut {NoStop}%
\bibitem [{\citenamefont {Deubel}\ \emph {et~al.}(2004)\citenamefont {Deubel},
  \citenamefont {von Freymann}, \citenamefont {Wegener}, \citenamefont
  {Pereira}, \citenamefont {Busch},\ and\ \citenamefont
  {Soukoulis}}]{deubel_direct_2004}%
  \BibitemOpen
  \bibfield  {author} {\bibinfo {author} {\bibnamefont {Deubel}, \bibfnamefont
  {M}}, \bibinfo {author} {\bibfnamefont {G.}~\bibnamefont {von Freymann}},
  \bibinfo {author} {\bibfnamefont {M.}~\bibnamefont {Wegener}}, \bibinfo
  {author} {\bibfnamefont {S.}~\bibnamefont {Pereira}}, \bibinfo {author}
  {\bibfnamefont {K.}~\bibnamefont {Busch}}, \ and\ \bibinfo {author}
  {\bibfnamefont {C.~M.}\ \bibnamefont {Soukoulis}}} (\bibinfo {year} {2004}),\
  \bibfield  {title} {\enquote {\bibinfo {title} {Direct laser writing of
  three-dimensional photonic-crystal templates for telecommunications},}\
  }\href@noop {} {\bibfield  {journal} {\bibinfo  {journal} {Nat. Mater.}\
  }\textbf {\bibinfo {volume} {3}},\ \bibinfo {pages} {444--447}}\BibitemShut
  {NoStop}%
\bibitem [{\citenamefont {Dietrich}\ \emph {et~al.}(2000)\citenamefont
  {Dietrich}, \citenamefont {Krausz},\ and\ \citenamefont
  {Corkum}}]{dietrich_determining_2000}%
  \BibitemOpen
  \bibfield  {author} {\bibinfo {author} {\bibnamefont {Dietrich},
  \bibfnamefont {P}}, \bibinfo {author} {\bibfnamefont {F.}~\bibnamefont
  {Krausz}}, \ and\ \bibinfo {author} {\bibfnamefont {P.~B.}\ \bibnamefont
  {Corkum}}} (\bibinfo {year} {2000}),\ \bibfield  {title} {\enquote {\bibinfo
  {title} {Determining the absolute carrier phase of a few-cycle laser
  pulse},}\ }\href@noop {} {\bibfield  {journal} {\bibinfo  {journal} {Opt.
  Lett.}\ }\textbf {\bibinfo {volume} {25}},\ \bibinfo {pages}
  {16--18}}\BibitemShut {NoStop}%
\bibitem [{\citenamefont {Ditmire}\ \emph
  {et~al.}(1997{\natexlab{a}})\citenamefont {Ditmire}, \citenamefont {Smith},
  \citenamefont {Tisch},\ and\ \citenamefont {Hutchinson}}]{Ditmire1997}%
  \BibitemOpen
  \bibfield  {author} {\bibinfo {author} {\bibnamefont {Ditmire}, \bibfnamefont
  {T}}, \bibinfo {author} {\bibfnamefont {R.~A.}\ \bibnamefont {Smith}},
  \bibinfo {author} {\bibfnamefont {J.~W.~G.}\ \bibnamefont {Tisch}}, \ and\
  \bibinfo {author} {\bibfnamefont {M.~H.~R.}\ \bibnamefont {Hutchinson}}}
  (\bibinfo {year} {1997}{\natexlab{a}}),\ \bibfield  {title} {\enquote
  {\bibinfo {title} {High intensity laser absorption by gases of atomic
  clusters},}\ }\href@noop {} {\bibfield  {journal} {\bibinfo  {journal} {Phys.
  Rev. Lett.}\ }\textbf {\bibinfo {volume} {78}},\ \bibinfo {pages}
  {3121}}\BibitemShut {NoStop}%
\bibitem [{\citenamefont {Ditmire}\ \emph
  {et~al.}(1997{\natexlab{b}})\citenamefont {Ditmire}, \citenamefont {Tisch},
  \citenamefont {Springate}, \citenamefont {Mason}, \citenamefont {Hay},
  \citenamefont {Smith}, \citenamefont {Marangos},\ and\ \citenamefont
  {Hutchinson}}]{DitmireNature}%
  \BibitemOpen
  \bibfield  {author} {\bibinfo {author} {\bibnamefont {Ditmire}, \bibfnamefont
  {T}}, \bibinfo {author} {\bibfnamefont {J.~W.~G.}\ \bibnamefont {Tisch}},
  \bibinfo {author} {\bibfnamefont {E.}~\bibnamefont {Springate}}, \bibinfo
  {author} {\bibfnamefont {M.~B.}\ \bibnamefont {Mason}}, \bibinfo {author}
  {\bibfnamefont {N.}~\bibnamefont {Hay}}, \bibinfo {author} {\bibfnamefont
  {R.~A.}\ \bibnamefont {Smith}}, \bibinfo {author} {\bibfnamefont
  {J.}~\bibnamefont {Marangos}}, \ and\ \bibinfo {author} {\bibfnamefont
  {M.~H.~R.}\ \bibnamefont {Hutchinson}}} (\bibinfo {year}
  {1997}{\natexlab{b}}),\ \bibfield  {title} {\enquote {\bibinfo {title}
  {High-energy ions produced in explosions of superheated atomic clusters},}\
  }\href@noop {} {\bibfield  {journal} {\bibinfo  {journal} {Nature (London)}\
  }\textbf {\bibinfo {volume} {386}},\ \bibinfo {pages} {54}}\BibitemShut
  {NoStop}%
\bibitem [{\citenamefont {Dombi}\ \emph {et~al.}(2013)\citenamefont {Dombi},
  \citenamefont {H\"orl}, \citenamefont {R\'acz}, \citenamefont {Marton},
  \citenamefont {Tr\"ugler}, \citenamefont {Krenn},\ and\ \citenamefont
  {Hohenester}}]{Dombi2013}%
  \BibitemOpen
  \bibfield  {author} {\bibinfo {author} {\bibnamefont {Dombi}, \bibfnamefont
  {P}}, \bibinfo {author} {\bibfnamefont {A.}~\bibnamefont {H\"orl}}, \bibinfo
  {author} {\bibfnamefont {P.}~\bibnamefont {R\'acz}}, \bibinfo {author}
  {\bibfnamefont {I.}~\bibnamefont {Marton}}, \bibinfo {author} {\bibfnamefont
  {A.}~\bibnamefont {Tr\"ugler}}, \bibinfo {author} {\bibfnamefont {J.~R.}\
  \bibnamefont {Krenn}}, \ and\ \bibinfo {author} {\bibfnamefont
  {U.}~\bibnamefont {Hohenester}}} (\bibinfo {year} {2013}),\ \bibfield
  {title} {\enquote {\bibinfo {title} {Ultrafast strong-field photoemission
  from plasmonic nanoparticles},}\ }\href@noop {} {\bibfield  {journal}
  {\bibinfo  {journal} {Nano Lett.}\ }\textbf {\bibinfo {volume} {13}},\
  \bibinfo {pages} {674--678}}\BibitemShut {NoStop}%
\bibitem [{\citenamefont {Dombi}\ \emph {et~al.}(2010)\citenamefont {Dombi},
  \citenamefont {Irvine}, \citenamefont {R{\'a}cz}, \citenamefont {Lenner},
  \citenamefont {Kro{\'o}}, \citenamefont {Farkas}, \citenamefont {Mitrofanov},
  \citenamefont {Baltu{\u{s}}ka}, \citenamefont {Fuji}, \citenamefont
  {Krausz},\ and\ \citenamefont {Elezzabi}}]{Dombi10}%
  \BibitemOpen
  \bibfield  {author} {\bibinfo {author} {\bibnamefont {Dombi}, \bibfnamefont
  {P}}, \bibinfo {author} {\bibfnamefont {S.~E.}\ \bibnamefont {Irvine}},
  \bibinfo {author} {\bibfnamefont {P.}~\bibnamefont {R{\'a}cz}}, \bibinfo
  {author} {\bibfnamefont {M.}~\bibnamefont {Lenner}}, \bibinfo {author}
  {\bibfnamefont {N.}~\bibnamefont {Kro{\'o}}}, \bibinfo {author}
  {\bibfnamefont {G.}~\bibnamefont {Farkas}}, \bibinfo {author} {\bibfnamefont
  {A.}~\bibnamefont {Mitrofanov}}, \bibinfo {author} {\bibfnamefont
  {A.}~\bibnamefont {Baltu{\u{s}}ka}}, \bibinfo {author} {\bibfnamefont
  {T.}~\bibnamefont {Fuji}}, \bibinfo {author} {\bibfnamefont {F.}~\bibnamefont
  {Krausz}}, \ and\ \bibinfo {author} {\bibfnamefont {A.~Y.}\ \bibnamefont
  {Elezzabi}}} (\bibinfo {year} {2010}),\ \bibfield  {title} {\enquote
  {\bibinfo {title} {Observation of few-cycle, strong-field phenomena in
  surface plasmon fields},}\ }\href@noop {} {\bibfield  {journal} {\bibinfo
  {journal} {Opt. Exp.}\ }\textbf {\bibinfo {volume} {18}},\ \bibinfo {pages}
  {24206--24212}}\BibitemShut {NoStop}%
\bibitem [{\citenamefont {Donnelly}\ \emph {et~al.}(2001)\citenamefont
  {Donnelly}, \citenamefont {Rust}, \citenamefont {Weiner}, \citenamefont
  {Allen}, \citenamefont {Smith}, \citenamefont {Steinke}, \citenamefont
  {Wilks}, \citenamefont {Zweiback}, \citenamefont {Cowan},\ and\ \citenamefont
  {Ditmire}}]{Donnelly2001}%
  \BibitemOpen
  \bibfield  {author} {\bibinfo {author} {\bibnamefont {Donnelly},
  \bibfnamefont {T~D}}, \bibinfo {author} {\bibfnamefont {M.}~\bibnamefont
  {Rust}}, \bibinfo {author} {\bibfnamefont {I.}~\bibnamefont {Weiner}},
  \bibinfo {author} {\bibfnamefont {M.}~\bibnamefont {Allen}}, \bibinfo
  {author} {\bibfnamefont {R.~A.}\ \bibnamefont {Smith}}, \bibinfo {author}
  {\bibfnamefont {C.~A.}\ \bibnamefont {Steinke}}, \bibinfo {author}
  {\bibfnamefont {S.}~\bibnamefont {Wilks}}, \bibinfo {author} {\bibfnamefont
  {J.}~\bibnamefont {Zweiback}}, \bibinfo {author} {\bibfnamefont {T.~E.}\
  \bibnamefont {Cowan}}, \ and\ \bibinfo {author} {\bibfnamefont
  {T.}~\bibnamefont {Ditmire}}} (\bibinfo {year} {2001}),\ \bibfield  {title}
  {\enquote {\bibinfo {title} {Hard x-ray and hot electron production from
  intense laser irradiation of wavelength-scale particles},}\ }\href@noop {}
  {\bibfield  {journal} {\bibinfo  {journal} {J. Phys. B}\ }\textbf {\bibinfo
  {volume} {34}},\ \bibinfo {pages} {L313}}\BibitemShut {NoStop}%
\bibitem [{\citenamefont {Drescher}\ \emph {et~al.}(2001)\citenamefont
  {Drescher}, \citenamefont {Hentschel}, \citenamefont {Kienberger},
  \citenamefont {Tempea}, \citenamefont {Spielmann}, \citenamefont {Reider},
  \citenamefont {Corkum},\ and\ \citenamefont {Krausz}}]{drescher2001x}%
  \BibitemOpen
  \bibfield  {author} {\bibinfo {author} {\bibnamefont {Drescher},
  \bibfnamefont {M}}, \bibinfo {author} {\bibfnamefont {M.}~\bibnamefont
  {Hentschel}}, \bibinfo {author} {\bibfnamefont {R.}~\bibnamefont
  {Kienberger}}, \bibinfo {author} {\bibfnamefont {G.}~\bibnamefont {Tempea}},
  \bibinfo {author} {\bibfnamefont {C.}~\bibnamefont {Spielmann}}, \bibinfo
  {author} {\bibfnamefont {G.~A.}\ \bibnamefont {Reider}}, \bibinfo {author}
  {\bibfnamefont {P.~B.}\ \bibnamefont {Corkum}}, \ and\ \bibinfo {author}
  {\bibfnamefont {F.}~\bibnamefont {Krausz}}} (\bibinfo {year} {2001}),\
  \bibfield  {title} {\enquote {\bibinfo {title} {X-ray pulses approaching the
  attosecond frontier},}\ }\href@noop {} {\bibfield  {journal} {\bibinfo
  {journal} {Science}\ }\textbf {\bibinfo {volume} {291}},\ \bibinfo {pages}
  {1923--1927}}\BibitemShut {NoStop}%
\bibitem [{\citenamefont {Drescher}\ \emph {et~al.}(2002)\citenamefont
  {Drescher}, \citenamefont {Hentschel}, \citenamefont {Kienberger},
  \citenamefont {Uiberacker}, \citenamefont {Yakovlev}, \citenamefont
  {Scrinzi}, \citenamefont {Westerwalbesloh}, \citenamefont {Kleineberg},
  \citenamefont {Heinzmann},\ and\ \citenamefont {Krausz}}]{drescher2002time}%
  \BibitemOpen
  \bibfield  {author} {\bibinfo {author} {\bibnamefont {Drescher},
  \bibfnamefont {M}}, \bibinfo {author} {\bibfnamefont {M.}~\bibnamefont
  {Hentschel}}, \bibinfo {author} {\bibfnamefont {R.}~\bibnamefont
  {Kienberger}}, \bibinfo {author} {\bibfnamefont {M.}~\bibnamefont
  {Uiberacker}}, \bibinfo {author} {\bibfnamefont {V.}~\bibnamefont
  {Yakovlev}}, \bibinfo {author} {\bibfnamefont {A.}~\bibnamefont {Scrinzi}},
  \bibinfo {author} {\bibfnamefont {Th.}\ \bibnamefont {Westerwalbesloh}},
  \bibinfo {author} {\bibfnamefont {U.}~\bibnamefont {Kleineberg}}, \bibinfo
  {author} {\bibfnamefont {U.}~\bibnamefont {Heinzmann}}, \ and\ \bibinfo
  {author} {\bibfnamefont {F.}~\bibnamefont {Krausz}}} (\bibinfo {year}
  {2002}),\ \bibfield  {title} {\enquote {\bibinfo {title} {Time-resolved
  atomic inner-shell spectroscopy},}\ }\href@noop {} {\bibfield  {journal}
  {\bibinfo  {journal} {Nature}\ }\textbf {\bibinfo {volume} {419}},\ \bibinfo
  {pages} {803--807}}\BibitemShut {NoStop}%
\bibitem [{\citenamefont {Ebadi}(2014)}]{Ebadi14A}%
  \BibitemOpen
  \bibfield  {author} {\bibinfo {author} {\bibnamefont {Ebadi}, \bibfnamefont
  {H}}} (\bibinfo {year} {2014}),\ \bibfield  {title} {\enquote {\bibinfo
  {title} {Interferences induced by spatially nonhomogeneous fields in
  high-harmonic generation},}\ }\href@noop {} {\bibfield  {journal} {\bibinfo
  {journal} {Phys. Rev. A}\ }\textbf {\bibinfo {volume} {89}},\ \bibinfo
  {pages} {053413}}\BibitemShut {NoStop}%
\bibitem [{\citenamefont {Ehberger}\ \emph {et~al.}(2015)\citenamefont
  {Ehberger}, \citenamefont {Hammer}, \citenamefont {Eisele}, \citenamefont
  {Kr\"uger}, \citenamefont {Noe}, \citenamefont {H\"ogele},\ and\
  \citenamefont {Hommelhoff}}]{Ehberger2015}%
  \BibitemOpen
  \bibfield  {author} {\bibinfo {author} {\bibnamefont {Ehberger},
  \bibfnamefont {D}}, \bibinfo {author} {\bibfnamefont {J.}~\bibnamefont
  {Hammer}}, \bibinfo {author} {\bibfnamefont {M.}~\bibnamefont {Eisele}},
  \bibinfo {author} {\bibfnamefont {M.}~\bibnamefont {Kr\"uger}}, \bibinfo
  {author} {\bibfnamefont {J.}~\bibnamefont {Noe}}, \bibinfo {author}
  {\bibfnamefont {A.}~\bibnamefont {H\"ogele}}, \ and\ \bibinfo {author}
  {\bibfnamefont {P.}~\bibnamefont {Hommelhoff}}} (\bibinfo {year} {2015}),\
  \bibfield  {title} {\enquote {\bibinfo {title} {Highly coherent electron beam
  from a laser-triggered tungsten needle tip},}\ }\href@noop {} {\bibfield
  {journal} {\bibinfo  {journal} {Phys. Rev. Lett.}\ }\textbf {\bibinfo
  {volume} {114}},\ \bibinfo {pages} {227601}}\BibitemShut {NoStop}%
\bibitem [{\citenamefont {Einstein}(1905)}]{einstein1905erzeugung}%
  \BibitemOpen
  \bibfield  {author} {\bibinfo {author} {\bibnamefont {Einstein},
  \bibfnamefont {A}}} (\bibinfo {year} {1905}),\ \bibfield  {title} {\enquote
  {\bibinfo {title} {{\"U}ber einen die erzeugung und verwandlung des lichtes
  betreffenden heuristischen gesichtspunkt},}\ }\href@noop {} {\bibfield
  {journal} {\bibinfo  {journal} {Annalen der Physik}\ }\textbf {\bibinfo
  {volume} {322}},\ \bibinfo {pages} {132--148}}\BibitemShut {NoStop}%
\bibitem [{\citenamefont {Esteban}\ \emph {et~al.}(2012)\citenamefont
  {Esteban}, \citenamefont {Borisov}, \citenamefont {Nordlander},\ and\
  \citenamefont {Aizpurua}}]{Esteban2012}%
  \BibitemOpen
  \bibfield  {author} {\bibinfo {author} {\bibnamefont {Esteban}, \bibfnamefont
  {R}}, \bibinfo {author} {\bibfnamefont {A.~G.}\ \bibnamefont {Borisov}},
  \bibinfo {author} {\bibfnamefont {P.}~\bibnamefont {Nordlander}}, \ and\
  \bibinfo {author} {\bibfnamefont {J.}~\bibnamefont {Aizpurua}}} (\bibinfo
  {year} {2012}),\ \bibfield  {title} {\enquote {\bibinfo {title} {Bridging
  quantum and classical plasmonics with a quantum-corrected model},}\
  }\href@noop {} {\bibfield  {journal} {\bibinfo  {journal} {Nat. Commun.}\
  }\textbf {\bibinfo {volume} {3}},\ \bibinfo {pages} {825}}\BibitemShut
  {NoStop}%
\bibitem [{\citenamefont {Fabris}\ \emph {et~al.}(2015)\citenamefont {Fabris},
  \citenamefont {Witting}, \citenamefont {Okell}, \citenamefont {Walke},
  \citenamefont {Matia-Hernando}, \citenamefont {Henkel}, \citenamefont
  {Barillot}, \citenamefont {Lein}, \citenamefont {Marangos},\ and\
  \citenamefont {Tisch}}]{fabris2015synchronized}%
  \BibitemOpen
  \bibfield  {author} {\bibinfo {author} {\bibnamefont {Fabris}, \bibfnamefont
  {D}}, \bibinfo {author} {\bibfnamefont {T.}~\bibnamefont {Witting}}, \bibinfo
  {author} {\bibfnamefont {W.~A.}\ \bibnamefont {Okell}}, \bibinfo {author}
  {\bibfnamefont {D.~J.}\ \bibnamefont {Walke}}, \bibinfo {author}
  {\bibfnamefont {P.}~\bibnamefont {Matia-Hernando}}, \bibinfo {author}
  {\bibfnamefont {J.}~\bibnamefont {Henkel}}, \bibinfo {author} {\bibfnamefont
  {T.~R.}\ \bibnamefont {Barillot}}, \bibinfo {author} {\bibfnamefont
  {M.}~\bibnamefont {Lein}}, \bibinfo {author} {\bibfnamefont {J.~P.}\
  \bibnamefont {Marangos}}, \ and\ \bibinfo {author} {\bibfnamefont {J.~W.~G.}\
  \bibnamefont {Tisch}}} (\bibinfo {year} {2015}),\ \bibfield  {title}
  {\enquote {\bibinfo {title} {Synchronized pulses generated at 20 ev and 90 ev
  for attosecond pump--probe experiments},}\ }\href@noop {} {\bibfield
  {journal} {\bibinfo  {journal} {Nat. Phot.}\ }\textbf {\bibinfo {volume}
  {9}},\ \bibinfo {pages} {383--387}}\BibitemShut {NoStop}%
\bibitem [{\citenamefont {Faisal}(1987)}]{FaisalBook}%
  \BibitemOpen
  \bibfield  {author} {\bibinfo {author} {\bibnamefont {Faisal}, \bibfnamefont
  {F~H~M}}} (\bibinfo {year} {1987}),\ \href@noop {} {\emph {\bibinfo {title}
  {Theory of Multiphoton processes}}}\ (\bibinfo  {publisher} {Springer},\
  \bibinfo {address} {New York})\BibitemShut {NoStop}%
\bibitem [{\citenamefont {Feist}\ \emph {et~al.}(2015)\citenamefont {Feist},
  \citenamefont {Echternkamp}, \citenamefont {Schauss}, \citenamefont
  {Yalunin}, \citenamefont {Sch\"afer},\ and\ \citenamefont
  {Ropers}}]{Feist2015}%
  \BibitemOpen
  \bibfield  {author} {\bibinfo {author} {\bibnamefont {Feist}, \bibfnamefont
  {A}}, \bibinfo {author} {\bibfnamefont {K.~E.}\ \bibnamefont {Echternkamp}},
  \bibinfo {author} {\bibfnamefont {J.}~\bibnamefont {Schauss}}, \bibinfo
  {author} {\bibfnamefont {S.~V.}\ \bibnamefont {Yalunin}}, \bibinfo {author}
  {\bibfnamefont {S.}~\bibnamefont {Sch\"afer}}, \ and\ \bibinfo {author}
  {\bibfnamefont {C.}~\bibnamefont {Ropers}}} (\bibinfo {year} {2015}),\
  \bibfield  {title} {\enquote {\bibinfo {title} {Quantum coherent optical
  phase modulation in an ultrafast transmission electron microscope},}\
  }\href@noop {} {\bibfield  {journal} {\bibinfo  {journal} {Nature}\ }\textbf
  {\bibinfo {volume} {521}},\ \bibinfo {pages} {200--203}}\BibitemShut
  {NoStop}%
\bibitem [{\citenamefont {Feng}\ and\ \citenamefont {Liu}(2015)}]{Feng15}%
  \BibitemOpen
  \bibfield  {author} {\bibinfo {author} {\bibnamefont {Feng}, \bibfnamefont
  {L}}, \ and\ \bibinfo {author} {\bibfnamefont {H.}~\bibnamefont {Liu}}}
  (\bibinfo {year} {2015}),\ \bibfield  {title} {\enquote {\bibinfo {title}
  {Attosecond extreme ultraviolet generation in cluster by using spatially
  inhomogeneous field},}\ }\href@noop {} {\bibfield  {journal} {\bibinfo
  {journal} {Phys. Plasmas}\ }\textbf {\bibinfo {volume} {22}},\ \bibinfo
  {pages} {013107}}\BibitemShut {NoStop}%
\bibitem [{\citenamefont {Feng}\ \emph {et~al.}(2013)\citenamefont {Feng},
  \citenamefont {Yuan},\ and\ \citenamefont {Chu}}]{Feng13}%
  \BibitemOpen
  \bibfield  {author} {\bibinfo {author} {\bibnamefont {Feng}, \bibfnamefont
  {L}}, \bibinfo {author} {\bibfnamefont {M.}~\bibnamefont {Yuan}}, \ and\
  \bibinfo {author} {\bibfnamefont {T.}~\bibnamefont {Chu}}} (\bibinfo {year}
  {2013}),\ \bibfield  {title} {\enquote {\bibinfo {title} {Attosecond x-ray
  source generation from two-color polarized gating plasmonic field
  enhancement},}\ }\href@noop {} {\bibfield  {journal} {\bibinfo  {journal}
  {Phys. Plasmas}\ }\textbf {\bibinfo {volume} {20}},\ \bibinfo {pages}
  {122307}}\BibitemShut {NoStop}%
\bibitem [{\citenamefont {Feng}\ \emph {et~al.}(2009)\citenamefont {Feng},
  \citenamefont {Gilbertson}, \citenamefont {Mashiko}, \citenamefont {Wang},
  \citenamefont {Khan}, \citenamefont {Chini}, \citenamefont {Wu},
  \citenamefont {Zhao},\ and\ \citenamefont {Chang}}]{feng2009generation}%
  \BibitemOpen
  \bibfield  {author} {\bibinfo {author} {\bibnamefont {Feng}, \bibfnamefont
  {X}}, \bibinfo {author} {\bibfnamefont {S.}~\bibnamefont {Gilbertson}},
  \bibinfo {author} {\bibfnamefont {H.}~\bibnamefont {Mashiko}}, \bibinfo
  {author} {\bibfnamefont {H.}~\bibnamefont {Wang}}, \bibinfo {author}
  {\bibfnamefont {S.~D.}\ \bibnamefont {Khan}}, \bibinfo {author}
  {\bibfnamefont {M.}~\bibnamefont {Chini}}, \bibinfo {author} {\bibfnamefont
  {Y.}~\bibnamefont {Wu}}, \bibinfo {author} {\bibfnamefont {K.}~\bibnamefont
  {Zhao}}, \ and\ \bibinfo {author} {\bibfnamefont {Z.}~\bibnamefont {Chang}}}
  (\bibinfo {year} {2009}),\ \bibfield  {title} {\enquote {\bibinfo {title}
  {Generation of isolated attosecond pulses with 20 to 28 femtosecond
  lasers},}\ }\href@noop {} {\bibfield  {journal} {\bibinfo  {journal} {Phys.
  Rev. Lett.}\ }\textbf {\bibinfo {volume} {103}},\ \bibinfo {pages}
  {183901}}\BibitemShut {NoStop}%
\bibitem [{\citenamefont {Fern\'andez-Garc\'ia}\ \emph
  {et~al.}(2014)\citenamefont {Fern\'andez-Garc\'ia}, \citenamefont
  {Sonnefraud}, \citenamefont {Fern\'andez-Dom\'inguez}, \citenamefont
  {Giannini},\ and\ \citenamefont {Maier}}]{fernandez-garcia_design_2014}%
  \BibitemOpen
  \bibfield  {author} {\bibinfo {author} {\bibnamefont {Fern\'andez-Garc\'ia},
  \bibfnamefont {R}}, \bibinfo {author} {\bibfnamefont {Y.}~\bibnamefont
  {Sonnefraud}}, \bibinfo {author} {\bibfnamefont {A.~I.}\ \bibnamefont
  {Fern\'andez-Dom\'inguez}}, \bibinfo {author} {\bibfnamefont
  {V.}~\bibnamefont {Giannini}}, \ and\ \bibinfo {author} {\bibfnamefont
  {S.~A.}\ \bibnamefont {Maier}}} (\bibinfo {year} {2014}),\ \bibfield  {title}
  {\enquote {\bibinfo {title} {Design considerations for near-field enhancement
  in optical antennas},}\ }\href@noop {} {\bibfield  {journal} {\bibinfo
  {journal} {Contemporary Physics}\ }\textbf {\bibinfo {volume} {55}},\
  \bibinfo {pages} {1--11}}\BibitemShut {NoStop}%
\bibitem [{\citenamefont {Fernandez-Varea}\ \emph {et~al.}(1993)\citenamefont
  {Fernandez-Varea}, \citenamefont {Mayol}, \citenamefont {Liljequist},\ and\
  \citenamefont {Salvat}}]{Fernandez-Varea1993}%
  \BibitemOpen
  \bibfield  {author} {\bibinfo {author} {\bibnamefont {Fernandez-Varea},
  \bibfnamefont {J~M}}, \bibinfo {author} {\bibfnamefont {R.}~\bibnamefont
  {Mayol}}, \bibinfo {author} {\bibfnamefont {D.}~\bibnamefont {Liljequist}}, \
  and\ \bibinfo {author} {\bibfnamefont {F.}~\bibnamefont {Salvat}}} (\bibinfo
  {year} {1993}),\ \bibfield  {title} {\enquote {\bibinfo {title} {Inelastic
  scattering of electrons in solids from a generalized oscillator strength
  model using optical and photoelectric data},}\ }\href@noop {} {\bibfield
  {journal} {\bibinfo  {journal} {J. Phys. Condens. Matter}\ }\textbf {\bibinfo
  {volume} {5}},\ \bibinfo {pages} {3593}}\BibitemShut {NoStop}%
\bibitem [{\citenamefont {Ferrari}\ \emph {et~al.}(2010)\citenamefont
  {Ferrari}, \citenamefont {Calegari}, \citenamefont {Lucchini}, \citenamefont
  {Vozzi}, \citenamefont {Stagira}, \citenamefont {Sansone},\ and\
  \citenamefont {Nisoli}}]{ferrari2010high}%
  \BibitemOpen
  \bibfield  {author} {\bibinfo {author} {\bibnamefont {Ferrari}, \bibfnamefont
  {F}}, \bibinfo {author} {\bibfnamefont {F.}~\bibnamefont {Calegari}},
  \bibinfo {author} {\bibfnamefont {M.}~\bibnamefont {Lucchini}}, \bibinfo
  {author} {\bibfnamefont {C.}~\bibnamefont {Vozzi}}, \bibinfo {author}
  {\bibfnamefont {S.}~\bibnamefont {Stagira}}, \bibinfo {author} {\bibfnamefont
  {G.}~\bibnamefont {Sansone}}, \ and\ \bibinfo {author} {\bibfnamefont
  {M.}~\bibnamefont {Nisoli}}} (\bibinfo {year} {2010}),\ \bibfield  {title}
  {\enquote {\bibinfo {title} {High-energy isolated attosecond pulses generated
  by above-saturation few-cycle fields},}\ }\href@noop {} {\bibfield  {journal}
  {\bibinfo  {journal} {Nat. Phot.}\ }\textbf {\bibinfo {volume} {4}},\
  \bibinfo {pages} {875--879}}\BibitemShut {NoStop}%
\bibitem [{\citenamefont {Ferr{\'e}}\ \emph {et~al.}(2015)\citenamefont
  {Ferr{\'e}}, \citenamefont {Handschin}, \citenamefont {Dumergue},
  \citenamefont {Burgy}, \citenamefont {Comby}, \citenamefont {Descamps},
  \citenamefont {Fabre}, \citenamefont {Garcia}, \citenamefont {G{\'e}neaux},
  \citenamefont {Merceron}, \citenamefont {M\'evel}, \citenamefont {Nahon},
  \citenamefont {Petit}, \citenamefont {Pons}, \citenamefont {Staedter},
  \citenamefont {Weber}, \citenamefont {Ruchon}, \citenamefont {Blanchet},\
  and\ \citenamefont {Mairesse}}]{ferre2015table}%
  \BibitemOpen
  \bibfield  {author} {\bibinfo {author} {\bibnamefont {Ferr{\'e}},
  \bibfnamefont {A}}, \bibinfo {author} {\bibfnamefont {C.}~\bibnamefont
  {Handschin}}, \bibinfo {author} {\bibfnamefont {M.}~\bibnamefont {Dumergue}},
  \bibinfo {author} {\bibfnamefont {F.}~\bibnamefont {Burgy}}, \bibinfo
  {author} {\bibfnamefont {A.}~\bibnamefont {Comby}}, \bibinfo {author}
  {\bibfnamefont {D.}~\bibnamefont {Descamps}}, \bibinfo {author}
  {\bibfnamefont {B.}~\bibnamefont {Fabre}}, \bibinfo {author} {\bibfnamefont
  {G.~A.}\ \bibnamefont {Garcia}}, \bibinfo {author} {\bibfnamefont
  {R.}~\bibnamefont {G{\'e}neaux}}, \bibinfo {author} {\bibfnamefont
  {L.}~\bibnamefont {Merceron}}, \bibinfo {author} {\bibfnamefont
  {E.}~\bibnamefont {M\'evel}}, \bibinfo {author} {\bibfnamefont
  {L.}~\bibnamefont {Nahon}}, \bibinfo {author} {\bibfnamefont
  {S.}~\bibnamefont {Petit}}, \bibinfo {author} {\bibfnamefont
  {B.}~\bibnamefont {Pons}}, \bibinfo {author} {\bibfnamefont {D}~\bibnamefont
  {Staedter}}, \bibinfo {author} {\bibfnamefont {S.}~\bibnamefont {Weber}},
  \bibinfo {author} {\bibfnamefont {T.}~\bibnamefont {Ruchon}}, \bibinfo
  {author} {\bibfnamefont {V.}~\bibnamefont {Blanchet}}, \ and\ \bibinfo
  {author} {\bibfnamefont {Y.}~\bibnamefont {Mairesse}}} (\bibinfo {year}
  {2015}),\ \bibfield  {title} {\enquote {\bibinfo {title} {A table-top
  ultrashort light source in the extreme ultraviolet for circular dichroism
  experiments},}\ }\href@noop {} {\bibfield  {journal} {\bibinfo  {journal}
  {Nat. Phot.}\ }\textbf {\bibinfo {volume} {9}},\ \bibinfo {pages}
  {93--98}}\BibitemShut {NoStop}%
\bibitem [{\citenamefont {Feti{\'c}}\ \emph {et~al.}(2012)\citenamefont
  {Feti{\'c}}, \citenamefont {Kalajd\u{z}i{\'c}},\ and\ \citenamefont
  {Milo\u{s}evi{\'c}}}]{Fetic12}%
  \BibitemOpen
  \bibfield  {author} {\bibinfo {author} {\bibnamefont {Feti{\'c}},
  \bibfnamefont {B}}, \bibinfo {author} {\bibfnamefont {K.}~\bibnamefont
  {Kalajd\u{z}i{\'c}}}, \ and\ \bibinfo {author} {\bibfnamefont {D.~B.}\
  \bibnamefont {Milo\u{s}evi{\'c}}}} (\bibinfo {year} {2012}),\ \bibfield
  {title} {\enquote {\bibinfo {title} {High-order harmonic generation by a
  spatially inhomogeneous field},}\ }\href@noop {} {\bibfield  {journal}
  {\bibinfo  {journal} {Ann. Phys. (Berlin)}\ }\textbf {\bibinfo {volume}
  {525}},\ \bibinfo {pages} {107--117}}\BibitemShut {NoStop}%
\bibitem [{\citenamefont {Fie{\ss}}\ \emph {et~al.}(2010)\citenamefont
  {Fie{\ss}}, \citenamefont {Schultze}, \citenamefont {Goulielmakis},
  \citenamefont {Dennhardt}, \citenamefont {Gagnon}, \citenamefont
  {Hofstetter}, \citenamefont {Kienberger},\ and\ \citenamefont
  {Krausz}}]{fiess2010versatile}%
  \BibitemOpen
  \bibfield  {author} {\bibinfo {author} {\bibnamefont {Fie{\ss}},
  \bibfnamefont {M}}, \bibinfo {author} {\bibfnamefont {M.}~\bibnamefont
  {Schultze}}, \bibinfo {author} {\bibfnamefont {E.}~\bibnamefont
  {Goulielmakis}}, \bibinfo {author} {\bibfnamefont {B.}~\bibnamefont
  {Dennhardt}}, \bibinfo {author} {\bibfnamefont {J.}~\bibnamefont {Gagnon}},
  \bibinfo {author} {\bibfnamefont {M.}~\bibnamefont {Hofstetter}}, \bibinfo
  {author} {\bibfnamefont {R.}~\bibnamefont {Kienberger}}, \ and\ \bibinfo
  {author} {\bibfnamefont {F.}~\bibnamefont {Krausz}}} (\bibinfo {year}
  {2010}),\ \bibfield  {title} {\enquote {\bibinfo {title} {Versatile apparatus
  for attosecond metrology and spectroscopy},}\ }\href@noop {} {\bibfield
  {journal} {\bibinfo  {journal} {Rev. of Sci. Instr.}\ }\textbf {\bibinfo
  {volume} {81}},\ \bibinfo {pages} {093103}}\BibitemShut {NoStop}%
\bibitem [{\citenamefont {Foreman}\ \emph {et~al.}(2013)\citenamefont
  {Foreman}, \citenamefont {Kealhofer}, \citenamefont {Skulason}, \citenamefont
  {Klopfer},\ and\ \citenamefont {Kasevich}}]{Foreman2013}%
  \BibitemOpen
  \bibfield  {author} {\bibinfo {author} {\bibnamefont {Foreman}, \bibfnamefont
  {S~M}}, \bibinfo {author} {\bibfnamefont {C.}~\bibnamefont {Kealhofer}},
  \bibinfo {author} {\bibfnamefont {G.~E.}\ \bibnamefont {Skulason}}, \bibinfo
  {author} {\bibfnamefont {B.~B.}\ \bibnamefont {Klopfer}}, \ and\ \bibinfo
  {author} {\bibfnamefont {M.~A.}\ \bibnamefont {Kasevich}}} (\bibinfo {year}
  {2013}),\ \bibfield  {title} {\enquote {\bibinfo {title} {Ultrafast
  microfocus x-ray source based on a femtosecond laser-triggered tip},}\
  }\href@noop {} {\bibfield  {journal} {\bibinfo  {journal} {Ann. Phys.
  (Berlin)}\ }\textbf {\bibinfo {volume} {525}},\ \bibinfo {pages}
  {L19--L22}}\BibitemShut {NoStop}%
\bibitem [{\citenamefont {F\"org}\ \emph {et~al.}(2016)\citenamefont {F\"org},
  \citenamefont {Sch\"otz}, \citenamefont {S{\"u}{\ss}mann}, \citenamefont
  {F\"orster}, \citenamefont {M.~Kr\"uger}, \citenamefont {Ahn}, \citenamefont
  {Okell}, \citenamefont {Wintersperger}, \citenamefont {Zherebtsov},
  \citenamefont {Guggenmos}, \citenamefont {Pervak}, \citenamefont {Kessel},
  \citenamefont {Trushin}, \citenamefont {Azzeer}, \citenamefont {Stockman},
  \citenamefont {Kim}, \citenamefont {Krausz}, \citenamefont {Hommelhoff},\
  and\ \citenamefont {Kling}}]{Foerg2016}%
  \BibitemOpen
  \bibfield  {author} {\bibinfo {author} {\bibnamefont {F\"org}, \bibfnamefont
  {B}}, \bibinfo {author} {\bibfnamefont {J.}~\bibnamefont {Sch\"otz}},
  \bibinfo {author} {\bibfnamefont {F.}~\bibnamefont {S{\"u}{\ss}mann}},
  \bibinfo {author} {\bibfnamefont {M.}~\bibnamefont {F\"orster}}, \bibinfo
  {author} {\bibfnamefont {M.}~\bibnamefont {M.~Kr\"uger}}, \bibinfo {author}
  {\bibfnamefont {B.}~\bibnamefont {Ahn}}, \bibinfo {author} {\bibfnamefont
  {W.~A.}\ \bibnamefont {Okell}}, \bibinfo {author} {\bibfnamefont
  {K.}~\bibnamefont {Wintersperger}}, \bibinfo {author} {\bibfnamefont
  {S.}~\bibnamefont {Zherebtsov}}, \bibinfo {author} {\bibfnamefont
  {A.}~\bibnamefont {Guggenmos}}, \bibinfo {author} {\bibfnamefont
  {V.}~\bibnamefont {Pervak}}, \bibinfo {author} {\bibfnamefont
  {A.}~\bibnamefont {Kessel}}, \bibinfo {author} {\bibfnamefont {S.~A.}\
  \bibnamefont {Trushin}}, \bibinfo {author} {\bibfnamefont {A.~M.}\
  \bibnamefont {Azzeer}}, \bibinfo {author} {\bibfnamefont {M.~I.}\
  \bibnamefont {Stockman}}, \bibinfo {author} {\bibfnamefont {D.}~\bibnamefont
  {Kim}}, \bibinfo {author} {\bibfnamefont {F.}~\bibnamefont {Krausz}},
  \bibinfo {author} {\bibfnamefont {P.}~\bibnamefont {Hommelhoff}}, \ and\
  \bibinfo {author} {\bibfnamefont {M.~F.}\ \bibnamefont {Kling}}} (\bibinfo
  {year} {2016}),\ \bibfield  {title} {\enquote {\bibinfo {title} {Attosecond
  nanoscale near-field sampling},}\ }\href@noop {} {\bibfield  {journal}
  {\bibinfo  {journal} {Nat. Comm.}\ }\textbf {\bibinfo {volume} {7}},\
  \bibinfo {pages} {11717}}\BibitemShut {NoStop}%
\bibitem [{\citenamefont {Fowler}\ and\ \citenamefont
  {Nordheim}(1928)}]{Fowler1928}%
  \BibitemOpen
  \bibfield  {author} {\bibinfo {author} {\bibnamefont {Fowler}, \bibfnamefont
  {R~H}}, \ and\ \bibinfo {author} {\bibfnamefont {L.}~\bibnamefont
  {Nordheim}}} (\bibinfo {year} {1928}),\ \bibfield  {title} {\enquote
  {\bibinfo {title} {Electron emission in intense electric fields},}\
  }\href@noop {} {\bibfield  {journal} {\bibinfo  {journal} {Proc. Roy. Soc.
  London Ser. A}\ }\textbf {\bibinfo {volume} {119}},\ \bibinfo {pages}
  {173--181}}\BibitemShut {NoStop}%
\bibitem [{\citenamefont {Frank}\ \emph {et~al.}(2012)\citenamefont {Frank},
  \citenamefont {Arrell}, \citenamefont {Witting}, \citenamefont {Okell},
  \citenamefont {McKenna}, \citenamefont {Robinson}, \citenamefont {Haworth},
  \citenamefont {Austin}, \citenamefont {Teng}, \citenamefont {Walmsley},
  \citenamefont {Marangos},\ and\ \citenamefont {Tisch}}]{frank2012invited}%
  \BibitemOpen
  \bibfield  {author} {\bibinfo {author} {\bibnamefont {Frank}, \bibfnamefont
  {F}}, \bibinfo {author} {\bibfnamefont {C.}~\bibnamefont {Arrell}}, \bibinfo
  {author} {\bibfnamefont {T.}~\bibnamefont {Witting}}, \bibinfo {author}
  {\bibfnamefont {W.~A.}\ \bibnamefont {Okell}}, \bibinfo {author}
  {\bibfnamefont {J.}~\bibnamefont {McKenna}}, \bibinfo {author} {\bibfnamefont
  {J.~S.}\ \bibnamefont {Robinson}}, \bibinfo {author} {\bibfnamefont {C.~A.}\
  \bibnamefont {Haworth}}, \bibinfo {author} {\bibfnamefont {D.}~\bibnamefont
  {Austin}}, \bibinfo {author} {\bibfnamefont {H.}~\bibnamefont {Teng}},
  \bibinfo {author} {\bibfnamefont {I.~A.}\ \bibnamefont {Walmsley}}, \bibinfo
  {author} {\bibfnamefont {J.~P}\ \bibnamefont {Marangos}}, \ and\ \bibinfo
  {author} {\bibfnamefont {J.~W.~G.}\ \bibnamefont {Tisch}}} (\bibinfo {year}
  {2012}),\ \bibfield  {title} {\enquote {\bibinfo {title} {Invited review
  article: technology for attosecond science},}\ }\href@noop {} {\bibfield
  {journal} {\bibinfo  {journal} {Rev. of Sci. Instr.}\ }\textbf {\bibinfo
  {volume} {83}},\ \bibinfo {pages} {071101}}\BibitemShut {NoStop}%
\bibitem [{\citenamefont {Frassetto}\ \emph {et~al.}(2014)\citenamefont
  {Frassetto}, \citenamefont {Trabattoni}, \citenamefont {Anumula},
  \citenamefont {Sansone}, \citenamefont {Calegari}, \citenamefont {Nisoli},\
  and\ \citenamefont {Poletto}}]{frassetto2014high}%
  \BibitemOpen
  \bibfield  {author} {\bibinfo {author} {\bibnamefont {Frassetto},
  \bibfnamefont {F}}, \bibinfo {author} {\bibfnamefont {A.}~\bibnamefont
  {Trabattoni}}, \bibinfo {author} {\bibfnamefont {S.}~\bibnamefont {Anumula}},
  \bibinfo {author} {\bibfnamefont {G.}~\bibnamefont {Sansone}}, \bibinfo
  {author} {\bibfnamefont {F.}~\bibnamefont {Calegari}}, \bibinfo {author}
  {\bibfnamefont {M.}~\bibnamefont {Nisoli}}, \ and\ \bibinfo {author}
  {\bibfnamefont {L.}~\bibnamefont {Poletto}}} (\bibinfo {year} {2014}),\
  \bibfield  {title} {\enquote {\bibinfo {title} {High-throughput beamline for
  attosecond pulses based on toroidal mirrors with microfocusing
  capabilities},}\ }\href@noop {} {\bibfield  {journal} {\bibinfo  {journal}
  {Rev. of Sci. Instr.}\ }\textbf {\bibinfo {volume} {85}},\ \bibinfo {pages}
  {103115}}\BibitemShut {NoStop}%
\bibitem [{\citenamefont {Fr\"ohlich}\ and\ \citenamefont
  {Pelzer}(1955)}]{Frohlich1955}%
  \BibitemOpen
  \bibfield  {author} {\bibinfo {author} {\bibnamefont {Fr\"ohlich},
  \bibfnamefont {H}}, \ and\ \bibinfo {author} {\bibfnamefont {H}~\bibnamefont
  {Pelzer}}} (\bibinfo {year} {1955}),\ \bibfield  {title} {\enquote {\bibinfo
  {title} {Plasma oscillations and energy loss of charged particles in
  solids},}\ }\href@noop {} {\bibfield  {journal} {\bibinfo  {journal} {Proc.
  Phys. Soc. A}\ }\textbf {\bibinfo {volume} {68}},\ \bibinfo {pages}
  {525}}\BibitemShut {NoStop}%
\bibitem [{\citenamefont {Fursey}(2005)}]{Fursey2005}%
  \BibitemOpen
  \bibfield  {author} {\bibinfo {author} {\bibnamefont {Fursey}, \bibfnamefont
  {G~N}}} (\bibinfo {year} {2005}),\ \href@noop {} {\emph {\bibinfo {title}
  {Field Emission in Vacuum Microelectronics}}}\ (\bibinfo  {publisher} {Kluwer
  Academic/Plenum},\ \bibinfo {address} {New York})\BibitemShut {NoStop}%
\bibitem [{\citenamefont {Ganter}\ \emph {et~al.}(2008)\citenamefont {Ganter},
  \citenamefont {Bakker}, \citenamefont {Gough}, \citenamefont {Leemann},
  \citenamefont {Paraliev}, \citenamefont {Pedrozzi}, \citenamefont
  {Le~Pimpec}, \citenamefont {Schlott}, \citenamefont {Rivkin},\ and\
  \citenamefont {Wrulich}}]{Ganter2008}%
  \BibitemOpen
  \bibfield  {author} {\bibinfo {author} {\bibnamefont {Ganter}, \bibfnamefont
  {R}}, \bibinfo {author} {\bibfnamefont {R.}~\bibnamefont {Bakker}}, \bibinfo
  {author} {\bibfnamefont {C.}~\bibnamefont {Gough}}, \bibinfo {author}
  {\bibfnamefont {S.~C.}\ \bibnamefont {Leemann}}, \bibinfo {author}
  {\bibfnamefont {M.}~\bibnamefont {Paraliev}}, \bibinfo {author}
  {\bibfnamefont {M.}~\bibnamefont {Pedrozzi}}, \bibinfo {author}
  {\bibfnamefont {F.}~\bibnamefont {Le~Pimpec}}, \bibinfo {author}
  {\bibfnamefont {V.}~\bibnamefont {Schlott}}, \bibinfo {author} {\bibfnamefont
  {L.}~\bibnamefont {Rivkin}}, \ and\ \bibinfo {author} {\bibfnamefont
  {A.}~\bibnamefont {Wrulich}}} (\bibinfo {year} {2008}),\ \bibfield  {title}
  {\enquote {\bibinfo {title} {Laser-photofield emission from needle cathodes
  for low-emittance electron beams},}\ }\href@noop {} {\bibfield  {journal}
  {\bibinfo  {journal} {Phys. Rev. Lett.}\ }\textbf {\bibinfo {volume} {100}},\
  \bibinfo {pages} {064801}}\BibitemShut {NoStop}%
\bibitem [{\citenamefont {Ghimire}\ \emph {et~al.}(2011)\citenamefont
  {Ghimire}, \citenamefont {DiChiara}, \citenamefont {Sistrunk}, \citenamefont
  {Agostini}, \citenamefont {DiMauro},\ and\ \citenamefont
  {Reis}}]{Ghimire2011}%
  \BibitemOpen
  \bibfield  {author} {\bibinfo {author} {\bibnamefont {Ghimire}, \bibfnamefont
  {S}}, \bibinfo {author} {\bibfnamefont {A.~D.}\ \bibnamefont {DiChiara}},
  \bibinfo {author} {\bibfnamefont {E.}~\bibnamefont {Sistrunk}}, \bibinfo
  {author} {\bibfnamefont {P.}~\bibnamefont {Agostini}}, \bibinfo {author}
  {\bibfnamefont {L.~F.}\ \bibnamefont {DiMauro}}, \ and\ \bibinfo {author}
  {\bibfnamefont {D.~A.}\ \bibnamefont {Reis}}} (\bibinfo {year} {2011}),\
  \bibfield  {title} {\enquote {\bibinfo {title} {Observation of high-order
  harmonic generation in a bulk crystal},}\ }\href@noop {} {\bibfield
  {journal} {\bibinfo  {journal} {Nat. Phys.}\ }\textbf {\bibinfo {volume}
  {7}},\ \bibinfo {pages} {138--141}}\BibitemShut {NoStop}%
\bibitem [{\citenamefont {Gong}\ \emph {et~al.}(2015)\citenamefont {Gong},
  \citenamefont {Joly}, \citenamefont {Hu}, \citenamefont {El-Khoury},\ and\
  \citenamefont {Hess}}]{Gong15}%
  \BibitemOpen
  \bibfield  {author} {\bibinfo {author} {\bibnamefont {Gong}, \bibfnamefont
  {Y}}, \bibinfo {author} {\bibfnamefont {A.~G.}\ \bibnamefont {Joly}},
  \bibinfo {author} {\bibfnamefont {D.}~\bibnamefont {Hu}}, \bibinfo {author}
  {\bibfnamefont {P.~Z.}\ \bibnamefont {El-Khoury}}, \ and\ \bibinfo {author}
  {\bibfnamefont {W.~P.}\ \bibnamefont {Hess}}} (\bibinfo {year} {2015}),\
  \bibfield  {title} {\enquote {\bibinfo {title} {Ultrafast imaging of surface
  plasmons propagating on a gold surface},}\ }\href@noop {} {\bibfield
  {journal} {\bibinfo  {journal} {Nano Lett.}\ }\textbf {\bibinfo {volume}
  {15}},\ \bibinfo {pages} {3472--3478}}\BibitemShut {NoStop}%
\bibitem [{\citenamefont {Goulielmakis}\ \emph {et~al.}(2008)\citenamefont
  {Goulielmakis}, \citenamefont {Schultze}, \citenamefont {Hofstetter},
  \citenamefont {Yakovlev}, \citenamefont {Gagnon}, \citenamefont {Uiberacker},
  \citenamefont {Aquila}, \citenamefont {Gullikson}, \citenamefont {Attwood},
  \citenamefont {Kienberger}, \citenamefont {Krausz},\ and\ \citenamefont
  {Kleineberg}}]{goulielmakis2008single}%
  \BibitemOpen
  \bibfield  {author} {\bibinfo {author} {\bibnamefont {Goulielmakis},
  \bibfnamefont {E}}, \bibinfo {author} {\bibfnamefont {M.}~\bibnamefont
  {Schultze}}, \bibinfo {author} {\bibfnamefont {M.}~\bibnamefont
  {Hofstetter}}, \bibinfo {author} {\bibfnamefont {V.~S.}\ \bibnamefont
  {Yakovlev}}, \bibinfo {author} {\bibfnamefont {J.}~\bibnamefont {Gagnon}},
  \bibinfo {author} {\bibfnamefont {M.}~\bibnamefont {Uiberacker}}, \bibinfo
  {author} {\bibfnamefont {A.~L.}\ \bibnamefont {Aquila}}, \bibinfo {author}
  {\bibfnamefont {E.~M.}\ \bibnamefont {Gullikson}}, \bibinfo {author}
  {\bibfnamefont {D.~T.}\ \bibnamefont {Attwood}}, \bibinfo {author}
  {\bibfnamefont {R.}~\bibnamefont {Kienberger}}, \bibinfo {author}
  {\bibfnamefont {F.}~\bibnamefont {Krausz}}, \ and\ \bibinfo {author}
  {\bibfnamefont {U.}~\bibnamefont {Kleineberg}}} (\bibinfo {year} {2008}),\
  \bibfield  {title} {\enquote {\bibinfo {title} {Single-cycle nonlinear
  optics},}\ }\href@noop {} {\bibfield  {journal} {\bibinfo  {journal}
  {Science}\ }\textbf {\bibinfo {volume} {320}},\ \bibinfo {pages}
  {1614--1617}}\BibitemShut {NoStop}%
\bibitem [{\citenamefont {Grobe}\ and\ \citenamefont
  {Eberly}(1993)}]{Grobe1993HHG}%
  \BibitemOpen
  \bibfield  {author} {\bibinfo {author} {\bibnamefont {Grobe}, \bibfnamefont
  {R}}, \ and\ \bibinfo {author} {\bibfnamefont {J.~H.}\ \bibnamefont
  {Eberly}}} (\bibinfo {year} {1993}),\ \bibfield  {title} {\enquote {\bibinfo
  {title} {One-dimensional model of a negative ion and its interaction with
  laser fields},}\ }\href@noop {} {\bibfield  {journal} {\bibinfo  {journal}
  {Phys. Rev. A}\ }\textbf {\bibinfo {volume} {48}},\ \bibinfo {pages}
  {4664}}\BibitemShut {NoStop}%
\bibitem [{\citenamefont {Guggenmos}\ \emph {et~al.}(2015)\citenamefont
  {Guggenmos}, \citenamefont {Jobst}, \citenamefont {Ossiander}, \citenamefont
  {Rad{\"u}nz}, \citenamefont {Riemensberger}, \citenamefont {Sch{\"a}ffer},
  \citenamefont {Akil}, \citenamefont {Jakubeit}, \citenamefont {B{\"o}hm},
  \citenamefont {Noever}, \citenamefont {Nickel}, \citenamefont {Kienberger},\
  and\ \citenamefont {Kleineberg}}]{guggenmos2015chromium}%
  \BibitemOpen
  \bibfield  {author} {\bibinfo {author} {\bibnamefont {Guggenmos},
  \bibfnamefont {A}}, \bibinfo {author} {\bibfnamefont {M.}~\bibnamefont
  {Jobst}}, \bibinfo {author} {\bibfnamefont {M.}~\bibnamefont {Ossiander}},
  \bibinfo {author} {\bibfnamefont {S.}~\bibnamefont {Rad{\"u}nz}}, \bibinfo
  {author} {\bibfnamefont {J.}~\bibnamefont {Riemensberger}}, \bibinfo {author}
  {\bibfnamefont {M.}~\bibnamefont {Sch{\"a}ffer}}, \bibinfo {author}
  {\bibfnamefont {A.}~\bibnamefont {Akil}}, \bibinfo {author} {\bibfnamefont
  {C.}~\bibnamefont {Jakubeit}}, \bibinfo {author} {\bibfnamefont
  {P.}~\bibnamefont {B{\"o}hm}}, \bibinfo {author} {\bibfnamefont
  {S.}~\bibnamefont {Noever}}, \bibinfo {author} {\bibfnamefont
  {B.}~\bibnamefont {Nickel}}, \bibinfo {author} {\bibfnamefont
  {R.}~\bibnamefont {Kienberger}}, \ and\ \bibinfo {author} {\bibfnamefont
  {U.}~\bibnamefont {Kleineberg}}} (\bibinfo {year} {2015}),\ \bibfield
  {title} {\enquote {\bibinfo {title} {Chromium/scandium multilayer mirrors for
  isolated attosecond pulses at 145 ev},}\ }\href@noop {} {\bibfield  {journal}
  {\bibinfo  {journal} {Opt. Lett.}\ }\textbf {\bibinfo {volume} {40}},\
  \bibinfo {pages} {2846--2849}}\BibitemShut {NoStop}%
\bibitem [{\citenamefont {Gulde}\ \emph {et~al.}(2014)\citenamefont {Gulde},
  \citenamefont {Schweda}, \citenamefont {Storeck}, \citenamefont {Maiti},
  \citenamefont {Yu}, \citenamefont {Wodtke}, \citenamefont {Sch{\"a}fer},\
  and\ \citenamefont {Ropers}}]{Gulde2014}%
  \BibitemOpen
  \bibfield  {author} {\bibinfo {author} {\bibnamefont {Gulde}, \bibfnamefont
  {M}}, \bibinfo {author} {\bibfnamefont {S.}~\bibnamefont {Schweda}}, \bibinfo
  {author} {\bibfnamefont {G.}~\bibnamefont {Storeck}}, \bibinfo {author}
  {\bibfnamefont {M.}~\bibnamefont {Maiti}}, \bibinfo {author} {\bibfnamefont
  {H.~K.}\ \bibnamefont {Yu}}, \bibinfo {author} {\bibfnamefont {A.~M.}\
  \bibnamefont {Wodtke}}, \bibinfo {author} {\bibfnamefont {S.}~\bibnamefont
  {Sch{\"a}fer}}, \ and\ \bibinfo {author} {\bibfnamefont {C.}~\bibnamefont
  {Ropers}}} (\bibinfo {year} {2014}),\ \bibfield  {title} {\enquote {\bibinfo
  {title} {Ultrafast low-energy electron diffraction in transmission resolves
  polymer/graphene superstructure dynamics},}\ }\href@noop {} {\bibfield
  {journal} {\bibinfo  {journal} {Science}\ }\textbf {\bibinfo {volume}
  {345}},\ \bibinfo {pages} {200--204}}\BibitemShut {NoStop}%
\bibitem [{\citenamefont {Gumbrell}\ \emph {et~al.}(2001)\citenamefont
  {Gumbrell}, \citenamefont {Comley}, \citenamefont {Hutchinson},\ and\
  \citenamefont {Smith}}]{Gumbrell2001}%
  \BibitemOpen
  \bibfield  {author} {\bibinfo {author} {\bibnamefont {Gumbrell},
  \bibfnamefont {E~T}}, \bibinfo {author} {\bibfnamefont {A.~J.}\ \bibnamefont
  {Comley}}, \bibinfo {author} {\bibfnamefont {M.~H.~R.}\ \bibnamefont
  {Hutchinson}}, \ and\ \bibinfo {author} {\bibfnamefont {R.~A.}\ \bibnamefont
  {Smith}}} (\bibinfo {year} {2001}),\ \bibfield  {title} {\enquote {\bibinfo
  {title} {Intense laser interactions with sprays of submicron droplets},}\
  }\href@noop {} {\bibfield  {journal} {\bibinfo  {journal} {Phys. Plasmas}\
  }\textbf {\bibinfo {volume} {8}},\ \bibinfo {pages} {1329}}\BibitemShut
  {NoStop}%
\bibitem [{\citenamefont {Habteyes}\ \emph {et~al.}(2012)\citenamefont
  {Habteyes}, \citenamefont {Dhuey}, \citenamefont {Wood}, \citenamefont
  {Gargas}, \citenamefont {Cabrini}, \citenamefont {Schuck}, \citenamefont
  {Alivisatos},\ and\ \citenamefont {Leone}}]{habteyes_metallic_2012}%
  \BibitemOpen
  \bibfield  {author} {\bibinfo {author} {\bibnamefont {Habteyes},
  \bibfnamefont {T~G}}, \bibinfo {author} {\bibfnamefont {S.}~\bibnamefont
  {Dhuey}}, \bibinfo {author} {\bibfnamefont {E.}~\bibnamefont {Wood}},
  \bibinfo {author} {\bibfnamefont {D.}~\bibnamefont {Gargas}}, \bibinfo
  {author} {\bibfnamefont {S.}~\bibnamefont {Cabrini}}, \bibinfo {author}
  {\bibfnamefont {P.~J.}\ \bibnamefont {Schuck}}, \bibinfo {author}
  {\bibfnamefont {A.~P.}\ \bibnamefont {Alivisatos}}, \ and\ \bibinfo {author}
  {\bibfnamefont {S.~R.}\ \bibnamefont {Leone}}} (\bibinfo {year} {2012}),\
  \bibfield  {title} {\enquote {\bibinfo {title} {Metallic adhesion layer
  induced plasmon damping and molecular linker as a nondamping alternative},}\
  }\href@noop {} {\bibfield  {journal} {\bibinfo  {journal} {ACS Nano}\
  }\textbf {\bibinfo {volume} {6}},\ \bibinfo {pages} {5702--5709}}\BibitemShut
  {NoStop}%
\bibitem [{\citenamefont {Hafner}(1999)}]{Hafner1999}%
  \BibitemOpen
  \bibfield  {author} {\bibinfo {author} {\bibnamefont {Hafner}, \bibfnamefont
  {C}}} (\bibinfo {year} {1999}),\ \href@noop {} {\emph {\bibinfo {title}
  {Post-modern Electromagnetics: Using Intelligent MaXwell Solvers}}}\
  (\bibinfo  {publisher} {Wiley})\BibitemShut {NoStop}%
\bibitem [{\citenamefont {Han}\ \emph {et~al.}(2016)\citenamefont {Han},
  \citenamefont {Kim}, \citenamefont {Kim}, \citenamefont {Kim}, \citenamefont
  {Kim}, \citenamefont {Park},\ and\ \citenamefont {Kim}}]{Han2016}%
  \BibitemOpen
  \bibfield  {author} {\bibinfo {author} {\bibnamefont {Han}, \bibfnamefont
  {S}}, \bibinfo {author} {\bibfnamefont {H.}~\bibnamefont {Kim}}, \bibinfo
  {author} {\bibfnamefont {Y.~W.}\ \bibnamefont {Kim}}, \bibinfo {author}
  {\bibfnamefont {Y.-J.}\ \bibnamefont {Kim}}, \bibinfo {author} {\bibfnamefont
  {S.}~\bibnamefont {Kim}}, \bibinfo {author} {\bibfnamefont {I.-Y.}\
  \bibnamefont {Park}}, \ and\ \bibinfo {author} {\bibfnamefont {S.-W.}\
  \bibnamefont {Kim}}} (\bibinfo {year} {2016}),\ \href@noop {} {}\bibinfo
  {note} {Submitted}\BibitemShut {NoStop}%
\bibitem [{\citenamefont {Hanke}\ \emph {et~al.}(2009)\citenamefont {Hanke},
  \citenamefont {Krauss}, \citenamefont {Tr{\"a}utlein}, \citenamefont {Wild},
  \citenamefont {Bratschitsch},\ and\ \citenamefont {Leitenstorfer}}]{Hanke}%
  \BibitemOpen
  \bibfield  {author} {\bibinfo {author} {\bibnamefont {Hanke}, \bibfnamefont
  {T}}, \bibinfo {author} {\bibfnamefont {G.}~\bibnamefont {Krauss}}, \bibinfo
  {author} {\bibfnamefont {D.}~\bibnamefont {Tr{\"a}utlein}}, \bibinfo {author}
  {\bibfnamefont {B.}~\bibnamefont {Wild}}, \bibinfo {author} {\bibfnamefont
  {R.}~\bibnamefont {Bratschitsch}}, \ and\ \bibinfo {author} {\bibfnamefont
  {A.}~\bibnamefont {Leitenstorfer}}} (\bibinfo {year} {2009}),\ \bibfield
  {title} {\enquote {\bibinfo {title} {Efficient nonlinear light emission of
  single gold optical antennas driven by few-cycle near-infrared pulses},}\
  }\href@noop {} {\bibfield  {journal} {\bibinfo  {journal} {Phys. Rev. Lett.}\
  }\textbf {\bibinfo {volume} {103}},\ \bibinfo {pages} {257404}}\BibitemShut
  {NoStop}%
\bibitem [{\citenamefont {Hansen}\ \emph {et~al.}(2005)\citenamefont {Hansen},
  \citenamefont {Bhatia}, \citenamefont {Harrit},\ and\ \citenamefont
  {Oddershede}}]{Hansen05}%
  \BibitemOpen
  \bibfield  {author} {\bibinfo {author} {\bibnamefont {Hansen}, \bibfnamefont
  {P~M}}, \bibinfo {author} {\bibfnamefont {V.~K.}\ \bibnamefont {Bhatia}},
  \bibinfo {author} {\bibfnamefont {N.}~\bibnamefont {Harrit}}, \ and\ \bibinfo
  {author} {\bibfnamefont {L.}~\bibnamefont {Oddershede}}} (\bibinfo {year}
  {2005}),\ \bibfield  {title} {\enquote {\bibinfo {title} {Expanding the
  optical trapping range of gold nanoparticles},}\ }\href@noop {} {\bibfield
  {journal} {\bibinfo  {journal} {Nano Lett.}\ }\textbf {\bibinfo {volume}
  {5}},\ \bibinfo {pages} {1937--1942}}\BibitemShut {NoStop}%
\bibitem [{\citenamefont {Hartschuh}(2008)}]{Hartschuh2008}%
  \BibitemOpen
  \bibfield  {author} {\bibinfo {author} {\bibnamefont {Hartschuh},
  \bibfnamefont {A}}} (\bibinfo {year} {2008}),\ \bibfield  {title} {\enquote
  {\bibinfo {title} {Tip-enhanced near-field optical microscopy},}\ }\href@noop
  {} {\bibfield  {journal} {\bibinfo  {journal} {Angewandte Chemie
  International Edition}\ }\textbf {\bibinfo {volume} {47}},\ \bibinfo {pages}
  {8178--8191}}\BibitemShut {NoStop}%
\bibitem [{\citenamefont {Hassan}\ \emph {et~al.}(2016)\citenamefont {Hassan},
  \citenamefont {Luu}, \citenamefont {Moulet}, \citenamefont {Raskazovskaya},
  \citenamefont {Zhokhov}, \citenamefont {Garg}, \citenamefont {Karpowicz},
  \citenamefont {Zheltikov}, \citenamefont {Pervak}, \citenamefont {Krausz},\
  and\ \citenamefont {Goulielmakis}}]{hassan_optical_2016}%
  \BibitemOpen
  \bibfield  {author} {\bibinfo {author} {\bibnamefont {Hassan}, \bibfnamefont
  {M~Th}}, \bibinfo {author} {\bibfnamefont {T.~T.}\ \bibnamefont {Luu}},
  \bibinfo {author} {\bibfnamefont {A.}~\bibnamefont {Moulet}}, \bibinfo
  {author} {\bibfnamefont {O.}~\bibnamefont {Raskazovskaya}}, \bibinfo {author}
  {\bibfnamefont {P.}~\bibnamefont {Zhokhov}}, \bibinfo {author} {\bibfnamefont
  {M.}~\bibnamefont {Garg}}, \bibinfo {author} {\bibfnamefont {N.}~\bibnamefont
  {Karpowicz}}, \bibinfo {author} {\bibfnamefont {A.~M.}\ \bibnamefont
  {Zheltikov}}, \bibinfo {author} {\bibfnamefont {V.}~\bibnamefont {Pervak}},
  \bibinfo {author} {\bibfnamefont {F.}~\bibnamefont {Krausz}}, \ and\ \bibinfo
  {author} {\bibfnamefont {E.}~\bibnamefont {Goulielmakis}}} (\bibinfo {year}
  {2016}),\ \bibfield  {title} {\enquote {\bibinfo {title} {Optical attosecond
  pulses and tracking the nonlinear response of bound electrons},}\ }\href@noop
  {} {\bibfield  {journal} {\bibinfo  {journal} {Nature}\ }\textbf {\bibinfo
  {volume} {530}},\ \bibinfo {pages} {66--70}}\BibitemShut {NoStop}%
\bibitem [{\citenamefont {He}\ \emph {et~al.}(2013)\citenamefont {He},
  \citenamefont {Wang}, \citenamefont {Li}, \citenamefont {Zhang},
  \citenamefont {Lan},\ and\ \citenamefont {Lu}}]{He13}%
  \BibitemOpen
  \bibfield  {author} {\bibinfo {author} {\bibnamefont {He}, \bibfnamefont
  {L}}, \bibinfo {author} {\bibfnamefont {Z.}~\bibnamefont {Wang}}, \bibinfo
  {author} {\bibfnamefont {Y.}~\bibnamefont {Li}}, \bibinfo {author}
  {\bibfnamefont {Q.}~\bibnamefont {Zhang}}, \bibinfo {author} {\bibfnamefont
  {P.}~\bibnamefont {Lan}}, \ and\ \bibinfo {author} {\bibfnamefont
  {P.}~\bibnamefont {Lu}}} (\bibinfo {year} {2013}),\ \bibfield  {title}
  {\enquote {\bibinfo {title} {Wavelength dependence of high-order-harmonic
  yield in inhomogeneous fields},}\ }\href@noop {} {\bibfield  {journal}
  {\bibinfo  {journal} {Phys. Rev. A}\ }\textbf {\bibinfo {volume} {88}},\
  \bibinfo {pages} {053404}}\BibitemShut {NoStop}%
\bibitem [{\citenamefont {Hemmers}\ \emph {et~al.}(1998)\citenamefont
  {Hemmers}, \citenamefont {Whitfield}, \citenamefont {Glans}, \citenamefont
  {Wang}, \citenamefont {Lindle}, \citenamefont {Wehlitz},\ and\ \citenamefont
  {Sellin}}]{hemmers1998high}%
  \BibitemOpen
  \bibfield  {author} {\bibinfo {author} {\bibnamefont {Hemmers}, \bibfnamefont
  {O}}, \bibinfo {author} {\bibfnamefont {S.~B.}\ \bibnamefont {Whitfield}},
  \bibinfo {author} {\bibfnamefont {P.}~\bibnamefont {Glans}}, \bibinfo
  {author} {\bibfnamefont {H.}~\bibnamefont {Wang}}, \bibinfo {author}
  {\bibfnamefont {D.~W.}\ \bibnamefont {Lindle}}, \bibinfo {author}
  {\bibfnamefont {R.}~\bibnamefont {Wehlitz}}, \ and\ \bibinfo {author}
  {\bibfnamefont {I.~A.}\ \bibnamefont {Sellin}}} (\bibinfo {year} {1998}),\
  \bibfield  {title} {\enquote {\bibinfo {title} {High-resolution electron
  time-of-flight apparatus for the soft x-ray region},}\ }\href@noop {}
  {\bibfield  {journal} {\bibinfo  {journal} {Rev. of Sci. Instr.}\ }\textbf
  {\bibinfo {volume} {69}},\ \bibinfo {pages} {3809--3817}}\BibitemShut
  {NoStop}%
\bibitem [{\citenamefont {Henkel}\ \emph {et~al.}(2013)\citenamefont {Henkel},
  \citenamefont {Witting}, \citenamefont {Fabris}, \citenamefont {Lein},
  \citenamefont {Knight}, \citenamefont {Tisch},\ and\ \citenamefont
  {Marangos}}]{henkel2013prediction}%
  \BibitemOpen
  \bibfield  {author} {\bibinfo {author} {\bibnamefont {Henkel}, \bibfnamefont
  {J}}, \bibinfo {author} {\bibfnamefont {T.}~\bibnamefont {Witting}}, \bibinfo
  {author} {\bibfnamefont {D.}~\bibnamefont {Fabris}}, \bibinfo {author}
  {\bibfnamefont {M.}~\bibnamefont {Lein}}, \bibinfo {author} {\bibfnamefont
  {P.~L.}\ \bibnamefont {Knight}}, \bibinfo {author} {\bibfnamefont {J.~W.~G.}\
  \bibnamefont {Tisch}}, \ and\ \bibinfo {author} {\bibfnamefont {J.~P.}\
  \bibnamefont {Marangos}}} (\bibinfo {year} {2013}),\ \bibfield  {title}
  {\enquote {\bibinfo {title} {Prediction of attosecond light pulses in the vuv
  range in a high-order-harmonic-generation regime},}\ }\href@noop {}
  {\bibfield  {journal} {\bibinfo  {journal} {Phys. Rev. A}\ }\textbf {\bibinfo
  {volume} {87}},\ \bibinfo {pages} {043818}}\BibitemShut {NoStop}%
\bibitem [{\citenamefont {Hentschel}\ \emph {et~al.}(2001)\citenamefont
  {Hentschel}, \citenamefont {Kienberger}, \citenamefont {Spielmann},
  \citenamefont {Reider}, \citenamefont {Milosevic}, \citenamefont {Brabec},
  \citenamefont {Corkum}, \citenamefont {Heinzmann}, \citenamefont {Drescher},\
  and\ \citenamefont {Krausz}}]{Krausz01}%
  \BibitemOpen
  \bibfield  {author} {\bibinfo {author} {\bibnamefont {Hentschel},
  \bibfnamefont {M}}, \bibinfo {author} {\bibfnamefont {R.}~\bibnamefont
  {Kienberger}}, \bibinfo {author} {\bibfnamefont {C.}~\bibnamefont
  {Spielmann}}, \bibinfo {author} {\bibfnamefont {G.~A.}\ \bibnamefont
  {Reider}}, \bibinfo {author} {\bibfnamefont {N.}~\bibnamefont {Milosevic}},
  \bibinfo {author} {\bibfnamefont {T.}~\bibnamefont {Brabec}}, \bibinfo
  {author} {\bibfnamefont {P.~B.}\ \bibnamefont {Corkum}}, \bibinfo {author}
  {\bibfnamefont {U.}~\bibnamefont {Heinzmann}}, \bibinfo {author}
  {\bibfnamefont {M.}~\bibnamefont {Drescher}}, \ and\ \bibinfo {author}
  {\bibfnamefont {F.}~\bibnamefont {Krausz}}} (\bibinfo {year} {2001}),\
  \bibfield  {title} {\enquote {\bibinfo {title} {Attosecond metrology},}\
  }\href@noop {} {\bibfield  {journal} {\bibinfo  {journal} {Nature}\ }\textbf
  {\bibinfo {volume} {414}},\ \bibinfo {pages} {509--513}}\BibitemShut
  {NoStop}%
\bibitem [{\citenamefont {Herink}\ \emph {et~al.}(2012)\citenamefont {Herink},
  \citenamefont {Solli}, \citenamefont {Gulde},\ and\ \citenamefont
  {Ropers}}]{Herink12}%
  \BibitemOpen
  \bibfield  {author} {\bibinfo {author} {\bibnamefont {Herink}, \bibfnamefont
  {G}}, \bibinfo {author} {\bibfnamefont {D.~R.}\ \bibnamefont {Solli}},
  \bibinfo {author} {\bibfnamefont {M.}~\bibnamefont {Gulde}}, \ and\ \bibinfo
  {author} {\bibfnamefont {C.}~\bibnamefont {Ropers}}} (\bibinfo {year}
  {2012}),\ \bibfield  {title} {\enquote {\bibinfo {title} {Field-driven
  photoemission from nanostructures quenches the quiver motion},}\ }\href@noop
  {} {\bibfield  {journal} {\bibinfo  {journal} {Nature}\ }\textbf {\bibinfo
  {volume} {483}},\ \bibinfo {pages} {190--193}}\BibitemShut {NoStop}%
\bibitem [{\citenamefont {Herink}\ \emph {et~al.}(2014)\citenamefont {Herink},
  \citenamefont {Wimmer},\ and\ \citenamefont {Ropers}}]{Herink2014}%
  \BibitemOpen
  \bibfield  {author} {\bibinfo {author} {\bibnamefont {Herink}, \bibfnamefont
  {G}}, \bibinfo {author} {\bibfnamefont {L.}~\bibnamefont {Wimmer}}, \ and\
  \bibinfo {author} {\bibfnamefont {C.}~\bibnamefont {Ropers}}} (\bibinfo
  {year} {2014}),\ \bibfield  {title} {\enquote {\bibinfo {title} {Field
  emission at terahertz frequencies: {AC}-tunneling and ultrafast carrier
  dynamics},}\ }\href@noop {} {\bibfield  {journal} {\bibinfo  {journal} {New
  J. Phys.}\ }\textbf {\bibinfo {volume} {16}},\ \bibinfo {pages}
  {123005}}\BibitemShut {NoStop}%
\bibitem [{\citenamefont {van Linden van~den Heuvell}\ and\ \citenamefont
  {Muller}(1988)}]{muller}%
  \BibitemOpen
  \bibfield  {author} {\bibinfo {author} {\bibnamefont {van Linden van~den
  Heuvell}, \bibfnamefont {H~B}}, \ and\ \bibinfo {author} {\bibfnamefont
  {H.~G.}\ \bibnamefont {Muller}}} (\bibinfo {year} {1988}),\ in\ \href@noop {}
  {\emph {\bibinfo {booktitle} {Multiphoton Processes}}},\ \bibinfo {editor}
  {edited by\ \bibinfo {editor} {\bibfnamefont {S.~J.}\ \bibnamefont {Smith}}\
  and\ \bibinfo {editor} {\bibfnamefont {P.~L.}\ \bibnamefont {Knight}}}\
  (\bibinfo  {publisher} {Cambridge University Press},\ \bibinfo {address}
  {Cambridge, England})\ p.~\bibinfo {pages} {25}\BibitemShut {NoStop}%
\bibitem [{\citenamefont {Higuchi}\ \emph {et~al.}(2015)\citenamefont
  {Higuchi}, \citenamefont {Maisenbacher}, \citenamefont {Liehl}, \citenamefont
  {Dombi},\ and\ \citenamefont {Hommelhoff}}]{Higuchi2015}%
  \BibitemOpen
  \bibfield  {author} {\bibinfo {author} {\bibnamefont {Higuchi}, \bibfnamefont
  {T}}, \bibinfo {author} {\bibfnamefont {L.}~\bibnamefont {Maisenbacher}},
  \bibinfo {author} {\bibfnamefont {A.}~\bibnamefont {Liehl}}, \bibinfo
  {author} {\bibfnamefont {P.}~\bibnamefont {Dombi}}, \ and\ \bibinfo {author}
  {\bibfnamefont {P.}~\bibnamefont {Hommelhoff}}} (\bibinfo {year} {2015}),\
  \bibfield  {title} {\enquote {\bibinfo {title} {A nanoscale vacuum-tube diode
  triggered by few-cycle laser pulses},}\ }\href@noop {} {\bibfield  {journal}
  {\bibinfo  {journal} {Appl. Phys. Lett.}\ }\textbf {\bibinfo {volume}
  {106}},\ \bibinfo {pages} {051109}}\BibitemShut {NoStop}%
\bibitem [{\citenamefont {Hilbert}\ \emph {et~al.}(2009)\citenamefont
  {Hilbert}, \citenamefont {Neukirch}, \citenamefont {Uiterwaal},\ and\
  \citenamefont {Batelaan}}]{Hilbert2009}%
  \BibitemOpen
  \bibfield  {author} {\bibinfo {author} {\bibnamefont {Hilbert}, \bibfnamefont
  {S~A}}, \bibinfo {author} {\bibfnamefont {A.}~\bibnamefont {Neukirch}},
  \bibinfo {author} {\bibfnamefont {C.~J.~G.~J.}\ \bibnamefont {Uiterwaal}}, \
  and\ \bibinfo {author} {\bibfnamefont {H.}~\bibnamefont {Batelaan}}}
  (\bibinfo {year} {2009}),\ \bibfield  {title} {\enquote {\bibinfo {title}
  {Exploring temporal and rate limits of laser-induced electron emission},}\
  }\href@noop {} {\bibfield  {journal} {\bibinfo  {journal} {J. Phys. B}\
  }\textbf {\bibinfo {volume} {42}},\ \bibinfo {pages} {141001}}\BibitemShut
  {NoStop}%
\bibitem [{\citenamefont {Hobbs}\ \emph {et~al.}(2014)\citenamefont {Hobbs},
  \citenamefont {Yang}, \citenamefont {Fallahi}, \citenamefont {Keathley},
  \citenamefont {Leo}, \citenamefont {K\"artner}, \citenamefont {Graves},\ and\
  \citenamefont {Berggren}}]{Hobbs2014}%
  \BibitemOpen
  \bibfield  {author} {\bibinfo {author} {\bibnamefont {Hobbs}, \bibfnamefont
  {R~G}}, \bibinfo {author} {\bibfnamefont {Y.}~\bibnamefont {Yang}}, \bibinfo
  {author} {\bibfnamefont {A.}~\bibnamefont {Fallahi}}, \bibinfo {author}
  {\bibfnamefont {P.~D.}\ \bibnamefont {Keathley}}, \bibinfo {author}
  {\bibfnamefont {E.~De}\ \bibnamefont {Leo}}, \bibinfo {author} {\bibfnamefont
  {F.~X.}\ \bibnamefont {K\"artner}}, \bibinfo {author} {\bibfnamefont {W.~S.}\
  \bibnamefont {Graves}}, \ and\ \bibinfo {author} {\bibfnamefont {K.~K.}\
  \bibnamefont {Berggren}}} (\bibinfo {year} {2014}),\ \bibfield  {title}
  {\enquote {\bibinfo {title} {High-yield, ultrafast, surface plasmon-enhanced,
  {Au} nanorod optical field electron emitter arrays},}\ }\href@noop {}
  {\bibfield  {journal} {\bibinfo  {journal} {{ACS} Nano}\ }\textbf {\bibinfo
  {volume} {8}},\ \bibinfo {pages} {11474--11482}}\BibitemShut {NoStop}%
\bibitem [{\citenamefont {Hoffrogge}\ \emph {et~al.}(2014)\citenamefont
  {Hoffrogge}, \citenamefont {Stein}, \citenamefont {Kr\"uger}, \citenamefont
  {F\"orster}, \citenamefont {Hammer}, \citenamefont {Ehberger}, \citenamefont
  {Baum},\ and\ \citenamefont {Hommelhoff}}]{Hoffrogge2014}%
  \BibitemOpen
  \bibfield  {author} {\bibinfo {author} {\bibnamefont {Hoffrogge},
  \bibfnamefont {J}}, \bibinfo {author} {\bibfnamefont {J.~P.}\ \bibnamefont
  {Stein}}, \bibinfo {author} {\bibfnamefont {M.}~\bibnamefont {Kr\"uger}},
  \bibinfo {author} {\bibfnamefont {M.}~\bibnamefont {F\"orster}}, \bibinfo
  {author} {\bibfnamefont {J.}~\bibnamefont {Hammer}}, \bibinfo {author}
  {\bibfnamefont {D.}~\bibnamefont {Ehberger}}, \bibinfo {author}
  {\bibfnamefont {P.}~\bibnamefont {Baum}}, \ and\ \bibinfo {author}
  {\bibfnamefont {P.}~\bibnamefont {Hommelhoff}}} (\bibinfo {year} {2014}),\
  \bibfield  {title} {\enquote {\bibinfo {title} {Tip-based source of
  femtosecond electron pulses at 30 {keV}},}\ }\href@noop {} {\bibfield
  {journal} {\bibinfo  {journal} {J. Appl. Phys.}\ }\textbf {\bibinfo {volume}
  {115}},\ \bibinfo {pages} {094506}}\BibitemShut {NoStop}%
\bibitem [{\citenamefont {Hommelhoff}\ \emph
  {et~al.}(2006{\natexlab{a}})\citenamefont {Hommelhoff}, \citenamefont
  {Kealhofer},\ and\ \citenamefont {Kasevich}}]{Hommelhoff2006b}%
  \BibitemOpen
  \bibfield  {author} {\bibinfo {author} {\bibnamefont {Hommelhoff},
  \bibfnamefont {P}}, \bibinfo {author} {\bibfnamefont {C.}~\bibnamefont
  {Kealhofer}}, \ and\ \bibinfo {author} {\bibfnamefont {M.~A.}\ \bibnamefont
  {Kasevich}}} (\bibinfo {year} {2006}{\natexlab{a}}),\ \bibfield  {title}
  {\enquote {\bibinfo {title} {Femtosecond laser meets field emission tip -- a
  sensor for the carrier envelope phase?}}\ }in\ \href@noop {} {\emph {\bibinfo
  {booktitle} {Proceedings of the 2006 IEEE International Frequency Control
  Symposium and Expositions, Vols. 1 and 2}}},\ pp.\ \bibinfo {pages}
  {470--474}\BibitemShut {NoStop}%
\bibitem [{\citenamefont {Hommelhoff}\ \emph
  {et~al.}(2006{\natexlab{b}})\citenamefont {Hommelhoff}, \citenamefont
  {Kealhofer},\ and\ \citenamefont {Kasevich}}]{Hommelhoff2006a}%
  \BibitemOpen
  \bibfield  {author} {\bibinfo {author} {\bibnamefont {Hommelhoff},
  \bibfnamefont {P}}, \bibinfo {author} {\bibfnamefont {C.}~\bibnamefont
  {Kealhofer}}, \ and\ \bibinfo {author} {\bibfnamefont {M.~A.}\ \bibnamefont
  {Kasevich}}} (\bibinfo {year} {2006}{\natexlab{b}}),\ \bibfield  {title}
  {\enquote {\bibinfo {title} {Ultrafast electron pulses from a tungsten tip
  triggered by low-power femtosecond laser pulses},}\ }\href@noop {} {\bibfield
   {journal} {\bibinfo  {journal} {Phys. Rev. Lett.}\ }\textbf {\bibinfo
  {volume} {97}},\ \bibinfo {pages} {247402}}\BibitemShut {NoStop}%
\bibitem [{\citenamefont {Hommelhoff}\ and\ \citenamefont
  {Kling}(2015)}]{Hommelhoff15}%
  \BibitemOpen
  \bibfield  {author} {\bibinfo {author} {\bibnamefont {Hommelhoff},
  \bibfnamefont {P}}, \ and\ \bibinfo {author} {\bibfnamefont {M.~F.}\
  \bibnamefont {Kling}}} (\bibinfo {year} {2015}),\ \href@noop {} {\emph
  {\bibinfo {title} {Attosecond Nanophysics: From Basic Science to
  Applications}}}\ (\bibinfo  {publisher} {Wiley-VCH},\ \bibinfo {address}
  {Berlin})\BibitemShut {NoStop}%
\bibitem [{\citenamefont {Hommelhoff}\ \emph
  {et~al.}(2006{\natexlab{c}})\citenamefont {Hommelhoff}, \citenamefont
  {Sortais}, \citenamefont {Aghajani-Talesh},\ and\ \citenamefont
  {Kasevich}}]{PeterH06}%
  \BibitemOpen
  \bibfield  {author} {\bibinfo {author} {\bibnamefont {Hommelhoff},
  \bibfnamefont {P}}, \bibinfo {author} {\bibfnamefont {Y.}~\bibnamefont
  {Sortais}}, \bibinfo {author} {\bibfnamefont {A.}~\bibnamefont
  {Aghajani-Talesh}}, \ and\ \bibinfo {author} {\bibfnamefont {M.~A.}\
  \bibnamefont {Kasevich}}} (\bibinfo {year} {2006}{\natexlab{c}}),\ \bibfield
  {title} {\enquote {\bibinfo {title} {Field emission tip as a nanometer source
  of free electron femtosecond pulses},}\ }\href@noop {} {\bibfield  {journal}
  {\bibinfo  {journal} {Phys. Rev. Lett.}\ }\textbf {\bibinfo {volume} {96}},\
  \bibinfo {pages} {077401}}\BibitemShut {NoStop}%
\bibitem [{\citenamefont {Huppert}\ \emph {et~al.}(2015)\citenamefont
  {Huppert}, \citenamefont {Jordan},\ and\ \citenamefont
  {W{\"o}rner}}]{huppert2015attosecond}%
  \BibitemOpen
  \bibfield  {author} {\bibinfo {author} {\bibnamefont {Huppert}, \bibfnamefont
  {M}}, \bibinfo {author} {\bibfnamefont {I.}~\bibnamefont {Jordan}}, \ and\
  \bibinfo {author} {\bibfnamefont {H.~J.}\ \bibnamefont {W{\"o}rner}}}
  (\bibinfo {year} {2015}),\ \bibfield  {title} {\enquote {\bibinfo {title}
  {Attosecond beamline with actively stabilized and spatially separated beam
  paths},}\ }\href@noop {} {\bibfield  {journal} {\bibinfo  {journal} {Rev. of
  Sci. Instr.}\ }\textbf {\bibinfo {volume} {86}},\ \bibinfo {pages}
  {123106}}\BibitemShut {NoStop}%
\bibitem [{\citenamefont {Husakou}\ and\ \citenamefont
  {Herrmann}(2014)}]{Husakou14A}%
  \BibitemOpen
  \bibfield  {author} {\bibinfo {author} {\bibnamefont {Husakou}, \bibfnamefont
  {A}}, \ and\ \bibinfo {author} {\bibfnamefont {J.}~\bibnamefont {Herrmann}}}
  (\bibinfo {year} {2014}),\ \bibfield  {title} {\enquote {\bibinfo {title}
  {Quasi-phase-matched high-harmonic generation in composites of metal
  nanoparticles and a noble gas},}\ }\href@noop {} {\bibfield  {journal}
  {\bibinfo  {journal} {Phys. Rev. A}\ }\textbf {\bibinfo {volume} {90}},\
  \bibinfo {pages} {023831}}\BibitemShut {NoStop}%
\bibitem [{\citenamefont {Husakou}\ \emph
  {et~al.}(2011{\natexlab{a}})\citenamefont {Husakou}, \citenamefont {Im},\
  and\ \citenamefont {Herrmann}}]{Husakou11A}%
  \BibitemOpen
  \bibfield  {author} {\bibinfo {author} {\bibnamefont {Husakou}, \bibfnamefont
  {A}}, \bibinfo {author} {\bibfnamefont {S-J.}\ \bibnamefont {Im}}, \ and\
  \bibinfo {author} {\bibfnamefont {J.}~\bibnamefont {Herrmann}}} (\bibinfo
  {year} {2011}{\natexlab{a}}),\ \bibfield  {title} {\enquote {\bibinfo {title}
  {Theory of plasmon-enhanced high-order harmonic generation in the vicinity of
  metal nanostructures in noble gases},}\ }\href@noop {} {\bibfield  {journal}
  {\bibinfo  {journal} {Phys. Rev. A}\ }\textbf {\bibinfo {volume} {83}},\
  \bibinfo {pages} {043839}}\BibitemShut {NoStop}%
\bibitem [{\citenamefont {Husakou}\ \emph {et~al.}(2015)\citenamefont
  {Husakou}, \citenamefont {Im}, \citenamefont {Kim},\ and\ \citenamefont
  {Herrmann}}]{Husakou15}%
  \BibitemOpen
  \bibfield  {author} {\bibinfo {author} {\bibnamefont {Husakou}, \bibfnamefont
  {A}}, \bibinfo {author} {\bibfnamefont {S-J.}\ \bibnamefont {Im}}, \bibinfo
  {author} {\bibfnamefont {K.H.}\ \bibnamefont {Kim}}, \ and\ \bibinfo {author}
  {\bibfnamefont {J.}~\bibnamefont {Herrmann}}} (\bibinfo {year} {2015}),\
  \bibfield  {title} {\enquote {\bibinfo {title} {High harmonic generation
  assisted by metal nanostructures and nanoparticles},}\ }in\ \href@noop {}
  {\emph {\bibinfo {booktitle} {{P}rogress in {N}onlinear {N}ano-{O}ptics}}},\
  \bibinfo {editor} {edited by\ \bibinfo {editor} {\bibfnamefont
  {S.}~\bibnamefont {Sakabe}}, \bibinfo {editor} {\bibfnamefont
  {C.}~\bibnamefont {Lienau}}, \ and\ \bibinfo {editor} {\bibfnamefont
  {R.}~\bibnamefont {Grunwald}}}\ (\bibinfo  {publisher} {Springer
  International Publishing})\ pp.\ \bibinfo {pages} {251--268}\BibitemShut
  {NoStop}%
\bibitem [{\citenamefont {Husakou}\ \emph
  {et~al.}(2011{\natexlab{b}})\citenamefont {Husakou}, \citenamefont
  {Kelkensberg}, \citenamefont {Herrmann},\ and\ \citenamefont
  {Vrakking}}]{Husakou11OE}%
  \BibitemOpen
  \bibfield  {author} {\bibinfo {author} {\bibnamefont {Husakou}, \bibfnamefont
  {A}}, \bibinfo {author} {\bibfnamefont {F.}~\bibnamefont {Kelkensberg}},
  \bibinfo {author} {\bibfnamefont {J.}~\bibnamefont {Herrmann}}, \ and\
  \bibinfo {author} {\bibfnamefont {M.~J.~J.}\ \bibnamefont {Vrakking}}}
  (\bibinfo {year} {2011}{\natexlab{b}}),\ \bibfield  {title} {\enquote
  {\bibinfo {title} {Polarization gating and circularly-polarized high harmonic
  generation using plasmonic enhancement in metal nanostructures},}\
  }\href@noop {} {\bibfield  {journal} {\bibinfo  {journal} {Opt. Exp.}\
  }\textbf {\bibinfo {volume} {19}},\ \bibinfo {pages}
  {25346--25354}}\BibitemShut {NoStop}%
\bibitem [{\citenamefont {Huth}\ \emph {et~al.}(2013)\citenamefont {Huth},
  \citenamefont {Chuvilin}, \citenamefont {Schnell}, \citenamefont {Amenabar},
  \citenamefont {Krutokhvostov}, \citenamefont {Lopatin},\ and\ \citenamefont
  {Hillenbrand}}]{Huth2013}%
  \BibitemOpen
  \bibfield  {author} {\bibinfo {author} {\bibnamefont {Huth}, \bibfnamefont
  {F}}, \bibinfo {author} {\bibfnamefont {A.}~\bibnamefont {Chuvilin}},
  \bibinfo {author} {\bibfnamefont {M.}~\bibnamefont {Schnell}}, \bibinfo
  {author} {\bibfnamefont {I.}~\bibnamefont {Amenabar}}, \bibinfo {author}
  {\bibfnamefont {R.}~\bibnamefont {Krutokhvostov}}, \bibinfo {author}
  {\bibfnamefont {S.}~\bibnamefont {Lopatin}}, \ and\ \bibinfo {author}
  {\bibfnamefont {R.}~\bibnamefont {Hillenbrand}}} (\bibinfo {year} {2013}),\
  \bibfield  {title} {\enquote {\bibinfo {title} {Resonant antenna probes for
  tip-enhanced infrared near-field microscopy},}\ }\href@noop {} {\bibfield
  {journal} {\bibinfo  {journal} {Nano Lett.}\ }\textbf {\bibinfo {volume}
  {13}},\ \bibinfo {pages} {1065--1072}}\BibitemShut {NoStop}%
\bibitem [{\citenamefont {Iaconis}\ and\ \citenamefont
  {Walmsley}(1998)}]{iaconis_spectral_1998}%
  \BibitemOpen
  \bibfield  {author} {\bibinfo {author} {\bibnamefont {Iaconis}, \bibfnamefont
  {C}}, \ and\ \bibinfo {author} {\bibfnamefont {I.~A.}\ \bibnamefont
  {Walmsley}}} (\bibinfo {year} {1998}),\ \bibfield  {title} {\enquote
  {\bibinfo {title} {Spectral phase interferometry for direct electric-field
  reconstruction of ultrashort optical pulses},}\ }\href@noop {} {\bibfield
  {journal} {\bibinfo  {journal} {Opt. Lett.}\ }\textbf {\bibinfo {volume}
  {23}},\ \bibinfo {pages} {792--794}}\BibitemShut {NoStop}%
\bibitem [{\citenamefont {Inouye}\ and\ \citenamefont
  {Kawata}(1994)}]{Inouye1994}%
  \BibitemOpen
  \bibfield  {author} {\bibinfo {author} {\bibnamefont {Inouye}, \bibfnamefont
  {Y}}, \ and\ \bibinfo {author} {\bibfnamefont {S.}~\bibnamefont {Kawata}}}
  (\bibinfo {year} {1994}),\ \bibfield  {title} {\enquote {\bibinfo {title}
  {Near-field scanning optical microscope with a metallic probe tip},}\
  }\href@noop {} {\bibfield  {journal} {\bibinfo  {journal} {Opt. Lett.}\
  }\textbf {\bibinfo {volume} {19}}~(\bibinfo {number} {3}),\ \bibinfo {pages}
  {159--161}}\BibitemShut {NoStop}%
\bibitem [{\citenamefont {Itatani}\ \emph {et~al.}(2002)\citenamefont
  {Itatani}, \citenamefont {Qu{\'e}r{\'e}}, \citenamefont {Yudin},
  \citenamefont {Ivanov}, \citenamefont {Krausz},\ and\ \citenamefont
  {Corkum}}]{itatani2002attosecond}%
  \BibitemOpen
  \bibfield  {author} {\bibinfo {author} {\bibnamefont {Itatani}, \bibfnamefont
  {J}}, \bibinfo {author} {\bibfnamefont {F.}~\bibnamefont {Qu{\'e}r{\'e}}},
  \bibinfo {author} {\bibfnamefont {G.~L.}\ \bibnamefont {Yudin}}, \bibinfo
  {author} {\bibfnamefont {M.~Yu.}\ \bibnamefont {Ivanov}}, \bibinfo {author}
  {\bibfnamefont {F.}~\bibnamefont {Krausz}}, \ and\ \bibinfo {author}
  {\bibfnamefont {P.~B.}\ \bibnamefont {Corkum}}} (\bibinfo {year} {2002}),\
  \bibfield  {title} {\enquote {\bibinfo {title} {Attosecond streak camera},}\
  }\href@noop {} {\bibfield  {journal} {\bibinfo  {journal} {Phys. Rev. Lett.}\
  }\textbf {\bibinfo {volume} {88}},\ \bibinfo {pages} {173903}}\BibitemShut
  {NoStop}%
\bibitem [{\citenamefont {Jackson}(1999)}]{Jackson1999}%
  \BibitemOpen
  \bibfield  {author} {\bibinfo {author} {\bibnamefont {Jackson}, \bibfnamefont
  {J~D}}} (\bibinfo {year} {1999}),\ \href@noop {} {\emph {\bibinfo {title}
  {Classical {E}lectrodynamics}}}\ (\bibinfo  {publisher} {Wiley})\BibitemShut
  {NoStop}%
\bibitem [{\citenamefont {Jain}\ \emph {et~al.}(2006)\citenamefont {Jain},
  \citenamefont {Eustis},\ and\ \citenamefont {El-Sayed}}]{jain_plasmon_2006}%
  \BibitemOpen
  \bibfield  {author} {\bibinfo {author} {\bibnamefont {Jain}, \bibfnamefont
  {P~K}}, \bibinfo {author} {\bibfnamefont {S.}~\bibnamefont {Eustis}}, \ and\
  \bibinfo {author} {\bibfnamefont {M.~A.}\ \bibnamefont {El-Sayed}}} (\bibinfo
  {year} {2006}),\ \bibfield  {title} {\enquote {\bibinfo {title} {Plasmon
  coupling in nanorod assemblies: Optical absorption, discrete dipole
  approximation simulation, and exciton-coupling model},}\ }\href@noop {}
  {\bibfield  {journal} {\bibinfo  {journal} {J. Phys. Chem. B}\ }\textbf
  {\bibinfo {volume} {110}},\ \bibinfo {pages} {18243--18253}}\BibitemShut
  {NoStop}%
\bibitem [{\citenamefont {Joachain}\ \emph {et~al.}(2012)\citenamefont
  {Joachain}, \citenamefont {Kylstra},\ and\ \citenamefont
  {Potvliege}}]{Joachain1}%
  \BibitemOpen
  \bibfield  {author} {\bibinfo {author} {\bibnamefont {Joachain},
  \bibfnamefont {C~J}}, \bibinfo {author} {\bibfnamefont {N.~J.}\ \bibnamefont
  {Kylstra}}, \ and\ \bibinfo {author} {\bibfnamefont {R.~M.}\ \bibnamefont
  {Potvliege}}} (\bibinfo {year} {2012}),\ \href@noop {} {\emph {\bibinfo
  {title} {Atoms in Intense Laser Fields}}}\ (\bibinfo  {publisher} {Cambridge
  University Press},\ \bibinfo {address} {Cambridge, England})\BibitemShut
  {NoStop}%
\bibitem [{\citenamefont {Jones}\ \emph {et~al.}(2000)\citenamefont {Jones},
  \citenamefont {Diddams}, \citenamefont {Ranka}, \citenamefont {Stentz},
  \citenamefont {Windeler}, \citenamefont {Hall},\ and\ \citenamefont
  {Cundiff}}]{jones2000carrier}%
  \BibitemOpen
  \bibfield  {author} {\bibinfo {author} {\bibnamefont {Jones}, \bibfnamefont
  {D~J}}, \bibinfo {author} {\bibfnamefont {S.~A.}\ \bibnamefont {Diddams}},
  \bibinfo {author} {\bibfnamefont {J.~K.}\ \bibnamefont {Ranka}}, \bibinfo
  {author} {\bibfnamefont {A.}~\bibnamefont {Stentz}}, \bibinfo {author}
  {\bibfnamefont {R.~S.}\ \bibnamefont {Windeler}}, \bibinfo {author}
  {\bibfnamefont {J.~L.}\ \bibnamefont {Hall}}, \ and\ \bibinfo {author}
  {\bibfnamefont {S.~T.}\ \bibnamefont {Cundiff}}} (\bibinfo {year} {2000}),\
  \bibfield  {title} {\enquote {\bibinfo {title} {Carrier-envelope phase
  control of femtosecond mode-locked lasers and direct optical frequency
  synthesis},}\ }\href@noop {} {\bibfield  {journal} {\bibinfo  {journal}
  {Science}\ }\textbf {\bibinfo {volume} {288}},\ \bibinfo {pages}
  {635--639}}\BibitemShut {NoStop}%
\bibitem [{\citenamefont {Juffmann}\ \emph {et~al.}(2015)\citenamefont
  {Juffmann}, \citenamefont {Klopfer}, \citenamefont {Skulason}, \citenamefont
  {Kealhofer}, \citenamefont {Xiao}, \citenamefont {Foreman},\ and\
  \citenamefont {Kasevich}}]{Juffmann2015}%
  \BibitemOpen
  \bibfield  {author} {\bibinfo {author} {\bibnamefont {Juffmann},
  \bibfnamefont {T}}, \bibinfo {author} {\bibfnamefont {B.~B.}\ \bibnamefont
  {Klopfer}}, \bibinfo {author} {\bibfnamefont {G.~E.}\ \bibnamefont
  {Skulason}}, \bibinfo {author} {\bibfnamefont {C.}~\bibnamefont {Kealhofer}},
  \bibinfo {author} {\bibfnamefont {F.}~\bibnamefont {Xiao}}, \bibinfo {author}
  {\bibfnamefont {S.~M.}\ \bibnamefont {Foreman}}, \ and\ \bibinfo {author}
  {\bibfnamefont {M.~A.}\ \bibnamefont {Kasevich}}} (\bibinfo {year} {2015}),\
  \bibfield  {title} {\enquote {\bibinfo {title} {Ultrafast time-resolved
  photoelectric emission},}\ }\href@noop {} {\bibfield  {journal} {\bibinfo
  {journal} {Phys. Rev. Lett.}\ }\textbf {\bibinfo {volume} {115}},\ \bibinfo
  {pages} {264803}}\BibitemShut {NoStop}%
\bibitem [{\citenamefont {Kameta}\ \emph {et~al.}(2002)\citenamefont {Kameta},
  \citenamefont {Kouchi}, \citenamefont {Ukai},\ and\ \citenamefont
  {Hatano}}]{kameta2002photoabsorption}%
  \BibitemOpen
  \bibfield  {author} {\bibinfo {author} {\bibnamefont {Kameta}, \bibfnamefont
  {K}}, \bibinfo {author} {\bibfnamefont {N.}~\bibnamefont {Kouchi}}, \bibinfo
  {author} {\bibfnamefont {M.}~\bibnamefont {Ukai}}, \ and\ \bibinfo {author}
  {\bibfnamefont {Y.}~\bibnamefont {Hatano}}} (\bibinfo {year} {2002}),\
  \bibfield  {title} {\enquote {\bibinfo {title} {Photoabsorption,
  photoionization, and neutral-dissociation cross sections of simple
  hydrocarbons in the vacuum ultraviolet range},}\ }\href@noop {} {\bibfield
  {journal} {\bibinfo  {journal} {Journal of Electron Spectroscopy and Related
  Phenomena}\ }\textbf {\bibinfo {volume} {123}},\ \bibinfo {pages}
  {225--238}}\BibitemShut {NoStop}%
\bibitem [{\citenamefont {Kane}(1999)}]{kane1999recent}%
  \BibitemOpen
  \bibfield  {author} {\bibinfo {author} {\bibnamefont {Kane}, \bibfnamefont
  {D~J}}} (\bibinfo {year} {1999}),\ \bibfield  {title} {\enquote {\bibinfo
  {title} {Recent progress toward real-time measurement of ultrashort laser
  pulses},}\ }\href@noop {} {\bibfield  {journal} {\bibinfo  {journal} {IEEE J.
  of Quant. Electron.}\ }\textbf {\bibinfo {volume} {35}},\ \bibinfo {pages}
  {421--431}}\BibitemShut {NoStop}%
\bibitem [{\citenamefont {Kane}\ \emph {et~al.}(1997)\citenamefont {Kane},
  \citenamefont {Rodr\'{\i}guez}, \citenamefont {Taylor},\ and\ \citenamefont
  {Clement}}]{kane1997simultaneous}%
  \BibitemOpen
  \bibfield  {author} {\bibinfo {author} {\bibnamefont {Kane}, \bibfnamefont
  {D~J}}, \bibinfo {author} {\bibfnamefont {G.}~\bibnamefont {Rodr\'{\i}guez}},
  \bibinfo {author} {\bibfnamefont {A.~J.}\ \bibnamefont {Taylor}}, \ and\
  \bibinfo {author} {\bibfnamefont {T.~S.}\ \bibnamefont {Clement}}} (\bibinfo
  {year} {1997}),\ \bibfield  {title} {\enquote {\bibinfo {title} {Simultaneous
  measurement of two ultrashort laser pulses from a single spectrogram in a
  single shot},}\ }\href@noop {} {\bibfield  {journal} {\bibinfo  {journal} {J.
  Opt. Soc. Am. B}\ }\textbf {\bibinfo {volume} {14}},\ \bibinfo {pages}
  {935--943}}\BibitemShut {NoStop}%
\bibitem [{\citenamefont {Kane}\ and\ \citenamefont
  {Trebino}(1993)}]{kane_characterization_1993}%
  \BibitemOpen
  \bibfield  {author} {\bibinfo {author} {\bibnamefont {Kane}, \bibfnamefont
  {D~J}}, \ and\ \bibinfo {author} {\bibfnamefont {R.}~\bibnamefont {Trebino}}}
  (\bibinfo {year} {1993}),\ \bibfield  {title} {\enquote {\bibinfo {title}
  {Characterization of arbitrary femtosecond pulses using frequency-resolved
  optical gating},}\ }\href@noop {} {\bibfield  {journal} {\bibinfo  {journal}
  {IEEE J. Quant. Electron.}\ }\textbf {\bibinfo {volume} {29}},\ \bibinfo
  {pages} {571--579}}\BibitemShut {NoStop}%
\bibitem [{\citenamefont {Kauranen}\ and\ \citenamefont
  {Zayats}(2012)}]{Kauranen2012}%
  \BibitemOpen
  \bibfield  {author} {\bibinfo {author} {\bibnamefont {Kauranen},
  \bibfnamefont {M}}, \ and\ \bibinfo {author} {\bibfnamefont {A.~V.}\
  \bibnamefont {Zayats}}} (\bibinfo {year} {2012}),\ \bibfield  {title}
  {\enquote {\bibinfo {title} {Nonlinear plasmonics},}\ }\href@noop {}
  {\bibfield  {journal} {\bibinfo  {journal} {Nat. Photon.}\ }\textbf {\bibinfo
  {volume} {9}},\ \bibinfo {pages} {737--748}}\BibitemShut {NoStop}%
\bibitem [{\citenamefont {Kazansky}\ and\ \citenamefont
  {Echenique}(2009)}]{kazansky2009one}%
  \BibitemOpen
  \bibfield  {author} {\bibinfo {author} {\bibnamefont {Kazansky},
  \bibfnamefont {A~K}}, \ and\ \bibinfo {author} {\bibfnamefont {P.~M.}\
  \bibnamefont {Echenique}}} (\bibinfo {year} {2009}),\ \bibfield  {title}
  {\enquote {\bibinfo {title} {One-electron model for the electronic response
  of metal surfaces to subfemtosecond photoexcitation},}\ }\href@noop {}
  {\bibfield  {journal} {\bibinfo  {journal} {Phys. Rev. Lett.}\ }\textbf
  {\bibinfo {volume} {102}},\ \bibinfo {pages} {177401}}\BibitemShut {NoStop}%
\bibitem [{\citenamefont {Kealhofer}\ \emph {et~al.}(2012)\citenamefont
  {Kealhofer}, \citenamefont {Foreman}, \citenamefont {Gerlich},\ and\
  \citenamefont {Kasevich}}]{Kealhofer2012}%
  \BibitemOpen
  \bibfield  {author} {\bibinfo {author} {\bibnamefont {Kealhofer},
  \bibfnamefont {C}}, \bibinfo {author} {\bibfnamefont {S.~M.}\ \bibnamefont
  {Foreman}}, \bibinfo {author} {\bibfnamefont {S.}~\bibnamefont {Gerlich}}, \
  and\ \bibinfo {author} {\bibfnamefont {M.~A.}\ \bibnamefont {Kasevich}}}
  (\bibinfo {year} {2012}),\ \bibfield  {title} {\enquote {\bibinfo {title}
  {Ultrafast laser-triggered emission from hafnium carbide tips},}\ }\href@noop
  {} {\bibfield  {journal} {\bibinfo  {journal} {Phys. Rev. B}\ }\textbf
  {\bibinfo {volume} {86}},\ \bibinfo {pages} {035405}}\BibitemShut {NoStop}%
\bibitem [{\citenamefont {Keathley}\ \emph {et~al.}(2013)\citenamefont
  {Keathley}, \citenamefont {Sell}, \citenamefont {Putnam}, \citenamefont
  {Guerrera}, \citenamefont {Velasquez-Garc\'{\i}a},\ and\ \citenamefont
  {K\"artner}}]{Keathley2013}%
  \BibitemOpen
  \bibfield  {author} {\bibinfo {author} {\bibnamefont {Keathley},
  \bibfnamefont {P~D}}, \bibinfo {author} {\bibfnamefont {A.}~\bibnamefont
  {Sell}}, \bibinfo {author} {\bibfnamefont {W.~P.}\ \bibnamefont {Putnam}},
  \bibinfo {author} {\bibfnamefont {S.}~\bibnamefont {Guerrera}}, \bibinfo
  {author} {\bibfnamefont {L.}~\bibnamefont {Velasquez-Garc\'{\i}a}}, \ and\
  \bibinfo {author} {\bibfnamefont {F.~X.}\ \bibnamefont {K\"artner}}}
  (\bibinfo {year} {2013}),\ \bibfield  {title} {\enquote {\bibinfo {title}
  {Strong-field photoemission from silicon field emitter arrays},}\ }\href@noop
  {} {\bibfield  {journal} {\bibinfo  {journal} {Ann. Phys. (Berlin)}\ }\textbf
  {\bibinfo {volume} {525}},\ \bibinfo {pages} {144--150}}\BibitemShut
  {NoStop}%
\bibitem [{\citenamefont {Keldysh}(1965)}]{Keldysh}%
  \BibitemOpen
  \bibfield  {author} {\bibinfo {author} {\bibnamefont {Keldysh}, \bibfnamefont
  {L~V}}} (\bibinfo {year} {1965}),\ \bibfield  {title} {\enquote {\bibinfo
  {title} {Ionization in the field of a strong electromagnetic wave},}\
  }\href@noop {} {\bibfield  {journal} {\bibinfo  {journal} {J. Expt. Theo.
  Phys.}\ }\textbf {\bibinfo {volume} {20}},\ \bibinfo {pages}
  {1307}}\BibitemShut {NoStop}%
\bibitem [{\citenamefont {Kelkensberg}\ \emph {et~al.}(2012)\citenamefont
  {Kelkensberg}, \citenamefont {Koenderink},\ and\ \citenamefont
  {Vrakking}}]{Kelkensberg12}%
  \BibitemOpen
  \bibfield  {author} {\bibinfo {author} {\bibnamefont {Kelkensberg},
  \bibfnamefont {F}}, \bibinfo {author} {\bibfnamefont {A.~F.}\ \bibnamefont
  {Koenderink}}, \ and\ \bibinfo {author} {\bibfnamefont {M.~J.~J.}\
  \bibnamefont {Vrakking}}} (\bibinfo {year} {2012}),\ \bibfield  {title}
  {\enquote {\bibinfo {title} {Attosecond streaking in a nano-plasmonic
  field},}\ }\href@noop {} {\bibfield  {journal} {\bibinfo  {journal} {New J.
  Phys.}\ }\textbf {\bibinfo {volume} {14}},\ \bibinfo {pages}
  {093034}}\BibitemShut {NoStop}%
\bibitem [{\citenamefont {Kienberger}\ \emph {et~al.}(2004)\citenamefont
  {Kienberger}, \citenamefont {Goulielmakis}, \citenamefont {Uiberacker},
  \citenamefont {Baltuska}, \citenamefont {Yakovlev}, \citenamefont {Bammer},
  \citenamefont {Scrinzi}, \citenamefont {Westerwalbesloh}, \citenamefont
  {Kleineberg}, \citenamefont {Heinzmann}, \citenamefont {Drescher},\ and\
  \citenamefont {Krausz}}]{kienberger2004atomic}%
  \BibitemOpen
  \bibfield  {author} {\bibinfo {author} {\bibnamefont {Kienberger},
  \bibfnamefont {R}}, \bibinfo {author} {\bibfnamefont {E.}~\bibnamefont
  {Goulielmakis}}, \bibinfo {author} {\bibfnamefont {M.}~\bibnamefont
  {Uiberacker}}, \bibinfo {author} {\bibfnamefont {A.}~\bibnamefont
  {Baltuska}}, \bibinfo {author} {\bibfnamefont {V.}~\bibnamefont {Yakovlev}},
  \bibinfo {author} {\bibfnamefont {F.}~\bibnamefont {Bammer}}, \bibinfo
  {author} {\bibfnamefont {A.}~\bibnamefont {Scrinzi}}, \bibinfo {author}
  {\bibfnamefont {Th.}\ \bibnamefont {Westerwalbesloh}}, \bibinfo {author}
  {\bibfnamefont {U.}~\bibnamefont {Kleineberg}}, \bibinfo {author}
  {\bibfnamefont {U.}~\bibnamefont {Heinzmann}}, \bibinfo {author}
  {\bibfnamefont {M.}~\bibnamefont {Drescher}}, \ and\ \bibinfo {author}
  {\bibfnamefont {F.}~\bibnamefont {Krausz}}} (\bibinfo {year} {2004}),\
  \bibfield  {title} {\enquote {\bibinfo {title} {Atomic transient recorder},}\
  }\href@noop {} {\bibfield  {journal} {\bibinfo  {journal} {Nature}\ }\textbf
  {\bibinfo {volume} {427}},\ \bibinfo {pages} {817--821}}\BibitemShut
  {NoStop}%
\bibitem [{\citenamefont {Kim}\ \emph {et~al.}(2013)\citenamefont {Kim},
  \citenamefont {Zhang}, \citenamefont {Shiner}, \citenamefont {Schmidt},
  \citenamefont {L\'egar\'e}, \citenamefont {Villeneuve},\ and\ \citenamefont
  {Corkum}}]{kim_petahertz_2013}%
  \BibitemOpen
  \bibfield  {author} {\bibinfo {author} {\bibnamefont {Kim}, \bibfnamefont
  {K~T}}, \bibinfo {author} {\bibfnamefont {C.}~\bibnamefont {Zhang}}, \bibinfo
  {author} {\bibfnamefont {A.~D.}\ \bibnamefont {Shiner}}, \bibinfo {author}
  {\bibfnamefont {B.~E.}\ \bibnamefont {Schmidt}}, \bibinfo {author}
  {\bibfnamefont {F.}~\bibnamefont {L\'egar\'e}}, \bibinfo {author}
  {\bibfnamefont {D.~M.}\ \bibnamefont {Villeneuve}}, \ and\ \bibinfo {author}
  {\bibfnamefont {P.~B.}\ \bibnamefont {Corkum}}} (\bibinfo {year} {2013}),\
  \bibfield  {title} {\enquote {\bibinfo {title} {Petahertz optical
  oscilloscope},}\ }\href@noop {} {\bibfield  {journal} {\bibinfo  {journal}
  {Nat. Phot.}\ }\textbf {\bibinfo {volume} {7}},\ \bibinfo {pages}
  {958--962}}\BibitemShut {NoStop}%
\bibitem [{\citenamefont {Kim}\ \emph {et~al.}(2008)\citenamefont {Kim},
  \citenamefont {Jin}, \citenamefont {Kim}, \citenamefont {Park}, \citenamefont
  {Kim},\ and\ \citenamefont {Kim}}]{Kim08}%
  \BibitemOpen
  \bibfield  {author} {\bibinfo {author} {\bibnamefont {Kim}, \bibfnamefont
  {S}}, \bibinfo {author} {\bibfnamefont {J.}~\bibnamefont {Jin}}, \bibinfo
  {author} {\bibfnamefont {Y-J.}\ \bibnamefont {Kim}}, \bibinfo {author}
  {\bibfnamefont {I-Y.}\ \bibnamefont {Park}}, \bibinfo {author} {\bibfnamefont
  {Y.}~\bibnamefont {Kim}}, \ and\ \bibinfo {author} {\bibfnamefont {S-W.}\
  \bibnamefont {Kim}}} (\bibinfo {year} {2008}),\ \bibfield  {title} {\enquote
  {\bibinfo {title} {High-harmonic generation by resonant plasmon field
  enhancement},}\ }\href@noop {} {\bibfield  {journal} {\bibinfo  {journal}
  {Nature}\ }\textbf {\bibinfo {volume} {453}},\ \bibinfo {pages}
  {757--760}}\BibitemShut {NoStop}%
\bibitem [{\citenamefont {Kim}\ \emph {et~al.}(2012)\citenamefont {Kim},
  \citenamefont {Jin}, \citenamefont {Kim}, \citenamefont {Park}, \citenamefont
  {Kim},\ and\ \citenamefont {Kim}}]{Kim12Reply}%
  \BibitemOpen
  \bibfield  {author} {\bibinfo {author} {\bibnamefont {Kim}, \bibfnamefont
  {S}}, \bibinfo {author} {\bibfnamefont {J.}~\bibnamefont {Jin}}, \bibinfo
  {author} {\bibfnamefont {Y-J.}\ \bibnamefont {Kim}}, \bibinfo {author}
  {\bibfnamefont {I-Y.}\ \bibnamefont {Park}}, \bibinfo {author} {\bibfnamefont
  {Y.}~\bibnamefont {Kim}}, \ and\ \bibinfo {author} {\bibfnamefont {S-W.}\
  \bibnamefont {Kim}}} (\bibinfo {year} {2012}),\ \bibfield  {title} {\enquote
  {\bibinfo {title} {Reply nature)},}\ }\href@noop {} {\bibfield  {journal}
  {\bibinfo  {journal} {Nature}\ }\textbf {\bibinfo {volume} {485}},\ \bibinfo
  {pages} {E1--E3}}\BibitemShut {NoStop}%
\bibitem [{\citenamefont {Koval}\ \emph {et~al.}(2007)\citenamefont {Koval},
  \citenamefont {Wilken}, \citenamefont {Bauer},\ and\ \citenamefont
  {Keitel}}]{Koval2007}%
  \BibitemOpen
  \bibfield  {author} {\bibinfo {author} {\bibnamefont {Koval}, \bibfnamefont
  {P}}, \bibinfo {author} {\bibfnamefont {F.}~\bibnamefont {Wilken}}, \bibinfo
  {author} {\bibfnamefont {D.}~\bibnamefont {Bauer}}, \ and\ \bibinfo {author}
  {\bibfnamefont {C.~H.}\ \bibnamefont {Keitel}}} (\bibinfo {year} {2007}),\
  \bibfield  {title} {\enquote {\bibinfo {title} {Nonsequential double
  recombination in intense laser fields},}\ }\href@noop {} {\bibfield
  {journal} {\bibinfo  {journal} {Phys. Rev. Lett.}\ }\textbf {\bibinfo
  {volume} {98}},\ \bibinfo {pages} {043904}}\BibitemShut {NoStop}%
\bibitem [{\citenamefont {Krasovskii}(2011)}]{krasovskii2011attosecond}%
  \BibitemOpen
  \bibfield  {author} {\bibinfo {author} {\bibnamefont {Krasovskii},
  \bibfnamefont {E~E}}} (\bibinfo {year} {2011}),\ \bibfield  {title} {\enquote
  {\bibinfo {title} {Attosecond spectroscopy of solids: streaking phase shift
  due to lattice scattering},}\ }\href@noop {} {\bibfield  {journal} {\bibinfo
  {journal} {Phys. Rev. B}\ }\textbf {\bibinfo {volume} {84}},\ \bibinfo
  {pages} {195106}}\BibitemShut {NoStop}%
\bibitem [{\citenamefont {Krausz}\ and\ \citenamefont
  {Ivanov}(2009)}]{Krausz09}%
  \BibitemOpen
  \bibfield  {author} {\bibinfo {author} {\bibnamefont {Krausz}, \bibfnamefont
  {F}}, \ and\ \bibinfo {author} {\bibfnamefont {M.}~\bibnamefont {Ivanov}}}
  (\bibinfo {year} {2009}),\ \bibfield  {title} {\enquote {\bibinfo {title}
  {Attosecond physics},}\ }\href@noop {} {\bibfield  {journal} {\bibinfo
  {journal} {Rev. Mod. Phys.}\ }\textbf {\bibinfo {volume} {81}},\ \bibinfo
  {pages} {163--234}}\BibitemShut {NoStop}%
\bibitem [{\citenamefont {Krausz}\ and\ \citenamefont
  {Stockman}(2014)}]{krausz2014attosecond}%
  \BibitemOpen
  \bibfield  {author} {\bibinfo {author} {\bibnamefont {Krausz}, \bibfnamefont
  {F}}, \ and\ \bibinfo {author} {\bibfnamefont {M.~I.}\ \bibnamefont
  {Stockman}}} (\bibinfo {year} {2014}),\ \bibfield  {title} {\enquote
  {\bibinfo {title} {Attosecond metrology: from electron capture to future
  signal processing},}\ }\href@noop {} {\bibfield  {journal} {\bibinfo
  {journal} {Nat. Phot.}\ }\textbf {\bibinfo {volume} {8}},\ \bibinfo {pages}
  {205--213}}\BibitemShut {NoStop}%
\bibitem [{\citenamefont {Kr{\"u}ger}\ \emph {et~al.}(2014)\citenamefont
  {Kr{\"u}ger}, \citenamefont {F{\"o}rster},\ and\ \citenamefont
  {Hommelhoff}}]{Kruger14}%
  \BibitemOpen
  \bibfield  {author} {\bibinfo {author} {\bibnamefont {Kr{\"u}ger},
  \bibfnamefont {M}}, \bibinfo {author} {\bibfnamefont {M.}~\bibnamefont
  {F{\"o}rster}}, \ and\ \bibinfo {author} {\bibfnamefont {P.}~\bibnamefont
  {Hommelhoff}}} (\bibinfo {year} {2014}),\ \bibfield  {title} {\enquote
  {\bibinfo {title} {Self-probing of metal nanotips by rescattered electrons
  reveals the nano-optical near-field},}\ }\href@noop {} {\bibfield  {journal}
  {\bibinfo  {journal} {J. Phys. B}\ }\textbf {\bibinfo {volume} {47}},\
  \bibinfo {pages} {124022}}\BibitemShut {NoStop}%
\bibitem [{\citenamefont {Kr{\"u}ger}\ \emph
  {et~al.}(2012{\natexlab{a}})\citenamefont {Kr{\"u}ger}, \citenamefont
  {Schenk}, \citenamefont {F{\"o}rster},\ and\ \citenamefont
  {Hommelhoff}}]{Kruger12B}%
  \BibitemOpen
  \bibfield  {author} {\bibinfo {author} {\bibnamefont {Kr{\"u}ger},
  \bibfnamefont {M}}, \bibinfo {author} {\bibfnamefont {M.}~\bibnamefont
  {Schenk}}, \bibinfo {author} {\bibfnamefont {M.}~\bibnamefont {F{\"o}rster}},
  \ and\ \bibinfo {author} {\bibfnamefont {P.}~\bibnamefont {Hommelhoff}}}
  (\bibinfo {year} {2012}{\natexlab{a}}),\ \bibfield  {title} {\enquote
  {\bibinfo {title} {Attosecond physics in photoemission from a metal
  nanotip},}\ }\href@noop {} {\bibfield  {journal} {\bibinfo  {journal} {J.
  Phys. B}\ }\textbf {\bibinfo {volume} {45}},\ \bibinfo {pages}
  {074006}}\BibitemShut {NoStop}%
\bibitem [{\citenamefont {Kr{\"u}ger}\ \emph {et~al.}(2011)\citenamefont
  {Kr{\"u}ger}, \citenamefont {Schenk},\ and\ \citenamefont
  {Hommelhoff}}]{Kruger11}%
  \BibitemOpen
  \bibfield  {author} {\bibinfo {author} {\bibnamefont {Kr{\"u}ger},
  \bibfnamefont {M}}, \bibinfo {author} {\bibfnamefont {M.}~\bibnamefont
  {Schenk}}, \ and\ \bibinfo {author} {\bibfnamefont {P.}~\bibnamefont
  {Hommelhoff}}} (\bibinfo {year} {2011}),\ \bibfield  {title} {\enquote
  {\bibinfo {title} {Attosecond control of electrons emitted from a nanoscale
  metal tip},}\ }\href@noop {} {\bibfield  {journal} {\bibinfo  {journal}
  {Nature}\ }\textbf {\bibinfo {volume} {475}},\ \bibinfo {pages}
  {78--81}}\BibitemShut {NoStop}%
\bibitem [{\citenamefont {Kr{\"u}ger}\ \emph
  {et~al.}(2012{\natexlab{b}})\citenamefont {Kr{\"u}ger}, \citenamefont
  {Schenk}, \citenamefont {Hommelhoff}, \citenamefont {Wachter}, \citenamefont
  {Lemell},\ and\ \citenamefont {Burgd{\"o}rfer}}]{Kruger12N}%
  \BibitemOpen
  \bibfield  {author} {\bibinfo {author} {\bibnamefont {Kr{\"u}ger},
  \bibfnamefont {M}}, \bibinfo {author} {\bibfnamefont {M.}~\bibnamefont
  {Schenk}}, \bibinfo {author} {\bibfnamefont {P.}~\bibnamefont {Hommelhoff}},
  \bibinfo {author} {\bibfnamefont {G.}~\bibnamefont {Wachter}}, \bibinfo
  {author} {\bibfnamefont {C.}~\bibnamefont {Lemell}}, \ and\ \bibinfo {author}
  {\bibfnamefont {J.}~\bibnamefont {Burgd{\"o}rfer}}} (\bibinfo {year}
  {2012}{\natexlab{b}}),\ \bibfield  {title} {\enquote {\bibinfo {title}
  {Interaction of ultrashort laser pulses with metal nanotips: a model system
  for strong-field phenomena},}\ }\href@noop {} {\bibfield  {journal} {\bibinfo
   {journal} {New J. Phys.}\ }\textbf {\bibinfo {volume} {14}},\ \bibinfo
  {pages} {085019}}\BibitemShut {NoStop}%
\bibitem [{\citenamefont {K\"ubel}\ \emph {et~al.}(2016)\citenamefont
  {K\"ubel}, \citenamefont {Siemering}, \citenamefont {Burger}, \citenamefont
  {Kling}, \citenamefont {Li}, \citenamefont {Alnaser}, \citenamefont
  {Bergues}, \citenamefont {Zherebtsov}, \citenamefont {Azzeer}, \citenamefont
  {Ben-Itzhak}, \citenamefont {Moshammer}, \citenamefont {de~Vivie-Riedle},\
  and\ \citenamefont {Kling}}]{kuebel2015}%
  \BibitemOpen
  \bibfield  {author} {\bibinfo {author} {\bibnamefont {K\"ubel}, \bibfnamefont
  {M}}, \bibinfo {author} {\bibfnamefont {R.}~\bibnamefont {Siemering}},
  \bibinfo {author} {\bibfnamefont {C.}~\bibnamefont {Burger}}, \bibinfo
  {author} {\bibfnamefont {N.~G.}\ \bibnamefont {Kling}}, \bibinfo {author}
  {\bibfnamefont {H.}~\bibnamefont {Li}}, \bibinfo {author} {\bibfnamefont
  {A.S.}\ \bibnamefont {Alnaser}}, \bibinfo {author} {\bibfnamefont
  {B.}~\bibnamefont {Bergues}}, \bibinfo {author} {\bibfnamefont
  {S.}~\bibnamefont {Zherebtsov}}, \bibinfo {author} {\bibfnamefont {A.~M.}\
  \bibnamefont {Azzeer}}, \bibinfo {author} {\bibfnamefont {I.}~\bibnamefont
  {Ben-Itzhak}}, \bibinfo {author} {\bibfnamefont {R.}~\bibnamefont
  {Moshammer}}, \bibinfo {author} {\bibfnamefont {R.}~\bibnamefont
  {de~Vivie-Riedle}}, \ and\ \bibinfo {author} {\bibfnamefont {M.~F.}\
  \bibnamefont {Kling}}} (\bibinfo {year} {2016}),\ \bibfield  {title}
  {\enquote {\bibinfo {title} {Steering proton migration in hydrocarbons using
  intense few-cycle laser fields},}\ }\href@noop {} {\bibfield  {journal}
  {\bibinfo  {journal} {Phys. Rev. Lett.}\ }\textbf {\bibinfo {volume} {116}},\
  \bibinfo {pages} {193001}}\BibitemShut {NoStop}%
\bibitem [{\citenamefont {Kuchiev}(1987)}]{kuchiev1987}%
  \BibitemOpen
  \bibfield  {author} {\bibinfo {author} {\bibnamefont {Kuchiev}, \bibfnamefont
  {M~Yu}}} (\bibinfo {year} {1987}),\ \bibfield  {title} {\enquote {\bibinfo
  {title} {Atomic antenna},}\ }\href@noop {} {\bibfield  {journal} {\bibinfo
  {journal} {Sov.~Phys.~JETP}\ }\textbf {\bibinfo {volume} {45}},\ \bibinfo
  {pages} {404--406}}\BibitemShut {NoStop}%
\bibitem [{\citenamefont {Kulander}\ \emph {et~al.}(1993)\citenamefont
  {Kulander}, \citenamefont {Schafer},\ and\ \citenamefont
  {Krause}}]{kulander}%
  \BibitemOpen
  \bibfield  {author} {\bibinfo {author} {\bibnamefont {Kulander},
  \bibfnamefont {K~C}}, \bibinfo {author} {\bibfnamefont {K.~J.}\ \bibnamefont
  {Schafer}}, \ and\ \bibinfo {author} {\bibfnamefont {J.~L.}\ \bibnamefont
  {Krause}}} (\bibinfo {year} {1993}),\ \bibfield  {title} {\enquote {\bibinfo
  {title} {Dynamics of short-pulse excitation, ionization and harmonic
  conversion},}\ }in\ \href@noop {} {\emph {\bibinfo {booktitle} {Super-Intense
  Laser-Atom Physics}}},\ \bibinfo {editor} {edited by\ \bibinfo {editor}
  {\bibfnamefont {B.}~\bibnamefont {Piraux}}, \bibinfo {editor} {\bibfnamefont
  {A.}~\bibnamefont {L'~Huillier}}, \ and\ \bibinfo {editor} {\bibfnamefont
  {K.}~\bibnamefont {Rzazewski}}}\ (\bibinfo  {publisher} {Plenum},\ \bibinfo
  {address} {New York})\ pp.\ \bibinfo {pages} {95--110}\BibitemShut {NoStop}%
\bibitem [{\citenamefont {Kusa}\ \emph {et~al.}(2015)\citenamefont {Kusa},
  \citenamefont {Echternkamp}, \citenamefont {Herink}, \citenamefont {Ropers},\
  and\ \citenamefont {Ashihara}}]{Kusa2015}%
  \BibitemOpen
  \bibfield  {author} {\bibinfo {author} {\bibnamefont {Kusa}, \bibfnamefont
  {F}}, \bibinfo {author} {\bibfnamefont {K.~E.}\ \bibnamefont {Echternkamp}},
  \bibinfo {author} {\bibfnamefont {G.}~\bibnamefont {Herink}}, \bibinfo
  {author} {\bibfnamefont {C.}~\bibnamefont {Ropers}}, \ and\ \bibinfo {author}
  {\bibfnamefont {S.}~\bibnamefont {Ashihara}}} (\bibinfo {year} {2015}),\
  \bibfield  {title} {\enquote {\bibinfo {title} {Optical field emission from
  resonant gold nanorods driven by femtosecond mid-infrared pulses},}\
  }\href@noop {} {\bibfield  {journal} {\bibinfo  {journal} {AIP Advances}\
  }\textbf {\bibinfo {volume} {5}},\ \bibinfo {pages} {077138}}\BibitemShut
  {NoStop}%
\bibitem [{\citenamefont {Lafrate}\ \emph {et~al.}(1980)\citenamefont
  {Lafrate}, \citenamefont {Ziegler},\ and\ \citenamefont
  {Nass}}]{Iafrate1980}%
  \BibitemOpen
  \bibfield  {author} {\bibinfo {author} {\bibnamefont {Lafrate}, \bibfnamefont
  {G~J}}, \bibinfo {author} {\bibfnamefont {J.~F.}\ \bibnamefont {Ziegler}}, \
  and\ \bibinfo {author} {\bibfnamefont {M.~J.}\ \bibnamefont {Nass}}}
  (\bibinfo {year} {1980}),\ \bibfield  {title} {\enquote {\bibinfo {title}
  {Application of lindhard's dielectric theory to the stopping of ions in
  solids},}\ }\href@noop {} {\bibfield  {journal} {\bibinfo  {journal} {J.
  Appl. Phys.}\ }\textbf {\bibinfo {volume} {51}},\ \bibinfo {pages}
  {984--987}}\BibitemShut {NoStop}%
\bibitem [{\citenamefont {Landsman}\ and\ \citenamefont
  {Keller}(2015)}]{AttoTunnelTime}%
  \BibitemOpen
  \bibfield  {author} {\bibinfo {author} {\bibnamefont {Landsman},
  \bibfnamefont {A~S}}, \ and\ \bibinfo {author} {\bibfnamefont
  {U.}~\bibnamefont {Keller}}} (\bibinfo {year} {2015}),\ \bibfield  {title}
  {\enquote {\bibinfo {title} {Attosecond science and the tunnelling time
  problem},}\ }\href@noop {} {\bibfield  {journal} {\bibinfo  {journal} {Phys.
  Rep.}\ }\textbf {\bibinfo {volume} {547}},\ \bibinfo {pages}
  {1--24}}\BibitemShut {NoStop}%
\bibitem [{\citenamefont {Lappas}\ \emph {et~al.}(1996)\citenamefont {Lappas},
  \citenamefont {Sanpera}, \citenamefont {Watson}, \citenamefont {Burnett},
  \citenamefont {Knight}, \citenamefont {Grobe},\ and\ \citenamefont
  {Eberly}}]{ASanpera1}%
  \BibitemOpen
  \bibfield  {author} {\bibinfo {author} {\bibnamefont {Lappas}, \bibfnamefont
  {D~G}}, \bibinfo {author} {\bibfnamefont {A.}~\bibnamefont {Sanpera}},
  \bibinfo {author} {\bibfnamefont {J.~B.}\ \bibnamefont {Watson}}, \bibinfo
  {author} {\bibfnamefont {K}~\bibnamefont {Burnett}}, \bibinfo {author}
  {\bibfnamefont {P.~L.}\ \bibnamefont {Knight}}, \bibinfo {author}
  {\bibfnamefont {R.}~\bibnamefont {Grobe}}, \ and\ \bibinfo {author}
  {\bibfnamefont {J.~H}\ \bibnamefont {Eberly}}} (\bibinfo {year} {1996}),\
  \bibfield  {title} {\enquote {\bibinfo {title} {Two-electron effects in
  harmonic generation and ionization from a model he atom},}\ }\href@noop {}
  {\bibfield  {journal} {\bibinfo  {journal} {J. Phys. B}\ }\textbf {\bibinfo
  {volume} {29}},\ \bibinfo {pages} {L619--L627}}\BibitemShut {NoStop}%
\bibitem [{\citenamefont {Lemell}\ \emph {et~al.}(2009)\citenamefont {Lemell},
  \citenamefont {Solleder}, \citenamefont {T\"ok\'esi},\ and\ \citenamefont
  {Burgd\"orfer}}]{lemell_simulation_2009}%
  \BibitemOpen
  \bibfield  {author} {\bibinfo {author} {\bibnamefont {Lemell}, \bibfnamefont
  {C}}, \bibinfo {author} {\bibfnamefont {B.}~\bibnamefont {Solleder}},
  \bibinfo {author} {\bibfnamefont {K.}~\bibnamefont {T\"ok\'esi}}, \ and\
  \bibinfo {author} {\bibfnamefont {J.}~\bibnamefont {Burgd\"orfer}}} (\bibinfo
  {year} {2009}),\ \bibfield  {title} {\enquote {\bibinfo {title} {Simulation
  of attosecond streaking of electrons emitted from a tungsten surface},}\
  }\href@noop {} {\bibfield  {journal} {\bibinfo  {journal} {Phys. Rev. A}\
  }\textbf {\bibinfo {volume} {79}},\ \bibinfo {pages} {062901}}\BibitemShut
  {NoStop}%
\bibitem [{\citenamefont {Lewenstein}\ \emph {et~al.}(1994)\citenamefont
  {Lewenstein}, \citenamefont {Balcou}, \citenamefont {Ivanov}, \citenamefont
  {L'Huillier},\ and\ \citenamefont {Corkum}}]{Lewenstein94}%
  \BibitemOpen
  \bibfield  {author} {\bibinfo {author} {\bibnamefont {Lewenstein},
  \bibfnamefont {M}}, \bibinfo {author} {\bibfnamefont {P.}~\bibnamefont
  {Balcou}}, \bibinfo {author} {\bibfnamefont {M.~Y.}\ \bibnamefont {Ivanov}},
  \bibinfo {author} {\bibfnamefont {A.}~\bibnamefont {L'Huillier}}, \ and\
  \bibinfo {author} {\bibfnamefont {P.~B.}\ \bibnamefont {Corkum}}} (\bibinfo
  {year} {1994}),\ \bibfield  {title} {\enquote {\bibinfo {title} {Theory of
  high-harmonic generation by low-frequency laser fields},}\ }\href@noop {}
  {\bibfield  {journal} {\bibinfo  {journal} {Phys. Rev. A}\ }\textbf {\bibinfo
  {volume} {49}},\ \bibinfo {pages} {2117}}\BibitemShut {NoStop}%
\bibitem [{\citenamefont {Lewenstein}\ and\ \citenamefont
  {L'Huillier}(2009)}]{MaciejChapter}%
  \BibitemOpen
  \bibfield  {author} {\bibinfo {author} {\bibnamefont {Lewenstein},
  \bibfnamefont {M}}, \ and\ \bibinfo {author} {\bibfnamefont {A.}~\bibnamefont
  {L'Huillier}}} (\bibinfo {year} {2009}),\ \bibfield  {title} {\enquote
  {\bibinfo {title} {Principles of single atom physics: High-order harmonic
  generation, above-threshold ionization and non-sequential ionization},}\ }in\
  \href@noop {} {\emph {\bibinfo {booktitle} {Strong Field Laser Physics}}},\
  \bibinfo {editor} {edited by\ \bibinfo {editor} {\bibfnamefont
  {T.}~\bibnamefont {Brabec}}}\ (\bibinfo  {publisher} {Springer},\ \bibinfo
  {address} {New York})\ pp.\ \bibinfo {pages} {147--183}\BibitemShut {NoStop}%
\bibitem [{\citenamefont {L'Huiller}\ \emph {et~al.}(1993)\citenamefont
  {L'Huiller}, \citenamefont {Lewenstein}, \citenamefont {Sali\`eres},
  \citenamefont {Balcou}, \citenamefont {Ivanov}, \citenamefont {Larsson},\
  and\ \citenamefont {Wahlstr\"om}}]{AnneHHG}%
  \BibitemOpen
  \bibfield  {author} {\bibinfo {author} {\bibnamefont {L'Huiller},
  \bibfnamefont {A}}, \bibinfo {author} {\bibfnamefont {M.}~\bibnamefont
  {Lewenstein}}, \bibinfo {author} {\bibfnamefont {P.}~\bibnamefont
  {Sali\`eres}}, \bibinfo {author} {\bibfnamefont {Ph.}\ \bibnamefont
  {Balcou}}, \bibinfo {author} {\bibfnamefont {M.~Yu.}\ \bibnamefont {Ivanov}},
  \bibinfo {author} {\bibfnamefont {J.}~\bibnamefont {Larsson}}, \ and\
  \bibinfo {author} {\bibfnamefont {C.~G.}\ \bibnamefont {Wahlstr\"om}}}
  (\bibinfo {year} {1993}),\ \bibfield  {title} {\enquote {\bibinfo {title}
  {High-order harmonic-generation cutoff},}\ }\href@noop {} {\bibfield
  {journal} {\bibinfo  {journal} {Phys. Rev. A}\ }\textbf {\bibinfo {volume}
  {48}},\ \bibinfo {pages} {R3433}}\BibitemShut {NoStop}%
\bibitem [{\citenamefont {L'Huiller}\ \emph {et~al.}(1983)\citenamefont
  {L'Huiller}, \citenamefont {Lompre}, \citenamefont {Mainfray},\ and\
  \citenamefont {Manus}}]{Anne-knee}%
  \BibitemOpen
  \bibfield  {author} {\bibinfo {author} {\bibnamefont {L'Huiller},
  \bibfnamefont {A}}, \bibinfo {author} {\bibfnamefont {L.~A.}\ \bibnamefont
  {Lompre}}, \bibinfo {author} {\bibfnamefont {G.}~\bibnamefont {Mainfray}}, \
  and\ \bibinfo {author} {\bibfnamefont {C.}~\bibnamefont {Manus}}} (\bibinfo
  {year} {1983}),\ \bibfield  {title} {\enquote {\bibinfo {title} {Multiply
  charged ions induced by multiphoton absorption in rate gases at 0.53
  $\mu$m},}\ }\href@noop {} {\bibfield  {journal} {\bibinfo  {journal} {Phys.
  Rev. A}\ }\textbf {\bibinfo {volume} {27}},\ \bibinfo {pages}
  {2503}}\BibitemShut {NoStop}%
\bibitem [{\citenamefont {L'Huillier}\ \emph {et~al.}(1991)\citenamefont
  {L'Huillier}, \citenamefont {Schafer},\ and\ \citenamefont
  {Kulander}}]{l1991theoretical}%
  \BibitemOpen
  \bibfield  {author} {\bibinfo {author} {\bibnamefont {L'Huillier},
  \bibfnamefont {A}}, \bibinfo {author} {\bibfnamefont {K.~J.}\ \bibnamefont
  {Schafer}}, \ and\ \bibinfo {author} {\bibfnamefont {K.~C.}\ \bibnamefont
  {Kulander}}} (\bibinfo {year} {1991}),\ \bibfield  {title} {\enquote
  {\bibinfo {title} {Theoretical aspects of intense field harmonic
  generation},}\ }\href@noop {} {\bibfield  {journal} {\bibinfo  {journal} {J.
  of Phys. B}\ }\textbf {\bibinfo {volume} {24}},\ \bibinfo {pages}
  {3315}}\BibitemShut {NoStop}%
\bibitem [{\citenamefont {Li}\ \emph {et~al.}(2015)\citenamefont {Li},
  \citenamefont {Mignolet}, \citenamefont {Wachter}, \citenamefont
  {Skruszewicz}, \citenamefont {Zherebtsov}, \citenamefont {S\"u{\ss}mann},
  \citenamefont {Kessel}, \citenamefont {Trushin}, \citenamefont {Kling},
  \citenamefont {K\"ubel}, \citenamefont {Ahn}, \citenamefont {Kim},
  \citenamefont {Ben-Itzhak}, \citenamefont {Cocke}, \citenamefont {Fennel},
  \citenamefont {Tiggesb\"aumker}, \citenamefont {Meiwes-Broer}, \citenamefont
  {Lemell}, \citenamefont {Burgd\"orfer}, \citenamefont {Levine}, \citenamefont
  {Remacle},\ and\ \citenamefont {Kling}}]{li_etal2015}%
  \BibitemOpen
  \bibfield  {author} {\bibinfo {author} {\bibnamefont {Li}, \bibfnamefont
  {H}}, \bibinfo {author} {\bibfnamefont {B.}~\bibnamefont {Mignolet}},
  \bibinfo {author} {\bibfnamefont {G.}~\bibnamefont {Wachter}}, \bibinfo
  {author} {\bibfnamefont {S.}~\bibnamefont {Skruszewicz}}, \bibinfo {author}
  {\bibfnamefont {S.}~\bibnamefont {Zherebtsov}}, \bibinfo {author}
  {\bibfnamefont {F.}~\bibnamefont {S\"u{\ss}mann}}, \bibinfo {author}
  {\bibfnamefont {A.}~\bibnamefont {Kessel}}, \bibinfo {author} {\bibfnamefont
  {S.~A.}\ \bibnamefont {Trushin}}, \bibinfo {author} {\bibfnamefont {N.~G.}\
  \bibnamefont {Kling}}, \bibinfo {author} {\bibfnamefont {M.}~\bibnamefont
  {K\"ubel}}, \bibinfo {author} {\bibfnamefont {B.}~\bibnamefont {Ahn}},
  \bibinfo {author} {\bibfnamefont {D.}~\bibnamefont {Kim}}, \bibinfo {author}
  {\bibfnamefont {I.}~\bibnamefont {Ben-Itzhak}}, \bibinfo {author}
  {\bibfnamefont {C.~L.}\ \bibnamefont {Cocke}}, \bibinfo {author}
  {\bibfnamefont {T.}~\bibnamefont {Fennel}}, \bibinfo {author} {\bibfnamefont
  {J.}~\bibnamefont {Tiggesb\"aumker}}, \bibinfo {author} {\bibfnamefont
  {K.-H.}\ \bibnamefont {Meiwes-Broer}}, \bibinfo {author} {\bibfnamefont
  {C.}~\bibnamefont {Lemell}}, \bibinfo {author} {\bibfnamefont
  {J.}~\bibnamefont {Burgd\"orfer}}, \bibinfo {author} {\bibfnamefont {R.D.}\
  \bibnamefont {Levine}}, \bibinfo {author} {\bibfnamefont {F.}~\bibnamefont
  {Remacle}}, \ and\ \bibinfo {author} {\bibfnamefont {M.~F.}\ \bibnamefont
  {Kling}}} (\bibinfo {year} {2015}),\ \bibfield  {title} {\enquote {\bibinfo
  {title} {Coherent electronic wave packet motion in {C}$_{60}$ controlled by
  the waveform and polarization of few-cycle laser fields},}\ }\href@noop {}
  {\bibfield  {journal} {\bibinfo  {journal} {Phys. Rev. Lett.}\ }\textbf
  {\bibinfo {volume} {114}},\ \bibinfo {pages} {123004}}\BibitemShut {NoStop}%
\bibitem [{\citenamefont {Liao}\ and\ \citenamefont
  {Thumm}(2014)}]{liao_attosecond_2014}%
  \BibitemOpen
  \bibfield  {author} {\bibinfo {author} {\bibnamefont {Liao}, \bibfnamefont
  {Q}}, \ and\ \bibinfo {author} {\bibfnamefont {U.}~\bibnamefont {Thumm}}}
  (\bibinfo {year} {2014}),\ \bibfield  {title} {\enquote {\bibinfo {title}
  {Attosecond {Time}-{Resolved} {Photoelectron} {Dispersion} and
  {Photoemission} {Time} {Delays}},}\ }\href@noop {} {\bibfield  {journal}
  {\bibinfo  {journal} {Phys. Rev. Lett.}\ }\textbf {\bibinfo {volume} {112}},\
  \bibinfo {pages} {023602}}\BibitemShut {NoStop}%
\bibitem [{\citenamefont {Limpert}\ \emph {et~al.}(2011)\citenamefont
  {Limpert}, \citenamefont {H\"adrich}, \citenamefont {Rothhardt},
  \citenamefont {Krebs}, \citenamefont {Eidam}, \citenamefont {Schreiber},\
  and\ \citenamefont {T\"unnermann}}]{limpert_ultrafast_2011}%
  \BibitemOpen
  \bibfield  {author} {\bibinfo {author} {\bibnamefont {Limpert}, \bibfnamefont
  {J}}, \bibinfo {author} {\bibfnamefont {S.}~\bibnamefont {H\"adrich}},
  \bibinfo {author} {\bibfnamefont {J.}~\bibnamefont {Rothhardt}}, \bibinfo
  {author} {\bibfnamefont {M.}~\bibnamefont {Krebs}}, \bibinfo {author}
  {\bibfnamefont {T.}~\bibnamefont {Eidam}}, \bibinfo {author} {\bibfnamefont
  {T.}~\bibnamefont {Schreiber}}, \ and\ \bibinfo {author} {\bibfnamefont
  {A.}~\bibnamefont {T\"unnermann}}} (\bibinfo {year} {2011}),\ \bibfield
  {title} {\enquote {\bibinfo {title} {Ultrafast fiber lasers for strong-field
  physics experiments},}\ }\href@noop {} {\bibfield  {journal} {\bibinfo
  {journal} {Laser \& Photonics Reviews}\ }\textbf {\bibinfo {volume} {5}},\
  \bibinfo {pages} {634--646}}\BibitemShut {NoStop}%
\bibitem [{\citenamefont {Lindhard}(1954)}]{Lindhard1954}%
  \BibitemOpen
  \bibfield  {author} {\bibinfo {author} {\bibnamefont {Lindhard},
  \bibfnamefont {J}}} (\bibinfo {year} {1954}),\ \href@noop {} {\bibfield
  {journal} {\bibinfo  {journal} {K. Dan. Vidensk. Selsk. Mat. Fys. Medd.}\
  }\textbf {\bibinfo {volume} {28}},\ \bibinfo {pages} {1}}\BibitemShut
  {NoStop}%
\bibitem [{\citenamefont {Lindner}\ \emph {et~al.}(2005)\citenamefont
  {Lindner}, \citenamefont {Sch\"atzel}, \citenamefont {Walther}, \citenamefont
  {Baltu\v{s}ka}, \citenamefont {Goulielmakis}, \citenamefont {Krausz},
  \citenamefont {Milo\v{s}evi\'{c}}, \citenamefont {Bauer}, \citenamefont
  {Becker},\ and\ \citenamefont {Paulus}}]{Lindner2005}%
  \BibitemOpen
  \bibfield  {author} {\bibinfo {author} {\bibnamefont {Lindner}, \bibfnamefont
  {F}}, \bibinfo {author} {\bibfnamefont {M.~G.}\ \bibnamefont {Sch\"atzel}},
  \bibinfo {author} {\bibfnamefont {H.}~\bibnamefont {Walther}}, \bibinfo
  {author} {\bibfnamefont {A.}~\bibnamefont {Baltu\v{s}ka}}, \bibinfo {author}
  {\bibfnamefont {E.}~\bibnamefont {Goulielmakis}}, \bibinfo {author}
  {\bibfnamefont {F.}~\bibnamefont {Krausz}}, \bibinfo {author} {\bibfnamefont
  {D.~B.}\ \bibnamefont {Milo\v{s}evi\'{c}}}, \bibinfo {author} {\bibfnamefont
  {D.}~\bibnamefont {Bauer}}, \bibinfo {author} {\bibfnamefont
  {W.}~\bibnamefont {Becker}}, \ and\ \bibinfo {author} {\bibfnamefont {G.~G.}\
  \bibnamefont {Paulus}}} (\bibinfo {year} {2005}),\ \bibfield  {title}
  {\enquote {\bibinfo {title} {Attosecond double-slit experiment},}\
  }\href@noop {} {\bibfield  {journal} {\bibinfo  {journal} {Phys. Rev. Lett.}\
  }\textbf {\bibinfo {volume} {95}},\ \bibinfo {pages} {040401}}\BibitemShut
  {NoStop}%
\bibitem [{\citenamefont {Liu}\ \emph {et~al.}(2011)\citenamefont {Liu},
  \citenamefont {Tang}, \citenamefont {Hentschel},\ and\ \citenamefont
  {Giessen}}]{Liu11}%
  \BibitemOpen
  \bibfield  {author} {\bibinfo {author} {\bibnamefont {Liu}, \bibfnamefont
  {N}}, \bibinfo {author} {\bibfnamefont {M.~L.}\ \bibnamefont {Tang}},
  \bibinfo {author} {\bibfnamefont {M.}~\bibnamefont {Hentschel}}, \ and\
  \bibinfo {author} {\bibfnamefont {A.~P.}\ \bibnamefont {Giessen},
  \bibfnamefont {H.~Alivisatos}}} (\bibinfo {year} {2011}),\ \bibfield  {title}
  {\enquote {\bibinfo {title} {Nanoantenna-enhanced gas sensing in a single
  tailored nanofocus},}\ }\href@noop {} {\bibfield  {journal} {\bibinfo
  {journal} {Nat. Mater.}\ }\textbf {\bibinfo {volume} {10}},\ \bibinfo {pages}
  {631--636}}\BibitemShut {NoStop}%
\bibitem [{\citenamefont {Locher}\ \emph {et~al.}(2015)\citenamefont {Locher},
  \citenamefont {Castiglioni}, \citenamefont {Lucchini}, \citenamefont {Greif},
  \citenamefont {Gallmann}, \citenamefont {Osterwalder}, \citenamefont
  {Hengsberger},\ and\ \citenamefont {Keller}}]{locher2015}%
  \BibitemOpen
  \bibfield  {author} {\bibinfo {author} {\bibnamefont {Locher}, \bibfnamefont
  {R}}, \bibinfo {author} {\bibfnamefont {L.}~\bibnamefont {Castiglioni}},
  \bibinfo {author} {\bibfnamefont {M.}~\bibnamefont {Lucchini}}, \bibinfo
  {author} {\bibfnamefont {M.}~\bibnamefont {Greif}}, \bibinfo {author}
  {\bibfnamefont {L.}~\bibnamefont {Gallmann}}, \bibinfo {author}
  {\bibfnamefont {J.}~\bibnamefont {Osterwalder}}, \bibinfo {author}
  {\bibfnamefont {M.}~\bibnamefont {Hengsberger}}, \ and\ \bibinfo {author}
  {\bibfnamefont {U.}~\bibnamefont {Keller}}} (\bibinfo {year} {2015}),\
  \bibfield  {title} {\enquote {\bibinfo {title} {Energy-dependent
  photoemission delays from noble metal surfaces by attosecond
  interferometry},}\ }\href@noop {} {\bibfield  {journal} {\bibinfo  {journal}
  {Optica}\ }\textbf {\bibinfo {volume} {2}},\ \bibinfo {pages}
  {405}}\BibitemShut {NoStop}%
\bibitem [{\citenamefont {Locher}\ \emph {et~al.}(2014)\citenamefont {Locher},
  \citenamefont {Lucchini}, \citenamefont {Herrmann}, \citenamefont {Sabbar},
  \citenamefont {Weger}, \citenamefont {Ludwig}, \citenamefont {Castiglioni},
  \citenamefont {Greif}, \citenamefont {Hengsberger}, \citenamefont {Gallmann}
  \emph {et~al.}}]{locher2014versatile}%
  \BibitemOpen
  \bibfield  {author} {\bibinfo {author} {\bibnamefont {Locher}, \bibfnamefont
  {R}}, \bibinfo {author} {\bibfnamefont {M.}~\bibnamefont {Lucchini}},
  \bibinfo {author} {\bibfnamefont {J.}~\bibnamefont {Herrmann}}, \bibinfo
  {author} {\bibfnamefont {M.}~\bibnamefont {Sabbar}}, \bibinfo {author}
  {\bibfnamefont {M.}~\bibnamefont {Weger}}, \bibinfo {author} {\bibfnamefont
  {A.}~\bibnamefont {Ludwig}}, \bibinfo {author} {\bibfnamefont
  {L.}~\bibnamefont {Castiglioni}}, \bibinfo {author} {\bibfnamefont
  {M.}~\bibnamefont {Greif}}, \bibinfo {author} {\bibfnamefont
  {M.}~\bibnamefont {Hengsberger}}, \bibinfo {author} {\bibfnamefont
  {L.}~\bibnamefont {Gallmann}},  \emph {et~al.}} (\bibinfo {year} {2014}),\
  \bibfield  {title} {\enquote {\bibinfo {title} {Versatile attosecond beamline
  in a two-foci configuration for simultaneous time-resolved measurements},}\
  }\href@noop {} {\bibfield  {journal} {\bibinfo  {journal} {Rev. of Sci.
  Instr.}\ }\textbf {\bibinfo {volume} {85}},\ \bibinfo {pages}
  {013113}}\BibitemShut {NoStop}%
\bibitem [{\citenamefont {Lorek}\ \emph {et~al.}(2015)\citenamefont {Lorek},
  \citenamefont {M{\aa}rsell}, \citenamefont {Losquin}, \citenamefont
  {Miranda}, \citenamefont {Harth}, \citenamefont {Guo}, \citenamefont
  {Sv\"{a}rd}, \citenamefont {Arnold}, \citenamefont {L'Huiller}, \citenamefont
  {Mikkelsen},\ and\ \citenamefont {Mauritsson}}]{anneoptexp2015}%
  \BibitemOpen
  \bibfield  {author} {\bibinfo {author} {\bibnamefont {Lorek}, \bibfnamefont
  {E}}, \bibinfo {author} {\bibfnamefont {E.}~\bibnamefont {M{\aa}rsell}},
  \bibinfo {author} {\bibfnamefont {A.}~\bibnamefont {Losquin}}, \bibinfo
  {author} {\bibfnamefont {M.}~\bibnamefont {Miranda}}, \bibinfo {author}
  {\bibfnamefont {A.}~\bibnamefont {Harth}}, \bibinfo {author} {\bibfnamefont
  {C.}~\bibnamefont {Guo}}, \bibinfo {author} {\bibfnamefont {R.}~\bibnamefont
  {Sv\"{a}rd}}, \bibinfo {author} {\bibfnamefont {C.~L.}\ \bibnamefont
  {Arnold}}, \bibinfo {author} {\bibfnamefont {A.}~\bibnamefont {L'Huiller}},
  \bibinfo {author} {\bibfnamefont {A.}~\bibnamefont {Mikkelsen}}, \ and\
  \bibinfo {author} {\bibfnamefont {J.}~\bibnamefont {Mauritsson}}} (\bibinfo
  {year} {2015}),\ \bibfield  {title} {\enquote {\bibinfo {title} {Size and
  shape dependent few-cycle near-field dynamics of bowtie nanoantennas},}\
  }\href@noop {} {\bibfield  {journal} {\bibinfo  {journal} {Opt. Exp.}\
  }\textbf {\bibinfo {volume} {23}},\ \bibinfo {pages}
  {31460--31471}}\BibitemShut {NoStop}%
\bibitem [{\citenamefont {Lotz}(1967)}]{Lotz1967}%
  \BibitemOpen
  \bibfield  {author} {\bibinfo {author} {\bibnamefont {Lotz}, \bibfnamefont
  {W}}} (\bibinfo {year} {1967}),\ \bibfield  {title} {\enquote {\bibinfo
  {title} {An empirical formula for the electron-impact ionization
  cross-section},}\ }\href@noop {} {\bibfield  {journal} {\bibinfo  {journal}
  {Zeitschrift f{\"u}r Physik}\ }\textbf {\bibinfo {volume} {206}},\ \bibinfo
  {pages} {205--211}}\BibitemShut {NoStop}%
\bibitem [{\citenamefont {Luecking}\ \emph {et~al.}(2012)\citenamefont
  {Luecking}, \citenamefont {Assion}, \citenamefont {Apolonski}, \citenamefont
  {Krausz},\ and\ \citenamefont {Steinmeyer}}]{luecking_long_term_2012}%
  \BibitemOpen
  \bibfield  {author} {\bibinfo {author} {\bibnamefont {Luecking},
  \bibfnamefont {F}}, \bibinfo {author} {\bibfnamefont {A.}~\bibnamefont
  {Assion}}, \bibinfo {author} {\bibfnamefont {A.}~\bibnamefont {Apolonski}},
  \bibinfo {author} {\bibfnamefont {F.}~\bibnamefont {Krausz}}, \ and\ \bibinfo
  {author} {\bibfnamefont {G.}~\bibnamefont {Steinmeyer}}} (\bibinfo {year}
  {2012}),\ \bibfield  {title} {\enquote {\bibinfo {title} {Long-term
  carrier-envelope-phase-stable few-cycle pulses by use of the feed-forward
  method},}\ }\href@noop {} {\bibfield  {journal} {\bibinfo  {journal} {Opt.
  Lett.}\ }\textbf {\bibinfo {volume} {37}},\ \bibinfo {pages}
  {2076--2078}}\BibitemShut {NoStop}%
\bibitem [{\citenamefont {L\"uneburg}\ \emph {et~al.}(2013)\citenamefont
  {L\"uneburg}, \citenamefont {M\"uller}, \citenamefont {Paarmann},\ and\
  \citenamefont {Ernstorfer}}]{Luneburg2013}%
  \BibitemOpen
  \bibfield  {author} {\bibinfo {author} {\bibnamefont {L\"uneburg},
  \bibfnamefont {S}}, \bibinfo {author} {\bibfnamefont {M.}~\bibnamefont
  {M\"uller}}, \bibinfo {author} {\bibfnamefont {A.}~\bibnamefont {Paarmann}},
  \ and\ \bibinfo {author} {\bibfnamefont {R.}~\bibnamefont {Ernstorfer}}}
  (\bibinfo {year} {2013}),\ \bibfield  {title} {\enquote {\bibinfo {title}
  {Microelectrode for energy and current control of nanotip field electron
  emitters},}\ }\href@noop {} {\bibfield  {journal} {\bibinfo  {journal} {Appl.
  Phys. Lett.}\ }\textbf {\bibinfo {volume} {103}},\ \bibinfo {pages}
  {213506}}\BibitemShut {NoStop}%
\bibitem [{\citenamefont {Luo}\ \emph {et~al.}(2013{\natexlab{a}})\citenamefont
  {Luo}, \citenamefont {Li}, \citenamefont {Wang}, \citenamefont {He},
  \citenamefont {Zhang},\ and\ \citenamefont {Lu}}]{Lu13A}%
  \BibitemOpen
  \bibfield  {author} {\bibinfo {author} {\bibnamefont {Luo}, \bibfnamefont
  {J}}, \bibinfo {author} {\bibfnamefont {Y.}~\bibnamefont {Li}}, \bibinfo
  {author} {\bibfnamefont {Z.}~\bibnamefont {Wang}}, \bibinfo {author}
  {\bibfnamefont {L.}~\bibnamefont {He}}, \bibinfo {author} {\bibfnamefont
  {Q.}~\bibnamefont {Zhang}}, \ and\ \bibinfo {author} {\bibfnamefont
  {P.}~\bibnamefont {Lu}}} (\bibinfo {year} {2013}{\natexlab{a}}),\ \bibfield
  {title} {\enquote {\bibinfo {title} {Efficient supercontinuum generation by
  uv-assisted midinfrared plasmonic fields},}\ }\href@noop {} {\bibfield
  {journal} {\bibinfo  {journal} {Phys. Rev. A}\ }\textbf {\bibinfo {volume}
  {89}},\ \bibinfo {pages} {023405}}\BibitemShut {NoStop}%
\bibitem [{\citenamefont {Luo}\ \emph {et~al.}(2013{\natexlab{b}})\citenamefont
  {Luo}, \citenamefont {Li}, \citenamefont {Wang}, \citenamefont {Zhang},
  \citenamefont {Lan},\ and\ \citenamefont {Lu}}]{Luo13JOSAB}%
  \BibitemOpen
  \bibfield  {author} {\bibinfo {author} {\bibnamefont {Luo}, \bibfnamefont
  {J}}, \bibinfo {author} {\bibfnamefont {Y.}~\bibnamefont {Li}}, \bibinfo
  {author} {\bibfnamefont {Z.}~\bibnamefont {Wang}}, \bibinfo {author}
  {\bibfnamefont {Q.}~\bibnamefont {Zhang}}, \bibinfo {author} {\bibfnamefont
  {P.}~\bibnamefont {Lan}}, \ and\ \bibinfo {author} {\bibfnamefont
  {P.}~\bibnamefont {Lu}}} (\bibinfo {year} {2013}{\natexlab{b}}),\ \bibfield
  {title} {\enquote {\bibinfo {title} {Wavelength dependence of
  high-order-harmonic yield in inhomogeneous fields},}\ }\href@noop {}
  {\bibfield  {journal} {\bibinfo  {journal} {J. Opt. Soc. Am. B}\ }\textbf
  {\bibinfo {volume} {30}},\ \bibinfo {pages} {2469--2475}}\BibitemShut
  {NoStop}%
\bibitem [{\citenamefont {Luo}\ \emph {et~al.}(2013{\natexlab{c}})\citenamefont
  {Luo}, \citenamefont {Li}, \citenamefont {Wang}, \citenamefont {Zhang},\ and\
  \citenamefont {Lu}}]{Luo13}%
  \BibitemOpen
  \bibfield  {author} {\bibinfo {author} {\bibnamefont {Luo}, \bibfnamefont
  {J}}, \bibinfo {author} {\bibfnamefont {Y.}~\bibnamefont {Li}}, \bibinfo
  {author} {\bibfnamefont {Z.}~\bibnamefont {Wang}}, \bibinfo {author}
  {\bibfnamefont {Q.}~\bibnamefont {Zhang}}, \ and\ \bibinfo {author}
  {\bibfnamefont {P.}~\bibnamefont {Lu}}} (\bibinfo {year}
  {2013}{\natexlab{c}}),\ \bibfield  {title} {\enquote {\bibinfo {title}
  {Ultra-short isolated attosecond emission in mid-infrared inhomogeneous
  fields without cep stabilization},}\ }\href@noop {} {\bibfield  {journal}
  {\bibinfo  {journal} {J. Phys. B}\ }\textbf {\bibinfo {volume} {46}},\
  \bibinfo {pages} {145602}}\BibitemShut {NoStop}%
\bibitem [{\citenamefont {Luu}\ \emph {et~al.}(2015)\citenamefont {Luu},
  \citenamefont {Garg}, \citenamefont {Kruchinin}, \citenamefont {Moulet},
  \citenamefont {Hassan},\ and\ \citenamefont {Goulielmakis}}]{Trang2015}%
  \BibitemOpen
  \bibfield  {author} {\bibinfo {author} {\bibnamefont {Luu}, \bibfnamefont
  {T~T}}, \bibinfo {author} {\bibfnamefont {M.}~\bibnamefont {Garg}}, \bibinfo
  {author} {\bibfnamefont {S.~Yu.}\ \bibnamefont {Kruchinin}}, \bibinfo
  {author} {\bibfnamefont {A.}~\bibnamefont {Moulet}}, \bibinfo {author}
  {\bibfnamefont {M.~Th.}\ \bibnamefont {Hassan}}, \ and\ \bibinfo {author}
  {\bibfnamefont {E.}~\bibnamefont {Goulielmakis}}} (\bibinfo {year} {2015}),\
  \bibfield  {title} {\enquote {\bibinfo {title} {Extreme ultraviolet
  high-harmonic spectroscopy of solids},}\ }\href@noop {} {\bibfield  {journal}
  {\bibinfo  {journal} {Nature}\ }\textbf {\bibinfo {volume} {521}},\ \bibinfo
  {pages} {498--502}}\BibitemShut {NoStop}%
\bibitem [{\citenamefont {Maier}(2007)}]{maier_plasmonics:_2007}%
  \BibitemOpen
  \bibfield  {author} {\bibinfo {author} {\bibnamefont {Maier}, \bibfnamefont
  {S~A}}} (\bibinfo {year} {2007}),\ \href@noop {} {\emph {\bibinfo {title}
  {Plasmonics: {Fundamentals} and {Applications}}}}\ (\bibinfo  {publisher}
  {Springer},\ \bibinfo {address} {New York})\BibitemShut {NoStop}%
\bibitem [{\citenamefont {Mairesse}\ \emph {et~al.}(2003)\citenamefont
  {Mairesse}, \citenamefont {de~Bohan}, \citenamefont {Frasinski},
  \citenamefont {Merdji}, \citenamefont {Dinu}, \citenamefont {Monchicourt},
  \citenamefont {Breger}, \citenamefont {Kova{\v{c}}ev}, \citenamefont
  {Ta{\"\i}eb}, \citenamefont {Carr{\'e}}, \citenamefont {Muller},
  \citenamefont {Agostini},\ and\ \citenamefont
  {Sali\'eres}}]{mairesse2003attosecond}%
  \BibitemOpen
  \bibfield  {author} {\bibinfo {author} {\bibnamefont {Mairesse},
  \bibfnamefont {Y}}, \bibinfo {author} {\bibfnamefont {A.}~\bibnamefont
  {de~Bohan}}, \bibinfo {author} {\bibfnamefont {L.~J.}\ \bibnamefont
  {Frasinski}}, \bibinfo {author} {\bibfnamefont {H.}~\bibnamefont {Merdji}},
  \bibinfo {author} {\bibfnamefont {L.~C.}\ \bibnamefont {Dinu}}, \bibinfo
  {author} {\bibfnamefont {P.}~\bibnamefont {Monchicourt}}, \bibinfo {author}
  {\bibfnamefont {P.}~\bibnamefont {Breger}}, \bibinfo {author} {\bibfnamefont
  {M.}~\bibnamefont {Kova{\v{c}}ev}}, \bibinfo {author} {\bibfnamefont
  {R.}~\bibnamefont {Ta{\"\i}eb}}, \bibinfo {author} {\bibfnamefont
  {B.}~\bibnamefont {Carr{\'e}}}, \bibinfo {author} {\bibfnamefont {H.~G.}\
  \bibnamefont {Muller}}, \bibinfo {author} {\bibfnamefont {P}~\bibnamefont
  {Agostini}}, \ and\ \bibinfo {author} {\bibfnamefont {P.}~\bibnamefont
  {Sali\'eres}}} (\bibinfo {year} {2003}),\ \bibfield  {title} {\enquote
  {\bibinfo {title} {Attosecond synchronization of high-harmonic soft
  x-rays},}\ }\href@noop {} {\bibfield  {journal} {\bibinfo  {journal}
  {Science}\ }\textbf {\bibinfo {volume} {302}},\ \bibinfo {pages}
  {1540--1543}}\BibitemShut {NoStop}%
\bibitem [{\citenamefont {Mairesse}\ and\ \citenamefont
  {Qu{\'e}r{\'e}}(2005)}]{mairesse2005frequency}%
  \BibitemOpen
  \bibfield  {author} {\bibinfo {author} {\bibnamefont {Mairesse},
  \bibfnamefont {Y}}, \ and\ \bibinfo {author} {\bibfnamefont {F.}~\bibnamefont
  {Qu{\'e}r{\'e}}}} (\bibinfo {year} {2005}),\ \bibfield  {title} {\enquote
  {\bibinfo {title} {Frequency-resolved optical gating for complete
  reconstruction of attosecond bursts},}\ }\href@noop {} {\bibfield  {journal}
  {\bibinfo  {journal} {Phys. Rev. A}\ }\textbf {\bibinfo {volume} {71}},\
  \bibinfo {pages} {011401}}\BibitemShut {NoStop}%
\bibitem [{\citenamefont {Marangos}(2011)}]{marangos2011introduction}%
  \BibitemOpen
  \bibfield  {author} {\bibinfo {author} {\bibnamefont {Marangos},
  \bibfnamefont {J~P}}} (\bibinfo {year} {2011}),\ \bibfield  {title} {\enquote
  {\bibinfo {title} {Introduction to the new science with x-ray free electron
  lasers},}\ }\href@noop {} {\bibfield  {journal} {\bibinfo  {journal}
  {Contemporary Physics}\ }\textbf {\bibinfo {volume} {52}},\ \bibinfo {pages}
  {551--569}}\BibitemShut {NoStop}%
\bibitem [{\citenamefont {Marangos}(2016)}]{marangos2016Review}%
  \BibitemOpen
  \bibfield  {author} {\bibinfo {author} {\bibnamefont {Marangos},
  \bibfnamefont {J~P}}} (\bibinfo {year} {2016}),\ \bibfield  {title} {\enquote
  {\bibinfo {title} {Development of high harmonic generation spectroscopy of
  organic molecules and biomolecules},}\ }\href@noop {} {\bibfield  {journal}
  {\bibinfo  {journal} {J. Phys. B}\ }\textbf {\bibinfo {volume} {49}},\
  \bibinfo {pages} {132001}}\BibitemShut {NoStop}%
\bibitem [{\citenamefont {Marinica}\ \emph {et~al.}(2015)\citenamefont
  {Marinica}, \citenamefont {Zapata}, \citenamefont {Nordlander}, \citenamefont
  {Kazansky}, \citenamefont {Echenique}, \citenamefont {Aizpurua},\ and\
  \citenamefont {Borisov}}]{Marinica2015}%
  \BibitemOpen
  \bibfield  {author} {\bibinfo {author} {\bibnamefont {Marinica},
  \bibfnamefont {D~C}}, \bibinfo {author} {\bibfnamefont {M.}~\bibnamefont
  {Zapata}}, \bibinfo {author} {\bibfnamefont {P.}~\bibnamefont {Nordlander}},
  \bibinfo {author} {\bibfnamefont {A.~K.}\ \bibnamefont {Kazansky}}, \bibinfo
  {author} {\bibfnamefont {P.}~\bibnamefont {Echenique}}, \bibinfo {author}
  {\bibfnamefont {J.}~\bibnamefont {Aizpurua}}, \ and\ \bibinfo {author}
  {\bibfnamefont {A.~G.}\ \bibnamefont {Borisov}}} (\bibinfo {year} {2015}),\
  \bibfield  {title} {\enquote {\bibinfo {title} {Active quantum plasmonics},}\
  }\href@noop {} {\bibfield  {journal} {\bibinfo  {journal} {Sci. Adv.}\
  }\textbf {\bibinfo {volume} {1}},\ \bibinfo {pages} {e1501095}}\BibitemShut
  {NoStop}%
\bibitem [{\citenamefont {Martin}\ and\ \citenamefont
  {Girard}(1997)}]{Martin1997}%
  \BibitemOpen
  \bibfield  {author} {\bibinfo {author} {\bibnamefont {Martin}, \bibfnamefont
  {O~J~F}}, \ and\ \bibinfo {author} {\bibfnamefont {C.}~\bibnamefont
  {Girard}}} (\bibinfo {year} {1997}),\ \bibfield  {title} {\enquote {\bibinfo
  {title} {Controlling and tuning strong optical field gradients at a local
  probe microscope tip apex},}\ }\href@noop {} {\bibfield  {journal} {\bibinfo
  {journal} {Appl.~Phys.~Lett.}\ }\textbf {\bibinfo {volume} {70}}~(\bibinfo
  {number} {6}),\ \bibinfo {pages} {705--707}}\BibitemShut {NoStop}%
\bibitem [{\citenamefont {Martin}\ \emph {et~al.}(2001)\citenamefont {Martin},
  \citenamefont {Hamann},\ and\ \citenamefont {Wickramasinghe}}]{Martin2001}%
  \BibitemOpen
  \bibfield  {author} {\bibinfo {author} {\bibnamefont {Martin}, \bibfnamefont
  {Y~C}}, \bibinfo {author} {\bibfnamefont {H.~F.}\ \bibnamefont {Hamann}}, \
  and\ \bibinfo {author} {\bibfnamefont {H.~K.}\ \bibnamefont
  {Wickramasinghe}}} (\bibinfo {year} {2001}),\ \bibfield  {title} {\enquote
  {\bibinfo {title} {Strength of the electric field in apertureless near-field
  optical microscopy},}\ }\href@noop {} {\bibfield  {journal} {\bibinfo
  {journal} {J. Appl. Phys.}\ }\textbf {\bibinfo {volume} {89}}~(\bibinfo
  {number} {10}),\ \bibinfo {pages} {5774--5778}}\BibitemShut {NoStop}%
\bibitem [{\citenamefont {Mashiko}\ \emph {et~al.}(2010)\citenamefont
  {Mashiko}, \citenamefont {Bell}, \citenamefont {Beck}, \citenamefont {Abel},
  \citenamefont {Nagel}, \citenamefont {Steiner}, \citenamefont {Robinson},
  \citenamefont {Neumark},\ and\ \citenamefont {Leone}}]{mashiko2010tunable}%
  \BibitemOpen
  \bibfield  {author} {\bibinfo {author} {\bibnamefont {Mashiko}, \bibfnamefont
  {H}}, \bibinfo {author} {\bibfnamefont {M.~J.}\ \bibnamefont {Bell}},
  \bibinfo {author} {\bibfnamefont {A.~R.}\ \bibnamefont {Beck}}, \bibinfo
  {author} {\bibfnamefont {M.~J.}\ \bibnamefont {Abel}}, \bibinfo {author}
  {\bibfnamefont {P.~M.}\ \bibnamefont {Nagel}}, \bibinfo {author}
  {\bibfnamefont {C.~P.}\ \bibnamefont {Steiner}}, \bibinfo {author}
  {\bibfnamefont {J.}~\bibnamefont {Robinson}}, \bibinfo {author}
  {\bibfnamefont {D.~M.}\ \bibnamefont {Neumark}}, \ and\ \bibinfo {author}
  {\bibfnamefont {S.~R.}\ \bibnamefont {Leone}}} (\bibinfo {year} {2010}),\
  \bibfield  {title} {\enquote {\bibinfo {title} {Tunable frequency-controlled
  isolated attosecond pulses characterized by either 750 nm or 400 nm
  wavelength streak fields},}\ }\href@noop {} {\bibfield  {journal} {\bibinfo
  {journal} {Opt. Exp.}\ }\textbf {\bibinfo {volume} {18}},\ \bibinfo {pages}
  {25887--25895}}\BibitemShut {NoStop}%
\bibitem [{\citenamefont {Mashiko}\ \emph {et~al.}(2008)\citenamefont
  {Mashiko}, \citenamefont {Gilbertson}, \citenamefont {Li}, \citenamefont
  {Khan}, \citenamefont {Shakya}, \citenamefont {Moon},\ and\ \citenamefont
  {Chang}}]{mashiko2008double}%
  \BibitemOpen
  \bibfield  {author} {\bibinfo {author} {\bibnamefont {Mashiko}, \bibfnamefont
  {H}}, \bibinfo {author} {\bibfnamefont {S.}~\bibnamefont {Gilbertson}},
  \bibinfo {author} {\bibfnamefont {C.}~\bibnamefont {Li}}, \bibinfo {author}
  {\bibfnamefont {S.~D.}\ \bibnamefont {Khan}}, \bibinfo {author}
  {\bibfnamefont {M.~M.}\ \bibnamefont {Shakya}}, \bibinfo {author}
  {\bibfnamefont {E.}~\bibnamefont {Moon}}, \ and\ \bibinfo {author}
  {\bibfnamefont {Z.}~\bibnamefont {Chang}}} (\bibinfo {year} {2008}),\
  \bibfield  {title} {\enquote {\bibinfo {title} {Double optical gating of
  high-order harmonic generation with carrier-envelope phase stabilized
  lasers},}\ }\href@noop {} {\bibfield  {journal} {\bibinfo  {journal} {Phys.
  Rev. Lett.}\ }\textbf {\bibinfo {volume} {100}},\ \bibinfo {pages}
  {103906}}\BibitemShut {NoStop}%
\bibitem [{\citenamefont {Mermin}(1970)}]{Mermin1970}%
  \BibitemOpen
  \bibfield  {author} {\bibinfo {author} {\bibnamefont {Mermin}, \bibfnamefont
  {N~D}}} (\bibinfo {year} {1970}),\ \bibfield  {title} {\enquote {\bibinfo
  {title} {Lindhard dielectric function in the relaxation-time
  approximation},}\ }\href@noop {} {\bibfield  {journal} {\bibinfo  {journal}
  {Phys. Rev. B}\ }\textbf {\bibinfo {volume} {1}},\ \bibinfo {pages}
  {2362--2363}}\BibitemShut {NoStop}%
\bibitem [{\citenamefont {Miaja-Avila}\ \emph {et~al.}(2006)\citenamefont
  {Miaja-Avila}, \citenamefont {Lei}, \citenamefont {Aeschlimann},
  \citenamefont {Gland}, \citenamefont {Murnane}, \citenamefont {Kapteyn},\
  and\ \citenamefont {Saathoff}}]{miaja2006laser}%
  \BibitemOpen
  \bibfield  {author} {\bibinfo {author} {\bibnamefont {Miaja-Avila},
  \bibfnamefont {L}}, \bibinfo {author} {\bibfnamefont {C.}~\bibnamefont
  {Lei}}, \bibinfo {author} {\bibfnamefont {M.}~\bibnamefont {Aeschlimann}},
  \bibinfo {author} {\bibfnamefont {J.~L.}\ \bibnamefont {Gland}}, \bibinfo
  {author} {\bibfnamefont {M.~M.}\ \bibnamefont {Murnane}}, \bibinfo {author}
  {\bibfnamefont {H.~C.}\ \bibnamefont {Kapteyn}}, \ and\ \bibinfo {author}
  {\bibfnamefont {G.}~\bibnamefont {Saathoff}}} (\bibinfo {year} {2006}),\
  \bibfield  {title} {\enquote {\bibinfo {title} {Laser-assisted photoelectric
  effect from surfaces},}\ }\href@noop {} {\bibfield  {journal} {\bibinfo
  {journal} {Phys. Rev. Lett.}\ }\textbf {\bibinfo {volume} {97}},\ \bibinfo
  {pages} {113604}}\BibitemShut {NoStop}%
\bibitem [{\citenamefont {Mikkelsen}\ \emph {et~al.}(2009)\citenamefont
  {Mikkelsen}, \citenamefont {Schwenke}, \citenamefont {Fordell}, \citenamefont
  {Luo}, \citenamefont {Kl{\"u}nder}, \citenamefont {Hilner}, \citenamefont
  {Anttu}, \citenamefont {Zakharov}, \citenamefont {Lundgren}, \citenamefont
  {Mauritsson}, \citenamefont {Andersen}, \citenamefont {Xu},\ and\
  \citenamefont {L'Huillier}}]{Mikkelsen09}%
  \BibitemOpen
  \bibfield  {author} {\bibinfo {author} {\bibnamefont {Mikkelsen},
  \bibfnamefont {A}}, \bibinfo {author} {\bibfnamefont {J.}~\bibnamefont
  {Schwenke}}, \bibinfo {author} {\bibfnamefont {T.}~\bibnamefont {Fordell}},
  \bibinfo {author} {\bibfnamefont {G.}~\bibnamefont {Luo}}, \bibinfo {author}
  {\bibfnamefont {K.}~\bibnamefont {Kl{\"u}nder}}, \bibinfo {author}
  {\bibfnamefont {E.}~\bibnamefont {Hilner}}, \bibinfo {author} {\bibfnamefont
  {N.}~\bibnamefont {Anttu}}, \bibinfo {author} {\bibfnamefont {A.~A.}\
  \bibnamefont {Zakharov}}, \bibinfo {author} {\bibfnamefont {E.}~\bibnamefont
  {Lundgren}}, \bibinfo {author} {\bibfnamefont {J.}~\bibnamefont
  {Mauritsson}}, \bibinfo {author} {\bibfnamefont {J.~N.}\ \bibnamefont
  {Andersen}}, \bibinfo {author} {\bibfnamefont {H.~Q.}\ \bibnamefont {Xu}}, \
  and\ \bibinfo {author} {\bibfnamefont {A.}~\bibnamefont {L'Huillier}}}
  (\bibinfo {year} {2009}),\ \bibfield  {title} {\enquote {\bibinfo {title}
  {Photoemission electron microscopy using extreme ultraviolet attosecond pulse
  trains},}\ }\href@noop {} {\bibfield  {journal} {\bibinfo  {journal} {Rev.
  Sci. Instr.}\ }\textbf {\bibinfo {volume} {80}},\ \bibinfo {pages}
  {123703}}\BibitemShut {NoStop}%
\bibitem [{\citenamefont {Milo\u{s}evi\'c}\ \emph {et~al.}(2006)\citenamefont
  {Milo\u{s}evi\'c}, \citenamefont {Paulus}, \citenamefont {Bauer},\ and\
  \citenamefont {Becker}}]{Milosevic06}%
  \BibitemOpen
  \bibfield  {author} {\bibinfo {author} {\bibnamefont {Milo\u{s}evi\'c},
  \bibfnamefont {D~B}}, \bibinfo {author} {\bibfnamefont {G.G.}\ \bibnamefont
  {Paulus}}, \bibinfo {author} {\bibfnamefont {D.}~\bibnamefont {Bauer}}, \
  and\ \bibinfo {author} {\bibfnamefont {W.}~\bibnamefont {Becker}}} (\bibinfo
  {year} {2006}),\ \bibfield  {title} {\enquote {\bibinfo {title}
  {Above-threshold ionization by few-cycle pulses},}\ }\href@noop {} {\bibfield
   {journal} {\bibinfo  {journal} {J. Phys. B}\ }\textbf {\bibinfo {volume}
  {39}},\ \bibinfo {pages} {R203--R262}}\BibitemShut {NoStop}%
\bibitem [{\citenamefont {Miranda}\ \emph {et~al.}(2012)\citenamefont
  {Miranda}, \citenamefont {Fordell}, \citenamefont {Arnold}, \citenamefont
  {L'Huillier},\ and\ \citenamefont {Crespo}}]{miranda_simultaneous_2012}%
  \BibitemOpen
  \bibfield  {author} {\bibinfo {author} {\bibnamefont {Miranda}, \bibfnamefont
  {M}}, \bibinfo {author} {\bibfnamefont {T.}~\bibnamefont {Fordell}}, \bibinfo
  {author} {\bibfnamefont {C.}~\bibnamefont {Arnold}}, \bibinfo {author}
  {\bibfnamefont {A.}~\bibnamefont {L'Huillier}}, \ and\ \bibinfo {author}
  {\bibfnamefont {H.}~\bibnamefont {Crespo}}} (\bibinfo {year} {2012}),\
  \bibfield  {title} {\enquote {\bibinfo {title} {Simultaneous compression and
  characterization of ultrashort laser pulses using chirped mirrors and glass
  wedges},}\ }\href@noop {} {\bibfield  {journal} {\bibinfo  {journal} {Opt.
  Expr.}\ }\textbf {\bibinfo {volume} {20}},\ \bibinfo {pages}
  {688--697}}\BibitemShut {NoStop}%
\bibitem [{\citenamefont {Mountford}\ \emph {et~al.}(1998)\citenamefont
  {Mountford}, \citenamefont {Smith},\ and\ \citenamefont
  {Hutchinson}}]{Mountford1998}%
  \BibitemOpen
  \bibfield  {author} {\bibinfo {author} {\bibnamefont {Mountford},
  \bibfnamefont {L~C}}, \bibinfo {author} {\bibfnamefont {R.~A.}\ \bibnamefont
  {Smith}}, \ and\ \bibinfo {author} {\bibfnamefont {M.~H.~R.}\ \bibnamefont
  {Hutchinson}}} (\bibinfo {year} {1998}),\ \bibfield  {title} {\enquote
  {\bibinfo {title} {Characterization of a sub-micron liquid spray for
  laser-plasma x-ray generation},}\ }\href@noop {} {\bibfield  {journal}
  {\bibinfo  {journal} {Rev. Sci. Instrum.}\ }\textbf {\bibinfo {volume}
  {69}},\ \bibinfo {pages} {3780}}\BibitemShut {NoStop}%
\bibitem [{\citenamefont {Muller}\ \emph {et~al.}(1986)\citenamefont {Muller},
  \citenamefont {van Linden van~den Heuvell},\ and\ \citenamefont {van~der
  Wiel}}]{VanderWiel}%
  \BibitemOpen
  \bibfield  {author} {\bibinfo {author} {\bibnamefont {Muller}, \bibfnamefont
  {H~G}}, \bibinfo {author} {\bibfnamefont {H.~B.}\ \bibnamefont {van Linden
  van~den Heuvell}}, \ and\ \bibinfo {author} {\bibfnamefont {M.~J.}\
  \bibnamefont {van~der Wiel}}} (\bibinfo {year} {1986}),\ \bibfield  {title}
  {\enquote {\bibinfo {title} {Experiments on "above-threshold ionization" of
  atomic hydrogen},}\ }\href@noop {} {\bibfield  {journal} {\bibinfo  {journal}
  {Phys. Rev. A}\ }\textbf {\bibinfo {volume} {34}},\ \bibinfo {pages}
  {236}}\BibitemShut {NoStop}%
\bibitem [{\citenamefont {M{\"u}ller}\ \emph {et~al.}(2016)\citenamefont
  {M{\"u}ller}, \citenamefont {Kravtsov}, \citenamefont {Paarmann},
  \citenamefont {Raschke},\ and\ \citenamefont {Ernstorfer}}]{Muller2016}%
  \BibitemOpen
  \bibfield  {author} {\bibinfo {author} {\bibnamefont {M{\"u}ller},
  \bibfnamefont {M}}, \bibinfo {author} {\bibfnamefont {V.}~\bibnamefont
  {Kravtsov}}, \bibinfo {author} {\bibfnamefont {A.}~\bibnamefont {Paarmann}},
  \bibinfo {author} {\bibfnamefont {M.~B.}\ \bibnamefont {Raschke}}, \ and\
  \bibinfo {author} {\bibfnamefont {R.}~\bibnamefont {Ernstorfer}}} (\bibinfo
  {year} {2016}),\ \bibfield  {title} {\enquote {\bibinfo {title} {Nanofocused
  plasmon-driven sub-10 fs electron point source},}\ }\href@noop {} {\bibfield
  {journal} {\bibinfo  {journal} {ACS Photonics}\ }\textbf {\bibinfo {volume}
  {3}},\ \bibinfo {pages} {611--619}}\BibitemShut {NoStop}%
\bibitem [{\citenamefont {M\"uller}\ \emph {et~al.}(2015)\citenamefont
  {M\"uller}, \citenamefont {Paarmann},\ and\ \citenamefont
  {Ernstorfer}}]{Muller2014}%
  \BibitemOpen
  \bibfield  {author} {\bibinfo {author} {\bibnamefont {M\"uller},
  \bibfnamefont {M}}, \bibinfo {author} {\bibfnamefont {A.}~\bibnamefont
  {Paarmann}}, \ and\ \bibinfo {author} {\bibfnamefont {R.}~\bibnamefont
  {Ernstorfer}}} (\bibinfo {year} {2015}),\ \bibfield  {title} {\enquote
  {\bibinfo {title} {Femtosecond electrons probing currents and atomic
  structure in nanomaterials},}\ }\href@noop {} {\bibfield  {journal} {\bibinfo
   {journal} {Nat. Comm.}\ }\textbf {\bibinfo {volume} {5}},\ \bibinfo {pages}
  {5292}}\BibitemShut {NoStop}%
\bibitem [{\citenamefont {Mustonen}\ \emph {et~al.}(2011)\citenamefont
  {Mustonen}, \citenamefont {Beaud}, \citenamefont {Kirk}, \citenamefont
  {Feurer},\ and\ \citenamefont {Tsujino}}]{Mustonen2011}%
  \BibitemOpen
  \bibfield  {author} {\bibinfo {author} {\bibnamefont {Mustonen},
  \bibfnamefont {A}}, \bibinfo {author} {\bibfnamefont {P.}~\bibnamefont
  {Beaud}}, \bibinfo {author} {\bibfnamefont {E.}~\bibnamefont {Kirk}},
  \bibinfo {author} {\bibfnamefont {T.}~\bibnamefont {Feurer}}, \ and\ \bibinfo
  {author} {\bibfnamefont {S.}~\bibnamefont {Tsujino}}} (\bibinfo {year}
  {2011}),\ \bibfield  {title} {\enquote {\bibinfo {title} {Five picocoulomb
  electron bunch generation by ultrafast laser-induced field emission from
  metallic nano-tip arrays},}\ }\href@noop {} {\bibfield  {journal} {\bibinfo
  {journal} {Appl. Phys. Lett.}\ }\textbf {\bibinfo {volume} {99}},\ \bibinfo
  {pages} {103504}}\BibitemShut {NoStop}%
\bibitem [{\citenamefont {Mustonen}\ \emph {et~al.}(2012)\citenamefont
  {Mustonen}, \citenamefont {Beaud}, \citenamefont {Kirk}, \citenamefont
  {Feurer},\ and\ \citenamefont {Tsujino}}]{Mustonen2012}%
  \BibitemOpen
  \bibfield  {author} {\bibinfo {author} {\bibnamefont {Mustonen},
  \bibfnamefont {A}}, \bibinfo {author} {\bibfnamefont {P.}~\bibnamefont
  {Beaud}}, \bibinfo {author} {\bibfnamefont {E.}~\bibnamefont {Kirk}},
  \bibinfo {author} {\bibfnamefont {T.}~\bibnamefont {Feurer}}, \ and\ \bibinfo
  {author} {\bibfnamefont {S.}~\bibnamefont {Tsujino}}} (\bibinfo {year}
  {2012}),\ \bibfield  {title} {\enquote {\bibinfo {title} {Efficient light
  coupling for optically excited high-density metallic nanotip arrays},}\
  }\href@noop {} {\bibfield  {journal} {\bibinfo  {journal} {Sci. Rep.}\
  }\textbf {\bibinfo {volume} {2}},\ \bibinfo {pages} {915}}\BibitemShut
  {NoStop}%
\bibitem [{\citenamefont {Nagel}\ \emph {et~al.}(2013)\citenamefont {Nagel},
  \citenamefont {Robinson}, \citenamefont {Harteneck}, \citenamefont {Pfeifer},
  \citenamefont {Abel}, \citenamefont {Prell}, \citenamefont {Neumark},
  \citenamefont {Kaindl},\ and\ \citenamefont {Leone}}]{Nagel2013}%
  \BibitemOpen
  \bibfield  {author} {\bibinfo {author} {\bibnamefont {Nagel}, \bibfnamefont
  {P~M}}, \bibinfo {author} {\bibfnamefont {J.~S.}\ \bibnamefont {Robinson}},
  \bibinfo {author} {\bibfnamefont {B.~D.}\ \bibnamefont {Harteneck}}, \bibinfo
  {author} {\bibfnamefont {T.}~\bibnamefont {Pfeifer}}, \bibinfo {author}
  {\bibfnamefont {M.~J.}\ \bibnamefont {Abel}}, \bibinfo {author}
  {\bibfnamefont {J.~S.}\ \bibnamefont {Prell}}, \bibinfo {author}
  {\bibfnamefont {D.~M.}\ \bibnamefont {Neumark}}, \bibinfo {author}
  {\bibfnamefont {R.~A.}\ \bibnamefont {Kaindl}}, \ and\ \bibinfo {author}
  {\bibfnamefont {S.~R.}\ \bibnamefont {Leone}}} (\bibinfo {year} {2013}),\
  \bibfield  {title} {\enquote {\bibinfo {title} {Surface plasmon assisted
  electron acceleration in photoemission from gold nanopillars},}\ }\href@noop
  {} {\bibfield  {journal} {\bibinfo  {journal} {Chem. Phys.}\ }\textbf
  {\bibinfo {volume} {414}},\ \bibinfo {pages} {106--111}}\BibitemShut
  {NoStop}%
\bibitem [{\citenamefont {Naik}\ \emph {et~al.}(2013)\citenamefont {Naik},
  \citenamefont {Shalaev},\ and\ \citenamefont
  {Boltasseva}}]{naik_alternative_2013}%
  \BibitemOpen
  \bibfield  {author} {\bibinfo {author} {\bibnamefont {Naik}, \bibfnamefont
  {G~V}}, \bibinfo {author} {\bibfnamefont {V.~M.}\ \bibnamefont {Shalaev}}, \
  and\ \bibinfo {author} {\bibfnamefont {A.}~\bibnamefont {Boltasseva}}}
  (\bibinfo {year} {2013}),\ \bibfield  {title} {\enquote {\bibinfo {title}
  {Alternative {Plasmonic} {Materials}: {Beyond} {Gold} and {Silver}},}\
  }\href@noop {} {\bibfield  {journal} {\bibinfo  {journal} {Adv. Mater.}\
  }\textbf {\bibinfo {volume} {25}},\ \bibinfo {pages}
  {3264--3294}}\BibitemShut {NoStop}%
\bibitem [{\citenamefont {Nakamura}\ and\ \citenamefont
  {Yamada}(1981)}]{nakamura_fundamental_1981}%
  \BibitemOpen
  \bibfield  {author} {\bibinfo {author} {\bibnamefont {Nakamura},
  \bibfnamefont {A}}, \ and\ \bibinfo {author} {\bibfnamefont {S.}~\bibnamefont
  {Yamada}}} (\bibinfo {year} {1981}),\ \bibfield  {title} {\enquote {\bibinfo
  {title} {Fundamental absorption edge of evaporated amorphous {WO}3 films},}\
  }\href@noop {} {\bibfield  {journal} {\bibinfo  {journal} {Appl. Phys.}\
  }\textbf {\bibinfo {volume} {24}},\ \bibinfo {pages} {55--59}}\BibitemShut
  {NoStop}%
\bibitem [{\citenamefont {Neacsu}\ \emph {et~al.}(2005)\citenamefont {Neacsu},
  \citenamefont {Reider},\ and\ \citenamefont {Raschke}}]{Neacsu2005}%
  \BibitemOpen
  \bibfield  {author} {\bibinfo {author} {\bibnamefont {Neacsu}, \bibfnamefont
  {C~C}}, \bibinfo {author} {\bibfnamefont {G.~A.}\ \bibnamefont {Reider}}, \
  and\ \bibinfo {author} {\bibfnamefont {M.~B.}\ \bibnamefont {Raschke}}}
  (\bibinfo {year} {2005}),\ \bibfield  {title} {\enquote {\bibinfo {title}
  {Second-harmonic generation from nanoscopic metal tips: {S}ymmetry selection
  rules for single asymmetric nanostructures},}\ }\href@noop {} {\bibfield
  {journal} {\bibinfo  {journal} {Phys. Rev. B}\ }\textbf {\bibinfo {volume}
  {71}},\ \bibinfo {pages} {201402}}\BibitemShut {NoStop}%
\bibitem [{\citenamefont {Nelayah}\ \emph {et~al.}(2007)\citenamefont
  {Nelayah}, \citenamefont {Kociak}, \citenamefont {St{\'e}phan}, \citenamefont
  {Garc{\'\i}a~de Abajo}, \citenamefont {Tenc{\'e}}, \citenamefont {Henrard},
  \citenamefont {Taverna}, \citenamefont {Pastoriza-Santos}, \citenamefont
  {Liz-Marz{\'a}n},\ and\ \citenamefont {Colliex}}]{Nelayah2007}%
  \BibitemOpen
  \bibfield  {author} {\bibinfo {author} {\bibnamefont {Nelayah}, \bibfnamefont
  {J}}, \bibinfo {author} {\bibfnamefont {M.}~\bibnamefont {Kociak}}, \bibinfo
  {author} {\bibfnamefont {O.}~\bibnamefont {St{\'e}phan}}, \bibinfo {author}
  {\bibfnamefont {F.~J.}\ \bibnamefont {Garc{\'\i}a~de Abajo}}, \bibinfo
  {author} {\bibfnamefont {M.}~\bibnamefont {Tenc{\'e}}}, \bibinfo {author}
  {\bibfnamefont {L.}~\bibnamefont {Henrard}}, \bibinfo {author} {\bibfnamefont
  {D.}~\bibnamefont {Taverna}}, \bibinfo {author} {\bibfnamefont
  {I.}~\bibnamefont {Pastoriza-Santos}}, \bibinfo {author} {\bibfnamefont
  {L.~M.}\ \bibnamefont {Liz-Marz{\'a}n}}, \ and\ \bibinfo {author}
  {\bibfnamefont {C.}~\bibnamefont {Colliex}}} (\bibinfo {year} {2007}),\
  \bibfield  {title} {\enquote {\bibinfo {title} {Mapping surface plasmons on a
  single metallic nanoparticle},}\ }\href@noop {} {\bibfield  {journal}
  {\bibinfo  {journal} {Nat. Phys.}\ }\textbf {\bibinfo {volume} {3}}~(\bibinfo
  {number} {5}),\ \bibinfo {pages} {348--353}}\BibitemShut {NoStop}%
\bibitem [{\citenamefont {Neppl}\ \emph {et~al.}(2012)\citenamefont {Neppl},
  \citenamefont {Ernstorfer}, \citenamefont {Bothschafter}, \citenamefont
  {Cavalieri}, \citenamefont {Menzel}, \citenamefont {Barth}, \citenamefont
  {Krausz}, \citenamefont {Kienberger},\ and\ \citenamefont
  {Feulner}}]{neppl2012attosecond}%
  \BibitemOpen
  \bibfield  {author} {\bibinfo {author} {\bibnamefont {Neppl}, \bibfnamefont
  {S}}, \bibinfo {author} {\bibfnamefont {R.}~\bibnamefont {Ernstorfer}},
  \bibinfo {author} {\bibfnamefont {E.~M.}\ \bibnamefont {Bothschafter}},
  \bibinfo {author} {\bibfnamefont {A.~L.}\ \bibnamefont {Cavalieri}}, \bibinfo
  {author} {\bibfnamefont {D.}~\bibnamefont {Menzel}}, \bibinfo {author}
  {\bibfnamefont {J.~V.}\ \bibnamefont {Barth}}, \bibinfo {author}
  {\bibfnamefont {F.}~\bibnamefont {Krausz}}, \bibinfo {author} {\bibfnamefont
  {R.}~\bibnamefont {Kienberger}}, \ and\ \bibinfo {author} {\bibfnamefont
  {P.}~\bibnamefont {Feulner}}} (\bibinfo {year} {2012}),\ \bibfield  {title}
  {\enquote {\bibinfo {title} {Attosecond time-resolved photoemission from core
  and valence states of magnesium},}\ }\href@noop {} {\bibfield  {journal}
  {\bibinfo  {journal} {Phys. Rev. Lett.}\ }\textbf {\bibinfo {volume} {109}},\
  \bibinfo {pages} {087401}}\BibitemShut {NoStop}%
\bibitem [{\citenamefont {Neppl}(2012)}]{neppl_attosecond_2012}%
  \BibitemOpen
  \bibfield  {author} {\bibinfo {author} {\bibnamefont {Neppl}, \bibfnamefont
  {Stefan}}} (\bibinfo {year} {2012}),\ \emph {\bibinfo {title} {Attosecond
  {Time}-{Resolved} {Photoemission} from {Surfaces} and {Interfaces}}},\
  \href@noop {} {Ph.D. thesis}\ (\bibinfo  {school} {LMU Munich})\BibitemShut
  {NoStop}%
\bibitem [{\citenamefont {Novotny}\ and\ \citenamefont
  {Hecht}(2012)}]{Novotny2012}%
  \BibitemOpen
  \bibfield  {author} {\bibinfo {author} {\bibnamefont {Novotny}, \bibfnamefont
  {L}}, \ and\ \bibinfo {author} {\bibfnamefont {B.}~\bibnamefont {Hecht}}}
  (\bibinfo {year} {2012}),\ \href@noop {} {\emph {\bibinfo {title} {Principles
  of {N}ano-{O}ptics}}}\ (\bibinfo  {publisher} {Cambridge University
  Press})\BibitemShut {NoStop}%
\bibitem [{\citenamefont {Novotny}\ and\ \citenamefont {{van
  Hulst}}(2011)}]{Novotny2011}%
  \BibitemOpen
  \bibfield  {author} {\bibinfo {author} {\bibnamefont {Novotny}, \bibfnamefont
  {L}}, \ and\ \bibinfo {author} {\bibfnamefont {N.}~\bibnamefont {{van
  Hulst}}}} (\bibinfo {year} {2011}),\ \bibfield  {title} {\enquote {\bibinfo
  {title} {Antennas for light},}\ }\href@noop {} {\bibfield  {journal}
  {\bibinfo  {journal} {Nat. Phot.}\ }\textbf {\bibinfo {volume} {5}},\
  \bibinfo {pages} {83--90}}\BibitemShut {NoStop}%
\bibitem [{\citenamefont {Okell}\ \emph {et~al.}(2015)\citenamefont {Okell},
  \citenamefont {Witting}, \citenamefont {Fabris}, \citenamefont {Arrell},
  \citenamefont {Hengster}, \citenamefont {Ibrahimkutty}, \citenamefont
  {Seiler}, \citenamefont {Barthelmess}, \citenamefont {Stankov}, \citenamefont
  {Lei}, \citenamefont {Sonnefraud}, \citenamefont {Rahmani}, \citenamefont
  {Uphues}, \citenamefont {Maier}, \citenamefont {Marangos},\ and\
  \citenamefont {Tisch}}]{okell_temporal_2015}%
  \BibitemOpen
  \bibfield  {author} {\bibinfo {author} {\bibnamefont {Okell}, \bibfnamefont
  {W~A}}, \bibinfo {author} {\bibfnamefont {T.}~\bibnamefont {Witting}},
  \bibinfo {author} {\bibfnamefont {D.}~\bibnamefont {Fabris}}, \bibinfo
  {author} {\bibfnamefont {C.~A.}\ \bibnamefont {Arrell}}, \bibinfo {author}
  {\bibfnamefont {J.}~\bibnamefont {Hengster}}, \bibinfo {author}
  {\bibfnamefont {S.}~\bibnamefont {Ibrahimkutty}}, \bibinfo {author}
  {\bibfnamefont {A.}~\bibnamefont {Seiler}}, \bibinfo {author} {\bibfnamefont
  {M.}~\bibnamefont {Barthelmess}}, \bibinfo {author} {\bibfnamefont
  {S.}~\bibnamefont {Stankov}}, \bibinfo {author} {\bibfnamefont {D.~Y.}\
  \bibnamefont {Lei}}, \bibinfo {author} {\bibfnamefont {Y.}~\bibnamefont
  {Sonnefraud}}, \bibinfo {author} {\bibfnamefont {M.}~\bibnamefont {Rahmani}},
  \bibinfo {author} {\bibfnamefont {T.}~\bibnamefont {Uphues}}, \bibinfo
  {author} {\bibfnamefont {S.~A.}\ \bibnamefont {Maier}}, \bibinfo {author}
  {\bibfnamefont {J.~P.}\ \bibnamefont {Marangos}}, \ and\ \bibinfo {author}
  {\bibfnamefont {J.~W.~G.}\ \bibnamefont {Tisch}}} (\bibinfo {year} {2015}),\
  \bibfield  {title} {\enquote {\bibinfo {title} {Temporal broadening of
  attosecond photoelectron wavepackets from solid surfaces},}\ }\href@noop {}
  {\bibfield  {journal} {\bibinfo  {journal} {Optica}\ }\textbf {\bibinfo
  {volume} {2}},\ \bibinfo {pages} {383--387}}\BibitemShut {NoStop}%
\bibitem [{\citenamefont {Okell}\ \emph {et~al.}(2013)\citenamefont {Okell},
  \citenamefont {Witting}, \citenamefont {Fabris}, \citenamefont {Austin},
  \citenamefont {Bocoum}, \citenamefont {Frank}, \citenamefont {Ricci},
  \citenamefont {Jullien}, \citenamefont {Walke}, \citenamefont {Marangos},
  \citenamefont {L\'opez-Martens},\ and\ \citenamefont
  {Tisch}}]{okell_carrier_envelope_2013}%
  \BibitemOpen
  \bibfield  {author} {\bibinfo {author} {\bibnamefont {Okell}, \bibfnamefont
  {W~A}}, \bibinfo {author} {\bibfnamefont {T.}~\bibnamefont {Witting}},
  \bibinfo {author} {\bibfnamefont {D.}~\bibnamefont {Fabris}}, \bibinfo
  {author} {\bibfnamefont {D.}~\bibnamefont {Austin}}, \bibinfo {author}
  {\bibfnamefont {M.}~\bibnamefont {Bocoum}}, \bibinfo {author} {\bibfnamefont
  {F.}~\bibnamefont {Frank}}, \bibinfo {author} {\bibfnamefont
  {A.}~\bibnamefont {Ricci}}, \bibinfo {author} {\bibfnamefont
  {A.}~\bibnamefont {Jullien}}, \bibinfo {author} {\bibfnamefont
  {D.}~\bibnamefont {Walke}}, \bibinfo {author} {\bibfnamefont {J.~P.}\
  \bibnamefont {Marangos}}, \bibinfo {author} {\bibfnamefont {R.}~\bibnamefont
  {L\'opez-Martens}}, \ and\ \bibinfo {author} {\bibfnamefont {J.~W.~G.}\
  \bibnamefont {Tisch}}} (\bibinfo {year} {2013}),\ \bibfield  {title}
  {\enquote {\bibinfo {title} {Carrier-envelope phase stability of hollow
  fibers used for high-energy few-cycle pulse generation},}\ }\href@noop {}
  {\bibfield  {journal} {\bibinfo  {journal} {Opt. Lett.}\ }\textbf {\bibinfo
  {volume} {38}},\ \bibinfo {pages} {3918--3921}}\BibitemShut {NoStop}%
\bibitem [{\citenamefont {Paarmann}\ \emph {et~al.}(2012)\citenamefont
  {Paarmann}, \citenamefont {Gulde}, \citenamefont {M{\"u}ller}, \citenamefont
  {Sch{\"a}fer}, \citenamefont {Schweda}, \citenamefont {Maiti}, \citenamefont
  {Xu}, \citenamefont {Hohage}, \citenamefont {Schenk}, \citenamefont
  {Ropers},\ and\ \citenamefont {Ernstorfer}}]{Paarmann2012}%
  \BibitemOpen
  \bibfield  {author} {\bibinfo {author} {\bibnamefont {Paarmann},
  \bibfnamefont {A}}, \bibinfo {author} {\bibfnamefont {M.}~\bibnamefont
  {Gulde}}, \bibinfo {author} {\bibfnamefont {M.}~\bibnamefont {M{\"u}ller}},
  \bibinfo {author} {\bibfnamefont {S.}~\bibnamefont {Sch{\"a}fer}}, \bibinfo
  {author} {\bibfnamefont {S.}~\bibnamefont {Schweda}}, \bibinfo {author}
  {\bibfnamefont {M.}~\bibnamefont {Maiti}}, \bibinfo {author} {\bibfnamefont
  {C.}~\bibnamefont {Xu}}, \bibinfo {author} {\bibfnamefont {T.}~\bibnamefont
  {Hohage}}, \bibinfo {author} {\bibfnamefont {F.}~\bibnamefont {Schenk}},
  \bibinfo {author} {\bibfnamefont {C.}~\bibnamefont {Ropers}}, \ and\ \bibinfo
  {author} {\bibfnamefont {R.}~\bibnamefont {Ernstorfer}}} (\bibinfo {year}
  {2012}),\ \bibfield  {title} {\enquote {\bibinfo {title} {Coherent
  femtosecond low-energy single-electron pulses for time-resolved diffraction
  and imaging: A numerical study},}\ }\href@noop {} {\bibfield  {journal}
  {\bibinfo  {journal} {J. Appl. Phys.}\ }\textbf {\bibinfo {volume} {112}},\
  \bibinfo {pages} {113109}}\BibitemShut {NoStop}%
\bibitem [{\citenamefont {Park}\ \emph {et~al.}(2013)\citenamefont {Park},
  \citenamefont {Piglosiewicz}, \citenamefont {Schmidt}, \citenamefont
  {Kollmann}, \citenamefont {Mascheck}, \citenamefont {Gro{\ss}},\ and\
  \citenamefont {Lienau}}]{Park13}%
  \BibitemOpen
  \bibfield  {author} {\bibinfo {author} {\bibnamefont {Park}, \bibfnamefont
  {D~J}}, \bibinfo {author} {\bibfnamefont {B.}~\bibnamefont {Piglosiewicz}},
  \bibinfo {author} {\bibfnamefont {S.}~\bibnamefont {Schmidt}}, \bibinfo
  {author} {\bibfnamefont {H.}~\bibnamefont {Kollmann}}, \bibinfo {author}
  {\bibfnamefont {M.}~\bibnamefont {Mascheck}}, \bibinfo {author}
  {\bibfnamefont {P.}~\bibnamefont {Gro{\ss}}}, \ and\ \bibinfo {author}
  {\bibfnamefont {C.}~\bibnamefont {Lienau}}} (\bibinfo {year} {2013}),\
  \bibfield  {title} {\enquote {\bibinfo {title} {Characterizing the optical
  near-field in the vicinity of a sharp metallic nanoprobe by angle-resolved
  electron kinetic energy spectroscopy},}\ }\href@noop {} {\bibfield  {journal}
  {\bibinfo  {journal} {Ann. Phys. (Berlin)}\ }\textbf {\bibinfo {volume}
  {525}},\ \bibinfo {pages} {135--142}}\BibitemShut {NoStop}%
\bibitem [{\citenamefont {Park}\ \emph {et~al.}(2012)\citenamefont {Park},
  \citenamefont {Piglosiewicz}, \citenamefont {Schmidt}, \citenamefont
  {Kollmann}, \citenamefont {Mascheck},\ and\ \citenamefont
  {Lienau}}]{Park2012}%
  \BibitemOpen
  \bibfield  {author} {\bibinfo {author} {\bibnamefont {Park}, \bibfnamefont
  {D~J}}, \bibinfo {author} {\bibfnamefont {B.}~\bibnamefont {Piglosiewicz}},
  \bibinfo {author} {\bibfnamefont {S.}~\bibnamefont {Schmidt}}, \bibinfo
  {author} {\bibfnamefont {H.}~\bibnamefont {Kollmann}}, \bibinfo {author}
  {\bibfnamefont {M.}~\bibnamefont {Mascheck}}, \ and\ \bibinfo {author}
  {\bibfnamefont {C.}~\bibnamefont {Lienau}}} (\bibinfo {year} {2012}),\
  \bibfield  {title} {\enquote {\bibinfo {title} {Strong field acceleration and
  steering of ultrafast electron pulses from a sharp metallic nanotip},}\
  }\href@noop {} {\bibfield  {journal} {\bibinfo  {journal} {Phys. Rev. Lett.}\
  }\textbf {\bibinfo {volume} {109}},\ \bibinfo {pages} {244803}}\BibitemShut
  {NoStop}%
\bibitem [{\citenamefont {Park}\ \emph {et~al.}(2011)\citenamefont {Park},
  \citenamefont {Kim}, \citenamefont {Choi}, \citenamefont {Lee}, \citenamefont
  {Kim}, \citenamefont {Kling}, \citenamefont {Stockman},\ and\ \citenamefont
  {Kim}}]{Park11}%
  \BibitemOpen
  \bibfield  {author} {\bibinfo {author} {\bibnamefont {Park}, \bibfnamefont
  {I-Y}}, \bibinfo {author} {\bibfnamefont {S.}~\bibnamefont {Kim}}, \bibinfo
  {author} {\bibfnamefont {J.}~\bibnamefont {Choi}}, \bibinfo {author}
  {\bibfnamefont {D-H.}\ \bibnamefont {Lee}}, \bibinfo {author} {\bibfnamefont
  {Y.~J.}\ \bibnamefont {Kim}}, \bibinfo {author} {\bibfnamefont {M.~F.}\
  \bibnamefont {Kling}}, \bibinfo {author} {\bibfnamefont {M.~I.}\ \bibnamefont
  {Stockman}}, \ and\ \bibinfo {author} {\bibfnamefont {S-W.}\ \bibnamefont
  {Kim}}} (\bibinfo {year} {2011}),\ \bibfield  {title} {\enquote {\bibinfo
  {title} {Plasmonic generation of ultrashort extreme-ultraviolet light
  pulses},}\ }\href@noop {} {\bibfield  {journal} {\bibinfo  {journal} {Nat.
  Phot.}\ }\textbf {\bibinfo {volume} {5}},\ \bibinfo {pages}
  {677}}\BibitemShut {NoStop}%
\bibitem [{\citenamefont {Paulus}\ \emph {et~al.}(2001)\citenamefont {Paulus},
  \citenamefont {Grasbon}, \citenamefont {Walther}, \citenamefont {Villoresi},
  \citenamefont {Nisoli}, \citenamefont {Stagira}, \citenamefont {Priori},\
  and\ \citenamefont {De~Silvestri}}]{paulus2001}%
  \BibitemOpen
  \bibfield  {author} {\bibinfo {author} {\bibnamefont {Paulus}, \bibfnamefont
  {G~G}}, \bibinfo {author} {\bibfnamefont {F.}~\bibnamefont {Grasbon}},
  \bibinfo {author} {\bibfnamefont {H.}~\bibnamefont {Walther}}, \bibinfo
  {author} {\bibfnamefont {P.}~\bibnamefont {Villoresi}}, \bibinfo {author}
  {\bibfnamefont {M.}~\bibnamefont {Nisoli}}, \bibinfo {author} {\bibfnamefont
  {S.}~\bibnamefont {Stagira}}, \bibinfo {author} {\bibfnamefont
  {E.}~\bibnamefont {Priori}}, \ and\ \bibinfo {author} {\bibfnamefont
  {S.}~\bibnamefont {De~Silvestri}}} (\bibinfo {year} {2001}),\ \bibfield
  {title} {\enquote {\bibinfo {title} {Absolute-phase phenomena in
  photoionization with few-cycle laser pulses},}\ }\href@noop {} {\bibfield
  {journal} {\bibinfo  {journal} {Nature}\ }\textbf {\bibinfo {volume} {414}},\
  \bibinfo {pages} {182--184}}\BibitemShut {NoStop}%
\bibitem [{\citenamefont {Paulus}\ \emph {et~al.}(2003)\citenamefont {Paulus},
  \citenamefont {Lindner}, \citenamefont {Walther}, \citenamefont
  {Baltu\u{s}ka}, \citenamefont {Goulielmakis}, \citenamefont {Lezius},\ and\
  \citenamefont {Krausz}}]{paulus_measurement_2003}%
  \BibitemOpen
  \bibfield  {author} {\bibinfo {author} {\bibnamefont {Paulus}, \bibfnamefont
  {G~G}}, \bibinfo {author} {\bibfnamefont {F.}~\bibnamefont {Lindner}},
  \bibinfo {author} {\bibfnamefont {H.}~\bibnamefont {Walther}}, \bibinfo
  {author} {\bibfnamefont {A.}~\bibnamefont {Baltu\u{s}ka}}, \bibinfo {author}
  {\bibfnamefont {E.}~\bibnamefont {Goulielmakis}}, \bibinfo {author}
  {\bibfnamefont {M.}~\bibnamefont {Lezius}}, \ and\ \bibinfo {author}
  {\bibfnamefont {F.}~\bibnamefont {Krausz}}} (\bibinfo {year} {2003}),\
  \bibfield  {title} {\enquote {\bibinfo {title} {Measurement of the phase of
  few-cycle laser pulses},}\ }\href@noop {} {\bibfield  {journal} {\bibinfo
  {journal} {Phys. Rev. Lett.}\ }\textbf {\bibinfo {volume} {91}},\ \bibinfo
  {pages} {253004}}\BibitemShut {NoStop}%
\bibitem [{\citenamefont {Paulus}\ \emph {et~al.}(2004)\citenamefont {Paulus},
  \citenamefont {Lindner}, \citenamefont {Walther},\ and\ \citenamefont
  {Milo{\v s}evi{\'c}}}]{Paulus2004}%
  \BibitemOpen
  \bibfield  {author} {\bibinfo {author} {\bibnamefont {Paulus}, \bibfnamefont
  {G~G}}, \bibinfo {author} {\bibfnamefont {F.}~\bibnamefont {Lindner}},
  \bibinfo {author} {\bibfnamefont {H.}~\bibnamefont {Walther}}, \ and\
  \bibinfo {author} {\bibfnamefont {D.}~\bibnamefont {Milo{\v s}evi{\'c}}}}
  (\bibinfo {year} {2004}),\ \bibfield  {title} {\enquote {\bibinfo {title}
  {Phase-controlled single-cycle strong-field photoionization},}\ }\href@noop
  {} {\bibfield  {journal} {\bibinfo  {journal} {Physica Scripta}\ }\textbf
  {\bibinfo {volume} {T110}},\ \bibinfo {pages} {120--125}}\BibitemShut
  {NoStop}%
\bibitem [{\citenamefont {Paulus}\ \emph {et~al.}(1994)\citenamefont {Paulus},
  \citenamefont {Nicklich}, \citenamefont {Huale}, \citenamefont
  {Lambropoulus},\ and\ \citenamefont {Walter}}]{Paulusplateau}%
  \BibitemOpen
  \bibfield  {author} {\bibinfo {author} {\bibnamefont {Paulus}, \bibfnamefont
  {G~G}}, \bibinfo {author} {\bibfnamefont {W.}~\bibnamefont {Nicklich}},
  \bibinfo {author} {\bibfnamefont {X.}~\bibnamefont {Huale}}, \bibinfo
  {author} {\bibfnamefont {P.}~\bibnamefont {Lambropoulus}}, \ and\ \bibinfo
  {author} {\bibfnamefont {H.}~\bibnamefont {Walter}}} (\bibinfo {year}
  {1994}),\ \bibfield  {title} {\enquote {\bibinfo {title} {Plateau in above
  threshold ionization spectra},}\ }\href@noop {} {\bibfield  {journal}
  {\bibinfo  {journal} {Phys. Rev. Lett.}\ }\textbf {\bibinfo {volume} {72}},\
  \bibinfo {pages} {2851}}\BibitemShut {NoStop}%
\bibitem [{\citenamefont {Pazourek}\ \emph {et~al.}(2015)\citenamefont
  {Pazourek}, \citenamefont {Nagele},\ and\ \citenamefont
  {Burgd\"orfer}}]{pazourek15}%
  \BibitemOpen
  \bibfield  {author} {\bibinfo {author} {\bibnamefont {Pazourek},
  \bibfnamefont {R}}, \bibinfo {author} {\bibfnamefont {S.}~\bibnamefont
  {Nagele}}, \ and\ \bibinfo {author} {\bibfnamefont {J.}~\bibnamefont
  {Burgd\"orfer}}} (\bibinfo {year} {2015}),\ \bibfield  {title} {\enquote
  {\bibinfo {title} {Attosecond chronoscopy of photoemission},}\ }\href@noop {}
  {\bibfield  {journal} {\bibinfo  {journal} {Rev. Mod. Phys.}\ }\textbf
  {\bibinfo {volume} {87}},\ \bibinfo {pages} {765}}\BibitemShut {NoStop}%
\bibitem [{\citenamefont {Perelomov}\ \emph {et~al.}(1966)\citenamefont
  {Perelomov}, \citenamefont {Popov},\ and\ \citenamefont
  {Terentev}}]{PPT1966}%
  \BibitemOpen
  \bibfield  {author} {\bibinfo {author} {\bibnamefont {Perelomov},
  \bibfnamefont {A~M}}, \bibinfo {author} {\bibfnamefont {V.~S.}\ \bibnamefont
  {Popov}}, \ and\ \bibinfo {author} {\bibfnamefont {M.~V.}\ \bibnamefont
  {Terentev}}} (\bibinfo {year} {1966}),\ \bibfield  {title} {\enquote
  {\bibinfo {title} {Ionization of atoms in an alternating electric field},}\
  }\href@noop {} {\bibfield  {journal} {\bibinfo  {journal} {Sov. Phys. JETP}\
  }\textbf {\bibinfo {volume} {23}},\ \bibinfo {pages} {924}}\BibitemShut
  {NoStop}%
\bibitem [{\citenamefont {P{\'e}rez-Hern{\'a}ndez}\ \emph
  {et~al.}(2013)\citenamefont {P{\'e}rez-Hern{\'a}ndez}, \citenamefont
  {Ciappina}, \citenamefont {Lewenstein}, \citenamefont {Roso},\ and\
  \citenamefont {Za{\"i}r}}]{Jose13}%
  \BibitemOpen
  \bibfield  {author} {\bibinfo {author} {\bibnamefont
  {P{\'e}rez-Hern{\'a}ndez}, \bibfnamefont {J~A}}, \bibinfo {author}
  {\bibfnamefont {M.~F.}\ \bibnamefont {Ciappina}}, \bibinfo {author}
  {\bibfnamefont {M.}~\bibnamefont {Lewenstein}}, \bibinfo {author}
  {\bibfnamefont {L.}~\bibnamefont {Roso}}, \ and\ \bibinfo {author}
  {\bibfnamefont {A.}~\bibnamefont {Za{\"i}r}}} (\bibinfo {year} {2013}),\
  \bibfield  {title} {\enquote {\bibinfo {title} {{B}eyond {C}arbon {K}-{E}dge
  {H}armonic {E}mission {U}sing a {S}patial and {T}emporal {S}ynthesized
  {L}aser {F}ield},}\ }\href@noop {} {\bibfield  {journal} {\bibinfo  {journal}
  {Phys. Rev. Lett.}\ }\textbf {\bibinfo {volume} {110}},\ \bibinfo {pages}
  {053001}}\BibitemShut {NoStop}%
\bibitem [{\citenamefont {P{\'e}rez-Hern{\'a}ndez}\ \emph
  {et~al.}(2009)\citenamefont {P{\'e}rez-Hern{\'a}ndez}, \citenamefont
  {Hoffmann}, \citenamefont {Za{\"i}r}, \citenamefont {Chipperfield},
  \citenamefont {Plaja}, \citenamefont {Ruiz}, \citenamefont {Marangos},\ and\
  \citenamefont {Roso}}]{Jose09A}%
  \BibitemOpen
  \bibfield  {author} {\bibinfo {author} {\bibnamefont
  {P{\'e}rez-Hern{\'a}ndez}, \bibfnamefont {J~A}}, \bibinfo {author}
  {\bibfnamefont {D.~J.}\ \bibnamefont {Hoffmann}}, \bibinfo {author}
  {\bibfnamefont {A.}~\bibnamefont {Za{\"i}r}}, \bibinfo {author}
  {\bibfnamefont {L.~E.}\ \bibnamefont {Chipperfield}}, \bibinfo {author}
  {\bibfnamefont {L.}~\bibnamefont {Plaja}}, \bibinfo {author} {\bibfnamefont
  {C.}~\bibnamefont {Ruiz}}, \bibinfo {author} {\bibfnamefont {J.~P.}\
  \bibnamefont {Marangos}}, \ and\ \bibinfo {author} {\bibfnamefont
  {L.}~\bibnamefont {Roso}}} (\bibinfo {year} {2009}),\ \bibfield  {title}
  {\enquote {\bibinfo {title} {Extension of the cut-off in high-harmonic
  generation using two delayed pulses of the same colour},}\ }\href@noop {}
  {\bibfield  {journal} {\bibinfo  {journal} {J. Phys. B}\ }\textbf {\bibinfo
  {volume} {42}},\ \bibinfo {pages} {134004}}\BibitemShut {NoStop}%
\bibitem [{\citenamefont {P\'erez-Hern\'andez}\ \emph
  {et~al.}(2011)\citenamefont {P\'erez-Hern\'andez}, \citenamefont {Roso},
  \citenamefont {Za\"ir},\ and\ \citenamefont {Plaja}}]{JoseNat}%
  \BibitemOpen
  \bibfield  {author} {\bibinfo {author} {\bibnamefont {P\'erez-Hern\'andez},
  \bibfnamefont {J~A}}, \bibinfo {author} {\bibfnamefont {L.}~\bibnamefont
  {Roso}}, \bibinfo {author} {\bibfnamefont {A.}~\bibnamefont {Za\"ir}}, \ and\
  \bibinfo {author} {\bibfnamefont {L.}~\bibnamefont {Plaja}}} (\bibinfo {year}
  {2011}),\ \bibfield  {title} {\enquote {\bibinfo {title} {Valley in the
  efficiency of the high-order harmonic yield at ultra-high laser
  intrensities},}\ }\href@noop {} {\bibfield  {journal} {\bibinfo  {journal}
  {Opt. Exp.}\ }\textbf {\bibinfo {volume} {19}},\ \bibinfo {pages}
  {19430--19439}}\BibitemShut {NoStop}%
\bibitem [{\citenamefont {Pfullmann}\ \emph {et~al.}(2013)\citenamefont
  {Pfullmann}, \citenamefont {Waltermann}, \citenamefont {Noack}, \citenamefont
  {Rausch}, \citenamefont {Nagy}, \citenamefont {Reinhardt}, \citenamefont
  {Kova\u{c}ev}, \citenamefont {Knittel}, \citenamefont {Bratschitsch},
  \citenamefont {Akemeier}, \citenamefont {H{\"u}tten}, \citenamefont
  {Leitenstorfer},\ and\ \citenamefont {Morgner}}]{Kovacev13NJP}%
  \BibitemOpen
  \bibfield  {author} {\bibinfo {author} {\bibnamefont {Pfullmann},
  \bibfnamefont {N}}, \bibinfo {author} {\bibfnamefont {C.}~\bibnamefont
  {Waltermann}}, \bibinfo {author} {\bibfnamefont {M.}~\bibnamefont {Noack}},
  \bibinfo {author} {\bibfnamefont {S.}~\bibnamefont {Rausch}}, \bibinfo
  {author} {\bibfnamefont {T.}~\bibnamefont {Nagy}}, \bibinfo {author}
  {\bibfnamefont {C.}~\bibnamefont {Reinhardt}}, \bibinfo {author}
  {\bibfnamefont {M.}~\bibnamefont {Kova\u{c}ev}}, \bibinfo {author}
  {\bibfnamefont {V.}~\bibnamefont {Knittel}}, \bibinfo {author} {\bibfnamefont
  {R.}~\bibnamefont {Bratschitsch}}, \bibinfo {author} {\bibfnamefont
  {D.}~\bibnamefont {Akemeier}}, \bibinfo {author} {\bibfnamefont
  {A.}~\bibnamefont {H{\"u}tten}}, \bibinfo {author} {\bibfnamefont
  {A.}~\bibnamefont {Leitenstorfer}}, \ and\ \bibinfo {author} {\bibfnamefont
  {U.}~\bibnamefont {Morgner}}} (\bibinfo {year} {2013}),\ \bibfield  {title}
  {\enquote {\bibinfo {title} {Bow-tie nano-antenna assisted generation of
  extreme ultraviolet radiation},}\ }\href@noop {} {\bibfield  {journal}
  {\bibinfo  {journal} {New J. Phys.}\ }\textbf {\bibinfo {volume} {15}},\
  \bibinfo {pages} {093027}}\BibitemShut {NoStop}%
\bibitem [{\citenamefont {Piglosiewicz}\ \emph {et~al.}(2014)\citenamefont
  {Piglosiewicz}, \citenamefont {Schmidt}, \citenamefont {Park}, \citenamefont
  {Vogelsang}, \citenamefont {Gross}, \citenamefont {Manzoni}, \citenamefont
  {Farinello}, \citenamefont {Cerullo},\ and\ \citenamefont
  {Lienau}}]{Piglosiewicz2014}%
  \BibitemOpen
  \bibfield  {author} {\bibinfo {author} {\bibnamefont {Piglosiewicz},
  \bibfnamefont {B}}, \bibinfo {author} {\bibfnamefont {S.}~\bibnamefont
  {Schmidt}}, \bibinfo {author} {\bibfnamefont {D.~J.}\ \bibnamefont {Park}},
  \bibinfo {author} {\bibfnamefont {J.}~\bibnamefont {Vogelsang}}, \bibinfo
  {author} {\bibfnamefont {P.}~\bibnamefont {Gross}}, \bibinfo {author}
  {\bibfnamefont {C.}~\bibnamefont {Manzoni}}, \bibinfo {author} {\bibfnamefont
  {P.}~\bibnamefont {Farinello}}, \bibinfo {author} {\bibfnamefont
  {G.}~\bibnamefont {Cerullo}}, \ and\ \bibinfo {author} {\bibfnamefont
  {C.}~\bibnamefont {Lienau}}} (\bibinfo {year} {2014}),\ \bibfield  {title}
  {\enquote {\bibinfo {title} {Carrier-envelope phase effects on the
  strong-field photoemission of electrons from metallic nanostructures},}\
  }\href@noop {} {\bibfield  {journal} {\bibinfo  {journal} {Nat. Phot.}\
  }\textbf {\bibinfo {volume} {8}},\ \bibinfo {pages} {37--42}}\BibitemShut
  {NoStop}%
\bibitem [{\citenamefont {Prinz}\ \emph {et~al.}(2015)\citenamefont {Prinz},
  \citenamefont {Haefner}, \citenamefont {Teisset}, \citenamefont {Bessing},
  \citenamefont {Michel}, \citenamefont {Lee}, \citenamefont {Geng},
  \citenamefont {Kim}, \citenamefont {Kim}, \citenamefont {Metzger},\ and\
  \citenamefont {Schultze}}]{prinz_cep_stable_2015}%
  \BibitemOpen
  \bibfield  {author} {\bibinfo {author} {\bibnamefont {Prinz}, \bibfnamefont
  {S}}, \bibinfo {author} {\bibfnamefont {M.}~\bibnamefont {Haefner}}, \bibinfo
  {author} {\bibfnamefont {C.~Y.}\ \bibnamefont {Teisset}}, \bibinfo {author}
  {\bibfnamefont {R.}~\bibnamefont {Bessing}}, \bibinfo {author} {\bibfnamefont
  {K.}~\bibnamefont {Michel}}, \bibinfo {author} {\bibfnamefont
  {Y.}~\bibnamefont {Lee}}, \bibinfo {author} {\bibfnamefont {X.~T.}\
  \bibnamefont {Geng}}, \bibinfo {author} {\bibfnamefont {S.}~\bibnamefont
  {Kim}}, \bibinfo {author} {\bibfnamefont {D.~E.}\ \bibnamefont {Kim}},
  \bibinfo {author} {\bibfnamefont {T.}~\bibnamefont {Metzger}}, \ and\
  \bibinfo {author} {\bibfnamefont {M.}~\bibnamefont {Schultze}}} (\bibinfo
  {year} {2015}),\ \bibfield  {title} {\enquote {\bibinfo {title} {Cep-stable,
  sub-6 fs, 300-khz opcpa system with more than 15 w of average power},}\
  }\href@noop {} {\bibfield  {journal} {\bibinfo  {journal} {Opt. Exp.}\
  }\textbf {\bibinfo {volume} {23}},\ \bibinfo {pages}
  {1388--1394}}\BibitemShut {NoStop}%
\bibitem [{\citenamefont {Pullen}\ \emph {et~al.}(2015)\citenamefont {Pullen},
  \citenamefont {Wolter}, \citenamefont {Le}, \citenamefont {Baudisch},
  \citenamefont {Hemmer}, \citenamefont {Senftleben}, \citenamefont
  {Schr\"oter}, \citenamefont {Ullrich}, \citenamefont {Moshammer},
  \citenamefont {Lin},\ and\ \citenamefont {Biegert}}]{mick2015}%
  \BibitemOpen
  \bibfield  {author} {\bibinfo {author} {\bibnamefont {Pullen}, \bibfnamefont
  {M~G}}, \bibinfo {author} {\bibfnamefont {B.}~\bibnamefont {Wolter}},
  \bibinfo {author} {\bibfnamefont {A-T.}\ \bibnamefont {Le}}, \bibinfo
  {author} {\bibfnamefont {M.}~\bibnamefont {Baudisch}}, \bibinfo {author}
  {\bibfnamefont {M.}~\bibnamefont {Hemmer}}, \bibinfo {author} {\bibfnamefont
  {A.}~\bibnamefont {Senftleben}}, \bibinfo {author} {\bibfnamefont {C.~D.}\
  \bibnamefont {Schr\"oter}}, \bibinfo {author} {\bibfnamefont
  {J.}~\bibnamefont {Ullrich}}, \bibinfo {author} {\bibfnamefont
  {R.}~\bibnamefont {Moshammer}}, \bibinfo {author} {\bibfnamefont {C.~D.}\
  \bibnamefont {Lin}}, \ and\ \bibinfo {author} {\bibfnamefont
  {J.}~\bibnamefont {Biegert}}} (\bibinfo {year} {2015}),\ \bibfield  {title}
  {\enquote {\bibinfo {title} {Imaging an aligned polyatomic molecule with
  laser-induced electron diffraction},}\ }\href@noop {} {\bibfield  {journal}
  {\bibinfo  {journal} {Nat. Comm.}\ }\textbf {\bibinfo {volume} {6}},\
  \bibinfo {pages} {7262}}\BibitemShut {NoStop}%
\bibitem [{\citenamefont {Qin}\ \emph {et~al.}(2010)\citenamefont {Qin},
  \citenamefont {Xia},\ and\ \citenamefont {Whitesides}}]{qin_soft_2010}%
  \BibitemOpen
  \bibfield  {author} {\bibinfo {author} {\bibnamefont {Qin}, \bibfnamefont
  {D}}, \bibinfo {author} {\bibfnamefont {Y.}~\bibnamefont {Xia}}, \ and\
  \bibinfo {author} {\bibfnamefont {G.~M.}\ \bibnamefont {Whitesides}}}
  (\bibinfo {year} {2010}),\ \bibfield  {title} {\enquote {\bibinfo {title}
  {Soft lithography for micro- and nanoscale patterning},}\ }\href@noop {}
  {\bibfield  {journal} {\bibinfo  {journal} {Nat. Prot.}\ }\textbf {\bibinfo
  {volume} {5}},\ \bibinfo {pages} {491--502}}\BibitemShut {NoStop}%
\bibitem [{\citenamefont {Quinonez}\ \emph {et~al.}(2013)\citenamefont
  {Quinonez}, \citenamefont {Handali},\ and\ \citenamefont
  {Barwick}}]{Quinonez2013}%
  \BibitemOpen
  \bibfield  {author} {\bibinfo {author} {\bibnamefont {Quinonez},
  \bibfnamefont {E}}, \bibinfo {author} {\bibfnamefont {J.}~\bibnamefont
  {Handali}}, \ and\ \bibinfo {author} {\bibfnamefont {B.}~\bibnamefont
  {Barwick}}} (\bibinfo {year} {2013}),\ \bibfield  {title} {\enquote {\bibinfo
  {title} {Femtosecond photoelectron point projection microscope},}\
  }\href@noop {} {\bibfield  {journal} {\bibinfo  {journal} {Rev. Sci. Instr.}\
  }\textbf {\bibinfo {volume} {84}},\ \bibinfo {pages} {103710}}\BibitemShut
  {NoStop}%
\bibitem [{\citenamefont {Rathje}\ \emph {et~al.}(2012)\citenamefont {Rathje},
  \citenamefont {Johnson}, \citenamefont {M\"uller}, \citenamefont
  {S{\"u}{\ss}mann}, \citenamefont {Adolph}, \citenamefont {K{\"u}bel},
  \citenamefont {Kienberger}, \citenamefont {Kling}, \citenamefont {Paulus},\
  and\ \citenamefont {Sayler}}]{Rathje12}%
  \BibitemOpen
  \bibfield  {author} {\bibinfo {author} {\bibnamefont {Rathje}, \bibfnamefont
  {T}}, \bibinfo {author} {\bibfnamefont {N.~G.}\ \bibnamefont {Johnson}},
  \bibinfo {author} {\bibfnamefont {M.}~\bibnamefont {M\"uller}}, \bibinfo
  {author} {\bibfnamefont {F.}~\bibnamefont {S{\"u}{\ss}mann}}, \bibinfo
  {author} {\bibfnamefont {D.}~\bibnamefont {Adolph}}, \bibinfo {author}
  {\bibfnamefont {M.}~\bibnamefont {K{\"u}bel}}, \bibinfo {author}
  {\bibfnamefont {R.}~\bibnamefont {Kienberger}}, \bibinfo {author}
  {\bibfnamefont {M.~F.}\ \bibnamefont {Kling}}, \bibinfo {author}
  {\bibfnamefont {G.~G.}\ \bibnamefont {Paulus}}, \ and\ \bibinfo {author}
  {\bibfnamefont {A.~M.}\ \bibnamefont {Sayler}}} (\bibinfo {year} {2012}),\
  \bibfield  {title} {\enquote {\bibinfo {title} {Review of attosecond resolved
  measurement and control via carrier-envelope phase tagging with
  above-threshold ionization},}\ }\href@noop {} {\bibfield  {journal} {\bibinfo
   {journal} {J. Phys. B}\ }\textbf {\bibinfo {volume} {45}},\ \bibinfo {pages}
  {074003}}\BibitemShut {NoStop}%
\bibitem [{\citenamefont {Reiss}(1980)}]{Reiss}%
  \BibitemOpen
  \bibfield  {author} {\bibinfo {author} {\bibnamefont {Reiss}, \bibfnamefont
  {H~R}}} (\bibinfo {year} {1980}),\ \bibfield  {title} {\enquote {\bibinfo
  {title} {Effect of an intense electromagnetic field on a weakly bound
  system},}\ }\href@noop {} {\bibfield  {journal} {\bibinfo  {journal} {Phys.
  Rev. A}\ }\textbf {\bibinfo {volume} {22}},\ \bibinfo {pages}
  {1786}}\BibitemShut {NoStop}%
\bibitem [{\citenamefont {Rewitz}\ \emph {et~al.}(2012)\citenamefont {Rewitz},
  \citenamefont {Keitzl}, \citenamefont {Tuchscherer}, \citenamefont {Huang},
  \citenamefont {Geisler}, \citenamefont {Razinskas}, \citenamefont {Hecht},\
  and\ \citenamefont {Brixner}}]{Rewitz12}%
  \BibitemOpen
  \bibfield  {author} {\bibinfo {author} {\bibnamefont {Rewitz}, \bibfnamefont
  {C}}, \bibinfo {author} {\bibfnamefont {T.}~\bibnamefont {Keitzl}}, \bibinfo
  {author} {\bibfnamefont {P.}~\bibnamefont {Tuchscherer}}, \bibinfo {author}
  {\bibfnamefont {J.-S.}\ \bibnamefont {Huang}}, \bibinfo {author}
  {\bibfnamefont {P.}~\bibnamefont {Geisler}}, \bibinfo {author} {\bibfnamefont
  {G.}~\bibnamefont {Razinskas}}, \bibinfo {author} {\bibfnamefont
  {B.}~\bibnamefont {Hecht}}, \ and\ \bibinfo {author} {\bibfnamefont
  {T.}~\bibnamefont {Brixner}}} (\bibinfo {year} {2012}),\ \bibfield  {title}
  {\enquote {\bibinfo {title} {Few-femtosecond plasmon dephasing of a single
  metallic nanostructure from optical response function reconstruction by
  interferometric frequency resolved optical gating},}\ }\href@noop {}
  {\bibfield  {journal} {\bibinfo  {journal} {Nano Lett.}\ }\textbf {\bibinfo
  {volume} {12}},\ \bibinfo {pages} {45--49}}\BibitemShut {NoStop}%
\bibitem [{\citenamefont {Robinson}\ \emph {et~al.}(2006)\citenamefont
  {Robinson}, \citenamefont {Haworth}, \citenamefont {Teng}, \citenamefont
  {Smith}, \citenamefont {Marangos},\ and\ \citenamefont
  {Tisch}}]{robinson_generation_2006}%
  \BibitemOpen
  \bibfield  {author} {\bibinfo {author} {\bibnamefont {Robinson},
  \bibfnamefont {J~S}}, \bibinfo {author} {\bibfnamefont {C.~A.}\ \bibnamefont
  {Haworth}}, \bibinfo {author} {\bibfnamefont {H.}~\bibnamefont {Teng}},
  \bibinfo {author} {\bibfnamefont {R.~A.}\ \bibnamefont {Smith}}, \bibinfo
  {author} {\bibfnamefont {J.~P.}\ \bibnamefont {Marangos}}, \ and\ \bibinfo
  {author} {\bibfnamefont {J.~W.~G.}\ \bibnamefont {Tisch}}} (\bibinfo {year}
  {2006}),\ \bibfield  {title} {\enquote {\bibinfo {title} {The generation of
  intense, transform-limited laser pulses with tunable duration from 6 to 30 fs
  in a differentially pumped hollow fibre},}\ }\href@noop {} {\bibfield
  {journal} {\bibinfo  {journal} {Appl. Phys. B}\ }\textbf {\bibinfo {volume}
  {85}},\ \bibinfo {pages} {525--529}}\BibitemShut {NoStop}%
\bibitem [{\citenamefont {Ropers}\ \emph
  {et~al.}(2007{\natexlab{a}})\citenamefont {Ropers}, \citenamefont
  {Els{\"a}sser}, \citenamefont {Cerullo}, \citenamefont {Zavelani-Rossi},\
  and\ \citenamefont {Lienau}}]{Ropers07}%
  \BibitemOpen
  \bibfield  {author} {\bibinfo {author} {\bibnamefont {Ropers}, \bibfnamefont
  {C}}, \bibinfo {author} {\bibfnamefont {T.}~\bibnamefont {Els{\"a}sser}},
  \bibinfo {author} {\bibfnamefont {G.}~\bibnamefont {Cerullo}}, \bibinfo
  {author} {\bibfnamefont {M.}~\bibnamefont {Zavelani-Rossi}}, \ and\ \bibinfo
  {author} {\bibfnamefont {C.}~\bibnamefont {Lienau}}} (\bibinfo {year}
  {2007}{\natexlab{a}}),\ \bibfield  {title} {\enquote {\bibinfo {title}
  {Ultrafast optical excitations of metallic nanostructures: From light
  confinement to a novel electron source},}\ }\href@noop {} {\bibfield
  {journal} {\bibinfo  {journal} {New J. Phys.}\ }\textbf {\bibinfo {volume}
  {9}},\ \bibinfo {pages} {397}}\BibitemShut {NoStop}%
\bibitem [{\citenamefont {Ropers}\ \emph
  {et~al.}(2007{\natexlab{b}})\citenamefont {Ropers}, \citenamefont {Solli},
  \citenamefont {Schulz}, \citenamefont {Lienau},\ and\ \citenamefont
  {Els{\"a}sser}}]{Ropers07PRL}%
  \BibitemOpen
  \bibfield  {author} {\bibinfo {author} {\bibnamefont {Ropers}, \bibfnamefont
  {C}}, \bibinfo {author} {\bibfnamefont {D.~R.}\ \bibnamefont {Solli}},
  \bibinfo {author} {\bibfnamefont {C-P.}\ \bibnamefont {Schulz}}, \bibinfo
  {author} {\bibfnamefont {C.}~\bibnamefont {Lienau}}, \ and\ \bibinfo {author}
  {\bibfnamefont {T.}~\bibnamefont {Els{\"a}sser}}} (\bibinfo {year}
  {2007}{\natexlab{b}}),\ \bibfield  {title} {\enquote {\bibinfo {title}
  {Localized multiphoton emission of femtosecond electron pulses from metal
  nanotips},}\ }\href@noop {} {\bibfield  {journal} {\bibinfo  {journal} {Phys.
  Rev. Lett.}\ }\textbf {\bibinfo {volume} {98}},\ \bibinfo {pages}
  {043907}}\BibitemShut {NoStop}%
\bibitem [{\citenamefont {Sali\`eres}\ \emph {et~al.}(2001)\citenamefont
  {Sali\`eres}, \citenamefont {Carr\'e}, \citenamefont {Le~D\'eroff},
  \citenamefont {Grasbon}, \citenamefont {Paulus}, \citenamefont {Walther},
  \citenamefont {Kopold}, \citenamefont {Becker}, \citenamefont {Milosevic},
  \citenamefont {Sanpera},\ and\ \citenamefont {Lewenstein}}]{salieres2001}%
  \BibitemOpen
  \bibfield  {author} {\bibinfo {author} {\bibnamefont {Sali\`eres},
  \bibfnamefont {P}}, \bibinfo {author} {\bibfnamefont {B.}~\bibnamefont
  {Carr\'e}}, \bibinfo {author} {\bibfnamefont {L.}~\bibnamefont
  {Le~D\'eroff}}, \bibinfo {author} {\bibfnamefont {F.}~\bibnamefont
  {Grasbon}}, \bibinfo {author} {\bibfnamefont {G.~G.}\ \bibnamefont {Paulus}},
  \bibinfo {author} {\bibfnamefont {H.}~\bibnamefont {Walther}}, \bibinfo
  {author} {\bibfnamefont {R.}~\bibnamefont {Kopold}}, \bibinfo {author}
  {\bibfnamefont {W.}~\bibnamefont {Becker}}, \bibinfo {author} {\bibfnamefont
  {D.~B.}\ \bibnamefont {Milosevic}}, \bibinfo {author} {\bibfnamefont
  {A.}~\bibnamefont {Sanpera}}, \ and\ \bibinfo {author} {\bibfnamefont
  {M.}~\bibnamefont {Lewenstein}}} (\bibinfo {year} {2001}),\ \bibfield
  {title} {\enquote {\bibinfo {title} {Feynman's path-integral approach for
  intense-laser-atom interactions},}\ }\href@noop {} {\bibfield  {journal}
  {\bibinfo  {journal} {Science}\ }\textbf {\bibinfo {volume} {292}},\ \bibinfo
  {pages} {902--905}}\BibitemShut {NoStop}%
\bibitem [{\citenamefont {Sali\`eres}\ \emph {et~al.}(1999)\citenamefont
  {Sali\`eres}, \citenamefont {L'Huillier}, \citenamefont {Antoine},\ and\
  \citenamefont {Lewenstein}}]{Salieres-adv}%
  \BibitemOpen
  \bibfield  {author} {\bibinfo {author} {\bibnamefont {Sali\`eres},
  \bibfnamefont {P}}, \bibinfo {author} {\bibfnamefont {A.}~\bibnamefont
  {L'Huillier}}, \bibinfo {author} {\bibfnamefont {P.}~\bibnamefont {Antoine}},
  \ and\ \bibinfo {author} {\bibfnamefont {M.}~\bibnamefont {Lewenstein}}}
  (\bibinfo {year} {1999}),\ \bibfield  {title} {\enquote {\bibinfo {title}
  {Study of the spatial and temporal coherence of high-order harmonics},}\ }in\
  \href@noop {} {\emph {\bibinfo {booktitle} {Advances in Atomic, Molecular and
  Optical Physics. Vol. 41}}},\ \bibinfo {editor} {edited by\ \bibinfo {editor}
  {\bibfnamefont {B.}~\bibnamefont {Bederson}}\ and\ \bibinfo {editor}
  {\bibfnamefont {H.}~\bibnamefont {Walter}}}\ (\bibinfo  {publisher} {Academic
  Press},\ \bibinfo {address} {San Diego})\ pp.\ \bibinfo {pages}
  {83--142}\BibitemShut {NoStop}%
\bibitem [{\citenamefont {Salvat}\ \emph {et~al.}(2005)\citenamefont {Salvat},
  \citenamefont {Jablonski},\ and\ \citenamefont {Powell}}]{Salvat2005}%
  \BibitemOpen
  \bibfield  {author} {\bibinfo {author} {\bibnamefont {Salvat}, \bibfnamefont
  {F}}, \bibinfo {author} {\bibfnamefont {A.}~\bibnamefont {Jablonski}}, \ and\
  \bibinfo {author} {\bibfnamefont {C.~J.}\ \bibnamefont {Powell}}} (\bibinfo
  {year} {2005}),\ \bibfield  {title} {\enquote {\bibinfo {title}
  {elsepa---dirac partial-wave calculation of elastic scattering of electrons
  and positrons by atoms, positive ions and molecules},}\ }\href@noop {}
  {\bibfield  {journal} {\bibinfo  {journal} {Comput. Phys. Commun.}\ }\textbf
  {\bibinfo {volume} {165}},\ \bibinfo {pages} {157 -- 190}}\BibitemShut
  {NoStop}%
\bibitem [{\citenamefont {Sansone}\ \emph {et~al.}(2006)\citenamefont
  {Sansone}, \citenamefont {Benedetti}, \citenamefont {Calegari}, \citenamefont
  {Vozzi}, \citenamefont {Avaldi}, \citenamefont {Flammini}, \citenamefont
  {Poletto}, \citenamefont {Villoresi}, \citenamefont {Altucci}, \citenamefont
  {Velotta}, \citenamefont {Stagira}, \citenamefont {De~Silvestri},\ and\
  \citenamefont {Nisoli}}]{sansone2006isolated}%
  \BibitemOpen
  \bibfield  {author} {\bibinfo {author} {\bibnamefont {Sansone}, \bibfnamefont
  {G}}, \bibinfo {author} {\bibfnamefont {E.}~\bibnamefont {Benedetti}},
  \bibinfo {author} {\bibfnamefont {F.}~\bibnamefont {Calegari}}, \bibinfo
  {author} {\bibfnamefont {C.}~\bibnamefont {Vozzi}}, \bibinfo {author}
  {\bibfnamefont {L.}~\bibnamefont {Avaldi}}, \bibinfo {author} {\bibfnamefont
  {R.}~\bibnamefont {Flammini}}, \bibinfo {author} {\bibfnamefont
  {L.}~\bibnamefont {Poletto}}, \bibinfo {author} {\bibfnamefont
  {P.}~\bibnamefont {Villoresi}}, \bibinfo {author} {\bibfnamefont
  {C.}~\bibnamefont {Altucci}}, \bibinfo {author} {\bibfnamefont
  {R.}~\bibnamefont {Velotta}}, \bibinfo {author} {\bibfnamefont
  {S.}~\bibnamefont {Stagira}}, \bibinfo {author} {\bibfnamefont
  {S.}~\bibnamefont {De~Silvestri}}, \ and\ \bibinfo {author} {\bibfnamefont
  {M.}~\bibnamefont {Nisoli}}} (\bibinfo {year} {2006}),\ \bibfield  {title}
  {\enquote {\bibinfo {title} {Isolated single-cycle attosecond pulses},}\
  }\href@noop {} {\bibfield  {journal} {\bibinfo  {journal} {Science}\ }\textbf
  {\bibinfo {volume} {314}},\ \bibinfo {pages} {443--446}}\BibitemShut
  {NoStop}%
\bibitem [{\citenamefont {Santra}\ and\ \citenamefont
  {Gordon}(2006)}]{Santra2006}%
  \BibitemOpen
  \bibfield  {author} {\bibinfo {author} {\bibnamefont {Santra}, \bibfnamefont
  {R}}, \ and\ \bibinfo {author} {\bibfnamefont {A.}~\bibnamefont {Gordon}}}
  (\bibinfo {year} {2006}),\ \bibfield  {title} {\enquote {\bibinfo {title}
  {Three-step model for high-harmonic generation in many-electron systems},}\
  }\href@noop {} {\bibfield  {journal} {\bibinfo  {journal} {Phys. Rev. Lett.}\
  }\textbf {\bibinfo {volume} {96}},\ \bibinfo {pages} {073906}}\BibitemShut
  {NoStop}%
\bibitem [{\citenamefont {Sarid}\ and\ \citenamefont
  {Challener}(2010)}]{Sarid2010}%
  \BibitemOpen
  \bibfield  {author} {\bibinfo {author} {\bibnamefont {Sarid}, \bibfnamefont
  {Dror}}, \ and\ \bibinfo {author} {\bibfnamefont {William}\ \bibnamefont
  {Challener}}} (\bibinfo {year} {2010}),\ \href@noop {} {\emph {\bibinfo
  {title} {Modern {I}ntroduction to {S}urface {P}lasmons}}}\ (\bibinfo
  {publisher} {Cambridge University Press})\BibitemShut {NoStop}%
\bibitem [{\citenamefont {Sau}\ and\ \citenamefont {Murphy}(2004)}]{Sau04}%
  \BibitemOpen
  \bibfield  {author} {\bibinfo {author} {\bibnamefont {Sau}, \bibfnamefont
  {T~K}}, \ and\ \bibinfo {author} {\bibfnamefont {C.~J.}\ \bibnamefont
  {Murphy}}} (\bibinfo {year} {2004}),\ \bibfield  {title} {\enquote {\bibinfo
  {title} {Room temperature, high-yield synthesis of multiple shapes of gold
  nanoparticles in aqueous solution},}\ }\href@noop {} {\bibfield  {journal}
  {\bibinfo  {journal} {J. Am. Chem. Soc.}\ }\textbf {\bibinfo {volume}
  {126}},\ \bibinfo {pages} {8648--8649}}\BibitemShut {NoStop}%
\bibitem [{\citenamefont {Savage}\ \emph {et~al.}(2012)\citenamefont {Savage},
  \citenamefont {Hawkeye}, \citenamefont {Esteban}, \citenamefont {Borisov},
  \citenamefont {Aizpurua},\ and\ \citenamefont {Baumberg}}]{Savage2012}%
  \BibitemOpen
  \bibfield  {author} {\bibinfo {author} {\bibnamefont {Savage}, \bibfnamefont
  {K~J}}, \bibinfo {author} {\bibfnamefont {M.~M.}\ \bibnamefont {Hawkeye}},
  \bibinfo {author} {\bibfnamefont {R.}~\bibnamefont {Esteban}}, \bibinfo
  {author} {\bibfnamefont {A.~G.}\ \bibnamefont {Borisov}}, \bibinfo {author}
  {\bibfnamefont {J.}~\bibnamefont {Aizpurua}}, \ and\ \bibinfo {author}
  {\bibfnamefont {J.~J.}\ \bibnamefont {Baumberg}}} (\bibinfo {year} {2012}),\
  \bibfield  {title} {\enquote {\bibinfo {title} {Revealing the quantum regime
  in tunnelling plasmonics},}\ }\href@noop {} {\bibfield  {journal} {\bibinfo
  {journal} {Nature}\ }\textbf {\bibinfo {volume} {491}},\ \bibinfo {pages}
  {574--577}}\BibitemShut {NoStop}%
\bibitem [{\citenamefont {Sayler}\ \emph {et~al.}(2011)\citenamefont {Sayler},
  \citenamefont {Rathje}, \citenamefont {M\"uller}, \citenamefont {R\"uhle},
  \citenamefont {Kienberger},\ and\ \citenamefont {Paulus}}]{paulus2011}%
  \BibitemOpen
  \bibfield  {author} {\bibinfo {author} {\bibnamefont {Sayler}, \bibfnamefont
  {A~M}}, \bibinfo {author} {\bibfnamefont {T.}~\bibnamefont {Rathje}},
  \bibinfo {author} {\bibfnamefont {W.}~\bibnamefont {M\"uller}}, \bibinfo
  {author} {\bibfnamefont {K.}~\bibnamefont {R\"uhle}}, \bibinfo {author}
  {\bibfnamefont {R.}~\bibnamefont {Kienberger}}, \ and\ \bibinfo {author}
  {\bibfnamefont {G.~G.}\ \bibnamefont {Paulus}}} (\bibinfo {year} {2011}),\
  \bibfield  {title} {\enquote {\bibinfo {title} {Precise, real-time,
  every-single-shot, carrier-envelope phase measurement of ultrashort laser
  pulses},}\ }\href@noop {} {\bibfield  {journal} {\bibinfo  {journal} {Opt.
  Lett}\ }\textbf {\bibinfo {volume} {36}},\ \bibinfo {pages}
  {1--3}}\BibitemShut {NoStop}%
\bibitem [{\citenamefont {Schafer}(1991)}]{schaferwop1}%
  \BibitemOpen
  \bibfield  {author} {\bibinfo {author} {\bibnamefont {Schafer}, \bibfnamefont
  {K~J}}} (\bibinfo {year} {1991}),\ \bibfield  {title} {\enquote {\bibinfo
  {title} {The energy analysis of time-dependent numerical wave functions},}\
  }\href@noop {} {\bibfield  {journal} {\bibinfo  {journal} {Comp. Phys.
  Comm.}\ }\textbf {\bibinfo {volume} {63}},\ \bibinfo {pages}
  {427--434}}\BibitemShut {NoStop}%
\bibitem [{\citenamefont {Schafer}(2009)}]{schaferwop2}%
  \BibitemOpen
  \bibfield  {author} {\bibinfo {author} {\bibnamefont {Schafer}, \bibfnamefont
  {K~J}}} (\bibinfo {year} {2009}),\ \bibfield  {title} {\enquote {\bibinfo
  {title} {Numerical methods in strong field physics},}\ }in\ \href@noop {}
  {\emph {\bibinfo {booktitle} {Strong Field Laser Physics}}},\ \bibinfo
  {editor} {edited by\ \bibinfo {editor} {\bibfnamefont {T.}~\bibnamefont
  {Brabec}}}\ (\bibinfo  {publisher} {Springer},\ \bibinfo {address} {New
  York})\ pp.\ \bibinfo {pages} {111--145}\BibitemShut {NoStop}%
\bibitem [{\citenamefont {Schafer}\ \emph {et~al.}(1993)\citenamefont
  {Schafer}, \citenamefont {B.Yang}, \citenamefont {DiMauro},\ and\
  \citenamefont {Kulander}}]{schafer93}%
  \BibitemOpen
  \bibfield  {author} {\bibinfo {author} {\bibnamefont {Schafer}, \bibfnamefont
  {K~J}}, \bibinfo {author} {\bibnamefont {B.Yang}}, \bibinfo {author}
  {\bibfnamefont {L.~F.}\ \bibnamefont {DiMauro}}, \ and\ \bibinfo {author}
  {\bibfnamefont {K.~C.}\ \bibnamefont {Kulander}}} (\bibinfo {year} {1993}),\
  \bibfield  {title} {\enquote {\bibinfo {title} {Above threshold ionization
  beyond the high harmonic cutoff},}\ }\href@noop {} {\bibfield  {journal}
  {\bibinfo  {journal} {Phys. Rev. Lett.}\ }\textbf {\bibinfo {volume} {70}},\
  \bibinfo {pages} {1599}}\BibitemShut {NoStop}%
\bibitem [{\citenamefont {Schafer}\ and\ \citenamefont
  {Kulander}(1990)}]{schaferwop}%
  \BibitemOpen
  \bibfield  {author} {\bibinfo {author} {\bibnamefont {Schafer}, \bibfnamefont
  {K~J}}, \ and\ \bibinfo {author} {\bibfnamefont {K.~C.}\ \bibnamefont
  {Kulander}}} (\bibinfo {year} {1990}),\ \bibfield  {title} {\enquote
  {\bibinfo {title} {Energy analysis of time-dependent wave functions:
  Application to above-threshold ionization},}\ }\href@noop {} {\bibfield
  {journal} {\bibinfo  {journal} {Phys. Rev. A}\ }\textbf {\bibinfo {volume}
  {42}},\ \bibinfo {pages} {5794(R)}}\BibitemShut {NoStop}%
\bibitem [{\citenamefont {Schenk}\ \emph {et~al.}(2010)\citenamefont {Schenk},
  \citenamefont {Kr{\"u}ger},\ and\ \citenamefont {Hommelhoff}}]{Schenk10}%
  \BibitemOpen
  \bibfield  {author} {\bibinfo {author} {\bibnamefont {Schenk}, \bibfnamefont
  {M}}, \bibinfo {author} {\bibfnamefont {M.}~\bibnamefont {Kr{\"u}ger}}, \
  and\ \bibinfo {author} {\bibfnamefont {P.}~\bibnamefont {Hommelhoff}}}
  (\bibinfo {year} {2010}),\ \bibfield  {title} {\enquote {\bibinfo {title}
  {Strong-field above-threshold photoemission from sharp metal tips},}\
  }\href@noop {} {\bibfield  {journal} {\bibinfo  {journal} {Phys. Rev. Lett.}\
  }\textbf {\bibinfo {volume} {105}},\ \bibinfo {pages} {257601}}\BibitemShut
  {NoStop}%
\bibitem [{\citenamefont {Schenk}\ \emph {et~al.}(2011)\citenamefont {Schenk},
  \citenamefont {Kr\"uger},\ and\ \citenamefont {Hommelhoff}}]{Schenk2011}%
  \BibitemOpen
  \bibfield  {author} {\bibinfo {author} {\bibnamefont {Schenk}, \bibfnamefont
  {M}}, \bibinfo {author} {\bibfnamefont {M.}~\bibnamefont {Kr\"uger}}, \ and\
  \bibinfo {author} {\bibfnamefont {P.}~\bibnamefont {Hommelhoff}}} (\bibinfo
  {year} {2011}),\ \bibfield  {title} {\enquote {\bibinfo {title}
  {Carrier-envelope phase dependent photoemission from a nanometric metal
  tip},}\ }in\ \href@noop {} {\emph {\bibinfo {booktitle} {Frequency Control
  and the European Frequency and Time Forum (FCS), 2011 Joint Conference of the
  IEEE International}}}\ (\bibinfo  {publisher} {IEEE})\ pp.\ \bibinfo {pages}
  {1--3}\BibitemShut {NoStop}%
\bibitem [{\citenamefont {Schertz}\ \emph {et~al.}(2012)\citenamefont
  {Schertz}, \citenamefont {Schmelzeisen}, \citenamefont {Kreiter},
  \citenamefont {Elmers},\ and\ \citenamefont {Sch\"onhense}}]{Schertz2012}%
  \BibitemOpen
  \bibfield  {author} {\bibinfo {author} {\bibnamefont {Schertz}, \bibfnamefont
  {F}}, \bibinfo {author} {\bibfnamefont {M.}~\bibnamefont {Schmelzeisen}},
  \bibinfo {author} {\bibfnamefont {M.}~\bibnamefont {Kreiter}}, \bibinfo
  {author} {\bibfnamefont {H.-J.}\ \bibnamefont {Elmers}}, \ and\ \bibinfo
  {author} {\bibfnamefont {G.}~\bibnamefont {Sch\"onhense}}} (\bibinfo {year}
  {2012}),\ \bibfield  {title} {\enquote {\bibinfo {title} {Field emission of
  electrons generated by the near field of strongly coupled plasmons},}\
  }\href@noop {} {\bibfield  {journal} {\bibinfo  {journal} {Phys. Rev. Lett}\
  }\textbf {\bibinfo {volume} {108}},\ \bibinfo {pages} {237602}}\BibitemShut
  {NoStop}%
\bibitem [{\citenamefont {Schiffrin}\ \emph {et~al.}(2013)\citenamefont
  {Schiffrin}, \citenamefont {Paasch-Colberg}, \citenamefont {Karpowicz},
  \citenamefont {Apalkov}, \citenamefont {Gerster}, \citenamefont
  {M\"uhlbrandt}, \citenamefont {Korbman}, \citenamefont {Reichert},
  \citenamefont {Schultze}, \citenamefont {Holzner}, \citenamefont {Barth},
  \citenamefont {Kienberger}, \citenamefont {Ernstorfer}, \citenamefont
  {Yakovlev}, \citenamefont {Stockman},\ and\ \citenamefont
  {Krausz}}]{schiffrin2013}%
  \BibitemOpen
  \bibfield  {author} {\bibinfo {author} {\bibnamefont {Schiffrin},
  \bibfnamefont {A}}, \bibinfo {author} {\bibfnamefont {T.}~\bibnamefont
  {Paasch-Colberg}}, \bibinfo {author} {\bibfnamefont {N.}~\bibnamefont
  {Karpowicz}}, \bibinfo {author} {\bibfnamefont {V.}~\bibnamefont {Apalkov}},
  \bibinfo {author} {\bibfnamefont {D.}~\bibnamefont {Gerster}}, \bibinfo
  {author} {\bibfnamefont {S.}~\bibnamefont {M\"uhlbrandt}}, \bibinfo {author}
  {\bibfnamefont {M.}~\bibnamefont {Korbman}}, \bibinfo {author} {\bibfnamefont
  {J.}~\bibnamefont {Reichert}}, \bibinfo {author} {\bibfnamefont
  {M.}~\bibnamefont {Schultze}}, \bibinfo {author} {\bibfnamefont
  {S.}~\bibnamefont {Holzner}}, \bibinfo {author} {\bibfnamefont {J.~V.}\
  \bibnamefont {Barth}}, \bibinfo {author} {\bibfnamefont {R.}~\bibnamefont
  {Kienberger}}, \bibinfo {author} {\bibfnamefont {R.}~\bibnamefont
  {Ernstorfer}}, \bibinfo {author} {\bibfnamefont {V.~S.}\ \bibnamefont
  {Yakovlev}}, \bibinfo {author} {\bibfnamefont {M.~I.}\ \bibnamefont
  {Stockman}}, \ and\ \bibinfo {author} {\bibfnamefont {F.}~\bibnamefont
  {Krausz}}} (\bibinfo {year} {2013}),\ \bibfield  {title} {\enquote {\bibinfo
  {title} {Optical-field-induced current in dielectrics},}\ }\href@noop {}
  {\bibfield  {journal} {\bibinfo  {journal} {Nature}\ }\textbf {\bibinfo
  {volume} {493}},\ \bibinfo {pages} {70--74}}\BibitemShut {NoStop}%
\bibitem [{\citenamefont {Schmidt}\ \emph {et~al.}(2011)\citenamefont
  {Schmidt}, \citenamefont {Shiner}, \citenamefont {Lassonde}, \citenamefont
  {Kieffer}, \citenamefont {Corkum}, \citenamefont {Villeneuve},\ and\
  \citenamefont {L\'egar\'e}}]{schmidt_cep_2011}%
  \BibitemOpen
  \bibfield  {author} {\bibinfo {author} {\bibnamefont {Schmidt}, \bibfnamefont
  {B~E}}, \bibinfo {author} {\bibfnamefont {A.~D.}\ \bibnamefont {Shiner}},
  \bibinfo {author} {\bibfnamefont {P.}~\bibnamefont {Lassonde}}, \bibinfo
  {author} {\bibfnamefont {J.-C.}\ \bibnamefont {Kieffer}}, \bibinfo {author}
  {\bibfnamefont {P.~B.}\ \bibnamefont {Corkum}}, \bibinfo {author}
  {\bibfnamefont {D.~M.}\ \bibnamefont {Villeneuve}}, \ and\ \bibinfo {author}
  {\bibfnamefont {F.}~\bibnamefont {L\'egar\'e}}} (\bibinfo {year} {2011}),\
  \bibfield  {title} {\enquote {\bibinfo {title} {Cep stable 1.6 cycle laser
  pulses at 1.8 $\mu$m},}\ }\href@noop {} {\bibfield  {journal} {\bibinfo
  {journal} {Opt. Exp.}\ }\textbf {\bibinfo {volume} {19}},\ \bibinfo {pages}
  {6858--6864}}\BibitemShut {NoStop}%
\bibitem [{\citenamefont {Scholl}\ \emph {et~al.}(2013)\citenamefont {Scholl},
  \citenamefont {Garc{\'i}a-Etxarri}, \citenamefont {Koh},\ and\ \citenamefont
  {Dionne}}]{Scholl2013}%
  \BibitemOpen
  \bibfield  {author} {\bibinfo {author} {\bibnamefont {Scholl}, \bibfnamefont
  {J~A}}, \bibinfo {author} {\bibfnamefont {A.}~\bibnamefont
  {Garc{\'i}a-Etxarri}}, \bibinfo {author} {\bibfnamefont {A.~L.}\ \bibnamefont
  {Koh}}, \ and\ \bibinfo {author} {\bibfnamefont {J.~A.}\ \bibnamefont
  {Dionne}}} (\bibinfo {year} {2013}),\ \bibfield  {title} {\enquote {\bibinfo
  {title} {Observation of quantum tunneling between two plasmonic
  nanoparticles},}\ }\href@noop {} {\bibfield  {journal} {\bibinfo  {journal}
  {Nano Lett.}\ }\textbf {\bibinfo {volume} {13}},\ \bibinfo {pages}
  {564--569}}\BibitemShut {NoStop}%
\bibitem [{\citenamefont {Scholz}\ \emph {et~al.}(2013)\citenamefont {Scholz},
  \citenamefont {Himmel}, \citenamefont {Heinemann}, \citenamefont {Schleyer},
  \citenamefont {Meyer},\ and\ \citenamefont {Krossing}}]{scholz2013crystal}%
  \BibitemOpen
  \bibfield  {author} {\bibinfo {author} {\bibnamefont {Scholz}, \bibfnamefont
  {F}}, \bibinfo {author} {\bibfnamefont {D.}~\bibnamefont {Himmel}}, \bibinfo
  {author} {\bibfnamefont {F.~W.}\ \bibnamefont {Heinemann}}, \bibinfo {author}
  {\bibfnamefont {P.~v.~R.}\ \bibnamefont {Schleyer}}, \bibinfo {author}
  {\bibfnamefont {K.}~\bibnamefont {Meyer}}, \ and\ \bibinfo {author}
  {\bibfnamefont {I.}~\bibnamefont {Krossing}}} (\bibinfo {year} {2013}),\
  \bibfield  {title} {\enquote {\bibinfo {title} {Crystal structure
  determination of the nonclassical 2-norbornyl cation},}\ }\href@noop {}
  {\bibfield  {journal} {\bibinfo  {journal} {Science}\ }\textbf {\bibinfo
  {volume} {341}},\ \bibinfo {pages} {62--64}}\BibitemShut {NoStop}%
\bibitem [{\citenamefont {Schr\"oder}\ \emph
  {et~al.}(2015{\natexlab{a}})\citenamefont {Schr\"oder}, \citenamefont
  {Sivis}, \citenamefont {Bormann}, \citenamefont {Sch\"afer},\ and\
  \citenamefont {Ropers}}]{Schroder2015}%
  \BibitemOpen
  \bibfield  {author} {\bibinfo {author} {\bibnamefont {Schr\"oder},
  \bibfnamefont {B}}, \bibinfo {author} {\bibfnamefont {M.}~\bibnamefont
  {Sivis}}, \bibinfo {author} {\bibfnamefont {R.}~\bibnamefont {Bormann}},
  \bibinfo {author} {\bibfnamefont {S.}~\bibnamefont {Sch\"afer}}, \ and\
  \bibinfo {author} {\bibfnamefont {C.}~\bibnamefont {Ropers}}} (\bibinfo
  {year} {2015}{\natexlab{a}}),\ \bibfield  {title} {\enquote {\bibinfo {title}
  {An ultrafast nanotip electron gun triggered by grating-coupled surface
  plasmons},}\ }\href@noop {} {\bibfield  {journal} {\bibinfo  {journal} {Appl.
  Phys. Lett.}\ }\textbf {\bibinfo {volume} {107}},\ \bibinfo {pages}
  {231105}}\BibitemShut {NoStop}%
\bibitem [{\citenamefont {Schr\"oder}\ \emph
  {et~al.}(2015{\natexlab{b}})\citenamefont {Schr\"oder}, \citenamefont
  {Weber}, \citenamefont {Yalunin}, \citenamefont {Kiel}, \citenamefont
  {Matyssek}, \citenamefont {Sivis}, \citenamefont {Sch\"afer}, \citenamefont
  {von Cube}, \citenamefont {Irsen}, \citenamefont {Busch}, \citenamefont
  {Ropers},\ and\ \citenamefont {Linden}}]{Schroder2015a}%
  \BibitemOpen
  \bibfield  {author} {\bibinfo {author} {\bibnamefont {Schr\"oder},
  \bibfnamefont {B}}, \bibinfo {author} {\bibfnamefont {T.}~\bibnamefont
  {Weber}}, \bibinfo {author} {\bibfnamefont {S.~V.}\ \bibnamefont {Yalunin}},
  \bibinfo {author} {\bibfnamefont {T.}~\bibnamefont {Kiel}}, \bibinfo {author}
  {\bibfnamefont {C.}~\bibnamefont {Matyssek}}, \bibinfo {author}
  {\bibfnamefont {M.}~\bibnamefont {Sivis}}, \bibinfo {author} {\bibfnamefont
  {S.}~\bibnamefont {Sch\"afer}}, \bibinfo {author} {\bibfnamefont
  {F.}~\bibnamefont {von Cube}}, \bibinfo {author} {\bibfnamefont
  {S.}~\bibnamefont {Irsen}}, \bibinfo {author} {\bibfnamefont
  {K.}~\bibnamefont {Busch}}, \bibinfo {author} {\bibfnamefont
  {C.}~\bibnamefont {Ropers}}, \ and\ \bibinfo {author} {\bibfnamefont
  {S.}~\bibnamefont {Linden}}} (\bibinfo {year} {2015}{\natexlab{b}}),\
  \bibfield  {title} {\enquote {\bibinfo {title} {Real-space imaging of nanotip
  plasmons using electron energy loss spectroscopy},}\ }\href@noop {}
  {\bibfield  {journal} {\bibinfo  {journal} {Phys. Rev. B}\ }\textbf {\bibinfo
  {volume} {92}},\ \bibinfo {pages} {085411}}\BibitemShut {NoStop}%
\bibitem [{\citenamefont {Schubert}\ \emph {et~al.}(2014)\citenamefont
  {Schubert}, \citenamefont {Hohenleutner}, \citenamefont {Langer},
  \citenamefont {Urbanek}, \citenamefont {Lange}, \citenamefont {Huttner},
  \citenamefont {Golde}, \citenamefont {Meier}, \citenamefont {Kira},
  \citenamefont {Koch},\ and\ \citenamefont {R.~Huber}}]{Huber2014}%
  \BibitemOpen
  \bibfield  {author} {\bibinfo {author} {\bibnamefont {Schubert},
  \bibfnamefont {O}}, \bibinfo {author} {\bibfnamefont {M.}~\bibnamefont
  {Hohenleutner}}, \bibinfo {author} {\bibfnamefont {F.}~\bibnamefont
  {Langer}}, \bibinfo {author} {\bibfnamefont {B.}~\bibnamefont {Urbanek}},
  \bibinfo {author} {\bibfnamefont {C.}~\bibnamefont {Lange}}, \bibinfo
  {author} {\bibfnamefont {U.}~\bibnamefont {Huttner}}, \bibinfo {author}
  {\bibfnamefont {D.}~\bibnamefont {Golde}}, \bibinfo {author} {\bibfnamefont
  {T.}~\bibnamefont {Meier}}, \bibinfo {author} {\bibfnamefont
  {M.}~\bibnamefont {Kira}}, \bibinfo {author} {\bibfnamefont {S.~W.}\
  \bibnamefont {Koch}}, \ and\ \bibinfo {author} {\bibfnamefont
  {R.}~\bibnamefont {R.~Huber}}} (\bibinfo {year} {2014}),\ \bibfield  {title}
  {\enquote {\bibinfo {title} {Sub-cycle control of terahertz high-harmonic
  generation by dynamical bloch oscillations},}\ }\href@noop {} {\bibfield
  {journal} {\bibinfo  {journal} {Nat. Phot.}\ }\textbf {\bibinfo {volume}
  {8}},\ \bibinfo {pages} {119--123}}\BibitemShut {NoStop}%
\bibitem [{\citenamefont {Schultze}\ \emph {et~al.}(2013)\citenamefont
  {Schultze}, \citenamefont {Bothschafter}, \citenamefont {Sommer},
  \citenamefont {Holzner}, \citenamefont {Schweinberger}, \citenamefont
  {Fiess}, \citenamefont {Hofstetter}, \citenamefont {Kienberger},
  \citenamefont {Apalkov}, \citenamefont {Yakovlev}, \citenamefont {Stockman},\
  and\ \citenamefont {Krausz}}]{schultze2013controlling}%
  \BibitemOpen
  \bibfield  {author} {\bibinfo {author} {\bibnamefont {Schultze},
  \bibfnamefont {M}}, \bibinfo {author} {\bibfnamefont {E.~M}\ \bibnamefont
  {Bothschafter}}, \bibinfo {author} {\bibfnamefont {A.}~\bibnamefont
  {Sommer}}, \bibinfo {author} {\bibfnamefont {S.}~\bibnamefont {Holzner}},
  \bibinfo {author} {\bibfnamefont {W.}~\bibnamefont {Schweinberger}}, \bibinfo
  {author} {\bibfnamefont {M.}~\bibnamefont {Fiess}}, \bibinfo {author}
  {\bibfnamefont {M.}~\bibnamefont {Hofstetter}}, \bibinfo {author}
  {\bibfnamefont {R.}~\bibnamefont {Kienberger}}, \bibinfo {author}
  {\bibfnamefont {V.}~\bibnamefont {Apalkov}}, \bibinfo {author} {\bibfnamefont
  {V.~S.}\ \bibnamefont {Yakovlev}}, \bibinfo {author} {\bibfnamefont {M.~I.}\
  \bibnamefont {Stockman}}, \ and\ \bibinfo {author} {\bibfnamefont
  {F.}~\bibnamefont {Krausz}}} (\bibinfo {year} {2013}),\ \bibfield  {title}
  {\enquote {\bibinfo {title} {Controlling dielectrics with the electric field
  of light},}\ }\href@noop {} {\bibfield  {journal} {\bibinfo  {journal}
  {Nature}\ }\textbf {\bibinfo {volume} {493}},\ \bibinfo {pages}
  {75--78}}\BibitemShut {NoStop}%
\bibitem [{\citenamefont {Schultze}\ \emph {et~al.}(2010)\citenamefont
  {Schultze}, \citenamefont {Fie{\ss}}, \citenamefont {Karpowicz},
  \citenamefont {Gagnon}, \citenamefont {Korbman}, \citenamefont {Hofstetter},
  \citenamefont {Neppl}, \citenamefont {Cavalieri}, \citenamefont {Komninos},
  \citenamefont {Mercouris}, \citenamefont {Nicolaides}, \citenamefont
  {Pazourek}, \citenamefont {Nagele}, \citenamefont {Feist}, \citenamefont
  {Burgd\"orfer}, \citenamefont {Azzeer}, \citenamefont {Ernstorfer},
  \citenamefont {Kienberger}, \citenamefont {Kleineberg}, \citenamefont
  {Goulielmakis}, \citenamefont {Krausz},\ and\ \citenamefont
  {Yakovlev}}]{schultze2010delay}%
  \BibitemOpen
  \bibfield  {author} {\bibinfo {author} {\bibnamefont {Schultze},
  \bibfnamefont {M}}, \bibinfo {author} {\bibfnamefont {M.}~\bibnamefont
  {Fie{\ss}}}, \bibinfo {author} {\bibfnamefont {N.}~\bibnamefont {Karpowicz}},
  \bibinfo {author} {\bibfnamefont {J.}~\bibnamefont {Gagnon}}, \bibinfo
  {author} {\bibfnamefont {M.}~\bibnamefont {Korbman}}, \bibinfo {author}
  {\bibfnamefont {M.}~\bibnamefont {Hofstetter}}, \bibinfo {author}
  {\bibfnamefont {S.}~\bibnamefont {Neppl}}, \bibinfo {author} {\bibfnamefont
  {A.~L.}\ \bibnamefont {Cavalieri}}, \bibinfo {author} {\bibfnamefont
  {Y.}~\bibnamefont {Komninos}}, \bibinfo {author} {\bibfnamefont {Th.}\
  \bibnamefont {Mercouris}}, \bibinfo {author} {\bibfnamefont {C.~A.}\
  \bibnamefont {Nicolaides}}, \bibinfo {author} {\bibfnamefont
  {R.}~\bibnamefont {Pazourek}}, \bibinfo {author} {\bibfnamefont
  {S.}~\bibnamefont {Nagele}}, \bibinfo {author} {\bibfnamefont
  {J.}~\bibnamefont {Feist}}, \bibinfo {author} {\bibfnamefont
  {J.}~\bibnamefont {Burgd\"orfer}}, \bibinfo {author} {\bibfnamefont {A.~M.}\
  \bibnamefont {Azzeer}}, \bibinfo {author} {\bibfnamefont {R.}~\bibnamefont
  {Ernstorfer}}, \bibinfo {author} {\bibfnamefont {R.}~\bibnamefont
  {Kienberger}}, \bibinfo {author} {\bibfnamefont {U.}~\bibnamefont
  {Kleineberg}}, \bibinfo {author} {\bibfnamefont {E.}~\bibnamefont
  {Goulielmakis}}, \bibinfo {author} {\bibfnamefont {F.}~\bibnamefont
  {Krausz}}, \ and\ \bibinfo {author} {\bibfnamefont {V.~S.}\ \bibnamefont
  {Yakovlev}}} (\bibinfo {year} {2010}),\ \bibfield  {title} {\enquote
  {\bibinfo {title} {Delay in photoemission},}\ }\href@noop {} {\bibfield
  {journal} {\bibinfo  {journal} {Science}\ }\textbf {\bibinfo {volume}
  {328}},\ \bibinfo {pages} {1658--1662}}\BibitemShut {NoStop}%
\bibitem [{\citenamefont {Schweinberger}\ \emph {et~al.}(2012)\citenamefont
  {Schweinberger}, \citenamefont {Sommer}, \citenamefont {Bothschafter},
  \citenamefont {Li}, \citenamefont {Krausz}, \citenamefont {Kienberger},\ and\
  \citenamefont {Schultze}}]{schweinberger_waveform_controlled_2012}%
  \BibitemOpen
  \bibfield  {author} {\bibinfo {author} {\bibnamefont {Schweinberger},
  \bibfnamefont {W}}, \bibinfo {author} {\bibfnamefont {A.}~\bibnamefont
  {Sommer}}, \bibinfo {author} {\bibfnamefont {E.}~\bibnamefont
  {Bothschafter}}, \bibinfo {author} {\bibfnamefont {J.}~\bibnamefont {Li}},
  \bibinfo {author} {\bibfnamefont {F.}~\bibnamefont {Krausz}}, \bibinfo
  {author} {\bibfnamefont {R.}~\bibnamefont {Kienberger}}, \ and\ \bibinfo
  {author} {\bibfnamefont {M.}~\bibnamefont {Schultze}}} (\bibinfo {year}
  {2012}),\ \bibfield  {title} {\enquote {\bibinfo {title} {Waveform-controlled
  near-single-cycle milli-joule laser pulses generate sub-10 nm extreme
  ultraviolet continua},}\ }\href@noop {} {\bibfield  {journal} {\bibinfo
  {journal} {Opt. Lett.}\ }\textbf {\bibinfo {volume} {37}},\ \bibinfo {pages}
  {3573--3575}}\BibitemShut {NoStop}%
\bibitem [{\citenamefont {Scrinzi}\ \emph {et~al.}(2006)\citenamefont
  {Scrinzi}, \citenamefont {Ivanov}, \citenamefont {Kienberger},\ and\
  \citenamefont {Villeneuve}}]{Scrinzi06}%
  \BibitemOpen
  \bibfield  {author} {\bibinfo {author} {\bibnamefont {Scrinzi}, \bibfnamefont
  {A}}, \bibinfo {author} {\bibfnamefont {M.~Y.}\ \bibnamefont {Ivanov}},
  \bibinfo {author} {\bibfnamefont {R.}~\bibnamefont {Kienberger}}, \ and\
  \bibinfo {author} {\bibfnamefont {D.~M.}\ \bibnamefont {Villeneuve}}}
  (\bibinfo {year} {2006}),\ \bibfield  {title} {\enquote {\bibinfo {title}
  {Attosecond physics},}\ }\href@noop {} {\bibfield  {journal} {\bibinfo
  {journal} {J. Phys. B}\ }\textbf {\bibinfo {volume} {39}},\ \bibinfo {pages}
  {R1--R37}}\BibitemShut {NoStop}%
\bibitem [{\citenamefont {Seiffert}\ \emph {et~al.}(2016)\citenamefont
  {Seiffert}, \citenamefont {S{\"u}{\ss}mann}, \citenamefont {Zherebtsov},
  \citenamefont {Rupp}, \citenamefont {Peltz}, \citenamefont {R{\"u}hl},
  \citenamefont {Kling},\ and\ \citenamefont {Fennel}}]{Seiffert15}%
  \BibitemOpen
  \bibfield  {author} {\bibinfo {author} {\bibnamefont {Seiffert},
  \bibfnamefont {L}}, \bibinfo {author} {\bibfnamefont {F.}~\bibnamefont
  {S{\"u}{\ss}mann}}, \bibinfo {author} {\bibfnamefont {S.}~\bibnamefont
  {Zherebtsov}}, \bibinfo {author} {\bibfnamefont {P.}~\bibnamefont {Rupp}},
  \bibinfo {author} {\bibfnamefont {C.}~\bibnamefont {Peltz}}, \bibinfo
  {author} {\bibfnamefont {E.}~\bibnamefont {R{\"u}hl}}, \bibinfo {author}
  {\bibfnamefont {M.~F.}\ \bibnamefont {Kling}}, \ and\ \bibinfo {author}
  {\bibfnamefont {T.}~\bibnamefont {Fennel}}} (\bibinfo {year} {2016}),\
  \bibfield  {title} {\enquote {\bibinfo {title} {Competition of single and
  double rescattering in the strong-field photoemission from dielectric
  nanospheres},}\ }\href@noop {} {\bibfield  {journal} {\bibinfo  {journal}
  {Appl. Phys. B}\ }\textbf {\bibinfo {volume} {122}},\ \bibinfo {pages}
  {101}}\BibitemShut {NoStop}%
\bibitem [{\citenamefont {Shaaran}\ \emph
  {et~al.}(2013{\natexlab{a}})\citenamefont {Shaaran}, \citenamefont
  {Ciappina}, \citenamefont {Guichard}, \citenamefont
  {P{\'e}rez-Hern{\'a}ndez}, \citenamefont {Roso}, \citenamefont {Arnold},
  \citenamefont {Siegel}, \citenamefont {Za{\"i}r},\ and\ \citenamefont
  {Lewenstein}}]{Marcelo13AR}%
  \BibitemOpen
  \bibfield  {author} {\bibinfo {author} {\bibnamefont {Shaaran}, \bibfnamefont
  {T}}, \bibinfo {author} {\bibfnamefont {M.~F.}\ \bibnamefont {Ciappina}},
  \bibinfo {author} {\bibfnamefont {R.}~\bibnamefont {Guichard}}, \bibinfo
  {author} {\bibfnamefont {J.~A.}\ \bibnamefont {P{\'e}rez-Hern{\'a}ndez}},
  \bibinfo {author} {\bibfnamefont {L.}~\bibnamefont {Roso}}, \bibinfo {author}
  {\bibfnamefont {M.}~\bibnamefont {Arnold}}, \bibinfo {author} {\bibfnamefont
  {T.}~\bibnamefont {Siegel}}, \bibinfo {author} {\bibfnamefont
  {A.}~\bibnamefont {Za{\"i}r}}, \ and\ \bibinfo {author} {\bibfnamefont
  {M.}~\bibnamefont {Lewenstein}}} (\bibinfo {year} {2013}{\natexlab{a}}),\
  \bibfield  {title} {\enquote {\bibinfo {title} {High-order-harmonic
  generation by enhanced plasmonic near-fields in metal nanoparticles},}\
  }\href@noop {} {\bibfield  {journal} {\bibinfo  {journal} {Phys. Rev. A}\
  }\textbf {\bibinfo {volume} {87}},\ \bibinfo {pages} {041402(R)}}\BibitemShut
  {NoStop}%
\bibitem [{\citenamefont {Shaaran}\ \emph
  {et~al.}(2012{\natexlab{a}})\citenamefont {Shaaran}, \citenamefont
  {Ciappina},\ and\ \citenamefont {Lewenstein}}]{Marcelo12JMO}%
  \BibitemOpen
  \bibfield  {author} {\bibinfo {author} {\bibnamefont {Shaaran}, \bibfnamefont
  {T}}, \bibinfo {author} {\bibfnamefont {M.~F.}\ \bibnamefont {Ciappina}}, \
  and\ \bibinfo {author} {\bibfnamefont {M.}~\bibnamefont {Lewenstein}}}
  (\bibinfo {year} {2012}{\natexlab{a}}),\ \bibfield  {title} {\enquote
  {\bibinfo {title} {Estimating the plasmonic field enhancement using
  high-order harmonic generation: the role of the field inhomogeneity},}\
  }\href@noop {} {\bibfield  {journal} {\bibinfo  {journal} {J. Mod. Opt.}\
  }\textbf {\bibinfo {volume} {86}},\ \bibinfo {pages}
  {1634--1639}}\BibitemShut {NoStop}%
\bibitem [{\citenamefont {Shaaran}\ \emph
  {et~al.}(2012{\natexlab{b}})\citenamefont {Shaaran}, \citenamefont
  {Ciappina},\ and\ \citenamefont {Lewenstein}}]{Marcelo12AA}%
  \BibitemOpen
  \bibfield  {author} {\bibinfo {author} {\bibnamefont {Shaaran}, \bibfnamefont
  {T}}, \bibinfo {author} {\bibfnamefont {M.~F.}\ \bibnamefont {Ciappina}}, \
  and\ \bibinfo {author} {\bibfnamefont {M.}~\bibnamefont {Lewenstein}}}
  (\bibinfo {year} {2012}{\natexlab{b}}),\ \bibfield  {title} {\enquote
  {\bibinfo {title} {Quantum-orbit analysis of high-order-harmonic generation
  by resonant plasmon field enhancement},}\ }\href@noop {} {\bibfield
  {journal} {\bibinfo  {journal} {Phys. Rev. A}\ }\textbf {\bibinfo {volume}
  {86}},\ \bibinfo {pages} {023408}}\BibitemShut {NoStop}%
\bibitem [{\citenamefont {Shaaran}\ \emph
  {et~al.}(2013{\natexlab{b}})\citenamefont {Shaaran}, \citenamefont
  {Ciappina},\ and\ \citenamefont {Lewenstein}}]{Marcelo13AA}%
  \BibitemOpen
  \bibfield  {author} {\bibinfo {author} {\bibnamefont {Shaaran}, \bibfnamefont
  {T}}, \bibinfo {author} {\bibfnamefont {M.~F.}\ \bibnamefont {Ciappina}}, \
  and\ \bibinfo {author} {\bibfnamefont {M.}~\bibnamefont {Lewenstein}}}
  (\bibinfo {year} {2013}{\natexlab{b}}),\ \bibfield  {title} {\enquote
  {\bibinfo {title} {Quantum-orbit analysis of above-threshold ionization
  driven by an intense spatially inhomogeneous field},}\ }\href@noop {}
  {\bibfield  {journal} {\bibinfo  {journal} {Phys. Rev. A}\ }\textbf {\bibinfo
  {volume} {87}},\ \bibinfo {pages} {053415}}\BibitemShut {NoStop}%
\bibitem [{\citenamefont {Shao}\ \emph {et~al.}(1996)\citenamefont {Shao},
  \citenamefont {Ditmire}, \citenamefont {Tisch}, \citenamefont {Springate},
  \citenamefont {Marangos},\ and\ \citenamefont {Hutchinson}}]{Shao1996}%
  \BibitemOpen
  \bibfield  {author} {\bibinfo {author} {\bibnamefont {Shao}, \bibfnamefont
  {Y~L}}, \bibinfo {author} {\bibfnamefont {T.}~\bibnamefont {Ditmire}},
  \bibinfo {author} {\bibfnamefont {J.~W.~G.}\ \bibnamefont {Tisch}}, \bibinfo
  {author} {\bibfnamefont {E.}~\bibnamefont {Springate}}, \bibinfo {author}
  {\bibfnamefont {J.~P.}\ \bibnamefont {Marangos}}, \ and\ \bibinfo {author}
  {\bibfnamefont {M.~H.~R.}\ \bibnamefont {Hutchinson}}} (\bibinfo {year}
  {1996}),\ \bibfield  {title} {\enquote {\bibinfo {title} {Multi-kev electron
  generation in the interaction of intense laser pulses with xe clusters},}\
  }\href@noop {} {\bibfield  {journal} {\bibinfo  {journal} {Phys. Rev. Lett.}\
  }\textbf {\bibinfo {volume} {77}},\ \bibinfo {pages} {3343}}\BibitemShut
  {NoStop}%
\bibitem [{\citenamefont {Shi-Lin}\ and\ \citenamefont
  {Ting-Yun}(2013)}]{ShiLin2013}%
  \BibitemOpen
  \bibfield  {author} {\bibinfo {author} {\bibnamefont {Shi-Lin}, \bibfnamefont
  {H}}, \ and\ \bibinfo {author} {\bibfnamefont {S.}~\bibnamefont {Ting-Yun}}}
  (\bibinfo {year} {2013}),\ \bibfield  {title} {\enquote {\bibinfo {title}
  {Effect of electron correlation on high-order harmonic generation in helium
  model atom},}\ }\href@noop {} {\bibfield  {journal} {\bibinfo  {journal}
  {Chin. Phys. B}\ }\textbf {\bibinfo {volume} {22}},\ \bibinfo {pages}
  {013101}}\BibitemShut {NoStop}%
\bibitem [{\citenamefont {Shiner}\ \emph {et~al.}(2011)\citenamefont {Shiner},
  \citenamefont {Schmidt}, \citenamefont {Trallero-Herrero}, \citenamefont
  {W\"orner}, \citenamefont {Patchkovskii}, \citenamefont {Corkum},
  \citenamefont {Kieffer}, \citenamefont {L\'egar\'e},\ and\ \citenamefont
  {Villeneuve}}]{Corkumcollective}%
  \BibitemOpen
  \bibfield  {author} {\bibinfo {author} {\bibnamefont {Shiner}, \bibfnamefont
  {A~D}}, \bibinfo {author} {\bibfnamefont {B.~E.}\ \bibnamefont {Schmidt}},
  \bibinfo {author} {\bibfnamefont {C.}~\bibnamefont {Trallero-Herrero}},
  \bibinfo {author} {\bibfnamefont {H.~J.}\ \bibnamefont {W\"orner}}, \bibinfo
  {author} {\bibfnamefont {S.}~\bibnamefont {Patchkovskii}}, \bibinfo {author}
  {\bibfnamefont {P.~B.}\ \bibnamefont {Corkum}}, \bibinfo {author}
  {\bibfnamefont {J-C.}\ \bibnamefont {Kieffer}}, \bibinfo {author}
  {\bibfnamefont {F.}~\bibnamefont {L\'egar\'e}}, \ and\ \bibinfo {author}
  {\bibfnamefont {D.~M.}\ \bibnamefont {Villeneuve}}} (\bibinfo {year}
  {2011}),\ \bibfield  {title} {\enquote {\bibinfo {title} {Probing collective
  multi-electron dynamics in xenon with high-harmonic spectroscopy},}\
  }\href@noop {} {\bibfield  {journal} {\bibinfo  {journal} {Nat. Phys.}\
  }\textbf {\bibinfo {volume} {7}},\ \bibinfo {pages} {464--467}}\BibitemShut
  {NoStop}%
\bibitem [{\citenamefont {Sivis}\ \emph {et~al.}(2012)\citenamefont {Sivis},
  \citenamefont {Duwe}, \citenamefont {Abel},\ and\ \citenamefont
  {Ropers}}]{Sivis12A}%
  \BibitemOpen
  \bibfield  {author} {\bibinfo {author} {\bibnamefont {Sivis}, \bibfnamefont
  {M}}, \bibinfo {author} {\bibfnamefont {M.}~\bibnamefont {Duwe}}, \bibinfo
  {author} {\bibfnamefont {B.}~\bibnamefont {Abel}}, \ and\ \bibinfo {author}
  {\bibfnamefont {C.}~\bibnamefont {Ropers}}} (\bibinfo {year} {2012}),\
  \bibfield  {title} {\enquote {\bibinfo {title} {Nanostructure-enhanced atomic
  line emission},}\ }\href@noop {} {\bibfield  {journal} {\bibinfo  {journal}
  {Nature}\ }\textbf {\bibinfo {volume} {485}},\ \bibinfo {pages}
  {E1--E3}}\BibitemShut {NoStop}%
\bibitem [{\citenamefont {Sivis}\ \emph {et~al.}(2013)\citenamefont {Sivis},
  \citenamefont {Duwe}, \citenamefont {Abel},\ and\ \citenamefont
  {Ropers}}]{Sivis13}%
  \BibitemOpen
  \bibfield  {author} {\bibinfo {author} {\bibnamefont {Sivis}, \bibfnamefont
  {M}}, \bibinfo {author} {\bibfnamefont {M.}~\bibnamefont {Duwe}}, \bibinfo
  {author} {\bibfnamefont {B.}~\bibnamefont {Abel}}, \ and\ \bibinfo {author}
  {\bibfnamefont {C.}~\bibnamefont {Ropers}}} (\bibinfo {year} {2013}),\
  \bibfield  {title} {\enquote {\bibinfo {title} {Extreme-ultraviolet light
  generation in plasmonic nanostructures},}\ }\href@noop {} {\bibfield
  {journal} {\bibinfo  {journal} {Nat. Phys.}\ }\textbf {\bibinfo {volume}
  {9}},\ \bibinfo {pages} {304--309}}\BibitemShut {NoStop}%
\bibitem [{\citenamefont {Sivis}\ and\ \citenamefont
  {Ropers}(2013)}]{Sivis13PRL}%
  \BibitemOpen
  \bibfield  {author} {\bibinfo {author} {\bibnamefont {Sivis}, \bibfnamefont
  {M}}, \ and\ \bibinfo {author} {\bibfnamefont {C.}~\bibnamefont {Ropers}}}
  (\bibinfo {year} {2013}),\ \bibfield  {title} {\enquote {\bibinfo {title}
  {Generation and bistability of a waveguide nanoplasma observed by enhanced
  extreme-ultraviolet fluorescence},}\ }\href@noop {} {\bibfield  {journal}
  {\bibinfo  {journal} {Phys. Rev. Lett.}\ }\textbf {\bibinfo {volume} {111}},\
  \bibinfo {pages} {085001}}\BibitemShut {NoStop}%
\bibitem [{\citenamefont {Skopalov\'a}\ \emph {et~al.}(2010)\citenamefont
  {Skopalov\'a}, \citenamefont {El-Taha}, \citenamefont {Za\"ir}, \citenamefont
  {Hohenberger}, \citenamefont {Springate}, \citenamefont {Tisch},
  \citenamefont {Smith},\ and\ \citenamefont {Marangos}}]{Skopalova2010}%
  \BibitemOpen
  \bibfield  {author} {\bibinfo {author} {\bibnamefont {Skopalov\'a},
  \bibfnamefont {E}}, \bibinfo {author} {\bibfnamefont {Y.~C.}\ \bibnamefont
  {El-Taha}}, \bibinfo {author} {\bibfnamefont {A.}~\bibnamefont {Za\"ir}},
  \bibinfo {author} {\bibfnamefont {M.}~\bibnamefont {Hohenberger}}, \bibinfo
  {author} {\bibfnamefont {E.}~\bibnamefont {Springate}}, \bibinfo {author}
  {\bibfnamefont {J.~W.~G.}\ \bibnamefont {Tisch}}, \bibinfo {author}
  {\bibfnamefont {R.~A.}\ \bibnamefont {Smith}}, \ and\ \bibinfo {author}
  {\bibfnamefont {J.~P.}\ \bibnamefont {Marangos}}} (\bibinfo {year} {2010}),\
  \bibfield  {title} {\enquote {\bibinfo {title} {Pulse-length dependence of
  the anisotropy of laser-driven cluster explosions: Transition to the
  impulsive regime for pulses approaching the few-cycle limit},}\ }\href@noop
  {} {\bibfield  {journal} {\bibinfo  {journal} {Phys. Rev. Lett.}\ }\textbf
  {\bibinfo {volume} {104}},\ \bibinfo {pages} {203401}}\BibitemShut {NoStop}%
\bibitem [{\citenamefont {Skopalov\'a}\ \emph {et~al.}(2011)\citenamefont
  {Skopalov\'a}, \citenamefont {Lei}, \citenamefont {Witting}, \citenamefont
  {Arrell}, \citenamefont {Frank}, \citenamefont {Sonnefraud}, \citenamefont
  {Maier}, \citenamefont {Tisch},\ and\ \citenamefont
  {Marangos}}]{Skopalova11}%
  \BibitemOpen
  \bibfield  {author} {\bibinfo {author} {\bibnamefont {Skopalov\'a},
  \bibfnamefont {E}}, \bibinfo {author} {\bibfnamefont {D.~Y.}\ \bibnamefont
  {Lei}}, \bibinfo {author} {\bibfnamefont {T.}~\bibnamefont {Witting}},
  \bibinfo {author} {\bibfnamefont {C.}~\bibnamefont {Arrell}}, \bibinfo
  {author} {\bibfnamefont {F.}~\bibnamefont {Frank}}, \bibinfo {author}
  {\bibfnamefont {Y.}~\bibnamefont {Sonnefraud}}, \bibinfo {author}
  {\bibfnamefont {S.~A.}\ \bibnamefont {Maier}}, \bibinfo {author}
  {\bibfnamefont {J.~W.~G.}\ \bibnamefont {Tisch}}, \ and\ \bibinfo {author}
  {\bibfnamefont {J.~P.}\ \bibnamefont {Marangos}}} (\bibinfo {year} {2011}),\
  \bibfield  {title} {\enquote {\bibinfo {title} {Numerical simulation of
  attosecond nanoplasmonic streaking},}\ }\href@noop {} {\bibfield  {journal}
  {\bibinfo  {journal} {New J. Phys.}\ }\textbf {\bibinfo {volume} {13}},\
  \bibinfo {pages} {083003}}\BibitemShut {NoStop}%
\bibitem [{\citenamefont {Smirnova}\ \emph {et~al.}(2009)\citenamefont
  {Smirnova}, \citenamefont {Mairesse}, \citenamefont {Patchkovskii},
  \citenamefont {Dudovich}, \citenamefont {Villeneuve}, \citenamefont
  {Corkum},\ and\ \citenamefont {Ivanov}}]{olgaHHG}%
  \BibitemOpen
  \bibfield  {author} {\bibinfo {author} {\bibnamefont {Smirnova},
  \bibfnamefont {O}}, \bibinfo {author} {\bibfnamefont {Y.}~\bibnamefont
  {Mairesse}}, \bibinfo {author} {\bibfnamefont {S.}~\bibnamefont
  {Patchkovskii}}, \bibinfo {author} {\bibfnamefont {N.}~\bibnamefont
  {Dudovich}}, \bibinfo {author} {\bibfnamefont {D.}~\bibnamefont
  {Villeneuve}}, \bibinfo {author} {\bibfnamefont {P.~B.}\ \bibnamefont
  {Corkum}}, \ and\ \bibinfo {author} {\bibfnamefont {M.~Yu.}\ \bibnamefont
  {Ivanov}}} (\bibinfo {year} {2009}),\ \bibfield  {title} {\enquote {\bibinfo
  {title} {High harmonic interferometry of multi-electron dynamics in
  molecules},}\ }\href@noop {} {\bibfield  {journal} {\bibinfo  {journal}
  {Nature}\ }\textbf {\bibinfo {volume} {460}},\ \bibinfo {pages}
  {972--977}}\BibitemShut {NoStop}%
\bibitem [{\citenamefont {Smith}\ \emph {et~al.}(1998)\citenamefont {Smith},
  \citenamefont {Ditmire},\ and\ \citenamefont {Tisch}}]{Smith1998}%
  \BibitemOpen
  \bibfield  {author} {\bibinfo {author} {\bibnamefont {Smith}, \bibfnamefont
  {R~A}}, \bibinfo {author} {\bibfnamefont {T.}~\bibnamefont {Ditmire}}, \ and\
  \bibinfo {author} {\bibfnamefont {J.~W.~G.}\ \bibnamefont {Tisch}}} (\bibinfo
  {year} {1998}),\ \bibfield  {title} {\enquote {\bibinfo {title}
  {Characterization of a cryogenically cooled high-pressure gas jet for
  laser/cluster interaction experiments},}\ }\href@noop {} {\bibfield
  {journal} {\bibinfo  {journal} {Rev. Sci. Instrum.}\ }\textbf {\bibinfo
  {volume} {69}},\ \bibinfo {pages} {3798}}\BibitemShut {NoStop}%
\bibitem [{\citenamefont {Sola}\ \emph {et~al.}(2006)\citenamefont {Sola},
  \citenamefont {M{\'e}vel}, \citenamefont {Elouga}, \citenamefont {Constant},
  \citenamefont {Strelkov}, \citenamefont {Poletto}, \citenamefont {Villoresi},
  \citenamefont {Benedetti}, \citenamefont {Caumes}, \citenamefont {Stagira},
  \citenamefont {Vozzi}, \citenamefont {Sansone},\ and\ \citenamefont
  {Nisoli}}]{sola2006controlling}%
  \BibitemOpen
  \bibfield  {author} {\bibinfo {author} {\bibnamefont {Sola}, \bibfnamefont
  {I~J}}, \bibinfo {author} {\bibfnamefont {E.}~\bibnamefont {M{\'e}vel}},
  \bibinfo {author} {\bibfnamefont {L.}~\bibnamefont {Elouga}}, \bibinfo
  {author} {\bibfnamefont {E.}~\bibnamefont {Constant}}, \bibinfo {author}
  {\bibfnamefont {V.}~\bibnamefont {Strelkov}}, \bibinfo {author}
  {\bibfnamefont {L.}~\bibnamefont {Poletto}}, \bibinfo {author} {\bibfnamefont
  {P.}~\bibnamefont {Villoresi}}, \bibinfo {author} {\bibfnamefont
  {E.}~\bibnamefont {Benedetti}}, \bibinfo {author} {\bibfnamefont {J.-P.}\
  \bibnamefont {Caumes}}, \bibinfo {author} {\bibfnamefont {S.}~\bibnamefont
  {Stagira}}, \bibinfo {author} {\bibfnamefont {C.}~\bibnamefont {Vozzi}},
  \bibinfo {author} {\bibfnamefont {G.}~\bibnamefont {Sansone}}, \ and\
  \bibinfo {author} {\bibfnamefont {M.}~\bibnamefont {Nisoli}}} (\bibinfo
  {year} {2006}),\ \bibfield  {title} {\enquote {\bibinfo {title} {Controlling
  attosecond electron dynamics by phase-stabilized polarization gating},}\
  }\href@noop {} {\bibfield  {journal} {\bibinfo  {journal} {Nat. Phys.}\
  }\textbf {\bibinfo {volume} {2}},\ \bibinfo {pages} {319--322}}\BibitemShut
  {NoStop}%
\bibitem [{\citenamefont {Solleder}\ \emph {et~al.}(2007)\citenamefont
  {Solleder}, \citenamefont {Lemell}, \citenamefont
  {T\ifmmode~\mbox{\H{o}}\else \H{o}\fi{}k\'esi}, \citenamefont {Hatcher},\
  and\ \citenamefont {Burgd\"orfer}}]{Solleder2007}%
  \BibitemOpen
  \bibfield  {author} {\bibinfo {author} {\bibnamefont {Solleder},
  \bibfnamefont {B}}, \bibinfo {author} {\bibfnamefont {C.}~\bibnamefont
  {Lemell}}, \bibinfo {author} {\bibfnamefont {K.}~\bibnamefont
  {T\ifmmode~\mbox{\H{o}}\else \H{o}\fi{}k\'esi}}, \bibinfo {author}
  {\bibfnamefont {N.}~\bibnamefont {Hatcher}}, \ and\ \bibinfo {author}
  {\bibfnamefont {J.}~\bibnamefont {Burgd\"orfer}}} (\bibinfo {year} {2007}),\
  \bibfield  {title} {\enquote {\bibinfo {title} {Spin-dependent low-energy
  electron transport in metals},}\ }\href@noop {} {\bibfield  {journal}
  {\bibinfo  {journal} {Phys. Rev. B}\ }\textbf {\bibinfo {volume} {76}},\
  \bibinfo {pages} {075115}}\BibitemShut {NoStop}%
\bibitem [{\citenamefont {Sonnefraud}\ \emph {et~al.}(2012)\citenamefont
  {Sonnefraud}, \citenamefont {Leen~Koh}, \citenamefont {McComb},\ and\
  \citenamefont {Maier}}]{Sonnefraud2012}%
  \BibitemOpen
  \bibfield  {author} {\bibinfo {author} {\bibnamefont {Sonnefraud},
  \bibfnamefont {Y}}, \bibinfo {author} {\bibfnamefont {A.}~\bibnamefont
  {Leen~Koh}}, \bibinfo {author} {\bibfnamefont {D.W.}\ \bibnamefont {McComb}},
  \ and\ \bibinfo {author} {\bibfnamefont {S.A.}\ \bibnamefont {Maier}}}
  (\bibinfo {year} {2012}),\ \bibfield  {title} {\enquote {\bibinfo {title}
  {Nanoplasmonics: {E}ngineering and observation of localized plasmon modes},}\
  }\href@noop {} {\bibfield  {journal} {\bibinfo  {journal} {Laser Photon.
  Rev.}\ }\textbf {\bibinfo {volume} {6}}~(\bibinfo {number} {3}),\ \bibinfo
  {pages} {277--295}}\BibitemShut {NoStop}%
\bibitem [{\citenamefont {S\"onnichsen}\ \emph {et~al.}(2002)\citenamefont
  {S\"onnichsen}, \citenamefont {Franzl}, \citenamefont {Wilk}, \citenamefont
  {von Plessen}, \citenamefont {Feldmann}, \citenamefont {Wilson},\ and\
  \citenamefont {Mulvaney}}]{Sonnichsen2002}%
  \BibitemOpen
  \bibfield  {author} {\bibinfo {author} {\bibnamefont {S\"onnichsen},
  \bibfnamefont {C}}, \bibinfo {author} {\bibfnamefont {T.}~\bibnamefont
  {Franzl}}, \bibinfo {author} {\bibfnamefont {T.}~\bibnamefont {Wilk}},
  \bibinfo {author} {\bibfnamefont {G.}~\bibnamefont {von Plessen}}, \bibinfo
  {author} {\bibfnamefont {J.}~\bibnamefont {Feldmann}}, \bibinfo {author}
  {\bibfnamefont {O.}~\bibnamefont {Wilson}}, \ and\ \bibinfo {author}
  {\bibfnamefont {P.}~\bibnamefont {Mulvaney}}} (\bibinfo {year} {2002}),\
  \bibfield  {title} {\enquote {\bibinfo {title} {Drastic reduction of plasmon
  damping in gold nanorods},}\ }\href@noop {} {\bibfield  {journal} {\bibinfo
  {journal} {Phys. Rev. Lett.}\ }\textbf {\bibinfo {volume} {88}},\ \bibinfo
  {pages} {077402}}\BibitemShut {NoStop}%
\bibitem [{\citenamefont {Stebbings}\ \emph {et~al.}(2011)\citenamefont
  {Stebbings}, \citenamefont {S{\"u}{\ss}mann}, \citenamefont {Yang},
  \citenamefont {Scrinzi}, \citenamefont {Durach}, \citenamefont {Rusina},
  \citenamefont {Stockman},\ and\ \citenamefont {Kling}}]{Stebbings11}%
  \BibitemOpen
  \bibfield  {author} {\bibinfo {author} {\bibnamefont {Stebbings},
  \bibfnamefont {S~L}}, \bibinfo {author} {\bibfnamefont {F.}~\bibnamefont
  {S{\"u}{\ss}mann}}, \bibinfo {author} {\bibfnamefont {Y.-Y.}\ \bibnamefont
  {Yang}}, \bibinfo {author} {\bibfnamefont {A.}~\bibnamefont {Scrinzi}},
  \bibinfo {author} {\bibfnamefont {M.}~\bibnamefont {Durach}}, \bibinfo
  {author} {\bibfnamefont {A.}~\bibnamefont {Rusina}}, \bibinfo {author}
  {\bibfnamefont {M.~I.}\ \bibnamefont {Stockman}}, \ and\ \bibinfo {author}
  {\bibfnamefont {M.~F.}\ \bibnamefont {Kling}}} (\bibinfo {year} {2011}),\
  \bibfield  {title} {\enquote {\bibinfo {title} {Generation of isolated
  attosecond extreme ultraviolet pulses employing nanoplasmonic field
  enhancement: optimization of coupled ellipsoids},}\ }\href@noop {} {\bibfield
   {journal} {\bibinfo  {journal} {New J. Phys.}\ }\textbf {\bibinfo {volume}
  {13}},\ \bibinfo {pages} {073010}}\BibitemShut {NoStop}%
\bibitem [{\citenamefont {St{\"o}ber}\ \emph {et~al.}(1968)\citenamefont
  {St{\"o}ber}, \citenamefont {Fink},\ and\ \citenamefont {Bohn}}]{Stoeber68}%
  \BibitemOpen
  \bibfield  {author} {\bibinfo {author} {\bibnamefont {St{\"o}ber},
  \bibfnamefont {W}}, \bibinfo {author} {\bibfnamefont {A.}~\bibnamefont
  {Fink}}, \ and\ \bibinfo {author} {\bibfnamefont {E.}~\bibnamefont {Bohn}}}
  (\bibinfo {year} {1968}),\ \bibfield  {title} {\enquote {\bibinfo {title}
  {Controlled growth of monodisperse silica spheres in the micron size
  range},}\ }\href@noop {} {\bibfield  {journal} {\bibinfo  {journal} {J. Coll.
  Int. Sci.}\ }\textbf {\bibinfo {volume} {26}},\ \bibinfo {pages}
  {62--69}}\BibitemShut {NoStop}%
\bibitem [{\citenamefont {St\"ockle}\ \emph {et~al.}(2000)\citenamefont
  {St\"ockle}, \citenamefont {Suh}, \citenamefont {Deckert},\ and\
  \citenamefont {Zenobi}}]{Stockle2000}%
  \BibitemOpen
  \bibfield  {author} {\bibinfo {author} {\bibnamefont {St\"ockle},
  \bibfnamefont {Raoul~M}}, \bibinfo {author} {\bibfnamefont {Yung~Doug}\
  \bibnamefont {Suh}}, \bibinfo {author} {\bibfnamefont {Volker}\ \bibnamefont
  {Deckert}}, \ and\ \bibinfo {author} {\bibfnamefont {Renato}\ \bibnamefont
  {Zenobi}}} (\bibinfo {year} {2000}),\ \bibfield  {title} {\enquote {\bibinfo
  {title} {Nanoscale chemical analysis by tip-enhanced {R}aman spectroscopy},}\
  }\href@noop {} {\bibfield  {journal} {\bibinfo  {journal} {Chem. Phys.
  Lett.}\ }\textbf {\bibinfo {volume} {318}}~(\bibinfo {number} {1-3}),\
  \bibinfo {pages} {131--136}}\BibitemShut {NoStop}%
\bibitem [{\citenamefont {Stockman}(2011)}]{stockmanreview}%
  \BibitemOpen
  \bibfield  {author} {\bibinfo {author} {\bibnamefont {Stockman},
  \bibfnamefont {M~I}}} (\bibinfo {year} {2011}),\ \bibfield  {title} {\enquote
  {\bibinfo {title} {Nanoplasmonics: past, present, and glimpse into future},}\
  }\href@noop {} {\bibfield  {journal} {\bibinfo  {journal} {Opt. Exp.}\
  }\textbf {\bibinfo {volume} {19}},\ \bibinfo {pages}
  {22029--22106}}\BibitemShut {NoStop}%
\bibitem [{\citenamefont {Stockman}\ \emph {et~al.}(2007)\citenamefont
  {Stockman}, \citenamefont {Kling}, \citenamefont {Kleineberg},\ and\
  \citenamefont {Krausz}}]{Stockman07}%
  \BibitemOpen
  \bibfield  {author} {\bibinfo {author} {\bibnamefont {Stockman},
  \bibfnamefont {M~I}}, \bibinfo {author} {\bibfnamefont {M.~F.}\ \bibnamefont
  {Kling}}, \bibinfo {author} {\bibfnamefont {U.}~\bibnamefont {Kleineberg}}, \
  and\ \bibinfo {author} {\bibfnamefont {F.}~\bibnamefont {Krausz}}} (\bibinfo
  {year} {2007}),\ \bibfield  {title} {\enquote {\bibinfo {title} {Attosecond
  nanoplasmonic-field microscope},}\ }\href@noop {} {\bibfield  {journal}
  {\bibinfo  {journal} {Nat. Photon.}\ }\textbf {\bibinfo {volume} {1}},\
  \bibinfo {pages} {539--544}}\BibitemShut {NoStop}%
\bibitem [{\citenamefont {Sumeruk}\ \emph
  {et~al.}(2007{\natexlab{a}})\citenamefont {Sumeruk}, \citenamefont {Kneip},
  \citenamefont {Symes}, \citenamefont {Churina}, \citenamefont {Belolipetski},
  \citenamefont {Donnelly},\ and\ \citenamefont {Ditmire}}]{Sumeruk2007}%
  \BibitemOpen
  \bibfield  {author} {\bibinfo {author} {\bibnamefont {Sumeruk}, \bibfnamefont
  {H~A}}, \bibinfo {author} {\bibfnamefont {S.}~\bibnamefont {Kneip}}, \bibinfo
  {author} {\bibfnamefont {D.~R.}\ \bibnamefont {Symes}}, \bibinfo {author}
  {\bibfnamefont {I.~V.}\ \bibnamefont {Churina}}, \bibinfo {author}
  {\bibfnamefont {A.~V.}\ \bibnamefont {Belolipetski}}, \bibinfo {author}
  {\bibfnamefont {T.~D.}\ \bibnamefont {Donnelly}}, \ and\ \bibinfo {author}
  {\bibfnamefont {T.}~\bibnamefont {Ditmire}}} (\bibinfo {year}
  {2007}{\natexlab{a}}),\ \bibfield  {title} {\enquote {\bibinfo {title}
  {Control of strong-laser-field coupling to electrons in solid targets with
  wavelength-scale spheres},}\ }\href@noop {} {\bibfield  {journal} {\bibinfo
  {journal} {Phys. Rev. Lett.}\ }\textbf {\bibinfo {volume} {98}},\ \bibinfo
  {pages} {045001}}\BibitemShut {NoStop}%
\bibitem [{\citenamefont {Sumeruk}\ \emph
  {et~al.}(2007{\natexlab{b}})\citenamefont {Sumeruk}, \citenamefont {Kneip},
  \citenamefont {Symes}, \citenamefont {Churina}, \citenamefont {Belolipetski},
  \citenamefont {Dyer}, \citenamefont {Landry}, \citenamefont {Bansal},
  \citenamefont {Bernstein}, \citenamefont {Donnelly}, \citenamefont
  {Karmakar}, \citenamefont {Pukhov},\ and\ \citenamefont
  {Ditmire}}]{Sumeruk2007Plasmas}%
  \BibitemOpen
  \bibfield  {author} {\bibinfo {author} {\bibnamefont {Sumeruk}, \bibfnamefont
  {H~A}}, \bibinfo {author} {\bibfnamefont {S.}~\bibnamefont {Kneip}}, \bibinfo
  {author} {\bibfnamefont {D.~R.}\ \bibnamefont {Symes}}, \bibinfo {author}
  {\bibfnamefont {I.~V.}\ \bibnamefont {Churina}}, \bibinfo {author}
  {\bibfnamefont {A.~V.}\ \bibnamefont {Belolipetski}}, \bibinfo {author}
  {\bibfnamefont {G.}~\bibnamefont {Dyer}}, \bibinfo {author} {\bibfnamefont
  {J.}~\bibnamefont {Landry}}, \bibinfo {author} {\bibfnamefont
  {G.}~\bibnamefont {Bansal}}, \bibinfo {author} {\bibfnamefont
  {A.}~\bibnamefont {Bernstein}}, \bibinfo {author} {\bibfnamefont {T.~D.}\
  \bibnamefont {Donnelly}}, \bibinfo {author} {\bibfnamefont {A.}~\bibnamefont
  {Karmakar}}, \bibinfo {author} {\bibfnamefont {A.}~\bibnamefont {Pukhov}}, \
  and\ \bibinfo {author} {\bibfnamefont {T.}~\bibnamefont {Ditmire}}} (\bibinfo
  {year} {2007}{\natexlab{b}}),\ \bibfield  {title} {\enquote {\bibinfo {title}
  {Hot electron and x-ray production from intense laser irradiation of
  wavelength-scale polystyrene spheres},}\ }\href@noop {} {\bibfield  {journal}
  {\bibinfo  {journal} {Phys. Plasmas}\ }\textbf {\bibinfo {volume} {14}},\
  \bibinfo {pages} {062704}}\BibitemShut {NoStop}%
\bibitem [{\citenamefont {S{\"u}{\ss}mann}(2013)}]{SuessmannPhD}%
  \BibitemOpen
  \bibfield  {author} {\bibinfo {author} {\bibnamefont {S{\"u}{\ss}mann},
  \bibfnamefont {F}}} (\bibinfo {year} {2013}),\ \emph {\bibinfo {title}
  {Attosecond dynamics of nano-localized fields probed by photoelectron
  spectroscopy}},\ \href@noop {} {Ph.D. thesis}\ (\bibinfo  {school} {LMU
  Munich})\BibitemShut {NoStop}%
\bibitem [{\citenamefont {S{\"u}{\ss}mann}\ and\ \citenamefont
  {Kling}(2011{\natexlab{a}})}]{SuessmannSPIE11}%
  \BibitemOpen
  \bibfield  {author} {\bibinfo {author} {\bibnamefont {S{\"u}{\ss}mann},
  \bibfnamefont {F}}, \ and\ \bibinfo {author} {\bibfnamefont {M.~F.}\
  \bibnamefont {Kling}}} (\bibinfo {year} {2011}{\natexlab{a}}),\ \bibfield
  {title} {\enquote {\bibinfo {title} {Attosecond measurement of petahertz
  plasmonic near-fields},}\ }\href@noop {} {\bibfield  {journal} {\bibinfo
  {journal} {Proc. of SPIE}\ }\textbf {\bibinfo {volume} {8096}},\ \bibinfo
  {pages} {80961C}}\BibitemShut {NoStop}%
\bibitem [{\citenamefont {S{\"u}{\ss}mann}\ and\ \citenamefont
  {Kling}(2011{\natexlab{b}})}]{SuessmannPRB11}%
  \BibitemOpen
  \bibfield  {author} {\bibinfo {author} {\bibnamefont {S{\"u}{\ss}mann},
  \bibfnamefont {F}}, \ and\ \bibinfo {author} {\bibfnamefont {M.~F.}\
  \bibnamefont {Kling}}} (\bibinfo {year} {2011}{\natexlab{b}}),\ \bibfield
  {title} {\enquote {\bibinfo {title} {Attosecond nanoplasmonic streaking of
  localized fields near metal nanospheres},}\ }\href@noop {} {\bibfield
  {journal} {\bibinfo  {journal} {Phys. Rev. B}\ }\textbf {\bibinfo {volume}
  {84}},\ \bibinfo {pages} {121406(R)}}\BibitemShut {NoStop}%
\bibitem [{\citenamefont {S{\"u}{\ss}mann}\ \emph {et~al.}(2015)\citenamefont
  {S{\"u}{\ss}mann}, \citenamefont {Seiffert}, \citenamefont {Zherebtsov},
  \citenamefont {Mondes}, \citenamefont {Stierle}, \citenamefont {Arbeiter},
  \citenamefont {Plenge}, \citenamefont {Rupp}, \citenamefont {Peltz},
  \citenamefont {Kessel}, \citenamefont {Trushin}, \citenamefont {Ahn},
  \citenamefont {Kim}, \citenamefont {Graf}, \citenamefont {R{\"u}hl},
  \citenamefont {Kling},\ and\ \citenamefont {Fennel}}]{Suessmann15}%
  \BibitemOpen
  \bibfield  {author} {\bibinfo {author} {\bibnamefont {S{\"u}{\ss}mann},
  \bibfnamefont {F}}, \bibinfo {author} {\bibfnamefont {L.}~\bibnamefont
  {Seiffert}}, \bibinfo {author} {\bibfnamefont {S.}~\bibnamefont
  {Zherebtsov}}, \bibinfo {author} {\bibfnamefont {V.}~\bibnamefont {Mondes}},
  \bibinfo {author} {\bibfnamefont {J.}~\bibnamefont {Stierle}}, \bibinfo
  {author} {\bibfnamefont {M.}~\bibnamefont {Arbeiter}}, \bibinfo {author}
  {\bibfnamefont {J.}~\bibnamefont {Plenge}}, \bibinfo {author} {\bibfnamefont
  {P.}~\bibnamefont {Rupp}}, \bibinfo {author} {\bibfnamefont {C.}~\bibnamefont
  {Peltz}}, \bibinfo {author} {\bibfnamefont {A.}~\bibnamefont {Kessel}},
  \bibinfo {author} {\bibfnamefont {S.~A.}\ \bibnamefont {Trushin}}, \bibinfo
  {author} {\bibfnamefont {B.}~\bibnamefont {Ahn}}, \bibinfo {author}
  {\bibfnamefont {D.}~\bibnamefont {Kim}}, \bibinfo {author} {\bibfnamefont
  {C.}~\bibnamefont {Graf}}, \bibinfo {author} {\bibfnamefont {E.}~\bibnamefont
  {R{\"u}hl}}, \bibinfo {author} {\bibfnamefont {M.~F.}\ \bibnamefont {Kling}},
  \ and\ \bibinfo {author} {\bibfnamefont {T.}~\bibnamefont {Fennel}}}
  (\bibinfo {year} {2015}),\ \bibfield  {title} {\enquote {\bibinfo {title}
  {Field propagation-induced directionality of carrier-envelope
  phase-controlled photoemission from nanospheres},}\ }\href@noop {} {\bibfield
   {journal} {\bibinfo  {journal} {Nat. Comm.}\ }\textbf {\bibinfo {volume}
  {6}},\ \bibinfo {pages} {7944}}\BibitemShut {NoStop}%
\bibitem [{\citenamefont {S\"u{\ss}mann}\ \emph {et~al.}(2014)\citenamefont
  {S\"u{\ss}mann}, \citenamefont {Stebbings}, \citenamefont {Zherebtsov},
  \citenamefont {Chew}, \citenamefont {Stockman}, \citenamefont {R\"uhl},
  \citenamefont {Kleineberg}, \citenamefont {Fennel},\ and\ \citenamefont
  {Kling}}]{Sussmann2014}%
  \BibitemOpen
  \bibfield  {author} {\bibinfo {author} {\bibnamefont {S\"u{\ss}mann},
  \bibfnamefont {F}}, \bibinfo {author} {\bibfnamefont {S.~L.}\ \bibnamefont
  {Stebbings}}, \bibinfo {author} {\bibfnamefont {S.}~\bibnamefont
  {Zherebtsov}}, \bibinfo {author} {\bibfnamefont {S.~H.}\ \bibnamefont
  {Chew}}, \bibinfo {author} {\bibfnamefont {M.~I.}\ \bibnamefont {Stockman}},
  \bibinfo {author} {\bibfnamefont {E.}~\bibnamefont {R\"uhl}}, \bibinfo
  {author} {\bibfnamefont {U.}~\bibnamefont {Kleineberg}}, \bibinfo {author}
  {\bibfnamefont {T.}~\bibnamefont {Fennel}}, \ and\ \bibinfo {author}
  {\bibfnamefont {M.~F.}\ \bibnamefont {Kling}}} (\bibinfo {year} {2014}),\
  \bibfield  {title} {\enquote {\bibinfo {title} {Attosecond nanophysics},}\
  }in\ \href {\doibase 10.1002/9783527677689.ch14} {\emph {\bibinfo {booktitle}
  {Attosecond and {XUV} {P}hysics}}}\ (\bibinfo  {publisher} {Wiley-VCH})\ pp.\
  \bibinfo {pages} {421--462}\BibitemShut {NoStop}%
\bibitem [{\citenamefont {S{\"u}{\ss}mann}\ \emph {et~al.}(2011)\citenamefont
  {S{\"u}{\ss}mann}, \citenamefont {Zherebtsov}, \citenamefont {Plenge},
  \citenamefont {Johnson}, \citenamefont {K{\"u}bel}, \citenamefont {Sayler},
  \citenamefont {Mondes}, \citenamefont {Graf}, \citenamefont {R{\"u}hl},
  \citenamefont {Paulus}, \citenamefont {Schmischke}, \citenamefont
  {Swrschek},\ and\ \citenamefont {Kling}}]{Suessmann11}%
  \BibitemOpen
  \bibfield  {author} {\bibinfo {author} {\bibnamefont {S{\"u}{\ss}mann},
  \bibfnamefont {F}}, \bibinfo {author} {\bibfnamefont {S.}~\bibnamefont
  {Zherebtsov}}, \bibinfo {author} {\bibfnamefont {J.}~\bibnamefont {Plenge}},
  \bibinfo {author} {\bibfnamefont {Nora~G.}\ \bibnamefont {Johnson}}, \bibinfo
  {author} {\bibfnamefont {M.}~\bibnamefont {K{\"u}bel}}, \bibinfo {author}
  {\bibfnamefont {A.~M.}\ \bibnamefont {Sayler}}, \bibinfo {author}
  {\bibfnamefont {V.}~\bibnamefont {Mondes}}, \bibinfo {author} {\bibfnamefont
  {C.}~\bibnamefont {Graf}}, \bibinfo {author} {\bibfnamefont {E.}~\bibnamefont
  {R{\"u}hl}}, \bibinfo {author} {\bibfnamefont {G.~G.}\ \bibnamefont
  {Paulus}}, \bibinfo {author} {\bibfnamefont {D.}~\bibnamefont {Schmischke}},
  \bibinfo {author} {\bibfnamefont {P.}~\bibnamefont {Swrschek}}, \ and\
  \bibinfo {author} {\bibfnamefont {M.~F.}\ \bibnamefont {Kling}}} (\bibinfo
  {year} {2011}),\ \bibfield  {title} {\enquote {\bibinfo {title} {Single-shot
  velocity-map imaging of attosecond light-field control at kilohertz rate},}\
  }\href@noop {} {\bibfield  {journal} {\bibinfo  {journal} {Rev. Sci. Instr.}\
  }\textbf {\bibinfo {volume} {82}},\ \bibinfo {pages} {093109}}\BibitemShut
  {NoStop}%
\bibitem [{\citenamefont {Swanwick}\ \emph {et~al.}(2014)\citenamefont
  {Swanwick}, \citenamefont {Keathley}, \citenamefont {Fallahi}, \citenamefont
  {Krogen}, \citenamefont {Laurent}, \citenamefont {Moses}, \citenamefont
  {K\"artner},\ and\ \citenamefont {Vel\'asquez-Garc\'{\i}a}}]{Swanwick2014}%
  \BibitemOpen
  \bibfield  {author} {\bibinfo {author} {\bibnamefont {Swanwick},
  \bibfnamefont {M~E}}, \bibinfo {author} {\bibfnamefont {P.~D.}\ \bibnamefont
  {Keathley}}, \bibinfo {author} {\bibfnamefont {A.}~\bibnamefont {Fallahi}},
  \bibinfo {author} {\bibfnamefont {P.~R.}\ \bibnamefont {Krogen}}, \bibinfo
  {author} {\bibfnamefont {G.}~\bibnamefont {Laurent}}, \bibinfo {author}
  {\bibfnamefont {J.}~\bibnamefont {Moses}}, \bibinfo {author} {\bibfnamefont
  {F.~X.}\ \bibnamefont {K\"artner}}, \ and\ \bibinfo {author} {\bibfnamefont
  {L.~F.}\ \bibnamefont {Vel\'asquez-Garc\'{\i}a}}} (\bibinfo {year} {2014}),\
  \bibfield  {title} {\enquote {\bibinfo {title} {Nanostructured ultrafast
  silicon-tip optical field-emitter arrays},}\ }\href@noop {} {\bibfield
  {journal} {\bibinfo  {journal} {Nano Lett.}\ }\textbf {\bibinfo {volume}
  {14}},\ \bibinfo {pages} {5035--5043}}\BibitemShut {NoStop}%
\bibitem [{\citenamefont {Symes}\ \emph {et~al.}(2004)\citenamefont {Symes},
  \citenamefont {Comley},\ and\ \citenamefont {Smith}}]{Symes2004}%
  \BibitemOpen
  \bibfield  {author} {\bibinfo {author} {\bibnamefont {Symes}, \bibfnamefont
  {D~R}}, \bibinfo {author} {\bibfnamefont {A.~J.}\ \bibnamefont {Comley}}, \
  and\ \bibinfo {author} {\bibfnamefont {R.~A.}\ \bibnamefont {Smith}}}
  (\bibinfo {year} {2004}),\ \bibfield  {title} {\enquote {\bibinfo {title}
  {Fast-ion production from short-pulse irradiation of ethanol
  microdroplets},}\ }\href@noop {} {\bibfield  {journal} {\bibinfo  {journal}
  {Phys. Rev. Lett.}\ }\textbf {\bibinfo {volume} {93}},\ \bibinfo {pages}
  {145004}}\BibitemShut {NoStop}%
\bibitem [{\citenamefont {Taflove}\ and\ \citenamefont
  {Hagness}(2005)}]{Taflove2005}%
  \BibitemOpen
  \bibfield  {author} {\bibinfo {author} {\bibnamefont {Taflove}, \bibfnamefont
  {A}}, \ and\ \bibinfo {author} {\bibfnamefont {S.~C.}\ \bibnamefont
  {Hagness}}} (\bibinfo {year} {2005}),\ \href@noop {} {\emph {\bibinfo {title}
  {Computational {E}lectrodynamics: {T}he {F}inite-{D}ifference {T}ime-{D}omain
  {M}ethod}}}\ (\bibinfo  {publisher} {Artech House, Norwood})\BibitemShut
  {NoStop}%
\bibitem [{\citenamefont {Tanuma}\ \emph {et~al.}(2011)\citenamefont {Tanuma},
  \citenamefont {Powell},\ and\ \citenamefont
  {Penn}}]{tanuma_calculations_2011}%
  \BibitemOpen
  \bibfield  {author} {\bibinfo {author} {\bibnamefont {Tanuma}, \bibfnamefont
  {S}}, \bibinfo {author} {\bibfnamefont {C.~J.}\ \bibnamefont {Powell}}, \
  and\ \bibinfo {author} {\bibfnamefont {D.~R.}\ \bibnamefont {Penn}}}
  (\bibinfo {year} {2011}),\ \bibfield  {title} {\enquote {\bibinfo {title}
  {Calculations of electron inelastic mean free paths. {IX}. {Data} for 41
  elemental solids over the 50 {eV} to 30 {keV} range},}\ }\href@noop {}
  {\bibfield  {journal} {\bibinfo  {journal} {Surf. Interface Anal.}\ }\textbf
  {\bibinfo {volume} {43}},\ \bibinfo {pages} {689--713}}\BibitemShut {NoStop}%
\bibitem [{\citenamefont {Tanuma}\ \emph {et~al.}(2005)\citenamefont {Tanuma},
  \citenamefont {Shiratori}, \citenamefont {Kimura}, \citenamefont {Goto},
  \citenamefont {Ichimura},\ and\ \citenamefont
  {Powell}}]{tanuma_experimental_2005}%
  \BibitemOpen
  \bibfield  {author} {\bibinfo {author} {\bibnamefont {Tanuma}, \bibfnamefont
  {S}}, \bibinfo {author} {\bibfnamefont {T.}~\bibnamefont {Shiratori}},
  \bibinfo {author} {\bibfnamefont {T.}~\bibnamefont {Kimura}}, \bibinfo
  {author} {\bibfnamefont {K.}~\bibnamefont {Goto}}, \bibinfo {author}
  {\bibfnamefont {S.}~\bibnamefont {Ichimura}}, \ and\ \bibinfo {author}
  {\bibfnamefont {C.~J.}\ \bibnamefont {Powell}}} (\bibinfo {year} {2005}),\
  \bibfield  {title} {\enquote {\bibinfo {title} {Experimental determination of
  electron inelastic mean free paths in 13 elemental solids in the 50 to 5000
  {eV} energy range by elastic-peak electron spectroscopy},}\ }\href@noop {}
  {\bibfield  {journal} {\bibinfo  {journal} {Surf. Interface Anal.}\ }\textbf
  {\bibinfo {volume} {37}},\ \bibinfo {pages} {833--845}}\BibitemShut {NoStop}%
\bibitem [{\citenamefont {T\H{o}k\'{e}si}\ \emph {et~al.}(2001)\citenamefont
  {T\H{o}k\'{e}si}, \citenamefont {Wirtz}, \citenamefont {Lemell},\ and\
  \citenamefont {Burgd\"{o}rfer}}]{Toekesi2001}%
  \BibitemOpen
  \bibfield  {author} {\bibinfo {author} {\bibnamefont {T\H{o}k\'{e}si},
  \bibfnamefont {K}}, \bibinfo {author} {\bibfnamefont {L.}~\bibnamefont
  {Wirtz}}, \bibinfo {author} {\bibfnamefont {C.}~\bibnamefont {Lemell}}, \
  and\ \bibinfo {author} {\bibfnamefont {J.}~\bibnamefont {Burgd\"{o}rfer}}}
  (\bibinfo {year} {2001}),\ \bibfield  {title} {\enquote {\bibinfo {title}
  {Hollow-ion formation in microcapillaries},}\ }\href@noop {} {\bibfield
  {journal} {\bibinfo  {journal} {Phys. Rev. A}\ }\textbf {\bibinfo {volume}
  {64}},\ \bibinfo {pages} {042902}}\BibitemShut {NoStop}%
\bibitem [{\citenamefont {Thomas}\ \emph {et~al.}(2013)\citenamefont {Thomas},
  \citenamefont {Kr\"uger}, \citenamefont {F\"orster}, \citenamefont {Schenk},\
  and\ \citenamefont {Hommelhoff}}]{Thomas2013}%
  \BibitemOpen
  \bibfield  {author} {\bibinfo {author} {\bibnamefont {Thomas}, \bibfnamefont
  {S}}, \bibinfo {author} {\bibfnamefont {M.}~\bibnamefont {Kr\"uger}},
  \bibinfo {author} {\bibfnamefont {M.}~\bibnamefont {F\"orster}}, \bibinfo
  {author} {\bibfnamefont {M.}~\bibnamefont {Schenk}}, \ and\ \bibinfo {author}
  {\bibfnamefont {P.}~\bibnamefont {Hommelhoff}}} (\bibinfo {year} {2013}),\
  \bibfield  {title} {\enquote {\bibinfo {title} {Probing of optical
  near-fields by electron rescattering on the 1 nm scale},}\ }\href@noop {}
  {\bibfield  {journal} {\bibinfo  {journal} {Nano Lett.}\ }\textbf {\bibinfo
  {volume} {13}},\ \bibinfo {pages} {4790--4794}}\BibitemShut {NoStop}%
\bibitem [{\citenamefont {Thomas}\ \emph {et~al.}(2015)\citenamefont {Thomas},
  \citenamefont {Wachter}, \citenamefont {Lemell}, \citenamefont
  {Burgd\"orfer},\ and\ \citenamefont {Hommelhoff}}]{Thomas2015}%
  \BibitemOpen
  \bibfield  {author} {\bibinfo {author} {\bibnamefont {Thomas}, \bibfnamefont
  {S}}, \bibinfo {author} {\bibfnamefont {G.}~\bibnamefont {Wachter}}, \bibinfo
  {author} {\bibfnamefont {C.}~\bibnamefont {Lemell}}, \bibinfo {author}
  {\bibfnamefont {J.}~\bibnamefont {Burgd\"orfer}}, \ and\ \bibinfo {author}
  {\bibfnamefont {P.}~\bibnamefont {Hommelhoff}}} (\bibinfo {year} {2015}),\
  \bibfield  {title} {\enquote {\bibinfo {title} {Large optical field
  enhancement for nanotips with large opening angles},}\ }\href@noop {}
  {\bibfield  {journal} {\bibinfo  {journal} {New. J. Phys.}\ }\textbf
  {\bibinfo {volume} {17}},\ \bibinfo {pages} {063010}}\BibitemShut {NoStop}%
\bibitem [{\citenamefont {Tikman}\ \emph {et~al.}(2016)\citenamefont {Tikman},
  \citenamefont {Yavuz}, \citenamefont {Ciappina}, \citenamefont {Chac\'on},
  \citenamefont {Altun},\ and\ \citenamefont {Lewenstein}}]{rydberg}%
  \BibitemOpen
  \bibfield  {author} {\bibinfo {author} {\bibnamefont {Tikman}, \bibfnamefont
  {Y}}, \bibinfo {author} {\bibfnamefont {I.}~\bibnamefont {Yavuz}}, \bibinfo
  {author} {\bibfnamefont {M.~F.}\ \bibnamefont {Ciappina}}, \bibinfo {author}
  {\bibfnamefont {A.}~\bibnamefont {Chac\'on}}, \bibinfo {author}
  {\bibfnamefont {Z.}~\bibnamefont {Altun}}, \ and\ \bibinfo {author}
  {\bibfnamefont {M.}~\bibnamefont {Lewenstein}}} (\bibinfo {year} {2016}),\
  \bibfield  {title} {\enquote {\bibinfo {title} {High-order-harmonic
  generation from rydberg atoms driven by plasmon-enhanced laser fields},}\
  }\href@noop {} {\bibfield  {journal} {\bibinfo  {journal} {Phys. Rev. A}\
  }\textbf {\bibinfo {volume} {93}},\ \bibinfo {pages} {023410}}\BibitemShut
  {NoStop}%
\bibitem [{\citenamefont {Tisch}\ \emph {et~al.}(1997)\citenamefont {Tisch},
  \citenamefont {Ditmire}, \citenamefont {Fraser}, \citenamefont {Hay},
  \citenamefont {Mason}, \citenamefont {Springate}, \citenamefont {Marangos},\
  and\ \citenamefont {Hutchinson}}]{Tisch1997}%
  \BibitemOpen
  \bibfield  {author} {\bibinfo {author} {\bibnamefont {Tisch}, \bibfnamefont
  {J~W~G}}, \bibinfo {author} {\bibfnamefont {T.}~\bibnamefont {Ditmire}},
  \bibinfo {author} {\bibfnamefont {D.~J.}\ \bibnamefont {Fraser}}, \bibinfo
  {author} {\bibfnamefont {N.}~\bibnamefont {Hay}}, \bibinfo {author}
  {\bibfnamefont {M.~B.}\ \bibnamefont {Mason}}, \bibinfo {author}
  {\bibfnamefont {E.}~\bibnamefont {Springate}}, \bibinfo {author}
  {\bibfnamefont {J.~P.}\ \bibnamefont {Marangos}}, \ and\ \bibinfo {author}
  {\bibfnamefont {M.~H.~R.}\ \bibnamefont {Hutchinson}}} (\bibinfo {year}
  {1997}),\ \bibfield  {title} {\enquote {\bibinfo {title} {Investigation of
  high-harmonic generation from xenon atom clusters},}\ }\href@noop {}
  {\bibfield  {journal} {\bibinfo  {journal} {J. Phys. B}\ }\textbf {\bibinfo
  {volume} {30}},\ \bibinfo {pages} {L709}}\BibitemShut {NoStop}%
\bibitem [{\citenamefont {T{\'o}th}\ \emph {et~al.}(1991)\citenamefont
  {T{\'o}th}, \citenamefont {Farkas},\ and\ \citenamefont
  {Vodopyanov}}]{Toth1991}%
  \BibitemOpen
  \bibfield  {author} {\bibinfo {author} {\bibnamefont {T{\'o}th},
  \bibfnamefont {C}}, \bibinfo {author} {\bibfnamefont {G.}~\bibnamefont
  {Farkas}}, \ and\ \bibinfo {author} {\bibfnamefont {K.~L.}\ \bibnamefont
  {Vodopyanov}}} (\bibinfo {year} {1991}),\ \bibfield  {title} {\enquote
  {\bibinfo {title} {Laser-induced electron emission from an au surface
  irradiated bey single picosecond pulses at $\lambda = 2.94\,\mu$m. the
  intermediate region between multiphoton and tunneling effect},}\ }\href@noop
  {} {\bibfield  {journal} {\bibinfo  {journal} {Appl. Phys. B}\ }\textbf
  {\bibinfo {volume} {53}},\ \bibinfo {pages} {221--225}}\BibitemShut {NoStop}%
\bibitem [{\citenamefont {Trebino}\ \emph {et~al.}(1997)\citenamefont
  {Trebino}, \citenamefont {DeLong}, \citenamefont {Fittinghoff}, \citenamefont
  {Sweetser}, \citenamefont {Krumb{\"u}gel}, \citenamefont {Richman},\ and\
  \citenamefont {Kane}}]{trebino1997measuring}%
  \BibitemOpen
  \bibfield  {author} {\bibinfo {author} {\bibnamefont {Trebino}, \bibfnamefont
  {R}}, \bibinfo {author} {\bibfnamefont {K.~W.}\ \bibnamefont {DeLong}},
  \bibinfo {author} {\bibfnamefont {D.~N.}\ \bibnamefont {Fittinghoff}},
  \bibinfo {author} {\bibfnamefont {J.~N.}\ \bibnamefont {Sweetser}}, \bibinfo
  {author} {\bibfnamefont {M.~A.}\ \bibnamefont {Krumb{\"u}gel}}, \bibinfo
  {author} {\bibfnamefont {B.~A.}\ \bibnamefont {Richman}}, \ and\ \bibinfo
  {author} {\bibfnamefont {D.~J.}\ \bibnamefont {Kane}}} (\bibinfo {year}
  {1997}),\ \bibfield  {title} {\enquote {\bibinfo {title} {Measuring
  ultrashort laser pulses in the time-frequency domain using frequency-resolved
  optical gating},}\ }\href@noop {} {\bibfield  {journal} {\bibinfo  {journal}
  {Rev. of Sci. Instr.}\ }\textbf {\bibinfo {volume} {68}},\ \bibinfo {pages}
  {3277--3295}}\BibitemShut {NoStop}%
\bibitem [{\citenamefont {Tsujino}\ \emph {et~al.}(2008)\citenamefont
  {Tsujino}, \citenamefont {Beaud}, \citenamefont {Kirk}, \citenamefont
  {Vogel}, \citenamefont {Sehr}, \citenamefont {Gobrecht},\ and\ \citenamefont
  {Wrulich}}]{Tsujino2008}%
  \BibitemOpen
  \bibfield  {author} {\bibinfo {author} {\bibnamefont {Tsujino}, \bibfnamefont
  {S}}, \bibinfo {author} {\bibfnamefont {P.}~\bibnamefont {Beaud}}, \bibinfo
  {author} {\bibfnamefont {E.}~\bibnamefont {Kirk}}, \bibinfo {author}
  {\bibfnamefont {T.}~\bibnamefont {Vogel}}, \bibinfo {author} {\bibfnamefont
  {H.}~\bibnamefont {Sehr}}, \bibinfo {author} {\bibfnamefont {J.}~\bibnamefont
  {Gobrecht}}, \ and\ \bibinfo {author} {\bibfnamefont {A.}~\bibnamefont
  {Wrulich}}} (\bibinfo {year} {2008}),\ \bibfield  {title} {\enquote {\bibinfo
  {title} {Ultrafast electron emission from metallic nanotip arrays induced by
  near infrared femtosecond laser pulses},}\ }\href@noop {} {\bibfield
  {journal} {\bibinfo  {journal} {Appl. Phys. Lett.}\ }\textbf {\bibinfo
  {volume} {92}},\ \bibinfo {pages} {193501}}\BibitemShut {NoStop}%
\bibitem [{\citenamefont {Tsujino}\ \emph {et~al.}(2009)\citenamefont
  {Tsujino}, \citenamefont {le~Pimpec}, \citenamefont {Raabe}, \citenamefont
  {Buess}, \citenamefont {Dehler}, \citenamefont {Kirk}, \citenamefont
  {Gobrecht},\ and\ \citenamefont {Wrulich}}]{Tsujino2009}%
  \BibitemOpen
  \bibfield  {author} {\bibinfo {author} {\bibnamefont {Tsujino}, \bibfnamefont
  {S}}, \bibinfo {author} {\bibfnamefont {F.}~\bibnamefont {le~Pimpec}},
  \bibinfo {author} {\bibfnamefont {J.}~\bibnamefont {Raabe}}, \bibinfo
  {author} {\bibfnamefont {M.}~\bibnamefont {Buess}}, \bibinfo {author}
  {\bibfnamefont {M.}~\bibnamefont {Dehler}}, \bibinfo {author} {\bibfnamefont
  {E.}~\bibnamefont {Kirk}}, \bibinfo {author} {\bibfnamefont {J.}~\bibnamefont
  {Gobrecht}}, \ and\ \bibinfo {author} {\bibfnamefont {A.}~\bibnamefont
  {Wrulich}}} (\bibinfo {year} {2009}),\ \bibfield  {title} {\enquote {\bibinfo
  {title} {Static and optical field enhancement in metallic nanotips studied by
  two-photon photoemission microscopy and spectroscopy excited by picosecond
  laser pulses},}\ }\href@noop {} {\bibfield  {journal} {\bibinfo  {journal}
  {Appl. Phys. Lett.}\ }\textbf {\bibinfo {volume} {94}},\ \bibinfo {pages}
  {093508}}\BibitemShut {NoStop}%
\bibitem [{\citenamefont {Uiberacker}\ \emph {et~al.}(2007)\citenamefont
  {Uiberacker}, \citenamefont {Uphues}, \citenamefont {Schultze}, \citenamefont
  {Verhoef}, \citenamefont {Yakovlev}, \citenamefont {Kling}, \citenamefont
  {Rauschenberger}, \citenamefont {Kabachnik}, \citenamefont {Schr{\"o}der},
  \citenamefont {Lezius}, \citenamefont {Kompa}, \citenamefont {Muller},
  \citenamefont {Vrakking}, \citenamefont {Hendel}, \citenamefont {Kleineberg},
  \citenamefont {Heinzmann}, \citenamefont {Drescher},\ and\ \citenamefont
  {Krausz}}]{uiberacker2007attosecond}%
  \BibitemOpen
  \bibfield  {author} {\bibinfo {author} {\bibnamefont {Uiberacker},
  \bibfnamefont {M}}, \bibinfo {author} {\bibfnamefont {Th.}\ \bibnamefont
  {Uphues}}, \bibinfo {author} {\bibfnamefont {M.}~\bibnamefont {Schultze}},
  \bibinfo {author} {\bibfnamefont {A.~J.}\ \bibnamefont {Verhoef}}, \bibinfo
  {author} {\bibfnamefont {V.}~\bibnamefont {Yakovlev}}, \bibinfo {author}
  {\bibfnamefont {M.~F.}\ \bibnamefont {Kling}}, \bibinfo {author}
  {\bibfnamefont {J.}~\bibnamefont {Rauschenberger}}, \bibinfo {author}
  {\bibfnamefont {N.~M.}\ \bibnamefont {Kabachnik}}, \bibinfo {author}
  {\bibfnamefont {H.}~\bibnamefont {Schr{\"o}der}}, \bibinfo {author}
  {\bibfnamefont {M.}~\bibnamefont {Lezius}}, \bibinfo {author} {\bibfnamefont
  {K.~L.}\ \bibnamefont {Kompa}}, \bibinfo {author} {\bibfnamefont {H.-G.}\
  \bibnamefont {Muller}}, \bibinfo {author} {\bibfnamefont {M.~J.~J.}\
  \bibnamefont {Vrakking}}, \bibinfo {author} {\bibfnamefont {S.}~\bibnamefont
  {Hendel}}, \bibinfo {author} {\bibfnamefont {U.}~\bibnamefont {Kleineberg}},
  \bibinfo {author} {\bibfnamefont {U.}~\bibnamefont {Heinzmann}}, \bibinfo
  {author} {\bibfnamefont {M.}~\bibnamefont {Drescher}}, \ and\ \bibinfo
  {author} {\bibfnamefont {F.}~\bibnamefont {Krausz}}} (\bibinfo {year}
  {2007}),\ \bibfield  {title} {\enquote {\bibinfo {title} {Attosecond
  real-time observation of electron tunnelling in atoms},}\ }\href@noop {}
  {\bibfield  {journal} {\bibinfo  {journal} {Nature}\ }\textbf {\bibinfo
  {volume} {446}},\ \bibinfo {pages} {627--632}}\BibitemShut {NoStop}%
\bibitem [{\citenamefont {Uphues}\ \emph {et~al.}(2008)\citenamefont {Uphues},
  \citenamefont {Schultze}, \citenamefont {Kling}, \citenamefont {Uiberacker},
  \citenamefont {Hendel}, \citenamefont {Heinzmann}, \citenamefont
  {Kabachnik},\ and\ \citenamefont {Drescher}}]{uphues2008ion}%
  \BibitemOpen
  \bibfield  {author} {\bibinfo {author} {\bibnamefont {Uphues}, \bibfnamefont
  {Th}}, \bibinfo {author} {\bibfnamefont {M.}~\bibnamefont {Schultze}},
  \bibinfo {author} {\bibfnamefont {M.~F.}\ \bibnamefont {Kling}}, \bibinfo
  {author} {\bibfnamefont {M.}~\bibnamefont {Uiberacker}}, \bibinfo {author}
  {\bibfnamefont {S.}~\bibnamefont {Hendel}}, \bibinfo {author} {\bibfnamefont
  {U.}~\bibnamefont {Heinzmann}}, \bibinfo {author} {\bibfnamefont {N.~M.}\
  \bibnamefont {Kabachnik}}, \ and\ \bibinfo {author} {\bibfnamefont
  {M.}~\bibnamefont {Drescher}}} (\bibinfo {year} {2008}),\ \bibfield  {title}
  {\enquote {\bibinfo {title} {Ion-charge-state chronoscopy of cascaded atomic
  auger decay},}\ }\href@noop {} {\bibfield  {journal} {\bibinfo  {journal}
  {New J. of Phys.}\ }\textbf {\bibinfo {volume} {10}},\ \bibinfo {pages}
  {025009}}\BibitemShut {NoStop}%
\bibitem [{\citenamefont {Vampa}\ \emph {et~al.}(2015)\citenamefont {Vampa},
  \citenamefont {Hammond}, \citenamefont {Thir\'e}, \citenamefont {Schmidt},
  \citenamefont {Légar\'e}, \citenamefont {McDonald}, \citenamefont {Brabec},\
  and\ \citenamefont {Corkum}}]{Vampa2015}%
  \BibitemOpen
  \bibfield  {author} {\bibinfo {author} {\bibnamefont {Vampa}, \bibfnamefont
  {G}}, \bibinfo {author} {\bibfnamefont {T.~J.}\ \bibnamefont {Hammond}},
  \bibinfo {author} {\bibfnamefont {N.}~\bibnamefont {Thir\'e}}, \bibinfo
  {author} {\bibfnamefont {B.~E.}\ \bibnamefont {Schmidt}}, \bibinfo {author}
  {\bibfnamefont {F.}~\bibnamefont {Légar\'e}}, \bibinfo {author}
  {\bibfnamefont {C.~R.}\ \bibnamefont {McDonald}}, \bibinfo {author}
  {\bibfnamefont {T.}~\bibnamefont {Brabec}}, \ and\ \bibinfo {author}
  {\bibfnamefont {P.~B.}\ \bibnamefont {Corkum}}} (\bibinfo {year} {2015}),\
  \bibfield  {title} {\enquote {\bibinfo {title} {Linking high harmonics from
  gases and solids},}\ }\href@noop {} {\bibfield  {journal} {\bibinfo
  {journal} {Nature}\ }\textbf {\bibinfo {volume} {522}},\ \bibinfo {pages}
  {462--464}}\BibitemShut {NoStop}%
\bibitem [{\citenamefont {Veronis}\ and\ \citenamefont
  {Fan}(2007)}]{veronis_overview_2007}%
  \BibitemOpen
  \bibfield  {author} {\bibinfo {author} {\bibnamefont {Veronis}, \bibfnamefont
  {G}}, \ and\ \bibinfo {author} {\bibfnamefont {S.}~\bibnamefont {Fan}}}
  (\bibinfo {year} {2007}),\ \bibfield  {title} {\enquote {\bibinfo {title}
  {Overview of simulation techniques for plasmonic devices},}\ }in\ \href@noop
  {} {\emph {\bibinfo {booktitle} {Surface {Plasmon} {Nanophotonics}}}},\
  \bibinfo {series and number} {\bibinfo {series} {Springer {Series} in
  {Optical} {Sciences}}\ No.\ \bibinfo {number} {131}},\ \bibinfo {editor}
  {edited by\ \bibinfo {editor} {\bibfnamefont {M.~L.}\ \bibnamefont
  {Brongersma}}\ and\ \bibinfo {editor} {\bibfnamefont {P.~G.}\ \bibnamefont
  {Kik}}}\ (\bibinfo  {publisher} {Springer Netherlands})\ pp.\ \bibinfo
  {pages} {169--182}\BibitemShut {NoStop}%
\bibitem [{\citenamefont {Vesseur}\ \emph {et~al.}(2007)\citenamefont
  {Vesseur}, \citenamefont {de~Waele}, \citenamefont {Kuttge},\ and\
  \citenamefont {Polman}}]{Vesseur2007}%
  \BibitemOpen
  \bibfield  {author} {\bibinfo {author} {\bibnamefont {Vesseur}, \bibfnamefont
  {E~J~R}}, \bibinfo {author} {\bibfnamefont {R.}~\bibnamefont {de~Waele}},
  \bibinfo {author} {\bibfnamefont {M.}~\bibnamefont {Kuttge}}, \ and\ \bibinfo
  {author} {\bibfnamefont {A.}~\bibnamefont {Polman}}} (\bibinfo {year}
  {2007}),\ \bibfield  {title} {\enquote {\bibinfo {title} {Direct observation
  of plasmonic modes in au nanowires using high-resolution cathodoluminescence
  spectroscopy},}\ }\href@noop {} {\bibfield  {journal} {\bibinfo  {journal}
  {Nano Lett.}\ }\textbf {\bibinfo {volume} {7}},\ \bibinfo {pages}
  {2843--2846}}\BibitemShut {NoStop}%
\bibitem [{\citenamefont {Vogelsang}\ \emph {et~al.}(2015)\citenamefont
  {Vogelsang}, \citenamefont {Robin}, \citenamefont {Nagy}, \citenamefont
  {Dombi}, \citenamefont {Rosenkranz}, \citenamefont {Schiek}, \citenamefont
  {Gro{\ss}},\ and\ \citenamefont {Lienau}}]{Vogelsang15}%
  \BibitemOpen
  \bibfield  {author} {\bibinfo {author} {\bibnamefont {Vogelsang},
  \bibfnamefont {J}}, \bibinfo {author} {\bibfnamefont {J.}~\bibnamefont
  {Robin}}, \bibinfo {author} {\bibfnamefont {B.~J}\ \bibnamefont {Nagy}},
  \bibinfo {author} {\bibfnamefont {P.}~\bibnamefont {Dombi}}, \bibinfo
  {author} {\bibfnamefont {D.}~\bibnamefont {Rosenkranz}}, \bibinfo {author}
  {\bibfnamefont {M.}~\bibnamefont {Schiek}}, \bibinfo {author} {\bibfnamefont
  {P.}~\bibnamefont {Gro{\ss}}}, \ and\ \bibinfo {author} {\bibfnamefont
  {C.}~\bibnamefont {Lienau}}} (\bibinfo {year} {2015}),\ \bibfield  {title}
  {\enquote {\bibinfo {title} {Ultrafast electron emission from a sharp metal
  nanotaper driven by adiabatic nanofocusing of surface plasmons},}\
  }\href@noop {} {\bibfield  {journal} {\bibinfo  {journal} {Nano Lett.}\
  }\textbf {\bibinfo {volume} {15}},\ \bibinfo {pages}
  {4685--4691}}\BibitemShut {NoStop}%
\bibitem [{\citenamefont {Wachter}\ \emph {et~al.}(2012)\citenamefont
  {Wachter}, \citenamefont {Lemell}, \citenamefont {Burgd{\"o}rfer},
  \citenamefont {Schenk}, \citenamefont {Kr{\"u}ger},\ and\ \citenamefont
  {Hommelhoff}}]{Wachter12}%
  \BibitemOpen
  \bibfield  {author} {\bibinfo {author} {\bibnamefont {Wachter}, \bibfnamefont
  {G}}, \bibinfo {author} {\bibfnamefont {C.}~\bibnamefont {Lemell}}, \bibinfo
  {author} {\bibfnamefont {J.}~\bibnamefont {Burgd{\"o}rfer}}, \bibinfo
  {author} {\bibfnamefont {M.}~\bibnamefont {Schenk}}, \bibinfo {author}
  {\bibfnamefont {M.}~\bibnamefont {Kr{\"u}ger}}, \ and\ \bibinfo {author}
  {\bibfnamefont {P.}~\bibnamefont {Hommelhoff}}} (\bibinfo {year} {2012}),\
  \bibfield  {title} {\enquote {\bibinfo {title} {Electron rescattering at
  metal nanotips induced by ultrashort laser pulses},}\ }\href@noop {}
  {\bibfield  {journal} {\bibinfo  {journal} {Phys. Rev. B}\ }\textbf {\bibinfo
  {volume} {86}},\ \bibinfo {pages} {035402}}\BibitemShut {NoStop}%
\bibitem [{\citenamefont {Walker}\ \emph {et~al.}(1994)\citenamefont {Walker},
  \citenamefont {Sheehy}, \citenamefont {DiMauro}, \citenamefont {Agostini},
  \citenamefont {Schafer},\ and\ \citenamefont {Kulander}}]{Walker94}%
  \BibitemOpen
  \bibfield  {author} {\bibinfo {author} {\bibnamefont {Walker}, \bibfnamefont
  {B}}, \bibinfo {author} {\bibfnamefont {B.}~\bibnamefont {Sheehy}}, \bibinfo
  {author} {\bibfnamefont {L.~F.}\ \bibnamefont {DiMauro}}, \bibinfo {author}
  {\bibfnamefont {P.}~\bibnamefont {Agostini}}, \bibinfo {author}
  {\bibfnamefont {K.~J.}\ \bibnamefont {Schafer}}, \ and\ \bibinfo {author}
  {\bibfnamefont {K.~C.}\ \bibnamefont {Kulander}}} (\bibinfo {year} {1994}),\
  \bibfield  {title} {\enquote {\bibinfo {title} {Precision measurement of
  strong field double ionization of helium},}\ }\href@noop {} {\bibfield
  {journal} {\bibinfo  {journal} {Phys. Rev. Lett.}\ }\textbf {\bibinfo
  {volume} {73}},\ \bibinfo {pages} {1227}}\BibitemShut {NoStop}%
\bibitem [{\citenamefont {Walmsley}\ and\ \citenamefont
  {Dorrer}(2009)}]{walmsley_characterization_2009}%
  \BibitemOpen
  \bibfield  {author} {\bibinfo {author} {\bibnamefont {Walmsley},
  \bibfnamefont {I~A}}, \ and\ \bibinfo {author} {\bibfnamefont
  {C.}~\bibnamefont {Dorrer}}} (\bibinfo {year} {2009}),\ \bibfield  {title}
  {\enquote {\bibinfo {title} {Characterization of ultrashort electromagnetic
  pulses},}\ }\href@noop {} {\bibfield  {journal} {\bibinfo  {journal} {Adv.
  Opt. Photon.}\ }\textbf {\bibinfo {volume} {1}},\ \bibinfo {pages}
  {308--437}}\BibitemShut {NoStop}%
\bibitem [{\citenamefont {Wang}\ \emph {et~al.}(2014)\citenamefont {Wang},
  \citenamefont {He}, \citenamefont {Luo}, \citenamefont {Lan},\ and\
  \citenamefont {Lu}}]{Cao14a}%
  \BibitemOpen
  \bibfield  {author} {\bibinfo {author} {\bibnamefont {Wang}, \bibfnamefont
  {Z}}, \bibinfo {author} {\bibfnamefont {L.}~\bibnamefont {He}}, \bibinfo
  {author} {\bibfnamefont {J.}~\bibnamefont {Luo}}, \bibinfo {author}
  {\bibfnamefont {P.}~\bibnamefont {Lan}}, \ and\ \bibinfo {author}
  {\bibfnamefont {P.}~\bibnamefont {Lu}}} (\bibinfo {year} {2014}),\ \bibfield
  {title} {\enquote {\bibinfo {title} {High-order harmonic generation from
  rydberg atoms in inhomogeneous fields},}\ }\href@noop {} {\bibfield
  {journal} {\bibinfo  {journal} {Opt. Exp.}\ }\textbf {\bibinfo {volume}
  {22}},\ \bibinfo {pages} {25909--25922}}\BibitemShut {NoStop}%
\bibitem [{\citenamefont {Wang}\ \emph {et~al.}(2013)\citenamefont {Wang},
  \citenamefont {Lan}, \citenamefont {Luo}, \citenamefont {He}, \citenamefont
  {Zhang},\ and\ \citenamefont {Lu}}]{Wang13}%
  \BibitemOpen
  \bibfield  {author} {\bibinfo {author} {\bibnamefont {Wang}, \bibfnamefont
  {Z}}, \bibinfo {author} {\bibfnamefont {P.}~\bibnamefont {Lan}}, \bibinfo
  {author} {\bibfnamefont {J.}~\bibnamefont {Luo}}, \bibinfo {author}
  {\bibfnamefont {L.}~\bibnamefont {He}}, \bibinfo {author} {\bibfnamefont
  {Q.}~\bibnamefont {Zhang}}, \ and\ \bibinfo {author} {\bibfnamefont
  {P.}~\bibnamefont {Lu}}} (\bibinfo {year} {2013}),\ \bibfield  {title}
  {\enquote {\bibinfo {title} {Control of electron dynamics with a multicycle
  two-color spatially inhomogeneous field for efficient single-attosecond-pulse
  generation},}\ }\href@noop {} {\bibfield  {journal} {\bibinfo  {journal}
  {Phys. Rev. A}\ }\textbf {\bibinfo {volume} {88}},\ \bibinfo {pages}
  {063838}}\BibitemShut {NoStop}%
\bibitem [{\citenamefont {Wanzenboeck}\ and\ \citenamefont
  {Waid}(2011)}]{cui_focused_2011}%
  \BibitemOpen
  \bibfield  {author} {\bibinfo {author} {\bibnamefont {Wanzenboeck},
  \bibfnamefont {H~D}}, \ and\ \bibinfo {author} {\bibfnamefont
  {S.}~\bibnamefont {Waid}}} (\bibinfo {year} {2011}),\ \bibfield  {title}
  {\enquote {\bibinfo {title} {Focused {Ion} {Beam} {Lithography}},}\ }in\
  \href@noop {} {\emph {\bibinfo {booktitle} {Recent {Advances} in
  {Nanofabrication} {Techniques} and {Applications}}}},\ \bibinfo {editor}
  {edited by\ \bibinfo {editor} {\bibfnamefont {B.}~\bibnamefont {Cui}}}\
  (\bibinfo  {publisher} {InTech})\ pp.\ \bibinfo {pages} {27--50}\BibitemShut
  {NoStop}%
\bibitem [{\citenamefont {Weber}\ \emph {et~al.}(2015)\citenamefont {Weber},
  \citenamefont {Manschwetus}, \citenamefont {Billon}, \citenamefont
  {B{\"o}ttcher}, \citenamefont {Bougeard}, \citenamefont {Breger},
  \citenamefont {G{\'e}l{\'e}oc}, \citenamefont {Gruson}, \citenamefont
  {Huetz}, \citenamefont {Lin} \emph {et~al.}}]{weber2015flexible}%
  \BibitemOpen
  \bibfield  {author} {\bibinfo {author} {\bibnamefont {Weber}, \bibfnamefont
  {S~J}}, \bibinfo {author} {\bibfnamefont {B.}~\bibnamefont {Manschwetus}},
  \bibinfo {author} {\bibfnamefont {M.}~\bibnamefont {Billon}}, \bibinfo
  {author} {\bibfnamefont {M.}~\bibnamefont {B{\"o}ttcher}}, \bibinfo {author}
  {\bibfnamefont {M.}~\bibnamefont {Bougeard}}, \bibinfo {author}
  {\bibfnamefont {P.}~\bibnamefont {Breger}}, \bibinfo {author} {\bibfnamefont
  {M.}~\bibnamefont {G{\'e}l{\'e}oc}}, \bibinfo {author} {\bibfnamefont
  {V.}~\bibnamefont {Gruson}}, \bibinfo {author} {\bibfnamefont
  {A.}~\bibnamefont {Huetz}}, \bibinfo {author} {\bibfnamefont
  {N.}~\bibnamefont {Lin}},  \emph {et~al.}} (\bibinfo {year} {2015}),\
  \bibfield  {title} {\enquote {\bibinfo {title} {Flexible attosecond beamline
  for high harmonic spectroscopy and xuv/near-ir pump probe experiments
  requiring long acquisition times},}\ }\href@noop {} {\bibfield  {journal}
  {\bibinfo  {journal} {Rev. of Sci. Instr.}\ }\textbf {\bibinfo {volume}
  {86}},\ \bibinfo {pages} {033108}}\BibitemShut {NoStop}%
\bibitem [{\citenamefont {Wessel}(1985)}]{Wessel1985}%
  \BibitemOpen
  \bibfield  {author} {\bibinfo {author} {\bibnamefont {Wessel}, \bibfnamefont
  {J}}} (\bibinfo {year} {1985}),\ \bibfield  {title} {\enquote {\bibinfo
  {title} {Surface-enhanced optical microscopy},}\ }\href@noop {} {\bibfield
  {journal} {\bibinfo  {journal} {J. Opt. Soc. Am. B}\ }\textbf {\bibinfo
  {volume} {2}}~(\bibinfo {number} {9}),\ \bibinfo {pages}
  {1538--1541}}\BibitemShut {NoStop}%
\bibitem [{\citenamefont {Wimmer}\ \emph {et~al.}(2014)\citenamefont {Wimmer},
  \citenamefont {Herink}, \citenamefont {Solli}, \citenamefont {Yalunin},
  \citenamefont {Echternkamp},\ and\ \citenamefont {Ropers}}]{Wimmer2014}%
  \BibitemOpen
  \bibfield  {author} {\bibinfo {author} {\bibnamefont {Wimmer}, \bibfnamefont
  {L}}, \bibinfo {author} {\bibfnamefont {G.}~\bibnamefont {Herink}}, \bibinfo
  {author} {\bibfnamefont {D.~R.}\ \bibnamefont {Solli}}, \bibinfo {author}
  {\bibfnamefont {S.~V.}\ \bibnamefont {Yalunin}}, \bibinfo {author}
  {\bibfnamefont {K.~E.}\ \bibnamefont {Echternkamp}}, \ and\ \bibinfo {author}
  {\bibfnamefont {C.}~\bibnamefont {Ropers}}} (\bibinfo {year} {2014}),\
  \bibfield  {title} {\enquote {\bibinfo {title} {Terahertz control of nanotip
  photoemission},}\ }\href@noop {} {\bibfield  {journal} {\bibinfo  {journal}
  {Nat. Phys.}\ }\textbf {\bibinfo {volume} {10}},\ \bibinfo {pages}
  {432}}\BibitemShut {NoStop}%
\bibitem [{\citenamefont {Winkel}\ \emph {et~al.}(2012)\citenamefont {Winkel},
  \citenamefont {Speck}, \citenamefont {H\"ubner}, \citenamefont {Arnold},
  \citenamefont {Krause},\ and\ \citenamefont {Gibbon}}]{Winkel2012}%
  \BibitemOpen
  \bibfield  {author} {\bibinfo {author} {\bibnamefont {Winkel}, \bibfnamefont
  {M}}, \bibinfo {author} {\bibfnamefont {R.}~\bibnamefont {Speck}}, \bibinfo
  {author} {\bibfnamefont {H.}~\bibnamefont {H\"ubner}}, \bibinfo {author}
  {\bibfnamefont {L.}~\bibnamefont {Arnold}}, \bibinfo {author} {\bibfnamefont
  {R.}~\bibnamefont {Krause}}, \ and\ \bibinfo {author} {\bibfnamefont
  {P.}~\bibnamefont {Gibbon}}} (\bibinfo {year} {2012}),\ \bibfield  {title}
  {\enquote {\bibinfo {title} {A massively parallel, multi-disciplinary
  barnes-hut tree code for extreme-scale n-body simulations},}\ }\href@noop {}
  {\bibfield  {journal} {\bibinfo  {journal} {Comput. Phys. Commun.}\ }\textbf
  {\bibinfo {volume} {183}},\ \bibinfo {pages} {880 -- 889}}\BibitemShut
  {NoStop}%
\bibitem [{\citenamefont {Wirth}\ \emph {et~al.}(2011)\citenamefont {Wirth},
  \citenamefont {Hassan}, \citenamefont {Grgura\u{s}}, \citenamefont {Gagnon},
  \citenamefont {Moulet}, \citenamefont {Luu}, \citenamefont {Pabst},
  \citenamefont {Santra}, \citenamefont {Alahmed}, \citenamefont {Azzeer},
  \citenamefont {Yakovlev}, \citenamefont {Pervak}, \citenamefont {Krausz},\
  and\ \citenamefont {Goulielmakis}}]{wirth_synthesized_2011}%
  \BibitemOpen
  \bibfield  {author} {\bibinfo {author} {\bibnamefont {Wirth}, \bibfnamefont
  {A}}, \bibinfo {author} {\bibfnamefont {M.~Th.}\ \bibnamefont {Hassan}},
  \bibinfo {author} {\bibfnamefont {I.}~\bibnamefont {Grgura\u{s}}}, \bibinfo
  {author} {\bibfnamefont {J.}~\bibnamefont {Gagnon}}, \bibinfo {author}
  {\bibfnamefont {A.}~\bibnamefont {Moulet}}, \bibinfo {author} {\bibfnamefont
  {T.~T.}\ \bibnamefont {Luu}}, \bibinfo {author} {\bibfnamefont
  {S.}~\bibnamefont {Pabst}}, \bibinfo {author} {\bibfnamefont
  {R.}~\bibnamefont {Santra}}, \bibinfo {author} {\bibfnamefont {Z.~A.}\
  \bibnamefont {Alahmed}}, \bibinfo {author} {\bibfnamefont {A.~M.}\
  \bibnamefont {Azzeer}}, \bibinfo {author} {\bibfnamefont {V.~S.}\
  \bibnamefont {Yakovlev}}, \bibinfo {author} {\bibfnamefont {V.}~\bibnamefont
  {Pervak}}, \bibinfo {author} {\bibfnamefont {F.}~\bibnamefont {Krausz}}, \
  and\ \bibinfo {author} {\bibfnamefont {E.}~\bibnamefont {Goulielmakis}}}
  (\bibinfo {year} {2011}),\ \bibfield  {title} {\enquote {\bibinfo {title}
  {Synthesized light transients},}\ }\href@noop {} {\bibfield  {journal}
  {\bibinfo  {journal} {Science}\ }\textbf {\bibinfo {volume} {334}},\ \bibinfo
  {pages} {195--200}}\BibitemShut {NoStop}%
\bibitem [{\citenamefont {Witting}\ \emph {et~al.}(2011)\citenamefont
  {Witting}, \citenamefont {Frank}, \citenamefont {Arrell}, \citenamefont
  {Okell}, \citenamefont {Marangos},\ and\ \citenamefont
  {Tisch}}]{witting_characterization_2011}%
  \BibitemOpen
  \bibfield  {author} {\bibinfo {author} {\bibnamefont {Witting}, \bibfnamefont
  {T}}, \bibinfo {author} {\bibfnamefont {F.}~\bibnamefont {Frank}}, \bibinfo
  {author} {\bibfnamefont {C.~A.}\ \bibnamefont {Arrell}}, \bibinfo {author}
  {\bibfnamefont {W.~A.}\ \bibnamefont {Okell}}, \bibinfo {author}
  {\bibfnamefont {J.~P.}\ \bibnamefont {Marangos}}, \ and\ \bibinfo {author}
  {\bibfnamefont {J.~W.~G.}\ \bibnamefont {Tisch}}} (\bibinfo {year} {2011}),\
  \bibfield  {title} {\enquote {\bibinfo {title} {Characterization of
  high-intensity sub-4-fs laser pulses using spatially encoded spectral
  shearing interferometry.}}\ }\href@noop {} {\bibfield  {journal} {\bibinfo
  {journal} {Opt. Lett.}\ }\textbf {\bibinfo {volume} {36}},\ \bibinfo {pages}
  {1680--1682}}\BibitemShut {NoStop}%
\bibitem [{\citenamefont {Witting}\ \emph {et~al.}(2012)\citenamefont
  {Witting}, \citenamefont {Frank}, \citenamefont {Okell}, \citenamefont
  {Arrell}, \citenamefont {Marangos},\ and\ \citenamefont
  {Tisch}}]{witting2012sub}%
  \BibitemOpen
  \bibfield  {author} {\bibinfo {author} {\bibnamefont {Witting}, \bibfnamefont
  {T}}, \bibinfo {author} {\bibfnamefont {F.}~\bibnamefont {Frank}}, \bibinfo
  {author} {\bibfnamefont {W.~A.}\ \bibnamefont {Okell}}, \bibinfo {author}
  {\bibfnamefont {C.~A.}\ \bibnamefont {Arrell}}, \bibinfo {author}
  {\bibfnamefont {J.~P.}\ \bibnamefont {Marangos}}, \ and\ \bibinfo {author}
  {\bibfnamefont {J.~W.~G.}\ \bibnamefont {Tisch}}} (\bibinfo {year} {2012}),\
  \bibfield  {title} {\enquote {\bibinfo {title} {Sub-4-fs laser pulse
  characterization by spatially resolved spectral shearing interferometry and
  attosecond streaking},}\ }\href@noop {} {\bibfield  {journal} {\bibinfo
  {journal} {J. of Phys. B}\ }\textbf {\bibinfo {volume} {45}},\ \bibinfo
  {pages} {074014}}\BibitemShut {NoStop}%
\bibitem [{\citenamefont {Wittmann}\ \emph {et~al.}(2009)\citenamefont
  {Wittmann}, \citenamefont {Horvath}, \citenamefont {Helml}, \citenamefont
  {Sch\"atzel}, \citenamefont {Gu}, \citenamefont {Cavalieri}, \citenamefont
  {Paulus},\ and\ \citenamefont {R.~Kienberger}}]{Wittmann09}%
  \BibitemOpen
  \bibfield  {author} {\bibinfo {author} {\bibnamefont {Wittmann},
  \bibfnamefont {T}}, \bibinfo {author} {\bibfnamefont {B.}~\bibnamefont
  {Horvath}}, \bibinfo {author} {\bibfnamefont {W.}~\bibnamefont {Helml}},
  \bibinfo {author} {\bibfnamefont {M.~G.}\ \bibnamefont {Sch\"atzel}},
  \bibinfo {author} {\bibfnamefont {X.}~\bibnamefont {Gu}}, \bibinfo {author}
  {\bibfnamefont {A.~L.}\ \bibnamefont {Cavalieri}}, \bibinfo {author}
  {\bibfnamefont {G.~G.}\ \bibnamefont {Paulus}}, \ and\ \bibinfo {author}
  {\bibfnamefont {R.}~\bibnamefont {R.~Kienberger}}} (\bibinfo {year} {2009}),\
  \bibfield  {title} {\enquote {\bibinfo {title} {Single-shot carrier-envelope
  phase measurement of few-cycle laser pulses},}\ }\href@noop {} {\bibfield
  {journal} {\bibinfo  {journal} {Nat. Phys.}\ }\textbf {\bibinfo {volume}
  {5}},\ \bibinfo {pages} {357--362}}\BibitemShut {NoStop}%
\bibitem [{\citenamefont {Wolf}\ \emph {et~al.}(2016)\citenamefont {Wolf},
  \citenamefont {Schumacher},\ and\ \citenamefont {Lippitz}}]{Wolf2016}%
  \BibitemOpen
  \bibfield  {author} {\bibinfo {author} {\bibnamefont {Wolf}, \bibfnamefont
  {D}}, \bibinfo {author} {\bibfnamefont {T.}~\bibnamefont {Schumacher}}, \
  and\ \bibinfo {author} {\bibfnamefont {M.}~\bibnamefont {Lippitz}}} (\bibinfo
  {year} {2016}),\ \bibfield  {title} {\enquote {\bibinfo {title} {Shaping the
  nonlinear near field},}\ }\href@noop {} {\bibfield  {journal} {\bibinfo
  {journal} {Nat. Commun.}\ }\textbf {\bibinfo {volume} {7}},\ \bibinfo {pages}
  {10361}}\BibitemShut {NoStop}%
\bibitem [{\citenamefont {Wyatt}\ \emph {et~al.}(2016)\citenamefont {Wyatt},
  \citenamefont {Witting}, \citenamefont {Schiavi}, \citenamefont {Fabris},
  \citenamefont {Matia-Hernando}, \citenamefont {Walmsley}, \citenamefont
  {Marangos},\ and\ \citenamefont {Tisch}}]{wyatt_aries_2016}%
  \BibitemOpen
  \bibfield  {author} {\bibinfo {author} {\bibnamefont {Wyatt}, \bibfnamefont
  {A~S}}, \bibinfo {author} {\bibfnamefont {T.}~\bibnamefont {Witting}},
  \bibinfo {author} {\bibfnamefont {A.}~\bibnamefont {Schiavi}}, \bibinfo
  {author} {\bibfnamefont {D.}~\bibnamefont {Fabris}}, \bibinfo {author}
  {\bibfnamefont {P.}~\bibnamefont {Matia-Hernando}}, \bibinfo {author}
  {\bibfnamefont {I.~A.}\ \bibnamefont {Walmsley}}, \bibinfo {author}
  {\bibfnamefont {J.~P.}\ \bibnamefont {Marangos}}, \ and\ \bibinfo {author}
  {\bibfnamefont {J.~W.~G.}\ \bibnamefont {Tisch}}} (\bibinfo {year} {2016}),\
  \bibfield  {title} {\enquote {\bibinfo {title} {Attosecond sampling of
  arbitrary optical waveforms},}\ }\href@noop {} {\bibfield  {journal}
  {\bibinfo  {journal} {Optica}\ }\textbf {\bibinfo {volume} {3}},\ \bibinfo
  {pages} {303--310}}\BibitemShut {NoStop}%
\bibitem [{\citenamefont {Xu}\ \emph {et~al.}(2014)\citenamefont {Xu},
  \citenamefont {Blaga}, \citenamefont {Zhang}, \citenamefont {Lai},
  \citenamefont {Lin}, \citenamefont {Miller}, \citenamefont {Agostini},\ and\
  \citenamefont {DiMauro}}]{xu2014}%
  \BibitemOpen
  \bibfield  {author} {\bibinfo {author} {\bibnamefont {Xu}, \bibfnamefont
  {J}}, \bibinfo {author} {\bibfnamefont {C.~I.}\ \bibnamefont {Blaga}},
  \bibinfo {author} {\bibfnamefont {K.}~\bibnamefont {Zhang}}, \bibinfo
  {author} {\bibfnamefont {Y.~H.}\ \bibnamefont {Lai}}, \bibinfo {author}
  {\bibfnamefont {C.~D.}\ \bibnamefont {Lin}}, \bibinfo {author} {\bibfnamefont
  {T.~A.}\ \bibnamefont {Miller}}, \bibinfo {author} {\bibfnamefont
  {P.}~\bibnamefont {Agostini}}, \ and\ \bibinfo {author} {\bibfnamefont
  {L.~F.}\ \bibnamefont {DiMauro}}} (\bibinfo {year} {2014}),\ \bibfield
  {title} {\enquote {\bibinfo {title} {Diffraction using laser-driven broadband
  electron wave packets},}\ }\href@noop {} {\bibfield  {journal} {\bibinfo
  {journal} {Nat. Comm.}\ }\textbf {\bibinfo {volume} {5}},\ \bibinfo {pages}
  {4635}}\BibitemShut {NoStop}%
\bibitem [{\citenamefont {Xu}\ \emph {et~al.}(1996)\citenamefont {Xu},
  \citenamefont {Spielmann}, \citenamefont {Poppe}, \citenamefont {Brabec},
  \citenamefont {Krausz},\ and\ \citenamefont {H\"ansch}}]{xu_route_1996}%
  \BibitemOpen
  \bibfield  {author} {\bibinfo {author} {\bibnamefont {Xu}, \bibfnamefont
  {L}}, \bibinfo {author} {\bibfnamefont {Ch.}\ \bibnamefont {Spielmann}},
  \bibinfo {author} {\bibfnamefont {A.}~\bibnamefont {Poppe}}, \bibinfo
  {author} {\bibfnamefont {T.}~\bibnamefont {Brabec}}, \bibinfo {author}
  {\bibfnamefont {F.}~\bibnamefont {Krausz}}, \ and\ \bibinfo {author}
  {\bibfnamefont {T.~W.}\ \bibnamefont {H\"ansch}}} (\bibinfo {year} {1996}),\
  \bibfield  {title} {\enquote {\bibinfo {title} {Route to phase control of
  ultrashort light pulses},}\ }\href@noop {} {\bibfield  {journal} {\bibinfo
  {journal} {Opt. Lett.}\ }\textbf {\bibinfo {volume} {21}},\ \bibinfo {pages}
  {2008--2010}}\BibitemShut {NoStop}%
\bibitem [{\citenamefont {Yakovlev}\ \emph {et~al.}(2015)\citenamefont
  {Yakovlev}, \citenamefont {Stockman}, \citenamefont {Krausz},\ and\
  \citenamefont {Baum}}]{baum2015}%
  \BibitemOpen
  \bibfield  {author} {\bibinfo {author} {\bibnamefont {Yakovlev},
  \bibfnamefont {V}}, \bibinfo {author} {\bibfnamefont {M.~I.}\ \bibnamefont
  {Stockman}}, \bibinfo {author} {\bibfnamefont {F.}~\bibnamefont {Krausz}}, \
  and\ \bibinfo {author} {\bibfnamefont {P.}~\bibnamefont {Baum}}} (\bibinfo
  {year} {2015}),\ \bibfield  {title} {\enquote {\bibinfo {title} {Atomic-scale
  diffractive imaging of sub-cycle electron dynamics in condensed matter},}\
  }\href@noop {} {\bibfield  {journal} {\bibinfo  {journal} {Sci. Rep.}\
  }\textbf {\bibinfo {volume} {5}},\ \bibinfo {pages} {14581}}\BibitemShut
  {NoStop}%
\bibitem [{\citenamefont {Yalunin}\ \emph {et~al.}(2011)\citenamefont
  {Yalunin}, \citenamefont {Gulde},\ and\ \citenamefont
  {Ropers}}]{Yalunin2011}%
  \BibitemOpen
  \bibfield  {author} {\bibinfo {author} {\bibnamefont {Yalunin}, \bibfnamefont
  {S~V}}, \bibinfo {author} {\bibfnamefont {M.}~\bibnamefont {Gulde}}, \ and\
  \bibinfo {author} {\bibfnamefont {C.}~\bibnamefont {Ropers}}} (\bibinfo
  {year} {2011}),\ \bibfield  {title} {\enquote {\bibinfo {title} {Strong-field
  photoemission from surfaces: Theoretical approaches},}\ }\href@noop {}
  {\bibfield  {journal} {\bibinfo  {journal} {Phys. Rev. B}\ }\textbf {\bibinfo
  {volume} {84}},\ \bibinfo {pages} {195426}}\BibitemShut {NoStop}%
\bibitem [{\citenamefont {Yalunin}\ \emph {et~al.}(2013)\citenamefont
  {Yalunin}, \citenamefont {Herink}, \citenamefont {Solli}, \citenamefont
  {Kr{\"u}ger}, \citenamefont {Hommelhoff}, \citenamefont {Diehn},
  \citenamefont {Munk},\ and\ \citenamefont {Ropers}}]{Yalunin13}%
  \BibitemOpen
  \bibfield  {author} {\bibinfo {author} {\bibnamefont {Yalunin}, \bibfnamefont
  {S~V}}, \bibinfo {author} {\bibfnamefont {G.}~\bibnamefont {Herink}},
  \bibinfo {author} {\bibfnamefont {D.~R.}\ \bibnamefont {Solli}}, \bibinfo
  {author} {\bibfnamefont {M.}~\bibnamefont {Kr{\"u}ger}}, \bibinfo {author}
  {\bibfnamefont {P.}~\bibnamefont {Hommelhoff}}, \bibinfo {author}
  {\bibfnamefont {M.}~\bibnamefont {Diehn}}, \bibinfo {author} {\bibfnamefont
  {A.}~\bibnamefont {Munk}}, \ and\ \bibinfo {author} {\bibfnamefont
  {C.}~\bibnamefont {Ropers}}} (\bibinfo {year} {2013}),\ \bibfield  {title}
  {\enquote {\bibinfo {title} {Field localization and rescattering in tip-based
  photoemission},}\ }\href@noop {} {\bibfield  {journal} {\bibinfo  {journal}
  {Ann. Phys. (Berlin)}\ }\textbf {\bibinfo {volume} {525}},\ \bibinfo {pages}
  {L12--L18}}\BibitemShut {NoStop}%
\bibitem [{\citenamefont {Yanagisawa}\ \emph {et~al.}(2009)\citenamefont
  {Yanagisawa}, \citenamefont {Hafner}, \citenamefont {Don{\'a}}, \citenamefont
  {Kl{\"o}ckner}, \citenamefont {Leuenberger}, \citenamefont {Greber},
  \citenamefont {Hengsberger},\ and\ \citenamefont
  {Osterwalder}}]{Yanagisawa2009}%
  \BibitemOpen
  \bibfield  {author} {\bibinfo {author} {\bibnamefont {Yanagisawa},
  \bibfnamefont {H}}, \bibinfo {author} {\bibfnamefont {C.}~\bibnamefont
  {Hafner}}, \bibinfo {author} {\bibfnamefont {P.}~\bibnamefont {Don{\'a}}},
  \bibinfo {author} {\bibfnamefont {M.}~\bibnamefont {Kl{\"o}ckner}}, \bibinfo
  {author} {\bibfnamefont {D.}~\bibnamefont {Leuenberger}}, \bibinfo {author}
  {\bibfnamefont {T.}~\bibnamefont {Greber}}, \bibinfo {author} {\bibfnamefont
  {M.}~\bibnamefont {Hengsberger}}, \ and\ \bibinfo {author} {\bibfnamefont
  {J.}~\bibnamefont {Osterwalder}}} (\bibinfo {year} {2009}),\ \bibfield
  {title} {\enquote {\bibinfo {title} {Optical control of field-emission sites
  by femtosecond laser pulses},}\ }\href@noop {} {\bibfield  {journal}
  {\bibinfo  {journal} {Phys. Rev. Lett.}\ }\textbf {\bibinfo {volume} {103}},\
  \bibinfo {pages} {257603}}\BibitemShut {NoStop}%
\bibitem [{\citenamefont {Yanagisawa}\ \emph {et~al.}(2010)\citenamefont
  {Yanagisawa}, \citenamefont {Hafner}, \citenamefont {Don{\'a}}, \citenamefont
  {Kl{\"o}ckner}, \citenamefont {Leuenberger}, \citenamefont {Greber},
  \citenamefont {Osterwalder},\ and\ \citenamefont
  {Hengsberger}}]{Yanagisawa2010}%
  \BibitemOpen
  \bibfield  {author} {\bibinfo {author} {\bibnamefont {Yanagisawa},
  \bibfnamefont {H}}, \bibinfo {author} {\bibfnamefont {C.}~\bibnamefont
  {Hafner}}, \bibinfo {author} {\bibfnamefont {P.}~\bibnamefont {Don{\'a}}},
  \bibinfo {author} {\bibfnamefont {M.}~\bibnamefont {Kl{\"o}ckner}}, \bibinfo
  {author} {\bibfnamefont {D.}~\bibnamefont {Leuenberger}}, \bibinfo {author}
  {\bibfnamefont {T.}~\bibnamefont {Greber}}, \bibinfo {author} {\bibfnamefont
  {J.}~\bibnamefont {Osterwalder}}, \ and\ \bibinfo {author} {\bibfnamefont
  {M.}~\bibnamefont {Hengsberger}}} (\bibinfo {year} {2010}),\ \bibfield
  {title} {\enquote {\bibinfo {title} {Laser-induced field emission from a
  tungsten tip: Optical control of emission sites and the emission process},}\
  }\href@noop {} {\bibfield  {journal} {\bibinfo  {journal} {Phys. Rev. B}\
  }\textbf {\bibinfo {volume} {81}},\ \bibinfo {pages} {115429}}\BibitemShut
  {NoStop}%
\bibitem [{\citenamefont {Yanagisawa}\ \emph
  {et~al.}(2011{\natexlab{a}})\citenamefont {Yanagisawa}, \citenamefont
  {Hengsberger}, \citenamefont {Leuenberger}, \citenamefont {Kl{\"o}ckner},
  \citenamefont {Hafner}, \citenamefont {Greber},\ and\ \citenamefont
  {Osterwalder}}]{Yanagisawa2011a}%
  \BibitemOpen
  \bibfield  {author} {\bibinfo {author} {\bibnamefont {Yanagisawa},
  \bibfnamefont {H}}, \bibinfo {author} {\bibfnamefont {M.}~\bibnamefont
  {Hengsberger}}, \bibinfo {author} {\bibfnamefont {D.}~\bibnamefont
  {Leuenberger}}, \bibinfo {author} {\bibfnamefont {M.}~\bibnamefont
  {Kl{\"o}ckner}}, \bibinfo {author} {\bibfnamefont {C.}~\bibnamefont
  {Hafner}}, \bibinfo {author} {\bibfnamefont {T.}~\bibnamefont {Greber}}, \
  and\ \bibinfo {author} {\bibfnamefont {J.}~\bibnamefont {Osterwalder}}}
  (\bibinfo {year} {2011}{\natexlab{a}}),\ \bibfield  {title} {\enquote
  {\bibinfo {title} {Energy distribution curves of ultrafast laser-induced
  field emission and their implications for electron dynamics},}\ }\href@noop
  {} {\bibfield  {journal} {\bibinfo  {journal} {Phys. Rev. Lett.}\ }\textbf
  {\bibinfo {volume} {107}},\ \bibinfo {pages} {087601}}\BibitemShut {NoStop}%
\bibitem [{\citenamefont {Yanagisawa}\ \emph
  {et~al.}(2011{\natexlab{b}})\citenamefont {Yanagisawa}, \citenamefont
  {Hengsberger}, \citenamefont {Leuenberger}, \citenamefont {Kl\"ockner},
  \citenamefont {Hafner}, \citenamefont {Greber},\ and\ \citenamefont
  {Osterwalder}}]{Yanagisawa2011}%
  \BibitemOpen
  \bibfield  {author} {\bibinfo {author} {\bibnamefont {Yanagisawa},
  \bibfnamefont {H}}, \bibinfo {author} {\bibfnamefont {M.}~\bibnamefont
  {Hengsberger}}, \bibinfo {author} {\bibfnamefont {D.}~\bibnamefont
  {Leuenberger}}, \bibinfo {author} {\bibfnamefont {M.}~\bibnamefont
  {Kl\"ockner}}, \bibinfo {author} {\bibfnamefont {C.}~\bibnamefont {Hafner}},
  \bibinfo {author} {\bibfnamefont {T.}~\bibnamefont {Greber}}, \ and\ \bibinfo
  {author} {\bibfnamefont {J.}~\bibnamefont {Osterwalder}}} (\bibinfo {year}
  {2011}{\natexlab{b}}),\ \bibfield  {title} {\enquote {\bibinfo {title}
  {Energy distribution curves of ultrafast laser-induced field emission and
  their implications for electron dynamics},}\ }\href@noop {} {\bibfield
  {journal} {\bibinfo  {journal} {Phys. Rev. Lett.}\ }\textbf {\bibinfo
  {volume} {107}},\ \bibinfo {pages} {087601}}\BibitemShut {NoStop}%
\bibitem [{\citenamefont {Yanagisawa}\ \emph {et~al.}(2014)\citenamefont
  {Yanagisawa}, \citenamefont {Schnepp}, \citenamefont {Hafner}, \citenamefont
  {Hengsberger}, \citenamefont {Landsman}, \citenamefont {Gallmann},\ and\
  \citenamefont {Osterwalder}}]{Yanagisawa2014}%
  \BibitemOpen
  \bibfield  {author} {\bibinfo {author} {\bibnamefont {Yanagisawa},
  \bibfnamefont {H}}, \bibinfo {author} {\bibfnamefont {S.}~\bibnamefont
  {Schnepp}}, \bibinfo {author} {\bibfnamefont {C.}~\bibnamefont {Hafner}},
  \bibinfo {author} {\bibfnamefont {M.}~\bibnamefont {Hengsberger}}, \bibinfo
  {author} {\bibfnamefont {A.~S.}\ \bibnamefont {Landsman}}, \bibinfo {author}
  {\bibfnamefont {L.}~\bibnamefont {Gallmann}}, \ and\ \bibinfo {author}
  {\bibfnamefont {J.}~\bibnamefont {Osterwalder}}} (\bibinfo {year} {2014}),\
  \bibfield  {title} {\enquote {\bibinfo {title} {Temporal and spectral
  disentanglement of laser-driven electron tunneling emission from a solid},}\
  }\href@noop {} {\ }\Eprint {http://arxiv.org/abs/1405.0609} {arXiv:1405.0609
  [cond-mat.mes-hal]} \BibitemShut {NoStop}%
\bibitem [{\citenamefont {Yavuz}(2013)}]{Yavuz13}%
  \BibitemOpen
  \bibfield  {author} {\bibinfo {author} {\bibnamefont {Yavuz}, \bibfnamefont
  {I}}} (\bibinfo {year} {2013}),\ \bibfield  {title} {\enquote {\bibinfo
  {title} {Gas population effects in harmonic emission by plasmonic fields},}\
  }\href@noop {} {\bibfield  {journal} {\bibinfo  {journal} {Phys. Rev. A}\
  }\textbf {\bibinfo {volume} {87}},\ \bibinfo {pages} {053815}}\BibitemShut
  {NoStop}%
\bibitem [{\citenamefont {Yavuz}\ \emph {et~al.}(2012)\citenamefont {Yavuz},
  \citenamefont {Bleda}, \citenamefont {Altun},\ and\ \citenamefont
  {Topcu}}]{Yavuz12}%
  \BibitemOpen
  \bibfield  {author} {\bibinfo {author} {\bibnamefont {Yavuz}, \bibfnamefont
  {I}}, \bibinfo {author} {\bibfnamefont {E.~A.}\ \bibnamefont {Bleda}},
  \bibinfo {author} {\bibfnamefont {Z.}~\bibnamefont {Altun}}, \ and\ \bibinfo
  {author} {\bibfnamefont {T.}~\bibnamefont {Topcu}}} (\bibinfo {year}
  {2012}),\ \bibfield  {title} {\enquote {\bibinfo {title} {Generation of a
  broadband {XUV} continuum in high-order-harmonic generation by spatially
  inhomogeneous fields},}\ }\href@noop {} {\bibfield  {journal} {\bibinfo
  {journal} {Phys. Rev. A}\ }\textbf {\bibinfo {volume} {85}},\ \bibinfo
  {pages} {013416}}\BibitemShut {NoStop}%
\bibitem [{\citenamefont {Yavuz}\ \emph {et~al.}(2016)\citenamefont {Yavuz},
  \citenamefont {Ciappina}, \citenamefont {Chac\'on}, \citenamefont {Altun},
  \citenamefont {Kling},\ and\ \citenamefont {Lewenstein}}]{ilhan1}%
  \BibitemOpen
  \bibfield  {author} {\bibinfo {author} {\bibnamefont {Yavuz}, \bibfnamefont
  {I}}, \bibinfo {author} {\bibfnamefont {M.~F.}\ \bibnamefont {Ciappina}},
  \bibinfo {author} {\bibfnamefont {A.}~\bibnamefont {Chac\'on}}, \bibinfo
  {author} {\bibfnamefont {Z.}~\bibnamefont {Altun}}, \bibinfo {author}
  {\bibfnamefont {M.~F.}\ \bibnamefont {Kling}}, \ and\ \bibinfo {author}
  {\bibfnamefont {M.}~\bibnamefont {Lewenstein}}} (\bibinfo {year} {2016}),\
  \bibfield  {title} {\enquote {\bibinfo {title} {Controlling electron
  localization in {H$^{+}_{2}$} by intense plasmon-enhanced laser fields},}\
  }\href@noop {} {\bibfield  {journal} {\bibinfo  {journal} {Phys. Rev. A}\
  }\textbf {\bibinfo {volume} {93}},\ \bibinfo {pages} {033404}}\BibitemShut
  {NoStop}%
\bibitem [{\citenamefont {Yavuz}\ \emph {et~al.}(2015)\citenamefont {Yavuz},
  \citenamefont {Tikman},\ and\ \citenamefont {Altun}}]{Yavuz15}%
  \BibitemOpen
  \bibfield  {author} {\bibinfo {author} {\bibnamefont {Yavuz}, \bibfnamefont
  {I}}, \bibinfo {author} {\bibfnamefont {Y.}~\bibnamefont {Tikman}}, \ and\
  \bibinfo {author} {\bibfnamefont {Z.}~\bibnamefont {Altun}}} (\bibinfo {year}
  {2015}),\ \bibfield  {title} {\enquote {\bibinfo {title} {High-order-harmonic
  generation from {H$_2^+$} molecular ions near plasmon-enhanced laser
  fields},}\ }\href@noop {} {\bibfield  {journal} {\bibinfo  {journal} {Phys.
  Rev. A}\ }\textbf {\bibinfo {volume} {92}},\ \bibinfo {pages}
  {023413}}\BibitemShut {NoStop}%
\bibitem [{\citenamefont {Yu}\ \emph {et~al.}(2015)\citenamefont {Yu},
  \citenamefont {Wang}, \citenamefont {Cao}, \citenamefont {Jiang},\ and\
  \citenamefont {Lu}}]{Yu15}%
  \BibitemOpen
  \bibfield  {author} {\bibinfo {author} {\bibnamefont {Yu}, \bibfnamefont
  {C}}, \bibinfo {author} {\bibfnamefont {Y.}~\bibnamefont {Wang}}, \bibinfo
  {author} {\bibfnamefont {X.}~\bibnamefont {Cao}}, \bibinfo {author}
  {\bibfnamefont {S.}~\bibnamefont {Jiang}}, \ and\ \bibinfo {author}
  {\bibfnamefont {R.}~\bibnamefont {Lu}}} (\bibinfo {year} {2015}),\ \bibfield
  {title} {\enquote {\bibinfo {title} {Isolated few-attosecond emission in a
  multi-cycle asymmetrically nonhomogeneous two-color laser field},}\
  }\href@noop {} {\bibfield  {journal} {\bibinfo  {journal} {J. Phys. B}\
  }\textbf {\bibinfo {volume} {47}},\ \bibinfo {pages} {225602}}\BibitemShut
  {NoStop}%
\bibitem [{\citenamefont {Yudin}\ and\ \citenamefont
  {Ivanov}(2001)}]{Yudin2001}%
  \BibitemOpen
  \bibfield  {author} {\bibinfo {author} {\bibnamefont {Yudin}, \bibfnamefont
  {G~L}}, \ and\ \bibinfo {author} {\bibfnamefont {M.~Yu.}\ \bibnamefont
  {Ivanov}}} (\bibinfo {year} {2001}),\ \bibfield  {title} {\enquote {\bibinfo
  {title} {Nonadiabatic tunnel ionization: Looking inside a laser cycle},}\
  }\href@noop {} {\bibfield  {journal} {\bibinfo  {journal} {Phys. Rev. A}\
  }\textbf {\bibinfo {volume} {64}},\ \bibinfo {pages} {013409}}\BibitemShut
  {NoStop}%
\bibitem [{\citenamefont {Zagoya}\ \emph {et~al.}(2016)\citenamefont {Zagoya},
  \citenamefont {Bonner}, \citenamefont {Chomet}, \citenamefont {Slade},\ and\
  \citenamefont {Figueira~de Morisson~Faria}}]{carla2016}%
  \BibitemOpen
  \bibfield  {author} {\bibinfo {author} {\bibnamefont {Zagoya}, \bibfnamefont
  {C}}, \bibinfo {author} {\bibfnamefont {M.}~\bibnamefont {Bonner}}, \bibinfo
  {author} {\bibfnamefont {H.}~\bibnamefont {Chomet}}, \bibinfo {author}
  {\bibfnamefont {E.}~\bibnamefont {Slade}}, \ and\ \bibinfo {author}
  {\bibfnamefont {C.}~\bibnamefont {Figueira~de Morisson~Faria}}} (\bibinfo
  {year} {2016}),\ \bibfield  {title} {\enquote {\bibinfo {title} {Different
  time scales in plasmonically enhanced high-order-harmonic generation},}\
  }\href@noop {} {\bibfield  {journal} {\bibinfo  {journal} {Phys. Rev. A}\
  }\textbf {\bibinfo {volume} {93}},\ \bibinfo {pages} {053419}}\BibitemShut
  {NoStop}%
\bibitem [{\citenamefont {Zang}\ \emph {et~al.}(2013)\citenamefont {Zang},
  \citenamefont {Lui},\ and\ \citenamefont {Xu}}]{Zhang13}%
  \BibitemOpen
  \bibfield  {author} {\bibinfo {author} {\bibnamefont {Zang}, \bibfnamefont
  {C}}, \bibinfo {author} {\bibfnamefont {C.}~\bibnamefont {Lui}}, \ and\
  \bibinfo {author} {\bibfnamefont {Z.}~\bibnamefont {Xu}}} (\bibinfo {year}
  {2013}),\ \bibfield  {title} {\enquote {\bibinfo {title} {Control of higher
  spectral components by spatially inhomogeneous fields in quantum wells},}\
  }\href@noop {} {\bibfield  {journal} {\bibinfo  {journal} {Phys. Rev. A}\
  }\textbf {\bibinfo {volume} {88}},\ \bibinfo {pages} {035805}}\BibitemShut
  {NoStop}%
\bibitem [{\citenamefont {Zhang}\ and\ \citenamefont
  {Thumm}(2009)}]{zhang2009attosecond}%
  \BibitemOpen
  \bibfield  {author} {\bibinfo {author} {\bibnamefont {Zhang}, \bibfnamefont
  {C-H}}, \ and\ \bibinfo {author} {\bibfnamefont {U.}~\bibnamefont {Thumm}}}
  (\bibinfo {year} {2009}),\ \bibfield  {title} {\enquote {\bibinfo {title}
  {Attosecond photoelectron spectroscopy of metal surfaces},}\ }\href@noop {}
  {\bibfield  {journal} {\bibinfo  {journal} {Phys. Rev. Lett.}\ }\textbf
  {\bibinfo {volume} {102}},\ \bibinfo {pages} {123601}}\BibitemShut {NoStop}%
\bibitem [{\citenamefont {Zhang}\ and\ \citenamefont
  {Thumm}(2011{\natexlab{a}})}]{zhang_effect_2011}%
  \BibitemOpen
  \bibfield  {author} {\bibinfo {author} {\bibnamefont {Zhang}, \bibfnamefont
  {C-H}}, \ and\ \bibinfo {author} {\bibfnamefont {U.}~\bibnamefont {Thumm}}}
  (\bibinfo {year} {2011}{\natexlab{a}}),\ \bibfield  {title} {\enquote
  {\bibinfo {title} {Effect of wave-function localization on the time delay in
  photoemission from surfaces},}\ }\href@noop {} {\bibfield  {journal}
  {\bibinfo  {journal} {Phys. Rev. A}\ }\textbf {\bibinfo {volume} {84}},\
  \bibinfo {pages} {065403}}\BibitemShut {NoStop}%
\bibitem [{\citenamefont {Zhang}\ and\ \citenamefont
  {Thumm}(2011{\natexlab{b}})}]{zhang_streaking_2011}%
  \BibitemOpen
  \bibfield  {author} {\bibinfo {author} {\bibnamefont {Zhang}, \bibfnamefont
  {C-H}}, \ and\ \bibinfo {author} {\bibfnamefont {U.}~\bibnamefont {Thumm}}}
  (\bibinfo {year} {2011}{\natexlab{b}}),\ \bibfield  {title} {\enquote
  {\bibinfo {title} {Streaking and {Wigner} time delays in photoemission from
  atoms and surfaces},}\ }\href@noop {} {\bibfield  {journal} {\bibinfo
  {journal} {Phys. Rev. A}\ }\textbf {\bibinfo {volume} {84}},\ \bibinfo
  {pages} {033401}}\BibitemShut {NoStop}%
\bibitem [{\citenamefont {Zhang}\ \emph {et~al.}(2009)\citenamefont {Zhang},
  \citenamefont {Chen}, \citenamefont {Qi},\ and\ \citenamefont
  {Qing}}]{zhang2009four}%
  \BibitemOpen
  \bibfield  {author} {\bibinfo {author} {\bibnamefont {Zhang}, \bibfnamefont
  {G-Y}}, \bibinfo {author} {\bibfnamefont {Y.}~\bibnamefont {Chen}}, \bibinfo
  {author} {\bibfnamefont {W.-K.}\ \bibnamefont {Qi}}, \ and\ \bibinfo {author}
  {\bibfnamefont {S.-M.}\ \bibnamefont {Qing}}} (\bibinfo {year} {2009}),\
  \bibfield  {title} {\enquote {\bibinfo {title} {Four-state
  rock-paper-scissors games in constrained newman-watts networks},}\
  }\href@noop {} {\bibfield  {journal} {\bibinfo  {journal} {Phys. Rev. E}\
  }\textbf {\bibinfo {volume} {79}},\ \bibinfo {pages} {062901}}\BibitemShut
  {NoStop}%
\bibitem [{\citenamefont {Zhao}\ \emph {et~al.}(2012)\citenamefont {Zhao},
  \citenamefont {Zhang}, \citenamefont {Chini}, \citenamefont {Wu},
  \citenamefont {Wang},\ and\ \citenamefont {Chang}}]{zhao2012tailoring}%
  \BibitemOpen
  \bibfield  {author} {\bibinfo {author} {\bibnamefont {Zhao}, \bibfnamefont
  {K}}, \bibinfo {author} {\bibfnamefont {Q.}~\bibnamefont {Zhang}}, \bibinfo
  {author} {\bibfnamefont {M.}~\bibnamefont {Chini}}, \bibinfo {author}
  {\bibfnamefont {Y.}~\bibnamefont {Wu}}, \bibinfo {author} {\bibfnamefont
  {X.}~\bibnamefont {Wang}}, \ and\ \bibinfo {author} {\bibfnamefont
  {Z.}~\bibnamefont {Chang}}} (\bibinfo {year} {2012}),\ \bibfield  {title}
  {\enquote {\bibinfo {title} {Tailoring a 67 attosecond pulse through
  advantageous phase-mismatch},}\ }\href@noop {} {\bibfield  {journal}
  {\bibinfo  {journal} {Opt. Lett.}\ }\textbf {\bibinfo {volume} {37}},\
  \bibinfo {pages} {3891--3893}}\BibitemShut {NoStop}%
\bibitem [{\citenamefont {Zherebtsov}\ \emph {et~al.}(2011)\citenamefont
  {Zherebtsov}, \citenamefont {Fennel}, \citenamefont {Plenge}, \citenamefont
  {Antonsson}, \citenamefont {Znakovskaya}, \citenamefont {Wirth},
  \citenamefont {Herrwerth}, \citenamefont {S{\"u}{\ss}mann}, \citenamefont
  {Peltz}, \citenamefont {Ahmad}, \citenamefont {Trushin}, \citenamefont
  {Pervak}, \citenamefont {Karsch}, \citenamefont {Vrakking}, \citenamefont
  {Langer}, \citenamefont {Graf}, \citenamefont {Stockman}, \citenamefont
  {Krausz}, \citenamefont {R{\"u}hl},\ and\ \citenamefont
  {Kling}}]{Zherebtsov11}%
  \BibitemOpen
  \bibfield  {author} {\bibinfo {author} {\bibnamefont {Zherebtsov},
  \bibfnamefont {S}}, \bibinfo {author} {\bibfnamefont {T.}~\bibnamefont
  {Fennel}}, \bibinfo {author} {\bibfnamefont {J.}~\bibnamefont {Plenge}},
  \bibinfo {author} {\bibfnamefont {E.}~\bibnamefont {Antonsson}}, \bibinfo
  {author} {\bibfnamefont {I.}~\bibnamefont {Znakovskaya}}, \bibinfo {author}
  {\bibfnamefont {A.}~\bibnamefont {Wirth}}, \bibinfo {author} {\bibfnamefont
  {O.}~\bibnamefont {Herrwerth}}, \bibinfo {author} {\bibfnamefont
  {F.}~\bibnamefont {S{\"u}{\ss}mann}}, \bibinfo {author} {\bibfnamefont
  {C.}~\bibnamefont {Peltz}}, \bibinfo {author} {\bibfnamefont
  {I.}~\bibnamefont {Ahmad}}, \bibinfo {author} {\bibfnamefont {S.~A.}\
  \bibnamefont {Trushin}}, \bibinfo {author} {\bibfnamefont {V.}~\bibnamefont
  {Pervak}}, \bibinfo {author} {\bibfnamefont {S.}~\bibnamefont {Karsch}},
  \bibinfo {author} {\bibfnamefont {M.~J.~J.}\ \bibnamefont {Vrakking}},
  \bibinfo {author} {\bibfnamefont {B.}~\bibnamefont {Langer}}, \bibinfo
  {author} {\bibfnamefont {C.}~\bibnamefont {Graf}}, \bibinfo {author}
  {\bibfnamefont {M.~I.}\ \bibnamefont {Stockman}}, \bibinfo {author}
  {\bibfnamefont {F.}~\bibnamefont {Krausz}}, \bibinfo {author} {\bibfnamefont
  {E.}~\bibnamefont {R{\"u}hl}}, \ and\ \bibinfo {author} {\bibfnamefont
  {M.~F.}\ \bibnamefont {Kling}}} (\bibinfo {year} {2011}),\ \bibfield  {title}
  {\enquote {\bibinfo {title} {Controlled near-field enhanced electron
  acceleration from dielectric nanospheres with intense few-cycle laser
  fields},}\ }\href@noop {} {\bibfield  {journal} {\bibinfo  {journal} {Nat.
  Phys.}\ }\textbf {\bibinfo {volume} {7}},\ \bibinfo {pages}
  {656--662}}\BibitemShut {NoStop}%
\bibitem [{\citenamefont {Zherebtsov}\ \emph {et~al.}(2012)\citenamefont
  {Zherebtsov}, \citenamefont {S{\"u}{\ss}mann}, \citenamefont {Peltz},
  \citenamefont {Plenge}, \citenamefont {Betsch}, \citenamefont {Znakovskaya},
  \citenamefont {Alnaser}, \citenamefont {Johnson}, \citenamefont {K{\"u}bel},
  \citenamefont {Horn}, \citenamefont {Mondes}, \citenamefont {Graf},
  \citenamefont {Trushin}, \citenamefont {Azzeer}, \citenamefont {Vrakking},
  \citenamefont {Paulus}, \citenamefont {Krausz}, \citenamefont {R\"uhl},
  \citenamefont {Fennel},\ and\ \citenamefont {Kling}}]{Zherebtsov12}%
  \BibitemOpen
  \bibfield  {author} {\bibinfo {author} {\bibnamefont {Zherebtsov},
  \bibfnamefont {S}}, \bibinfo {author} {\bibfnamefont {F.}~\bibnamefont
  {S{\"u}{\ss}mann}}, \bibinfo {author} {\bibfnamefont {C.}~\bibnamefont
  {Peltz}}, \bibinfo {author} {\bibfnamefont {J}~\bibnamefont {Plenge}},
  \bibinfo {author} {\bibfnamefont {K.~J.}\ \bibnamefont {Betsch}}, \bibinfo
  {author} {\bibfnamefont {I.}~\bibnamefont {Znakovskaya}}, \bibinfo {author}
  {\bibfnamefont {A.~S.}\ \bibnamefont {Alnaser}}, \bibinfo {author}
  {\bibfnamefont {N.~G.}\ \bibnamefont {Johnson}}, \bibinfo {author}
  {\bibfnamefont {M.}~\bibnamefont {K{\"u}bel}}, \bibinfo {author}
  {\bibfnamefont {A.}~\bibnamefont {Horn}}, \bibinfo {author} {\bibfnamefont
  {V.}~\bibnamefont {Mondes}}, \bibinfo {author} {\bibfnamefont
  {C.}~\bibnamefont {Graf}}, \bibinfo {author} {\bibfnamefont {S.~A.}\
  \bibnamefont {Trushin}}, \bibinfo {author} {\bibfnamefont {A.}~\bibnamefont
  {Azzeer}}, \bibinfo {author} {\bibfnamefont {M.~J.~J.}\ \bibnamefont
  {Vrakking}}, \bibinfo {author} {\bibfnamefont {G.~G.}\ \bibnamefont
  {Paulus}}, \bibinfo {author} {\bibfnamefont {F.}~\bibnamefont {Krausz}},
  \bibinfo {author} {\bibfnamefont {E.}~\bibnamefont {R\"uhl}}, \bibinfo
  {author} {\bibfnamefont {T.}~\bibnamefont {Fennel}}, \ and\ \bibinfo {author}
  {\bibfnamefont {M.~F.}\ \bibnamefont {Kling}}} (\bibinfo {year} {2012}),\
  \bibfield  {title} {\enquote {\bibinfo {title} {Carrier-envelope phase-tagged
  imaging of the controlled electron acceleration from $\mathrm{SiO}_2$
  nanospheres in intense few-cycle laser fields},}\ }\href@noop {} {\bibfield
  {journal} {\bibinfo  {journal} {New J. Phys.}\ }\textbf {\bibinfo {volume}
  {14}},\ \bibinfo {pages} {075010}}\BibitemShut {NoStop}%
\bibitem [{\citenamefont {Zuloaga}\ \emph {et~al.}(2009)\citenamefont
  {Zuloaga}, \citenamefont {Prodan},\ and\ \citenamefont
  {Nordlander}}]{Zuloaga2009}%
  \BibitemOpen
  \bibfield  {author} {\bibinfo {author} {\bibnamefont {Zuloaga}, \bibfnamefont
  {J}}, \bibinfo {author} {\bibfnamefont {E.}~\bibnamefont {Prodan}}, \ and\
  \bibinfo {author} {\bibfnamefont {P.}~\bibnamefont {Nordlander}}} (\bibinfo
  {year} {2009}),\ \bibfield  {title} {\enquote {\bibinfo {title} {Quantum
  description of the plasmon resonances of a nanoparticle dimer},}\ }\href@noop
  {} {\bibfield  {journal} {\bibinfo  {journal} {Nano Lett.}\ }\textbf
  {\bibinfo {volume} {9}},\ \bibinfo {pages} {887--891}}\BibitemShut {NoStop}%
\bibitem [{\citenamefont {Zuloaga}\ \emph {et~al.}(2010)\citenamefont
  {Zuloaga}, \citenamefont {Prodan},\ and\ \citenamefont
  {Nordlander}}]{Zuloaga2010}%
  \BibitemOpen
  \bibfield  {author} {\bibinfo {author} {\bibnamefont {Zuloaga}, \bibfnamefont
  {J}}, \bibinfo {author} {\bibfnamefont {E.}~\bibnamefont {Prodan}}, \ and\
  \bibinfo {author} {\bibfnamefont {P.}~\bibnamefont {Nordlander}}} (\bibinfo
  {year} {2010}),\ \bibfield  {title} {\enquote {\bibinfo {title} {Quantum
  plasmonics: {O}ptical properties and tunability of metallic nanorods},}\
  }\href@noop {} {\bibfield  {journal} {\bibinfo  {journal} {ACS Nano}\
  }\textbf {\bibinfo {volume} {4}},\ \bibinfo {pages} {5269--5276}}\BibitemShut
  {NoStop}%
\end{thebibliography}

%merlin.mbs apsrmp4-1.bst 2010-07-25 4.21a (PWD, AO, DPC) hacked
%Control: author (3) reversed first dotless

\end{document}